\documentclass[a4paper, 11pt]{report}
\usepackage{my_jinstpub}
\usepackage{epsfig}
\usepackage[]{lineno}
\usepackage{hhline}

\usepackage{siunitx}
\DeclareSIUnit\gauss{G}
\DeclareSIUnit\bar{bar}
\DeclareSIUnit\barn{b}
\DeclareSIUnit\angstrom{\text{Å}}
\usepackage[version=3]{mhchem}
\usepackage{threeparttable}
\usepackage{hyperref}
\usepackage{subfigure}
\usepackage[section]{placeins} 
\usepackage{gensymb}
\usepackage{textcomp}
\usepackage{csquotes}
\usepackage{booktabs}
\usepackage{textgreek}

\usepackage{tikz}
\usetikzlibrary{calc, decorations, decorations.pathmorphing, decorations.markings , external}
\usepackage{pgfplots}
\pgfplotsset{compat=1.17}
\usepgfplotslibrary{groupplots, colorbrewer}
\pgfplotsset{
    cycle list/Dark2-4,
    cycle list/Dark2-8,
    cycle list/Set1-4,
    cycle list/Set1-9,
}
\pgfplotscreateplotcyclelist{mylinestyles}{solid, densely dashed, densely dotted, dashdotted, dashed, dotted}

\newcommand{\electron}{\ensuremath{\text{e}^{-}}}
\newcommand{\positron}{\ensuremath{\text{e}^{+}}}
\newcommand{\photon}{\text{\textgamma}}

\title{The Development of Energy-Recovery Linacs}

\author[t]{Chris Adolphsen,}
\author[d,i]{Kevin Andre,}
\author[f]{Deepa Angal-Kalinin,}
\author[g]{Michaela Arnold,}
\author[j]{Kurt Aulenbacher,}
\author[o]{Steve Benson,}
\author[m]{Jan Bernauer,}
\author[o]{Alex Bogacz,}
\author[l]{Maarten Boonekamp,}
\author{Reinhard Brinkmann,}
\author[o]{Max Bruker,}
\author[d]{Oliver Brüning,}
\author[p]{Camilla Curatolo,}
\author[k]{Patxi Duthill,}
\author[i]{Oliver Fischer,}
\author[e,c]{Georg Hoffstaetter,}
\author[d]{Bernhard Holzer,}
\author[k,i]{Ben Hounsell,}
\author[o,1]{Andrew Hutton,\note{Corresponding author.}}
\author[d]{Erk Jensen,}
\author[k]{Walid Kaabi,}
\author[c]{Dmitry Kayran,}
\author[i]{Max Klein,}
\author[a,s]{Jens Knobloch,}
\author[o]{Geoff Krafft,}
\author[a]{Julius Kühn,}
\author[a]{Bettina Kuske,}
\author[m]{Vladimir Litvinenko,}
\author[o]{Frank Marhauser,}
\author[f]{Boris Militsyn,}
\author[v]{Sergei Nagaitsev,}
\author[o]{George Neil,}
\author[a]{Axel Neumann,}
\author[g]{Norbert Pietralla,}
\author[o]{Bob Rimmer,}
\author[p]{Luca Serafini,}
\author[b]{Oleg A.~Shevchenko,}
\author[d,q]{Nick Shipman,}
\author[j]{Hubert Spiesberger,}
\author[n]{Olga Tanaka,}
\author[b,r]{Valery Telnov,}
\author[o]{Chris Tennant,}
\author[h]{Cristina Vaccarezza,}
\author[k]{David Verney,}
\author[b]{Nikolay Vinokurov,}
\author[f]{Peter Williams,}
\author[n]{Akira Yamamoto,}
\author[n]{Kaoru Yokoya,}
\author[d]{Frank Zimmermann}
\emailAdd{andrew@jlab.org}
\affiliation[a]{Helmholtz-Zentrum Berlin, Berlin, Germany}
\affiliation[b]{Budker Institute of Nuclear Physics, 630090, Novosibirsk, Russia}
\affiliation[c]{Brookhaven National Laboratory, Upton, NY, USA}
\affiliation[d]{CERN, Geneva, Switzerland}
\affiliation[e]{Cornell University, Ithaca, NY, USA}
\affiliation[f]{Daresbury Laboratory (STFC), Daresbury, UK}
\affiliation[g]{Technische Universität Darmstadt, Institute for Nuclear Physics, Darmstadt, Germany}
\affiliation[h]{INFN, Frascati, Italy}
\affiliation[i]{University of Liverpool, Liverpool, UK}
\affiliation[j]{University of Mainz, Mainz, Germany}
\affiliation[k]{IJCLab, Orsay, France}
\affiliation[l]{CEA Saclay, Saclay, France}
\affiliation[m]{Center for Frontiers in Nuclear Science, Department of Physics and Astronomy, Stony Brook University, Stony Brook, NY, USA, and RIKEN BNL Research Center, Brookhaven National Laboratory, Upton, NY, USA}
\affiliation[n]{KEK, Tsukuba, Japan}
\affiliation[o]{Thomas Jefferson National Accelerator Facility, Newport News, VA, USA}
\affiliation[p]{INFN, Milano, Italy, and LASA}
\affiliation[q]{Lancaster University, Lancaster, UK}
\affiliation[r]{Novosibirsk State University, 630090, Novosibirsk, Russia}
\affiliation[s]{University of Siegen, Siegen, Germany}
\affiliation[t]{SLAC, Menlo Park, CA, USA}
\affiliation[v]{Fermilab, Batavia, IL, USA}

\abstract{
Energy-recovery linacs (ERLs) have been emphasised by the recent (2020) update of the European
Strategy for Particle Physics as one of the most promising technologies for the accelerator base
of future high-energy physics. They are indeed 
beginning to assert their potential as game changers in the field of accelerators and their applications. Their unique combination of bright, linac-like beam quality with high average current and extremely flexible time structure, unprecedented operating efficiency, and compact footprint opens the door to previously unattainable performance regimes.

The current paper has been written as a base document to support and specify details of the recently published European roadmap for the development of energy-recovery linacs.
The paper summarises the previous achievements 
on ERLs and the status of the field and its basic technology items. The main possible future contributions and applications of ERLs to particle and nuclear physics as well as industrial developments 
are presented.   Many of  the single resulting requirements will be or have been already met
 in the ongoing concerted effort, which will move the
field forward with complementary facilities. A corresponding roadmap is established, with a European focus,  describing
major opportunities, new facilities, milestones, deliverables and necessary investments, as
a coherent global effort to meet expectations in  the next five years and further ahead. It is thus realistic to predict
that a viable technical ERL base will emerge in the not distant future, serving as a reliable input to 
strategic high-energy physics  decisions to come. 

The paper includes a vision for the further future, beyond 2030, as well as a comparative data base for the main existing and forthcoming ERL facilities. At hand is an unprecedented
technology combining strongly enhanced performance of electron- and photon-beam-based
physics with sustainable power consumption, by using the decelerated beam for
new acceleration, and with non-radiative waste, as the beam is dumped at injection energy.
A series of continuous innovations, such as on intense electron sources or high-quality superconducting 
cavity technology, will massively contribute to the development of accelerator physics at large. Industrial 
applications are potentially revolutionary and may carry the development of ERLs much further, 
establishing another shining example of the impact of particle physics on society and its technical foundation with a special view on sustaining nature.  
}
\begin{document}
\maketitle
\flushbottom
\newpage
\pagestyle{myplain}\pagenumbering{arabic}

%\tableofcontents{}
%\newpage
%
%
\chapter{Introduction}\label{sec:intro}

%\section{The Magic Principle of Energy Recovery, its Promises and Past}
%Andrew Hutton, Steve Benson, George Neil \\
%from Tigner~\cite{Tigner:1965wf} .. to high brightness, high energy e beams at hugely reduced power consumption.. 

\section{The Magic Principle of Energy Recovery, its Promises and Past}\label{sec:intro:magic}
%Andrew Hutton, Steve Benson, George Neil \\
\subsection{History}\label{sec:intro:magic:history}
The idea of an energy-recovery linac traces back to Maury Tigner~\cite{Tigner:1965wf} in 1965.
He was looking at ways to enhance the current in a collider for high-energy physics.
Accelerating two beams, colliding them, and then dumping them is extremely inefficient.
If one could recover the energy of the beams in the same cavities in which they were accelerated, then the efficiency of the machine could be greatly increased.
The design of the final dump also becomes much simpler.
Though the idea was sound, the implementation of an efficient solution relied on the development of reliable superconducting radiofrequency (SRF) accelerating cavities.
These were developed over the next decade.
The first large use of SRF cavities was at the High Energy Physics Lab (HEPL) at Stanford University.
% Sic. No hyphenation
Researchers there installed a recirculation loop with the capability of varying the path length so that the electrons in a second pass through the accelerating cavities could be either accelerated or decelerated.
Both options were demonstrated.
This was the first ERL with SRF cavities~\cite{SMITH19871}.
This type of ERL is called same-cell energy recovery.
The beam was not used for anything, and the current was pulsed, but evidence for energy recovery was clearly seen in the RF power requirements during the beam pulse.

Other demonstrations of energy recovery with room-temperature cavities were carried out at Chalk River~\cite{4328851} and Los Alamos National Lab~\cite{FELDMAN198726}. The Los Alamos demonstration used coupled accelerating and decelerating cavities, and it had an FEL in the beamline so the overall FEL efficiency could, in principle, be increased, but the cavity losses and the RF transport losses led to an overall increase in the RF power required, showing the advantage of using nearly lossless SRF cavities in the same-cell energy recovery mode.

During the early development of CEBAF at what is now Jefferson Lab, the ability to recirculate beam in the newly developed SRF cavities was tested in the Front End Test (FET)~\cite{PhysRevLett.84.662}, where the beam was recirculated in a fashion similar to the HEPL experiment. The current in this case, however, could be run continuously, and both recirculation (two accelerating passes) and an energy-recovery configuration were demonstrated. 

While all of this technology development work was taking place, several authors noted that the ERL was a natural way to increase the overall efficiency of a Free-Electron Laser (FEL) since the FEL usually only takes about \SI{1}{\percent} of the energy of the electron beam out as laser radiation and then dumps the rest. If one could recover most of the power from beam at the exit of the FEL, one could greatly enhance the overall efficiency of the laser. The Los Alamos experiment demonstrated some of the concepts of an ERL-based FEL but was a low-average-power, pulsed device.

This led to the development of an IR Demo project at Jefferson Lab~\cite{BENSON20021}, based on the same cryomodules that had been developed for CEBAF. This was a resounding success, exceeding all of the ambitious goals that had been established with a \SIrange{35}{48}{\mega\electronvolt}, \SI{5}{\milli\ampere} electron beam producing \SI{2.1}{\kilo\watt} of IR outcoupled to users.
This enabled the development of an even more ambitious goal: to increase the power levels by a factor of ten.
This was achieved by a rebuild of the recirculation arcs and an increase of the electron energy.
This facility circulated \SI{9}{\milli\ampere} at up to \SI{150}{\mega\electronvolt}, still the highest current that has been recirculated in an SRF ERL~\cite{4440128}.
There was a considerable amount of beam optics studies which laid the foundation for the design of later ERL facilities.

The ERLs at Jefferson Lab were important demonstrations that one can produce high beam power without a large installed RF power source.
The IR Upgrade ERL operated with over \SI{1.1}{\mega\watt} of beam power with only about \SI{300}{\kilo\watt} of installed RF, thus demonstrating the most basic reason for building an ERL.
Other devices were also built, however, which pushed other frontiers.
Novosibirsk has built two ERLs using room-temperature cavities~\cite{GAVRILOV200754}.
With the copper losses of the cavities, the efficiency is not high, but they were able to recirculate up to \SI{30}{\milli\ampere} of average current, still the record for recirculated current.
The two ERLs are used for far-infrared FELs in a very active user program. 

A group at JAERI built an ERL that used novel cryogenic cooling at long wavelengths to produce a very efficient ERL.
They also pushed the efficiency of the FEL to record levels for an ERL~\cite{HAJIMA2003115}.

The group at KEK commissioned a high-current ERL test machine that is designed for currents up to \SI{100}{\milli\ampere} and demonstrated \SI{1}{\milli\ampere} of beam recirculation.
The photocathode gun operates at \SI{500}{\kilo\volt}, the highest of any photocathode gun~\cite{PhotonFactoryReportJapanese}.
The KEK project lost funding and stopped, but recently the project restarted with a different funding source.

An ERL similar in design to the Jefferson Lab ERL, ALICE, was built at the Daresbury Laboratory (UK).
It operated pulsed due to radiation and refrigeration concerns but demonstrated both THz production and IR FEL operation~\cite{10yearsalice}.
ALICE was shut down after ten years of successful operation, having achieved its objectives.

As part of an ERL program for a light source, Cornell commissioned an injector with the highest average current demonstrated from a photocathode injector~\cite{cornell_pddr}.
Following this, they reused the gun, booster and a single cryomodule as the basis for CBETA.
The arcs that return the beam to the cryomodule used a novel technique, Fixed-Field Alternating-Gradient (FFAG) transport, to demonstrate the first multi-pass energy recovery in an SRF-based ERL~\cite{PhysRevLett.125.044803}.

\subsection{The Technology}\label{sec:intro:magic:technology}
Energy-Recovery Linacs are an extremely efficient technique for accelerating high-av\-er\-age-current electron beams.
In an ERL, a high-average-current electron beam is accelerated to relativistic energies in (typically) a superconducting RF CW linear accelerator.
The beam is then used for its intended purpose, i.e., providing a gain medium for a free-electron laser, synchrotron light production, a cooling source for ion beams, or a beam for colliding against ions.
The application usually creates a large increase in the energy spread or emittance of the electron beam, but the majority of the beam power remains.
To recover this power, the beam is then sent back through the accelerator again, only this time roughly \SI{180}{\degree} off the accelerating RF phase. The beam is therefore decelerated as it goes through the linac, putting its power back into the RF fields.
Eventually, the beam energy becomes so low that transport of the beam becomes awkward, so the beam is dumped with some (small) residual energy.

Three major system benefits accrue from this manipulation: the required RF power (and its capital cost and required electricity) is significantly reduced, the beam power that must be dissipated in the dump is reduced by a large factor, and often the electron beam dump energy can be reduced below the photo-neutron threshold, minimizing the activation of the dump region, so the required shielding of the facility can be reduced.
The cost savings associated with incorporation of energy recovery must be balanced against the need to provide a beam transport system to re-inject the beam to the linac for recovery.
If significant growth in the energy spread or emittance of the electron beam has occurred in the process of utilizing the beam, then this transport system can necessitate significant manipulation of the beam phase space.
These techniques are well understood by now, but a new machine requires considerable care in the design phase to minimize operational problems.

There are additional benefits that accrue from the geometry and physics of such a machine.
An ERL has the ability to supply extremely low emittances (of approximately equal value in both planes) for the production of synchrotron light with high peak and average brightness, or for electron beam cooling.
Additionally, the ERL has the advantage of being able to optimize beta functions independently without exceeding the dynamic aperture limitations that rings present.

Finally, the ability of the ERL to operate at low charges with small longitudinal emittances enables the production of very short electron pulses at extremely high repetition rates.
To achieve these benefits requires careful design, including answering a number of physics issues.

There are several hardware aspects that have been improved to enable the potential of ERLs, notably SRF cavity design to allow high currents, including damping of unwanted Higher Order Modes (HOMs) to avoid beam break-up issues.
However, the continual improvement in ERL capability is still pushing the technology limits in several areas, including SRF.
Another active research area is the development of a high-current, ultra-high-brightness, CW electron source.
Extensive development efforts for CW sources have been undertaken at many laboratories, and substantial efforts are also required for appropriate diagnostics. 

However, the following sections will show convincingly that the fundamental principles of ERLs have been successfully demonstrated, not just once, but across the globe.
There can no longer be any doubt that an ERL can be built and achieve its goals.
All of the subjects have now been addressed at some level, but not always simultaneously.
It is generally believed (and history bears this out) that progress in accelerator performance usually requires steps of about a factor of ten.
Less than this is usually a waste of valuable resources, more than this can lead to failure due to the unexpected collusion of multiple extensions of existing technology.
For ERLs to be adopted for larger machines, it will be necessary to have a demonstration ERL with parameters that require mastery of all the potential problem areas, with a beam power of $\sim\SI{10}{\mega\watt}$. PERLE~\cite{Angal_Kalinin_2018} is such a machine and opens the way for future large ERLs.

%\section{Science Goals and Requirements}
%a brief introduction to where ERL enters and what is required to be achieved
\section{Particle Physics and the Importance of ERLs}\label{sec:intro:importance}

For decades, a large community of particle physicists, theorists and experimentalists, in collaboration with ingenious engineers and technicians, has written history, opening a new chapter of physics and understanding of nature following the birth of quantum theory a hundred years ago. The  weak, the electromagnetic and the strong interactions could be  described very successfully by  an SU(2)$_L$ $\times$ U(1) $\times$ SU(3)$_c$ gauge field theory. Following the example of QED, renormalisation led to calculable predictions, while the principle of spontaneous symmetry breaking was confirmed with the discovery of the Higgs Boson by the ATLAS and CMS experiments at the LHC about 10 years ago. 

% Paragraph edited by Andrew on 3/7/22
Over time, the roles of experiment and theory had been of varying---and often alternating---importance.
Prior to the discovery of quark substructure at Stanford in 1968, there was strong agreement on the basic principles of particle physics, except for, e.g., Young-Mills theories, which were perceived to be rather abstract, or S-Matrix theory and a few invariance principles for the strong interaction, which evaded a perturbative description.  
While there was not much theoretical guidance in the fifties, experimenters found an increasing number of particles and began to look more closely at lepton-lepton and lepton-hadron collisions.
By the end of the seventies, which Weinberg describes as otherwise a ``most miserable peace time''\,\cite{WeinbergCC17}, the main elements of the Standard Model had emerged and the theory had become so predictive that for the subsequent decades of experimentation at hadron-hadron, lepton-hadron, and electron-positron colliders, essentially all results confirmed the Standard Model---with discoveries (gluons, W, Z bosons, and heavier quarks), and with ever more precise measurements, accompanied by continuing searches for phenomena beyond the Standard Model (SM).

Now that we have the SM and the Higgs Boson,  theory has been confirmed in its simplest configuration: doublets, triplets, mixing etc., albeit neutrinos oscillate. ``Changing the point of view of 
physicists''\,\cite{WeinbergCC17} is a due task set by Weinberg, as we leave the decades of confirming our theoretical base. 

Currently, the LHC experiments annually produce hundreds  of first-class publications, non-collider experiments search for BSM physics occurring in loops, with recent puzzles such as that of $g-2$ of the muon, and the LHC is preparing for a more intense luminosity phase (HL-LHC). For the future of particle physics, given the long lead time of its accelerator projects, two questions are slowly becoming pressing: i) could the SM be the end of insight, the end of particle physics? and ii), how can we proceed most sensibly, meaning physics reach, diversity and resources, in probing its consistency to look beyond? 

As to the first question, much now resembles the fifties: theory provides questions,  but no firm answers. Specifically, the SM has known, fundamental deficiencies: a proliferation of too many parameters, a missing explanation of the repetitive quark and lepton family pattern, an unresolved left-right asymmetry in the neutrino sector related to lepton-flavour non-conservation, an unexplained flavour hierarchy, the intriguing question of parton confinement, and others. The Standard Model carries the boson-fermion asymmetry, it mixes the three interactions but has no grand unification, the proton is stable, it needs experiments to determine the parton dynamics inside the proton, has no prediction for the existence of a yet lower layer of substructure, and it does not explain the difference between leptons and quarks. Moreover,  the SM has missing links to Dark Matter, possibly through Axions, and Quantum Gravity, while string theory still resides apart. The Standard Model is a phenomenologically successful theory, fine tuned to describe a possibly metastable 
universe\,\cite{Bednyakov:2015sca}.
Hardly is it the end.

Principally new theories would be required to ``turn the SM on its head'' while, as Steven Weinberg also stated not long ago: ``There isn't a clear idea to break into the future beyond the Standard Model''\,\cite{WeinbergCC17}, it remains the conviction, as Gian Giudice described it in his eloquent ``imaginary conversation'' with the late Guido Altarelli, that ``A new paradigm change seems to be necessary''\,\cite{Giudice:2017pzm} in the ``Dawn of the post naturalness era''.

Apparently, particle physics is as interesting, challenging, and far-reaching as it ever was in recent history.
But it needs revolutionary advances in insight, observation and technologies, not least for its accelerator base. It demands that new generation hadron-hadron, electron-hadron and
pure lepton colliders be developed and realised. A new paradigm can hardly be established
with just one type of collider in the future.
The field needs global cooperation, trust,
and complementarity of its techniques, a lesson  learned from the 
exploration of the Fermi scale with the Tevatron, HERA and LEP/SLC.
Similarly, the S$\mathrm{p}\bar{\mathrm{p}}$S collider, fixed-target muon-hadron and neutrino-hadron scattering experiments and PETRA/PEP/Tristan established the Standard Model, together with subsequent lower-energy experiments such as Babar, BELLE and BES-II. 

%As to the second question about how we shall proceed, one knows that
%progress in previous decades was achieved largely owing to major accelerator developments, for example the introduction of colliders for $e^+e^-$ (first with ADONE in Frascati), of hadron colliders such as the ISR, the Tevatron and LHC, and HERA as the first electron-proton collider. Following Kerst's innovation of colliders, using storage rings, new, far reaching concepts and technical innvoations, such as the stochastic cooling at the S$p\bar{p}S$ collider, for example, have paved the way to dramatic improvements in conquering the accelerator energy frontier with steadily rising luminosity. It has also been a tradition, and possible lesson, that a global distribution of accelerator frontiers was key to the rise of the SM and for the attractiveness of the field of particle physics. Given the absence of reliable theoretical guidance, the new frontiers of energy and intensity need to be evaluated and moved to new levels, while fixed target experiments, such as DUNE and HyperK, and lower energy colliders, such as BELLE,  advance.

Energy Recovery, described in the previous section and throughout this paper, is at the threshold of becoming one of the main means for the advancement of accelerators in a dramatic way.
Recycling  the kinetic energy of a used beam for accelerating a newly injected beam, i.e., minimising the power consumption, avoiding the emittance growth of storage rings, and dumping at injection energy---these are the key elements of a novel accelerator concept, invented half a century ago\,\cite{Tigner:1965wf}, which is almost ripe for renewing our field. The potential of this technique may indeed be compared with the finest
innovations of accelerator technology such as by Widerøe, Lawrence, Veksler, Kerst, van der Meer and others during the twentieth century.
While muon colliders radiate heavily and, like plasma wakefield accelerators, are rather far from being ready for deployment,
ERL is a green technology close to being exploited.
It also corresponds to the prediction of F.\,Bordry, expressed recently,  according to which
``there will be no future large-scale science project without an energy management component, an incentive for energy efficiency and energy recovery among the 
major objectives''\,\cite{freddy19}.

% MB: Suggested change "relies" to "rely" discouraged because it would be wrong ;)
While the future of hadron colliders, such as FCC-hh or HE-LHC, relies on a considerable extrapolation of superconducting, high-field dipole magnet technology, new ERL proposals are close to becoming the base of future energy frontier electron-hadron\,\cite{AbelleiraFernandez:2012cc,LHeC:2020oyt} 
and $\text{e}^+\text{e}^-$\,\cite{Litvinenko:2019txu,Telnov21} 
colliders with luminosities enhanced by orders of magnitude, extended kinematic reach and reduced power consumption.

 It is the key purpose of the now established European roadmap, as well as of ERL facilites and experiments elsewhere, to evaluate the performance prospects and to characterise the required R\&D that would be necessary for ERLs to become a reliable technology. When supported and successful, this will  open
  new avenues not only for the future of energy-frontier particle physics
but similarly for new generations of low-energy particle and nuclear physics experiments
and novel, potentially revolutionary industrial developments through the application of ERL based techniques, as are also described in this text.

%\section{Requirements to the Next Generation of ERLs}
%This is better written when the preceding section on ERLs is in 

\section{Outline}\label{sec:intro:outline}
%Chapter \ref{sec:intro} focuses on the science goals and the requirements that derive from them.
Chapter \ref{sec:current_facilities} describes ERLs that have closed down but still hold records in some technology, as well as those currently in operation or forthcoming, while Chapter \ref{sec:new_facilities} presents new facilities or upgrades that are being proposed.
Appendix~\ref{sec:appendix:facilities} lists the parameters of these facilities, showing the steady improvement in performance attained and still to be demonstrated. It also includes parameter lists for
the future high-energy accelerators using ERLs that are currently being considered.
Chapter \ref{sec:key_challenges} describes the various key technology challenges, and Chapter \ref{sec:frontier} describes the future uses of the ERL facilities for particle-physics research (both high-energy physics and nuclear physics).
Chapter \ref{sec:applications} describes the major industrial uses of ERLs.
Chapter \ref{sec:sustainability} addresses the special role that ERLs can play in sustainability.
A concluding Chapter \ref{sec:conclusions} describes the roadmap that the Panel recommends to enable ERLs to take their rightful place on the world stage. Finally, Appendix~\ref{sec:appendix:electronpositron} presents a detailed report prepared by a group of experts which evaluates two recent proposals for electron-positron colliders: the CERC, an ERL-based version of FCC-ee, and the ERLC, an ERL configuration of the ILC. 
%\section{Science Goals and Requirements}
%a brief introduction to where ERL enters and what is required to be achieved
%Max Klein, Frank Zimmermann

%text guiding through the past, current and forthcoming facilities in the Appendix \\
%will be a longer chapter around Andrews parameter table
%

%Intro - Andrew Hutton
\chapter{ERL---Facilities and  Current Status}\label{sec:current_facilities}
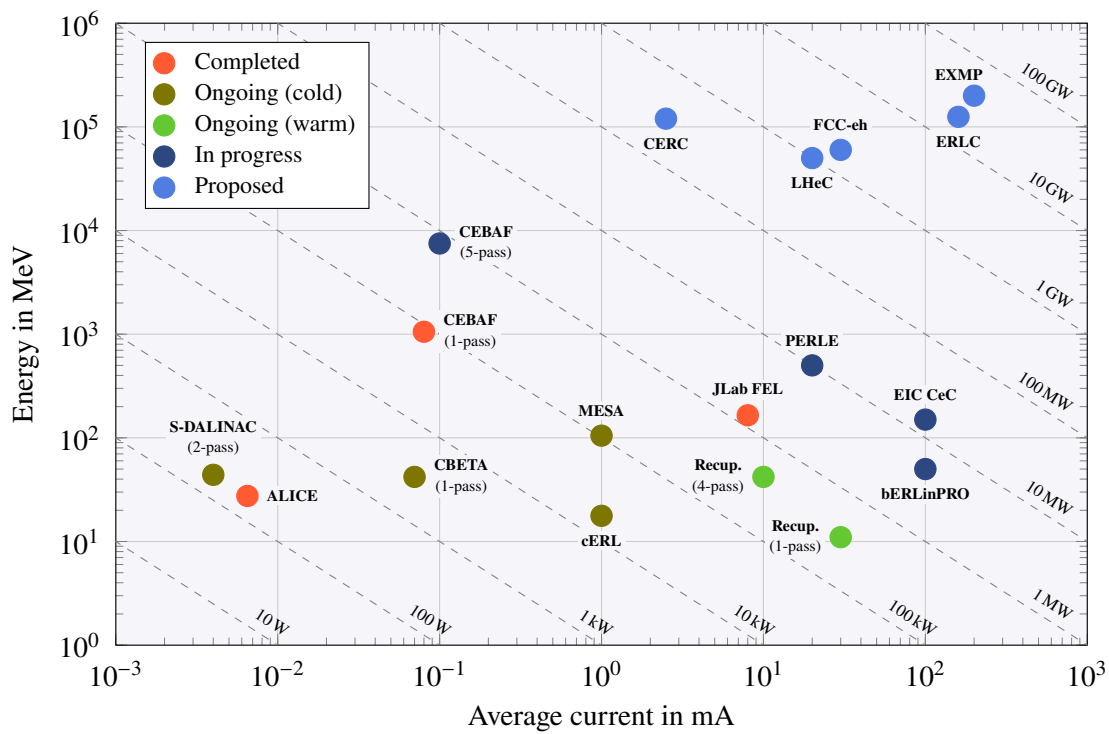
\begin{figure}[ht]\centering
\definecolor{ERL_Landscape_Background}{HTML}{EBEBF5}
\definecolor{ERL_Landscape_Legacy}{HTML}{FF5A32}
\definecolor{ERL_Landscape_Operational}{HTML}{7D7600}
\definecolor{ERL_Landscape_OperationalWarm}{HTML}{64C832}
\definecolor{ERL_Landscape_Construction}{HTML}{507DE1}
\definecolor{ERL_Landscape_Proposed}{HTML}{507DE1}
\definecolor{ERL_Landscape_Progress}{HTML}{2D4780}
\tikzsetnextfilename{erl_landscape}
\begin{tikzpicture}
[
    mymarks/.style={mark size=4pt, only marks, mark=*},
    marks_legacy/.style={mymarks, ERL_Landscape_Legacy},
    marks_operational/.style={mymarks, ERL_Landscape_Operational},
    marks_operational_warm/.style={mymarks, ERL_Landscape_OperationalWarm},
    marks_proposed/.style={mymarks, ERL_Landscape_Proposed},
    marks_progress/.style={mymarks, ERL_Landscape_Progress},
    powerline/.style={dashed, black!50!white},
    powernode/.style={anchor=south east, sloped, black, font=\tiny, inner xsep=1em},
    erl label/.style={font=\tiny, black, align=center, fill=ERL_Landscape_Background!50!white, inner sep=0.5mm},
]
\begin{axis}
[
    xmode=log,
    ymode=log,
    xmin=1e-3,
    xmax=1000,
    ymin=1,
    ymax=1e6,
    width=.95\linewidth,
    height=0.65\linewidth,
    xlabel={Average current in \si{\milli\ampere}},
    ylabel={Energy in \si{\mega\electronvolt}},
    axis background/.style={
        fill=ERL_Landscape_Background!50!white,
    },
    xmajorgrids=true,
    ymajorgrids=true,
    major grid style={black!20!white},
    legend pos=north west,
    legend cell align=left,
    legend style={
        fill=white,
        font=\footnotesize,
        /tikz/every odd column/.append style={column sep=.5em},
    },
    ticklabel shift=2pt,
]
\addplot[
    nodes near coords={\l},
    visualization depends on={value \thisrow{label} \as \l},
    marks_legacy,
    coordinate style/.from=\thisrow{style},
    coordinate style/.condition={1 < 2}{erl label}
] table[x=current, y=energy] {figures/erl_landscape_legacy.dat}; \addlegendentry{Completed}
\addplot[
    nodes near coords={\l},
    visualization depends on={value \thisrow{label} \as \l},
    marks_operational,
    coordinate style/.from=\thisrow{style},
    coordinate style/.condition={1 < 2}{erl label}
] table[x=current, y=energy] {figures/erl_landscape_ongoing_cold.dat}; \addlegendentry{Ongoing (cold)}
\addplot[
    nodes near coords={\l},
    visualization depends on={value \thisrow{label} \as \l},
    marks_operational_warm,
    coordinate style/.from=\thisrow{style},
    coordinate style/.condition={1 < 2}{erl label}
] table[x=current, y=energy] {figures/erl_landscape_ongoing_warm.dat}; \addlegendentry{Ongoing (warm)}
\addplot[
    nodes near coords={\l},
    visualization depends on={value \thisrow{label} \as \l},
    marks_progress,
    coordinate style/.from=\thisrow{style},
    coordinate style/.condition={1 < 2}{erl label}
] table[x=current, y=energy] {figures/erl_landscape_progress.dat}; \addlegendentry{In progress}
\addplot[
    nodes near coords={\l},
    visualization depends on={value \thisrow{label} \as \l},
    marks_proposed,
    coordinate style/.from=\thisrow{style},
    coordinate style/.condition={1 < 2}{erl label}
] table[x=current, y=energy] {figures/erl_landscape_proposed.dat}; \addlegendentry{Proposed}

% Dirty positioning? Yes, but works for now...
%\draw[powerline] (axis cs:1e-13, 1e+8) -- (axis cs:1e-5, 1e0) node[powernode, pos=1.02, ] {\SI{0.01}{\watt}};
%\draw[powerline] (axis cs:1e-12, 1e+8) -- (axis cs:1e-4, 1e0) node[powernode, pos=1.02, ] {\SI{0.1}{\watt}};
%\draw[powerline] (axis cs:1e-11, 1e+8) -- (axis cs:1e-3, 1e0) node[powernode, pos=1.02, ] {\SI{1}{\watt}};
\draw[powerline] (axis cs:1e-10, 1e+8) -- (axis cs:1e-2, 1e0) node[powernode, pos=1.02, ] {\SI{10}{\watt}};
\draw[powerline] (axis cs:1e-9, 1e+8) -- (axis cs:1e-1, 1e0) node[powernode, pos=1.02, ] {\SI{100}{\watt}};
\draw[powerline] (axis cs:1e-8, 1e+8) -- (axis cs:1e+0, 1e0) node[powernode, pos=1.02, ] {\SI{1}{\kilo\watt}};
\draw[powerline] (axis cs:1e-7, 1e+8) -- (axis cs:1e+1, 1e0) node[powernode, pos=1.02, ] {\SI{10}{\kilo\watt}};
\draw[powerline] (axis cs:1e-6, 1e+8) -- (axis cs:1e+2, 1e0) node[powernode, pos=1.02, ] {\SI{100}{\kilo\watt}};
\draw[powerline] (axis cs:1e-6, 1e+9) -- (axis cs:1e+3, 1e0) node[powernode, pos=1, ] {\SI{1}{\mega\watt}};
\draw[powerline] (axis cs:1e-6, 1e+10) -- (axis cs:1e+3, 1e1) node[powernode, pos=1, ] {\SI{10}{\mega\watt}};
\draw[powerline] (axis cs:1e-6, 1e+11) -- (axis cs:1e+3, 1e2) node[powernode, pos=1, ] {\SI{100}{\mega\watt}};
\draw[powerline] (axis cs:1e-6, 1e+12) -- (axis cs:1e+3, 1e3) node[powernode, pos=1, ] {\SI{1}{\giga\watt}};
\draw[powerline] (axis cs:1e-6, 1e+13) -- (axis cs:1e+3, 1e4) node[powernode, pos=1, ] {\SI{10}{\giga\watt}};
\draw[powerline] (axis cs:1e-6, 1e+14) -- (axis cs:1e+3, 1e5) node[powernode, pos=1, ] {\SI{100}{\giga\watt}};

%\draw (axis cs:1e-3, 1e3) node[fill=white, draw, align=center, font=\footnotesize] {DRAFT\\7/15/2021};
%
\end{axis}
\end{tikzpicture}
\caption{The landscape of past, present, and proposed ERLs. The dashed lines are contours of constant beam power.}\label{fig:current_facilities:erl_landscape}
\end{figure}

This section addresses the ERLs that have closed, those that are active, and those that are fully funded and under construction.
In order to limit the history, only those ERLs that still hold a record for at least one parameter have been retained.
Figure \ref{fig:current_facilities:erl_landscape} shows where all the facilities lie on a plot of energy versus circulating current.

To date, CW facilities have been limited to $\lesssim\SI{2}{\mega\watt}$ (the Jefferson Lab FEL) for a single-pass ERL.
BINP has pursued a different strategy with a pulsed, normal-conducting acceleration system.
They have achieved \SI{5}{\mega\watt} of peak pulse power in a four-pass ERL, the highest power achieved anywhere.
Normal-conducting ERLs may have a place in the future landscape, but probably not for high-energy colliders.
The highest energy recirculated, \SI{1}{\giga\electronvolt}, is the one-pass ERL test at CEBAF, with a five-pass test being planned.
PERLE is the only proposal for a multi-pass CW ERL with \SI{10}{\mega\watt} circulating beam power, a necessary first step towards LHeC.

However, the ERLs currently operating are pushing the limits in a variety of technologies, and are invaluable in moving the field forward.
The efforts to increase the gun current with small emittance have been successful (\SI{75}{\milli\ampere} at CBETA with a DC gun, \SI{100}{\milli\ampere} with an SRF gun), but a current this high has not been recirculated at this time.
These technology challenges will be addressed in Chapter~\ref{sec:key_challenges}.

\section{Completed Facilities}\label{sec:current_facilities:completed}

\subsection{ALICE at Daresbury}%
\label{sec:current_facilities:completed:alice}

%Deepa Angal-Kalinin, Peter Williams
% MB 07/17/2021: rough edit
% consistent use of siunitx
% some minor punctuation and stuff
% enlarged the pictures
 
\begin{figure}[htb]
    \centering
    \includegraphics[width=\linewidth]{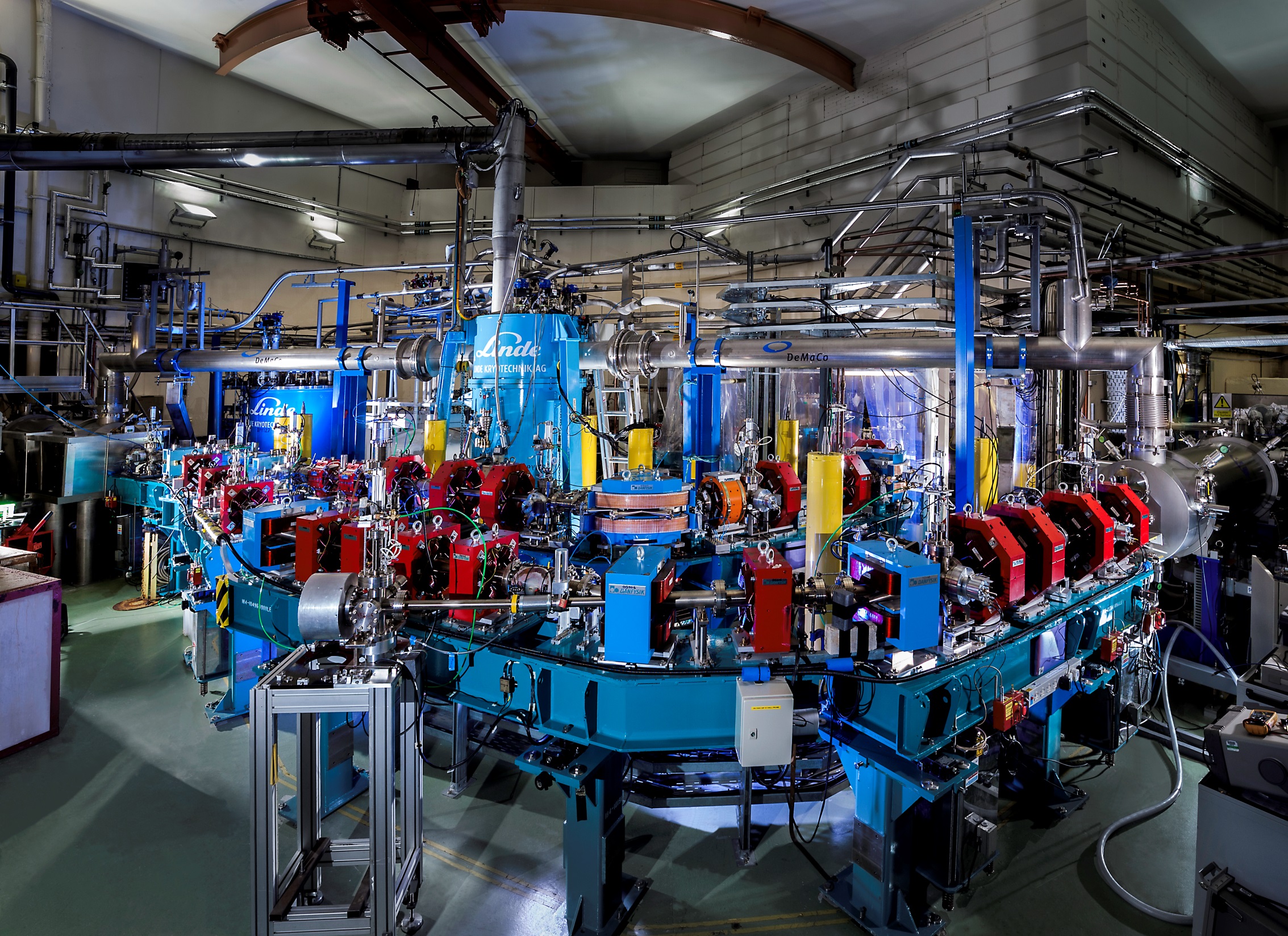}
    \caption{ALICE photographed in 2014 looking over the injection line and return arc to the \SI{2}{\kelvin} cryogenic coldbox and distribution to the booster module (right) and linac (left). The gun is behind the booster, and the chicane and FEL lie behind the cryosystem. The outward arc and dump lie beyond the left wall in a second bunker.}
    \label{fig:current_facilities:completed:alice:photo}
\end{figure}

Accelerators and Lasers in Combined Experiments (ALICE) was an ERL operational from 2005 to 2016 at STFC Daresbury Laboratory in the United Kingdom. Originally conceived as a test facility and technology demonstrator for FELs, it matured over its life into a round-the-clock operational facility for user experiments in the life sciences. Figure~\ref{fig:current_facilities:completed:alice:photo} shows a photograph of the machine.
\subsubsection{Conception}
Daresbury Laboratory was the site of the world's first dedicated X-ray Synchrotron Radiation Source (SRS, 1980--2008). A third-generation replacement for the SRS, the DIAMOND light source, had been designed by the team at Daresbury, but a decision was made to site this at Harwell in 2000. Daresbury was then tasked to develop ideas for the next generation of light sources, strongly felt to be Free-Electron Lasers (FELs).

The UK synchrotron user community were loath to lose the capability of high average power in the development of X-ray FELs for the UK, it being necessary to drive them with a linac rather than a storage ring.
Therefore, the possibility of utilising an ERL to drive an FEL was deemed attractive.
This led to the founding of the 4GLS project in 2000~\cite{ClarkeTheLaboratoryb}.
The 4GLS concept included a \SI{100}{\milli\ampere}, \SI{600}{\mega\electronvolt} ERL driving an EUV to soft X-ray FEL.
Given the multiple new technologies required for such a machine, for example DC photocathode guns, CW superconducting RF,  energy-recovery transport and beam instrumentation, of which there was no UK expertise, it was viewed as essential to construct a test facility to develop skills.
This prototype was called the Energy Recovery Linac Prototype (ERLP, later renamed ALICE, Accelerators and Lasers in Combined Experiments).
At this time, the success of the Jefferson Lab IR-DEMO FEL was recognised~\cite{Chattopadhyay2001ICFA26}, and a collaboration was established between JLab and Daresbury Laboratory to aid the development of ALICE.

\subsubsection{Construction}
Construction commenced in 2003 in a repurposed shielded bunker that had formerly been the experimental area for the defunct Nuclear Structure Facility.
The major components were: a \SI{350}{\kilo\volt} DC photocathode gun based on the JLab design with GaAs cathodes with internal re-caesiation system~\cite{Chattopadhyay2010ICFANewsletter}, a \SI{2}{\kelvin} LHe cryosystem with \SI{120}{\watt} capacity, and two Stanford / Rossendorf type cryomodules, each containing two 9-cell TESLA \SI{1.3}{\giga\hertz} cavities~\cite{Goulden2008INSTALLATIONERLP}.
The design beam energy after the booster was \SI{8}{\mega\electronvolt} and after the linac \SI{35}{\mega\electronvolt}.
The installed RF power was \SI{52}{\kilo\watt} in the booster and \SI{13}{\kilo\watt} in the linac.
The IR-Demo wiggler re-engineered for variable gap and a \SI{9}{\meter} optical cavity working at the 5th subharmonic of the \SI{81.25}{\mega\hertz} bunch repetition rate comprised the oscillator FEL.

%\begin{figure}[h]
%    \centering
%    \includegraphics[width=0.45\textwidth]{ALICEcartoon.png}
%    \caption{The ALICE layout. The arcs were 5m across and X apart leading to path length Y. The long injection line was forced by the existing shape of the shielded bunker.}
%    \label{fig:current_facilities:completed:alice:cartoon}
%\end{figure}
\begin{figure}[htb]
    \centering
    \includegraphics[width=0.8\linewidth]{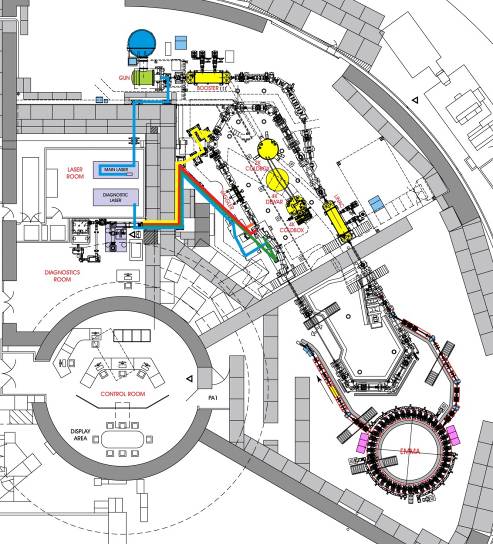}
    \caption{The ALICE layout within the ex-NSF bunker. The gun and booster at the top, outward arc and EMMA extraction at the bottom, laser, diagnostics and control rooms on the left. The arcs were \SI{6.5}{\meter} across and \SI{28}{\meter} apart, leading to $\sim \SI{180}{\nano\second}$ flight time between acceleration and deceleration.}
    \label{fig:current_facilities:completed:alice:layout}
\end{figure}
Various options were considered for the lattice design, coalescing on a racetrack with triple-bend achromat arcs at either end with independently variable first- and second-order longitudinal compaction, one of which was mounted on a movable trombone table to vary the total path length over one RF wavelength. The ALICE layout is shown in Fig.~\ref{fig:current_facilities:completed:alice:layout}.
The injection line was forced to be overly long due to layout restrictions in the bunker, causing difficulties during commissioning; this was exacerbated by a lack of diagnostics within the straight immediately prior to injection.
By 2005, all equipment specifications were complete, the photoinjector laser had been installed and commissioned and the magnets, IOTs, gun, SF6 vessel, and DC HV assembled.
Problems with the gun ceramic and buncher delayed beam commissioning, but first beam was achieved in August 2006.
2007 was plagued by issues with the cryogenic system and gun ceramic; it also became apparent that there were faults in the manufacturing of the cryomodules.
These were some of the very first TESLA-type cryomodules to be produced by ACCEL (now Research Instruments), and they had become badly contaminated during integration.
This led to significant field emission, which limited the linac gradient throughout the life of ALICE.
Additional shielding needed to be installed to protect the RF control racks from radiation damage, and ALICE operated at \SI{27}{\mega\electronvolt} rather than the intended \SI{35}{\mega\electronvolt}.
This situation was later somewhat alleviated after pulsed He processsing in 2013.
Poor performance of the boosted IOT co-axial couplers precluded the intended eventual move from a macropulsed system to CW operation.

\subsubsection{Operational Working Point}
\begin{figure}[htb]\centering%
\includegraphics[width=\linewidth]{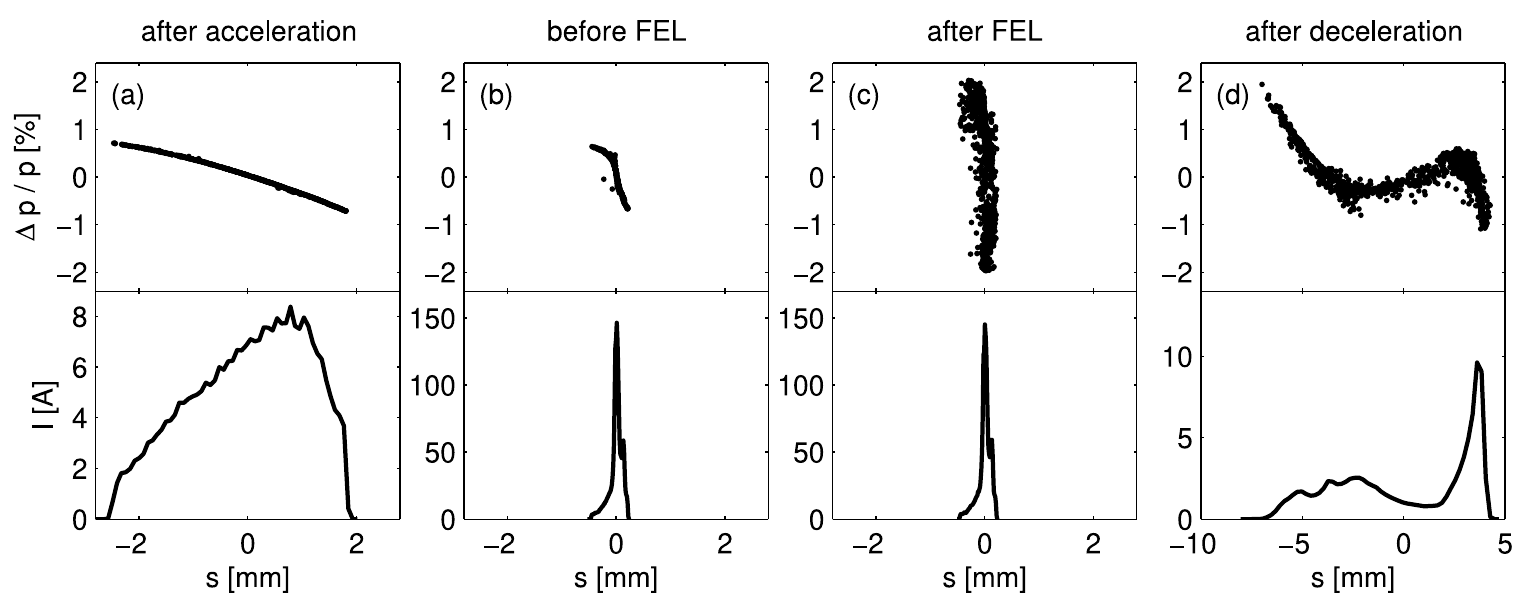}\\[1ex]
\begin{tikzpicture}
\path (0, 0) node (pic) {\includegraphics[width=.45\linewidth]{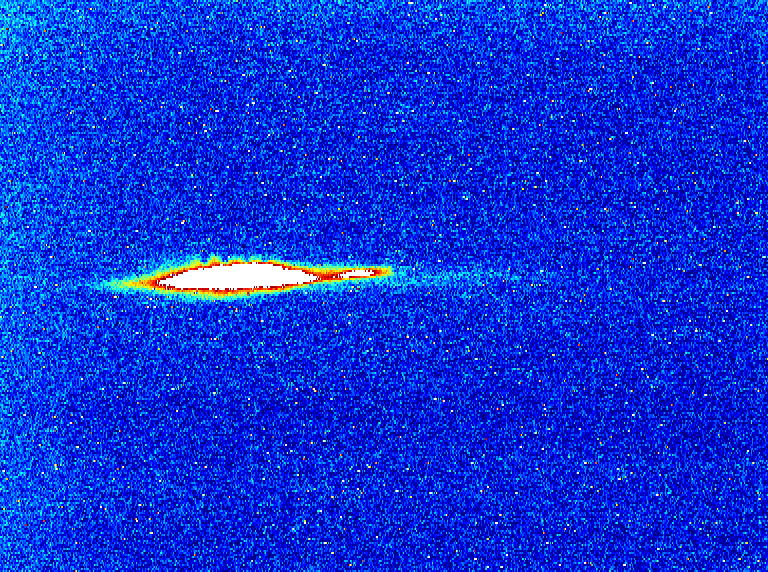}};
\draw (pic.north east) node[fill=white, anchor=north east, yshift=-1ex, xshift=-1ex] {not lasing};
\end{tikzpicture}%
\hfill
\begin{tikzpicture}
\path (0, 0) node (pic) {\includegraphics[width=.45\linewidth]{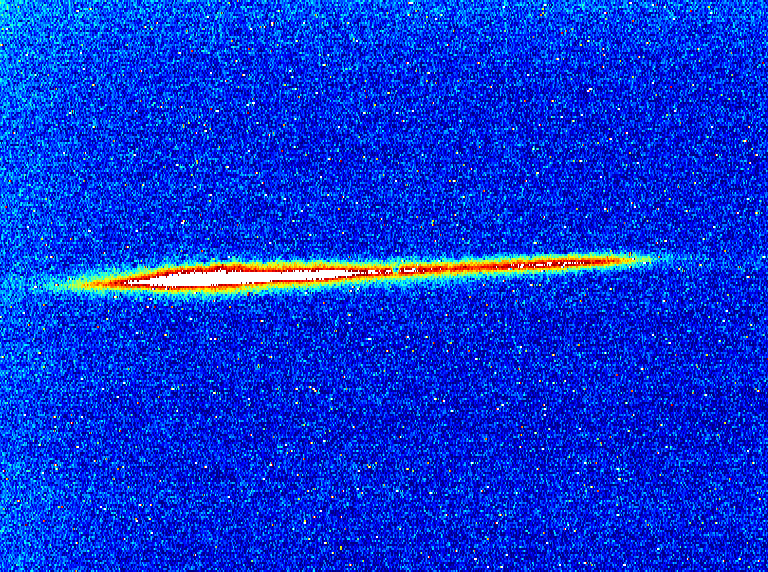}};
\draw (pic.north east) node[fill=white, anchor=north east, yshift=-1ex, xshift=-1ex] {lasing};
\end{tikzpicture}%
\caption{Simulations of the ALICE longitudinal phase space showing the point-to-parallel match (top) and bunch current profile at each stage (middle). Measured OTR images of the beam from the return arc showing FEL-lasing induced energy spread increase and mean energy drop (bottom).}%
\label{fig:current_facilities:completed:alice:lps}%
\end{figure}

The longitudinal match was a point-to-parallel longitudinal phase space double shear and reverse. The bunch was injected at \SI{6.5}{\mega\electronvolt}, then chirped and accelerated at nominal \ang{-8} (rising side of crest) to \SI{27}{\mega\electronvolt}, followed by linearisation and compression in the outward arc and chicane, respectively.
The longitudinal transport parameters were $R_{56} = \SI{0}{\meter}$ (arc) and \SI{+0.028}{\meter} (chicane), $T_{566} = \SI{-2.9}{\meter}$ (arc) and $\SI{-0.4}{\meter}$ (chicane).
The FEL lasing increased the energy spread from \SI{0.4}{\percent} to \SI{5}{\percent} FW and decreased the mean energy by \SI{2}{\percent}.
After lasing, the bunch was decompressed and de-linearised in the return arc ($R_{56} = \SI{-0.028}{\meter}$, $T_{566} = \SI{+2.9}{\meter}$) to ``paint'' the bunch back onto the RF waveform for deceleration. On re-entering the linac, the bunch was dechirped and decelerated on \ang{-8} (falling side of trough).
It was then dumped below the injection energy at \SI{6.0}{\mega\electronvolt}, as energy is lost to the FEL lasing. This scheme is illustrated in Fig.~\ref{fig:current_facilities:completed:alice:lps}.

There were various modes of transverse match, developed through a combination of design and operational learning, each tailored to the application under use. For example, during Compton backscattering runs, a \SI{30}{\micro\meter} round spot with zero divergence was created after the FEL on entrance to the return arc for interaction with the TW laser.
During FEL runs, a few mm elliptical waist was created along the wiggler.

\subsubsection{Commissioning \& User Exploitation}
\begin{figure}[htb]
    \centering
    \includegraphics[width=0.8\textwidth]{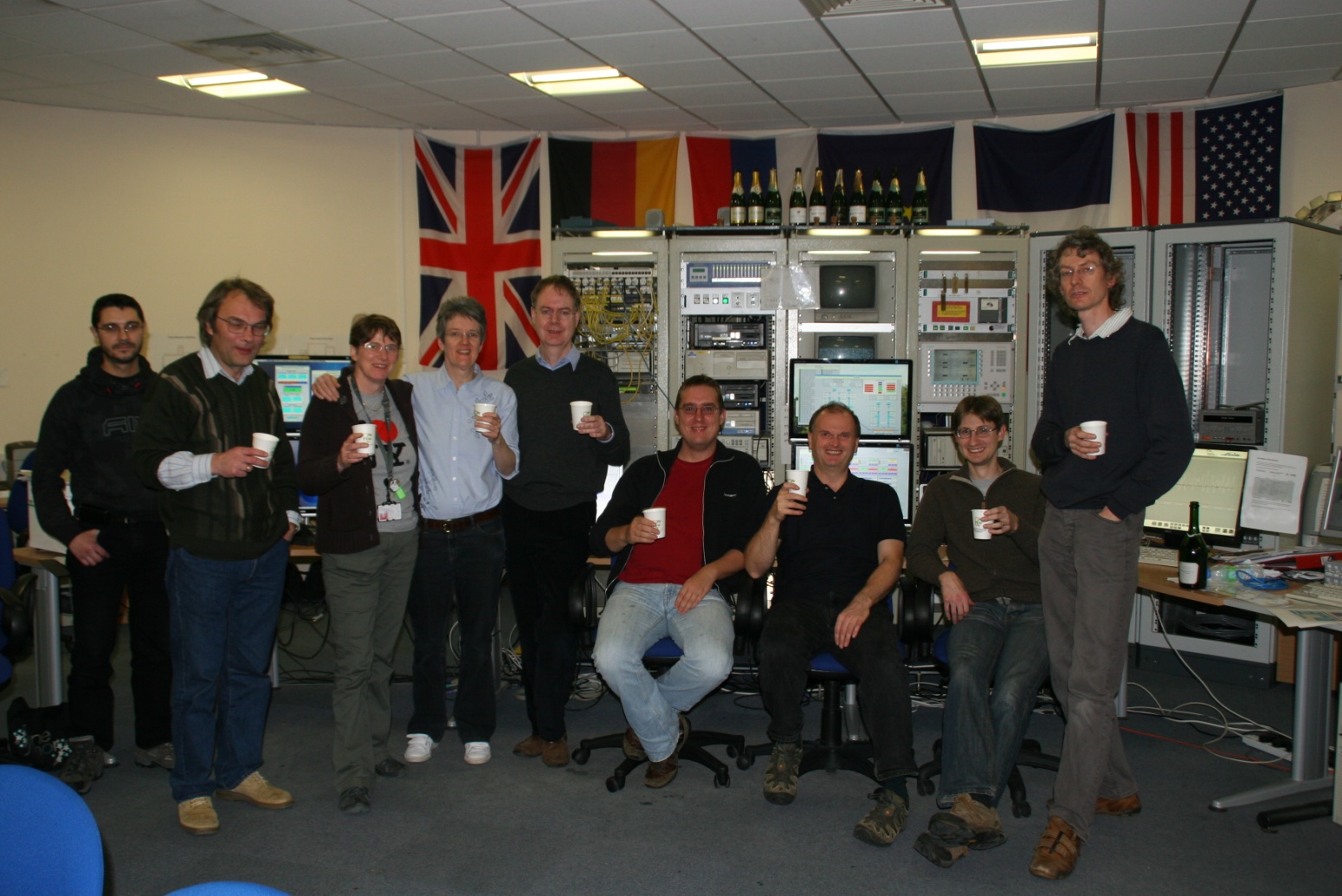}
    \caption{ALICE commissioning team celebrating first energy recovery in the control room in 2008.}
    \label{fig:current_facilities:completed:alice:team}
\end{figure}

\begin{figure}[htb]
    \centering
    \includegraphics[width=0.4\textwidth]{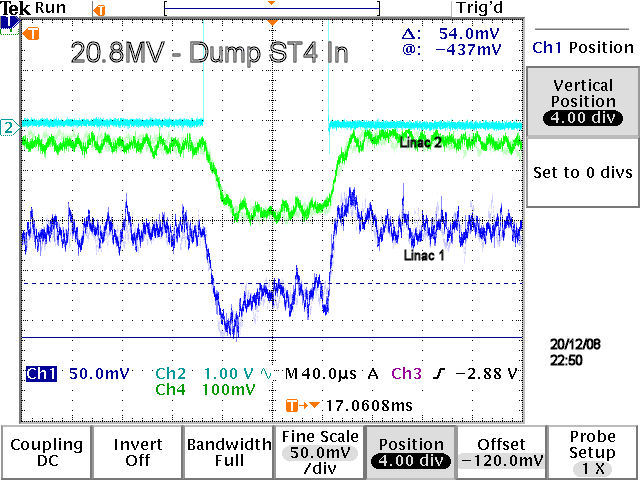} 
    \includegraphics[width=0.4\textwidth]{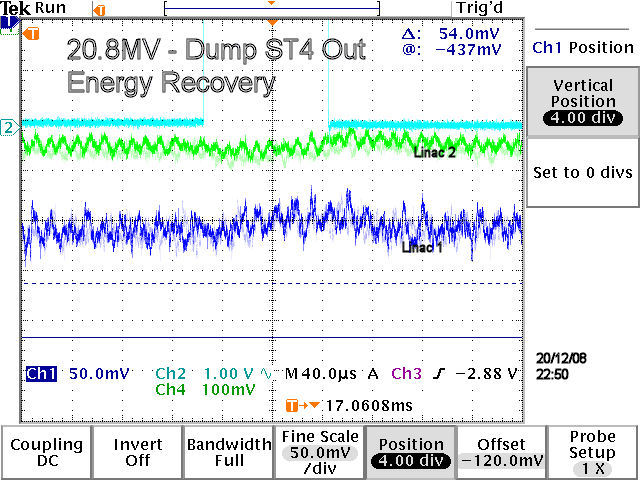}
    \caption{ALICE \SI{100}{\percent} energy recovery at \SI{20.8}{\mega\electronvolt} with \SI{80}{\pico\coulomb} bunch charge in 2009. Traces show the RF demand from the two linac cavities. Left: A movable dump is inserted to block the return beam before re-entering the linac; we see the \SI{100}{\micro\second} bunch train drawing power to accelerate. Right: On removal of the dump, we see the RF demand fall to zero.}
    \label{fig:current_facilities:completed:alice:er}
\end{figure}

In 2008, the booster needed to be sent to ACCEL for repair, and a smaller-diameter ceramic from Stanford was installed temporarily in the gun, allowing commissioning at a reduced voltage of \SI{230}{\kilo\volt}.
After this, fast progress was made: first beam through the booster in October and linac in December.
Just before Christmas 2008, full energy recovery was achieved at \SI{20.8}{\mega\electronvolt} with \SI{10}{\pico\coulomb} charge (Fig.~\ref{fig:current_facilities:completed:alice:team}), followed by \SI{100}{\percent} energy recovery at \SI{80}{\pico\coulomb} bunch charge in 2009. Figure~\ref{fig:current_facilities:completed:alice:er} shows the RF demand falling to zero on recovery.

Coinciding with the onset of user operations, the 4GLS project was cancelled, and ERLP was renamed to ALICE.
The first beamline exploited by users was the broadband THz source in 2009, which utilised a pick-off mirror within the vacuum chamber of the fourth dipole in the compression chicane.
The coherent radiation of the $\sim \SI{1}{\pico\second}$ FW long bunches peaking at \SI{0.3}{\tera\hertz} was transported to the diagnostic room located $\sim \SI{10}{\meter}$ away.
The source had \SI{70}{\kilo\watt} peak power and \SI{23}{\milli\watt} average power~\cite{Weightman2014Investigation.,Williams2013TheDifferentiation}.
The same year, an experiment demonstrating inverse Compton scattering (or Compton backscattering) was performed using a dedicated \SI{10}{\tera\watt} laser system.
This experiment produced $\sim \SI{30}{\kilo\electronvolt}$ X-rays in a head-on configuration~\cite{Priebe2008InverseDaresbury,Laundy2012ResultsSource}.

The main purpose of ALICE was to drive the IR-FEL, the commissioning of which was undertaken in 2010, with first lasing achieved in October~\cite{Thompson2012FirstLaser}.
Good operational reliability of the FEL was achieved in 2012 after the photocathode gun had been reworked to include a re-manufactured ceramic at the original diameter specification.
This allowed \SI{325}{\kilo\volt} operation and consequent reduction of the beam emittance to \SI{2}{\milli\meter\milli\radian}. The output radiation had a peak power of \SI{3}{\mega\watt} with a single-pass gain of \SI{25}{\percent} and was continuously tunable in the range of \SIrange{5}{10}{\micro\meter}.
It operated with user experiments in the application of FEL-beam-illuminated microscopy to cancer diagnosis for five years~\cite{Smith2013Near-fieldDiagnosis,Ingham2018AnImaging,Ingham2019SubmicronIR-FEL}.

Machine studies interleaved with the user programme established a detailed understanding of the beam dynamics, in particular the performance of the gun and the longitudinal behaviour of the TBA arcs~\cite{Saveliev2016ElectronGun,Jackson2016LongitudinalArcs}.
\subsubsection{Summary}
By the final run, ALICE successfully operated 24 hours per day, 7 days per week for an uninterrupted 3 months for external users.
However, ALICE had originally been intended as a short-lived (18 months) test bed and learning tool.
As a consequence of this history, by 2015 many components had gone well beyond their intended life, in particular the cryosystem was now of relatively primitive design and needed major overhaul.
ALICE had showed the potential of ERLs as user facilities, but ALICE itself had gone as far as it could. In 2019, as part of decommissioning, the ALICE injector was donated to IJCLab to form the basis for the PERLE injector.

ALICE constituted the first superconducting RF linac and photoinjector gun in the UK, the first ERL in Europe, the first FEL driven by an ERL in Europe, and the first IR-SNOM on a FEL.

%\subsection{ALICE at Daresbury}
%Deepa Angal-Kalinin, Peter Williams

\subsection{JLab FEL}\label{sec:current_facilities:completed:jlab_fel}
%George Neil, Steve Benson
\subsubsection{Early ERL work at Jefferson Lab}
As mentioned in section \ref{sec:intro:magic:history}, CEBAF needed to test the effects of recirculation on the Cornell SRF cavity design that they had adopted, so they repeated the earlier HEPL experiment, this time with CW beam.  This so-called Front-End Test demonstrated both energy recovery and recirculation, and it permitted the measurement of the beam break-up instability threshold in the Cornell cavity.
Some valuable lessons concerning ERLs were learned in operating this device.

\subsubsection{The IR Demo FEL}
In 1995, The Navy, prodded by Bill Colson of the Naval Postgraduate School, became interested in the possibility of building a high-power laser that could be tuned to atmospheric windows and used for shipboard defense~\cite{LYON1995ABS81}.
The highest FEL power to date at that time was \SI{11}{\watt} from a pulsed, room-temperature-linac-based FEL at Vanderbilt University~\cite{BRAU199238}.
The Navy agreed to fund the development of a \SI{1}{\kilo\watt} FEL in the infrared to demonstrate that an ERL using SRF cavities could greatly increase the maximum power from an FEL. This was the IR Demo project.

\begin{figure}[htb]\centering
\includegraphics[width=\linewidth]{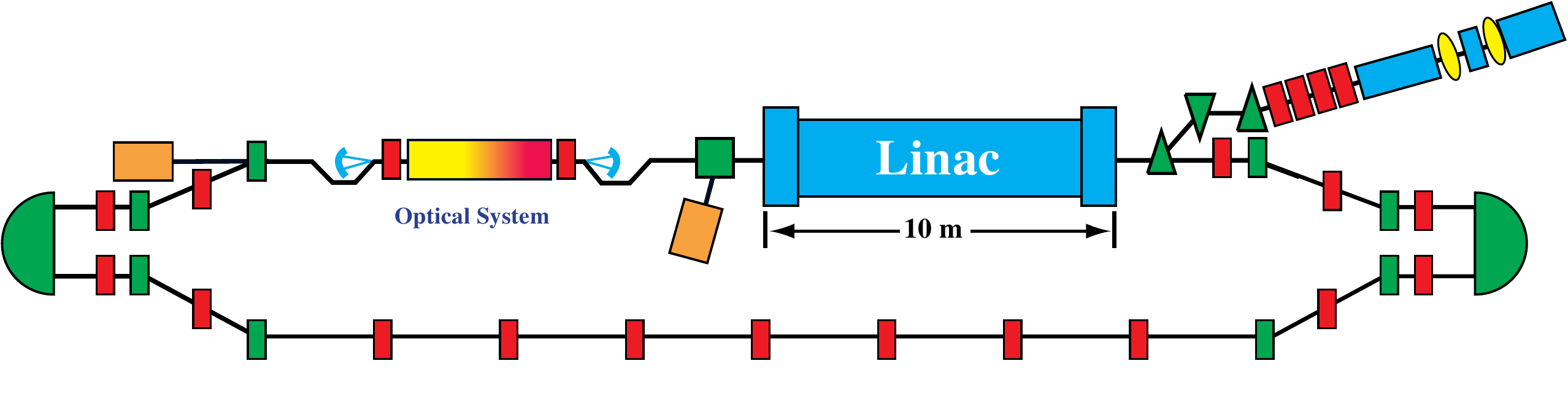}
\caption{IR Demo schematic layout. The photocathode injector is in the upper right. The beam is then merged with the recirculated beam and accelerated to full energy in a single cryomodule. The FEL is between two chicanes that give room for the two cavity mirrors of the resonator. The exhaust beam is transported through two Bates $180^{\circ}$ bends and decelerated to the injection energy. It is then dumped in a high-power dump.}%
\label{fig:current_facilities:completed:jlab_fel:2d}
\end{figure}

Since the IR Demo performance was a factor of 100 higher than existing FELs and since no FEL had every operated on a CW ERL, the JLab group produced a very conservative design.
The design of the FEL systems actually started with the formation of a team that worked on the design of an industrial UV laser system in 1995~\cite{NEIL1995159}.
This team learned a great deal by carefully thinking about the design of a recirculating, energy-recovering driver accelerator.
It also developed a good design for an injector, eventually used in the IR Demo~\cite{osti_754372}.
The first recirculation loop for the UV design, based on a proven $180^{\circ}$ bend used at the Bates lab at MIT~\cite{FLANZ1985325}, was used, with some modifications, for the energy-recovery transport for the IR Demo.
The geometry chosen, with the FEL preceding the energy-recovery loop, was a reflection of the fact that Coherent Synchrotron Radiation (CSR) effects were not completely understood at that time.
By putting most of the bending after the FEL, most of the CSR effects were avoided.
The design parameters and the final, as-built parameters are listed in Table \ref{tab:current_facilities:completed:jlab_fel:parameters}; a sketch of the machine is shown in Fig.~\ref{fig:current_facilities:completed:jlab_fel:2d}.

\begin{table}[htb]\centering
\caption{Design and As-built parameters for the IR Demo FEL.}\label{tab:current_facilities:completed:jlab_fel:parameters}
\begin{tabular}{lrr}
\toprule
Parameter & Design & As built \\
\midrule
Energy (MeV) & 41 & 47.3 \\
Current (mA) & 5 & 4.5 \\
Charge (nC) & 0.135 & 0.060 \\
Energy spread & \SI{0.2}{\percent} & \SI{0.15}{\percent} \\
$\epsilon_x$ (\si{\micro\meter}) & 8 & 6 \\
Wavelength (\si{\micro\meter}) & 3.2 & 3.2 \\
Wiggler $K_\text{rms}$ & 0.7 & 0.99 \\
Wiggler gap (cm) & 1.0 & 1.0 \\
Number of periods & 40 & 40 \\
Output coupling & \SI{10}{\percent} & \SI{10}{\percent} \\
Laser power (kW) & 1.0 & 2.1 \\
\bottomrule
\end{tabular}
\end{table}

As the machine was commissioned, the conservative design choices proved to be very advantageous.
The wiggler was stronger than expected and the maximum energy higher, which increased the gain.
To save commissioning time, the charge was reduced to half the design and the repetition rate doubled.
The measured beam parameters matched the simulation values fairly closely, so the projected gain was over \SI{80}{\percent} per pass, providing a comfortable gain margin for a \SI{10}{\percent} output coupler.

With the large gain margin, the laser lased surprisingly easily.
Initial lasing was carried out in a straight-ahead mode, which limited the current to \SI{1.1}{\milli\ampere} and the laser power to \SI{500}{\watt}.
First light was achieved with a \SI{42}{\mega\electronvolt} beam at \SI{5}{\micro\meter} on June 15, 1998, and \SI{300}{\watt} was achieved on July 27~\cite{BENSON199927} with a \SI{10}{\percent} output coupler.
The laser was quite stable and could operate for long periods of time at this power level, 30 times the previous world record power.

Once lasing had been achieved, the task of recovering the beam was next.
The design of the Bates bends used to return the beam to the linac worked well, and the calculated matches allowed transport of the beam back through the linac to the energy-recovery dump without much difficulty.
The relatively large apertures (another conservative design feature) meant that the losses in the transport were minimal.
Once energy recovery had been achieved, the current and laser power were slowly ramped up.
With the final beam dump energy less than \SI{10}{\mega\electronvolt}, the level of activation and neutron production went down enormously, and it was possible to enter the vault immediately after running for hours with milliamperes of recirculated beam.
This turned out to be one of the largest benefits of energy recovery.

Recirculating the beam turned out to be straightforward, but the FEL power was limited by mirror heating.
The theory of high power FEL operation with absorptive mirrors had been worked out during the design stage~\cite{BENSON1998401}.
Heat-induced distortion reduces the FEL gain and ends up clamping the power at a level proportional to the wavelength and a figure of merit (FOM) for the mirrors determined by their thermal characteristics.
Once this power is reached, it cannot be exceeded with any beam current.
The calcium fluoride mirrors used in the first lasing had a relatively poor thermal FOM, so the power clamped at about \SI{500}{\watt}. When sapphire mirrors, with a much higher FOM, were used, the power was ramped up to \SI{1720}{\watt} with \SI{4.4}{\milli\ampere} of beam current in July 1999.
The limitation now was just the efficiency of the laser and the available current~\cite{PhysRevLett.84.662}.
Many of the beam physics issues that were feared to limit the performance, such as RF instabilities, beam breakup, and halo, turned out not to be a problem~\cite{BENSON20021}.

Once the goal of \SI{1000}{\watt} from the IR Demo had been reached, a user program began.
Machine studies and optimization were carried out during off shifts.
The electron beam quality was sufficient to lase at not only the third harmonic, already demonstrated on several FELs, but also at the second and fifth harmonic, which had never previously been demonstrated~\cite{NEIL2002119}.
With more operating experience, \SI{2.1}{\kilo\watt} of laser power could be provided at \SI{3.2}{\micro\meter}, more than a factor of two over the design power.

Other wavelength ranges could be produced using parasitic processes.
Short, high-charge bunches radiate copious CSR and can ruin the beam quality.
The degradation from the chicane before the FEL was reasonable, however, and the THz radiation produced in this process could, in principle, be used.
Researchers had been working for years to produce THz radiation in a similar manner by using short-pulsed lasers to produce very short THz pulses.
The radiation in the \SI{48}{\mega\electronvolt} beam was enhanced by the cube of the ratio of the electron energy to the electron rest mass energy, which is about 100 for this case.
Using this simple scaling, it is easy to show that the \SI{100}{\micro\watt} of THz power in the laser-based sources could be increased by a million times to the \SI{100}{\watt} level.
These levels were demonstrated in the chicane magnet just upstream of the FEL~\cite{osti_842288}.

The second parasitic radiation source is Thomson backscattering.
The circulating light in the optical cavity collides with the electrons at the waist of the cavity where both beams are very small.
The infrared photons can then scatter off the electrons and produce X-rays along the direction of the electron beam.
The IR Demo produced copious amounts of these Thomson-scattered X-rays, and the center of the cavity was placed at a location where the wiggling electrons were angled upward so the X-rays could escape the optical cavity.
Careful measurements of the X-ray flux and spectrum were carried out while running the laser for other users~\cite{Boyce:748655}.

\subsubsection{The IR Upgrade FEL}
The next step in the evolution of the design was to produce a \SI{10}{\kilo\watt} FEL.
The plan was to triple the electron beam energy, double the electron current and keep the FEL efficiency the same as it was with the IR Demo.
This should produce over \SI{12}{\kilo\watt} of laser light.

Initially, the third cryomodule necessary to reach \SI{150}{\mega\electronvolt} was not available, limiting the energy to \SI{80}{\mega\electronvolt}. With \SI{10}{\milli\ampere} of beam current and \SI{1.25}{\percent} efficiency (\SI{1.5}{\percent} had been reached with the IR Demo) there was a chance of reaching \SI{10}{\kilo\watt}.

Unlike the IR Demo, however, the FEL was placed after the first \ang{180} bend, so the CSR in that first arc had to be dealt with.
The Bates bend allowed one to get even shorter bunches than in the IR Demo since it was possible to use the sextupoles in the arc to correct for the RF curvature in the linac.
This increased the CSR even more.
It was found that using a slightly under-bunched beam in the FEL produced the best FEL performance.

Initial lasing was attempted with an optical klystron operated with small dispersion~\cite{BENSON200340}.
As with the IR Demo, initial laser commissioning took place with a pulsed beam at low duty cycle.
Lasing at \SI{6}{\micro\meter} using zinc selenide mirrors was obtained on June 17, 2003.
This system was far more flexible than the IR Demo.
Up to four mirror sets could be installed in the vacuum chamber at one time, and the wiggler could be tuned in real time~\cite{SHINN2003196}.
Lasing over the full reflectivity range of the mirrors was demonstrated, but the gain was much smaller than expected.
This was due to a fourfold blow-up in the longitudinal emittance caused by longitudinal space charge~\cite{HernndezGarca2004LONGITUDINALSC}.
With this poor beam quality, the \SI{1.25}{\percent} efficiency was out of reach and had to await the addition of the third cryomodule in the linac.

While the third cryomodule was being completed, the problem of recirculating the beam was worked out.
With the longer transport and the need to transport through three cryomodules instead of one, this was not as easy as in the IR Demo but was eased somewhat by the increased flexibility of the lattice.
The shorter bunch lengths also led to increased CSR, which led to more mirror heating in the downstream mirror.
A bunch-lengthening chicane and THz-absorbing traps were therefore added to reduce the mirror heating~\cite{patent9590384, patent9209587}.

Once the beam energy could be increased to \SI{150}{\mega\electronvolt} and the optical klystron was replaced with a variable permanent-magnet wiggler, strong lasing could be achieved at short wavelengths and multikilowatt lasing became common, though the power was still limited by mirror heating.
With the stronger lasing, however, the concept of incomplete energy recovery was developed, where the energy acceptance of the ERL could be increased by a factor of two over the complete energy-recovery state~\cite{osti_921655}.
The new cryomodule also had a lower beam breakup (BBU) threshold, which led to the development of techniques to reduce BBU and carefully characterize the physics of this instability~\cite{PhysRevSTAB.9.064403}.

The solution to mirror heating was to enhance the thermal FOM by cooling the sapphire mirrors to cryogenic temperatures. The FOM at liquid nitrogen temperature is at least a factor of 200 larger than at room temperature. With these mirrors, the efficiency of the laser, which could routinely exceed \SI{1.6}{\percent} at low power, was now independent of power or average electron current.
The FEL could now lase at \SI{1.6}{\micro\meter} with a power output of up to \SI{14.3}{\kilo\watt}, easily exceeding the \SI{10}{\kilo\watt} goal~\cite{pac07_benson}.
The ERL provided \SI{8.5}{\milli\ampere} of current at \SI{115}{\mega\electronvolt} for this power level.

\subsubsection{THz Operations, UV Demo, and CEBAF ER}
Even more power was possible with further efforts, but the focus then shifted to going to other wavelengths.
In parallel with the IR Upgrade operations, an optical transport line was installed providing THz transport to one of the upstairs labs from the last bend before the FEL.
The vacuum system of the accelerator was isolated from the THz transport using diamond windows.
As noted above, this radiation can be well over \SI{100}{\watt}, a unique radiation source for users.
The higher energy and current of the IR upgrade enabled the device to produce a factor of ten more THz radiation than in the IR Demo.
This was used to make THz movies using THz detector arrays~\cite{patent8362430}.

\begin{figure}[htb]\centering
\includegraphics[trim=3cm 1.1cm 0.6cm 1.7cm, clip, width=.85\linewidth]{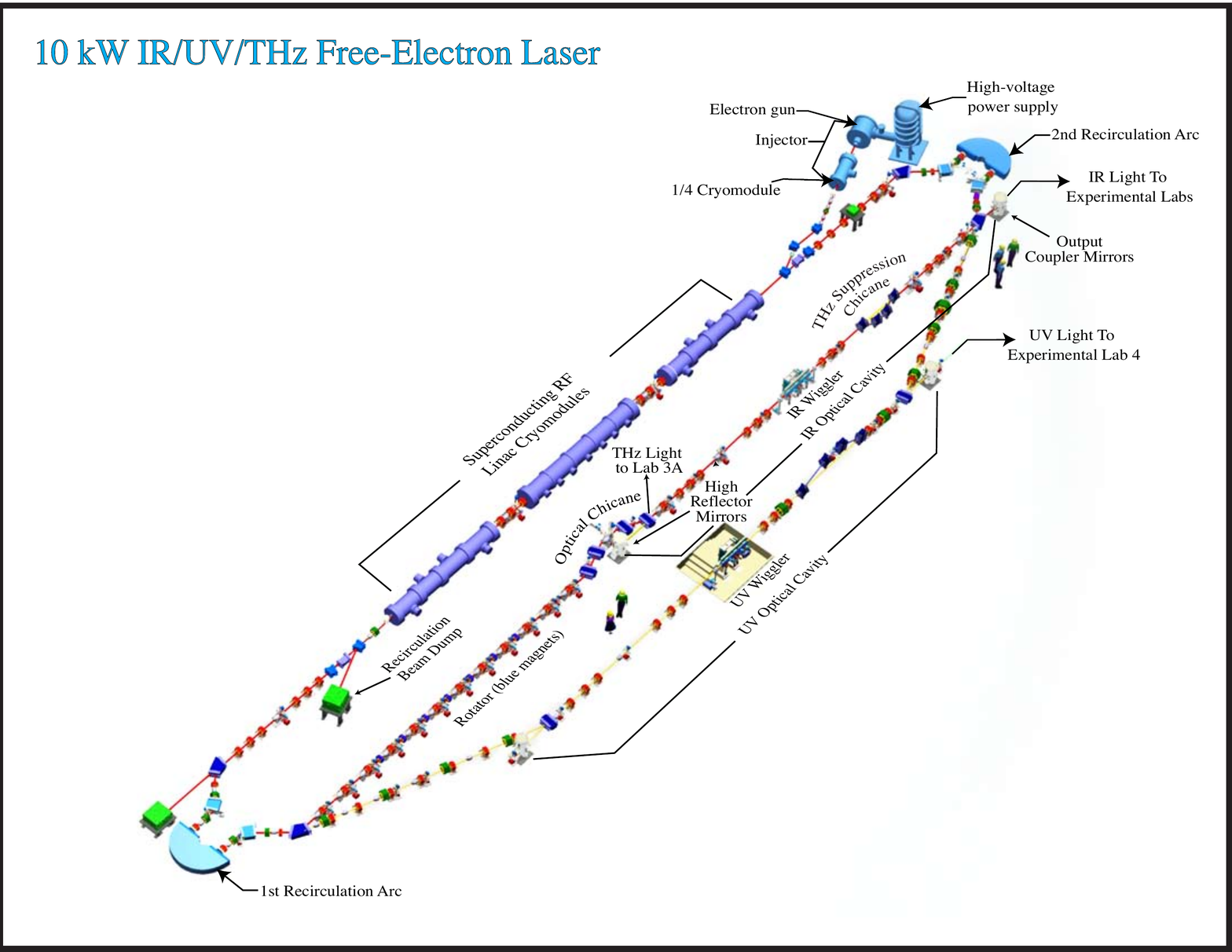}
\caption{Layout of the JLab FEL facility, showing the injector at the top right, the superconducting linac on the upper left, with 2 recirculation paths each containing an oscillator-based FEL. The inner path included the IRFEL of the IR Upgrade project, while the outer path contained the UVFEL of the UV Demo project. The recirculation arcs were mostly the same except the outer dipoles, which could be halved in strength to send the beam into the UV line. Each optical resonator was \SI{32}{\meter} in length and could use any of four sets of mirrors without breaking vacuum.}%
\label{fig:current_facilities:completed:jlab_fel:3d}
\end{figure}

Along with the IR Upgrade, the lab was funded to build another electron beam transport with an ultraviolet laser in it~\cite{napac11_douglas}.
The final FEL layout including this addition is shown in Fig.~\ref{fig:current_facilities:completed:jlab_fel:3d}.
The UV design benefited from all the lessons learned from the IR machine, though the electron beam requirements differed from the IR machine.
The UVFEL requires a smaller transverse emittance and a smaller energy spread, so the charge was halved to match the smaller acceptance.
With a proper longitudinal match, the gain can be substantial well into the ultraviolet.
The real challenge in the UVFEL is the optical resonator.
Operation at high power is much more of a challenge than in the infrared, both because the absorption is much higher and because the allowed distortion is proportional to the wavelength.
Kilowatt operation is not possible without cryogenic mirrors.
However, the pulse lasing performance was very strong once achieved.
A gain of over \SI{200}{\percent} per pass was seen, exceeding even optimistic projections.
The power in the near UV was limited by mirror heating to \SI{150}{\watt}~\cite{osti_1020755}.

One final ERL at Jefferson Lab was demonstrated at CEBAF.
This will be described in the next section. The peak current in CEBAF is too low for an FEL, but the experiment did demonstrate that a GeV FEL using an ERL is feasible.
Jefferson Lab has now proposed building a 5-pass recirculation system operating at up to \SI{8}{\giga\electronvolt} to demonstrate even higher energies.
This is much harder than a \SI{1}{\giga\electronvolt} system since incoherent synchrotron radiation must be compensated.

\subsubsection{Summary}
Taking advantage of the deep knowledge of SRF acceleration available at Jefferson Lab, the FEL group developed many of the techniques required to operate a high-power ERL.
The ERLs at Jefferson Lab continue to be the only ones that demonstrated more electron beam power than installed RF power, which is the main reason to build an ERL.
Other advantages of the ERL, like reduced activation in the dump and more stable RF operation, were found as well.
The group also definitively demonstrated that FELs were indeed capable of high power.

%\subsection{JLab FEL}
%George Neil, Steve Benson

\subsection{CEBAF Single-pass Energy Recovery Experiment (CEBAF-ER)}\label{sec:current_facilities:completed:cebaf_1pass}
%Alex Bogacz and Chris Tennant

% MB 7/18/21
% Edit for punctuation/hyphenation, minor details like passive voice, etc.
% Relabeled figures and replaced three of them with TikZ versions.

\subsubsection{Experimental Setup}
\begin{figure}[htb]
  \centering
  \includegraphics[width=0.7\textwidth]{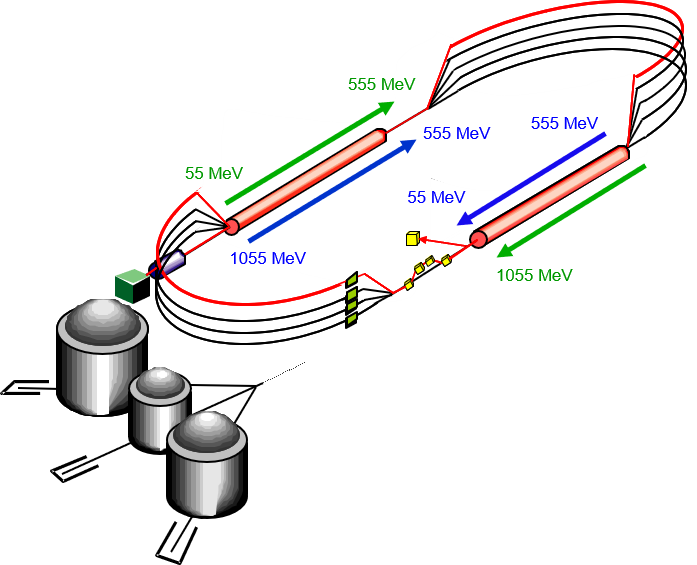}
  \caption{Schematic layout of ER-CEBAF.}
  \label{fig:current_facilities:completed:cebaf_1pass:schematic}
\end{figure}

The \SI{6}{\giga\electronvolt} CEBAF accelerator was a five-pass recirculating SRF (superconducting radio frequency) based linac capable of simultaneous delivery to three end stations of CW beam for nuclear physics experiments.
The CEBAF energy-recovery experiment was carried out in  March of 2003~\cite{Bogacz:pac03} with the goal of demonstrating the energy recovery of a \SI{1}{\giga\electronvolt} beam while characterizing the beam phase space at various points in the machine and measuring the RF system’s response to energy recovery. 
In order to perform the energy-recovery experiment, two major components had to be installed in the CEBAF accelerator: a phase delay chicane and a beam dump line.
A schematic representation of the CEBAF-ER experiment is shown in Fig.~\ref{fig:current_facilities:completed:cebaf_1pass:schematic}.
The beam is injected into the North Linac at \SI{55}{\mega\electronvolt}, where it is accelerated to \SI{555}{\mega\electronvolt}.
The beam traverses the first (East) arc and begins acceleration through the South Linac, where it reaches a maximum energy of \SI{1055}{\mega\electronvolt}.
Following the South Linac, the beam passes through the newly installed magnetic phase-delay chicane.
The chicane was designed to create a path length difference of exactly \textonehalf~RF wavelength so that upon re-entry into the North Linac, the beam will be \ang{180} out of phase with respect to the cavity RF fields and subsequently be decelerated to \SI{555}{\mega\electronvolt}.
After traversing the East Arc, the beam enters the South Linac and is decelerated to \SI{55}{\mega\electronvolt}, at which point the energy-recovered electron beam is sent to a dump.
Issues related to beam quality preservation in systems with a large energy ratio between final and injected beams (a factor of 20) were addressed. 

\subsubsection{RF Challenges}
Whereas Jefferson Lab’s IR FEL demonstrated the energy recovery of a \SI{45}{\mega\electronvolt} beam through a single cryomodule (8 CEBAF-type 5-cell cavities), the CEBAF-ER experiment energy recovered a beam through 39 cryomodules. Consequently, any harmful effects induced into the electron beam by the RF system were greatly enhanced.

The most apparent RF-induced effect observed was the coupling of the transverse planes.
It is known that the higher-order mode (HOM) coupler on each cavity introduces a phase-dependent skew quadrupole which couples the horizontal and vertical oscillations.
In nominal CEBAF operation, this effect is mitigated by the use of a DC magnetic skew quadrupole between cryomodules to produce a compensatory gradient integral.
However, in CEBAF-ER operation, the sign of the induced skew quadrupole changes since the second pass is \ang{180} out of phase with the first pass.
Therefore, although the external DC skew quadrupoles can locally correct for a single pass through the linac, the coupling will be doubled on the other pass.
During the experiment, we used an ``up-down'' correction scheme in which the accelerating pass through the North Linac and the decelerating pass through the South Linac were corrected using the DC skew quadrupoles.
Although the coupling is not suppressed using this configuration, it was the most attractive solution based on simulations, which indicated that the initial projected emittances would be recovered after energy recovery.

Additionally, a transverse electric-field gradient that exists in the cavity fundamental power couplers (FPC) produces a transverse deflection which leads to differential head-tail steering of an electron bunch.
The magnitude of the effect can be minimized with an appropriate choice of RF feed geometry~\cite{York}.
Simulations using the present feed configuration in CEBAF suggest that the projected transverse emittance could conceivably be degraded by a factor of 2 from passage through the acceleration pass due to the effects of the dipole-mode-driven head-tail steering.

\subsubsection{Phase Space Measurements}
\begin{figure}[htb]\centering
\tikzsetnextfilename{cebaf_1pass_bothbeams}
\begin{tikzpicture}
[
    annotation/.style={fill=white, draw, font=\footnotesize},
    myarrow/.style={-latex},
]
\begin{axis}
[
    width=0.9\linewidth,
    height=.6\linewidth,
    ymin=0,
    ymax=560,
    ylabel={Intensity (arb.~units)},
    xlabel={Wire scanner position (mm)},
    xmin=0,
    xmax=38,
    legend pos=north east,
    xmajorgrids=true,
    ymajorgrids=true,
    legend style={font=\footnotesize},
]
\addplot+[mark=none] table[x index=0, y index=1] {figures/cebaf_1pass_bothbeams.csv}; \addlegendentry{Wire scan at 2L22: $x$, $x/y$, $y$}
\draw (axis cs:4.8, 490) node[annotation] (highbeam) {\SI{1056}{\mega\electronvolt} beam};
\draw[myarrow] (highbeam) -- (axis cs: 10.7, 400);
\draw[myarrow] (highbeam) -- (axis cs: 19.8, 425);
\draw[myarrow] (highbeam) -- (axis cs: 28, 450);
\draw (axis cs:33, 200) node[annotation] (lowbeam) {\SI{56}{\mega\electronvolt} beam};
\draw[myarrow] (lowbeam) -- (axis cs: 7.3, 150);
\draw[myarrow] (lowbeam) -- (axis cs: 15.5, 60);
\draw[myarrow] (lowbeam) -- (axis cs: 22.8, 60);
\end{axis}
\end{tikzpicture}
\caption{Wire scan with fully accelerated and energy-recovered beams at the same time.}\label{fig:current_facilities:completed:cebaf_1pass:bothbeams}
\end{figure}
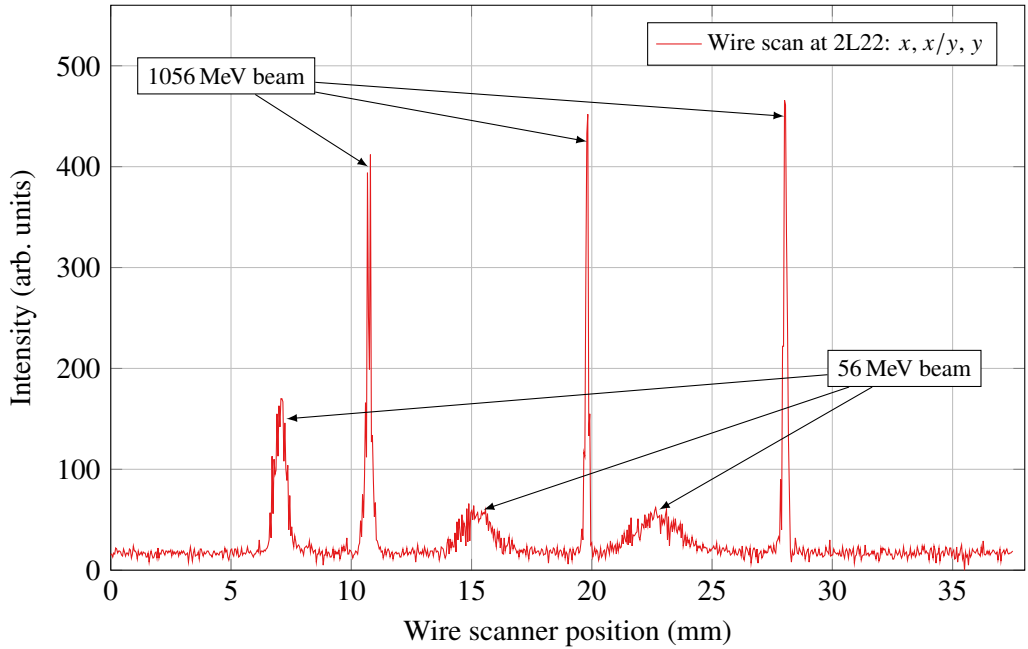

To gain a quantitative understanding of the beam behavior throughout the machine, an intense effort was made to characterize the 6D phase space during the CEBAF-ER experimental run.
A scheme was implemented to measure the geometric emittance of the energy-recovered beam prior to sending it to the dump, as well as in the injector and in each arc. 
In this way, one can understand how the emittance evolves throughout the machine.
In addition to describing the transverse phase space, the fractional momentum spread was measured in the injector and arcs to characterize the longitudinal phase space as illustrated in Fig.~\ref{fig:current_facilities:completed:cebaf_1pass:bothbeams}.

The emittance and momentum spread of the first-pass beam were measured in the injector, Arc 1, and Arc 2 utilizing a scheme involving multiple optics and multiple wire scanners.
Two wire scanners were placed in each arc, one at the beginning of the arc in a non-dispersive region and the second in the middle of the arc at a point of high dispersion (\SI{6}{\meter}).
The emittance in the injector was measured using five wire scanners along the injector line.
One of the unresolved difficulties with this measurement was finding a scheme for which the emittance and momentum spread of the recirculated beam could be measured in Arc 1.
During the measurement, an insertable, downstream dump was used to prohibit the transport of a recirculated beam.
But it is unclear how to resolve each beam from a wire scanner that is sampling two co-propagating beams; even more so in the case of Arc 1 where, notionally, both the first-pass and second-pass energy-recovered beam have the same energy.
This is not an issue for Arc 2; since the energy-recovered beam is sent to the dump immediately upon exiting the South Linac, there is only one beam being transported through Arc 2 at all times.

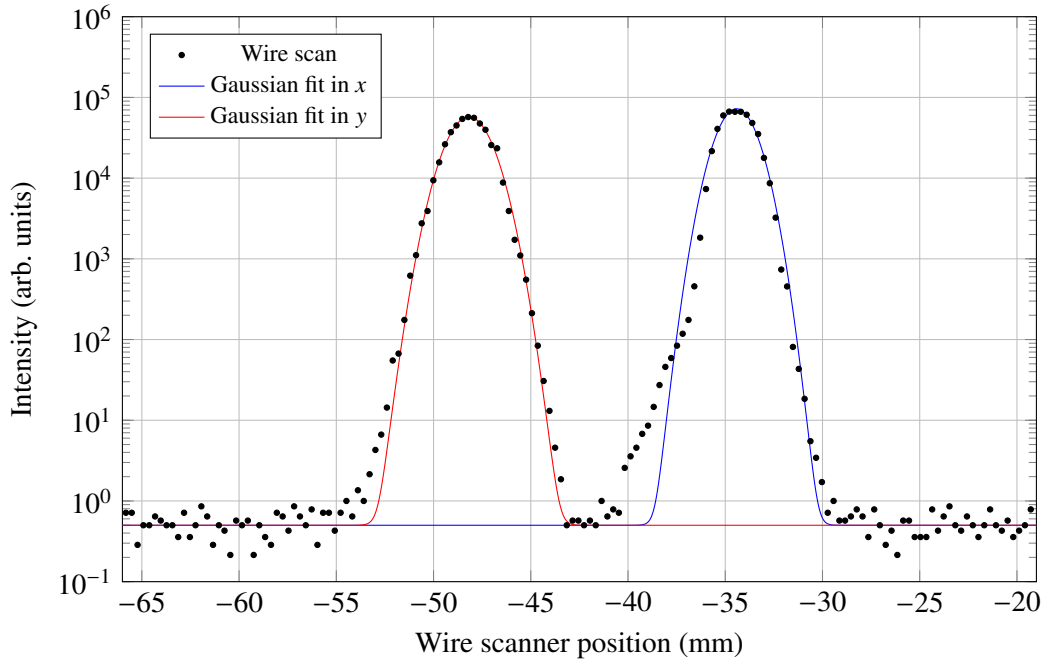
\begin{figure}[htb]\centering
\tikzsetnextfilename{cebaf_1pass_large_dynamic_range}
\begin{tikzpicture}
\begin{axis}
[
    width=0.9\linewidth,
    height=.6\linewidth,
    ymode=log,
    ymin=0.1,
    ymax=1000000,
    ylabel={Intensity (arb.~units)},
    xlabel={Wire scanner position (mm)},
    xmin=-66,
    xmax=-19,
    legend pos=north west,
    xmajorgrids=true,
    ymajorgrids=true,
    legend style={font=\footnotesize},
]
\addplot+[black, only marks, mark size=1pt] table[x index=0, y index=1] {figures/cebaf_1pass_55mev_wirescan.csv}; \addlegendentry{Wire scan}
\addplot+[blue, mark=none] table[x index=0, y index=1] {figures/cebaf_1pass_55mev_wirescan.f2}; \addlegendentry{Gaussian fit in $x$}
\addplot+[red, mark=none] table[x index=0, y index=1] {figures/cebaf_1pass_55mev_wirescan.f1}; \addlegendentry{Gaussian fit in $y$}
\end{axis}
\end{tikzpicture}
\caption{Large-dynamic-range $x$ and $y$ beam profile measurement of the energy-recovered beam with $E=\SI{55}{\mega\electronvolt}$. The beam profiles after energy recovery do not show any significant distortion; in this configuration, the profile was close to Gaussian over multiple orders of magnitude.}\label{fig:current_facilities:completed:cebaf_1pass:large_dynamic_range}
\end{figure}

To improve on the dynamic range of the wire scanner for beam profile measurements of the energy-recovered beam, instrumentation was added to the wire scanner just upstream of the beam dump.
This instrumentation relies on photomultiplier tubes (PMTs) to detect the scattered electron or the subsequent shower from the incident beam intercepting the wire.
The data from the wire scanner and PMTs are combined to yield a beam profile with two to three times greater dynamic range than one would obtain using a single photomultiplier or by measuring the induced current on the wire~\cite{Freyberger} as illustrated in Fig.~\ref{fig:current_facilities:completed:cebaf_1pass:large_dynamic_range}.

\subsubsection{RF Measurements}
\begin{figure}[htb]\centering
\tikzsetnextfilename{cebaf_1pass_rf_er}
\begin{tikzpicture}
\begin{axis}
[
    width=0.9\linewidth,
    height=.6\linewidth,
    ymin=-0.15,
    ymax=0.15,
    ylabel={Gradient modulator drive signal (\si{\volt})},
    xlabel={Time (\si{\micro\second})},
    xmin=0,
    xmax=350,
    xmajorgrids=true,
    ymajorgrids=true,
    legend pos=south west,
    y tick label style={/pgf/number format/fixed, /pgf/number format/precision=2},
]
\addplot+[mark=none] table[x index=0, y index=2] {figures/cebaf_1pass_drivesignal.csv}; \addlegendentry{with ER}
\addplot+[mark=none] table[x index=0, y index=1] {figures/cebaf_1pass_drivesignal.csv}; \addlegendentry{without ER}
\end{axis}
\end{tikzpicture}
\caption{Forward RF power for an RF cavity (SL20 cavity 8). The red data is for a single beam propagating through the cavity, whereas the blue data is for in-phase and \ang{180} out-of-phase beams copropagating through the cavity.}\label{fig:current_facilities:completed:cebaf_1pass:rf_er}
\end{figure}
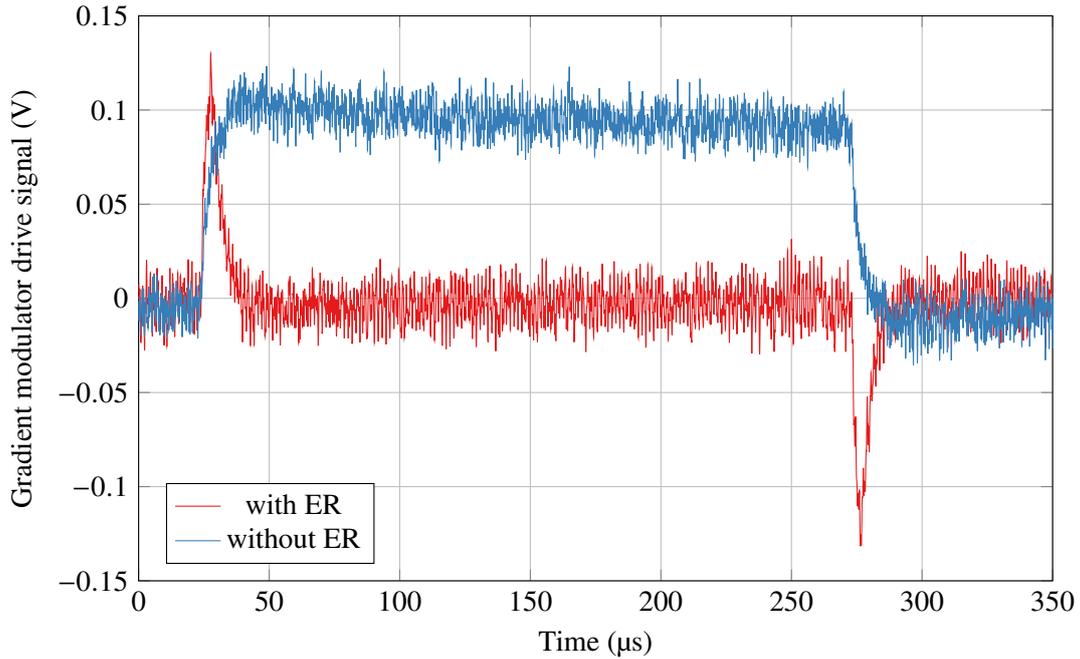

In addition to the beam-based measurements presented in the previous sections, another important class of measurements deals with the RF system’s response to energy recovery.
These measurements are intended to test the system’s response by measuring the gradient and phase stability with and without energy recovery in several cavities throughout the North and South Linac.
As an example, consider Fig.~\ref{fig:current_facilities:completed:cebaf_1pass:rf_er}, which illustrates the RF system gradient modulator drive signal during pulsed beam operation.
Without energy recovery, this signal is nonzero when a \SI{250}{\micro\second} beam pulse enters the RF cavity, indicating power is drawn from the cavity.
With energy recovery, the signal is zero once the initial transient passage of the leading edge of the pulse is over, indicating no additional power draw is required by the cavity.

\subsubsection{Conclusions}
The CEBAF-ER experiment has shown the feasibility of energy recovering a high-energy (\SI{1}{\giga\electronvolt}) beam through a large (about \SI{1}{\kilo\meter} circumference), superconducting accelerator with 39 cryomodules.
In doing so, sufficient operational control of two coupled beams of substantially different energies (up to a factor of 20 difference) was demonstrated in a common transport channel in the presence of steering and focusing errors.
In addition, the dynamic range of the system's performance was tested by demonstrating high final-to-injector energy ratios of 20:1.
With the injector set to provide \SI{55}{\mega\electronvolt} into the linac, \SI{80}{\micro\ampere} of CW beam, accelerated to \SI{1055}{\mega\electronvolt} and energy recovered at \SI{55}{\mega\electronvolt}, was steered to the energy-recovery dump. 
Measurements of the beam phase space show that energy
recovery does not introduce any substantial phase space
degradation. Beam profiles after energy recovery do not
show any significant distortion, and for the \SI{55}{\mega\electronvolt} configuration,
the profiles were close to Gaussian over multiple decades. 

%\subsection{CEBAF Single Pass}
% Alex Bogacz
%
%
%

\section{Ongoing Activities}\label{sec:current_facilities:ongoing}

\subsection{CBETA at Cornell}%
\label{sec:current_facilities:ongoing:cbeta}
% Georg Hoffstatter

% MB  7/15/2021
% Edit for minor language/typesetting oversights, consistent use of siunitx, figure/table labels.
% Merged bibliography into global bibtex file.
% Adopted present tense and passive voice where appropriate.

The Cornell-BNL Test Accelerator (CBETA) \cite{Mayes17} is the first multi-pass SRF accelerator operating in energy-recovery (ER) mode \cite{PhysRevLett.125.044803}, focusing on technologies for reduced energy consumption \cite{PhysRevAccelBeams.24.010101}.
The energy delivered to the beam during the first four passes through the accelerating structure is recovered during four subsequent decelerating passes.
In addition to ER, energy savings are achieved by using superconducting accelerating cavities and permanent magnets.
The permanent magnets are arranged in a  Fixed-Field Alternating-gradient (FFA) optical system to construct a single return loop that successfully transports electron bunches of 42, 78, 114, and \SI{150}{\mega\electronvolt} in one common vacuum chamber.
While beam loss and radiation limits only allowed commissioning at low currents, this new kind of accelerator, an 8-pass energy-recovery linac, has the potential to accelerate much higher current than existing linear accelerators.
Additionally, with its DC photoinjector,  CBETA is designed for high brightness while consuming much less energy per electron.
CBETA has also operated as a one-turn (i.e., two-pass) ERL to measure the recovery efficiency accurately~\cite{PhysRevAccelBeams.24.010101}.

After BNL had initiated ERL research to develop energy-efficient accelerator technology and Cornell had prototyped ERL components for a light source, a mutually beneficial collaboration was formed, BNL obtained funds from NYSERDA, and construction of CBETA at Cornell commenced.
Commissioning was completed in early 2020.
A large number of international collaborators helped during commissioning shifts, making it a joint effort of nearly all laboratories worldwide that pursue ERL technology.
Because recovering beam energy in SRF cavities was first proposed at Cornell~\cite{Tigner:1965wf}, it is pleasing that its first multi-pass system is constructed at the same university.

The FFA beam ERL return loop is also the first of its kind.
It is constructed of permanent magnets of the Halbach type \cite{Brooks:IPAC2017-THPIK007, Brooks:IPAC2019-THPTS088} and can simultaneously transport beams within an energy window that spans nearly a factor of 4, from somewhat below \SI{40}{\mega\electronvolt} to somewhat above \SI{150}{\mega\electronvolt}.
Having only one beamline for 7 different beams at 4 different energies saves construction and operation costs.
The permanent Halbach magnets contain several innovations: they are combined-function magnets, they were fine tuned to \SI{0.01}{\percent} accuracy by automated field shimming, and they provide an adiabatic transition between the arc and straight sections~\cite{Berg:ERL2017-TUIDCC004}.

Table \ref{tab:current_facilities:ongoing:cbeta:machine_settings} shows both the design and multi-turn commissioning parameters.
The commissioning period reported here established multi-turn energy recovery at low currents of about \SI{1}{\nano\ampere}.
A conservative, safe current level was used for equipment and personnel protection: to avoid radiation damage to the permanent magnets and to have an acceptable radiation level in areas adjacent to the accelerator.
A reduced bunch charge of \SI{5}{\pico\coulomb} was also used to avoid particle loss from Coherent Synchrotron Radiation (CSR).
A push to high current will be the next stage of this accelerator.

\begin{table}[!htb] 
\caption{CBETA Machine Parameters}
\begin{center}
\begin{tabular}{lcc}
\toprule
%>{\centering}m{0.8in}|
%>{\centering}m{0.42in}
%>{\centering\arraybackslash}m{0.42in}}
Parameter                   & Value & Units\\
\midrule
Bunch charge, design limit  & 125   & pC \\
Bunch charge, commissioning & 5     & pC \\
Bunch rate, design limit    & 325   & MHz \\
Bunch rate, commissioning   & $<1$  & kHz \\
Beam current, design limit  & 40    & mA \\
Beam current, commissioning & 1     & nA \\
Beam energy, injector       & 6     & MeV \\
Beam energy, peak           & 150   & MeV \\
\bottomrule
\end{tabular}
\label{tab:current_facilities:ongoing:cbeta:machine_settings}
\end{center}
\end{table}

%\section{Layout}

\begin{figure}[htb]
	\centering
		\includegraphics[width=\columnwidth]{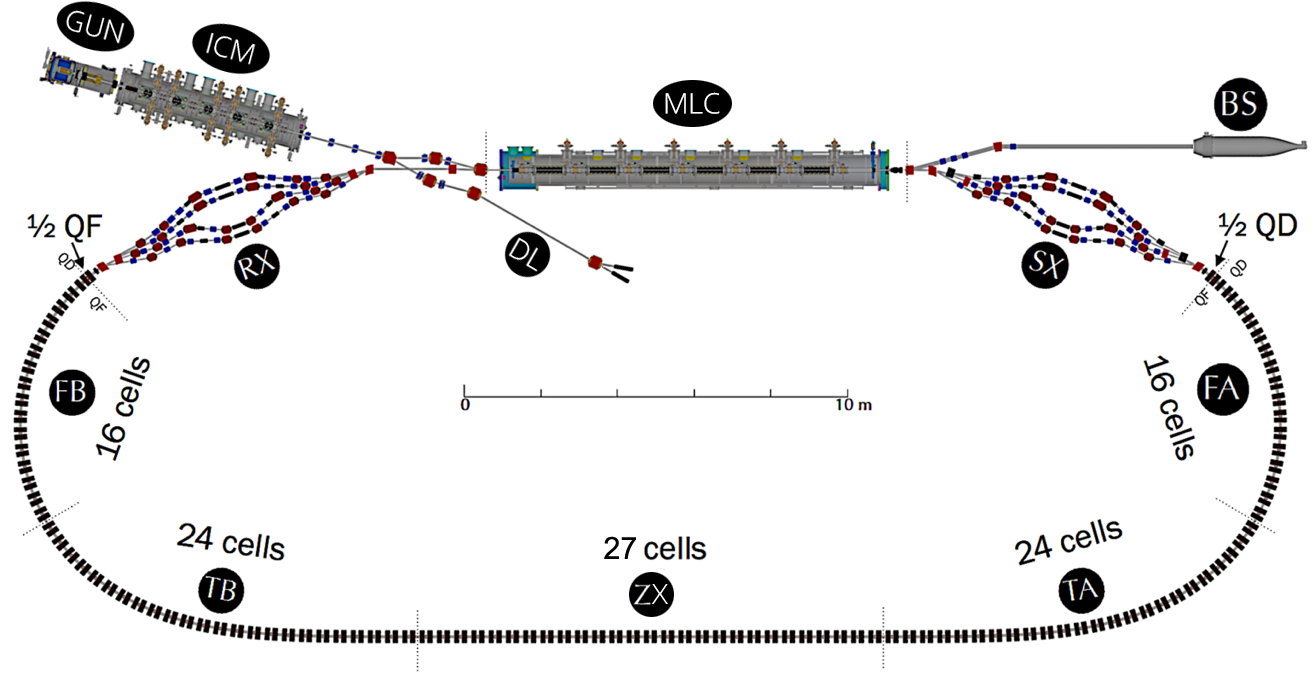}
	\caption{The major components of CBETA are the electron gun (GUN), the Injector Cryomodule (ICM), the Main Linac Cryomodule (MLC), the diagnostic line (DL), the four SX splitter/combiner lines,  the FFA arc consisting of the first arc (FA), first transition (TA), straight section (ZX), second transition (TB), second arc (FB), and the four RX splitter/combiner lines. The fully decelerated beam is absorbed in the beam stop (BS).}
	\label{fig:current_facilities:ongoing:cbeta:layout}
\end{figure}

The technical layout of the CBETA accelerator is shown in Fig.~\ref{fig:current_facilities:ongoing:cbeta:layout}.
The acceleration chain \cite{PhysRevSTAB.16.073401, doi:10.1063/1.4913678, PhysRevSTAB.18.083401} begins with a DC photoelectron gun operated at \SI{300}{\kilo\volt}, a pair of emittance-compensating solenoids, and a normal-conducting buncher cavity.
This is immediately followed by the superconducting injector cryomodule (ICM), accelerating the beam to the target injection energy of \SI{6}{\mega\electronvolt}.
The beam is then steered either to the left through a three-bend achromatic merger into the Main Linac Cryomodule (MLC) cavities, or to the right through a mirrored merger into a set of transverse and longitudinal diagnostics.
The layout of the mirror merger and the position of the diagnostics are chosen such that the bunch can be studied in a location equivalent to the beginning of the first MLC cavity, downstream of which the effects of space charge are greatly reduced.
The MLC is the first cryomodule that was custom designed for ERL applications, i.e., optimized for large beam powers but low input power; it consists of six cavities, providing a total energy gain of \SI{36}{\mega\electronvolt}.
The energy gain and phase of each cavity are not equal; instead, they are chosen to account for non-relativistic effects \cite{PhysRevAccelBeams.22.091602} and to minimize the growth of the energy spread throughout the machine. 

The higher energy beams downstream of the MLC are guided into the four SX ``splitter'' beamlines by a common electromagnet.
These beam lines serve to independently match the optics, the orbit, and the time of flight for each beam energy at the entrance of the return loop.
Each of the four beamlines contains 8 quadrupole magnets, up to 10 dipole magnets, and a motorized path-length adjusting chicane.
All magnets in the splitter beamlines are electromagnets.
The splitter lines feed into the FFA arc, which consists of periodic FODO cells with different periodic optics and orbits for different energies to which the merger beamlines are adjusted.

The path length is adjusted by motorized chicanes that are limited to 10--\ang{20} of RF phase for each beam line.
While all optical elements in the return loop are permanent magnets, each has either a vertical or horizontal dipole corrector with a strength corresponding to \SI{3}{\milli\meter} offset in an FFA quadrupole.

The return loop has three sections: arc, straight, and the transitions between them.
In the arc sections, the beam trajectories for the four energies are spatially separated, with the highest energy on the outside of the arc.
In the transition sections, the four orbits converge adiabatically towards the center of the pipe, and the periodic optical functions change adiabatically into those of the straight section~\cite{Berg:ERL2017-TUIDCC004}.
CBETA is the first accelerator to demonstrate this concept of adiabatic FFA transitions.

Downstream of the FFA, the four beams are separated into the RX splitter lines.
Their trajectory, optical functions, and path lengths again are individually tuned for further passes through the MLC.
Finally, the energy-recovered \SI{6}{\mega\electronvolt} beam is guided to the beam stop.

\begin{figure}[htb] 
\centering
    \begin{center}
         \subfigure[Horizontal FFA orbits]{%
            \label{fig:current_facilities:ongoing:cbeta:H_orbits}
            \includegraphics[width=0.48\columnwidth]{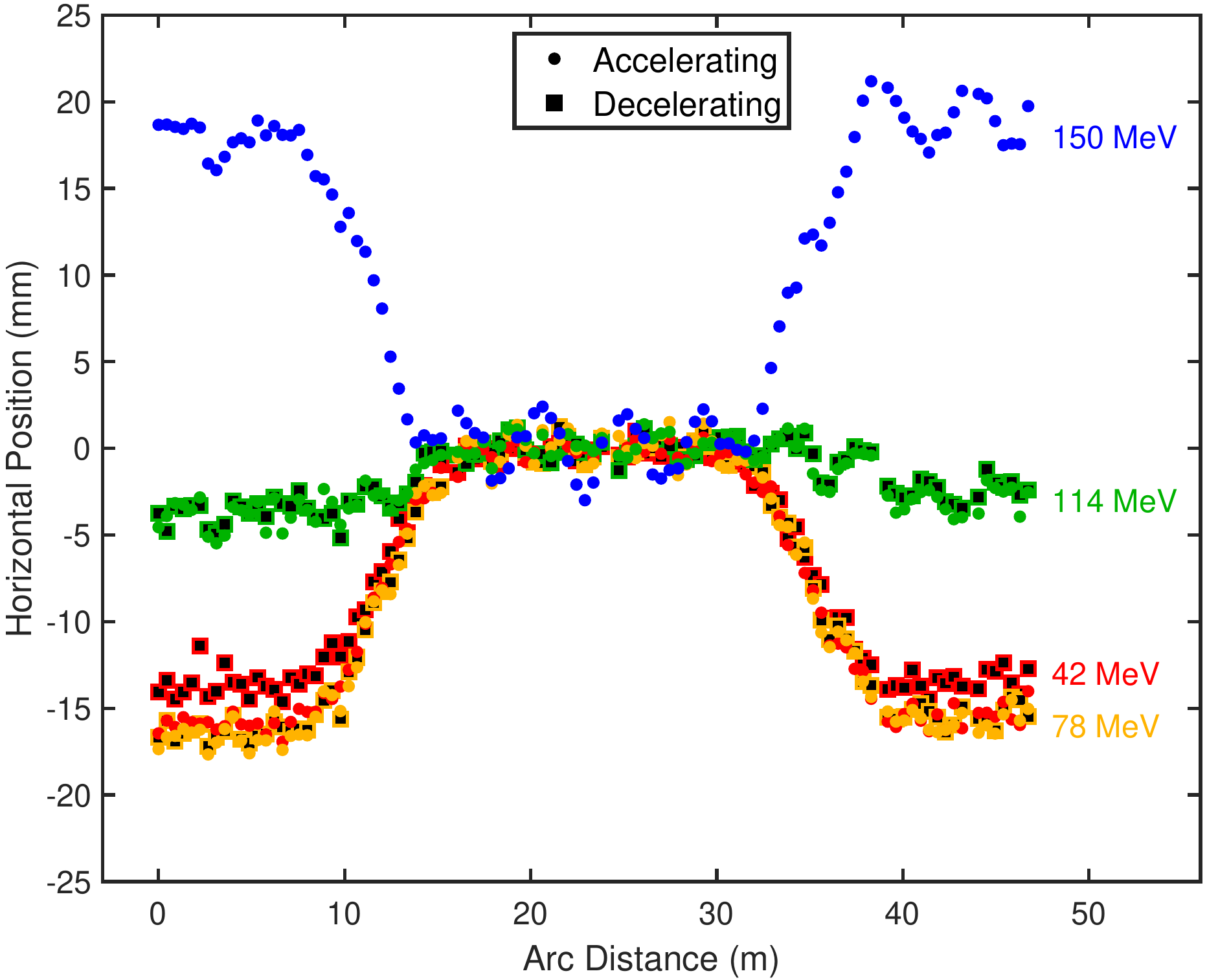}
        }%
        %\subfigure[Vertical FFA orbits]{%
        %   \label{fig:current_facilities:ongoing:cbeta:V_orbits}
        %   \includegraphics[width=0.5\columnwidth]{figs/v_orbits.pdf}
        %}\\
        \subfigure[Beam arrival phase]{%
           \label{fig:current_facilities:ongoing:cbeta:phases}
           \includegraphics[width=0.48\columnwidth]{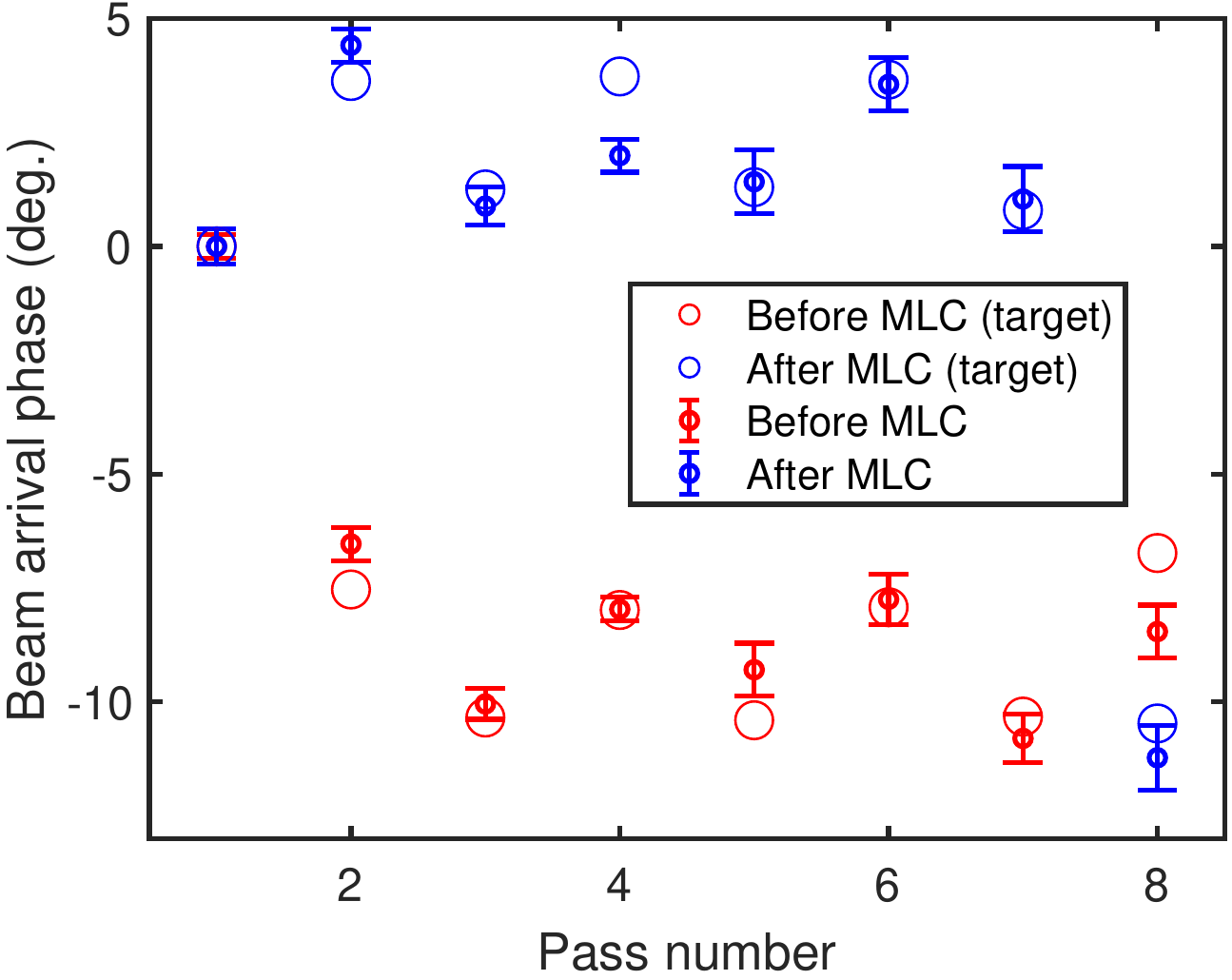}
        }
        \caption{Measured orbits through the FFA and arrival time before and after the MLC. Arrival time is shown in units of RF phase, and the phases of the decelerating passes are shown relative to \ang{180}.
        }
        \label{fig:current_facilities:ongoing:cbeta:orbits}
    \end{center}
\end{figure}

Figure~\ref{fig:current_facilities:ongoing:cbeta:H_orbits} %and \ref{fig:V_orbits} 
shows the measured beam orbits through the common FFA loop, demonstrating that the design trajectories were achieved.
Because beam position monitors (BPMs) are placed in periodic positions in the FFA loop, the design orbit maintains periodic values in the arc and straight sections, with adiabatic transitions between them. 
The beam arrival phases at the entrance and exit of the MLC are shown in Fig.~\ref{fig:current_facilities:ongoing:cbeta:phases} and compared to the target values from simulation.
All phases are shown with respect to their values from the first pass through the MLC, with the sign chosen such that negative phases indicate a later arrival time.
Compared to the first pass, higher passes show a systematically later arrival at the BPM before the MLC because the first-pass beam is the slower beam from the injector.
On top of that systematic offset, each pass is intentionally alternated slightly positive or negative to prevent growth in energy spread while maintaining energy balance.
To show that the full energy was recovered for each particle, the output energy was accurately adjusted to \SI{6}{\mega\electronvolt}.

\begin{figure}[htb] 
\centering
    \begin{center}
         \subfigure{%
            \label{fig:current_facilities:ongoing:cbeta:arc_tunes}
            \includegraphics[width=0.48\columnwidth]{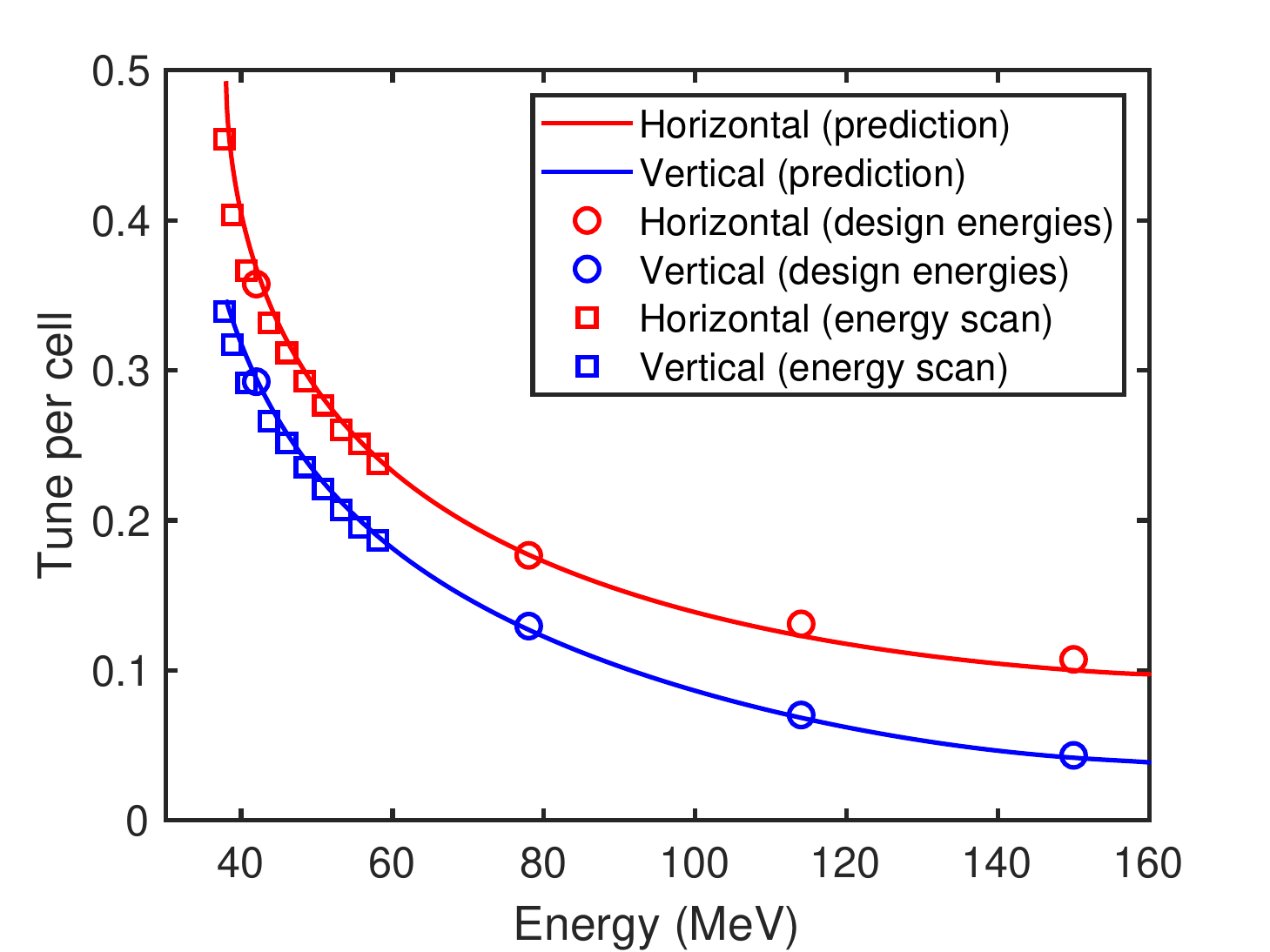}
        }%
        \subfigure{%
           \label{fig:current_facilities:ongoing:cbeta:straight_tunes}
           \includegraphics[width=0.48\columnwidth]{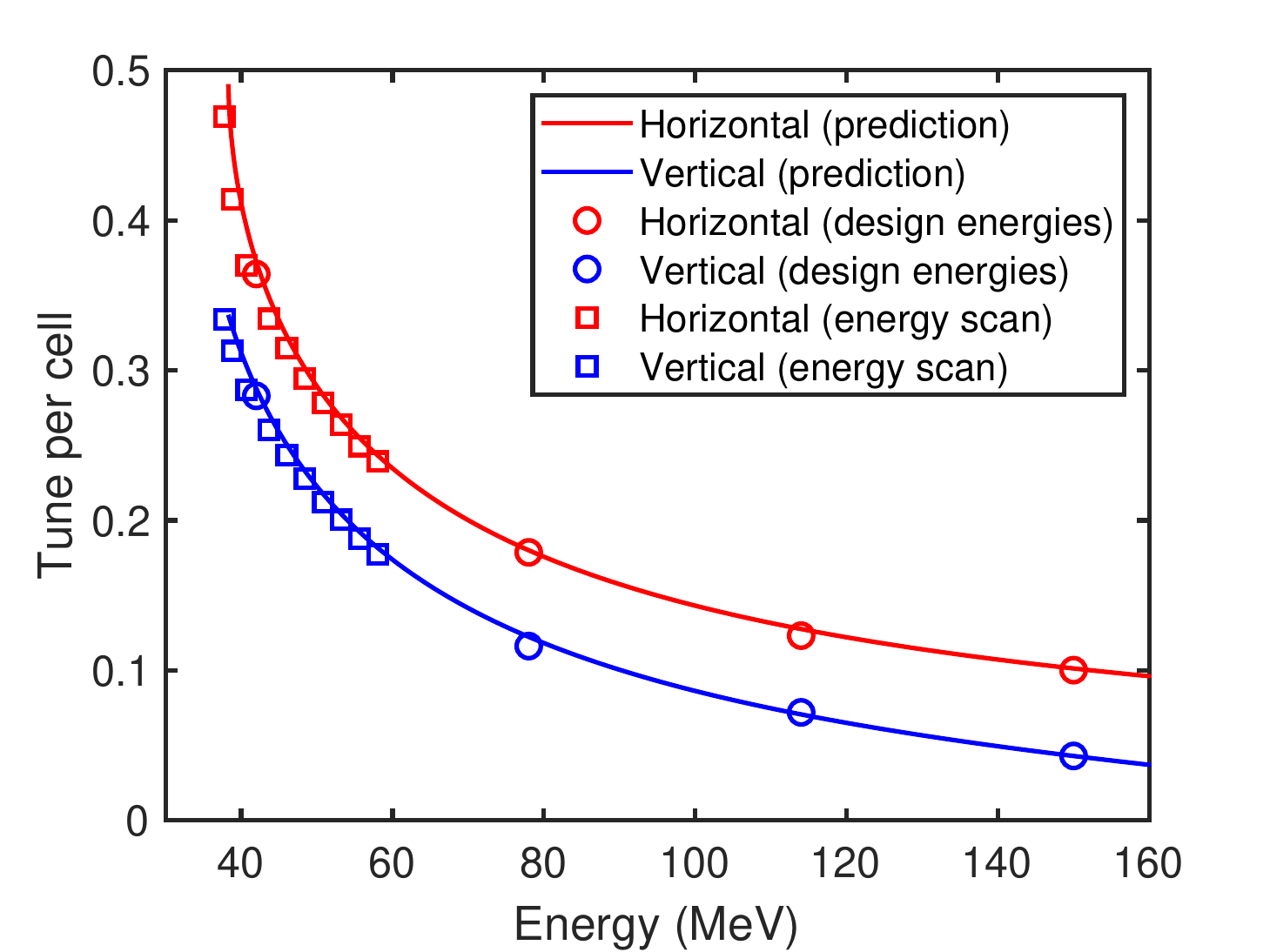}
        }
    \end{center}
    \caption{Measured tunes in the FFA arc sections (left) and straight section (right) as a function of beam energy. The lines show the result of a field-map-based model calculation.}
    \label{fig:current_facilities:ongoing:cbeta:tunes}
\end{figure}

To verify the optics models for the FFA section, Figure~\ref{fig:current_facilities:ongoing:cbeta:tunes} compares the measured and designed phase advance per FODO cell (tunes) as a function of the beam energy, showing that the orbits, RF phases, and optical functions are all as designed.
Tunes were measured by fitting sine functions to difference orbits.
In addition to the four design energies, a scan of the first-pass energy was performed from \SIrange{39}{60}{\mega\electronvolt} by varying the energy gain in the MLC.
The phase advances agree well with simulated predictions from field-map-based particle tracking.

\begin{figure}[!htbp] 
\centering
    \begin{center}
         \subfigure{%
            \label{fig:current_facilities:ongoing:cbeta:dump_beam}
            \includegraphics[width=0.45\columnwidth]{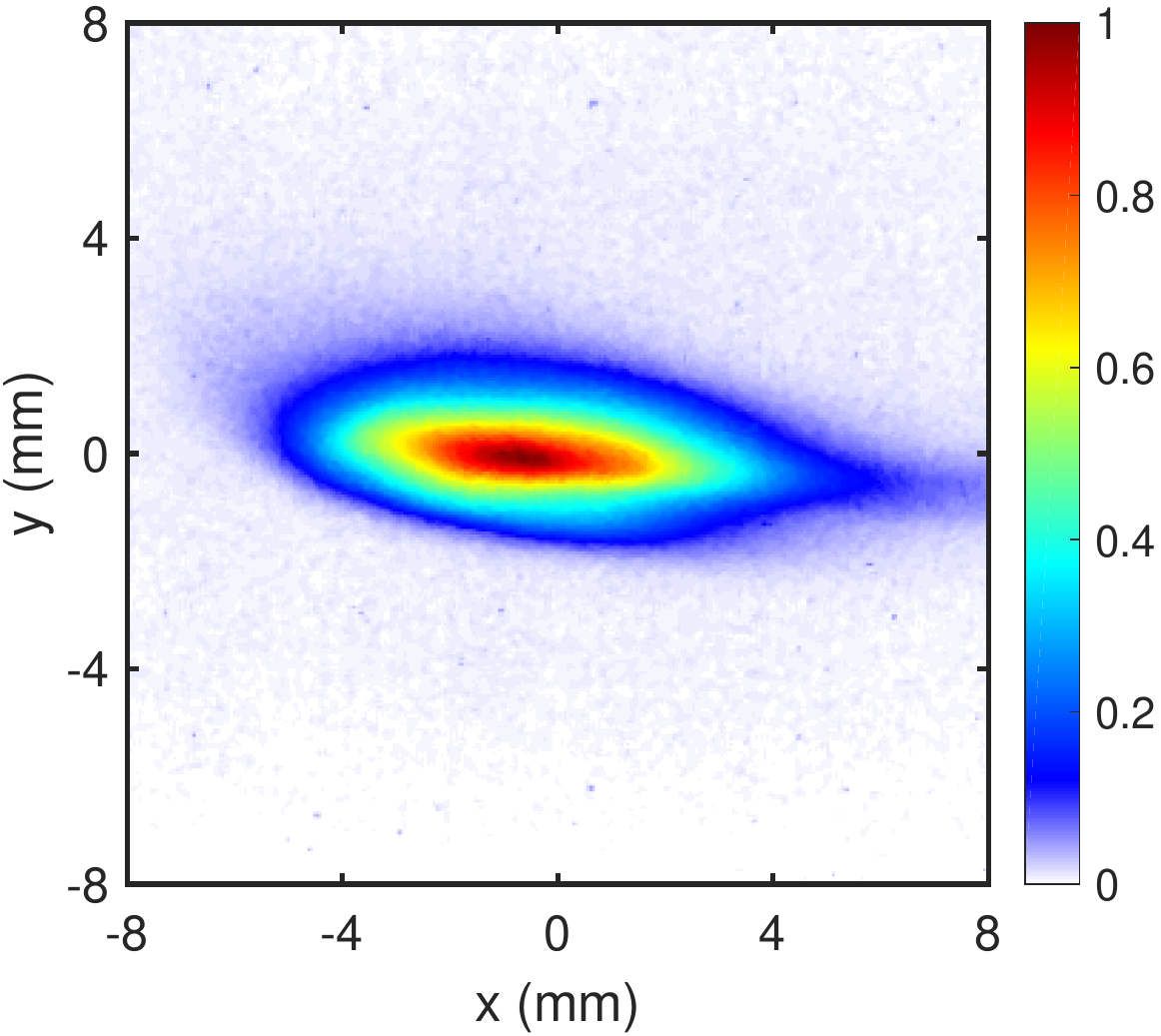}
        }%
        \subfigure{%
           \label{fig:current_facilities:ongoing:cbeta:transmission}
           \includegraphics[width=0.51\columnwidth]{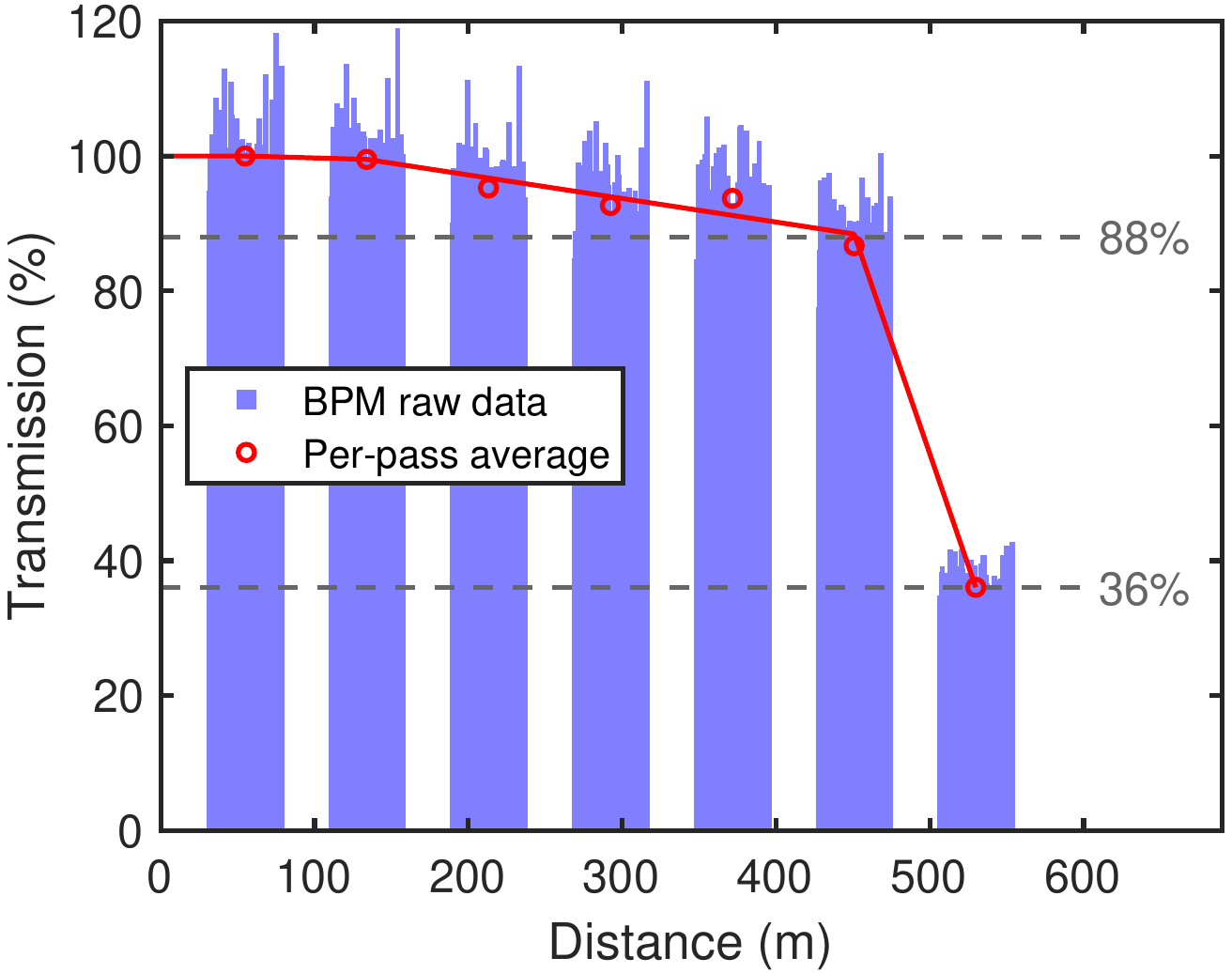}
        }
    \end{center}
    \caption{\label{fig:current_facilities:ongoing:cbeta:primary_results}%
(Left) Image of the beam on the view screen before the beam stop. (Right) Transmission for each of the seven passes through the FFA arc. Blue bars are a scaled reading of charge from individual BPMs, red circles are an average of that data over each pass. Red lines are included to guide the eye.}
\end{figure}

The primary result of the CBETA commissioning period can be summarized with Fig.~\ref{fig:current_facilities:ongoing:cbeta:dump_beam}, the image of the beam on the view screen in the beam stop beamline after 8 passes through the MLC.
The beam energy at this screen was measured to be the same as the injection energy (\SI{6}{\mega\electronvolt}), demonstrating that each particle on the view screen had its energy completely recovered.

While trajectories, RF phases, and optics propagation in the FFA are close to the design, and while each particle arriving at the beam stop has its energy recovered, not all particles made it through all eight passes.
Figure~\ref{fig:current_facilities:ongoing:cbeta:dump_beam} shows an image of the remaining beam on the view screen at the entrance to the beam stop, and Fig.~\ref{fig:current_facilities:ongoing:cbeta:transmission} shows a measure of the transmission throughout the machine. 

The team were able to recover the energy of about one third of the beam, the largest part of the loss occurring after the sixth pass.
The data suggest a slow loss of transmission, beginning as early as the second pass, accumulating to around \SI{10}{\percent} of total loss by the end of the sixth pass through the FFA, followed by a much larger drop in transmission before entering the seventh pass. 

Investigations into the source of these losses uncovered many small problems in optics settings, nonlinear stray fields, evidence of micro-bunching, and others; but these issues have not been fully investigated yet. Improving the transmission is the next focus of run plans for CBETA.

Beam dynamics in CBETA naturally splits into two separate regions.
In the low-energy injector, space-charge effects dominate, and obtaining the best beam quality relies on compensating their emittance-diluting effects.
The primary diagnostic is thus the Emittance Measurement System (EMS)~\cite{PhysRevSTAB.11.100703, heng} in the diagnostic line (DL).
Space-charge effects are relatively minor following the first-pass acceleration through the MLC.
Commissioning this section focuses on achieving the desired orbit, energy, and dispersion through the rest of the machine. 

For characterization of the injector beam, the beam is diverted into the DL, which is comprised of a suite of diagnostics, including the EMS, a vertical deflecting cavity~\cite{BELOMESTNYKH2010179}, and an energy spectrometer (dipole magnet) for measuring the longitudinal phase space of the beam.

To determine suitable machine settings for the low-energy injector, a Multi-Objective Genetic Algorithm optimization (MOGA) \cite{PhysRevSTAB.16.073401, doi:10.1063/1.4913678, PhysRevSTAB.18.083401}, is applied to 3D space-charge simulations of the beam passing through injector, merger, and MLC.

\begin{figure}[!htbp] 
\centering
    \begin{center}
         \subfigure{%
            \label{fig:current_facilities:ongoing:cbeta:meas_x_phase}
            \includegraphics[width=0.48\columnwidth]{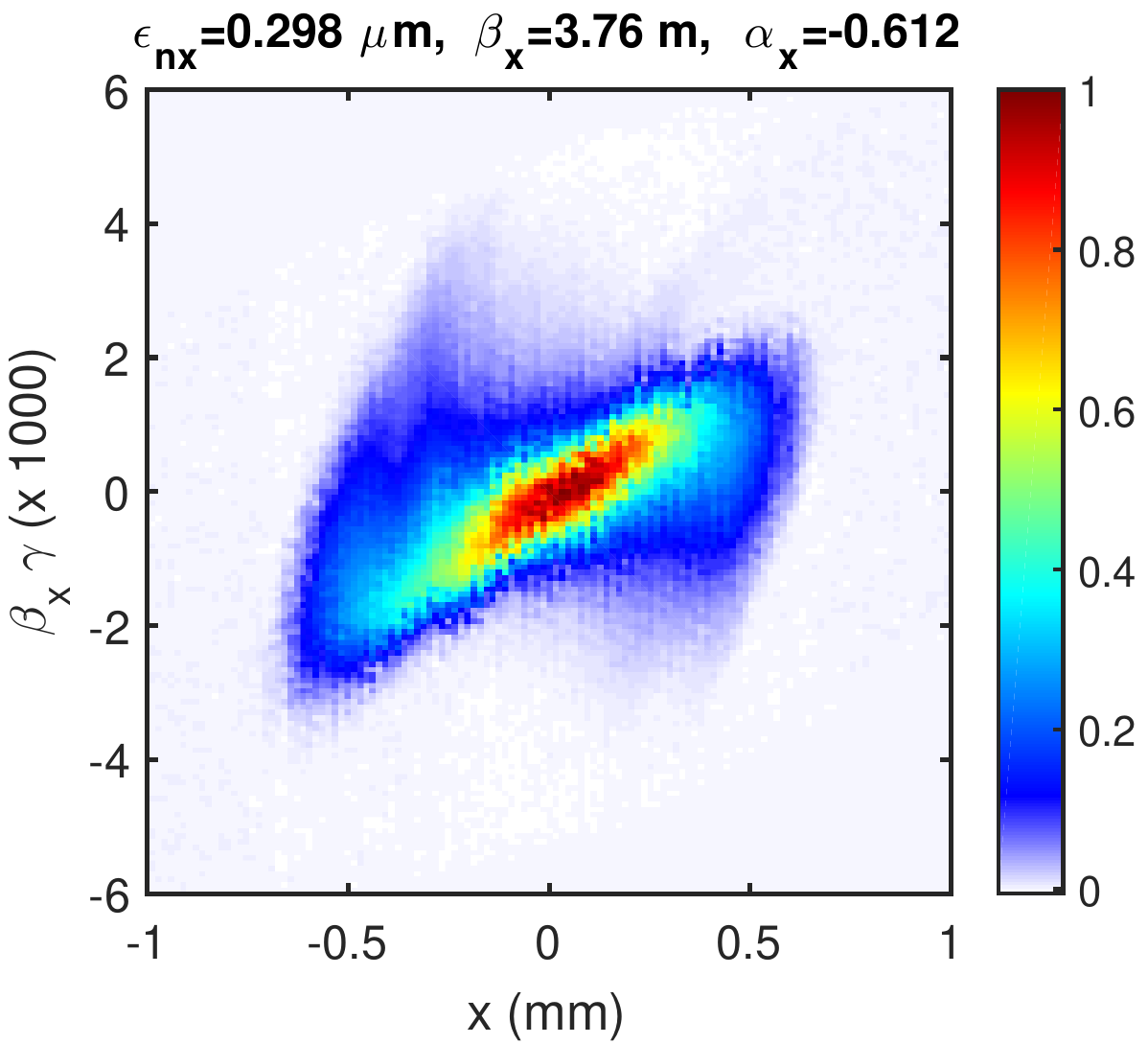}
        }%
        \subfigure{%
            \label{fig:current_facilities:ongoing:cbeta:sim_x_phase}
            \includegraphics[width=0.48\columnwidth]{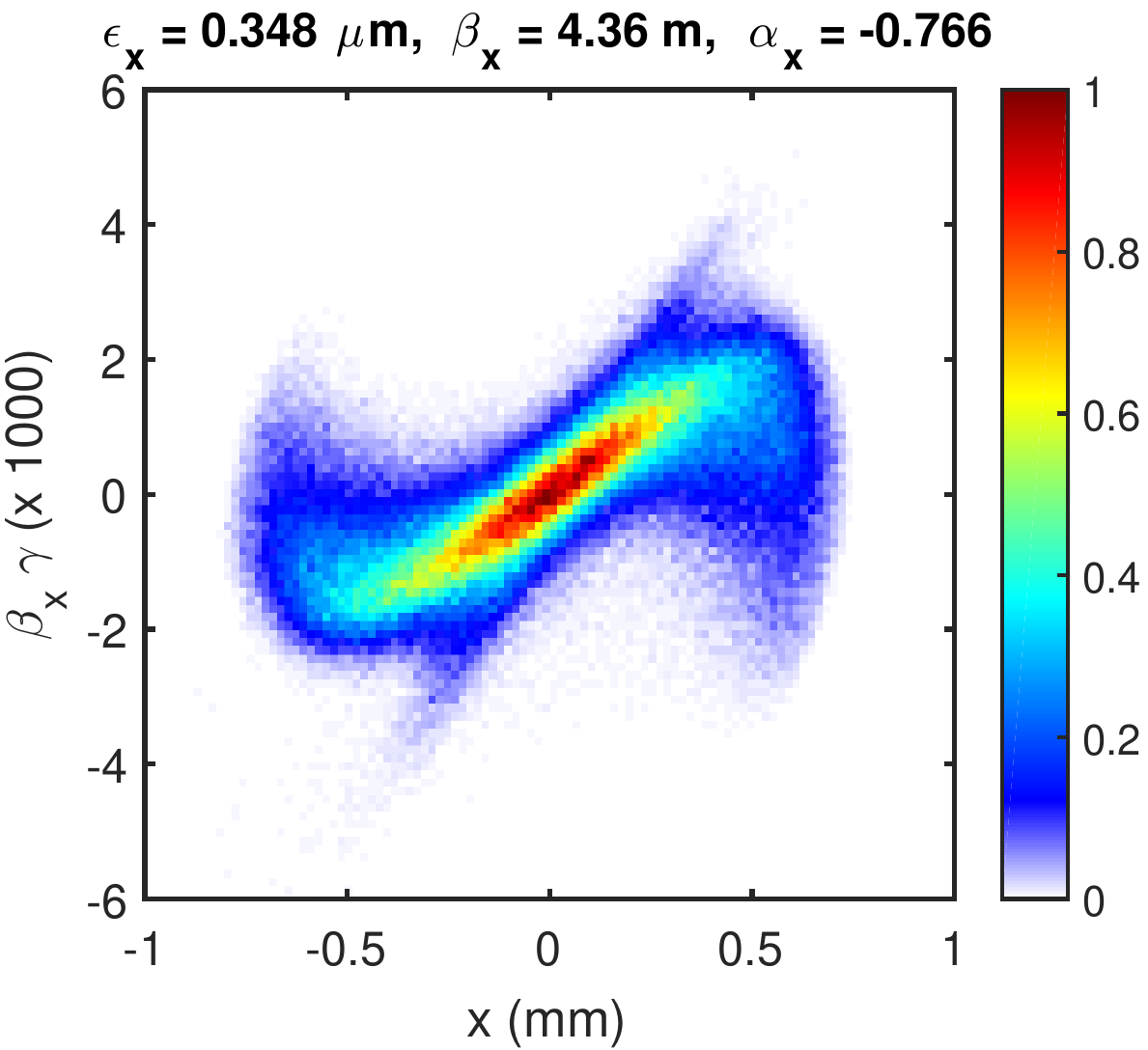}
        }
        %\subfigure[Measured y phase space.]{%
        %   \label{fig:current_facilities:ongoing:cbeta:meas_y_phase}
        %   \includegraphics[width=0.5\columnwidth]{figs/meas_y_phase.pdf}
        %}%
        %\subfigure[Simulated y phase space.]{%
        %   \label{fig:current_facilities:ongoing:cbeta:sim_y_phase}
        %   \includegraphics[width=0.5\columnwidth]{figs/sim_y_phase.pdf}
        %}
        %\subfigure[Measured transverse profile.]{%
        %   \label{fig:current_facilities:ongoing:cbeta:meas_profile}
        %   \includegraphics[width=0.3\textwidth]{figs/meas_profile.pdf}
        %}\\ %  ------- End of the first row ----------------------%
        %
        %\subfigure[Simulated transverse profile.]{%
        %   \label{fig:current_facilities:ongoing:cbeta:sim_profile}
        %   \includegraphics[width=0.3\textwidth]{figs/sim_profile.pdf}
        %}
    \end{center}
    \caption{\label{fig:current_facilities:ongoing:cbeta:injector_phase_space}%
Measured (left) and simulated (right) transverse phase spaces after the injector at 5 pC.}%  and beam profiles.
\end{figure}

Figure~\ref{fig:current_facilities:ongoing:cbeta:injector_phase_space} shows the measured horizontal phase space and the corresponding results of simulations at the operating setting.

\begin{figure}[!htbp] 
\centering
    \begin{center}
        %\subfigure{%
        %   \label{fig:current_facilities:ongoing:cbeta:three_pass_uncorrected_h}
        %   \includegraphics[width=0.5\columnwidth]{figs/3pass_uncorr_h.pdf}
        %}%
        %\subfigure{%
        %   \label{fig:current_facilities:ongoing:cbeta:three_pass_corrected_h}
        %   \includegraphics[width=0.5\columnwidth]{figs/3pass_corr_h.pdf}
        %}\\
        \subfigure{%
            \label{fig:current_facilities:ongoing:cbeta:three_pass_uncorrected_v}
            \includegraphics[width=0.48\columnwidth]{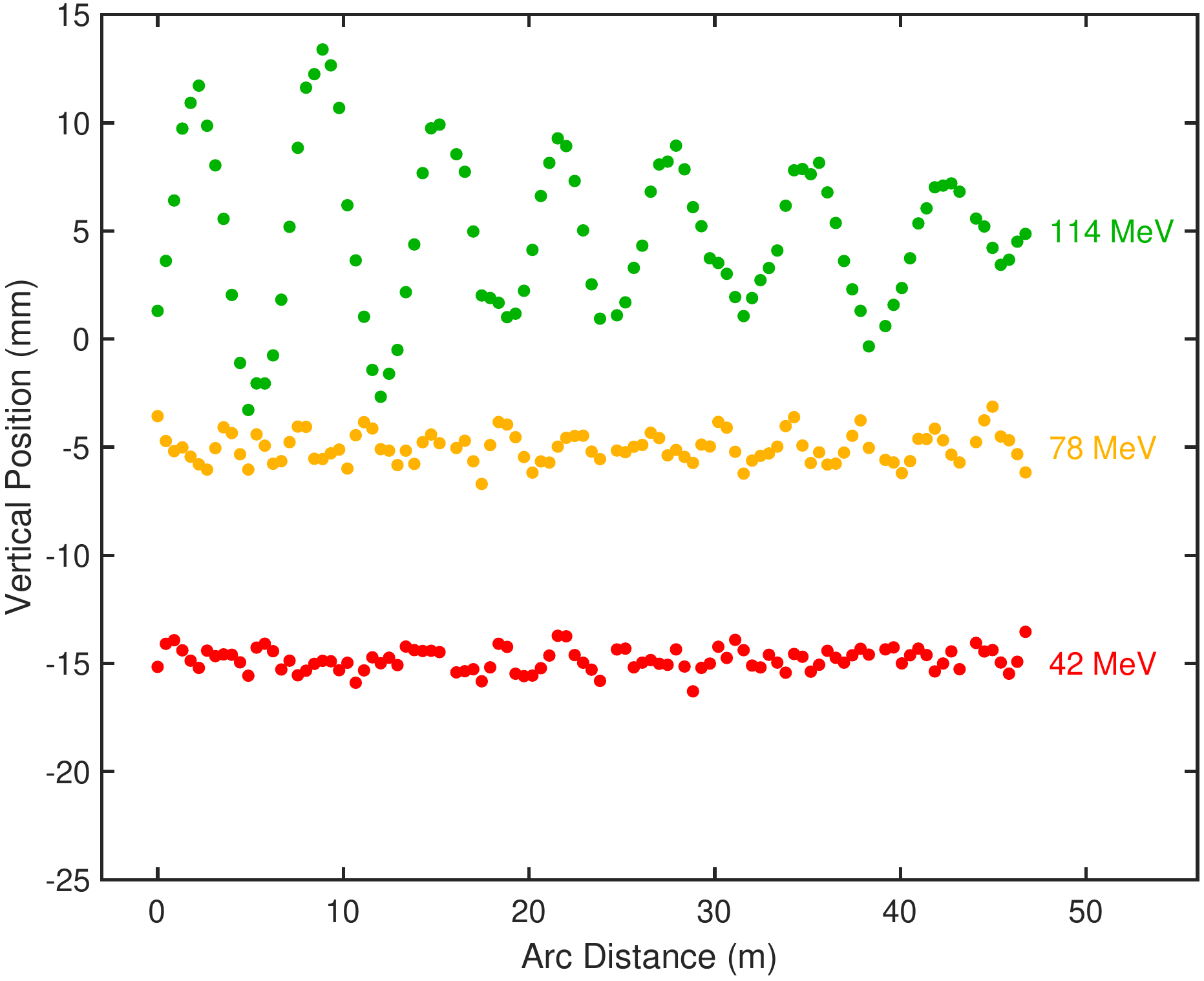}
        }%
        \subfigure{%
           \label{fig:current_facilities:ongoing:cbeta:three_pass_corrected_v}
           \includegraphics[width=0.48\columnwidth]{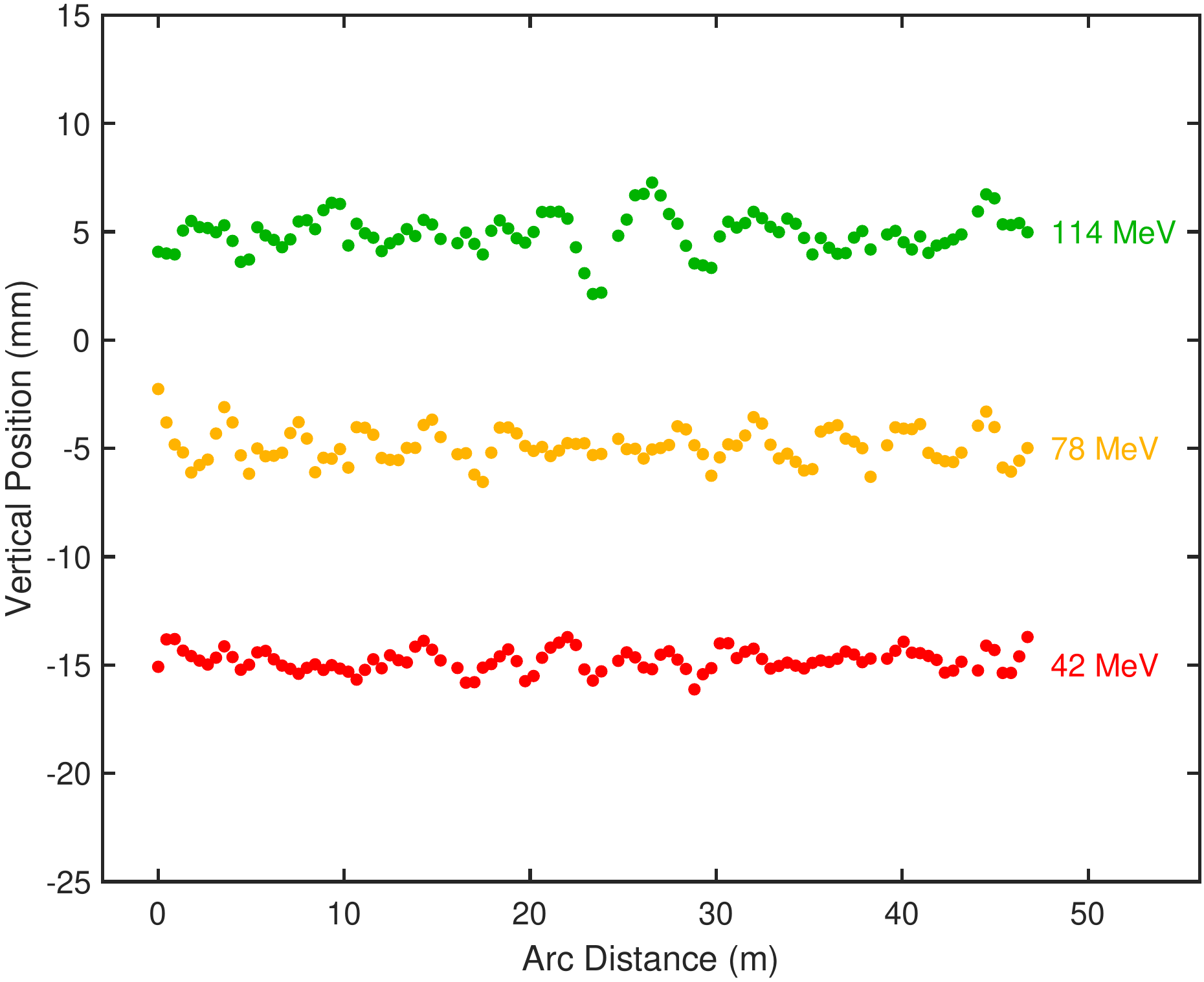}
        }
    \end{center}
    \caption{Measured vertical orbits of the first three passes through the FFA return loop before (left) and after (right) application of a simultaneous orbit correction algorithm. Orbits are offset for clarity.}
    \label{fig:current_facilities:ongoing:cbeta:three_pass_correction}
\end{figure}

Orbit correction for 7 simultaneous beams at 4 different energies in the same beam transport is not trivial.
The orbit correction methods therefore differ in each section of CBETA.
In the MLC, the beams are centered in the RF cavities; in the splitter sections, the beam is centered in the quadrupoles.
Particularly in the FFA return loop, the orbit correction is unconventional since the corrector coils act on all beams.
In general, the orbit correction uses a Singular-Value Decomposition (SVD) algorithm where the RMS orbit deviation of all beams is minimized using either a predicted or measured response matrix.
Predicted responses were effective only for single beams over short distances, with measured responses required for more complex corrections.
As an example, the orbits in the FFA are shown before and after simultaneously correcting the first three passes in Fig.~\ref{fig:current_facilities:ongoing:cbeta:three_pass_correction}.

In the splitters, BPMs were used to directly measure the orbit and path length, allowing the correction of these quantities.
In order to correct the beam optics functions, direct measurements are limited to the dispersion functions and the transport matrix element $R_{56}$, which describes the energy dependence of the position and arrival time of the beam.
Correcting these by adjusting quadrupoles also corrects other optical functions, which were checked by measuring the orbit response to magnet changes.
These procedures can only be used reliably for the accelerating pass because during deceleration, each splitter magnet (except the one at highest energy) is traversed by two beams, which simultaneously react to magnet changes.
For the decelerating passes, a manual, empirical tuning approach was used to maximize transmission into the beam stop. 
The next step is to improve transmission, which includes investigating better optics solutions, developing improved diagnostics for the decelerating passes, and reducing halo by using a low-halo cathode, possibly in conjunction with beam collimation.
% \subsection{CBETA at Cornell}
% Georg Hoffstatter

\subsection{S-DALINAC at Darmstadt}
%Michaela Arnold, Norbert Pietralla 

% MB 11/30/2021: Rough text edit
% MB 12/01/2021: Replaced beam loading figure with TikZ

    \subsubsection{Introduction}

\begin{figure}[htb]
    	    \centering
    	    \includegraphics[width=\linewidth]{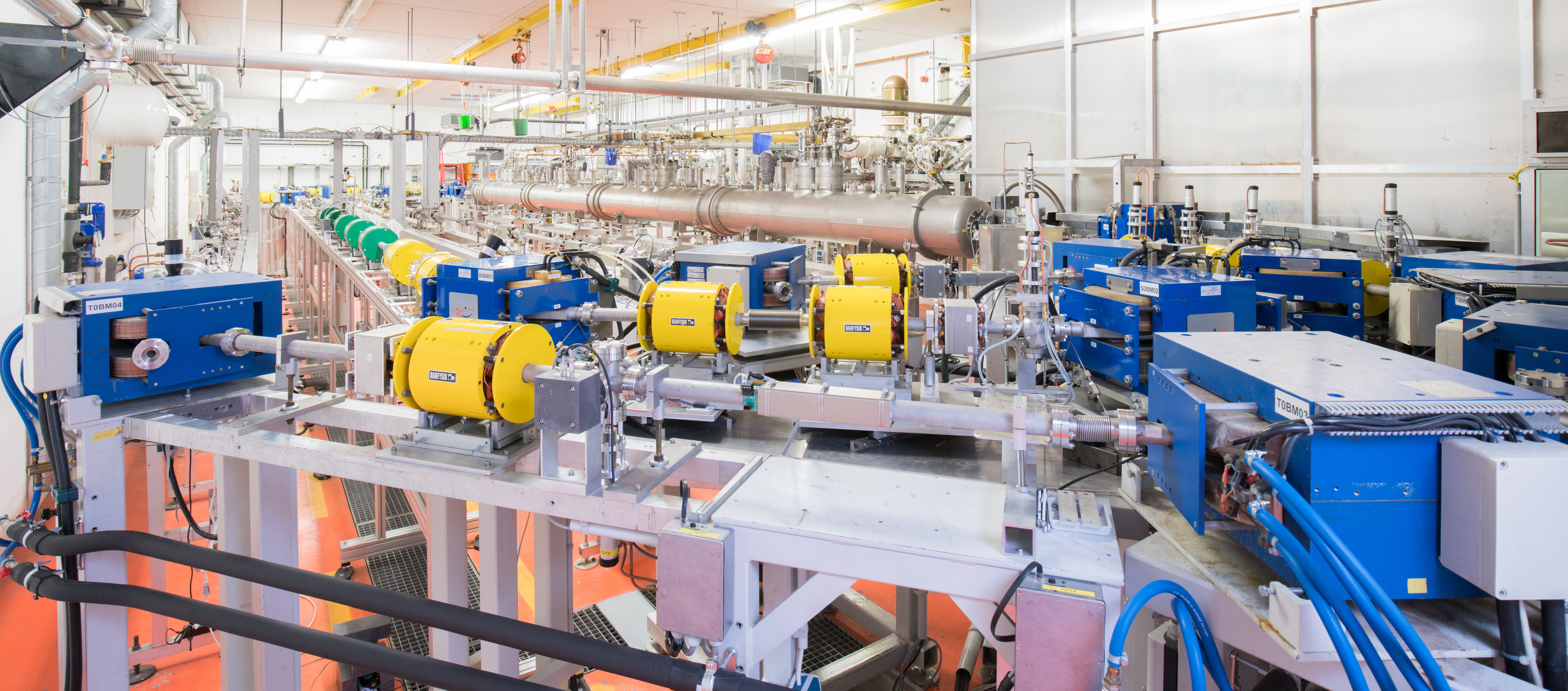}
    		\caption{Photograph of the \mbox{S-DALINAC} with its three recirculation beam lines (by Jan-Christoph Hartung).}
    		\label{fig:S-DALINAC_overview}
    	\end{figure}

\begin{figure}[htb]
    	    \centering
    		\includegraphics[width=\linewidth]{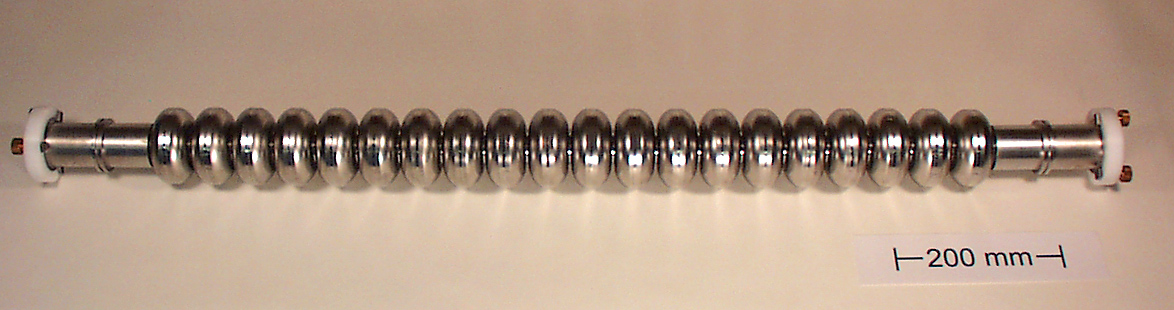}
    		\caption{Photograph of a 2.997\,GHz, 20-cell, $\beta=1$, niobium SRF cavity used in the \mbox{S-DALINAC}.}
    		\label{fig:S-DALINAC_cavity}
    	\end{figure}

\begin{figure}[htb]
        	\centering
        	\includegraphics[scale=0.4]{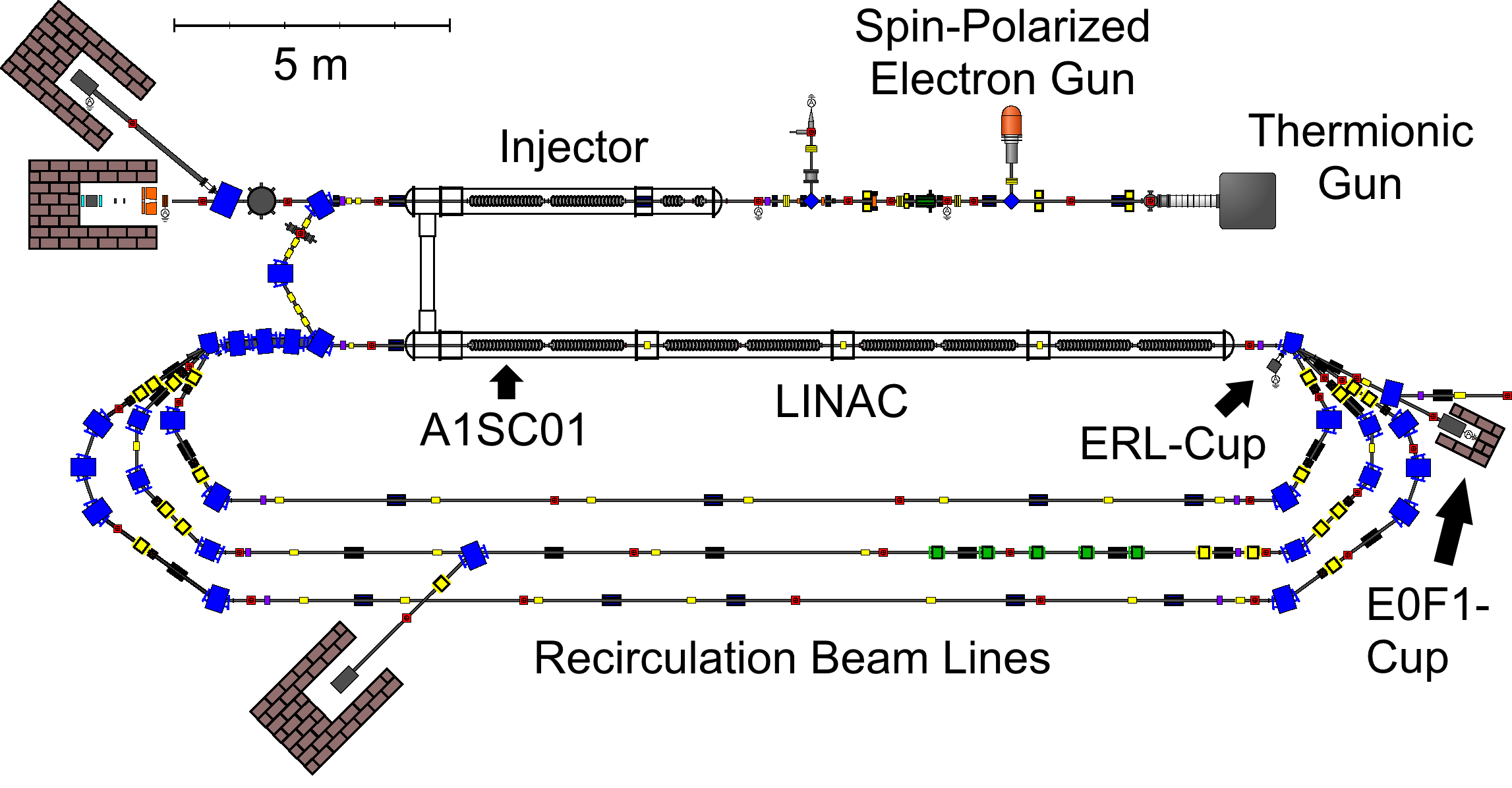}
        	\caption{Schematic floor plan of the \mbox{S-DALINAC}. It is operated in CW mode at 2.997\,GHz. The injector can accelerate the beam up to 10\,MeV (up to 7.6\,MeV is used for recirculating operation). The main accelerator can provide an energy gain of up to 30.4\,MeV. At maximum, an electron energy of 130\,MeV is possible in the conventional accelerating scheme. In ERL mode, energies of up to 68.4\,MeV (two acceleration passages through the main linac) or 34.2\,MeV (one acceleration passage) are feasible by design. In all recirculating modes, a time-averaged electron-beam current of up to \SI{20}{\micro\ampere} can be provided.}
        	\label{fig:S-DALINAC_floorplan}
        \end{figure}

        The superconducting Darmstadt electron linear accelerator (\mbox{S-DALINAC}) has been in operation at Technische Universität Darmstadt since 1991 \cite{Pietralla18}. It was initially built as a twice-recirculating machine. In 2015/2016, a new recirculation beam line was installed, allowing for a thrice-recirculating operation as well as for the operation as an ERL. In August 2017, the once-recirculating ERL operation was demonstrated \cite{Arnold:2020ubw}. Twice-recirculating ERL mode was achieved in August 2021 \cite{Schliessmann-prep}. Figures~\ref{fig:S-DALINAC_overview} and \ref{fig:S-DALINAC_cavity} show photographs of the current machine and one of its accelerating cavities, respectively, and Figure~\ref{fig:S-DALINAC_floorplan} gives an overview of the floor plan.
 

\begin{figure}
    	    \centering
    		\subfigure[Once-recirculating ERL mode.]{\includegraphics[width=.4\linewidth]{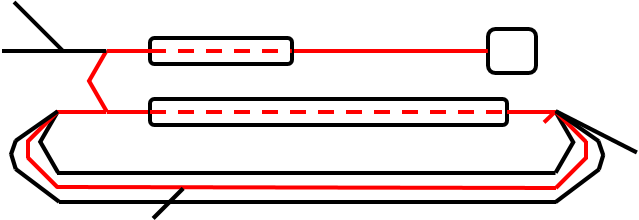}}\hspace{1cm}
    		\subfigure[Twice-recirculating ERL mode.]{\includegraphics[width=.4\linewidth]{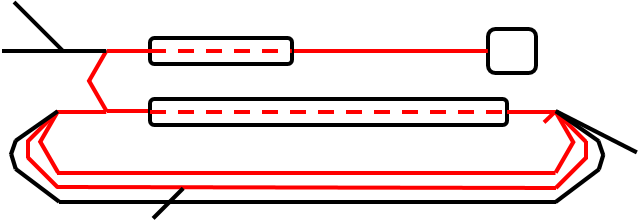}}
    		\caption{The lattice of the \mbox{S-DALINAC} is capable of once- or twice-recirculating ERL operation. The \ang{180} phase shift is applied in the second recirculation beam line.}
    		\label{fig:ERL_schemes_S-DALINAC}
    	\end{figure}

        The \mbox{S-DALINAC} has quite a versatile lattice.
        All recirculation beam lines include path-length adjustment systems.
        The path lengths of the recirculation beam lines can be changed by remotely adjusting the positions of the dipole magnets and the quadrupole magnets in the recirculation arcs, allowing for phase shifts of up to \ang{265} (first recirculation), \ang{360} (second), and \ang{205} (third) with respect to the accelerating phase on re-entry of the beam into the main linac. The following operation schemes are possible at the S-DALINAC:
        \begin{itemize}
        	\item Injector operation
        	\item Single-pass mode (one passage through the main linac and extraction to the experimental hall)
        	\item Once-recirculating mode (two passages through the main linac and extraction to the experimental hall)
        	\item Thrice-recirculating mode (four passages through the main linac and extraction to the experimental hall)		
        	\item Once-recirculating ERL mode (one accelerating and one decelerating passage through the main linac)
        	\item Twice-recirculating ERL mode (two accelerating and two decelerating passages through the main linac)
        \end{itemize}
        The beam path for both ERL modes is depicted in Fig.~\ref{fig:ERL_schemes_S-DALINAC}.

%    \FloatBarrier

    \subsubsection{Once-Recirculating ERL mode}

In August 2017, the once-recirculating ERL mode was first achieved \cite{Arnold:2020ubw} with the parameters shown in Table~\ref{tab:ERLparameters_S-DALINAC}. The injection energy was very low in this setting. For the first main accelerator cavity, this resulted in a combination of beams with $\gamma\approx5$ and with $\gamma\approx44$ after acceleration and on the way to deceleration, respectively.
A similar situation existed in the last main accelerator cavity. The combination of time-of-flight effects and phase slippage in the 20-cell SRF cavities, designed for ultrarelativistic particle velocities of $\beta=1$, was compensated by an additional \ang{6} detuning of the path-length adjustment system.
The RF-recovery effect ${\cal E}_{\text{RF}}$ \cite{Arnold:2020ubw}, the comparison of the RF beam loading for the two cases of either conventional operation or energy-recovery operation, was measured in the first main accelerator cavity.
In the meantime, a new RF power measurement system was installed and commissioned, capable of measuring all RF powers simultaneously \cite{Steinhorst21}. 
        
        \begin{table}[tb]\centering
            \caption{Main parameters of the once-recirculating ERL operation  \cite{Arnold:2020ubw}. The recovered RF power was measured in the cavity A1SC01 (see Fig.~\ref{fig:S-DALINAC_floorplan}).}
            \label{tab:ERLparameters_S-DALINAC}
            \begin{tabular}{lc}
                \toprule
                    Parameter & Value \\
                \midrule
                    Energy gain injector (setpoint) & \SI{2.5}{MeV} \\
                    Energy gain linac (setpoint) & \SI{20.0}{MeV} \\
                    Current (before injector, setpoint) & \SI{1.2}{\micro A} \\
                    Total change in phase (setpoint) & \SI{186}{\degree} \\
                    RF-recovery effect ${\cal E}_{\text{RF}}$ (measured) & \SI{90.1 +- 0.3}{\percent} \\
                \bottomrule
            \end{tabular}
        \end{table}        
        
        The measurement was separated into four different settings:
        \begin{enumerate}
            \item No beam in the main accelerator.
            \item Single pass: one beam is accelerated in the main accelerator.
            \item Once-recirculating mode: two beams are accelerated in the main accelerator.
            \item ERL mode: one beam is accelerated, another beam is decelerated in the main accelerator.
        \end{enumerate}
        The forward power and reverse power of the first main accelerator cavity as well as the beam current on the corresponding beam dumps were monitored. Table~\ref{tab:meanpowers_S-DALINAC} shows the mean values of the measured beam powers.
        The beam loading vanishes nearly completely in the case of ERL operation.  

        \begin{table}[htb]\centering
            \caption{Mean values of the beam power measured in the main-linac cavity A1SC01 for four different settings. The quoted uncertainties reflect the widths of the distributions of measured values \cite{Arnold:2020ubw}.}
            \label{tab:meanpowers_S-DALINAC}
            \begin{tabular}{lc}
            \toprule
        			Operation & Mean Beam Power (W)\\
        	\midrule
        			No Beam & $0.00 \pm 0.01$\\
        			ERL (acc. + dec.) & $0.45\pm 0.03$\\
        			One Beam (acc.) & $4.51 \pm 0.16$\\ 
        			Two Beams (acc.~+ acc.) & $8.59 \pm 0.01$\\ 
            \bottomrule
            \end{tabular}
        \end{table}
        
\begin{figure}[htb]
        	\centering
        	\includegraphics[scale=0.35]{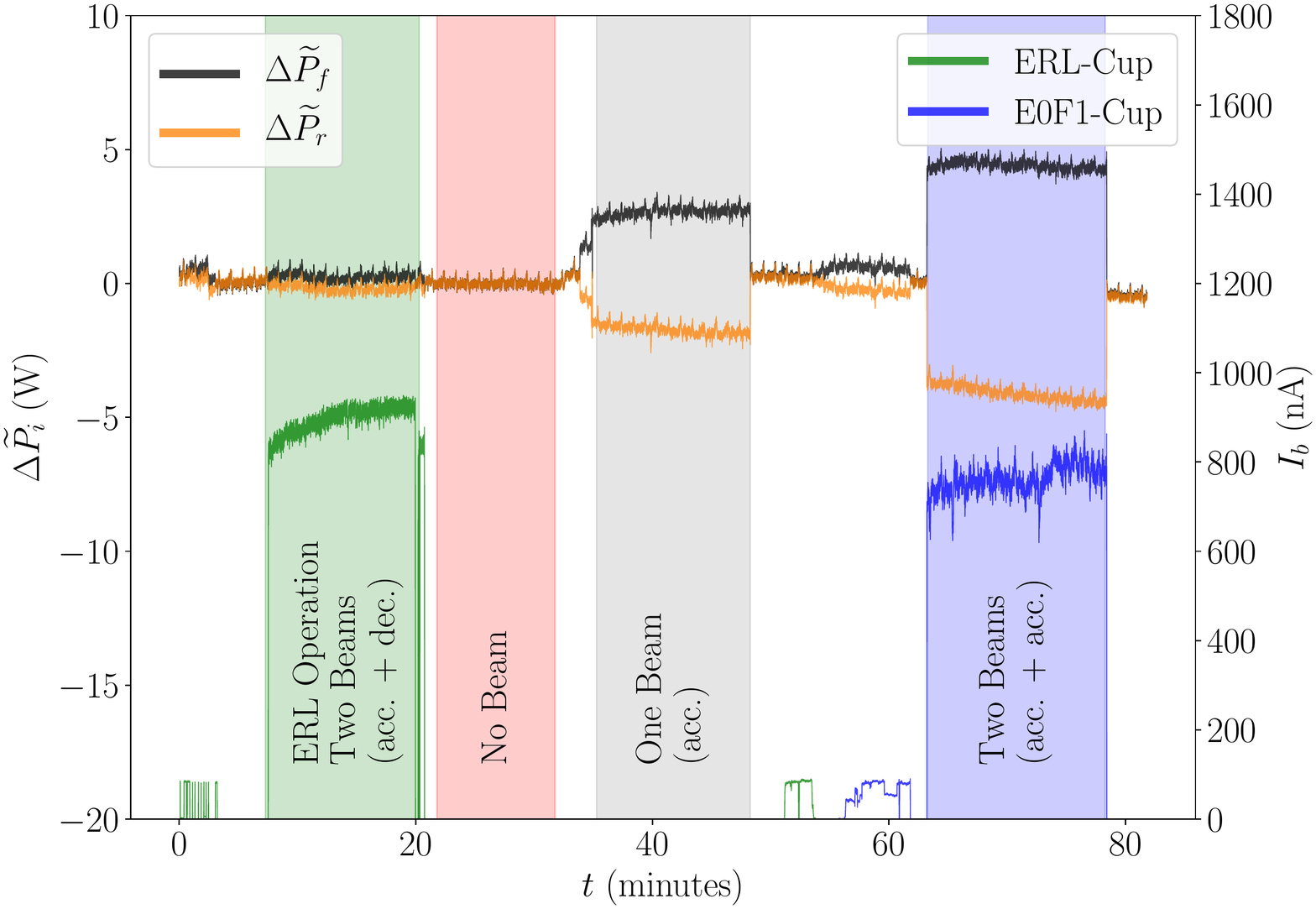}
        	\caption{The changes in forward RF power (black curve) and reverse  RF power (orange curve) of the first main accelerating cavity (A1SC01, see Fig.~\ref{fig:S-DALINAC_floorplan}) have been monitored during four different settings of the accelerator (ERL: green, no beam: red, single pass: grey, twice accelerating: blue). The beam current on the corresponding Faraday cups (ERL-Cup: green, E0F1-Cup: blue, see Fig.~\ref{fig:S-DALINAC_floorplan}) was measured simultaneously \cite{Arnold:2020ubw}.}
        	\label{fig:ERL-measurement-overview_S-DALINAC}
        \end{figure}

        Figure~\ref{fig:ERL-measurement-overview_S-DALINAC} gives an overview of the data obtained during the measurement. For more information on the once-recirculating ERL operation of the S-DALINAC, see \cite{Arnold:2020ubw}.

    \FloatBarrier

    \subsubsection{Twice-Recirculating ERL mode}

       In August 2021, the \mbox{S-DALINAC} was operated successfully in twice-recirculating ERL mode. The main parameters are listed in Table~\ref{tab:2xERLparameters_S-DALINAC}. 

        \begin{table}[htb]\centering
            \caption{Parameters used for twice-recirculating ERL operation. A detailed evaluation of the data is in preparation \cite{Schliessmann-prep}.}
            \label{tab:2xERLparameters_S-DALINAC}
            \begin{tabular}{lc}
            \toprule
                    Parameter & Value \\
            \midrule
                    Momentum downstream injector (setpoint) & $\SI{5.00}{MeV}/c$ \\
					Momentum downstream first main linac passage (setpoint) & $\SI{23.66}{MeV}/c$ \\
					Momentum downstream second main linac passage (setpoint) & $\SI{41.61}{MeV}/c$ \\
                    Maximum beam current used & \SI{7.1}{\micro\ampere} \\
					Maximum RF-recovery effect reached & \SI{86.7}{\percent} \\
            \bottomrule
            \end{tabular}
        \end{table}

        In this twice-recirculating ERL operation mode, indicated in  Fig.~\ref{fig:ERL_schemes_S-DALINAC} on the right-hand side, there are two beams transported in the first recirculation beam line: one beam on its way to the second acceleration and one on its way to the second deceleration.
        Thus, the measurement of both beams is complex.
        BeO targets with holes as well as beam loss monitors have proven their capabilities.
        Dedicated devices for the measurement of both beams are under investigation at the moment \cite{Dutine:2019mzh,Schliessmann:2019fvn}. 
		
		Beam dynamics simulations have shown the need for superior beam quality from injection \cite{Schliessmann:2020due}.
		Especially the bunch length is a very important parameter to mitigate phase slippage effects.
		A new capture cavity at the entrance of the injector linac was installed to improve the beam quality of the injected beam; this optimizes the initial emittance and energy spread.
		This six-cell cavity operates at a reduced beta of 0.86 \cite{Weih:2019unj}.

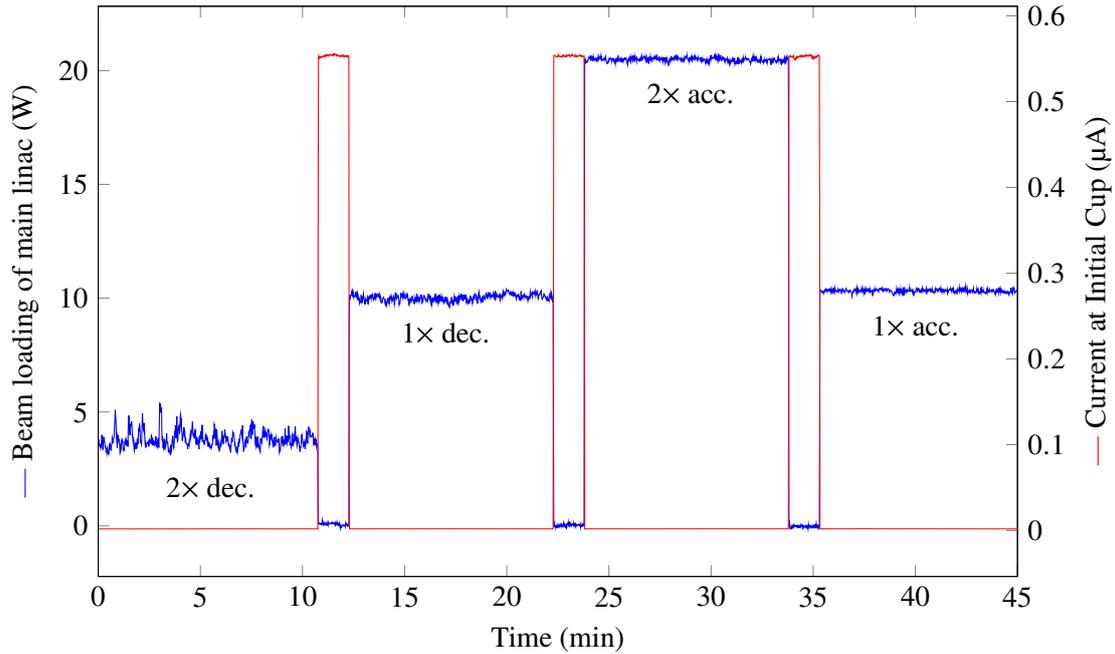
\begin{figure}
    		\centering
\tikzsetnextfilename{sdalinac_beamloading}
\begin{tikzpicture}
\begin{axis}
[
    width=.8\linewidth,
    height=.5\linewidth,
    scale only axis,
    ylabel={\raisebox{0.5ex}{\tikz{\draw[blue] (0, 0) -- (1em, 0);}}~Beam loading of main linac (\si{\watt})},
    xlabel={Time (min)},
    xmin=0,
    xmax=45,
    %ymin=0,
    %ymax=50,
    %legend pos=north west,
    ytick pos=left,
]
\addplot+[blue, mark=none] table[x index=0, y index=1] {figures/sdalinac_beamloading_power.dat};
%\addlegendentry{Beam loading}
\draw (axis cs:5.5, 1.7) node {$2\times$ dec.};
\draw (axis cs:17, 8.5) node {$1\times$ dec.};
\draw (axis cs:29, 19) node {$2\times$ acc.};
\draw (axis cs:40, 8.8) node {$1\times$ acc.};
\end{axis}
\begin{axis}
[
    width=.8\linewidth,
    height=.5\linewidth,
    scale only axis,
    ylabel={\raisebox{0.5ex}{\tikz{\draw[red] (0, 0) -- (1em, 0);}}~Current at Initial Cup (\si{\micro\ampere})},
    xmin=0,
    xmax=45,
    %ymin=0,
    %ymax=50,
    %legend pos=north east,
    ylabel near ticks, yticklabel pos=right, ytick pos=right,
    xticklabel=\empty,
    xtick=\empty,
]
\addplot+[red, mark=none] table[x index=0, y index=1] {figures/sdalinac_beamloading_current.dat};
%\addlegendentry{Current}
\end{axis}
\end{tikzpicture}
    		\caption{The beam loading of the full main accelerator is shown during four different operational phases in the twice-recirculating ERL run at a constant beam current of \SI{0.55}{\micro A}. During the first phase, the accelerator was running in full ERL mode. Only approx.~\SI{4}{W} was requested by the LLRF system. In the second phase, two accelerated and one decelerated beam had a total beam loading of approx.~\SI{10}{W}. For only one accelerated beam, the power consumption was nearly identical. This indicates a full recovery of the kinetic energy of the first decelerated beam. The third phase shows the beam loading needed with two accelerated beams with approx.~\SI{20}{W}. In between these four phases, the beam was blocked with an upstream cup to verify the beam current and to have a clear separation of the different phases. A detailed evaluation of the data is in preparation \cite{Schliessmann-prep}.}
    		\label{fig:2xERL_S-DALINAC}
    	\end{figure}

		In twice-recirculating ERL operation, the beam current was ramped up to \SI{7.1}{\micro A}.
		For each measurement point, the total beam loading of the main accelerator was measured during four phases:
		\begin{itemize}
			\item 2x dec.: two accelerated and two decelerated beams in the main accelerator
			\item 1x dec.: two accelerated and one decelerated beam in the main accelerator
			\item 2x acc.: two accelerated beams in the main accelerator
			\item 1x acc.: one accelerated beam in the main accelerator
		\end{itemize}
		Figure~\ref{fig:2xERL_S-DALINAC} shows a measurement with an initial beam current of \SI{0.55}{\micro A} as an example. A full discussion of the measurement is in preparation \cite{Schliessmann-prep}.

    \subsubsection{Summary and Outlook: \mbox{S-DALINAC}}

        Since its upgrade in 2015/2016, the \mbox{S-DALINAC} can be operated as a once- or twice-recirculating ERL. The lattice is quite versatile. In August 2017, the first successful once-recirculating ERL run with an RF-recovery effect of $(90.1 \pm 0.3\,)\%$ was achieved, the first operation of an ERL in Germany.
        The importance of high injection energies was clearly seen.
        Later, in August 2021, the first successful twice-recirculating ERL run was completed with beam currents of up to \SI{7.1}{\micro A} and a measured RF-recovery effect of up to \SI{86.7}{\%}.
        For the future, further work on dedicated beam diagnostics and a more detailed investigation on the impact of the phase-slippage effect are planned.

    \subsubsection{Outlook: DICE}

    The twice-recirculating ERL run at \mbox{S-DALINAC} demonstrated the feasibility of a multi-turn SRF-ERL with common beam transport for some sections of the accelerated and the decelerated beam. This scheme will be exploited in the MESA and PERLE ERLs that are currently under construction. At the same time, the S-DALINAC demonstrations emphasized the constraints due to a common beam transport of the accelerated and decelerated beam. Combined with the fact that high-energy and high-current multi-turn SRF ERLs require  high stability, reliability, and controllability, these observations make the planning for a future ERL using separated beam transport, besides those with common beam transport, desirable (see also Subsection~\ref{sec:key_challenges:art_of_arcs}).
    
    An external expert panel has reviewed the options for development of the physics department of TU Darmstadt in November 2020 and has encouraged its President to investigate concepts along these lines. 
    A more detailed design is being planned and referred to as DICE (Darmstadt Individually-recirculating Compact ERL). Separated beam transport for the accelerated and decelerated beams with a multi-bend-achromat arc design are intended. DICE may provide a top energy in excess of 520\,MeV with 20\,mA in CW operation in its final stage, using three recirculation beam lines. A double-sided linac is being considered. The SRF system may run at 801.58\,MHz in order to complement PERLE at Orsay (see Subsection \ref{sec:new_facilities:perle}) and to make a comparison as informative as possible. The pair of PERLE and DICE could then offer most valuable information about the pros and cons of combined versus individual recirculations in a multi-turn SRF-ERL. 
    
    TU Darmstadt is currently considering to establish and operate DICE near the international FAIR facility at Darmstadt.
    DICE is intended to be operable either in ERL mode or in conventional accelerating mode to deliver beam to fixed-target experiments.
    As ERLs would make for optimum colliders, foreseen applications of DICE's ERL mode include the production of brilliant MeV-range photon beams and even for an electron-ion collider on intense radioactive ion beams at FAIR.
    The MeV-range photon beams are intended to be obtained from laser Compton backscattering reactions of the high-current ERL beam at high repetition rate in a high-finesse optical resonator. They would ideally serve the internationally visible program on photonuclear reactions at TU Darmstadt (see \cite{Zilg_PPNP2022} and also Section \ref{sec:frontier:photonuclear}). Moreover, the establishment of DICE at FAIR would represent an additional investment of TU Darmstadt in the international FAIR facility for enabling electron scattering reactions on stored radioactive-ion beams at FAIR with unprecedented luminosity. The corresponding physics cases have previously been endorsed by international expert panels in the context of the planning of the ELISe experiment at FAIR.
    While ELISe could not be realized within FAIR's modularized start version, the establishment of DICE by TU Darmstadt at FAIR would make the ambitious research projects, initially intended by the ELISe proposal, possible. The corresponding design study of DICE is currently in preparation.

%\subsection{sDALINAC at Darmstadt}
%Michaela Arnold, Norbert Pietralla
%

\subsection{cERL at KEK}
\label{sec:cERL}
\subsubsection{Introduction}
\begin{figure}[htb]
\centering
\includegraphics[scale=0.33]{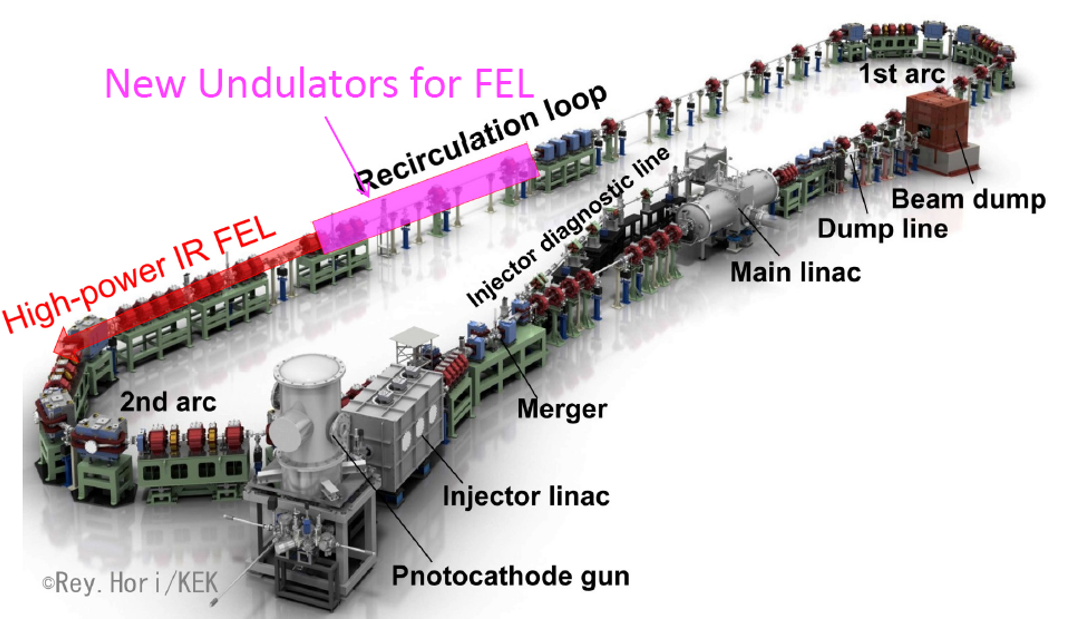}
\caption{Layout of the cERL.}
\label{fig:cerl}
\end{figure}

The Compact Energy-Recovery Linac (cERL) has been operating since 2013 at KEK.
It is a test accelerator to operate with a high average beam current and excellent beam quality.
The main purpose of the cERL is to develop key components required for a future high-average-current electron source with low emittance, such as a DC photocathode gun and cutting-edge superconducting cavity technologies. Figure~\ref{fig:cerl} shows the layout of the cERL.
An electron beam produced by the \SIrange{390}{500}{\kilo\volt} gun~\cite{doi:10.1063/1.4811158} is accelerated in the superconducting injector cavities to between 3 and \SI{5}{\mega\electronvolt}~\cite{WATANABE201367}, then accelerated in the SC main linac (ML) cavities up to \SIrange{17.6}{20}{\mega\electronvolt}~\cite{PhysRevAccelBeams.22.022002}.
The beam then travels around the re-circulation loop, is decelerated in the ML-SC down to the injection energy, and is delivered to the beam dump.

After the first beam commissioning in December 2013, the maximum beam current was increased in a step-by-step manner every year, namely, \SI{1}{\micro\ampere} in 2013, \SI{10}{\micro\ampere} in 2014, and \SI{100}{\micro\ampere} in 2015.
Details of the design and construction of the facility as well as the results of the initial commissioning were already reported in \cite{AKEMOTO2018197}.
In March 2016, high-current (\SI{1}{\milli\ampere}) CW operation with energy recovery was achieved \cite{Obina:IPAC2019-TUPGW036}.
Despite these successes, the future light source plan at KEK was shifted to a high-performance storage ring, and the ERL Project Office was closed at KEK in 2017.
However, the ``Utilization Promotion Team based on Superconducting Accelerator (SRF-application team)'' was kept because the KEK Directorates wanted to maintain the R\&D for industrial applications based on ERL technologies.
On this basis, cERL beam operation was restarted, but the objective changed from the future light source to industrial applications. 

\subsubsection{Industrial applications of the cERL}
To be able to realize such industrial applications as an Extreme Ultraviolet Free-Electron Laser (EUV-FEL) for lithography, high-intensity Laser-Compton-scattering (LSC) sources, a THz source, a Radio Isotope (RI) factory, etc., the following performance should be achieved: 
\begin{itemize}
 \item High-average-CW-current electron beam;
 \item High quality of the electron beam with high bunch charge. 
\end{itemize}

With these goals in mind, high-bunch-charge operation (max.~40\,pC/bunch) was launched in March 2017 to develop the beam handling techniques required for a high-average-current FEL.
Following the successful achievement of this goal, even higher-bunch-charge operation (max.~60\,pC/bunch) was achieved in March 2018.
Finally, in June 2018, stable \SI{1}{\milli\ampere} energy-recovery beam operation was achieved with small beam emittance, with minimal beam loss and halo.
A brief summary of this achievement is as follows: CW \SI{0.9}{\milli\ampere} operation with the recirculation loop energy of \SI{17.6}{\mega\electronvolt}.
It was stable in 2 hours after fine tuning to reduce beam loss.
To achieve stable CW operation, optics tuning and collimator tuning were very important.
The measured normalized emittances were close to the design values (design: $\epsilon_x / \epsilon_y = (0.34 / 0.24)\,\pi\,\si{\milli\meter\milli\radian}$; measured: $\epsilon_x / \epsilon_y = (0.29 / 0.26)\,\pi\,\si{\milli\meter\milli\radian}$). The energy recovery efficiency was $\SI{100 +- 0.03}{\percent}$ \cite{Obina:IPAC2019-TUPGW036}.

\begin{figure}[htb]
\centering
\includegraphics[scale=0.6]{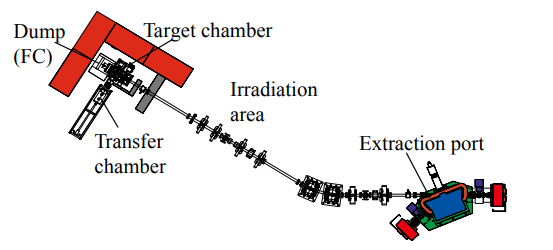}
\caption{Layout of the irradiation area.}
\label{fig:cERL:irradiation}
\end{figure}

In 2018, there were two focus topics of the cERL activity.
The first was a radioactive isotope manufacturing facility for nuclear medical isotope production of $^{99}$Mo/$^\text{99m}$Tc using an accelerator rather than a nuclear reactor, which should result in a more stable supply. For this, a new beam line for electron beam irradiation was constructed and successfully commissioned \cite{Morikawa:IPAC2019-THPMP012}.
This new beam line is only used for industrial applications: $^{99}$Mo production for nuclear medicine and asphalt modification for infrastructure sustainability (see Fig.~\ref{fig:cERL:irradiation}). The construction was finished in March 2019, and the first irradiation experiments were done in June 2019 \cite{sakai_srf2021}.

\begin{figure}[htb]
\centering
\includegraphics[width=10cm]{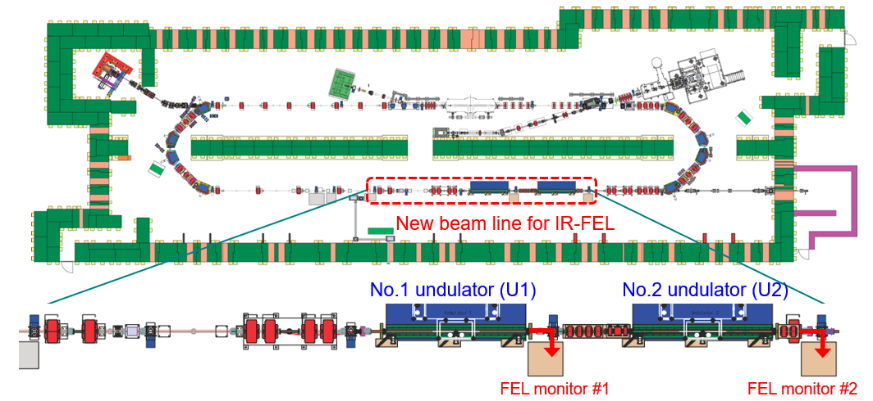}
\caption{Layout of the IR-FEL beam line in the cERL.}
\label{fig:cERL:ir_fel}
\end{figure}

Another industrial application of interest is the use of the EUV-FEL for Future Lithography, which is the prioritized target for the next years.
This design is discussed in Section~\ref{sec:applications:euv-fel}. For the EUV-FEL scheme, the realization of an ERL-based SASE-FEL with high-current beam is one of the key technologies.
In order to demonstrate ERL-FEL operation with SASE, two undulators, which produce Mid-Infrared (MIR) free-electron-laser light for high-efficiency laser processing to organic material, were installed at the south section of cERL in 2020 as shown in Fig.~\ref{fig:cERL:ir_fel}.
With stable beam operation involving AI beam tuning, IR-FEL light was successfully produced based on the SASE scheme \cite{sakai_srf2021, doi:10.1063/5.0072511}.

\begin{figure}[htb]
\centering
\includegraphics[height=8cm]{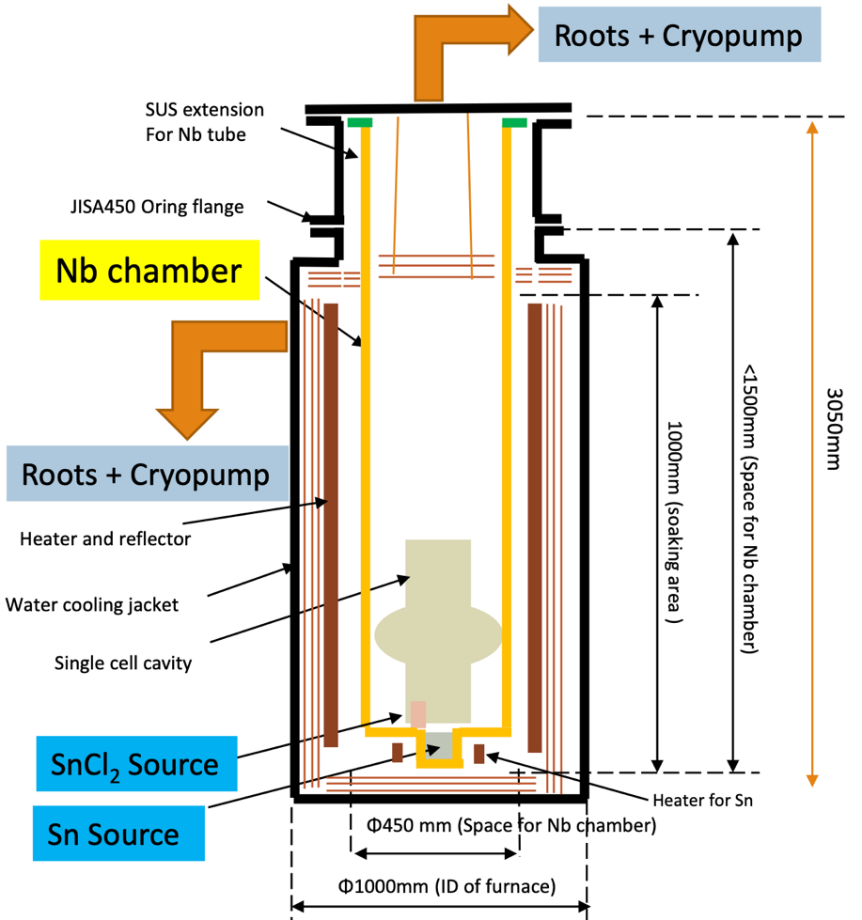}%
\quad%
\includegraphics[height=8cm]{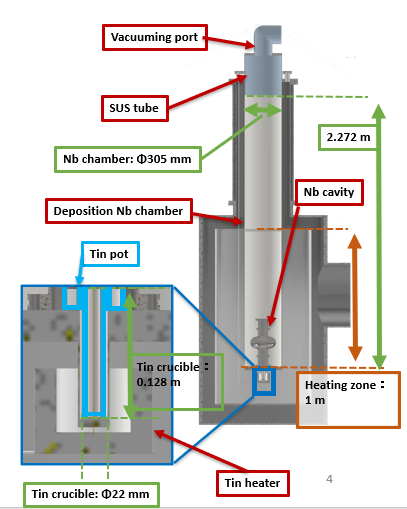}%
\caption{Schematic of the KEK Nb$_3$Sn furnace.}
\label{fig:cERL:furnace}
\end{figure}

An important factor for the commercialization of SRF is the ability to operate at \SI{4.5}{\kelvin}, and KEK has an aggressive development plan for this.
The first step is the fabrication of a furnace based on the Cornell furnace, large enough for a three-cell cavity at \SI{1300}{\mega\hertz} and capable of reaching \SI{1400}{\kelvin} (Fig.~\ref{fig:cERL:furnace}) \cite{takahashi_thinfilm2021}.
The tin is diffused into the interior of the cavity through a completely separate pumping system.
This furnace is part of a new research area which will focus on producing cavities coated with Nb$_3$Sn that can be cooled using commercial cryocoolers.

\subsection{Recuperator at Novosibirsk}%
\label{sec:current_facilities:recuperator}

% MB 07/17/2021: Rough edit:
% minor grammar and typesetting edits
% consistent use of siunitx
% adapted the parameter table to standard style

The Novosibirsk FEL facility~\cite{vinoerl} includes three FELs~\cite{vinoshev}
operating in the terahertz, far-, and mid-infrared spectral ranges.
Despite its rather long history, its potential has not been fully revealed so far.
The first FEL of this facility has been operating for users of terahertz radiation since 2004.
It remains the world's most powerful source of coherent narrow-band radiation in its wavelength range (\SIrange{90}{340}{\micro\meter}).
The second FEL was commissioned in 2009; now, it operates in the range of \SIrange{35}{80}{\micro\meter}, but it is planned to replace its undulator with a new one, shifting its short-wavelength boundary down to \SI{15}{\micro\meter}.
The average radiation power of the first and second FELs is up to \SI{0.5}{\kilo\watt}, and the peak power is about \SI{1}{\mega\watt}.
The third FEL was commissioned in 2015 to cover the wavelength range of \SIrange{5}{20}{\micro\meter}.
Its undulator is comprised of three separate sections.
Such a lattice is suited very well to demonstrate the new off-mirror way of radiation outcoupling in an FEL oscillator (so called electron outcoupling~\cite{matveenko}), 
which is also planned for the near future, along with improvements of the accelerator injection system.
As a result, the average electron beam current and, consequently, the radiation power of all the three FELs will increase.

\begin{figure}[ht]
\centering
\includegraphics[scale=1.0]{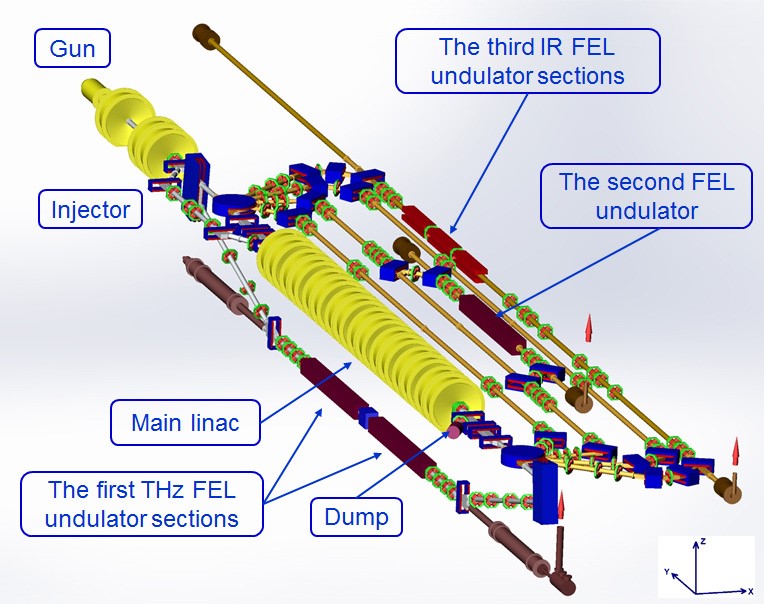}
\caption{The Novosibirsk ERL with three FELs (top view).}
\label{fig:current_facilities:recuperator:topview}
\end{figure}

The undulators of the FELs are installed on the first, second, and fourth orbits of the multi-turn ERL.
The scheme of the Novosibirsk ERL with three FELs is shown in Fig.~\ref{fig:current_facilities:recuperator:topview}.
The Novosibirsk ERL was the first multi-turn ERL in the world.
Its characteristic features include a normal-conducting \SI{180}{\mega\hertz} accelerating system, a DC electron gun with a grid-controlled thermionic cathode, three operation modes of the magnetic system, and a rather compact ($\SI{6}{\meter} \times \SI{40}{\meter}$) design.

The NovoFEL accelerator has a rather complex design.
One can treat it as three different ERLs that use the same injector and the same linac.
Starting from the low-energy injector, electrons pass through the accelerating radio-frequency (RF) structure (main linac) several times.
After that, they lose part of their energy in the FEL undulator.
The used electron beam is decelerated in the same RF structure, and the low-energy electrons are absorbed in the beam dump.

The first ERL of the facility has only one orbit.
It is stacked vertically (see Fig.~\ref{fig:current_facilities:recuperator:topview}).
The second and the third ERLs are two and four-turn ERLs, respectively.
Their beam lines are arranged horizontally.
The injector is common for all the ERLs.
It includes an electrostatic gun and one bunching and two accelerating cavities.
The gun voltage is \SI{300}{\kilo\volt}, which is applied to the grid-controlled thermionic cathode. The gun provides \SI{1}{\nano\second} bunches with a charge of up to \SI{1.5}{\nano\coulomb}, a normalized emittance of about \SI{20}{\micro\meter}, and a repetition rate of zero to \SI{22.5}{\mega\hertz}.
After the \SI{180.4}{\mega\hertz} bunching cavity, the bunches are compressed in the drift space (about \SI{3}{\meter} long), accelerated up to \SI{2}{\mega\electronvolt} (total energy) in the two \SI{180.4}{\mega\hertz} accelerating cavities, and injected through the injection beamline and the chicane into the main accelerating structure of the ERL (see Fig.~\ref{fig:current_facilities:recuperator:topview}).
The accelerating structure consists of 16 normal-conducting RF cavities, connected to two waveguides. The operating frequency is \SI{180.4}{\mega\hertz}.
Such a low frequency allows operation with long bunches and high currents.

The choice of the working ERL and corresponding FEL is determined by commutation of bending magnets.
The first FEL is installed underneath the accelerating RF structure.
Therefore, after the first pass through the RF structure, the electron beam with an energy of \SI{11}{\mega\electronvolt} is bent by \ang{180} into the vertical plane.
After being used in the FEL, the beam returns to the RF structure at the decelerating phase.
In this mode, the ERL operates as a single-orbit machine.

For operation with the second and third FELs, two round magnets (a spreader and a recombiner) are switched on.
They bend the beam in the horizontal plane as shown in Fig.~\ref{fig:current_facilities:recuperator:topview}.
After four passes through the RF accelerating structure, the electron beam enters the undulator of the third FEL. The energy of electrons in the third FEL is about \SI{42}{\mega\electronvolt}.
The used beam is decelerated four times and goes to the beam dump.

If the four magnets on the second track (see Fig.~\ref{fig:current_facilities:recuperator:topview}) are switched on, the beam with an energy of \SI{20}{\mega\electronvolt} passes through the second FEL.
After that, it enters the accelerating structure at the decelerating phase due to the choice of path length through the second FEL.
After two decelerating passes, the used beam is absorbed in the beam dump. 

The parameters of the operating modes are shown in Table~\ref{tab:current_facilities:recuperator:parameters}.

%%%%%%%%%%%%%%%%%%%% BINP starts here %%%%%%%%%%%%%%%%%%%%

\begin{table}[ht]\centering
\caption{Basic accelerator and FEL parameters}%
\label{tab:current_facilities:recuperator:parameters}
\begin{tabular}{lcccc}
\toprule
   & &  1st FEL & 2nd FEL & 3rd FEL \\
\midrule
Beam energy & MeV & \numrange{8.5}{13.4} & \numrange{21}{22.8} &   \numrange{39}{42} \\
 Peak Current & A    &    10         &  30          &  50            \\
 Average Current & mA &     30        &    10          &  4            \\
 Wavelength & \si{\micro\meter} &   \numrange{90}{340} & \numrange{37}{80} & \numrange{8}{11} \\
 Average radiation power & kW &     0.5      &    0.5          &    0.1        \\
\bottomrule
\end{tabular}
\end{table}

%%%%%%%%%%%%%%%%%%%%%%%%%%%%%%%%%%%%%
It is worth noting that all the \ang{180} bends are achromatic (even second-order achromatic on the first and second horizontal tracks) but non-isochronous.
That enables longitudinal ``gymnastics'' to increase the peak current in the FELs and to optimize the deceleration of the used beam.

The possibilities for users to conduct their experiments have been significantly expanded recently by implementation of the new operating mode~\cite{shevchenko:fel2019-tup024}.
In this mode, single or periodic radiation macropulses of duration of down to \SI{10}{\micro\second} can be obtained.
The radiation power modulation is done electronically by controlling the FEL lasing, and it can be triggered by an external signal.

The current of the Novosibirsk ERL is now limited by the electron gun.
A new RF gun was built and tested recently.
It operates at a frequency of \SI{90}{\mega\hertz}.
An average beam current of more than \SI{100}{\milli\ampere} was achieved~\cite{Volkov:2016cgy}.
It is planned to install this gun in the injector; the existing electrostatic gun will be kept there.
The RF gun beamline has already been manufactured and assembled on the test setup.
The beam parameters were measured after the first bending magnet and at the beamline exit.
% \subsection{Recuperator at Novosibirsk}
% Nikolay Vinokurov \\

\subsection{MESA}%
\label{sec:new_facilities:mesa}
%\chapterauthor{Kurt Aulenbacher}

% MB: rough typesetting edit, 06/11/2021
% MB: replaced overview figure by TikZ version, 06/14/2021

The Mainz Superconducting Energy-recovering Accelerator, MESA for short,  is going to be used for particle and nuclear physics experiments \cite{Hug:2020miu}.
The superconducting RF modules will allow CW operation at \SI{1300}{MHz}.
Two beam modes will be available:
On the one hand, the P2 experiment \cite{Becker:2018ggl} will use   an extracted spin-polarized beam (EB mode) of \SI{150}{\micro\ampere}.
On the other hand, the MAGIX experiment  will employ windowless targets using a gas-jet technique \cite{Schlimme:2021gjx}.
Because of the low areal density, the interaction of the beam with the target is minimal.
Hence, energy recovery of the beam after passing the target is efficient and  higher luminosities can be achieved with a given installed RF power.
For this \enquote{ER mode}, \SI{1}{mA} of beam intensity at \SI{105}{MeV} will be available in the first stage, which is planned to be increased to \SI{10}{mA} in a second stage.

\begin{figure}[htbp]\centering
\input{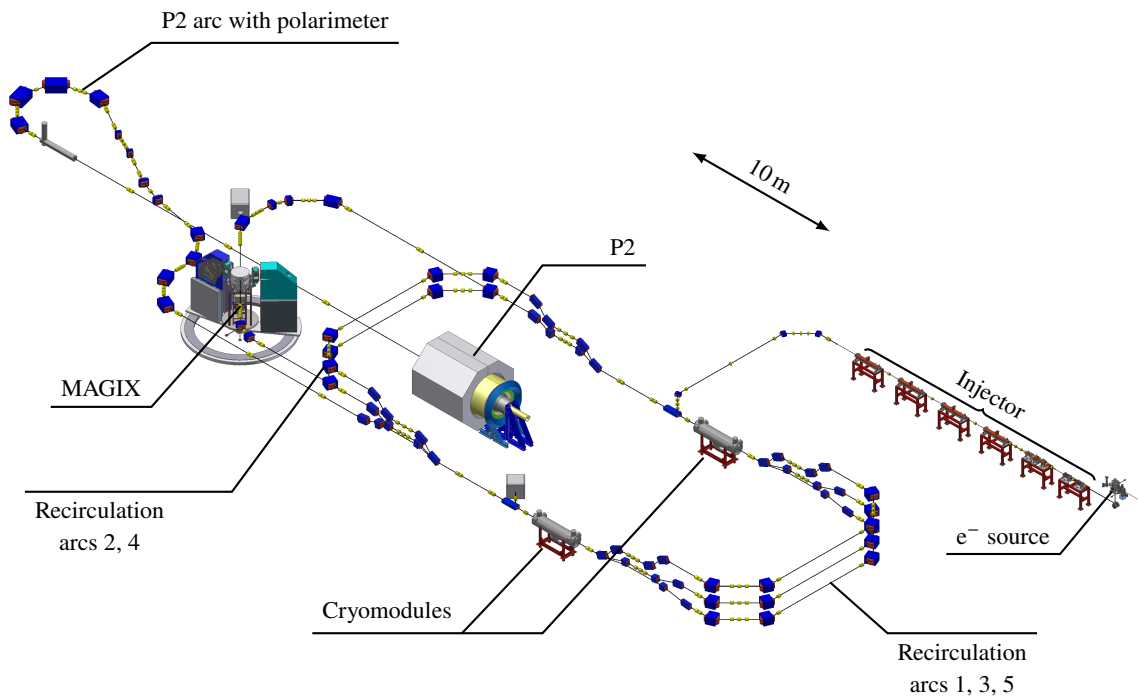}
\caption{Overview of the MESA accelerator components. Blue, yellow, and red items represent dipoles, quadrupoles, and beam position monitors, respectively.}%
\label{fig:new_facilities:mesa:view}%
\end{figure}

The recirculator magnet system is arranged in double-sided fashion with an accelerating cryomodule of the \enquote{ELBE} type  on each side (see Fig.~\ref{fig:new_facilities:mesa:view}).
The lattice offers high flexibility:
Firstly, very good energy spread, $\Delta E/E<10^{-4}$, can be  achieved  by non-isochronous acceleration~\cite{Hug:2020hwl}.
Secondly, by using path length compensation and variable $R_{56}$ in the arcs,  continuous variation of energy at the experimental sites from \SI{30}{MeV} up to the maximum energy  is possible~\cite{Simon:2021}.
This is important for the planned campaigns aiming at precision measurements of astrophysical S-factors and the charge radius of the proton.
The SRF cryomodules will operate at a gradient of \SI{12.5}{\mega\volt\per\meter}.
Four cavities of the TESLA type will then yield \SI{50}{MeV} per pass. The original ELBE design was supplemented by XFEL tuners for faster tuning. 
With five \ang{180} bending arcs,  three linac passages in EB mode and twice-accelerating and twice-decelerating operations in ER mode are possible.
In ER mode, a dedicated beam line can guide the beam towards the MAGIX experiment and back to the modules with the \ang{180} phase shift required for ER operation.
In EB mode, the beam is extracted after the third linac passage and guided through another external arc towards a long straight line in front of the P2 experiment.
The straight will incorporate an electron beam polarimeter \cite{Tyukin:2020efk} and 
the beam parameter stabilization for the P2 experiment.
This stabilization system is based on an arrangement of resonant $\text{TM}_{11}$ and $\text{TM}_{10}$ cavities.
It has already been tested under realistic conditions at the MAMI accelerator  and has demonstrated that the  bandwidth and sensitivity required for the P2 experiment can be achieved \cite{Kempf:2020}. 

The injection energy is \SI{5}{MeV}, which leads to  a beam energy of \SI{155}{MeV} in EB mode and \SI{105}{MeV} in ER mode.
The fundamental power couplers and the RF amplifiers are chosen in such a way that the beam power requirement of the P2 experiment (\SI{23.25}{kW}) can be met.
For ER mode, our investigations \cite{Stoll2019} have revealed that a beam current at the experiment above \SI{10}{mA}  can be obtained as far as BBU instability is concerned.
However, a more serious limitation seems to result from the HOM-damping antennas of the TESLA cavities.
The power handling capability of the HOM antennas is presently subject to large uncertainties.
Their thermal conductivity has been improved for the MESA cavities \cite{Stengler:2020vrh}.
Therefore, the  limitation for the beam current at the experiment is presently believed to be within the range of the design value of \SI{1}{mA}.
Further improvements, e.g., coating the antennas with material of higher $T_\text{C}$, are presently under investigation.
The MAGIX experiment will operate with a windowless gas jet target.
Together with the \SI{1}{mA} MESA beam, luminosities of $>\SI{e35}{\per\square\centi\meter\per\second}$  are possible.
This luminosity can be run permanently  while complying with radiation protection regulations valid for our institution (more details in section \ref{sec:key_challenges:interaction_region}), at least for targets with low nuclear charge like hydrogen or helium \cite{Ledroit2021}. 

A new building is presently being erected which will contain most of the recirculating part of MESA, whereas MAGIX and P2 as well as  the injector \cite{Heine:2021}  will be installed in old parts of the building complex at the Institut für Kernphysik at the Johannes Gutenberg-Universität in Mainz, Germany. 
The new building will become available in the second half of 2022.
Injector installation has already started, as it is located in the existing part of the site. Therefore, \SI{5}{MeV} beam could  become available as soon as the installation of the cryomodules in the new hall is completed. 

In a dedicated test set-up, the source and the longitudinal matching system were successfully operated at bunch charges up to \SI{0.77}{pC}, which yields the nominal \SI{1}{mA} at \SI{1300}{MHz} \cite{Heil:2021,Friederich:2019,Matejcek:2019}. All four cavities  of MESA have  been installed in their cryomodules, where they have achieved  $Q_0>\num{1e10}$ at the nominal field of \SI{12.5}{\mega\volt\per\meter}, though up to now without beam.
This is sufficient for the planned operations.
The cryogenic plant will be upgraded and  will afterwards 
have enough capacity to cool the cryomodules and auxiliary devices such as the superconducting solenoid for the P2 experiment  \cite{Simon:2021}.

After completion of civil construction   the installation of the main part of MESA will begin. Major components to be installed are   the cryogenic infrastructure and the recirculating magnet system.
Of course, there have been delays by the COVID-19 pandemic, but so far their impact only adds up to less than one year.
Therefore, start of beam commissioning for experiments is still envisaged in 2024, which of course depends on the further development of the pandemic crisis.

%--------------------------------- 
 
% \subsection{MESA at Mainz}
%Kurt Aulenbacher

\chapter{ERL---New Facilities in the Twenties}\label{sec:new_facilities}

There are two facilities in Europe that will be completed in this decade: bERLinPro, which is partially built and is ready to accept a cryomodule with the latest fundamental power couplers to dynamically match the loaded cavity to the klystron; and PERLE@Orsay, which has garnered a lot of international collaborators due to its pivotal position for the future of later ERLs.
This will be the only high-power, multi-turn ERL in operation anywhere in the world and will test all of the components needed for future ERLs simultaneously.  Completion of these two facilities and their successful commissioning to their full potential will set the stage for the large high-energy colliders to come.

In addition to these new facilities, there are important developments in the USA.
At Jefferson Lab, a funded project will enable CEBAF to operate as a five-pass ERL.
This will also be the highest-energy ERL being built and, more importantly, will demonstrate that it is possible to control a large number of cryomodules (50) in ERL mode.
At BNL, the Electron-Ion Collider (EIC) will require advanced electron cooling based on an ERL to minimize the power required.
The parameters are the most aggressive of the new facilities, and it must function correctly for the EIC to reach its performance specification.
The information obtained from bERLinPro and PERLE@Orsay in particular will be important for the success of the EIC Cooler.   

\section{bERLinPro}\label{sec:current_facilities:berlinpro}
% MB: Rough edit 6/2/21. Fixed duplicate figure label and a few hyphenation and punctuation issues.
% Also converted all numbers+units to \SI.
% The thing with the hyphens between number and unit so as to resemble a hyphenated compound adjective
% is meant well but incorrect, see sec. 5.4.3 in the SI brochure.
% Re Hyphens: Interesting: Wiley gave me a very hard time for not having used hyphens when writing the book on SRF and everything needed to be changed (Jens).
%Bettina Kuske, Axel Neumann, Jens Knobloch, Julius Kühn
\subsection{Goals and expectations}

The bERLinPro project at Helmholtz-Zentrum Berlin (HZB) officially started in 2011.
At the time, energy-recovery linacs (ERL) were considered an enabling accelerator concept to bridge the gap between the third generation \enquote{work horse} storage-ring-based light sources and newly developed X-ray FELs in terms of brightness, coherence,  pulse length, and number of user stations.
Several ERL facilities had successfully demonstrated the concept, which laid the basis for a number of proposals for ERL-based multi-user X-ray ERL facilities worldwide.

HZB, with its 3rd-generation storage ring BESSY II, has a long-standing tradition of supporting short-pulse experiments and  offering special and flexible timing options to users.
The femto-slicing facility, low-alpha-mode operation, and different fill patterns with single or few bunches in the gap of the continuous bunch train are examples of support for short-pulse and timing experiments.
However, these options provide low photon flux as they are limited to a fraction of the stored beam current.
TRIBs \cite{Goslawski:IPAC2019-THYYPLM2}, BESSY-VSR \cite{VSR@BESSY} are examples of ongoing developments with the intention to provide even more flexibility, in the case of VSR also in bunch length, while maintaining high-flux operation.

Therefore, an ERL-based light source was considered a promising alternative for a successor to BESSY II.
ERLs can provide
\begin{enumerate}
\item high-average flux without the peak being too high for experiments to cope with,
\item a very low emittance for a higher degree of coherence for users,
\item the option for short-pulse operation down into the \SI{100}{\femto\second} range, and
\item importantly, flexibility to tailor the bunch parameters and timing structure to address the wide variety of user demands.
\end{enumerate}
However, such a user facility would require improving on average current, brightness, beam loss, etc.~of the (then) to-date demonstrated ERLs by at least one order of magnitude.

\begin{figure}[htb]
    \centering
    \includegraphics[width=\textwidth]{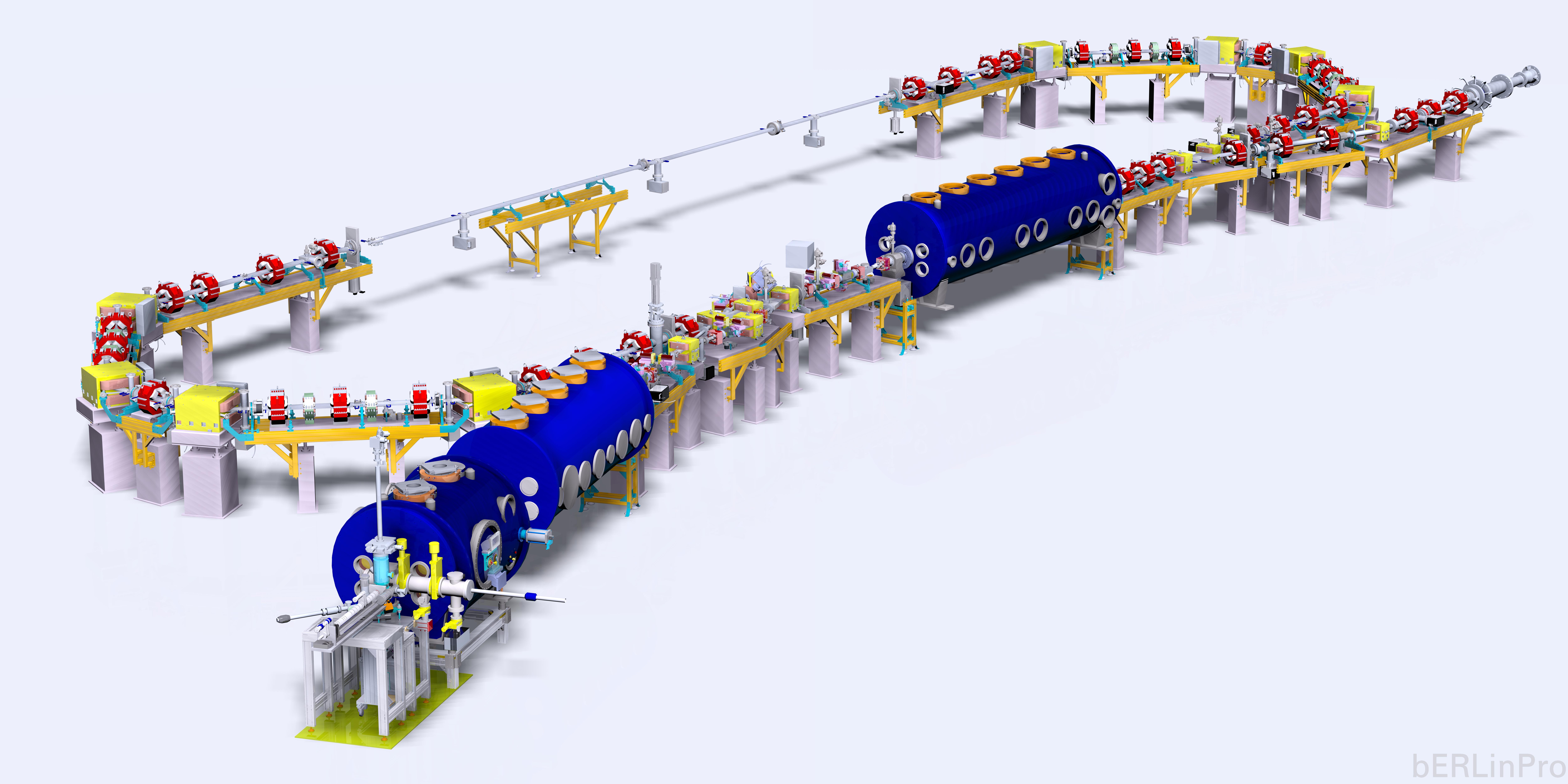}
    \caption{Schematic of bERLinPro. From bottom left to top right: SRF photoinjector, booster module, merger, main linac module, and beam dump.}
    \label{fig:current_facilities:berlinpro:3d}
\end{figure}

bERLinPro, whose layout is depicted in Figure~\ref{fig:current_facilities:berlinpro:3d}, was never intended as a user facility itself.
Rather, it was to serve as a demonstration experiment in accelerator science and technology, to push the electron-beam parameters to the levels needed for a future user light source.
In particular, the following questions were to be addressed: 
\begin{itemize}
  \item \textbf{High average current}: Storage rings run at currents of a few hundred mA. To achieve the same order of magnitude in flux with an ERL, high bunch charges and CW operation in the GHz range are required.
  \item \textbf{Flexibility in pulse length and timing structure}: Flexibility in bunch length was to be demonstrated using different machine configurations. The pulse timing structure is directly given by the flexibility of the cathode laser system. 
  \item \textbf{Emittance and coherence}: A normalized emittance more than an order of magnitude better than in BESSY II is needed to increase the radiation brilliance and coherence.
  \item \textbf{Stability}: The beam and bunch-parameter stability of storage rings, being equilibrium devices, is extraordinarily high and essential to many BESSY II users. bERLinPro was to investigate whether sufficient stability can also be achieved with a single-shot device.
  \item \textbf{Beam loss}: In a light-source facility, users must be able to conduct their experiments near the accelerator, placing stringent requirements on the radiation shielding and permitted beam loss. At \SI{300}{\milli\ampere}, the BESSY II losses are of the  order of \SI{20}{\pico\coulomb\per\second}. Given a \SI{100}{\milli\ampere} ERL, \SI{e11}{\pico\coulomb\per\second} are continuously generated and dumped. For a loss rate commensurate with BESSY II, fractional losses would need to be limited to $10^{-10}$. While this appears an unrealistic target, it must be demonstrated that losses can be tightly controlled and locally handled. This aspect is also of importance to maximize the energy efficiency of ERLs.
\end{itemize}

At a 2009 electron photoinjector workshop organised by HZB, international experts considered DC injectors to be near their performance limit. Such systems at Cornell and KEK were limited to about \SI{350}{\kilo\volt} due to repeated breakdown of the ceramics, a problem that was solved later \cite{Nagai500kVgun}.
Potentially, SRF photoinjectors can achieve fields and voltages significantly higher.
It was thus recommended to establish an SRF injector program, building upon the extensive experience at ELBE (Helmholtz-Zentrum Dresden-Rossendorf) \cite{ArnoldSRFGunHZDR} and  Brookhaven National Laboratory \cite{WencanXuBNLgunIPAC2013}.

\subsection{Revised Focus}
During the course of the project, two major developments shifted the focus of bERLinPro. 
\begin{enumerate}
  \item The construction of a multi-bend achromat based storage ring (MAX IV) demonstrated impressively that the emittance of GeV-class storage rings can be reduced to the \SI{1}{\kilo\electronvolt} diffraction limit, albeit at bunch lengths more than an order of magnitude longer than a standard ERL mode permits, to maintain reasonable lifetimes. 
  \item In 2015, the BESSY VSR project was started at HZB. It employs superconducting RF cavities, based on the bERLinPro LINAC design, to provide a large CW overvoltage. This system allows for bunch shortening in a storage ring such as BESSY II, while still permitting currents on par with that in ERLs. 
\end{enumerate}

Thus, \enquote{conventional} ring accelerators, albeit with technically challenging modifications, can be designed for highly coherent radiation with flexible pulse lengths at high average flux in a multiuser facility. This view was underscored by a DOE-BESAC subpanel report issued in 2013, which concluded that ERLs are technologically not yet sufficiently mature to be considered for light-source applications. Following further development, ERLs may be considered for upgrades of, for example, FEL facilities.

Given these developments, HZB decided not to pursue an ERL-based light source further.
The focus of bERLinPro shifted to studies of key challenges that must be addressed by all ERL facilities, independent of their application.
Many of these, of course, involve the targeted questions listed above.
bERLinPro now is considered a dedicated accelerator research facility, embedded in the Helmholtz Association’s Funding Research Topic 
\enquote{Accelerator Research \& Development.}
ERL-relevant research continues to be in the foreground, and bERLinPro is open for collaboration partners.
But the facility is also available for non-ERL accelerator research that takes advantage of the unique properties of SRF-based systems.
Examples presently include ultra-fast electron diffraction, potentially tests of the VSR SRF systems, and testing advanced artificial-intelligence and machine-learning algorithms in a test facility unconstrained by user operation.

\subsection{Current Status}

\subsubsection{Accelerator Installation}

\begin{figure}[htb]
    \centering
    \includegraphics[width=\textwidth]{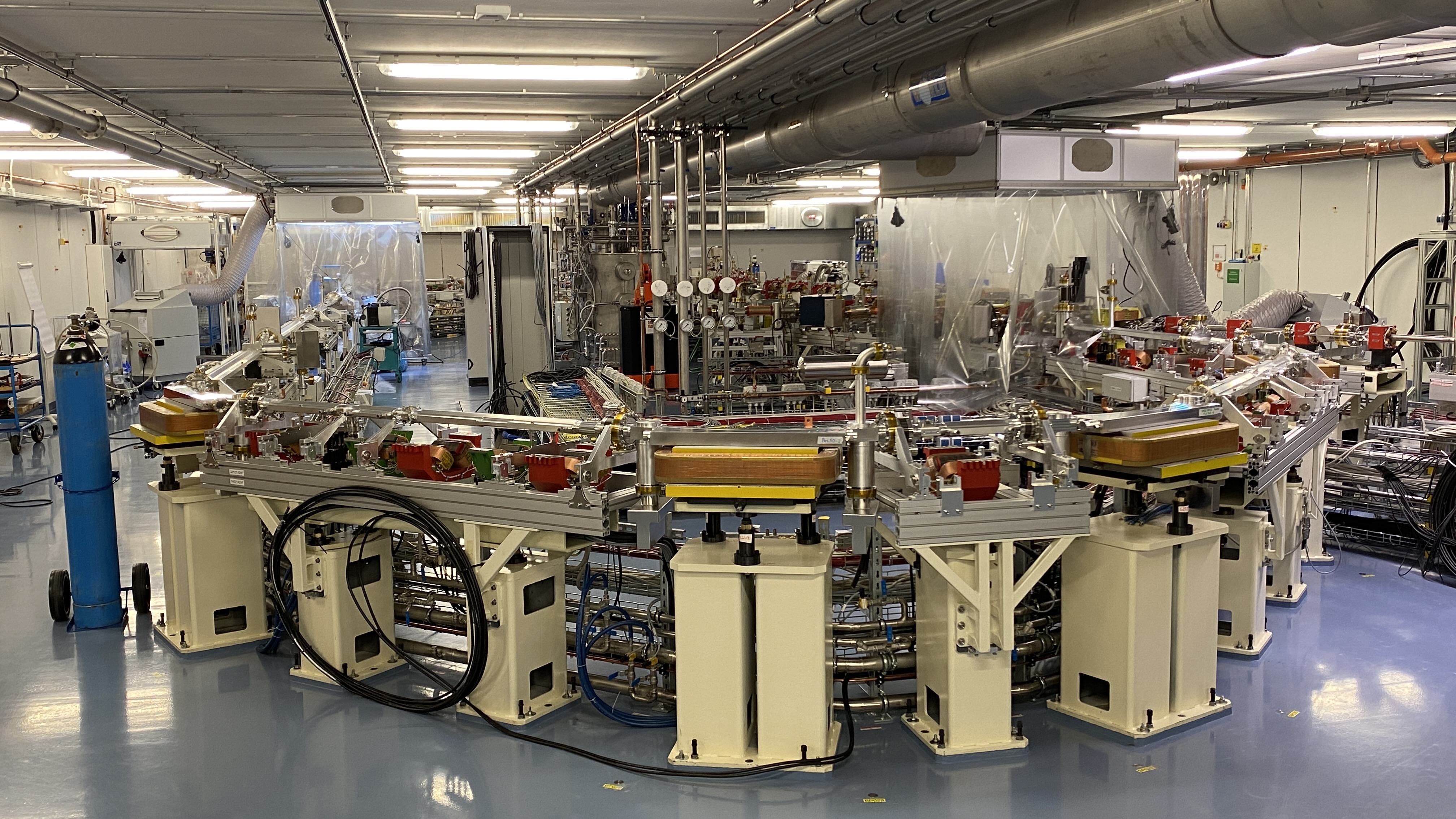}
    \caption{bERLinPro during the installation of the recirculator vacuum system. Two local clean rooms on a ceiling rail system are visible; they are used for particle-free installation (ISO-5) of components. The top half of the magnets of the return arc were removed for the installation of the vacuum system. The injection line comes from the right. }
    \label{fig:current_facilities:berlinpro:photo}
\end{figure}

The beam transport and vacuum system installed by ISO-5 clean room standards---see Figure~\ref{fig:current_facilities:berlinpro:photo}---as well as photocathode laser and laser beam line, beam diagnostics, and a \SI{600}{\kilo\watt} beam dump are installed and fully ready for operation with beam currents up to \SI{100}{\milli\ampere}.

A cathode production and transport system is operational and currently being used to develop Cs-K-Sb cathodes.
First commissioning of the SRF photoinjector started in 2018.
However, due to severe problems with the cathode implementation, a revamp and reassembly was required.
The installation of the SRF photoinjector as well as the 3-cavity booster module is now planned for 2022. The current is currently limited to \SI{10}{\milli\ampere} by the RF couplers in the photoinjector (thermal power limit).
While the couplers of the booster are capable of supporting \SI{100}{\milli\ampere} operation, their coupling strength is currently optimized by spacers for \SI{10}{\milli\ampere} beam current.

Funds are presently not available for the SRF LINAC module.
However, the perspective exists to employ a VSR SRF module in the future (2026) on a temporary basis, to provide approximately \SI{25}{\mega\electronvolt} total beam energy.  However, the accelerator is not designed for the VSR operating frequency (\SI{1.5}{\giga\hertz}), thus constraining the beam current to a few milliamperes and limiting the scope of studies, yet requiring considerable installation and adaptation efforts.

Present activities are focused on the high-current SRF photoinjector.
A dedicated diagnostic line capable of handling \SI{10}{\milli\ampere} is installed to characterize the beam.
Following the booster installation, the beam can be transported through the merger to the high-power beam dump following the splitter section, allowing studies of emittance preservation, beam loss, and bunch length manipulation.

\subsubsection{Technical Infrastructure} 

To date, all infrastructure needed to support the operation of an ERL facility up to \SI{50}{\mega\electronvolt} at \SI{100}{\milli\ampere} is in place. \SI{1.3}{\giga\hertz} high-power RF transmitters and cryogenics are installed for photoinjector, booster, and linac operation. 
In particular, a subterranean bunker provides sufficient radiation protection to handle \SI{30}{\kilo\watt} of continuous beam loss at \SI{50}{\mega\electronvolt}, thereby allowing extensive beam tests without these being overly constrained being by radiation-safety considerations. Dedicated cavity preparation facilities, including a \SI{130}{\square\meter} ISO-4 clean room are available to support the SRF program.

\subsubsection{Future Activities and Options}

\paragraph{100 mA operation}

Two changes to bERLinPro are planned to upgrade to \SI{100}{\milli\ampere} operation:
\begin{enumerate}
\item readjustment RF coupling strength of the booster module and
\item upgrade of the photoinjector module with a new cavity.  The \SI{600}{\kilo\watt} beam dump is already compatible with \SI{100}{\milli\ampere} operation at \SI{1.3}{\giga\hertz}.
\end{enumerate}
The booster module coupling is easily optimized for heavier beam loading by simply removing the presently installed coupler spacers. 

The photoinjector module was built to allow a retroactive upgrade of the module to full \SI{100}{\milli\ampere} operation. It can be re-equipped with a new cavity with dual power-coupler ports that accommodate the recently validated \SI{120}{\kilo\watt} high-power couplers \cite{Neumann:IPAC2021-MOPAB347}.\footnote{So far, the \SI{1.3}{\giga\hertz} coupler has been tested to \SI{120}{\kilo\watt} at \SI{50}{\percent} duty factor, the duty factor being currently constrained by an administrative limit. \SI{120}{\kilo\watt} CW operation appears feasible.} These couplers are essential to handle the ca.\ \SI{200}{\kilo\watt} beam loading at \SI{100}{\milli\ampere} operation. 

Ideally, contingent on funding, a second module would be constructed to avoid long ``dark times'', to mitigate risk, and to allow for operation and concurrent implementation of new injector improvements.
Past experience has underscored that efficient progress is severely hampered if one relies on a serial approach with a single system. In the case of bERLinPro, only a second cryostat and cavity body would need to be produced---the remaining components for a second cold string are already available. 

\paragraph{Robust, high-efficiency photocathodes}

The photocathode in the SRF injector is one of the most critical components. Especially when targeting \SI{100}{\milli\ampere}, high quantum efficiency, low thermal emittance, and longevity are essential. Cs-K-Sb-based systems currently are the state of the art, much effort being expended to devise coating and cathode transportation/insertion procedures that realize and maintain the intrinsic high performance of the material.
New developments using a triple evaporation chamber for co-deposition are already under way at HZB, with the goal to improve upon the performance of Cs-K-Sb using new Na-K-Sb cathodes. This system promises to be more robust against gas  contamination and ion pollution encountered in high-current operation. The ultimate goal is to employ such cathodes in the \SI{100}{\milli\ampere} injector.

\paragraph{Recirculation with VSR module}

Presently, the LINAC module required for full ERL operation is not funded. However, a \SI{1.5}{\giga\hertz} SRF module consisting of two four-cell waveguide HOM-damped cavities is under construction for the demonstration of VSR technology~\cite{Velez:IPAC2017-MOPVA053}. It is being considered for a \emph{temporary} LINAC substitute to accelerate the beam to about \SI{25}{\mega\electronvolt} for initial recirculation studies. While designed for \SI{300}{\milli\ampere} operation, its frequency is only compatible with a \SI{50}{\mega\hertz} bunch repetition rate in bERLinPro for an average current of about \SI{5}{\milli\ampere}, dependent on the achievable quantum efficiency of the injector photocathode.
The cryogenic distribution system is being designed to be compatible with this module.  However, the VSR module will not provide the design beam energy and requires a time-consuming readjustment of the beam transport to re-adapt the path length for recirculation.
So far, only preliminary feasibility studies were performed to determine whether this option works and justifies the effort. Further beam dynamics studies are mandatory for a conclusive answer.

\paragraph{Recirculation with full LINAC module}

For an in-depth ERL program, a \SI{1.3}{\giga\hertz} linac module with three seven-cell cavities can be used to accelerate the bunches to the design \SI{50}{\mega\electronvolt}. A new design with waveguide HOM absorbers and mechanical tuners is near ready for construction, contingent on funding.
Alternatively, it is being considered to adopt a lower-risk (and probably lower-cost) design such as the Cornell LINAC module with beam tube absorbers.
This module is compatible with full \SI{100}{\milli\ampere} operation once the high-current injector comes online.
This layout will enable high-current recirculation for studies of bunch length manipulation, emittance preservation, phase matching and beam stability for energy efficiency, and, critically, beam loss minimisation commensurate with high-current, energy-efficient ERL facilities.

\paragraph{Fast reactive tuners and microphonics compensation}

Contingent on the development of fast reactive tuners (FRTs) by partners such as CERN, this key component for efficient accelerator operations may be integrated in the LINAC module for testing with full beam.
The key here will be the study of microphonics compensation with FRTs in a real accelerator environment to reduce the required RF power for field stabilisation for drastically improved efficiency.
Addtionally, ``classical'' mechanical fast tuners can be incorporated to compare their effectiveness of compensation, as studied in the past~\cite{Neumann:microphonics},  including new Kalman-predictor methods currently under development at HZB~\cite{Ushakov:IPAC2018-WEPAK012}.

\paragraph{4 K cavity operation}

\SI{4}{\kelvin} (or higher) cavity operation has been identified to be mandatory for efficient future \emph{high-energy} ERL applications.
Presently, it is not yet clear what materials and cavity coating technique will be best suited for \SI{4}{\kelvin} operation and where the development will take place.
Still, it is planned to adapt the LINAC module's cryogenics in such a manner to optionally accommodate a ``\SI{4}{\kelvin} cavity'' once available to demonstrate reliable operation with beam.

\subsubsection{Contributions to ERL Development}
The theoretical, technical, and operational experiences developed in the last ten years are too numerous for an exhaustive list, and they are published in numerous reports. Some examples are given below.

\paragraph{Contributions to theory}
\begin{itemize}
    \item New shielding formulas \cite{OttShieldingFormula} and radiation detector design  \cite{OttCERNField}
    \item Optics code ‘OPAL’ modified for ERL start-to-end calculation  \cite{Metzger-Kraus:IPAC2016-WEPOY034}
    \item Detailed impedance studies for accelerator components \cite{Glock:IPAC2015-MOPWA018}
\end{itemize}

\paragraph{Technical and operational experience}

\begin{itemize}
    \item SRF photoinjector development  \cite{Neumann:IPAC2018-TUPML053}
    \item Cathode transfer system for integration of cathodes in an SRF injector \cite{Kuehn:IPAC2018-TUPMF002}
    \item Cathode production with high quantum efficiency and life time \cite{PhysRevAccelBeams.21.113401}
    \item Cathode laser development \cite{MBI_laser_development}
    \item Superconducting solenoid \cite{VoelkerSolenoid} with superconducting Nb magnetic shield \cite{VoelkerNB-Shield}
    \item \SI{100}{\kilo\watt} average-power class couplers  \cite{Neumann:IPAC2021-MOPAB347}
    \item HOM damped linac end cells \cite{NeumannLinacEndCell}
    \item \SI{600}{\kilo\watt} beam dump 
    \item Digital low-level RF systems for CW operation \cite{Echevarria:IPAC2016-TUPOW035}
    \item Repair procedure for damaged (scratched) superconducting cavities \cite{TamashevichScratchRepair}
    
\end{itemize}

\subsubsection{Lessons learned}

Ten years is little time to develop components that exceed the state of the art considerably.
This particularly applies to the photoinjector parameters: \SI{30}{\mega\volt\per\meter} and \SI{100}{\milli\ampere} was far beyond what had been built at the time and still is.
Correspondingly, both schedule and budget must include a large degree of contingency.
It must be communicated at all levels that the many unknowns of such a \emph{research} project (as opposed to an \emph{implementation} project) prevent precise time and budget planning, and much contingency must be included. 

\paragraph{Schedule}
Schedule contingency must accommodate technical setbacks that can have a severe impact: For example, a small scratch incurred during final cleaning of the SRF photoinjector cavity delayed bERLinPro by two years to develop a recovery procedure.
The schedule must also accommodate long times spent in recruiting and training personnel in light of the strong competition for a limited number of experts.
In the case of bERLinPro, this included personnel, for example for laser technology, clean room assembly of SRF components, digital low-level RF systems, and advanced engineering designs for complex accelerator and cryomodule components up to fully integrated systems.

\paragraph{Infrastructure}
Key infrastructure to support SRF development, in particular clean room and cavity handling and testing facilities, ideally should be available prior to the start of the project or at least must be an integral part of the project timing and budget.
The infrastructure must be sufficiently dimensioned to prevent time-consuming bottle necks from (at times unforeseen) parallel activities.

\paragraph{Budget}
Budgetary planning is complicated by the fact that many components are completely new developments, and no fiscal baseline experience exists.
Often, only few companies with the prerequisite production expertise exist worldwide, driving up costs further.
Tooling development results in high baseline costs for the many one-of-a-kind component orders.
Ideally, the budget should enable procurement of multiple systems of the most critical components to mitigate risk.
This allows manufacturers to \enquote{learn} as they produce, leaves room for modifications as know-how is acquired, and mitigates risks in case of system damage.

\paragraph{Collaborations and manufacturing}
The project faced significant hurdles stemming from inadequate, late, faulty, or damaged deliveries from industry as well as collaborating laboratories participating on a \enquote{best effort} basis.
This underscores that intense supervision and quality control throughout the production is essential---a time-consuming process that must be included in staffing plans.
The technical specifications of many components are highly specialized and the production techniques non-standard, even for expert manufacturers.
Supervision is complicated by frequently large distances to vendors, and even cultural differences can impact the communication.
Even \enquote{best-effort} cooperation contracts with laboratories proved insufficient at times.

%\subsection{bERLinPro}
% Jens Knobloch, Bettina Kuske

\section{PERLE at Orsay} \label{sec:new_facilities:perle}
%Oliver Bruening, Walid Kaabi

% MB 07/28/2021
% Minor spacing/hyphenation/punctuation consistency fixes, use siunitx throughout.
% Replaced includegraphics* by includegraphics.
% Allow more liberal float placement for now.
% Relabeled all floats for consistency.

% MB 08/05/2021: Replaced fig:injector_layout with new vector version from Benjamin

% MB 08/12/2021: Replace fig:multiturnoptics with new pgfplots version using Data from Kévin

\subsection{Facility Overview}
\begin{figure}[htb]
    \centering
    \includegraphics[width=.99\textwidth]{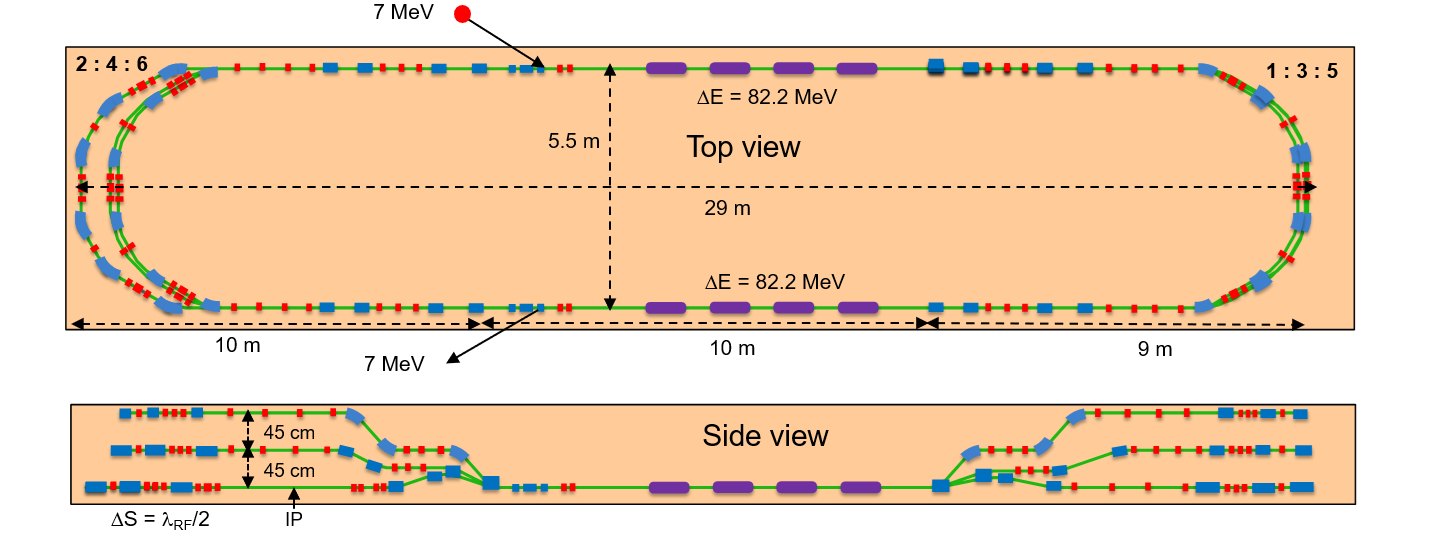}
    \caption{\label{fig:new_facilities:perle:layout} Top and side views of the PERLE facility
    planned to be built at IJClab Orsay.
    An electron energy of \SI{500}{\mega\electronvolt} is achieved in three turns passing through two 
    cryomodules, each housing four 5-cell cavities of \SI{802}{\mega\hertz} frequency.}
\end{figure}

PERLE, a Powerful Energy Recovery Linac for Experiments, emerged from the design of the 
Large Hadron Electron Collider as a 3-turn racetrack configuration with a linac
in each straight. Its principles were published first at the IPAC conference 
2014~\cite{Jensen:IPAC2014-TUOBA02}.
The CDR of the LHeC in 
2012~\cite{AbelleiraFernandez:2012cc} assumed an electron current of
 \SI{6.8}{\milli\ampere} to reach the initial design luminosity 
 of \SI{e33}{\per\square\centi\meter\per\second}, which was and still is considered to be large
 when compared to the HERA values, between 1 and \SI{4e31}{\per\square\centi\meter\per\second}.
The discovery of the Higgs boson  made it desirable 
to target a tenfold higher luminosity value than envisioned in that CDR. This was possible 
since the LHC reached a higher brightness than assumed:
the $\beta^*$ could possibly be reduced to below \SI{10}{\centi\meter} and  the electron current increased,
as was discussed in 2013~\cite{Bruening:2013bga,Zimmermann:2013aga}.  
The default electron beam current of
the LHeC became \SI{20}{\milli\ampere}~\cite{LHeC:2020jxs}, and this value has now been adopted
for PERLE. Emulating the LHeC configuration with one instead of about 50
cryomodules per linac determined the final electron energy to be \SI{500}{\mega\electronvolt}
and the footprint of the facility as shown in Fig.\,\ref{fig:new_facilities:perle:layout}.
In 2017, a group of accelerator, particle and nuclear physicists presented a PERLE design 
concept~\cite{Angal-Kalinin:2017iup}, together with detailed considerations for physics
and other applications, at that time still for \SI{1}{\giga\electronvolt} beam energy.
PERLE has now been established as a Collaboration of Institutes with mostly long 
experience on ERL, SRF and magnet technology as well as operation. The facility
will be hosted by Ir\`{e}ne Joliot Curie Laboratory at Orsay, and be built by
a collaboration of BINP Novosibirsk, CERN, University of Cornell, IJClab Orsay, Jefferson Laboratory Newport News, 
University of Liverpool, and STFC Daresbury including the Cockcroft Institute, with others expressing 
interest. Recently, 
an ambitious plan was endorsed aiming for first PERLE beam operation, with initially one linac, in the mid twenties. The Collaboration intends to use the ALICE gun, the JLEIC booster
and the SPL~\cite{SPL:2014} cryomodules as key components for an early start,  while the bulk funding is still to be realised.  
The importance of PERLE reaches far beyond the role it has for LHeC. Its parameters make it
very suitable for exploring ERLs in the new, \SI{100}{\milli\ampere} range, required also for further high-energy 
colliders, and its intense, low-emittance beam provides an ideal base for novel particle
and nuclear physics experiments.

\subsection{Injector}

In recent years the PERLE injector design has been pursued and  is approaching tentative conclusions.
A \SI{20}{\milli\ampere} current corresponds to \SI{500}{\pico\coulomb} bunch charge at \SI{40}{\mega\hertz} frequency as prescribed
by the LHC.
Delivery of such high-charge electron bunches into the main loop of an ERL while preserving the emittance is challenging. This is because at the typical injection momentum, space charge forces still have a significant effect on the beam dynamics. 
Simulations have shown that the baseline DC-gun-based injector can achieve the required emittance at the booster linac exit. The quality of the \SI{500}{\pico\coulomb} bunches must then be preserved with space charge through the merger at a total beam energy of \SI{7}{\mega\electronvolt} keeping the emittance below \SI{6}{\milli\meter\milli\radian}.
\begin{table}[htb]
   \centering
   \caption{PERLE merger specification}
   \begin{tabular}{lcc}
       \toprule
       \textbf{Parameter} & \textbf{Values} \\
       \midrule
           Bunch charge         & \SI{500}{\pico\coulomb} \\
           Emittance         & $\leq \SI{6}{\milli\meter\milli\radian}$ \\
           Total injection energy      & $\SI{7}{\mega\electronvolt}/c$ \\
           First arc energy        & \SI{89}{\mega\electronvolt} \\
           RMS bunch length       & \SI{3}{\milli\meter} \\
           Maximum RMS transverse beam size       & \SI{6}{\milli\meter} \\
           Twiss $\beta$ at 1st main linac pass exit & \SI{8.6}{\meter} \\
           Twiss $\alpha$ at 1st main linac pass exit & \num{-0.66}  \\
       \bottomrule
   \end{tabular}
   \label{tab:new_facilities:perle:specification}
\end{table}
The beam dynamics in the merger were simulated using the code OPAL and optimised using a genetic algorithm. Three possible merger schemes were investigated. The goal of the optimisation was to minimise the emittance growth while also achieving the required Twiss parameters to match onto the spreader at the main linac exit. A three-dipole solution was then examined in more detail.
Table~\ref{tab:new_facilities:perle:specification} shows the requirements on the beam at the exit of the main linac after the first pass.

\begin{figure}[htb]
    \centering
    \includegraphics[width=.8\linewidth]{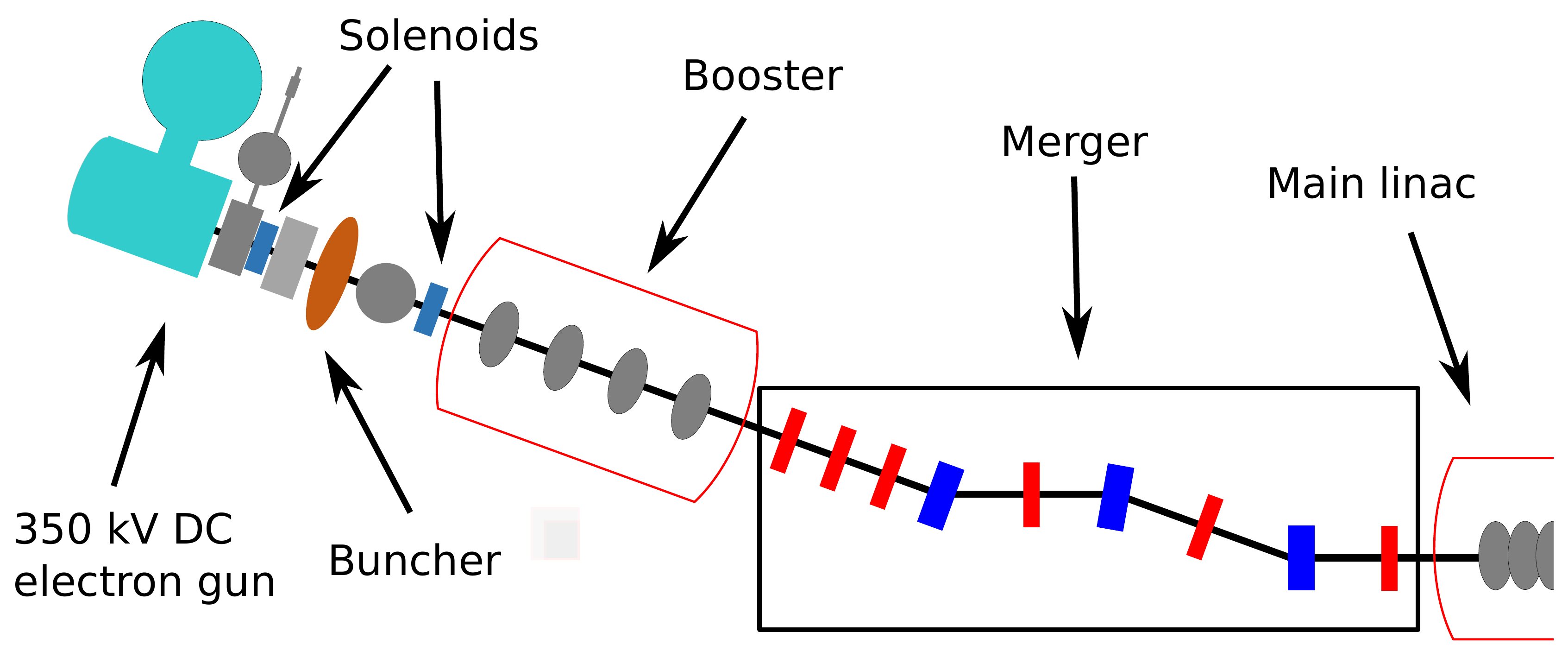}
    \caption{The layout of the injector with an example of a possible merger scheme.}
    \label{fig:new_facilities:perle:injector_layout}
\end{figure}
To achieve this low emittance with high average current, a DC-gun-based injector will be used.
This injector will consist of a \SI{350}{\kilo\volt} photocathode electron gun, a pair of solenoids for transverse beam size control and emittance compensation, an \SI{801.58}{\mega\hertz} buncher cavity, a booster linac consisting of four single cell \SI{801.58}{\mega\hertz} SRF cavities, and a merger to transport the beam into the main ERL loop.
The Twiss matching to the optics of the main ERL loop is also done in the merger.
The layout of the injector with a possible merger example can be seen in Fig.~\ref{fig:new_facilities:perle:injector_layout}. 
%After the injection line follows the main linac which consists of four five cell 801.58~MHz SRF cavities. 

\subsection{Accelerator Characteristics}
%
%PERLE is a compact three-pass ERL project based on SRF technology, pushing as a new generation machine the operational regime for multi-turn ERLs to around 10\,MW beam power level. Apart from the experiments it could host thanks to its beam characteristics, 
PERLE will serve as a hub for the validation and exploration of a broad range of accelerator phenomena in an unexplored operational power regime.
% serving for the development of ERL technology for future energy and intensity frontier machines. 
A summary of the design parameters is presented in Table~\ref{tab:new_facilities:perle:parameters}. 
\begin{table}[htb]
   \centering
   \begin{tabular}{lcc}
       \toprule
       Parameter & unit & value \\
       \midrule
       Injection beam energy & MeV  & 7\\
       Electron beam energy & MeV & 500 \\
       Norm. emittance $\gamma\varepsilon_{x,y}$ & \si{\milli\meter\milli\radian} & 6 \\
       Average beam current & mA & 20 \\
       Bunch charge & pC & 500 \\  
       Bunch length & mm & 3 \\
       Bunch spacing & ns & 24.95 \\
       RF frequency & MHz & 801.58 \\
       Duty factor & & CW \\
       \bottomrule
    \end{tabular}
    \caption{PERLE Beam Parameters}
    \label{tab:new_facilities:perle:parameters}
\end{table}
The bunch spacing in the ERL is assumed to be \SI{25}{\nano\second}; however, empty bunches might be required in the ERL for ion clearing gaps.
%It  features a 3-turn acceleration and 3-turn deceleration racetrack configuration, 802\,MHz SRF system and beam currents of up to 20\,mA (e.g. 2x3x20 mA = 120\,mA in the SRF cavities). 
PERLE will study important characteristics such as: CW operation, handling a high average beam current, low delivered beam energy spread and low delivered beam emittance. 
%The two linacs are composed of one Superconducting Proton Linac (SPL)\cite{SPL:2014} style cryomodule with four 5-cell 802\,MHz RF cavities. 

The linac optics design minimises the effect of wakefields such that the beta function must be minimised at low energy.
The ERL is operated on crest in order to benefit from the maximum voltage available in the cavity. 
%A schematic of the ERL geometry is shown in Fig.~\ref{fig:PERLElayout}. 
The spreaders/recombiners connect the linac structures to the arcs and route the electron bunches according to their energies. The design is a two-step achromatic vertical deflection system and features a specific magnet design in order to gain in compactness.

The three arcs on either side of the linacs are vertically stacked and composed of 6 dipoles instead of 4 dipoles with respect to the previous design \cite{Angal-Kalinin:2017iup}, reducing the effects of CSR. Moreover, the arc lattice is based on flexible-momentum-compaction optics such that the momentum compaction factor can be minimised but also adjusted if needed.
The low energy implies that the energy spread and emittance growth due to incoherent synchrotron radiation is negligible in the arcs.

The ERL lattice design provides a pair of low-beta insertions for experimental purposes, and the multi-pass optics optimisation gives a perfect transmission with the front-to-end tracking results including CSR.
Multi-bunch tracking has shown that instabilities from HOM can be damped with frequency detuning.
The optimal bunch recombination pattern gives some constraints on the length of the arcs. Furthermore, the arc with the low-beta insertions will provide the necessary shift to the decelerating phase in the RF cavities.
There are two chicanes in the lattice, located at the entrance of a linac and symmetrically at the exit of the other linac structure. They are needed to allow injection and extraction through a constant field. 
%The injection in the ERL takes place at 7\,MeV with an angle of around $20^{\circ}$, a precise description of the injector beam line can be found in~\cite{Ben:injector}.

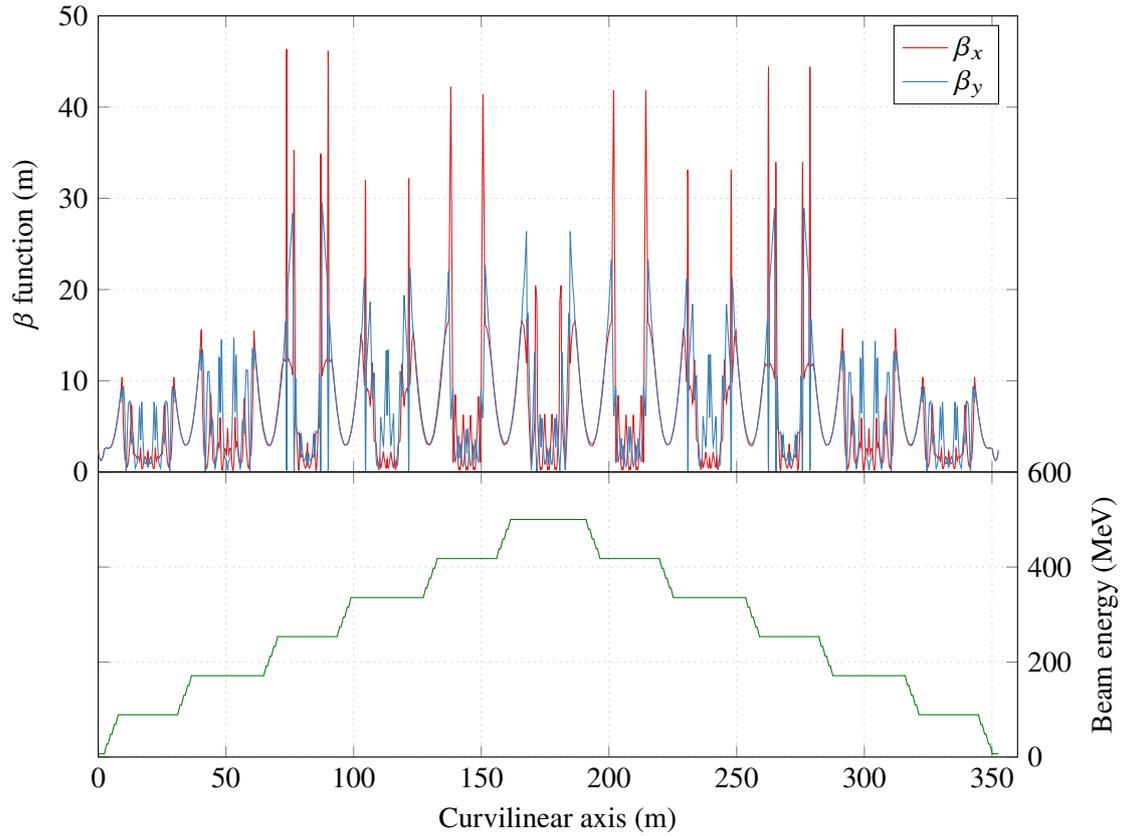
\begin{figure}[htb]\centering
\tikzsetnextfilename{perle_multiturnoptics}
\begin{tikzpicture}
\begin{groupplot}
[
    group style={
        group size=1 by 2,
        vertical sep=0,
    },
    width=.8\linewidth,
    height=.4\linewidth,
    scale only axis,
%    axis y line=left,
    ymajorgrids=true,
    xmajorgrids=true,
    major grid style={dotted},
    ylabel={$\beta$ function (m)},
    xmin=0,
    xmax=360,
    ymin=0,
    ymax=50,
]
\nextgroupplot[xticklabel=\empty]
\addplot+[mark=none] table[x index=0, y index=2] {figures/perle_multiturnoptics.csv};
\addlegendentry{$\beta_x$}
\addplot+[mark=none] table[x index=0, y index=3] {figures/perle_multiturnoptics.csv};
\addlegendentry{$\beta_y$}
\nextgroupplot[
    height=.25\linewidth,
    yticklabel pos=right,
    ymin=0,
    ymax=600,
    xlabel={Curvilinear axis (m)},
    ylabel={Beam energy (MeV)},
]
\addplot+[green!50!black, mark=none] table[x index=0, y expr={\thisrowno{1}*1e3}] {figures/perle_multiturnoptics.csv};
%\addlegendentry{Energy}
\end{groupplot}
\end{tikzpicture}
\caption{\label{fig:new_facilities:perle:multiturnoptics}Plot of the beta functions and the beam energy along the multi-turn ERL. PERLE has two linacs and 3 passes, which leads to a six-fold increase and subsequent decrease of the beam energy.}
\end{figure}

The optics design of the multi-turn ERL is shown in Fig.~\ref{fig:new_facilities:perle:multiturnoptics} and presents the sequence of linacs and arcs leading to the two interaction regions where experiment setups will be placed.
Note the relatively low values of the beta function along the ERL since the beam emittance is in fact larger at low energy.

\subsection{Prospect}
A vigorous R\&D program is being pursued to develop a Technical Design Report for
PERLE at Orsay within the next year. To achieve this goal, tentatively the
following sequence of accelerator design studies has been identified:
\begin{itemize}
\item Completion of the injector design
\item Momentum acceptance and longitudinal match
\item Start-to-end simulation with synchrotron radiation, CSR micro-bunching
\item HOM design and tests of a dressed cavity
\item Preparation of ALICE gun installation at Orsay
\item Multi-pass wake-field effects, BBU studies
\item Injection line/chicane design including space-charge studies at injection
\item Design of PERLE diagnostics
\item Preparation of facility infrastructure
\end{itemize}
The collaboration is aiming at the PERLE Technical Design Report to be concluded by end
of 2022, with the goal of achieving the first beam at PERLE by the mid-twenties. Important milestones will be the delivery and equipment of the JLEIC booster cryostat to Orsay and the production and test
of the complete linac cavity-cryomodule, as the first linac for 
PERLE and the 802\,MHz cryomodule demonstrator as part of the 
FCC-ee feasibility project. Further details on the current design 
of PERLE can be found in Ref.\,\cite{alexperledis21}.
The multi-turn,
high-current, small-emittance configuration
and the  time line of PERLE make it a central part of the European
plans for the development of energy-recovery linacs.

%\subsection{PERLE at Orsay} 
%Oliver Bruening, Walid Kaabi

%\section{Non-European Facilities}
\section{CEBAF 5-pass Energy Recovery Experiment} 
\label{sec:new_facilities:cebaf5}
 %Alex Bogacz
\subsection{Modifications to CEBAF}
\begin{figure}[tbh]
  \centering
  \includegraphics[width=0.7\textwidth]{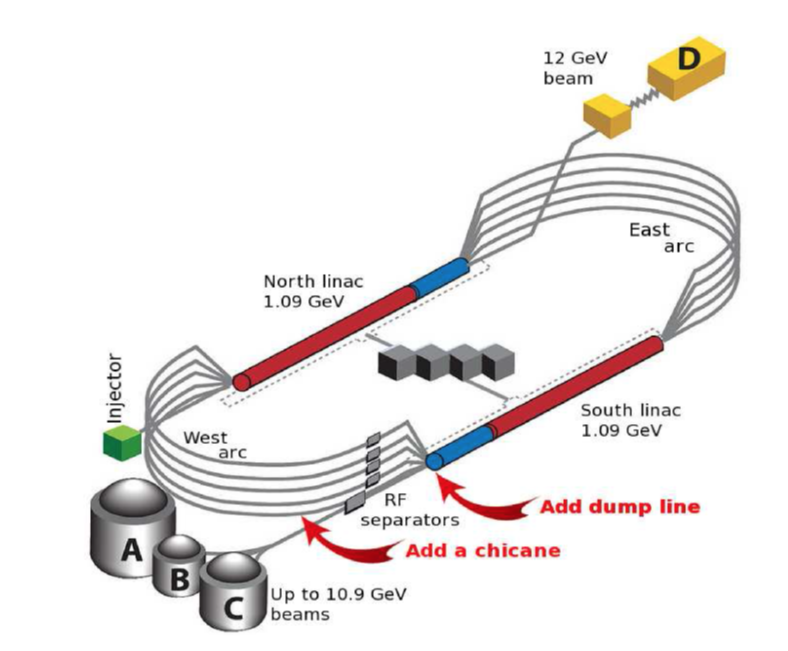}
  \caption{The CEBAF accelerator, with arrows indicating planned hardware installations for the ER@CEBAF experiment.}
  \label{fig:new_facilities:cebaf5:schematic}
\end{figure}

CEBAF presently accelerates CW beams for delivery to Hall D at \SI{12}{\giga\electronvolt} (11 linac passes) and Halls A--C at \SI{11}{\giga\electronvolt} and below (up to 10 linac passes).
After use in the halls, beam is delivered to the respective hall beam dumps at beam power up to \SI{1}{\mega\watt} for Halls A and C (high-current halls), \SI{60}{\kilo\watt} for Hall D, and \SI{1.5}{\kilo\watt} for Hall B (low-current halls).
All beam is accelerated on RF crest, and there is no energy recovery in routine CEBAF operations.
Energy recovery would be made feasible in CEBAF by the addition of two modest hardware sections: a path-length chicane insertion at the start of Arc A, and a low-power dump line at the end of the south linac (SL), before the first west spreader dipole magnet.
These areas are indicated in Fig.~\ref{fig:new_facilities:cebaf5:schematic}.
These alterations are designed to remain in place permanently and do not interfere with any capability of routine CEBAF \SI{12}{\giga\electronvolt} operations.
\par
\subsection{New path-length chicane}
\begin{figure}[tbh]
  \centering
  \includegraphics[width=\textwidth]{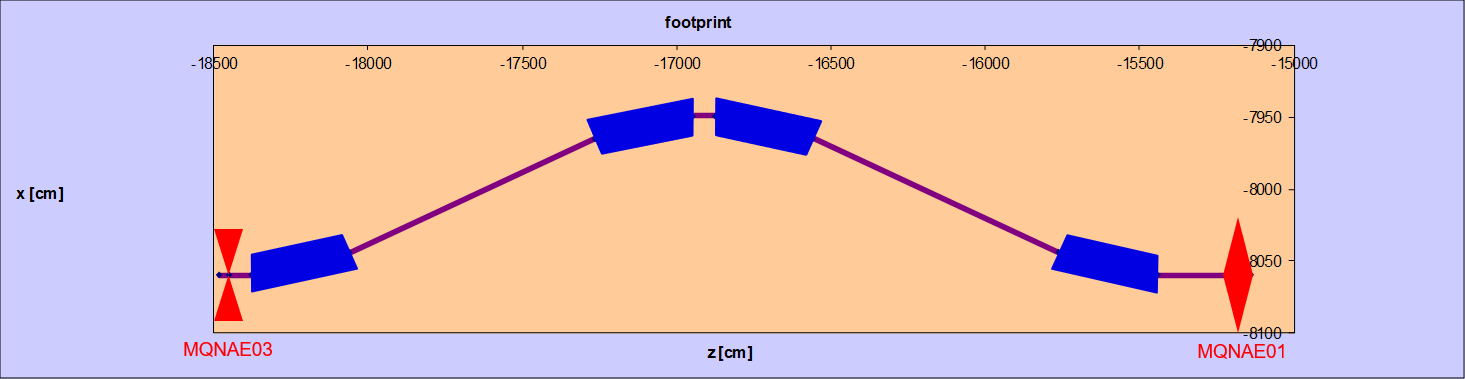}
  \includegraphics[width=\textwidth]{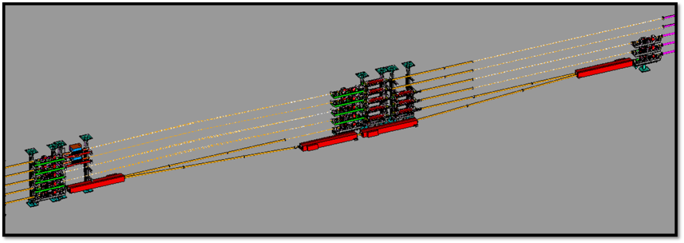}
  \caption{ER@CEBAF new AE-region path-length chicane with a bend angle of \ang{5} and a dipole length of \SI{3}{\meter}.}
  \label{fig:new_facilities:cebaf5:chicane}
\end{figure}

To invoke energy-recovery mode for ER@CEBAF, the beam must pass
through an additional half of the CEBAF RF wavelength (\SI{10}{\centi\meter}) in the
final arc (Arc A), which will be implemented with a new \SI{31}{\meter}-long
chicane in the matching straight just before the
arc. The chicane is configured with four standard CEBAF MBA 3-meter
40-turn dipole magnets as illustrated in
Fig.~\ref{fig:new_facilities:cebaf5:chicane}. The required maximum individual dipole
field at the peak beam energy of \SI{7073}{\mega\electronvolt} is about \SI{0.7}{\tesla}. 

Support of both ``straight through'' and ``chicane bypass'' modes
requires a special beampipe design with a \ang{5} split. This type
of vacuum chamber has already been used at CEBAF; there are no
additional vacuum concerns for this insertion.
An engineering perspective layout of the four chicane MBA dipoles, with the middle magnets extending
about 1 m into the aisle and additional lower-energy passes, is illustrated in Fig.~\ref{fig:new_facilities:cebaf5:chicane}. This chicane
would not interfere with tunnel passage or clearances in this area.

%%%%%%%%%%%%%%%%%%%%%%%%%%%%%%%%%%%%%%%%%%%%%%%%%%%%%%%%%%%%%%%%%%%%%%
\subsection{Longitudinal Match and Considerations}\label{S:longMatch}
%%%%%%%%%%%%%%%%%%%%%%%%%%%%%%%%%%%%%%%%%%%%%%%%%%%%%%%%%%%%%%%%%%%%%%

The longitudinal stability in ERLs is dependent upon judicious choices
for accelerating and decelerating RF phases, the initial bunch length
and energy spread of the beam, and the chromatic characteristics of
the return arcs.  ISR-driven energy loss in the
high-energy arcs, in both accelerating and decelerating passes,
results in two beams of different momentum traversing each arc.
Finally, as a consequence of anti-damping during deceleration, the
relative energy spread of the beam becomes larger during deceleration
and energy recovery. All of these effects can be mitigated by
performing the appropriate longitudinal phase space manipulations.

The injection chicane, located at the end of the injector, can be
used to perform initial bunch compression. We can also separately
adjust linac phases for acceleration and deceleration passes, and
control $M_{56}$ (linear momentum compaction) in each of the CEBAF
arcs separately.

\subsection{Momentum Acceptance}
The limiting factor for ER@CEBAF with 5 passes is the arc momentum
acceptance, which places a bound on the maximum energy gain one can
support in the linacs. Above that energy gain, the ISR losses are sufficiently large that the energy separation
between accelerated and decelerated beams in the last arcs is larger
than the arc momentum acceptance.

The standard configuration for the CEBAF ARC1 and ARC2 is high-dispersion optics
(8 and 6 meters peak, respectively). For the CEBAF ER
experiment, these optics were redesigned to be low dispersion like the
other arcs in the nominal CEBAF \SI{12}{\giga\electronvolt} design to give them larger
momentum acceptance. Changing these arcs to this configuration only
requires changes to quadrupole magnet setpoints. All arcs are first-order (but not second) achromats.
The initial estimates of the momentum acceptance were performed using a
simple spreadsheet and established ISR energy loss and energy spread formulas
from Sands~\cite{Sands}. From this, it was determined that the maximum
feasible energy gain for ER@CEBAF is \SI{700}{\mega\electronvolt} per linac, with a likely hard
bound somewhere below \SI{750}{\mega\electronvolt} per linac.

\begin{table}[htb]\centering
\caption{Pass-by-pass beam energy parameters for 5-pass ER@CEBAF at \SI{700}{\mega\electronvolt} per linac and phases as described in the text. $\Delta E$ is energy lost to SR in each arc. All energies are in MeV according to theory; the exit momentum at the end of each arc, $p_\text{exit}$, is obtained from tracking and includes small effects including spreader/recombiner and quadrupole SR energy losses.}
\label{TAB:passByPassEnergies}
\begin{tabular}{lccccc}
\hline\hline
Arc (accelerating) & $E_\text{entrance}$ & $E_\text{exit}$ & $\Delta E$ & $E_\text{loss, total}$ & $p_\text{exit}$ (tracking) \\
\hline
1 & 779.00   & 779.00  & 0.00 & 0.00 & 778.98 \\
2 & 1479.00  & 1478.98 & 0.02 & 0.02 & 1478.96 \\
3 & 2178.98  & 2178.88 & 0.10 & 0.12 & 2178.76 \\
4 & 2878.88  & 2878.73 & 0.15 & 0.27 & 2878.48 \\
5 & 3578.73  & 3578.37 & 0.36 & 0.63 & 3577.79 \\
6 & 4278.37  & 4277.89 & 0.49 & 1.11 & 4276.95 \\
7 & 4977.89  & 4977.00 & 0.89 & 2.00 & 4975.99 \\
8 & 5677.00  & 5675.49 & 1.51 & 3.51 & 5674.55 \\
9 & 6375.49  & 6373.10 & 2.40 & 5.90 & 6372.26 \\
10 & 7073.10 & 7070.37 & 2.72 & 8.63 & 7069.98 \\
\hline
Arc (decelerating) & $E_\text{entrance}$ & $E_\text{exit}$ & $\Delta E$ & $E_\text{loss, total}$ & $p_\text{exit}$ (tracking) \\
\hline
9 & 6371.83  & 6369.44 & 2.39 & 11.02 & 6370.19 \\
8 & 5670.90  & 5669.40 & 1.50 & 12.52 & 5672.07 \\
7 & 4970.85  & 4969.97 & 0.89 & 13.40 & 4966.02 \\
6 & 4271.42  & 4270.94 & 0.48 & 13.88 & 4966.61 \\
5 & 3572.39  & 3572.04 & 0.35 & 14.24 & 3568.62 \\
4 & 2873.50  & 2873.35 & 0.15 & 14.39 & 2871.69 \\
3 & 2174.80  & 2174.71 & 0.10 & 14.48 & 2174.30 \\
2 & 1476.16  & 1476.14 & 0.02 & 14.50 & 1477.65 \\
1 & 777.60   & 777.59  & 0.00 & 14.51 &  780.56 \\
dump & 79.04 & ---     & 0.00 & 14.51 &  83.50  \\
\hline
\end{tabular}
\end{table}

To maximize the momentum acceptance, the average momentum was calculated for the accelerated
and decelerated beams in each arc to obtain the resulting average momentum.

%
%%%%%%%%%%%%%%%%%%%%%%%%%%%%%%%%%%%%%%%%%%%%%%%%%%%%%%%%%%%%%%%%%%%%%%
\subsection{Multi-pass Linac Optics}
%%%%%%%%%%%%%%%%%%%%%%%%%%%%%%%%%%%%%%%%%%%%%%%%%%%%%%%%%%%%%%%%%%%%%%

\begin{figure}[tbh]
  \centering
  \includegraphics[width=\textwidth]{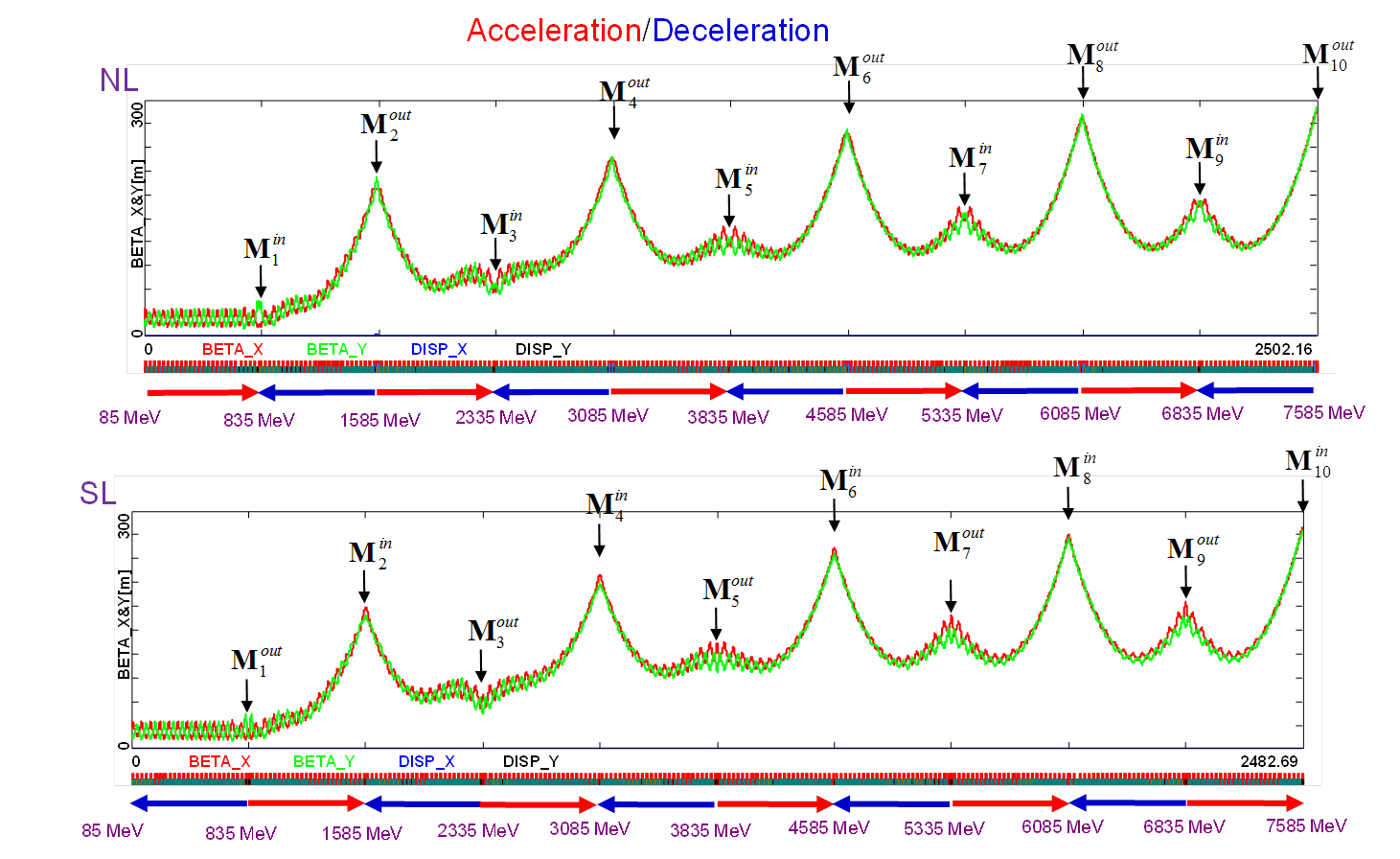}
  \caption{Complete multi-pass optics for both linacs optimized as slightly perturbed \ang{60} FODO}
  \label{fig:new_facilities:cebaf5:Linacs_Multipass}
\end{figure}

Energy recovery in a racetrack topology explicitly requires that both
the accelerating and decelerating beams share the individual return
arcs. This, in turn, imposes specific requirements for the Twiss functions
at the linacs ends: the Twiss functions have to be identical for both the accelerating and decelerating linac passes converging to the same
energy and thus entering the same arc, therefore requiring corresponding matching conditions at the linac ends.
To visualize the beta functions for multiple accelerating and decelerating
passes through a given linac, it is convenient to reverse the linac
direction for all decelerating passes and string them together with
the interleaved accelerating passes, as illustrated in
Fig.~\ref{fig:new_facilities:cebaf5:Linacs_Multipass}. This way, the corresponding
accelerating and decelerating passes are joined together at the
entrance/exit of the arc. Therefore, the matching conditions are automatically
built into the resulting multi-pass linac beamline. The figure
illustrates the optimum focusing profile for both North and South
linacs based on a FODO-like (\ang{60} phase advance) lattice. One can
see that both linacs uniquely define the Twiss functions for the arcs:
the NL fixes the input to all odd arcs and the output to all even arcs, while
the SL fixes the input to all even arcs and the output to all odd arcs.

To optimize the multi-pass linac optics, we conducted a thorough
exploratory study of the optimum phase advance in the linac FODO
structure; spanning from no-focusing drift linac to a strongly
focusing \ang{120} FODO (present CEBAF linac optics)
~\cite{Bogacz_Multipass_Optics}. A single-objective optimization was done to minimize
$\beta/E$ averaged over all accelerating and decelerating passes, where $\beta$ is the betatron function and $E$ the energy of the beam.
This quantity is a driving term for most collective phenomena in
recirculating linacs. As a result, the optimum value of the linac
phase advance was found to be \ang{60}. The optimum lowest-pass linac
optics was configured as a slightly perturbed \ang{60} FODO with
modified quads at the linac end.

%%%%%%%%%%%%%%%%%%%%%%%%%%%%%%%%%%%%%%%%%%%%%%%%%%%%%%%%%%%%%%%%%%%%%%
\subsection{New Arc Optics for ER}
%%%%%%%%%%%%%%%%%%%%%%%%%%%%%%%%%%%%%%%%%%%%%%%%%%%%%%%%%%%%%%%%%%%%%%
As concluded above, both linacs uniquely define the Twiss functions
for the arcs. Therefore, the optics settings of all 10 CEBAF arcs will have to be modified to meet the new values of the Twiss functions at both arc ends.
Since the linacs are almost identical, the required matching conditions are very close to mirror-symmetric for the arcs.
To redesign the arcs, we modified the spreader
and recombiner sections along their matching straights, leaving the
arc proper intact. This procedure was carried out for all 10 arcs. The acceleration process continues pass-by-pass through the SL pass after Arc 9.
Finally, the beam at top energy is transported through Arc A, where it gains an extra half-wavelength (via a path-length delay chicane), and it ends up being decelerated in the following passes through both the North and South linacs.
Similarly, the deceleration continues pass-by-pass through Arc 3, and it finally reaches the last decelerating pass. Finally, the beam is extracted into the dump at the end of the SL.%
This process results in the parameters listed in Table~\ref{TAB:passByPassEnergies}.

%%%%%%%%%%%%%%%%%%%%%%%%%%%%%%%%%%%%%%%%%%%%%%%%%%%%%%%%%%%%%%%%%%%%%%
\subsection{Operational Challenges---Diagnostics}
%%%%%%%%%%%%%%%%%%%%%%%%%%%%%%%%%%%%%%%%%%%%%%%%%%%%%%%%%%%%%%%%%%%%%%
Diagnostic measures, with potential calibration needs, are coupled with justification from procedural steps. This is likely
a minimally sufficient list since a successful demonstration of ER capability will illuminate issues, which drive later diagnostic development. They are aimed at supporting the following operational scenarios.

Once the beam is brought to the highest pass, delayed by a half-period of the accelerating RF, and directed into the first decelerating linac, it will undergo an almost unstoppable tumble down the accelerator
to its lowest energy and the extraction path. There is no straightforward mechanism with installed
beam-line components for differential steering control of the accelerating and decelerating passes. The RF separators provide the only obvious way to stop this \enquote{avalanche} at intermediate points, simultaneously providing a path for detailed circuit by circuit beam diagnostics on the decelerating path. Preparation may be made as outlined below to divert the beam on selected passes into the Hall-directed extraction lines for characterization.

Multi-energy steering of the beams had been problematic in CEBAF operation without compensation of the skew focusing inherent in the 5-cell accelerating cavities of the linacs. The cavity RF skew field
inverts on deceleration, unlike the DC magnetic compensation.
The presently employed skew compensation method works only on acceleration or on deceleration, but not both.
Recent accelerator tests have demonstrated
that with the \SI{12}{\giga\electronvolt} re-alignment of the linacs, and with linac steering guided by global analysis of the beam trajectories of differing energy, the RF skew focusing no longer poses a large problem. While this
compensation is normal operating procedure, it appears that the baseline for this ERL demo should omit skew compensation.

The CEBAF linac BPMs use time-of-flight separation of a \SI{4.2}{\micro\second} test pulse in a system with a round-trip circuit time of \SI{4.4}{\micro\second}.
During setup in \enquote{Tune Mode}, each of the various pass beam positions is identified by a time delay from a sync pulse. All of the \SI{1497}{\mega\hertz} BPMs fail in measuring concurrent accelerating and decelerating beam pulses because of destructive interference of the two signals.
The temporal separation of the accelerating and decelerating beam
macro-pulses allows the linac SEE BPMs to register all beam passes separately.
At the time of this writing, this system is configured only for 5 passes in the South Linac, extended to 6 passes (to support Hall D) in the North Linac. It was extremely useful during the 2003 single-pass ERL demonstration. It will be extended to 10 passes (5 up, 5 down) to provide adequate information for setup, operation, and fault detection for the ERL demo.

Furthermore, Arc 1 beam energy measurement of the decelerated beam requires knowledge of beam positions in Arc 1. This might be obtained with sufficient accuracy via OTR viewers installed at selected
locations in Arc 1, or alternatively by use of \SI{3}{\giga\hertz} BPM signal processing.
Another alternative, drilling holes into the existing arc viewers and directing the accelerated beam through the holes, is
not as viable a candidate because the full-energy, SR-affected configuration will not necessarily have the accelerated beam at the nominal matched arc energy. A minimal installation would monitor
two non-dispersive points and at least one following dispersive point. A parallel installation in Arc 9 determines the Arc 9
decelerating beam energy. All subsequent beam energies are determined by decisions made to this point, and it is reasonable to verify the result of these decisions in Arc 9. The Arc 1 BPMs also enable a precise comparison of the two beam energies in Arc 1 to test performance of longitudinal manipulations intended to compensate for SR-induced energy droop.

%%%%%%%%%%%%%%%%%%%%%%%%%%%%%%%%%%%%%%%%%%%%%%%%%%%%%%%%%%%%%%%%%%%%%%
\subsection{Measurements}
%%%%%%%%%%%%%%%%%%%%%%%%%%%%%%%%%%%%%%%%%%%%%%%%%%%%%%%%%%%%%%%%%%%%%%
In order to avoid head-tail emittance growth, only one RF separator system will be active at any given time. Normal CEBAF operation allows two (or more) RF separators to be active, as convenient for intermediate extraction while beam is directed to higher passes. This is not necessary for the ERL demo and, in principle, irrecoverably increases the beam emittance. 
A detailed outline of measurements has been generated using the ability to divert beam from its downward cascade into the successive extraction lines. The beam energy for the accelerating and decelerating beams will be measured using the well-calibrated Hall A dipole system (``9th dipole'' system), and the 4 meter dispersion optics configuration will allow the momentum spread to be measured with IHA1C12 and
the associated viewer.
At this same time, the machine protection system will provide a detailed comparison of the relative current transmission from the injector to the accelerated and decelerated beam passes.
The beam envelope parameters and emittances will be measured in the zero-dispersion 2C line, which is instrumented for such measurements.
Addition of thicker wires and/or plates to the wire scanner harp will enable high-resolution halo detection on a pass-dependent basis.
The diagnostic opportunities at the final energy-recovery dump are more restricted than in the 2C/1C lines, but emittances and momentum spread can also be measured to allow a comparison between an initial NL-accelerated, SL-decelerated beam and the final 5-pass up/down beam with its additional SR-driven energy spread and consequent chromatically driven emittance and halo growth.
The dependence of the beam parameters on the CW beam current is accessible only for the limited regions presently equipped with synchrotron light monitors. The possibility of adding more monitors is under consideration, but the Hall A beam line already has such a monitor, which is well-placed near the high-dispersion point, and the ER dump may also be so equipped.
%\subsection{CEBAF 5-Pass at Jlab} 
%Alex Bogacz

\section{Electron Cooler at BNL}\label{sec:new_facilities:bnl_electron_cooler}
%Dimitry Kayran, Vladimir Litvinenko, Erdong Wang

\begin{figure}[htb]\centering
\includegraphics[width=.4\linewidth]{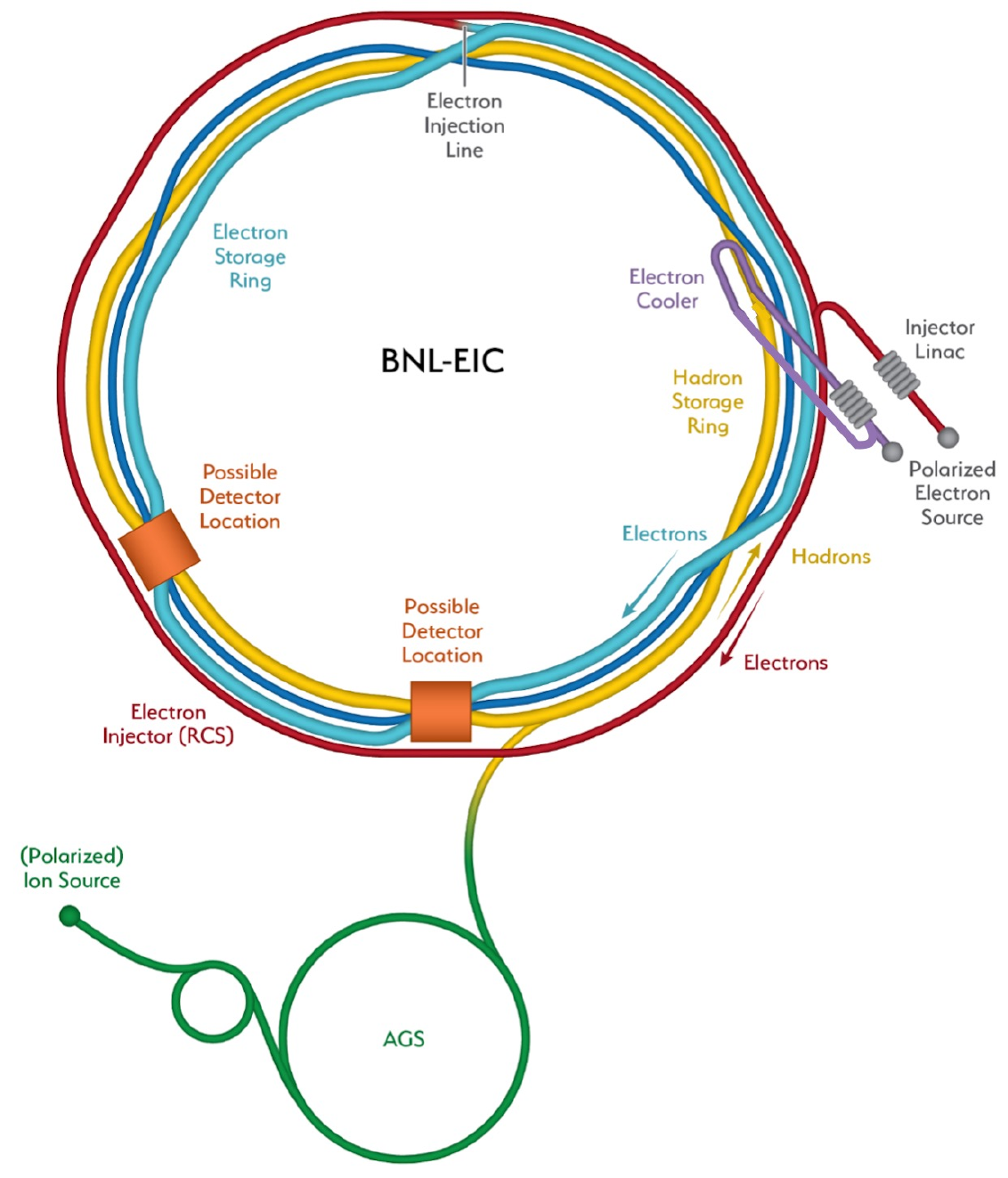}\quad%
\tikzsetnextfilename{bnl_electron_cooler_schematic}
\begin{tikzpicture}
\begin{axis}
[
    width=.5\linewidth,
    height=7.5cm,
    ymode=log,
    xmin=0,
    xmax=160,
    ymin=0.1,
    ymax=20,
    xlabel={Center-of-Mass Energy (GeV)},
    ylabel={Luminosity (\SI{e33}{\per\square\centi\meter\per\second})},
    ymajorgrids=true,
    yminorgrids=true,
    xmajorgrids=true,
    legend pos=north west,
    ytick={0.1, 1, 10},
    yticklabels={0.1, 1, 10},
    xtick={0, 40, 80, 120, 160},
]
\addplot+[thick, mark size=2pt] coordinates {
    (28.6    , 0.44)
    (44.7  , 3.68)
    (63.2  , 4.48)
    (104.9 , 10.0)
    (140.7 , 1.54)
};\addlegendentry{with SHC}
\addplot+[thick, mark size=2pt] coordinates {
    (28.6   , 0.14)
    (44.7  , 0.51)
    (63.2  , 0.64)
    (104.9 , 4.28)
    (140.7 , 0.54)
};\addlegendentry{w/o SHC}
\end{axis}
\end{tikzpicture}
\caption{Left: Schematic EIC layout including the electron cooler in a 200-meter-long straight section. Right: Maximum achievable luminosity in the EIC with and without strong hadron cooling (SHC).}%
\label{fig:new_facilities:bnl_electron_cooler:schematic}
\end{figure}
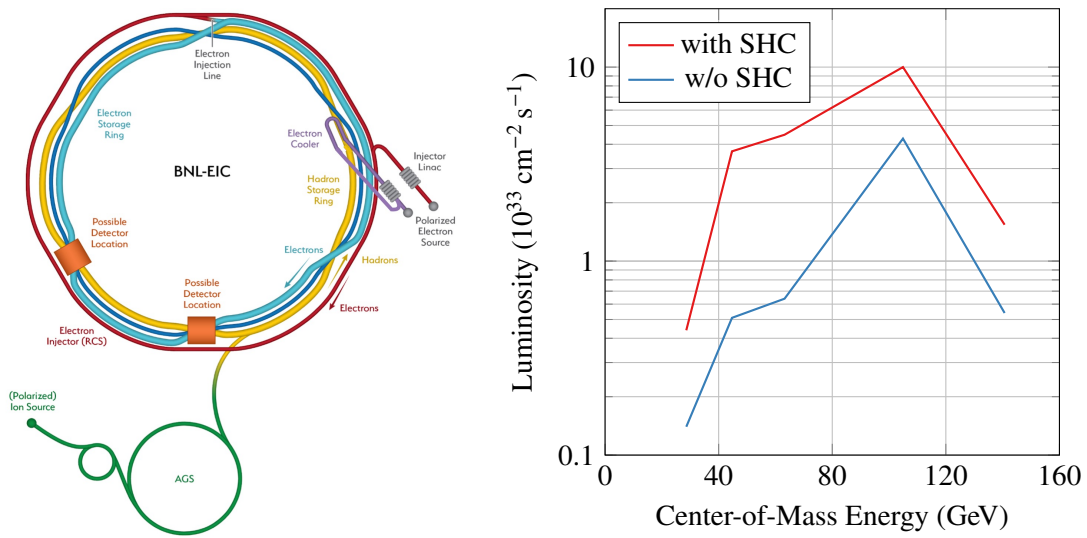

Brookhaven National Laboratory (BNL) has been selected to build the next US DOE nuclear physics facility called the Electron-Ion Collider or EIC.
A schematic view of the facility is shown in Fig.~\ref{fig:new_facilities:bnl_electron_cooler:schematic} (left).
The EIC will extend the existing Relativistic Heavy Ion Collider (RHIC) with an accelerator for polarized electrons to obtain electron-ion collisions~\cite{eic-cdr}.
A high luminosity of \SI{1e34}{\per\square\centi\meter\per\second} in the EIC can only be achieved using a strong beam cooling mechanism that counteracts IBS in the hadron bunches, which would otherwise cause a rapid increase of hadron emittance and a reduction of luminosity~\cite{eic-strong-hadron-cooling} as shown in Fig.~\ref{fig:new_facilities:bnl_electron_cooler:schematic} (right).
%In this paper, we present the requirements for an ERL which is required to drive the EIC cooler.

Electron cooling of hadron beams at the EIC top energy requires a \SI{150}{\mega\electronvolt} electron beam with an average power of \SI{15}{\mega\watt} or higher.
This task is a natural fit for an ERL driver, while being out of reach for DC accelerators~\cite{PhysRevLett.96.044801}.
BNL recently demonstrated successful e-cooling of a low-energy ion beam in RHIC using an RF accelerator~\cite{PhysRevLett.124.084801}.
However, the efficiency of traditional electron cooling---both magnetized and non-magnetized---falls as a high power of beam energy\footnote{The e-cooling time scales as $\gamma^{5/2}$, where $\gamma$ is the relativistic factor. For cooling to occur, the relativistic factors of electron and hadron beams must be equal.}.
For the EIC energy, a traditional e-cooler would require a multi-Ampere average current ERL, which is outside of the scope of this paper. 

Coherent Electron Cooling (CeC)~\cite{PhysRevLett.102.114801, PhysRevLett.111.084802, litvinenko2018plasmacascade, cool2013litvinenko} is a novel but untested technique with the potential to satisfy the stringent requirements for the EIC cooler.
CeC is a technique which uses an electron beam to perform all functions of a stochastic cooler~\cite{RevModPhys.57.689}: the pick-up, the amplifier, and the kicker.
The bandwidth of traditional RF stochastic cooling~\cite{PhysRevLett.100.174802} is limited to a few GHz.
In contrast, CeC amplifiers use the microbunching instability, which has a bandwidth of up to hundreds of THz~\cite{PhysRevLett.111.084802, litvinenko2018plasmacascade, PhysRevAccelBeams.24.014402, theory-of-plasma-cascade}, sufficient to cool the dense proton beam in the EIC.

\begin{figure}[htb]\centering
\includegraphics[width=\linewidth]{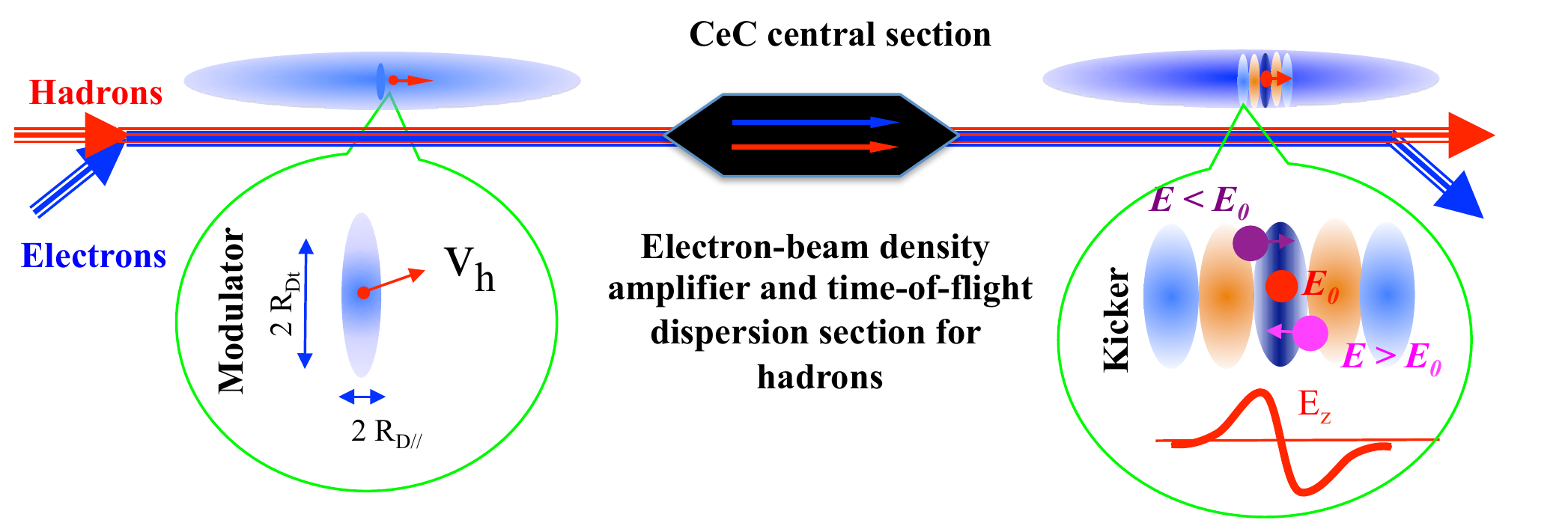}%
\caption{Schematic view of the principle of CeC. The device consists of three sections: a modulator, an amplifier plus a hadron dispersion section, and a kicker.}%
\label{fig:new_facilities:bnl_electron_cooler:cec}
\end{figure}

As shown in Fig.~\ref{fig:new_facilities:bnl_electron_cooler:cec}, the CeC consists of three sections: the modulator, the amplifier, and the kicker~\cite{PhysRevLett.102.114801, cool2013litvinenko}.
In the modulator, hadrons create a negative imprint of their density via a process known as Debye screening~\cite{PhysRevE.78.026413}.
The imprint is amplified by a selected microbunching instability in the central section, which also serves for the hadrons' time-of-flight dispersion.
In the kicker, the strongly---typically hundredfold---amplified imprint generates a longitudinal electric field, which is used to correct the energy of each individual hadron toward the central value.
Transverse cooling is accomplished by coupling between the longitudinal and transverse degrees of freedom~\cite{PhysRevLett.102.114801}.

\begin{figure}[htb]\centering
\includegraphics[width=\linewidth]{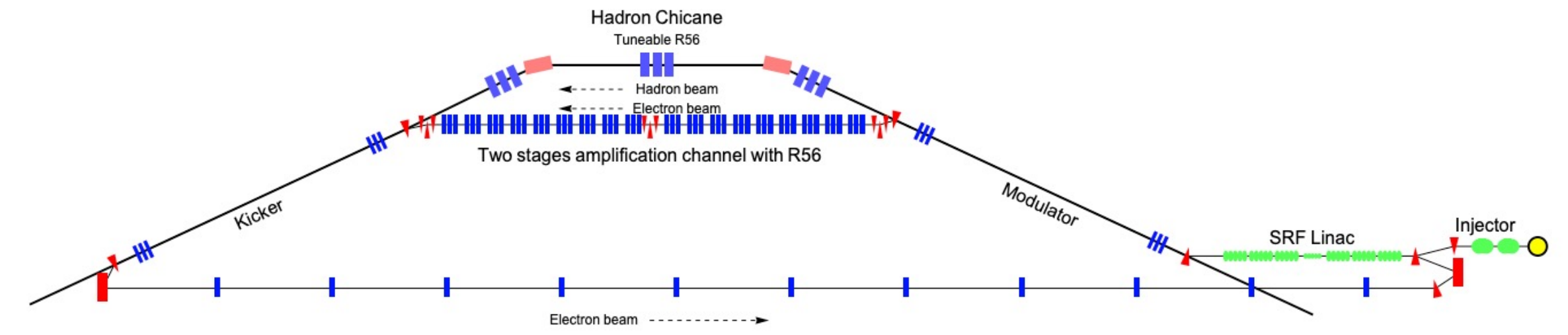}%
\caption{Layout of the chicane-based coherent electron cooler for the EIC. The hadron chicane for pathlength and $R_{56}$ adjustment is provided by using displaced superconducting dipole magnets at the IR2 straight section. Note that for a better view, the vertical scale is stretched by a factor of $\approx 50$.}%
\label{fig:new_facilities:bnl_electron_cooler:chicane}
\end{figure}

Currently, BNL is developing two CeC designs for EIC cooling.
The first CeC design is based on a conventional multi-chicane microbunching amplifier~\cite{PhysRevLett.111.084802, PhysRevAccelBeams.22.034401, PhysRevAccelBeams.22.081003}, which requires a modification of the RHIC accelerator to separate the electron and hadron beams~\cite{eic-strong-hadron-cooling, PhysRevAccelBeams.22.081003, bergan:ipac2021-tupab179}, as shown in Fig.~\ref{fig:new_facilities:bnl_electron_cooler:chicane}.
The amplifier in this CeC would have a bandwidth of \SI{30}{\tera\hertz}, and the system promises a cooling time of tens of minutes for a \SI{275}{\giga\electronvolt} proton beam in the EIC.

\begin{figure}[htb]\centering
\includegraphics[width=\linewidth]{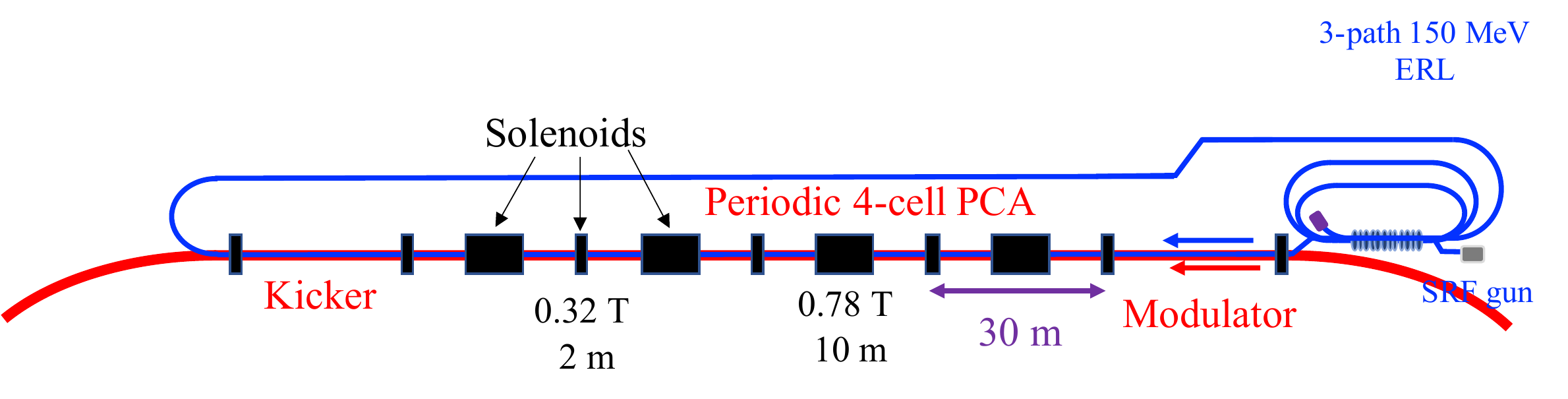}%
\caption{Layout of the CeC in the EIC with 4-cell PCA located at an existing 200-meter straight section in the RHIC tunnel. The system is driven by a 3-path \SI{150}{\mega\electronvolt} ERL with a high-quality electron beam generated by a QW SRF gun~\cite{PhysRevLett.124.244801}.}%
\label{fig:new_facilities:bnl_electron_cooler:plasma}
\end{figure}

The second CeC design, which is shown in Fig.~\ref{fig:new_facilities:bnl_electron_cooler:plasma}, is based on a plasma-cascade microbunching amplifier (PCA)~\cite{litvinenko2018plasmacascade, PhysRevAccelBeams.24.014402, theory-of-plasma-cascade}.
This system does not require separation of electron and ion bunches.
The PCA has a bandwidth of \SI{500}{\tera\hertz}, and this system promises to cool a \SI{275}{\giga\electronvolt} proton beam in the EIC in under 5 minutes.

In short, both CeC designs have the potential to cool hadron beams in the EIC to achieve its luminosity potential.
The team are working to identify the most promising as well as least expensive version of the CeC cooler.
Given that the technique is untested, the real cooler is expected to have a significantly longer cooling time that the estimation provided above.
This is the reason why CeC with 4-cell PCA is being tested experimentally at RHIC~\cite{litvinenko:cool2019-tuz01, doi:10.1142/S0217751X19420296, PhysRevAccelBeams.22.111002}.
This experiment has already discovered a number of additional limiting factors, such as excessive THz-scale noise in the electron beam.
At the same time, this experiment also established that the SRF gun built for CeC can generate an electron beam with the quality required for PCA-based CeC.

Table~\ref{tab:new_facilities:bnl_electron_cooler:requirements} summarizes the requirements for the ERL and the beam quality for both current CeC designs for a \SI{275}{\giga\electronvolt} EIC proton beam.

\begin{table}[htb]\centering
\caption{ERL and beam quality requirements for the \SI{275}{\giga\electronvolt} proton beam CeC cooler in the EIC. \enquote{CeC~1} and \enquote{CeC~2} refer to the two design options described in the text.}%
\label{tab:new_facilities:bnl_electron_cooler:requirements}
\begin{tabular}{lccc}
\toprule
\textbf{Parameter} & \textbf{Unit} & \textbf{CeC 1} & \textbf{CeC 2} \\
\midrule
Beam energy & MeV & 149.8 & 149.8 \\
\midrule
Injector & & & \\
\midrule
Electron gun & & DC, inverted & SRF, QW \\
Gun voltage & MV & 0.4 & 1.5~* \\
Charge per bunch & nC & 1 & 1.5~* \\
Bunch frequency, max & MHz & 98.5 & 49.25~** \\
Beam current & mA & 98.5 & 73.9 \\
Injection beam energy & MeV & 5.6 & 3.5 \\
\midrule
ERL & & &\\
\midrule
Number of passes & & 1 & 3 \\
ERL linac, fundamental & MV & 163 & 51 \\
Harmonic section & & 3\textsuperscript{rd} & 5\textsuperscript{th} \\
ERL linac, harmonic & MV & 18 & 2 \\
\midrule
Beam parameters in CeC & & & \\
\midrule
Peak current & A & \numrange{17}{30} & 150 \\
Norm.~emittance, rms & \si{\milli\meter\milli\radian} & 3 & 1.25~* \\
Bunch length, rms & mm & \numrange{3.5}{7} & 1.2 \\
Energy spread $\sigma_\gamma / \gamma$, rms & & $<\num{1e-4}$ & $<\num{2e-4}$~* \\
\bottomrule
\end{tabular}\\[1ex]
* Demonstrated parameters in the current CeC accelerator and SRF gun. In the current CeC accelerator, electron bunches with \SI{1.5}{\nano\coulomb} and \SI{1.75}{\mega\electronvolt} are compressed to \SIrange{50}{75}{\ampere} and accelerated to an energy of \SI{14.6}{\mega\electronvolt}. The beam has an RMS $\sigma_\gamma/\gamma$ of less then \num{2e-4} and a slice RMS normalized emittance of \SI{1}{\milli\meter\milli\radian}.\\
** For the same cooling time: with 98.5 MHz rep-rate cooling time will be 1/2 that of CeC 1.
\end{table}

Both CeC designs require an ERL operating with parameters beyond the state of the art.
Specifically, the average current exceeds that demonstrated in SRF ERLs by an order of magnitude or more, and further progress in ERL technology is needed for such a driver to become a reality.

%\subsection{Electron Cooler at BNL} 
%Dimitry Kayran, Vladimir Litvinenko  
%

 \chapter{Key Challenges---a Concerted Effort}\label{sec:key_challenges}

While ERLs have been extremely successful, the future lies with improved technology.
Rather than isolate these technology developments from the ongoing facility development, several of these improvements will be integrated into projects that have the technological skills as well as the need.
In particular, the further development of a DC photoelectron gun with low emittance and high current (initially \SI{8}{mA}, later \SI{20}{mA}) will be pursued at PERLE, based on the ALICE electron gun from STFC.
At bERLinPro, a superconducting RF gun has already been developed, but a new gun will be built to produce up to \SI{100}{mA} of beam current.
In parallel, in the USA, the EIC Cooler will use a DC photoelectron gun developed by Cornell, also with the goal of reaching \SI{100}{mA}.
Maintaining the small emittance through the booster and merger will also be demonstrated at these facilities.  
There is a dynamic effort world-wide to improve SRF for accelerators, but the ERLs have a couple of additional challenges that will be addressed.
The main focus will be to handle the high bunch currents and in particular managing the Higher-Order Modes (HOMs) for total linac currents up to \SI{200}{mA} (the sum of accelerating and decelerating beams).
ERLs, by design, have very low power requirements due to the energy-recovery process.
This makes matching the power source to the cavity extremely important.
Recent work on Fast Reactive Tuners (FRTs) will have a major impact on the cavity stability with low wasted power; these will be tested at bERLinPro.

There are two specific areas for the long-term development of ERLs: twin-axis cavities and the ability to operate at \SI{4.5}{K} with conventional cryogenics, rather than the sub-atmospheric plants required for \SI{2}{K}.
These areas will be developed separately from the facilities.

\section{Low-Emittance, High-Current Sources}
%  including polarisation, positrons \\
% Julius Kühn, Bettina Kuske, Boris Militsyn, Oleg Shevchenko, Olga Tanaka, Cristina Vaccarezza,

% MB: Rough edit for hyphenation, punctuation, word choices/grammar, and use of \SI and \SIrange. 06/14/2021%5 

% MB 12/20/2021: Since there have been a lot of changes, another edit for hyphenation, punctuation, and language.
%    Though it is not technically wrong, I removed the \cdot from unit products for consistency with the rest of the text; proper spacing makes it unnecessary.

%\subsection{Introduction}
The most critical component of an ERL is the electron injector, 
as it determines the limits for important beam parameters, 
such as temporal structure,  
bunch charge, and transverse and longitudinal emittance, which are critical for many applications.

Typical goal parameters for HEP applications are in the range of \SIrange{20}{25}{\milli\ampere} 
 average current, up to \SI{500}{\pico\coulomb} bunch charge, and emittance values of  
%$\approx\SI[inter-unit-product = \ensuremath{{}\cdot{}}]{30}{\milli\meter\milli\radian}$%
$\approx\SI{30}{\milli\meter\milli\radian}$
at the IP. The need to transport the beam through the ERL arcs  translates this emittance requirement
to $\approx\SI{5}{\milli\meter\milli\radian}$ at the entrance of the ERL.
In fact, light source applications might surpass the HEP demands on the injector.

Typical ERL injectors are comprised of an electron gun, 
which generates electrons with an energy in the range of a few 100's of keV up to the low MeV range, the
RF booster accelerator, which accelerates the electrons up to \SI{10}{MeV},
and a matching section or merger, which injects the electrons into the ERL loop. 
In injectors with low-energy guns, an RF buncher is installed between gun and 
booster to provide preliminary longitudinal compression of the bunches.   

\subsection{Electron guns}
The technological challenge within the injector is the electron gun. 
There are different  solutions for electron guns used in ERLs varying from traditional 
DC thermionic guns to state-of-the-art SRF photocathode guns. 
Thermionic injectors have the drawbacks of not easily providing an arbitrary temporal bunch structure,
which is necessary for ERLs, and of excluding any possibility to produce polarised electrons.
The most promising solutions are normal- or superconducting RF photocathode guns where the beam is extracted from a photocathode
illuminated by a pulsed laser. Although all parameters needed  by the HEP community have been met individually, to date, there is no gun available 
off the shelf which meets all the demands simultaneously. This goal requires significant further efforts.

The performance of  photocathode guns depends on appropriate photocathodes and  lasers for electron emission. 
Common efforts have been devoted to the development of 
high-quantum-efficiency (QE) photocathodes and their handling and transport; 
different laboratories concentrate on different cathode materials, e.g., 
Cs-Te (INFN-DESY), Cs-K-Sb (BNL, JLab, HZB), and Na-K-Sb (Cornell, HZB); QE values $>\SI{10}{\percent}$ 
at the desired wavelength have been achieved. 

Unfortunately, the existing technology of UV drive lasers does not allow for delivering the power 
 necessary to generate high average current in combination with the highly reliable Te-based photocathodes. 
Sufficient power is only provided by green lasers, restricting the applicability to more sensitive Sb-based photocathodes, 
which have a lower work function and lead to a higher risk of dark current. Further investigations in the field of 
high-QE photocathodes and associated lasers are still required.

Polarised beams can be produced with the mature technology of DC guns or potentially with the emerging SRF guns.
SRF guns hold the promise of low-emittance, high-current beams due to their high cathode field, 
and CW operation. 

The technology for normal-conducting RF buncher cavities and SRF booster linacs is well developed, and one 
of the existing designs may be used for HEP-ERL applications.

The unavoidable need to merge the injected beam with the recirculated beam is demanding, 
especially if multi-turn recirculation is considered.
First proof-of-principle experiments of injection into a multi-turn ERL were performed at Novosibirsk and Cornell.

\subsubsection{Thermionic Guns}
Thermionic guns are able to deliver high-average-current, high-brightness beams due to their relatively low effective cathode temperature \cite{PhysRevAccelBeams.22.123401}. 

Regular thermionic guns consist of a thermionic cathode, which serves as the electron emitter,
followed by an accelerating gap where the electrons are accelerated to energies varying from 10's to up to a few \SI{100}{keV}. 

As ERLs require bunched beams 
with a frequency varying from a few to 100's of MHz, the continuous beam 
needs to be separated into bunches.
%at the accelerating frequency
This beam modulation may be provided either at the exit of the gun with an RF deflector
and collimator or internally by modulating the cathode emission with a grid or a focusing electrode, which became more popular in  recent injectors \cite{2020.IEEE.TED.67.347}.
Unfortunately, the grid deteriorates the beam emittance; therefore, the applicability of grid-modulated guns is limited
to high-power, long-wavelength FELs and ERLs for high-energy physics applications such 
FELIX in Nijmegen, Netherlands, 
ELBE in Dresden, Germany, at FHI Berlin, Germany, and at Novosibirsk, Russia, where the beam is accelerated up to 
\SI{300}{\kilo\electronvolt} and the average electron current during normal operation in CW mode is about \SI{100}{\milli\ampere}.

Further acceleration of the beam in the grid-modulated guns is provided either with DC or RF accelerating fields that split thermionic guns into DC and RF guns, respectively.
A thermionic gun with acceleration in RF fields has been built in Novosibirsk to increase the average current (and, consequently, 
the power of the Novosibirsk FEL) \cite{Volkov2016}.
The cathode-grid assembly 
is mounted in a \SI{90}{\mega\hertz} RF cavity.
The gun has demonstrated an average current of more than \SI{100}{\milli\ampere} at an 
electron energy of about \SI{300}{\kilo\electronvolt}.

The measured normalized 
emittance of grid-modulated thermionic guns
is typically below \SI[%inter-unit-product=$\cdot$
]{20}{\milli\meter\milli\radian}, an order of magnitude larger than the corresponding thermal emittance of the cathode, which is explained by the inhomogeneity of the electric field caused by the presence of the grid and the charge of the bunch. The lifetime of thermionic cathodes is substantially longer than that of photocathodes, which is a decisive advantage for  user-facility applications. Additional efforts need to be applied to develop beam modulation schemes which preserve the beam emittance.

\subsubsection{DC Photocathode Guns}
An alternative to thermionic guns are photocathode guns; here, the beam modulation is provided by the photocathode drive laser without increasing the beam emittance. 

In these guns, the source of electrons is a metal or semiconductor crystal illuminated by laser pulses, which in addition provides flexibility, particularly in the temporal structure of the beam.
DC guns extract the charge from the cathode by using a DC-biased anode located a few \si{cm} from the cathode. 
The anode is biased to a voltage of up to \SI{500}{\kilo\volt}; a hole in the anode allows the electrons to be extracted.
The  electric field at the cathode is relatively low: $<\SI{5}{\mega\volt\per\meter}$.
DC-gun-based injectors are typically equipped with a normal-conducting (NC) RF buncher for longitudinal compression of the bunch.

DC guns were originally developed for operation 
with highly sensitive GaAs-based photocathodes for nuclear physics experiments requiring polarised beams. Vacuum pressures of better than \SI{e-11}{\milli\bar} provide relatively long dark lifetimes of the GaAs-based photocathodes, with a typical oxygen lifetime of \SI{2e-8}{\milli\bar\second}~\cite{Chanlek_2014}. Modern polarised GaAs-based photocathodes show a quantum efficiency of $\sim\SI{1}{\percent}$ 
and produce electron beams with a polarization of 
more than \SI{85}{\percent} \cite{Aulenbacher2011, doi:10.1063/1.4972180}.

Later on, DC guns, primarily with GaAs photocathodes, were utilized for FEL applications with low emittance and low bunch charge.

For use as a source of unpolarised electrons, DC guns also allow operation with antimonide-based photocathodes with an oxygen lifetime of \SI{e-5}{\milli\bar\second}. 

As the output beam energy from a DC gun is relatively low and to preserve the very low emittance, the high space-charge forces after compression have to be faced by properly matching and transporting the electron beam through the booster section. 

Record-level normalized transverse emittance values lower than \SI{1}{\milli\meter\milli\radian} 
have been measured with the  DC-gun-based Cornell ERL injector using a Na$_2$KSb photocathode at 
a wavelength $\lambda = \SI{520}{\nano\meter}$ (\SI{50}{\mega\hertz}) \cite{doi:10.1063/1.4913678}. 
This type of DC gun may be considered for a high-current ERL injector, though the longitudinal parameters 
of the beam may require further optimisation.

\subsubsection{Normal-Conducting RF Guns}
The continuing development of electron accelerators and their applications requires beams with higher brightness. This can be achieved using guns with higher cathode fields such as normal-conducting RF guns, where fields of \SI{100}{\mega\volt\per\meter} are demonstrated.

Unfortunately, high-field guns cannot operate in CW mode due to the very high heat load on the cavity surface, which cannot be removed by existing cooling systems.
The maximum RF frequency allowing CW operation is in the VHF range of up to \SI{300}{\mega\hertz} in quarter-wavelength cavities. 
These medium-to-low-field guns allow to reach cathode fields of up to \SI{20}{\mega\volt\per\meter}. 
The power dissipated in the cavity still reaches \SI{100}{\kilo\watt}, which makes cooling of the cavity a serious technical problem.

The lower electric field in VHF guns
%frequency 
requires longer laser pulses to preserve the low 
%thermal 
emittance defined by the cathode (\enquote{cigar emission}).
The beam needs to be 
%further 
compressed by a buncher cavity downstream to reach high peak currents and then 
be matched to a booster linac with a focusing solenoid for emittance compensation.
The APEX CW RF gun developed at LBNL \cite{doi:10.1063/1.5088521} works in the VHF frequency band with a peak electrode field of \SI{20}{\mega\volt\per\meter}, much higher than DC guns and 
therefore with reduced impact of space-charge forces.
The output beam energy is about \SI{800}{\kilo\electronvolt}, 
and the normalized transverse emittance measured with APEX showed values similar to the Cornell DC gun, 
i.e., \SI[parse-numbers=false]{0.2/0.4/0.6}{\milli\meter\milli\radian}
with \SI[parse-numbers=false]{20/100/300}{\pico\coulomb} electron bunches, respectively~\cite{doi:10.1063/1.4913678}.
Unlike Cornell’s experience, stable operation was demonstrated only at low average current (\SI{1}{\milli\ampere}) 
due to the limited repetition rate of the laser system (up to \SI{1}{\mega\hertz}). 
Ongoing studies on APEX2 show the possibility to reach \SI{0.1}{\milli\meter\milli\radian}
for \SI{100}{\pico\coulomb} bunches with
%\SI{0.6}[inter-unit-product=$\cdot$]{\milli\meter\milli\radian} emittance and 
about \SI{1.5}{\mega\electronvolt} final energy \cite{Mitchell:IPAC2019-TUPTS080}.

\subsubsection{Superconducting RF Guns}
As an alternative to NC-RF guns, SRF cavities can be used to allow for CW operation with cathode fields higher than \SI{20}{\mega\volt\per\meter}.

SRF guns have been developed at HZDR \cite{ArnoldSRFGunHZDR}, HZB \cite{KampsSRFGun}, BNL \cite{WencanXuBNLgunIPAC2013}, 
KEK \cite{Nishimori:IPAC2016-THPOW008}, and DESY \cite{vogel:srf2019-thp080}, to name but a few.
While the latest BNL Coherent electron Cooling (CeC) project uses a \SI{113}{\mega\hertz} quarter-wave cavity, 
the European \SI{1.3}{\giga\hertz} \enquote{TESLA}-type cavity is more commonly used.
Typical electric peak fields at the cathode lie at a few \SI{10}{\mega\volt\per\meter}.
Close to $\sim\SI{60}{\mega\volt\per\meter}$ was reached in vertical tests at KEK \cite{konomi:srf2019-frcab4}. The output beam energy is
up to \SIrange{3}{4}{\mega\electronvolt}.
Bunch charges extend up to a few \SI{}{\nano\coulomb} at very different bunch lengths and repetition rates.
ELBE at HZD-Rossendorf is the only facility operating an SRF gun in user operation since 2007.
The ELBE SRF gun II produces \SI{300}{\pico\coulomb} bunches with $\sim\SI{15}{\milli\meter\milli\radian}$
emittance and a few \si{\pico\second} of bunch length, quite close to the HEP needs.

The SRF technology itself, though, is a challenge: the theoretically high fields might be limited by field 
emission or multipacting, and the handling of the sensitive surfaces is critical during installation.
As in NCRF guns, dark current might be an issue.
The handling of high-QE cathodes is challenging, involving the danger of contamination during cathode exchange, overheating of the cathode by the drive laser, or low thermal conductivity.
Careful procedures are indispensable to mitigate the risks.

Once in operation, SRF technology runs smoothly and reliably, e.g., at the X-FEL at DESY.
SRF guns are suitable for high-power applications due to their potentially high RF field and the high 
repetition rate of the cathode laser up the \SI{}{\giga\hertz} level. 
Another big advantage of SRF guns is the confinement to a single technology, 
as most ERLs already utilize SRF technology in their linac designs.
The high cathode field and exit energy reduce the impact of space charge for high-density bunches.
No extra buncher cavity is needed. The RF field next to the cathode can provide longitudinal focusing. 
The typical bunch length is in the range of the pulse length of the drive laser.
A (mostly cold) solenoid focuses the beam into the booster linac, placed as close as possible to the gun cavity exit.
All these advantages make high-frequency SRF  guns a good candidate for  ERL injectors, though more operational experience is needed.

\subsection{High-Current Photocathodes}
Injectors for high-current operation (except the thermionic gun) rely on photocathodes, e.g., 
semiconductor materials based on metal, (multi)alkali, or GaAs-based systems for polarised beams, 
in combination with a photocathode drive laser. The quality of the photocathode is relevant for the performance 
of the photoinjector in terms of emittance, current, and  
lifetime, which 
are essential for photoinjector operation. 

For high-current operation, photocathodes with high quantum efficiency are necessary and are 
usually developed in-house \cite{doi:10.1063/1.4820132, PhysRevAccelBeams.21.113401, PANUGANTI2021164724}. 
Quantum efficiencies above \SI{10}{\percent} at the desired wavelength have been achieved in the laboratory.
Reproducible growth procedures have been developed, and months-long lifetime under operational conditions has been 
achieved as well~\cite{Wang2021}. 
One critical aspect is to preserve extremely high vacuum conditions ($<\SI{e-10}{\milli\bar}$) from the preparation system, 
throughout the complete transfer line to the photoinjector, and in the photoinjector itself. 

The photocathode substrates (called plugs or pucks,  usually made from molybdenum in RF guns) are optimised regarding 
their cleanliness and surface finish ($< \SI{10}{\nano\meter}$ RMS surface roughness) to achieve low emittance 
and to avoid field emission. Especially in SRF photoinjectors, the superconducting cavity is extremely sensitive 
to any kind of contamination; therefore, the photocathode exchange process is very critical.
The polarised electron beams needed for nuclear physics experiments demand GaAs photocathodes, but their lifetime has still to be improved, e.g., by multi-layer concepts \cite{PhysRevAccelBeams.23.023401}. 

Ongoing research topics in the field are the understanding of the photocathode materials (e.g., electronic properties), 
the photoemission process, and their intrinsic emittance \cite{Cocchi2019}. New growth procedures of high-quantum-efficiency, 
smooth, monocrystalline photocathodes or multi-layer systems and the screening of new photocathode materials are crucial 
for future electron accelerators.

\subsection{Buncher and Booster}
Any injector based on thermionic DC guns, DC photocathode guns, 
or low-frequency VHF guns necessarily has to include measures to 
%bunch 
chop the continuous beam and to longitudinally compress the bunch to the desired length.
The beam energy at the gun exit is in the order of few hundred~\SI{}{keV}, which leads to a strong impact of space-charge forces.
In order to bunch the beam and longitudinally compress the bunches, an RF buncher cavity is installed behind the gun. 

Further compression and acceleration of the bunches up to about \SI{10}{\mega\electronvolt} is usually achieved by SRF cavities 
grouped into the booster module. Typically, these cavities are independently fed and controlled to provide higher flexibility for 
longitudinal matching.

In high-field SRF guns, bunching is not necessary.
The operating point in the varying cathode field leads to velocity bunching so that the bunches leaving an SRF gun are already short, 
in the order of a few \SI{}{\pico\second}. The exit energy of SRF guns lies around a few \SI{}{MeV}. Space charge is still active, 
so the booster is placed as close as possible to the SRF gun, and care has to be taken to preserve the emittance 
by proper optical matching. Bunch compression can also be achieved in 
the booster by introducing an energy chirp in the bunch. 
Later, the dispersive  merger section may be used for further compression.

%An injector concept based on the characteristics of an electron beam generated by thermionic DC guns, DC photocathode guns, and low-frequency VHF guns, necessarily longitudinal compression of he bunch.
%The beam energy at the gun exit is of the order of few hundred \SI{}{keV}, which leads to a lengthening of the bunch by space charge forces. To longitudinally compress the bunch, an RF buncher is installed behind the gun. 
%Further compression and acceleration
%of the bunches up to about \SI{10}{\mega\electronvolt} is usually achieved by SC cavities grouped into the booster module. Typically, the cavities are independently fed and controlled to provide higher flexibility for longitudinal matching.

%In high-field SRF guns, bunching is not necessary.
%The operation point in the varying cathode field leads to velocity bunching, so that the bunches leaving an SRF gun are already short, in the order of a few \SI{}{ps}, and have a higher energy of a few \SI{}{MeV}. Space charge is still active, so the booster is placed as close as possible to the SRF gun, and care has to be taken to preserve the emittance by proper optical matching. Further bunch compression can be achieved in 
%the booster, by introducing an energetic chirp on the bunch. Later, the dispersive  merger section can be used for further compression.

\subsection{Merger}
The high-charge, low-emittance bunches that leave the booster at $\sim\SI{10}{\mega\electronvolt}$ 
have to be transported to and merged into the recirculator ring of the ERL.
The beam-optical section required to do so is called the merger.
Various merger schemes were realised. The most popular one is the
%dog-leg, 
S-shaped merger, exceptions being the vertical zigzag merger of the BNL ERL project \cite{BNLMerger}, 
or the Novosibirsk C-shape merger. 

The introduction of dipoles into the beam line necessarily implies an energy-dependent widening of the beam, 
which cannot fully be controlled due to the action of space-charge forces in the dispersive section.
A certain unavoidable emittance blow-up can be mitigated by careful design strategies that also need to 
take space-charge effects in the transport lines outside the dispersive merger section into account.

The injected low-energy bunches and the recirculated high-energy beam are merged in an interleaving scheme 
in front of the linac. This requires careful optics matching to secure the subsequent simultaneous transport of both beams 
through the linac, where the injected beam is accelerated and the recirculated beam is decelerated 
(single turn ERL) or accelerated further (multi-turn ERL). Usually, a small chicane is needed for the recirculated beam.
Besides the transverse optics, also the longitudinal phase space has to be matched---a precondition for high energy-recovery efficiency. 

The merger is a dense and complicated beam transport section. Up to now, mostly multi-objective optimization algorithms have been used in the development of the merger. The development of analytic approaches could significantly speed up the design.

\subsection{Conclusion}
In conclusion, we can state that low-emittance, high-current injectors are under development at many laboratories worldwide,
and much progress has been achieved during the last years. The parameters necessary for HEP were all met, 
although not simultaneously in one project. PERLE at Orsay sets out to achieve exactly that goal, and it can rely on 
many preparatory experiments and even hardware from the international community.

The gun remains the critical component of the injector. Different gun options were discussed in this chapter. 
PERLE relies on the well-established technology of DC guns. 
VHF NC and SRF guns can be promising alternatives, as APEX-II at LBNL and SLAC or the SRF guns developed at bERLinPro 
in Berlin and DESY in Hamburg. bERLinPro intends to conduct the proof-of-principle
experiment for high charge and low emittance in 2022. All these projects are indispensable for the further ERL development.

Technological challenges still cover the UV photocathode drive laser to allow delivery of the required average current from reliable Te-based photocathodes, 
though the operation of antimonide photocathode with a green laser is gaining importance and has already demonstrated photocurrents close to the required \SI{100}{mA}.

Further theoretical understanding is still required for the optimal transport of high-charge bunches, $>\SI{500}{pC}$, 
through the merger section with controlled emittance dilution. The current approach relies on time-consuming and CPU-intensive optimisation algorithms.

In summary, the injector development is considered ready for HEP ERL operation within the current time frame.

%In conclusion we can state, that low-emittance, high current sources are under development at many laboratories worldwide, and much progress has been achieved during the last years. The parameters necessary for HEP, where all achieved, although not simultaneously in one project. PERLE at Orsay sets out to achieve exactly that goal, and can rely on many preparatory experiments and even hardware of the international community.

%Different gun options were discussed in the chapter. PERLE relies on the well established technology of DC guns (??). SRF guns can be a promising alternative, and the SRF injector at bERLinPro in Berlin intends to conduct the prove of principle experiment for high charge and low emittance in 2022. Both projects are indispensable for the further ERL development.

%Technological challenges still cover the cathode laser... why?

%Further theoretical understanding is still required for the optimal transport of very high charge bunches \SI{>100}{pC} through the merger section with controlled emittance dilution. The current approach relies on time consuming and CPU intensive optimization algorithms.

%What else do you want to cover?

%\section{High Current Sources}
%  including polarisation, positrons \\
%  bERLinPro, ALICE Upgrade, cERL, cBETA 65mA,  eCOOLER, ..
%Boris Militsyn

%\input{s42.tex}
%\section{Low Emittance Injectors}
%  cBETA, cERL, ..
% Cristina Vaccarezza, Bettina Kuske, Olga Tanaka

\section{Challenges of SRF Cavities and Cryomodules}%
\label{sec:key_challenges:srf}

% MB: rough edit started on 06/11/2021, mostly punctuation/hyphenation, a few word choices, \SI usage, \labels. Finished up to line ~158
% MB: replaced some erroneous references to fig:key_challenges:srf:multi_kw_loads by fig:key_challenges:srf:blas

% MB: rough edit finished 10/5/2021

% Frank Marhauser, Erk Jensen
Many ERL facilities have been conceived worldwide since the conceptual proposal in 1965 \cite{Tigner:1965wf}, including the machines described in Section~\ref{sec:current_facilities}.
After the first reported energy recovery experiment at the Stanford accelerator-driven FEL (SCA/FEL) in 1987 \cite{SMITH19871}, experimental demonstrations of the energy recovery principle utilizing SRF cavities in a single acceleration and single deceleration pass progressed in the early 2000s at FEL facilities such as the Jefferson Lab IR FEL \cite{PhysRevLett.84.662} and the Japan Atomic Energy Research Institute (JAERI) FEL \cite{HAJIMA2003115}, albeit at rather low beam energies (\SI{100}{\mega\electronvolt} and \SI{17}{\mega\electronvolt}, respectively) and beam currents (\SI{5}{\milli\ampere} and \SI{10}{\milli\ampere}, respectively).
It was recognized that the beam power obtained was not necessarily higher than what could have been achieved in conventional linear accelerators.
Consequently, the interests spurred for higher energy (multi-GeV), higher beam current (typically \SI{100}{\milli\ampere}, but also higher), low emittance ($\sim\SI{1}{\milli\meter\milli\radian}$), and short bunch length ($\sim\SI{1}{\pico\second}$) ERLs with the ultimate goal to develop high-laser-power FELs, hard-x-ray sources, spontaneous-emission light sources attractive for synchrotron radiation (SR) users, as well as electron cooling machines, where large beam power is required for cooling ions, particularly useful for high-luminosity electron-ion colliders (see Section \ref{sec:frontier:high_energy_colliders}).
The potential use of ERL technology for energy-frontier accelerator physics has been realized.
Particularly for the conceived Large Hadron-electron Collider (LHeC), it is well acknowledged that an ERL is the only choice to achieve high luminosities with economic use of power \cite{Agostini20}.

The interest in ERL technology has lately been boosted by the advances made for SRF cavity and cryomodule technology, which is the key technology for enabling low-loss energy storage and highly efficient energy recovery. 

\begin{figure}[htp]
\centering
\includegraphics[width=0.8\textwidth]{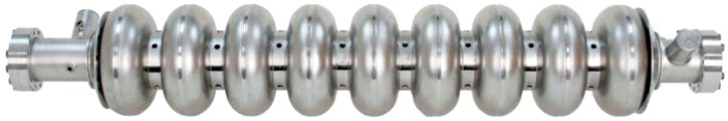}
\caption{TESLA 1.3 GHz nine-cell SRF cavity}
\label{fig:key_challenges:srf:tesla9cell}
\end{figure}

SRF cavities are conventionally made from fine-grain niobium being cooled by liquid helium below its critical temperature, $T_\text{c}$, of \SI{9.25}{K}, which is the highest of all pure metals.
The cavity operating temperature is typically $\sim\SI{2}{K}$.
At the typical operating temperatures and depending on the accelerating RF field, $E_\text{acc}$, cavity design, and resonant frequency $f$, the dynamic RF surface losses per resonator cell are on the order of \SIrange{1}{10}{\watt}.
This is achievable thanks to an about million-fold lower surface resistance, $R_\text{res}$, in the superconducting walls compared to normal conducting RF (NCRF) structures operating at room temperature.
It makes SRF technology the only viable option to allow operating in continuous wave (CW) while sustaining high $E_\text{acc}$ levels.
Fields around \SI{20}{\mega\volt\per\meter} have been envisaged for the main linac cavities in ERLs.
Already in the year 2000, industrially produced TESLA \footnote{The prominent TESLA SRF accelerator technology originated from the R\&D efforts for a \SI{1}{\tera\electronvolt} \positron\electron\ linear collider with an integrated X-ray laser facility as published in 1997 \cite{Brink07}.
The latter facility is fully operational and known as the European XFEL at DESY \cite{XFEL}, whereas the large-scale linear collider is still planned in the frame of the ILC global design effort \cite{ILC07}.} \SI{1.3}{GHz} nine-cell SRF cavities (see Fig.~\ref{fig:key_challenges:srf:tesla9cell}) have reached $E_\text{acc}$ = \SI{25}{\mega\volt\per\meter} on average with a high yield and at an unloaded quality factor, $Q_0$, of \num{5e9} \cite{Aune00}. 

Leveraging the progress of SRF accelerator technology and auxiliary RF systems dominantly developed at \SI{1.3}{GHz}, Cornell University in collaboration with JLab explored the potential of a hard X-ray user facility with a proposal in 2001 \cite{Gruner11}.
The study not only outlined the advantages of ERLs compared to state-of-the-art SR facilities but proposed a roadmap to first construct a \SI{100}{mA}, \SI{100}{MeV} demonstration ERL to experimentally test and develop ERL technology before trying to build a full-scale multi-GeV facility.
It was conceived that such an approach would have laid the groundwork for ERLs at other laboratories while serving as a vehicle for the training of accelerator physicists in ERL technology.
20 years later, this roadmap is yet to be realized.
Therefore, it is recommended to reinstate an ERL technology roadmap.
This directly addresses the 2020 European Strategy for Particle Physics having emphasized that the ERL technology is one of the most promising technologies.
Imminent questions are why the construction of multi-GeV, high beam current ERLs could not succeed within the last two decades despite the promising physics opportunities well outlined in the past, and why the progress of the ERL technology was curbed with only comparably small-scale ERL demonstration facilities being built.
This is related to the technical readiness level (TRL) of ERL key components.
We, therefore, need to also address the challenges that SRF cavities and cryomodules still face today. 

In this aspect, one should acknowledge that the advance of the SRF accelerator technology did not occur swiftly.
It relied on a concerted effort of many contributing laboratories worldwide, now covering more than five decades of funded R\&D with experimental trial and error.
Furthermore, the industrialization of SRF cavity production, specifically at two European vendor sites \footnote{RI Research Instruments GmbH (RI) in Germany (formerly ACCEL) and Zanon Research \& Innovation Srl (formerly Ettore Zanon S.p.A.)}, was vigorously supported with the knowledge transfer from DESY and INFN \cite{Singer16}.
Since then, TESLA cavities have been employed in many SRF accelerator facilities around the globe, and well over 1000 TESLA cavities have been produced by industry.
\SI{1.3}{GHz} SRF accelerator systems are already an integral part of the LCLS-II FEL project at SLAC \cite{Galayda14} as well as small-scale facilities like the ELBE accelerator in Dresden Rossendorf \cite{Buchner00}, the one-pass ERLs demonstrator facilities ALICE\cite{AboBakr09} (see Section~\ref{sec:current_facilities:completed:alice}) as well as the multipass ERLs CBETA \cite{Mayes17}  (see Section~\ref{sec:current_facilities:ongoing:cbeta}), and the proposed MESA facility \cite{Aulenbacher09} (see Section~\ref{sec:new_facilities:mesa}).

Consequently, it will be important that future large-scale ERLs can largely benefit from the industrialization of SRF cavity and cryomodule technology to reduce manufacturing costs and lead times.
Today, the two aforementioned European vendors have been elevated through infrastructure investments to a TRL that allows them to produce chemically cleaned, high-performance cavities preassembled in ISO-4 clean rooms, fully dressed in helium tanks, delivered under vacuum, and ready for vertical high-field RF tests in dewars before being assembled into cryomodules.
This manufacturing paradigm (``built-to-print'' approach) has first been practiced by DESY for the construction of the XFEL (800+ cavities) and subsequently adopted by the LCLS-II (300+ cavities), which included the continuing cavity procurement for its high energy upgrade.
Both projects were schedule-driven to fulfill the demands of prompt cryomodule assembly and delivery to the final accelerator sites.
The infrastructure investments at vendor sites were paramount, yielding unprecedented delivery rates of $\sim 7$ SRF cavities per week for the XFEL \cite{Walker:SRF2015-MOPB086}.
The requested throughput could however only be achieved by engaging two vendors concurrently. 

Given the arguments above, one might conclude that the SRF accelerator cavity and cryomodule technology is well matured for utilization in ERLs.
This however can be deceptive since the developments are predominantly based on \SI{1.3}{GHz} accelerator and auxiliary RF systems as emphasized above.
This concerns e.g.~DESY-owned production machines that are being utilized at vendor sites but are only tailored to TESLA cavities.
These are the automated cavity tuning machine and the semi-automated cavity sub-component RF measuring machine, respectively.
Such dedicated production tools are principally not usable for any other cavity design.
For designs at much lower frequencies, in particular, the infrastructure must be expanded or built from scratch to come to the same TRL levels as achieved for TESLA-type cavities.
This can, for example, mandate upgrades of buffered chemical polishing, electro-polishing, and ultra-pure water high-pressure rinse machines/cabinets, all of which are mandatory for cavity post-processing treatments of the superconducting surfaces.

Moreover, the decision of MESA to utilize TESLA cavities was not based on the reason that these cavities are best suitable for ERLs but based on the high TRL \cite{Stengler:SRF2015-THPB116}.
The required two main linac cryomodules are modified ELBE cryomodules each housing two TESLA cavities.
These cryomodules can be assembled and delivered by the European vendor RI \cite{RI}  as was the case in the past for the two ALICE main linac cryomodules \cite{doi:10.1063/1.3215597}.
The planned Polish FEL will benefit from the same development \cite{Szamota19} . 

In summary, TESLA-type cavities and cryomodules are matured technologies, but it cannot be assumed that this development will satisfy all the requirements of full-scale ERLs without modifications or major design changes.
The cavity and cryomodule designs shall rather adapt to the specific machine parameters and specifications, which can also mandate a different choice of RF frequency.
The continued knowledge transfer from SRF expert laboratories to industrial vendors is important since even the two most qualified SRF technology vendors in Europe show production discrepancies affecting geometrical tolerances and thus cavity performances, though similar, strict manufacturing protocols have been followed \cite{Marhauser18}.

\subsection{General Challenges and Concerns}
There are well-known challenges to overcome which can limit the performance of SRF cavities and auxiliary components.
These need to be addressed by appropriate design choices as best as practicably possible.
Some lingering key challenges are:  
\begin{enumerate}
    \item Thermal limitations in the fundamental power coupler (FPC) and higher-order mode (HOM) couplers,
    \item Resonant electron multipacting in cavity cells, the FPC, and HOM coupler,
    \item Electron field emission primarily depending on surface cleanliness,
    \item Microphonic detuning of cavities depending on the cavity mechanical design---including the cavity tuner and helium tank design---which is a more crucial challenge for the energy-recovery linac cavities driving residual RF power needs than it is for the heavily beam-loaded cavities in ERL injectors.
    % MB 3/7/22: Removed reference; the chapter being referred to is gone.
    %Details on this challenge are given in Section~\ref{sec:key_challenges:power}.
    This also includes challenges for transient effects, i.e., if the main linac cavities are filled and emptied or in case of a sudden beam loss. 
\end{enumerate}

In some cases, the performance degradation aggregates in CW operation and with the increase of the beam current, requiring thorough analysis.
Beam-dynamic issues such as transverse and longitudinal Beam Breakup (\emph{BBU}) instabilities and the beam-induced HOM power dissipation increase with the beam current in the machine.
This drives the choice of the required HOM coupler technology and the complexity of the cryomodule design.
Cavity damping concepts may vary but are principally based on coaxial and waveguide couplers as well as beam line absorbers (\emph{BLA}s) or a combination thereof.
Challenges for each of the coupler technologies are summarized below \footnote{A more detailed review of these HOM coupler technologies can be found in \cite{Marhauser17}.}.
Design modifications beyond the state of the art are likely necessary for ERLs.

\subsection{Coaxial HOM Dampers}
A good example of such failures is due to the simple adaption of TESLA-type coaxial HOM couplers for cavities resonating at different frequencies at JLab and SNS.
These couplers are made from solid niobium to remain superconducting during operation.
Each coupler consists of a hook-type inner HOM coupler housed in a cylindrical can to capture the beam-induced HOM energy via a coaxial cable so that it can be dissipated at room temperature in a standard \SI{50}{\ohm} load outside the cryomodule \cite{Sekutowicz93}.
A capacitive notch filter is included to reject the damping of the fundamental accelerating mode.
Note that these couplers were originally envisioned for pulsed machines operating at low duty cycles, thus not requiring active convection cooling but relying solely on thermal conduction to extract the RF heat load.
Therefore, their usability is limited in CW operation.
Thermal performance limitations were for instance encountered when JLab tried to utilize slightly scaled TESLA-type HOM couplers for their \SI{1497}{MHz} upgrade cavities in the Continuous Electron Beam Accelerator Facility (CEBAF), which is a \SI{12}{\giga\electronvolt} electron recirculator with up to 5.5 passes \cite{Harwood13}.
Premature quenches in prototype cavities were observed in the HOM end-groups being exposed to about \SI{10}{\percent} of the fundamental-mode peak magnetic field in the cavity.
This mandated improved thermal management including the replacement of material components in the RF ceramic feed-through with ones of higher thermal conductivity to eventually adapt these couplers to CW operation \cite{Reece05}.
Such coupler modifications were later adopted for LCLS-II operating with TESLA cavities in CW mode and more recently considered for the MESA cryomodule cavities.

\begin{figure}[htbp]
\centering
\includegraphics[width=0.8\textwidth]{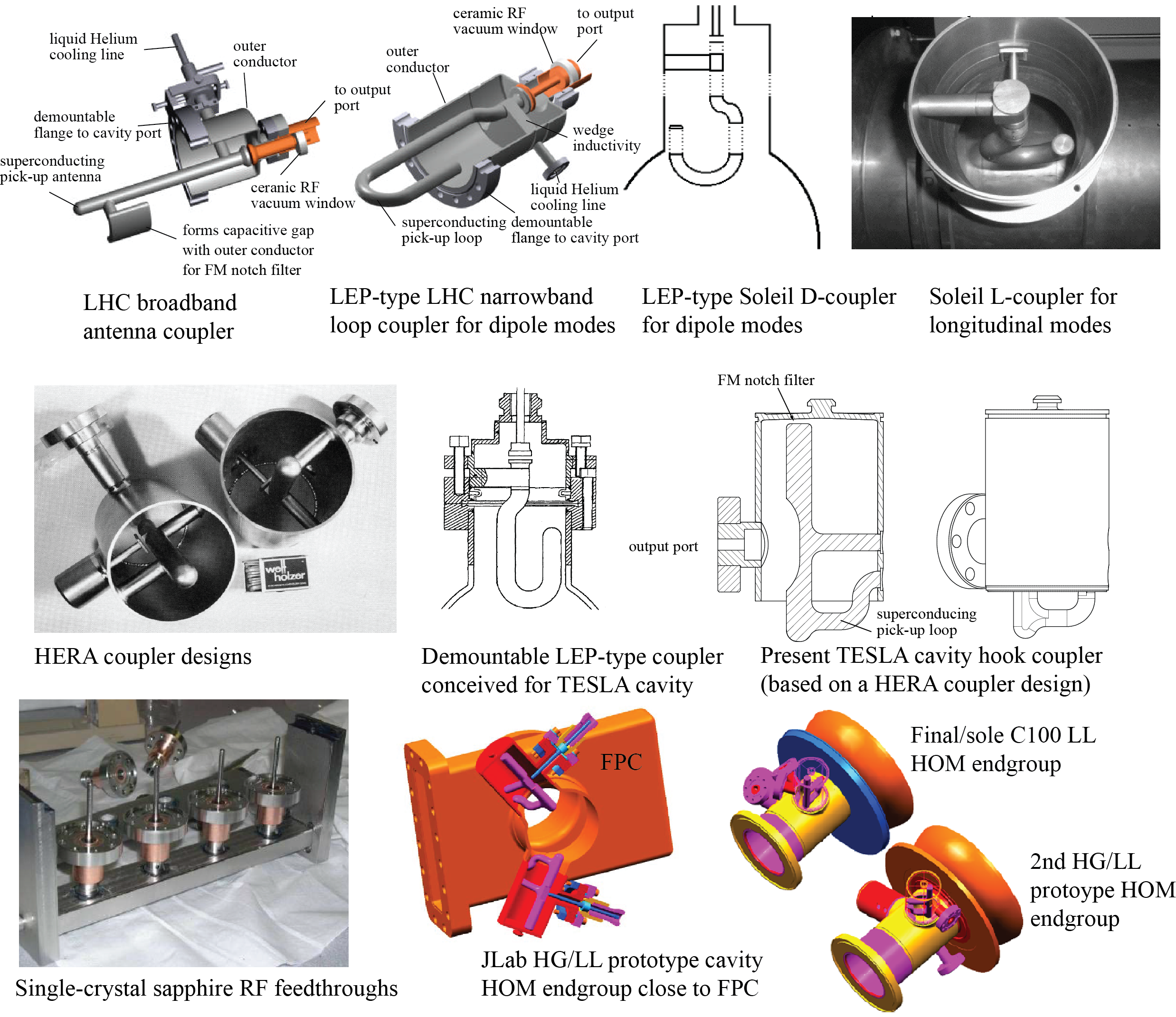}
\caption{Existing coaxial HOM coupler concepts for SRF cavities (figure taken from \cite{Marhauser17}).}
\label{fig:key_challenges:srf:coaxial_couplers}
\end{figure}

For higher-beam-current multi-pass ERLs, improved HOM coupler concepts are required to cope with the more stringent damping requirements and anticipated high HOM power levels.
Actively cooled HOM coaxial couplers that enhance their power capability have been in operation at CERN and Soleil for a long time (\cite{Boussard99, Mosnier98}).
The coaxial HOM couplers for the \SI{400}{\mega\hertz} LHC cavities, for instance, are being cooled via liquid helium circuits, and have been tested up to \SI{0.8}{kW} HOM power in the laboratory.
The conceptual designs of these couplers, however, are tailored for the specific application and cavities.
LHC cavities utilize both a narrow-band and broadband coupler to capture the most offending HOMs in specific frequency regions of interest.
Figure~\ref{fig:key_challenges:srf:coaxial_couplers} depicts different coaxial couplers in use as described above.

The more than 20-year-old coupler technologies need to be revisited and not just scaled to meet the more stringent requirements of modern ERLs.
Some conceptual design improvements for coaxial HOM dampers were explored more recently in the context of \SI{400}{\mega\hertz} crab cavity developments \cite{Angal_Kalinin_2018}, specifically for the high-luminosity upgrade of the LHC \cite{Carra15}.
The experiences made could become helpful for ERL R\&D.

\begin{figure}[htbp]
\centering
\includegraphics[width=0.8\textwidth]{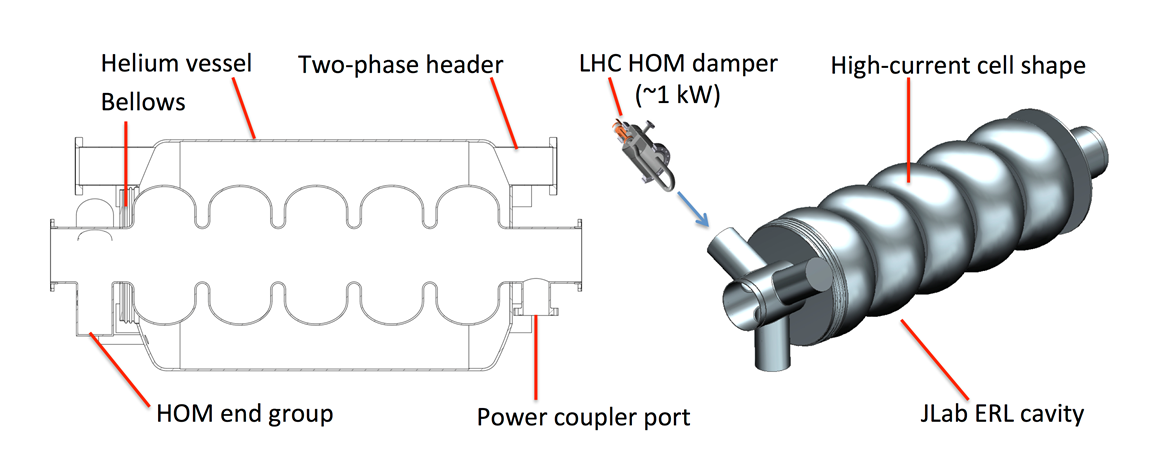}
\caption{Concept design of a cavity and helium vessel arrangement. The helium vessel may be made from titanium like the SNS helium tanks or stainless steel like the CEBAF \SI{12}{\giga\electronvolt} upgrade modules.}
\label{fig:key_challenges:srf:coax_y_concept}
\end{figure}

Moreover, most existing accelerating cavities employ two coaxial HOM couplers per cavity, positioned close to end cell irises.
In this case, one coupler can be located on each side of the cavity (e.g., TESLA cavities), or two on a single side (e.g., JLab upgrade cavities).
The importance is to rotate one coupler versus the other azimuthally to allow capturing both polarizations of dipole HOMs (and, to some extent, the polarizations of quadrupole HOMs, though of lesser concern).
Two coaxial couplers are therefore needed at least.
For more efficient damping but added complexity, one can increase the number of couplers to three for just for a single end-group (``Y end-group'') and allow for an FPC at the other end of the cavity.
This has been conceptually investigated at JLab.
Spacing the couplers by \ang{120} avoids noticeable variations in the damping of differently polarized modes, as evidenced experimentally for instance in TESLA cavities \cite{Marhauser18}.
An adequate HOM coupler technology for the cavity is yet to be developed.
A possible, comparably compact concept is depicted in Figure~\ref{fig:key_challenges:srf:coax_y_concept}, illustrating a five-cell cavity integrated into a helium vessel (left) with a single HOM end-group located outside the helium vessel but close to the cavity.
It accommodates three coaxial, actively cooled (LHC-style) HOM couplers. 

\subsection{Waveguide HOM Dampers}
JLab has prototyped Y-shaped end-groups in the past, but for waveguide HOM dampers.
In general, this coupler concept acknowledges the verified HOM damping efficiency of three-folded waveguide dampers, which have been successfully employed for NCRF cavities in high-beam current third-generation synchrotron light sources (e.g., BESSY II at HZB \cite{Marhauser01}, ALBA \cite{Perez11}, ESRF \cite{Guillotin06}), and \positron\electron\ colliders (e.g., PEP-II \cite{Rimmer92}, DAPHNE \cite{Bartalucci99}).
For NCRF cavities, the waveguides can however be put onto the cavity cells with minor consequences, whereas this is a major detriment for SRF cavities since increasing the magnetic fields at the cavity-waveguide intersection leads to a decrease of the quench field limit. 

\begin{figure}[htbp]
\centering
\includegraphics[width=0.8\textwidth]{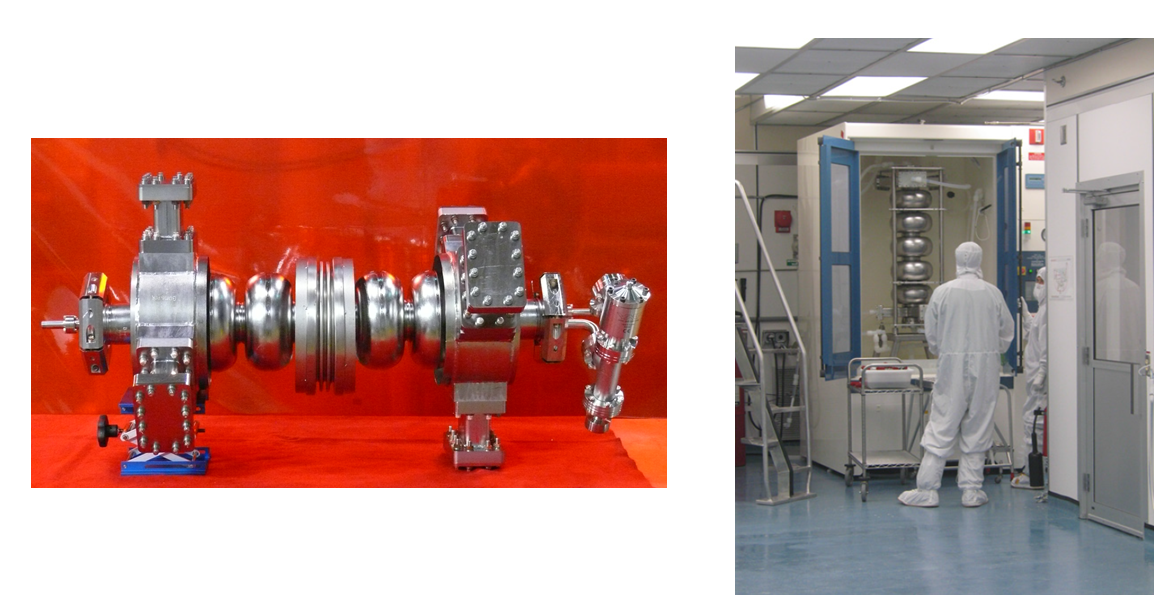}
\caption{JLab high-current ERL prototype cavities at \SI{1.5}{GHz} (left) and \SI{750}{MHz} during preparation (right) \cite{Rimmer10}.}
\label{fig:key_challenges:srf:High_Current_JLab_Cavities}
\end{figure}

Even two such HOM end-groups---one doing double duty as FPC---were considered in 2005 for Ampere-class cryomodules for future ERLs based on high-power compact FELs \cite{Rimmer05}.
By 2010, JLab had designed and fabricated three full-featured five-cell so-called high-current (HC) prototype cavities, two at \SI{1.5}{GHz} and one at \SI{750}{MHz} (see Fig.~\ref{fig:key_challenges:srf:High_Current_JLab_Cavities}) \cite{Rimmer10}.
All HC cavities ultimately exceeded the self-proclaimed goal of $E_\text{acc} = \SI{17}{\mega\volt\per\meter}$ in vertical tests and with a $Q_0$ typically around \num{1e10} at \SI{2}{K} (somewhat higher at \SI{750}{MHz}).
The HC cavity design features rather large boreholes to provide good cell-to-cell coupling, which reduces the probability of trapped HOMs.
This also leads to a smaller sensitivity to fabrication tolerances, minimizing concerns of accelerator-mode field-flatness distortions of other detuning effects as compared e.g.~to TESLA-cavities.

Furthermore, the comparably large boreholes along the cavity minimize beam halo interception, while three waveguides per end group could be better accommodated on a relatively large beam tube.
An advantage over coaxial HOM couplers is that the waveguides provide a natural cutoff (high-pass filter) for the fundamental mode, thus not requiring a delicate notch filter, while being broadband and thus capable of capturing very-high-frequency HOMs that short bunches will excite.
Broadband HOM absorbers can be placed at the end of waveguide dampers.
The consumption of beamline space can be minimized by folding the waveguides over towards the center of a cavity by using a \ang{90} bend inside the cryomodule, which can keep the cavity installation length (flange to flange) nearly as compact as it would be in the case of cavities with coaxial couplers.

\begin{figure}[htbp]
\centering
\includegraphics[width=0.8\textwidth]{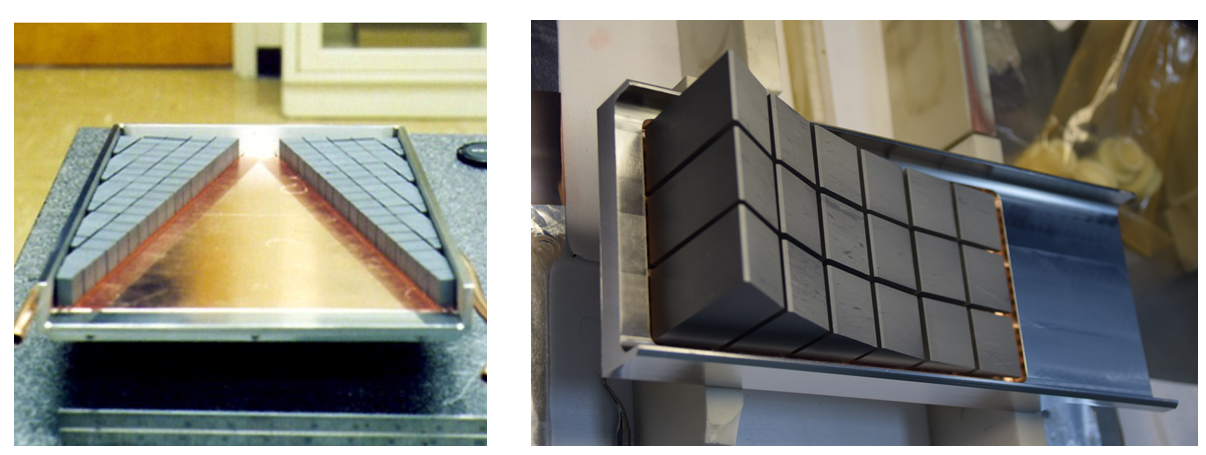}
\caption{Multi-kW PEP-II waveguide load developed in the mid-1990s \cite{Pendleton94} (left) and $\sim\SI{0.5}{\kilo\watt}$ VSR cavity waveguide load \cite{Guo19} (right).}
\label{fig:key_challenges:srf:Multi_kW_loads}
\end{figure}

The achievable power capability depends on the choice of material and the efficiency of extracting the HOM power deposition from the absorbers via heat conduction into a coolant.
The coolant can be nitrogen at medium power levels, but water-cooled absorbers placed at room temperature need to be considered in the kW range.
A variety of material choices, primarily lossy ceramics, are available commercially that provide broadband low-reflection properties at room temperature.
The HC cavity waveguide absorbers were conceived for the latter using lossy ceramic tiles brazed onto copper, which would have provided a power capability of \SI{4}{kW} per load as needed for a \SI{1}{\ampere} beam current \cite{Cheng10}.
The thermal conductivity and expansion coefficients of the lossy ceramics and the metal substrate play a crucial role in extracting the heat while minimizing thermal stress that could lead to a separation of tiles from the substrate and cracking.
An efficient concept is to braze the tiles to a copper pegboard as developed already in the mid-1990s for PEP-II cavity multi-kW loads \cite{Pendleton94} (see Fig.~\ref{fig:key_challenges:srf:Multi_kW_loads}).
A pegboard takes over some of the thermal stresses developing during brazing. The brazed assembly can be thermally shock-cycled before assembly to verify its integrity as well as tested under high heat load on the bench (e.g., via IR light).
An increased concern for SRF cavities compared to NCRF cavities, however, is the particulate outgassing from the tiles that could migrate into the cavity interior.
This must be minimized by proper surface cleaning procedures.
Such problems are being addressed by HZB Berlin in collaboration with JLab.
JLab has developed waveguide loads for two HZB projects, namely the variable-pulse-length storage ring (VSR) cavities (\SI{1.5}{\giga\hertz} and \SI{1.75}{\giga\hertz}) and the bERLinPro main ERL linac cavities (\SI{1.3}{GHz}). The HC cavity damping scheme developed at JLab was adopted for both projects (\cite{Neumann14, Guo19}).
As a single-pass energy-recovery machine, the seven-cell main linac cavities for bERLinPro have to account for \SI{200}{mA} of beam current, though the estimated power is still moderate ($\sim\SI{27}{\watt}$ per load), but higher for the VSR five-cell storage ring cavities accounting for \SI{300}{mA} ($\sim\SI{460}{\watt}$ per load).

It is in the eye of the beholder whether HOM waveguides add to the complexity of the cryomodule design compared to coaxial couplers.
The latter need numerous electron-beam welding (EBW) steps and are costly components.
Costs are compounded by strict dimensional tolerances.
An RF ceramic window feedthrough is required, which is limited in its transmission characteristics, and the RF power capability of coaxial cables and connectors have to be accounted for.
The required notch tuning of the fundamental filter is an added complexity for coaxial couplers.
This is done by mechanically deforming the HOM can, but detuning upon cool-down to cryogenic temperature is still a problem, necessitating deliberate work-hardening of the HOM can so that detuning during cool-down can be minimized.

Waveguides, on the other hand, can be produced by sheet metal deep-drawing and in fewer EBW steps, reducing process steps and costs, while fabrication tolerances are much more relaxed.
Costs will be added, however, for the development of the HOM loads, but this allows one to reach multi-kW power capability per waveguide load as required for operation at very high beam currents.
The increased static heat load from the waveguide into the helium bath can be managed by proper thermal heat interception.

\begin{figure}[htbp]
\centering
\includegraphics[width=0.8\textwidth]{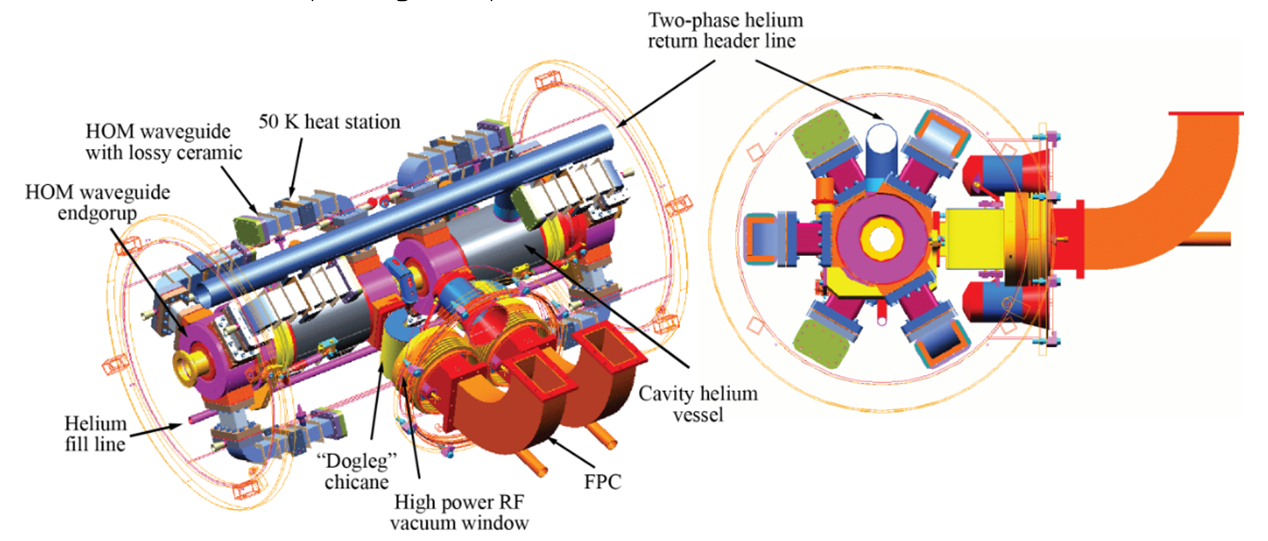}
\caption{Conceptual cryomodule design for high-current ERLs with a JLab HC cavity pair (figure taken from \cite{Marhauser17}). Each cavity employs six HOM waveguide dampers with one doing double duty as an FPC.}
\label{fig:key_challenges:srf:High_Current_JLab_Cryomodule}
\end{figure}

The funding for the HC SRF cavity and cryomodule R\&D at JLab ceased in 2010, which is unfortunate since it could have solved some of the most stringent challenges for high-beam-current ERLs.
At that time, the HC cavity design had already been prototyped at two different RF frequencies, and high-power FPC developments including high-power RF windows were ongoing, while a conceptual design of the cryomodule was being carried out (see Fig.~\ref{fig:key_challenges:srf:High_Current_JLab_Cryomodule}).

\subsection{Beam Line Absorbers}
\begin{figure}[htp]
\centering
\includegraphics[width=0.8\textwidth]{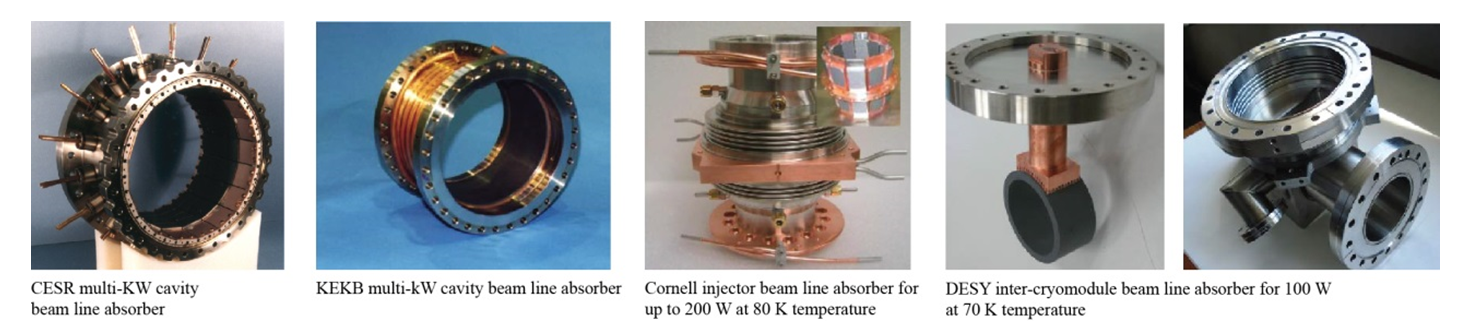}
\caption{From left to right: Room temperature multi-kW BLA for CESR and KEKB, respectively, \SI{200}{\watt} load for \SI{80}{\kelvin} operation at the Cornell injector, and DESY \SI{100}{\watt} load for \SI{70}{\kelvin} operation between cryomodules employed for the EU XFEL \cite{Altarelli07} and adapted for LCLS-II.  Each photograph falls under the Creative Commons Attribution 3.0 (CC-BY 3.0) license (\url{https://creativecommons.org/licenses/by/3.0/}).}
\label{fig:key_challenges:srf:BLAs}
\end{figure}

A proven alternative to coaxial and waveguide HOM dampers are cylindrical beam line absorbers (BLAs) with absorber material placed around a beam tube (see Fig.~\ref{fig:key_challenges:srf:BLAs}).

\begin{figure}[htp]
\centering
\includegraphics[width=0.8\textwidth]{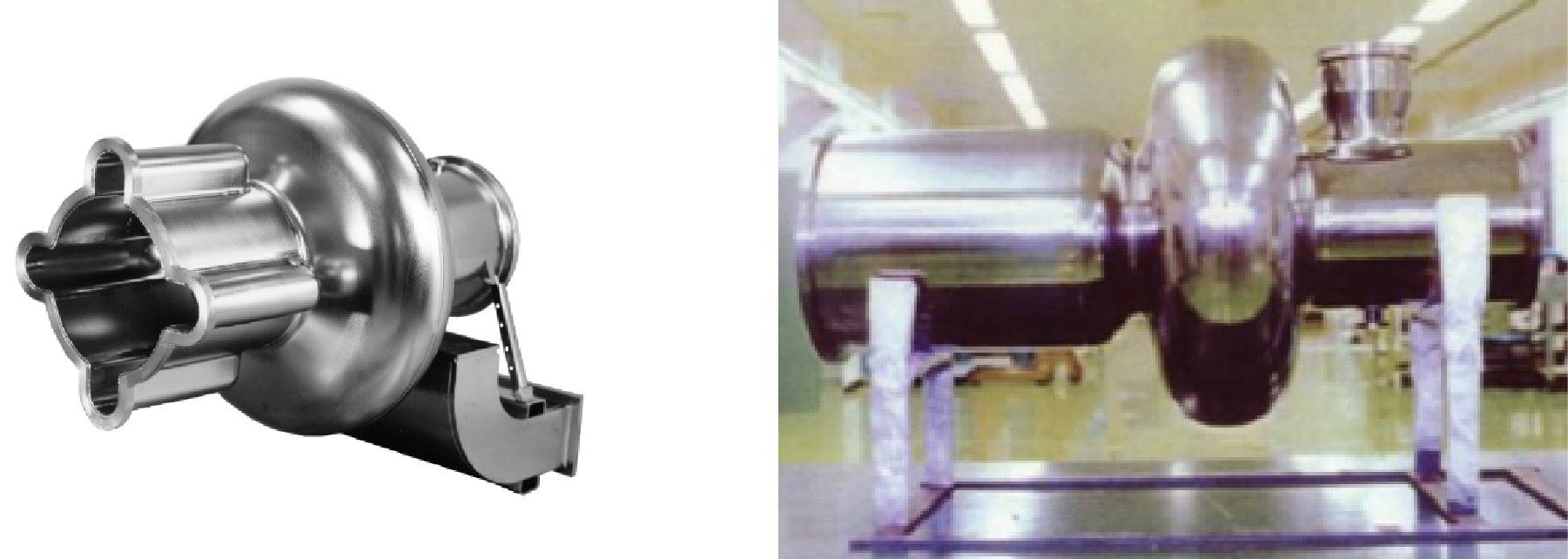}
\caption{The \SI{500}{MHz} CESR cavity (left) \cite{Vogel04}, and \SI{508}{MHz} KEKB cavity (right) \cite{Padamsee98}}
\label{fig:key_challenges:srf:CESR_KEKB_Cavities}
\end{figure}

Depending on the material properties, BLAs can yield relatively broadband damping.
To capture all critical HOMs, the cavity design must allow the lowest resonating modes to propagate out of the cavity above the mode-specific beam tube cutoff and towards a BLA.
Therefore, cavities using BLAs feature transitions to large beam tubes (see Fig.~\ref{fig:key_challenges:srf:CESR_KEKB_Cavities}).

This concept was first successfully employed in storage-ring cavities for average beam currents in the order of \SI{1}{\ampere}.
The first SRF cavities taking advantage of the concept were the CESR and KEKB single-cell cavities, followed later by various other cavities in storage-ring-based light sources \cite{Belomestnykh07}.
Most high-beam-current storage rings operate at \SI{500}{\mega\hertz}, but other RF frequencies have also been employed; for instance, the SR cryomodule cavities at SOLEIL resonating at \SI{352}{\mega\hertz} leveraged the RF technology formerly developed for LEP at CERN at the same frequency.
The SOLEIL cavities, however, rely on coaxial HOM coupler technology (see Fig.~\ref{fig:key_challenges:srf:coaxial_couplers}).

BLAs can be placed on either side of a cavity depending on HOM damping requirements.
Similar to waveguide absorbers, the BLAs can cope with multiple kW of induced HOM power.
Developments were made utilizing both lossy ferrite and ceramic materials.
Some absorbers more recently developed are for rather moderate power capability ($<\SI{1}{kW}$). This includes, e.g., the XFEL (or LCLS-II) inter-cryomodule BLAs that intercept HOMs not captured by the TESLA cavities and propagating through the beam tubes.

\begin{figure}[htbp]\centering
\includegraphics[width=0.8\textwidth]{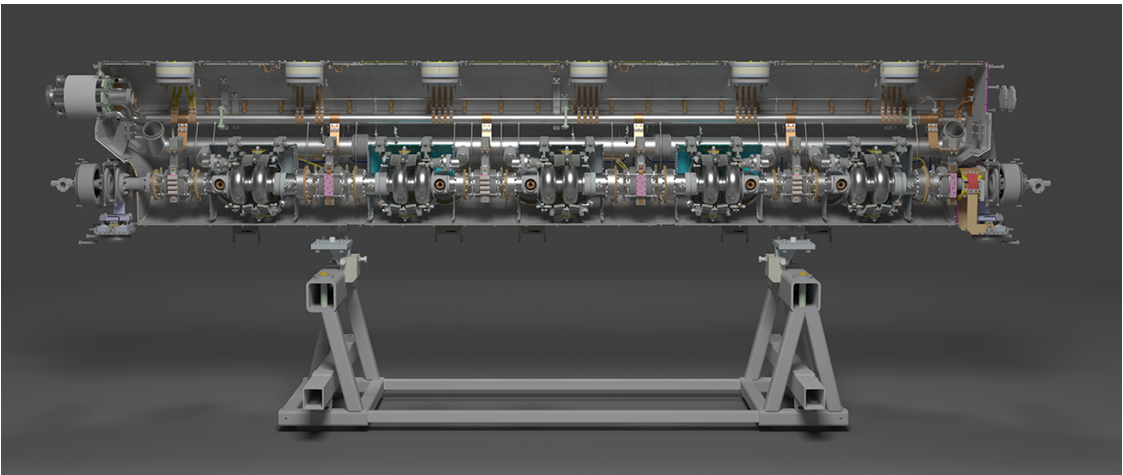}
\caption{Cutaway view revealing the inside of the CBETA injector cryomodule housing five two-cell \SI{1.3}{GHz} SRF cavities each equipped with two \SI{50}{\kilo\watt} fundamental power couplers. The cavities are powered by five \SI{130}{\kilo\watt} CW klystrons (figure taken from \cite{CornellInjector_link}).}
\label{fig:key_challenges:srf:cornell_injector}
\end{figure}

The highly complex CESR BLA was modified for the Cornell ERL injector cryomodule \cite{Shemelin06}.
Depending on the design, the complexity and thus costs of BLAs can be high.
High-power BLAs must be placed at room temperature and water-cooled.
They are placed sufficiently far away from the cavity to minimize the heat load into the cryogenic bath.
This has the disadvantage that the non-accelerating real estate becomes large compared to other designs.
For the Cornell ERL, BLAs conceived for up to \SI{200}{\watt} of HOM power load were placed between all cavities for aggressive HOM damping.
Here, the BLAs are cooled with helium gas at \SI{80}{\kelvin} operating temperature, which requires heat interception at the BLA flanges at \SI{5}{\kelvin}.
This concept could reduce the beamline space consumption, while the innermost cryomodule cavities can share the same BLA.
The cutaway of the Cornell ERL injector in Fig.~\ref{fig:key_challenges:srf:cornell_injector} reveals the five \SI{1.3}{GHz} SRF cavities and the BLAs.
In this design, the HOM power dissipation was estimated to be only \SI{26}{\watt} per load \cite{Chojnacki10}.
The number of cells per SRF cavity is only two compared to seven in the Cornell ERL main linac cryomodule, reducing the beam loading in the injector cavities and thus the FPC CW power capability, which is constrained by the choice of \SI{1.3}{GHz} compared to what is achievable at lower frequencies.
For the same reason, two \SI{50}{\kilo\watt} FPCs feed a single cavity.
This allows one to accelerate a beam current of \SI{100}{mA} at $E_\text{acc} = \SIrange{4.3}{5}{\mega\volt\per\meter}$ in the five-cavity cryomodule. 

The technical challenges of BLAs are similar to those of waveguide absorbers:
The absorbers are brazed to a metal substrate, and the outgassing of particulates from the lossy material has to be minimized; however, the risks are more severe since the BLAs are placed in rather close proximity to the cryogenic environment and directly in the beamline vacuum.
For very high power levels, sufficient heat management of the warm-to-cold transitions also becomes important to limit the heat leaking into the cryogenically cooled surfaces.
The cold-to-warm transitions can therefore occupy a significant length of beamline, which would not be the case when using waveguide dampers.

Furthermore, the electrostatic charge building up on lossy ceramics when the beam passes through can deflect the beam if the material has no residual DC conductivity or is not properly shielded.
In the latter case, some residual DC conductivity is required for charge drainage.
R\&D on this issue is still ongoing: For instance, one aim is to develop thinly coated materials (e.g., \ce{TiN}) on the BLA ceramics to provide DC conductivity while also creating a vacuum barrier to avoid particulate creation.
One challenge is that commercially available absorptive materials can be porous and are often found to be very dirty \cite{Eichhorn}.
Furthermore, batch-to-batch material variations from commercial vendors can be encountered, which can influence the damping efficiency.
R\&D to develop adequate coatings is beneficial for both BLAs and waveguide absorbers struggling with particulate creation and is therefore recommended. 

Instead of brazing individual absorber tiles, the concept of using a single fully cylindrical absorber ring has been considered in the past, e.g., at KEK for their multi-kW BLAs by hot isostatic pressing of a ferrite ring to a water-cooled copper tube \cite{Furuya10} (see Fig.~\ref{fig:key_challenges:srf:BLAs}).
This avoids the risk of individual tiles separating from the brazed substrate, which was experienced during early BLA development for the Cornell ERL \cite{Chojnacki10}.
Alternatively, DESY developed a concept to braze a cylindrical ceramic absorber to a copper stem utilizing a pegboard at its end.
These BLAs are in use in the XFEL and LCLS-II cryomodules (see Fig.~\ref{fig:key_challenges:srf:BLAs}), but for moderate power levels (\SI{100}{\watt} at \SI{70}{\kelvin} temperature).
More recently, a specific lossy \ce{SiC} material (CoorsTek SC-35) was considered at ANL, which had first been identified at JLab as a promising absorber \cite{Marhauser11} and later characterized at Cornell for use as BLA material.
HOM developments at ANL in the context of the Advanced Light Source Upgrade followed earlier studies performed at Cornell \cite{Eichhorn13} exploring the shrink-fit of a cylindrical absorber ring to a metal jacket with flanges \cite{Conway17}.
The production process has meanwhile been refined in an ANL/BNL collaboration for use in the proposed BNL Electron-Ion Collider, where the expected HOM power is as high as \SI{20}{kW} per BLA \cite{Holmes20}.
There is, however, numerical evidence that BLAs are self-heating devices since the beam excites numerous HOMs that are locally trapped due to the comparably high relative permittivity of the ceramic.
These trapped modes create significant, unwanted heating independent from the HOM excited in the SRF cavities; this issue warrants further investigation.
Additionally, the broadband damping characteristics of the BLAs should be characterized experimentally with a dedicated test setup, which can be done both at low and high power.
Corresponding efforts have started at BNL.
To contribute to these ERL-relevant developments in Europe, similar infrastructure should be set up at collaborating laboratories.

Overall, it is recommended to explore the conceptual improvement of all mentioned HOM coupler technologies beyond the present state of the art for potential use in ERLs.

At this point, it shall be mentioned that BNL was planning to build an R\&D ERL in the range of \SI{20}{MeV}, \SI{500}{mA} in the mid-2000s.
BNL also developed a series of high-current ERL cavity concepts over a long period, including different damping schemes.
A design of a high-current, strongly damped five-cell \SI{704}{MHz} cavity with enlarged beam tubes employing ferrite BLAs was conceived in 2003 \cite{Calaga03} and later built by industry (AES/US).
Tests of the cavity installed in the ERL by 2011 uncovered thermal issues in CW operation \cite{Sheehy11}.
These ultimately limited the CW operation to \SI{12}{\mega\volt\per\meter}.
The cause was attributed to increased heating in the \ce{NbTi} beam tube flange sealed by \ce{AlMg} seals, which is the preferred sealing technology for SRF cavities today. The reason was RF leaking into the gap that the seal leaves at the flange. 
The magnetic field of $\approx \SI{200}{\ampere\per\meter}$ at the seal and the rather poor thermal conduction of the \ce{NbTi} flange caused the flange temperature to rise beyond $T_\text{c}$, which caused the beam pipe to become normal-conducting, resulting in a thermal runaway and the cavity quenching prematurely.
Nevertheless, the cavity was commissioned in the BNL R\&D ERL in 2014.
Other issues, specifically related to the electron source, limited the average beam current to \SI{22}{\micro\ampere} \cite{Kayran14}.
Meanwhile, the project has been halted at BNL.
Such issues are relevant for ERLs but can be avoided by proper numerical design work requiring trained personnel.
It is therefore of high importance that the European roadmap planning for ERLs consider the investment in next-generation scientists and engineers to satisfy labor demands.

\subsection{Multipacting}
Multipacting (MP) refers to an unwanted self-amplification of secondary electron current that is emitted from the cavity walls and then impinges on them again in resonance with the RF field, multiplying as a result.
% MB: Rewrote the original sentence; you had to know what it was to begin with in order to understand the explanation.
%the strong augmentation of free electrons impacting cavity walls and being in resonance with the RF field.
As a consequence, MP can significantly drain the energy stored in the cavity and needs to be suppressed.
Fortunately, such resonances can be mitigated by design, which led to the typical elliptical cavity cell shape employed for medium-to-high-particle-velocity SRF cavities (see Fig.~\ref{fig:key_challenges:srf:tesla9cell}).
Yet, resonant conditions for MP still exist in every elliptical cell, though confined to the equator region.
The RF field regime in which resonant MP occurs is called a barrier and can cover several MV/m.
It can be readily understood that a resonant MP barrier in a cavity cell will shift to a lower regime if the resonant frequency is lowered.
In an elliptical cell, such a barrier is usually `soft', i.e., easy to push through when increasing the RF field levels---or not even observable---and resonance trajectories ultimately cease to exist once the barrier is overcome.
Sometimes, however, such MP barriers are harder to break through during high-field ramp-up, depending on the cleanliness at the impact sites.
A cleaner surface will lower the average secondary electron yield (SEY) per electron hitting the impact site.
The SEY is a function of the impact energy, which in turn can depend on the detailed surface topology at the resonant sites.
Prolonged RF field operation within the hard MP barrier can help to process the surface to lower the SEY (``RF conditioning'').

Today, MP is of lesser concern in elliptical cavity cells than it is in the auxiliary components.
For instance, simple scaling of coupler components to other frequencies can inadvertently create hard MP barriers.
In coaxial HOM-couplers, this does not only concern drainage of the stored energy but a detuning of the fundamental notch filter due to the resulting thermal deformations.
Such operational concerns were encountered at the \SI{805}{MHz} SRF proton cavities of the Spallation Neutron Source (SNS), in which scaled TESLA-type couplers were causing operational failures \cite{Howell14}.
The MP barrier occurring within the HOM coupler cans was unfortunately not identified until the cryomodules were installed in the machine; the issue was only verified by simulations in hindsight \cite{Ko10}.
These concerns were echoed for the SRF linac cavities at the European Spallation Source (ESS), but already at the design stage to avoid similar consequences \cite{Molley10}.
For the SNS, a program was initiated to remove the existing coaxial hooks from their HOM cans to eliminate the MP.
This was only justifiable since HOM damping for SNS proton cavities is not a stringent requirement as opposed to electron machines thanks to the larger mass of the protons.

Note that the use of two waveguide dampers in more than 300 CEBAF cavities showed no evidence of MP concerns during operation, and no evidence of MP was found in the \SI{1.5}{GHz} and \SI{0.75}{GHz} HC cavity prototype at JLab during vertical tests.
If improved coaxial HOM couplers are considered in future ERLs, leveraging the full potential of today's 3D RF simulation tools is highly recommended to eliminate the occurrence of hard MP barriers by design.
Such simulations can however be tedious depending on the problem size, and funding for expanding the capabilities of present-day software tools should be considered.

\subsection{Field Emission}
A general issue in all SRF cavities is the presence of \si{\micro\meter}- or sub-\si{\micro\meter}-size field emitters present on cold cavity surfaces in regions of high electric field.
The nature of these field emitters can be manifold but is usually attributed to the presence of particulate contamination and surface defects or irregularities.
These occurrences can cause significant local electric-field enhancements to enable the extraction of unwanted electrons via a quantum-mechanical process from the metal surface into the cavity vacuum, i.e., the tunneling through a potential barrier as first described by Fowler and Nordheim \cite{Fowler28}.
The local field enhancement leads to field emission (FE) to occur at RF peak field levels at merely a few MV/m to a few tens of MV/m, which otherwise would require fields in the GV/m range.
The RF field levels enabling FE are therefore well in the range at which SRF cavities have to operate in ERLs.
Since FE scales exponentially with increasing field, the concerns quickly compound for machine concepts that rely on high accelerating fields (\SI{20}{\mega\volt\per\meter} range or even less). 

Surface cleaning protocols are still being improved to mitigate the probability of field emitters.
Noticeable improvements were made at industrial vendor sites as well as expert laboratories---verified through vertical tests---by enforcing stricter cleanliness and assembly protocols.
This includes increased and repetitive cycles of ultra-pure high-pressure water rinsing (HPR) of the cavity interior during post-processing, which has been proven to be effective to wash out particulates loosely adhering to surfaces, but also stricter clean-room protocols to avoid recontamination of cavities during auxiliary hardware assembly in the clean room.
Such strategies were implemented and recorded for quality control measures for both the European X-ray XFEL (EXFEL) and the Linac Coherent Light Source (LCLS-II) FEL projects and more strictly than for any previous projects.

Despite these efforts, no guarantee can be provided to fully avoid the occurrence of field emitters in SRF cavities.
Sorting out cavities that carry stubborn field emitters before cryomodule installation and trying to chemically re-process and re-HPR the cavities can be required, affecting both the schedule and project costs.

The issue of field emission should not be underestimated since the consequences can be severe.
Note that field-emitted electrons can travel both upstream and downstream in the accelerator.
The electrons eventually hit either the cavity or other beamline components, losing their energy.
This can cause not only local heating effects but prompt gamma and neutron radiation depending on the impact energy.
As a consequence, cryomodules and beamline components can be highly activated after extended operational periods with both short- and long-lived isotopes being produced.
The resulting radiation damage to cryomodule and beamline components is a problem for SRF accelerators.
The severity of the radiation dose depends on the electron energy at impact and therefore the operating field strength of the cavities, but also on the materials involved. 

The concerns augment if field-emitted electrons are captured by the RF field to synchronously accelerate through neighboring cavities or even through a string of cryomodules with increasing energies.
Such issues are highly apparent at JLab's flagship \SI{12}{\giga\electronvolt} Continuous Electron Beam Accelerator Facility (CEBAF), a 5.5-pass electron recirculator operating at \SI{1.5}{GHz}~\cite{Harwood13}; they strongly affect reaching and maintaining its energy reach~\cite{benesch_fieldemission}.
Considering the large number of cavities---18 in the injector and 400 equally distributed in a north linac and south linac---data of statistical significance is still being gathered.
Currently, this focuses on the CEBAF upgrade cavities that shall operate at \SI{19.2}{\mega\volt\per\meter} nominally.
Immediately after the addition of the first CEBAF upgrade cryomodules, unusually high radiation levels were detected \cite{marhauser13}.
Nearly ten years after the installation of the first cryomodule, the consequences include the degradation of beamline and cryomodule materials, including leaking cryomodule gate valves and dramatic disintegration of magnetic superinsulation \cite{Reece21}.
Pinholes developed in the RF vacuum ceramics within the input power couplers have been identified recently, the cause of which is not fully understood but could be related to FE.
These consequences are highly relevant for future multi-GeV ERLs, influencing choices for cavity and cryomodule component design.
One should consider that the long-term effects are likely more severe in machines operating in CW mode compared to pulsed mode.

\begin{figure}[htbp]\centering%
\includegraphics[width=\linewidth]{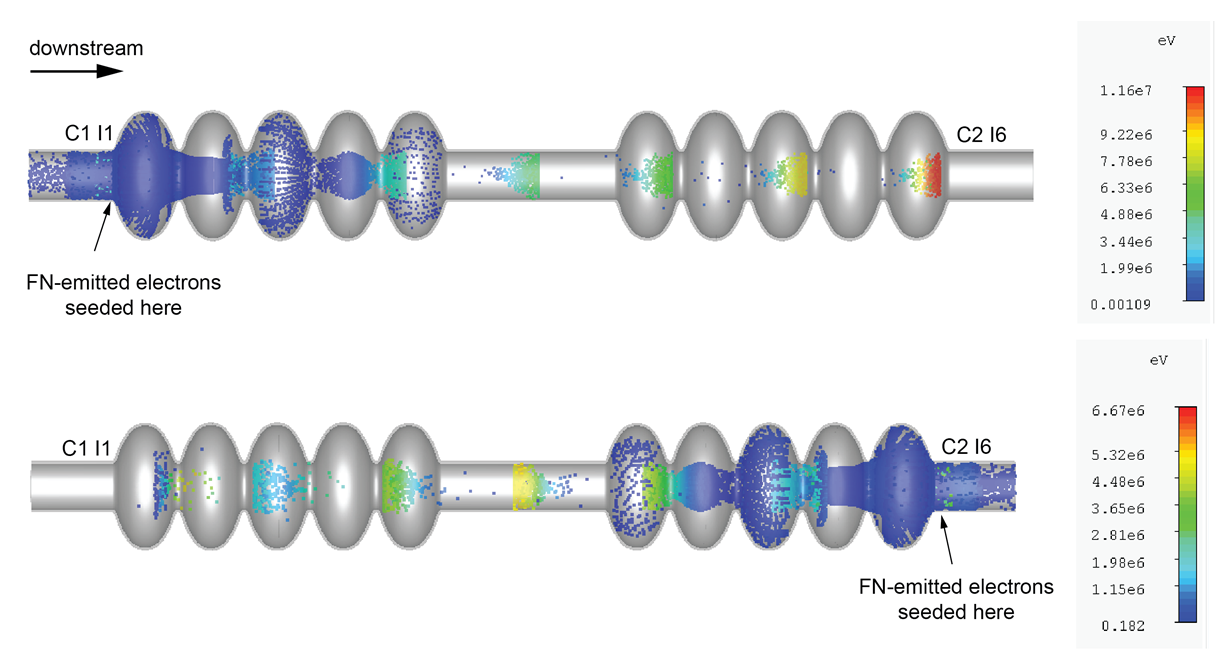}%
\caption{Fowler-Nordheim-emitted (FN) electrons traveling through two five-cell cavities, which are phased to provide maximum energy gain for the main beam. Top: Electrons are continuously field-emitted at the first iris of cavity~1 (C1~I1). Bottom: Electrons are continuously field-emitted at the last iris of cavity~2 (C2~I6).}
\label{fig:key_challenges:srf:Field_emission_trajetories}
\end{figure}

A concept has been outlined to suppress the upstream-directed field-emitted electrons in RF accelerators by design \cite{Marhauser:SRF2015-MOPB061}.
It merely relies on the implementation of cavity-interconnecting beam tubes that are a half-integer of the RF wavelength long. It is therefore not restricted to a certain number of cavity cells.
This concept will mitigate the synchronous acceleration of field-emitted electrons in the upstream direction.
This is illustrated in Fig.~\ref{fig:key_challenges:srf:Field_emission_trajetories}) for two five-cell cavities.
The figure shows the difference of the energy gain of field-emitted electrons in downstream (top) and upstream direction, respectively, as numerically calculated with the energy levels color-coded as quantified by the figure legends.

This concept ideally requests similar operating field levels in all cavities to efficiently annihilate the once accumulated energy.
Yet, even with some discrepancy in operating fields, one can expect a significant energy reduction for upstream-directed electrons within a relatively short distance.
Electrons will then impact on the surfaces at rather low energies, in turn reducing all FE issues described above.
The only implication is that the accelerator cannot be used for scenarios which envision the acceleration of beams in both directions.
Verifying such a concept experimentally would require the development of prototypes.
Such developments are hardly supported by funding agencies in the context of generic R\&D but instead require a strategic commitment to solid, continuous funding, otherwise only feasible in the context of larger, project-oriented R\&D.
It is however recommended that the conceptual planning of ERLs explore alternative ideas beyond state-of-the-art designs as early as possible. 

\subsection{Frequency Choice}
The choice of the cavity RF frequency and accelerating field affects all other systems and has to consider a wide range of other infrastructure.
For instance, the dynamic RF losses, $P_\text{RF}$, in SRF surfaces scale with $E_\text{acc}^2$.
This influences the choice of the optimum operating field and temperature apart from the RF frequency.
Capital and operational cost analyses for GeV-scale SRF linacs showed that operating fields above \SI{20}{\mega\volt\per\meter} are not favorable for CW machines---as opposed to pulsed machines such as the International Linear Collider (ILC), in which $E_\text{acc} > \SI{30}{\mega\volt\per\meter}$ is mandated \cite{ILC07}.
This statement applies to ERLs, which consequently benefit from moderate operational fields.
It is especially beneficial to mitigate field emission concerns as described above.  

A related study has been carried out in detail for a pre-conceptual design of a \SI{1}{\giga\electronvolt} proton linac that has shown that dynamic RF losses can be minimized by choosing lower operating frequencies than the predominant \SI{1.3}{\giga\hertz} for cost reasons, resulting in a rather broad minimum around \SI{800}{MHz} \cite{Marhauser:2014fya}.
The optimization studies started by assessing the optimum frequency to minimize $P_\text{RF}$ normalized per unit of active accelerating length, $L_\text{cell}$, and $E_\text{acc}^2$.
This normalization allowed evaluating cost optimum of the machines, while the frequency (thus $L_\text{cell}$) and $E_\text{acc}$ remained optimization parameters.
Since $P_\text{RF}$ depends on the superconducting properties of the materials, reference data have been used for fine-grain niobium that can be scaled to any frequency per BCS theory.
However, the residual resistance is also a function of the frequency---per present knowledge, while the scaling laws are still subject to R\&D---which has been assessed using experimental data for a large number of SNS, ILC, and CEBAF cavities.

\begin{figure}[htbp]\centering%
\includegraphics[width=\linewidth]{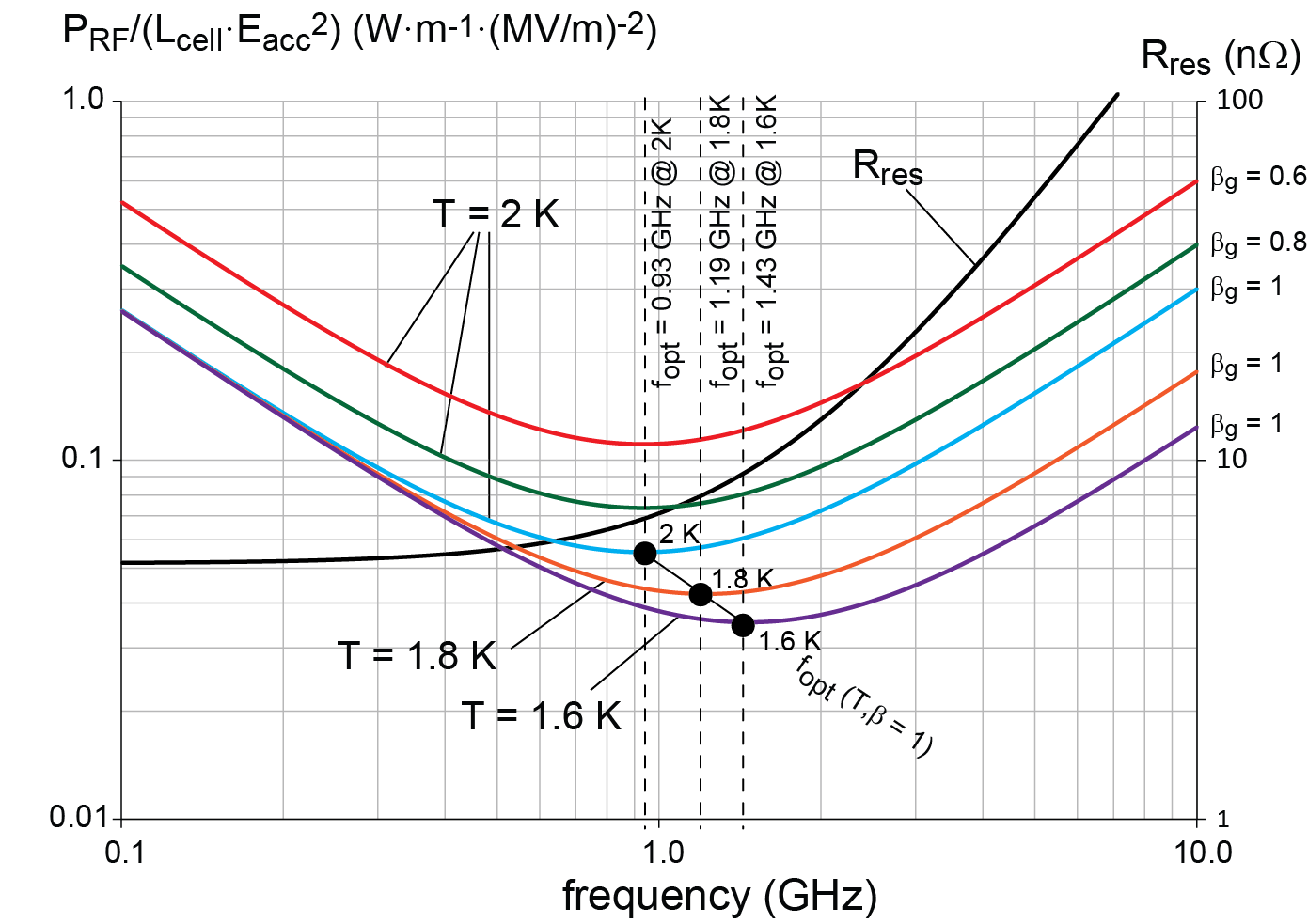}%
\caption{Normalized dynamic RF losses in elliptical SRF cavities (fine grain high-RRR Nb, BCP or EP, low-T baked) for different geometrical $\beta$-values and as a function of a frequency-dependent residual resistance. Realistic geometric parameters ($R/Q$ per accelerator cell, $G$) have been assumed depending on $\beta$. In this case, the optimum frequency is \SI{0.93}{GHz} at \SI{2}{K} and \SI{1.43}{GHz} at \SI{1.6}{K}, respectively.}%
\label{fig:key_challenges:srf:Dynamic_RF_losses}%
\end{figure}

\begin{figure}[htbp]\centering
\includegraphics[width=0.8\textwidth]{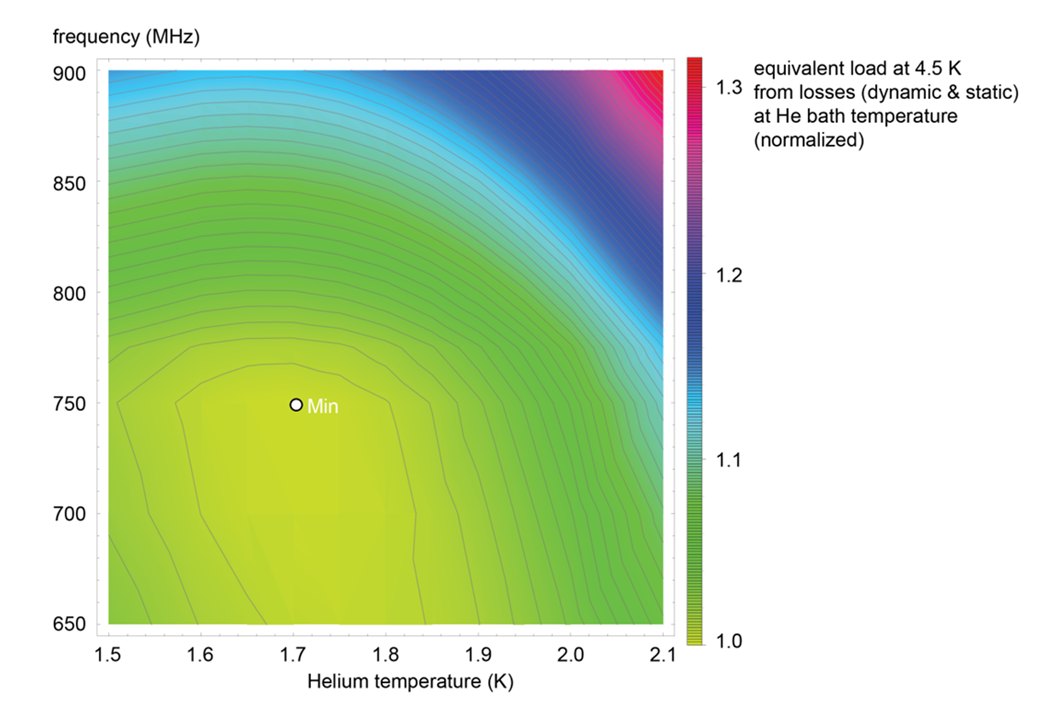}
\caption{Contour plot of the normalized equivalent load (dynamic and static) at  \SI{4.5}{K} in the SRF proton linac (\SIrange{0.1}{1}{GeV}) as a function of frequency and operating helium temperature.}
\label{fig:key_challenges:srf:Equivalent_load_4.5K}
\end{figure}

This resulted in the findings shown in Fig.~\ref{fig:key_challenges:srf:Dynamic_RF_losses}, which includes the velocity of light $\beta = 1$ and lower-$\beta$ cavities relevant for proton linacs.
Here, the RF losses increase with lower $\beta$ due to the squeezed cavities having inferior design characteristics lowering their shunt impedance, $R/Q$, and geometry factor, $G$.
For $\beta=1$, the normalized dynamic losses are evaluated at three temperature levels: \SI{1.6}{K}, \SI{1.8}{K}, and \SI{2}{K}.
For instance, the optimum frequency at \SI{2}{K} would be \SI{0.93}{GHz}, whereas it is \SI{1.43}{GHz} at \SI{1.6}{K}.
Lowering $T$ will lower the minimum dynamic losses achievable, but this comes with increased complexity of the helium plant.
The capital cost of the main helium cold box (at \SI{4.5}{K}) was assessed based on the equivalent refrigeration capacity at \SI{4.5}{K}, which also had to include the estimated static losses of the cryomodules.
The normalized equivalent loss at \SI{4.5}{K} is plotted in Fig.~\ref{fig:key_challenges:srf:Equivalent_load_4.5K} as a function of temperature and frequency.
This reveals that the optimum temperature to minimize the equivalent load would be  \SI{1.7}{K} with the overall optimum frequency to minimize the dynamic losses now at \SI{750}{MHz}.
The minimum is however rather broad.
Yet, the cost difference of operating around the typical \SI{2}{K} or at \SI{1.7}{K} was rather marginal.
Moreover, temperatures below  \SI{1.8}{K} have not been employed for SRF cavities so far.
Thus, an operating temperature between \SI{1.9}{K} and  \SI{2.0}{K} provides a good compromise without adding complexity to the cold compressors of the helium plant.

Note that the optimum frequency shifted to higher values (within \SIrange{800}{850}{MHz}) once the cost was allowed to scale with the cryomodule and overall tunnel length.
Choosing, e.g., \SI{805}{MHz} would take advantage of SNS RF system developments and operational experiences\footnote{The knowledge of the equivalent load allowed to assess the cost for the helium plant for various operating temperatures. Furthermore, with parametrized cost models for cavities and cryomodules components---as a function of frequency---the cryomodule hardware costs were evaluated and cross-checked with past expenses for SNS and CEBAF cryomodules to provide realistic estimates utilizing existing work breakdown structures. The linac tunnel costs were included, which linearly scale with the length. This allowed performing a capital and operational cost analysis of an SRF proton linac as a function of frequency and operating temperature.}. It was therefore concluded that a frequency around \SI{800}{MHz} at an operating temperature of \SI{2}{K} is a good compromise taking into account capital as well as long-term operational costs. 

\begin{figure}[htbp]\centering
\includegraphics[width=0.8\textwidth]{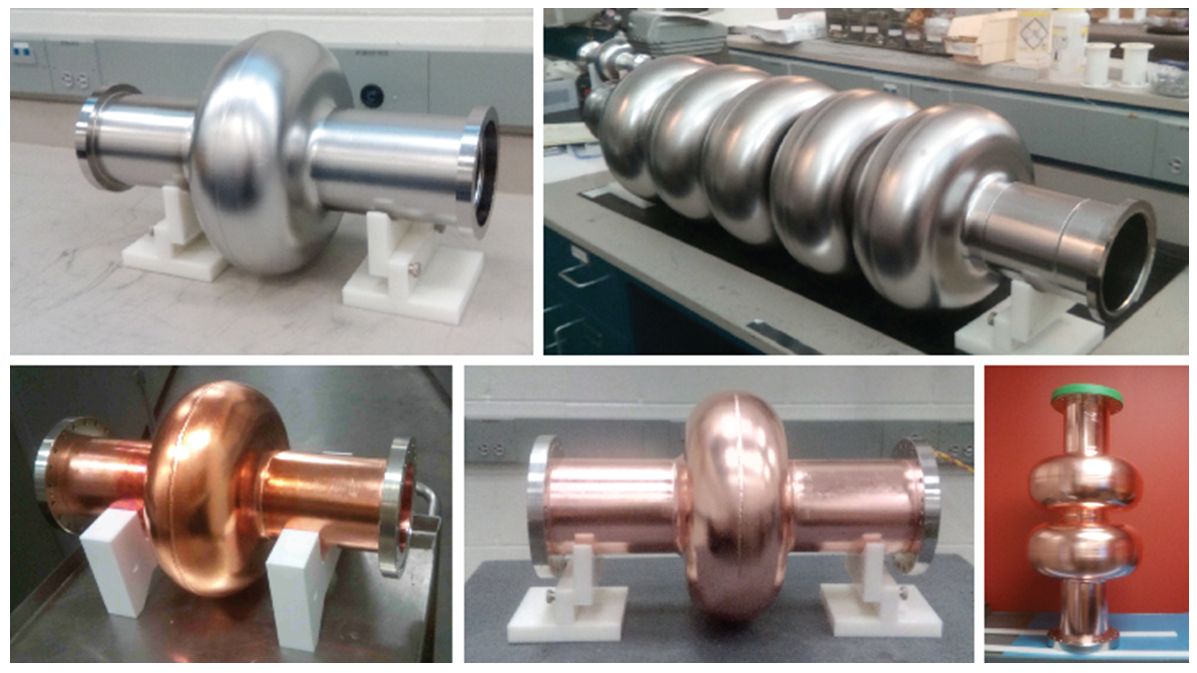}
\caption{\SI{802}{MHz} ERL cavity prototype development at JLab for CERN.}
\label{fig:key_challenges:srf:ERL_Cavities_Well_Balanced_Design}
\end{figure}

The outcome of this analysis was important for CERN's interests in a high-beam-current, multipass ERL, specifically for a \SI{80}{GeV} three-pass race-track ERL for the LHeC \cite{Calaga13}.
Taking into account the constraints of the LHC bunch repetition frequency of \SI{40.079}{MHz}, while allowing for a sufficiently high harmonic, $h$, for a flexible system, a frequency of \SI{801.58}{MHz} ($h = 20$) was chosen, which is very close to the SNS SRF linac frequency.
This choice of frequency would also enable nearly equal bunch spacing with three recirculating passes.
JLab and CERN then collaborated to develop an \SI{801.58}{\mega\hertz} five-cell SRF cavity made from fine-grain Nb.
Though the prototype efforts focused on the five-cell cavity, JLab also produced two OFHC copper cavities for Nb thin-film sputtering R\&D at CERN, a single-cell cavity for N-doping/infusion studies at FNAL, as well as a two-cell, low-power copper cavity for low-power bench measurements.
The ensemble of this cavity development is shown in Fig.~\ref{fig:key_challenges:srf:ERL_Cavities_Well_Balanced_Design}. 

\begin{figure}[htbp]\centering
\includegraphics[width=0.8\textwidth]{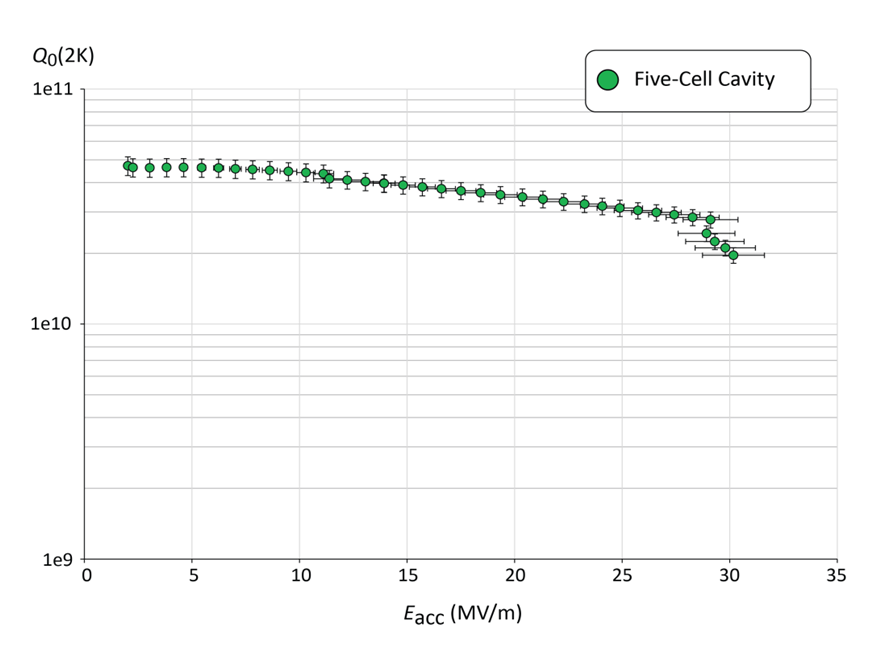}
\caption{Vertical test result of the five-cell \SI{802}{MHz} niobium cavity.}
\label{fig:key_challenges:srf:VTA_result}
\end{figure}

The Nb five-cell cavity was tested vertically at \SI{2}{\kelvin} at JLab with the result that a $Q_\text{0}$ well above \num{4e10} was achieved at low fields, while $Q_0$-values beyond \num{3e10} could be maintained for up to $\sim\SI{27}{\mega\volt\per\meter}$ (see Fig.~\ref{fig:key_challenges:srf:VTA_result}) before the cavity quenched \cite{Marhauser18FCC}. 

Overall, the results verified the prior theoretical assessment that a low surface resistance should be achievable around \SI{800}{MHz} accounting for the frequency dependency of both the BCS and residual resistance ($\approx\SI{3.2}{\nano\ohm}$).
This has the advantage that mature interior surface post-processing methods can be applied.
Newer treatment methods for achieving high-$Q_0$ cavities such as N-doping or N-infusion \cite{DHAKAL2020100034} are therefore not necessarily beneficial at this frequency since the BCS resistance is already small, while an increased residual resistance could be a drawback.
In this respect, high-$Q_0$ surface treatment methods are presently more beneficial for higher-frequency cavities such as TESLA cavities though it required adaptions of the recipe as detailed above.

\begin{figure}[htbp]\centering
\includegraphics[width=0.8\textwidth]{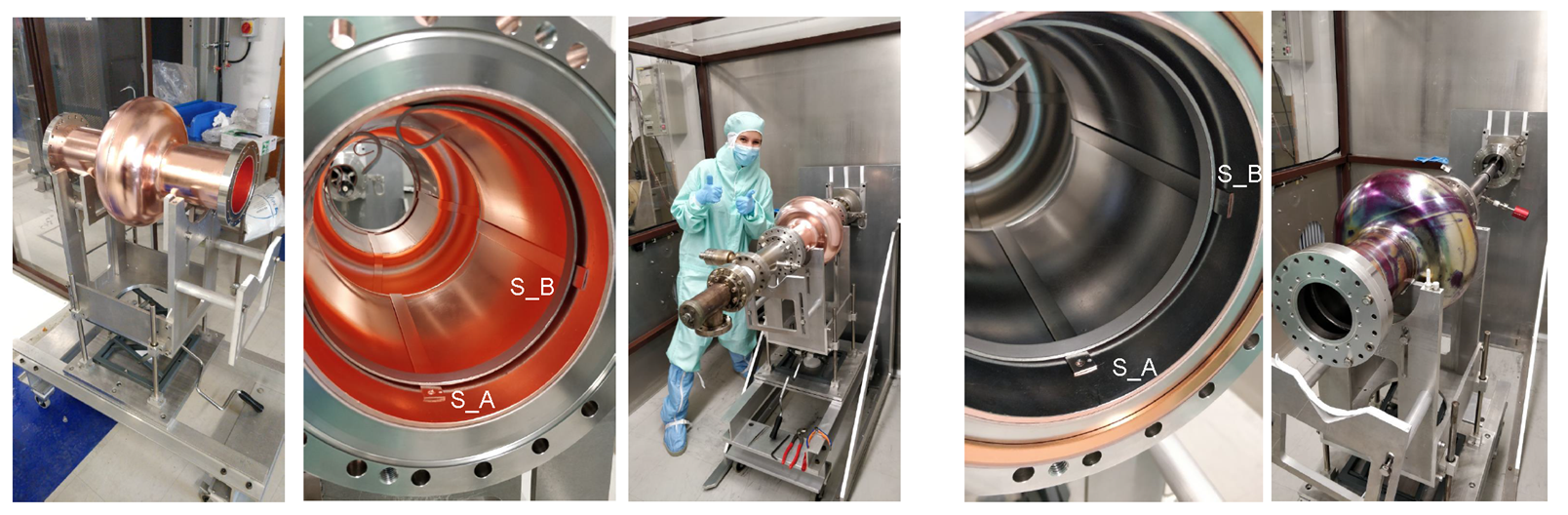}
\caption{First Nb thin-film-coated \SI{802}{MHz} cavity at CERN. The three leftmost figures show the copper cavity before coating including the clean-room assembly. The two rightmost figures show the cavity after the Nb coating \cite{Pereira21}.}
\label{fig:key_challenges:srf:Nb_on_Copper}
\end{figure}

Alternatively, the Nb thin sputtering technology on copper has been applied in the past for LEP and LHC cavities at CERN, but constraining cavities to reach rather low fields ($<\SI{10}{\mega\volt\per\meter}$).
More recent advances at CERN however show promising results to boost the $Q_0$ and $E_\text{acc}$ to higher levels \cite{Rosaz18}.
Meanwhile, the first Nb thin-film coating test on one of the \SI{802}{MHz} copper cavities has been carried out with encouraging results (see Fig.~\ref{fig:key_challenges:srf:Nb_on_Copper}).
Two witness samples placed in the cavity were analyzed after the coating process, verifying a rather uniform-thickness, dense, and high-quality Nb layer \cite{Pereira21}. RF high-field testing is planned in the future.

All currently existing ERL facilities and recently funded ERL projects serve as a proof of principle to validate the technology before building large-scale ERLs.
This is also true for the PERLE project \cite{Angal_Kalinin_2018}; PERLE, however, was conceived to demonstrate the technology viable for ERLs proposed for the LHeC and FCC-he, both hadron-electron colliders at CERN. For this reason, the \SI{802}{MHz} cavity developed at CERN has been chosen as a baseline design for the main linac. 

\subsection{Cavity Design Choice}
There are several design choices that have to be made for ERLs independent of the cavity frequency.
There is no single optimum cavity design, and beam dynamical aspects have to be considered at the same time.
The design starts with the choice of the number of cells.
Since the beam loading in the injector drives the RF power needs, a small number of cells are needed---as used for the Cornell Injector---due to the limited CW power capability of FPCs, either for coaxial couplers or waveguide couplers.
This immediately benefits from reducing the loaded $Q$ of HOMs \cite{Marhauser17}. 

Achieve higher power capability in FPCs is limited by the ceramic RF vacuum window and the overall challenge to extract high heat loads efficiently before reaching the cryogenic environment.
A reduced power density favors lower operating frequencies.
This implies that the cavities naturally have large boreholes reducing, e.g., beam halo interception but also providing a smaller bunch loss factor.
The latter scales linearly with frequency and reduces the HOM power losses accordingly for a given beam current.
Unfortunately, the concurrent optimization of many key design parameters is mutually exclusive.
This, for instance, concerns the electric, $E_\text{pk}$/$E_\text{acc}$, and magnetic, $B_\text{pk}$/$E_\text{acc}$, cavity surface peak field ratios, for which the desired minimum of one will lead to an increase of the other.
The so-called Low-Loss (LL) cavities aim for a small $B_\text{pk}$/$E_\text{acc}$, while High-Gradient (HG) designs seek a low $E_\text{pk}$/$E_\text{acc}$.
The LL cavities aim at lowering the cryogenic losses ($\propto R/Q G$) but mandate comparably small iris apertures, which is not favorable for HOM damping due to the probability of trapped modes, while the sensitivity to fabrication tolerances increases.
The HG cavities aim for higher-field operation as desired for ILC with the drawback of increased cryogenic losses.
Very high fields are not needed in ERLs based on the cost rationales elucidated above and the increased concerns about field emission.
On the other hand, the cavity walls are more inclined to provide better mechanical stability against Lorentz force and microphonic detuning effects.
This is a challenge in ERL main linac cavities since the energy recovery results in high external $Q_0$ of the FPC and thus a lower resonance bandwidth.
In this case, microphonic peak excursions must be well controllable to avoid cavity trips.
The mechanical stiffness of the cavities plays a role in this aspect.
By adding stiffening rings to the cavity cells, one can compensate for the lack of mechanical integrity of LL cavities.
The high-current (HC) cavities developed at JLab in the past are a trade-off between the two paradigms.
The recent development of the PERLE cavity (see Fig.~\ref{fig:key_challenges:srf:ERL_Cavities_Well_Balanced_Design}) refined the optimization strategy to provide a well-balanced set of all key parameters beneficial for ERLs \cite{Marhauser18FCC}.
The number of cells is five for use in ERL main linac cavities, which seeks a reasonably high ratio of active accelerating length to passive beamline length while preserving the chance of sufficient HOM-damping.
Reducing the impact energy for MP secondary electrons at the cell equators also influenced the design choices at the equator region.

\begin{figure}[htbp]\centering
\includegraphics[width=0.8\textwidth]{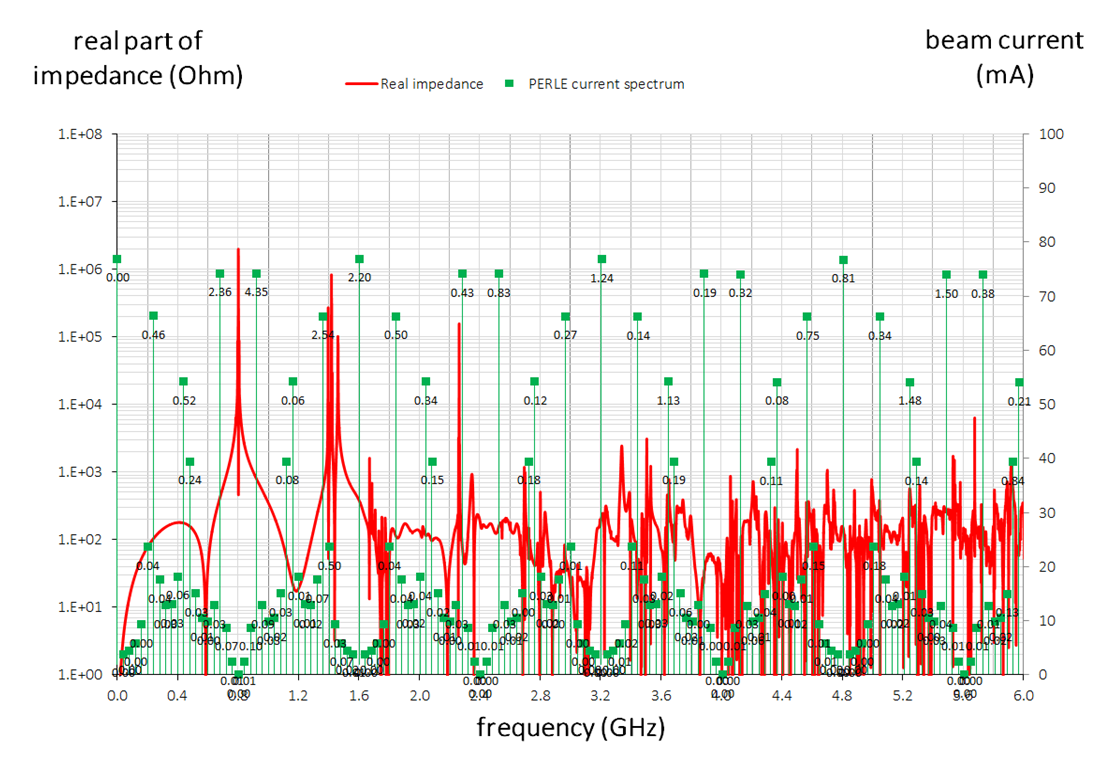}
\caption{Real monopole impedance spectrum of the five-cell \SI{802}{MHz} cavity prototype (red) together with beam current lines (green) for the baseline three-pass PERLE machine (\SI{25}{mA} injected current). The numbers associated with the spectral lines denote the power dissipation (in \si{\watt}), which yet need to be multiplied by two. Note that PERLE design parameters have changed meanwhile with the advance of a new lattice design \cite{Bogacz20}}
\label{fig:key_challenges:srf:PERLE_sprectrum_gaps}
\end{figure}

Moreover and specifically for the PERLE machine to be constructed at ICJLab in France, the most offending monopole and dipole HOM frequencies were cross-checked not to coincide with spectral lines of the beam.
If only every bucket at \SI{801.58}{MHz} were filled, the spectral lines would repeat at this interval.
However, with a more complex bunch pattern of accelerating and decelerating beams in a multipass machine---and not necessarily separated by equal bunch spacing---the beam current spectrum can become complicated.
Such a spectrum is shown in Fig.~\ref{fig:key_challenges:srf:PERLE_sprectrum_gaps} (green lines) for the initial design of PERLE; it is overlapped with the real part of the monopole mode spectrum of the PERLE cavity computed using a Y end-group with the three coaxial couplers.
The fundamental mode is not resolved, but one can see that the corresponding net beam current is zero in energy-recovery mode.
Even with these dense spectral lines, the high-impedance HOMs avoid the major spectral lines associated with high beam current.
Note that multiplying the real impedance with the square of the beam current lines provides the monopole HOM power per spectral line, and summing up the values yields the total power requirement that is shared among three couplers (transverse HOMs will not add dramatically to the assessed value).
In this case, the estimated power was tens of watts, shared among three couplers.
This heat load is deemed to be manageable with actively cooled coaxial couplers, but the appropriate coupler design still needs to be completed.
Depending on the HOMs escaping the cavities, BLAs might need to be employed as well, e.g., outside each cryomodule for their interception, thus not adding to the cryomodule design complexity. 

\begin{figure}[htbp]\centering
\includegraphics[width=0.8\textwidth]{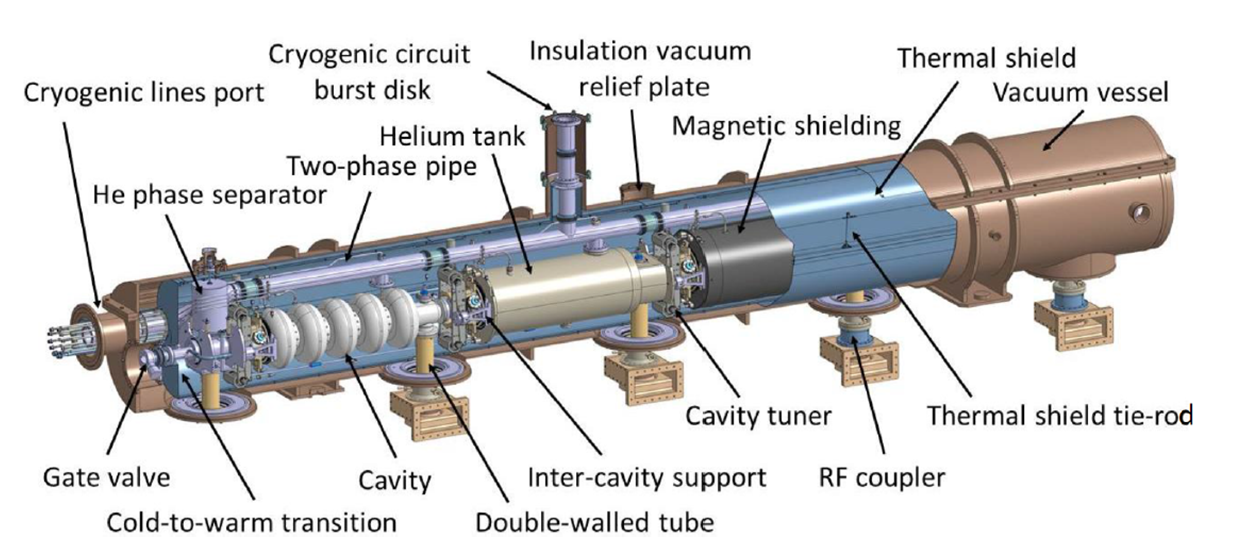}
\caption{General assembly view of an SPL cryomodule for $\beta=1$ five-cell SRF cavities.}
\label{fig:key_challenges:srf:SPL_Cryomodule}
\end{figure}

A prototype cryomodule concept has been worked out within the PERLE project trying to make use of an existing CERN SPL cryomodule vessel that would also accommodate four \SI{802}{MHz} five-cell cavities (see Fig.~\ref{fig:key_challenges:srf:SPL_Cryomodule}) \cite{Agostini20} but requires some modifications.

\begin{figure}[htbp]\centering
\includegraphics[width=0.8\textwidth]{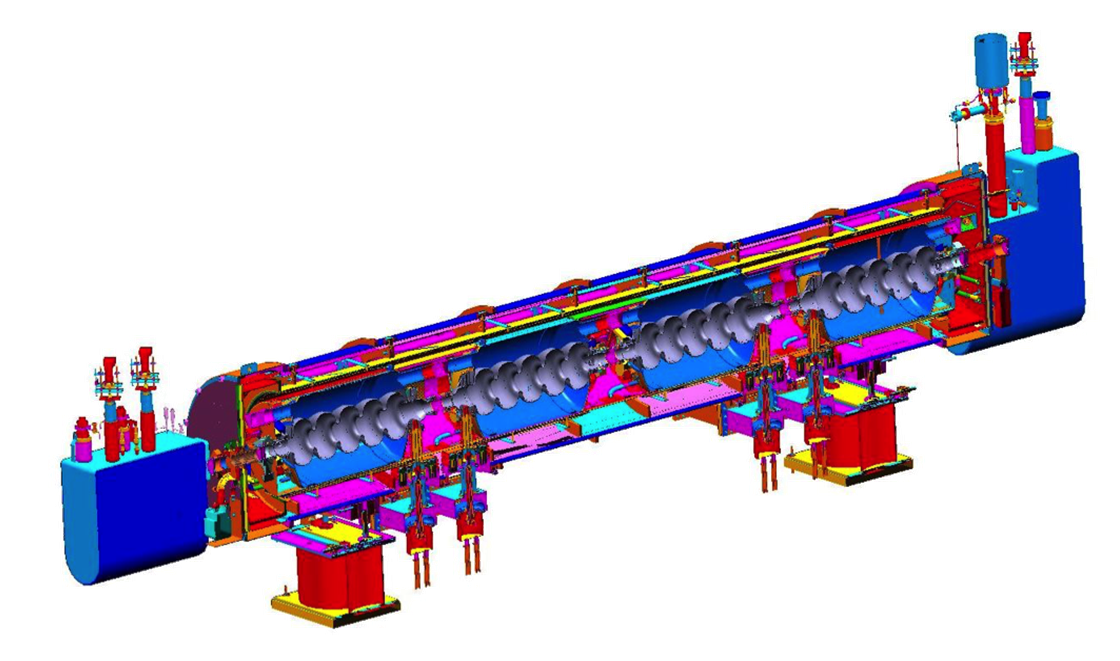}
\caption{General view of an SNS-type cryomodule accommodating PERLE five-cell \SI{802}{MHz} cavities.}
\label{fig:key_challenges:srf:SNS_type_Cryomodule}
\end{figure}

The SNS-type cryomodule has also been envisioned for PERLE based on experience made at JLab with this type of cryomodule for both SNS and CEBAF (see Fig.~\ref{fig:key_challenges:srf:SNS_type_Cryomodule}).
The static loads of the SNS-type cryomodule measured at \SI{2}{K} were typically less than the \SI{28}{W} budget, and the static load of the shield was less than the \SI{200}{\watt} budgeted at $\sim\SI{50}{K}$ (inlet is \SI{40}{K}, and outlet up to \SI{80}{K}).
For PERLE, the dynamic loads of the CW cavities will be much higher than for the pulsed SNS cavities as elucidated earlier.
With a $Q_0$ of \num{2e10} at \SI{2}{K}, the dynamic heat load is already \SI{28}{W} per cavity at the desired $E_\text{acc}$ of \SI{18.7}{\mega\volt\per\meter}.
The maximum dynamic load per module is thus $\approx\SI{112}{W}$, with a total \SI{2}{K} load estimated to be less than \SI{150}{W}.
This is well within the capacity of the helium circuit and end cans.
The choice of \SI{802}{MHz} should allow reaching a $Q_0$ of \num{2e10} at \SI{18.7}{\mega\volt\per\meter} given the prototype results, though $Q_0$ must be preserved after cryomodule installation.

The funding support for PERLE is highly important to drive this technology to a high TRL level while breaking the paradigm of using the more inadequate \SI{1.3}{GHz} SRF technology for high-beam-current ERLs. 

Finally, while the overarching cost optimization, especially for large-scale machines, is often central for funding success, it might not serve well for providing the major performance goals of an ERL.
It might therefore be conceived that the stability of operating an ERL can benefit from not operating too close to performance thresholds (e.g., for $E_\text{acc}$) and particularly in CW operation just for the sake of reducing overall cost.
The ERL inherently provides the chance to recover operational costs that over time could become a cost driver apart from initial capital costs.
More relaxed parameters might then be favorable to achieve success in a faster time scale with fewer technological setbacks.

\subsection{Prospects of New Cavity Surface Treatments and Recipes}
Steady performance improvements of SRF cavities in cryomodules have been made ever since the beginning: by streamlining surface treatments of the delicate SRF interior surfaces and by applying stricter quality controls throughout manufacturing, surface post-processing, and eventually cryomodule assembly.
Several electro-polished SRF cavities have already achieved $E_\text{acc} \approx \SI{50}{\mega\volt\per\meter}$ at various laboratories in vertical tests when cavities were fully immersed in liquid helium.
While such high fields have not been obtained with a high yield, they demonstrate that one can approach the theoretical limit of the peak surface magnetic flux density in pure niobium at which the cavity will ultimately quench.
The pursuit of robustly sustaining higher RF fields at higher $Q_0$-factors continues.
Both figures of merit drive the dynamic RF losses dissipated in the liquid helium and thus the requirements for the cryogenic helium plant.
These losses are consequently directly linked to the capital and operational expenses of the accelerator facility.
As such, the reduction of the dynamic RF losses in SRF cavities is highly relevant for large-scale ERLs.
Note that the static heat load leaking into a cryomodule in CW operation can be kept at much smaller levels by design than the dynamic RF heat load.
This is in contrast to cryomodules operating in pulsed mode at small duty cycles (few percent), where the dynamic RF losses are about an order of magnitude smaller than in CW machines.
Usually, it suffices to install one thermal shield in a CW-operating cryomodule versus two, when the static heat load is comparably high.
In CW, however, other adaptions must be made such as employing larger cavity helium risers capable of extracting the higher RF heat loads.

Reducing the heat load by elevating the cavity $Q_0$ is therefore highly desirable.
For that reason, high-$Q_0$ R\&D has flourished in recent years, pioneered at US laboratories.
Activities included new surface techniques that introduce non-magnetic impurities to the Nb surface as part of the SRF cavity post-processing, namely during standard high-temperature annealing in a vacuum furnace needed conventionally for hydrogen degassing.
A new phenomenon, i.e., the continuous decrease of the resistive losses with increasing field levels (anti-$Q_0$ slope), was first observed by chance at JLab with the diffusion (`doping') of a small amount of titanium during heat treatment of a \SI{1.47}{GHz} single-cell cavity \cite{Dhakal12}.
Doping with nitrogen rather than titanium became more prominent at FNAL thereafter \cite{Grassellino_2013} and was carried out primarily with \SI{1.3}{GHz} cavities.
Those treatments successfully reduce the temperature-dependent BCS resistance, $R_\text{BCS}$, of the superconductor.
A recipe for N-doping was quickly developed and chosen as the basis for the treatment of TESLA cavities for the LCLS-II FEL project at SLAC \cite{Galayda14}.
The N-doping was industrialized for LCLS-II at two aforementioned European vendor sites spearheaded by JLab \cite{Marhauser:IPAC2017-MOPVA131}.
The chosen recipe promised a threefold increase of $Q_0$ at moderate fields ($\sim\num{3e10}$ at \SI{16}{\mega\volt\per\meter} and \SI{2}{K}) compared to undoped cavities as was demonstrated with prototype cavities during the R\&D phase.
However, this recipe did not come without repercussions for LCLS-II. The applied recipe generally lowers the quench field of the cavities compared to undoped cavities, though the achieved fields were predominantly acceptable for LCLS-II per specification.
Secondly, $Q_0$ depends sensitively on the Nb properties such as the grain size.
This resulted in noticeable variations of the residual magnetic flux trapping in the cavity cells, which in turn affect temperature-independent residual losses, $R_\text{res}$.
It was experienced that the severity of the measured $Q_0$-degradation depended on the pedigree of the Nb sheets, i.e., the specific treatments of the sheets at the Nb suppliers.
This problem required continuing R\&D activity during LCLS-II cavity production to adapt the N-doping recipe so that a higher $Q_0$ could be more robustly achieved at the nominal operating field.
To develop a more robust recipe, a promising alternate recipe was explored at the same time, termed N-infusion.
It technically differs from the N-doping by exposing the cavities to nitrogen at lower temperatures in the furnace after high-temperature annealing, while not requiring a final electro-polishing treatment \cite{DHAKAL2020100034}.
It is possible that the TRL of these surface treatments can be further increased to be of use in ERL cavities.
Furthering the R\&D is however still recommended.
The pros and cons concerning the cryomodule complexity need to be evaluated at the same time.
The LCLS-II cryomodules, e.g., are designed to provide fast cool-down for effective magnetic flux expulsion during cool-down, while enhanced magnetic shielding is required to keep the residual magnetic field around the cavities smaller than in conventional cryomodules.
Moreover, the usefulness of the N-doping or N-infusion also depends on the choice of frequency as discussed above.

Alternatively, materials other than bulk niobium are still being explored worldwide \cite{Valente16}.
This includes the use of thin films, either on copper or on Nb, and the use of higher-$T_\text{c}$ materials, the latter promising higher sustainable fields than bulk Nb.
Candidate materials are A15 compounds (like \ce{Nb3Sn} or \ce{V3Si}), \ce{MgB2}, and oxypnictides.
Sputtering a thin film of Nb on copper cavities has already been practiced at CERN for LEP and LHC cavities in the past.
Using copper cavities as a substrate has the advantage to significantly increase the thermal conductivity, making the heat transfer to the helium bath more effective compared to bulk Nb.
Furthermore, fabrication costs can be limited. The challenge is the quality of the sputtered Nb film, which currently limits the reach of high RF fields routinely achievable in bulk Nb cavities.
The creation of high-$T_\text{c}$ compounds on cavity surfaces is technically more complex.
A significant focus in recent years has been the vapor diffusion deposition of \si{\micro\meter}-thick \ce{Nb3Sn} on SRF surfaces, which theoretically yields about twice the $T_\text{c}$ compared to pure Nb.
A detailed review is given in \cite{Posen17}.
The \ce{Nb3Sn} deposition technique is not novel, but was revived at Cornell University in 2009 after pioneering studies had already been conducted over a prolonged time from the 1970s into the 1990s, especially in Germany (e.g., \cite{Hillenbrand75, Heinrichs84}) but also at other institutions, but ceased in 2000.
JLab and FNAL have joined Cornell's renewed interests in \ce{Nb3Sn}-coated cavities thereafter.
A major benefit of \ce{Nb3Sn} cavities is the achievement of $Q_0$-values at \SI{4.2}{K} that bulk Nb cavities can only achieve around \SI{2}{K}.
This boosts the cryogenic efficiency by a factor of 3 to 4 and allows considering chilling the cavities with cryocoolers instead of liquid helium.
Yet, the repeatability of the coating process, especially for multi-cell cavities, and the brittleness of the \ce{Nb3Sn} layer are still major technical challenges to address before planning to apply \ce{Nb3Sn} cavities in ERLs. 

%\section{High Quality SRF: Cavity and Cryomodules}
%  $Q_0 \geq 10^{10}$, including power coupling, HOMs.
% Frank Marhauser, Erk Jensen, Bob Rimmer

\section{Multi-turn ERL Operation and the Art of Arcs}
\label{sec:key_challenges:art_of_arcs}
% CEBAF, cBETA
%Alex Bogacz and Peter Williams
\subsection{Multi-turn Recirculating Linacs and their Extension to Multi-turn ERLs}
The primary motivation for a multi-pass Recirculating Linear Accelerator (RLA) is its efficient usage of expensive superconducting linac structures. CEBAF is the prime example of such a machine. When contemplating the extension of an RLA to propose a multi-turn ERL, many techniques developed for RLAs can be directly transposed. One important example is an appropriate choice of linac optics. The focusing profile along the linac (quadrupole gradients) needs to be set (and then stay constant) so that multiple beams with differing energies may be transported simultaneously without loss.
The main constraint is that adequate transverse focusing must be provided for a given linac aperture. A robust solution is that used routinely in CEBAF: The optics of the two racetrack linacs are symmetric, the first being matched to the first accelerating passage and the second to the last decelerating one. In order to maximize the BBU threshold current, the optics is tuned so that the integral~$\int \frac{\beta}{E}\,\text{d}s$~along a linac is minimized.
This can be implemented by setting up periodic linac optics for the lowest-energy accelerating pass in the first and for the last decelerating pass in the second linac. Higher-energy passes naturally assume more ``caternary''-like optics.

\subsection{Topology and Recovery Transport Choices}
\subsubsection{'Dogbone' ERL}
\begin{figure}[tbh]
  \centering
  \includegraphics[width=0.7\textwidth]{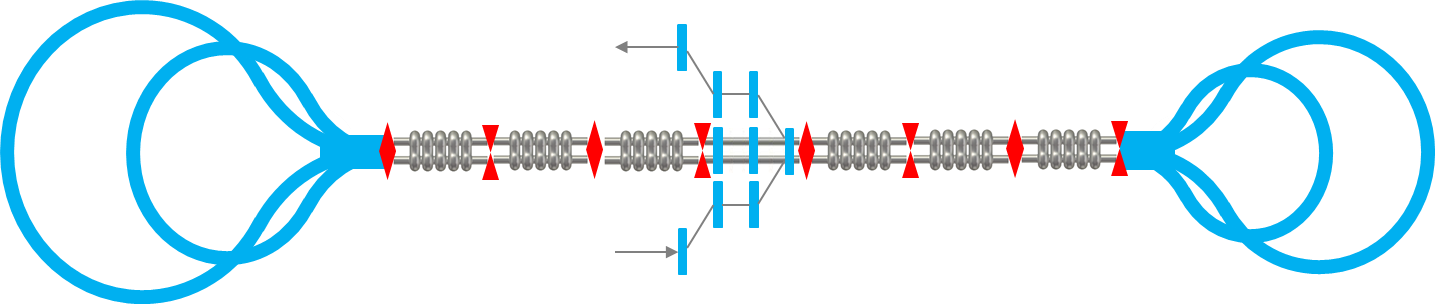}
  \caption{Multi-pass ‘dogbone’ ERL: Schematic view of the accelerator layout; featuring a single SRF linac based on elliptical twin-axis cavities, four return ‘droplet’ arcs and a pair of injection/extraction chicanes.}
  \label{fig:key_challenges:art_of_arcs:Dogbone_ER}
\end{figure}

Efficient usage of linac structures, promised by multi-pass recirculation, can be further enhanced by configuring an RLA in a ‘dogbone’ topology, which boosts the RF efficiency by a factor close to two (compare to a corresponding racetrack)~\cite{bogacz:erl2019-tucoxbs04}.
Such a structure is shown in Fig.~\ref{fig:key_challenges:art_of_arcs:Dogbone_ER}.

\begin{figure}[tbh]
  \centering
  \includegraphics[width=0.7\textwidth]{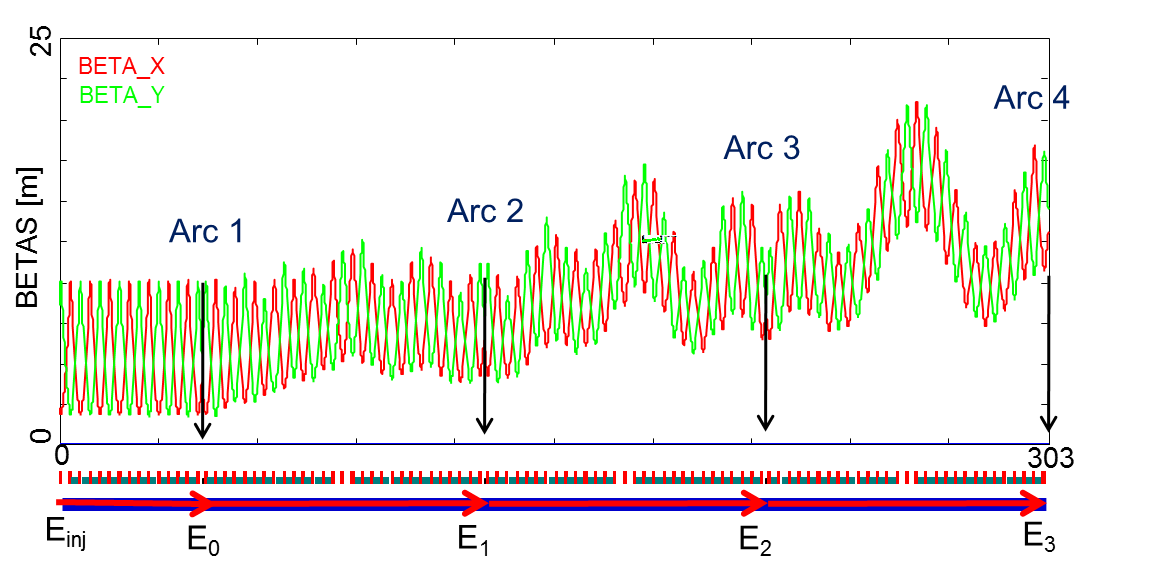}
  \caption{Multi-pass linac optics for all passes, with mirror-symmetric arcs inserted as point matrices (arrows).}
  \label{fig:key_challenges:art_of_arcs:Bisected_Linac}
\end{figure}

A dog-bone-shaped RLA was first considered for rapid acceleration of fast decaying muons, as part of a Neutrino Factory design~\cite{Dogbone_Concept}.  However, the ‘dogbone’ configuration requires the beam to traverse the linac in both directions while being accelerated. This can be facilitated by special ‘bisected’ linac optics~\cite{PhysRevSTAB.12.070101}, where the quadrupole gradients scale up with momentum to maintain a periodic FODO structure for the lowest-energy pass in the first half of the linac; then, the quadrupole strengths are mirrored in the second linac half.
The virtue of these optics is the appearance of distinct nodes in the beta beat-wave at the ends of each pass (where the droplet arcs begin), which limits the growth of initial betas at the beginning of each subsequent droplet arc, easing linac-to-arc matching.
Furthermore, ‘bisected’ linac optics naturally support energy recovery in a ’dogbone’ configuration.
Fig.~\ref{fig:key_challenges:art_of_arcs:Bisected_Linac} illustrates multi-pass linac optics for all passes, with mirror-symmetric arcs inserted as point matrices (arrows).

It would be necessary to configure a dogbone multi-pass electron ERL with elliptical twin-axis cavities~\cite{Twin_Axis}, which are capable of accelerating (or decelerating) beams in two separate beam pipes, avoiding parasitic collisions within the linac which have large beam-beam strength due to the similar betas in both planes.
It also allows for the acceleration of oppositely charged species in each part, enabling an $\positron{}\electron{}$\ collider: such cavities feature opposite longitudinal electric fields in the two halves of the cavity.

A dogbone ERL is attractive in the regime of hundreds of GeV and low current due to the more efficient use of RF.
However, because of the counter-propagating beams, there has been no proposed method to implement ion clearing gaps, precluding its use when a beam current of tens of mA is required.
As the motivation for ERLs is to enable high average beam power, we therefore concentrate on schemes where all beams co-propagate.

\subsubsection{Racetrack ERL}
The simplest layout to utilise a linac as an ERL is to link the end of the linac to its start via a long bypass\footnote{This was actually considered for the SLC in 1968, before the discovery of RF pulse compression~\cite{Hermannsfeldt:1971zz}}. However, this is very inefficient in terms of acceleration per unit beamline / tunnel length (packing fraction). The minimal modification to this is to split the linac symmetrically into a racetrack, leaving only the two \ang{180} arcs at each end being devoid of linac. This is essentially the layout of CEBAF and is described in detail in~\cite{jlab_tn_17_061}.
On implementation of energy recovery, one is faced with the choice of common vs.~separate recovery transport.
%\subsubsection{Common vs Separate Recovery Transport in Racetracks}
%\label{subsec:transport}

\begin{figure}[htb]
  \centering
  \includegraphics[width=0.7\textwidth]{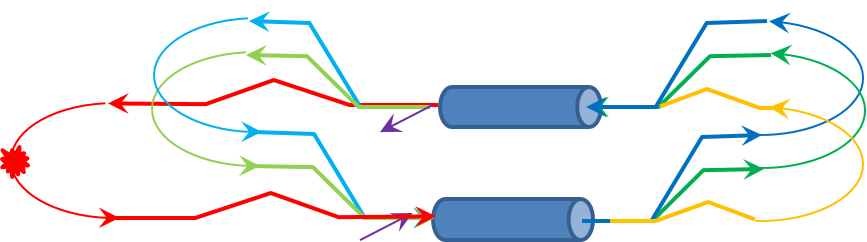}
  \caption{Common recovery transport: The spent beam is re-injected into the injection linac.}
  \label{fig:key_challenges:art_of_arcs:commontransport}
\end{figure}

If we choose to re-inject the spent beam into the injection linac, we select common transport. This means that, to first order, the pass-by-pass accelerating and decelerating beams at all locations outside the linacs have the same energy. They must therefore traverse the same arc beamline, so each arc carries two beams---one accelerating and one decelerating. The advantage of this is that in an $n$-pass ERL, one only needs $n$ arcs at either end. This is illustrated in Fig.~\ref{fig:key_challenges:art_of_arcs:commontransport}.
\begin{figure}[htb]
  \centering
  \includegraphics[width=0.7\textwidth]{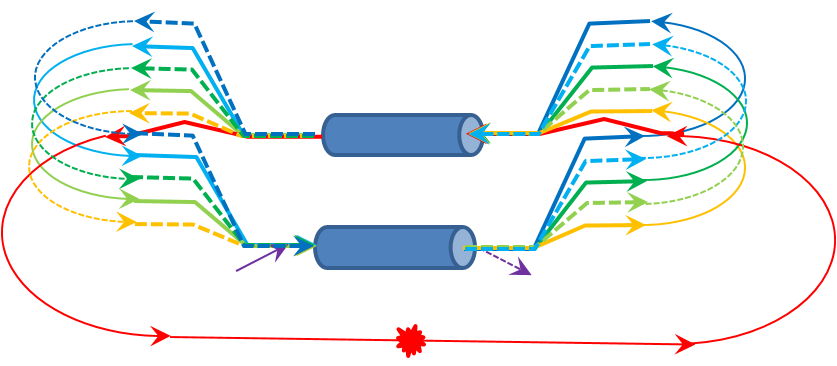}
  \caption{Separate recovery transport: The spent beam is re-injected into the non-injection linac.}
  \label{fig:key_challenges:art_of_arcs:separatetransport}
\end{figure}

Alternatively, if we choose to re-inject the spent beam into the non-injection linac, we select separate transport. Now, all beams outside the linac have well-separated energies and can traverse separate arc beamlines\footnote{This is ``can'' rather than ``must''. For example, if the arcs are FFA, some could physically occupy the same beamline, although the centroid and therefore optics will differ.}. This is illustrated in Fig.~\ref{fig:key_challenges:art_of_arcs:separatetransport}.
At first glance, common transport seems superior as it eliminates the need for $n$ additional arcs, thereby having significant cost implications. However, it also imposes additional constraints on the beam dynamics due to the fact that one loses independent control of the transverse optics and, likely more importantly, the longitudinal phase space. The ability to locally compensate may be of crucial operational importance and requires further study.

At multi-GeV scales, this choice also affects how one deals with synchrotron radiation energy loss: in common transport, one must physically replace the lost beam energy with additional RF structures before each arc in order to match the decelerating beam energy to the accelerating beam. This RF power cannot be recovered. In separate transport, there is no requirement to replace energy lost to SR in order to fit in the transport; however, of course, one cannot decelerate the beam energy to negative values. The energy lost to SR can instead be provided at injection, with the difference between injection energy and dump energy being that lost to SR. This preserves ER within the main linacs.

Common transport also imposes specific requirements on the Twiss function at the linacs ends in that they have to be identical for both the accelerating and decelerating linac passes converging to the same energy and therefore entering the same arc.

%There is an alternative scheme, pointing out that it would be beneficial to separate the accelerating and decelerating arcs. This would simplify energy compensation systems and linac-to-arc matching, but at an higher cost of the magnetic system of the arcs. However, doubling number of arcs is a  costly proposition. On the other hand, the CBETA experiment is pioneering a multi-pass arcs to transport a vast energy range through the same beam-line and it still intends to use them for energy recovery. 

\subsection{Arc Lattice Choices}
\subsubsection{‘Droplet’ Arcs}
At the ends of the RLA linac, the beams need to be directed into the appropriate energy-dependent (pass-dependent) droplet arc for recirculation. The entire droplet-arc architecture \cite{PhysRevSTAB.12.070101} is based on \ang{90}-phase-advance cells with periodic beta functions. For practical reasons, horizontal rather than vertical beam separation has been chosen. Rather than suppressing the horizontal dispersion created by the spreader, it has been matched to that of the outward arc. 
\begin{figure}[tbh]
  \centering
  \includegraphics[width=0.7\textwidth]{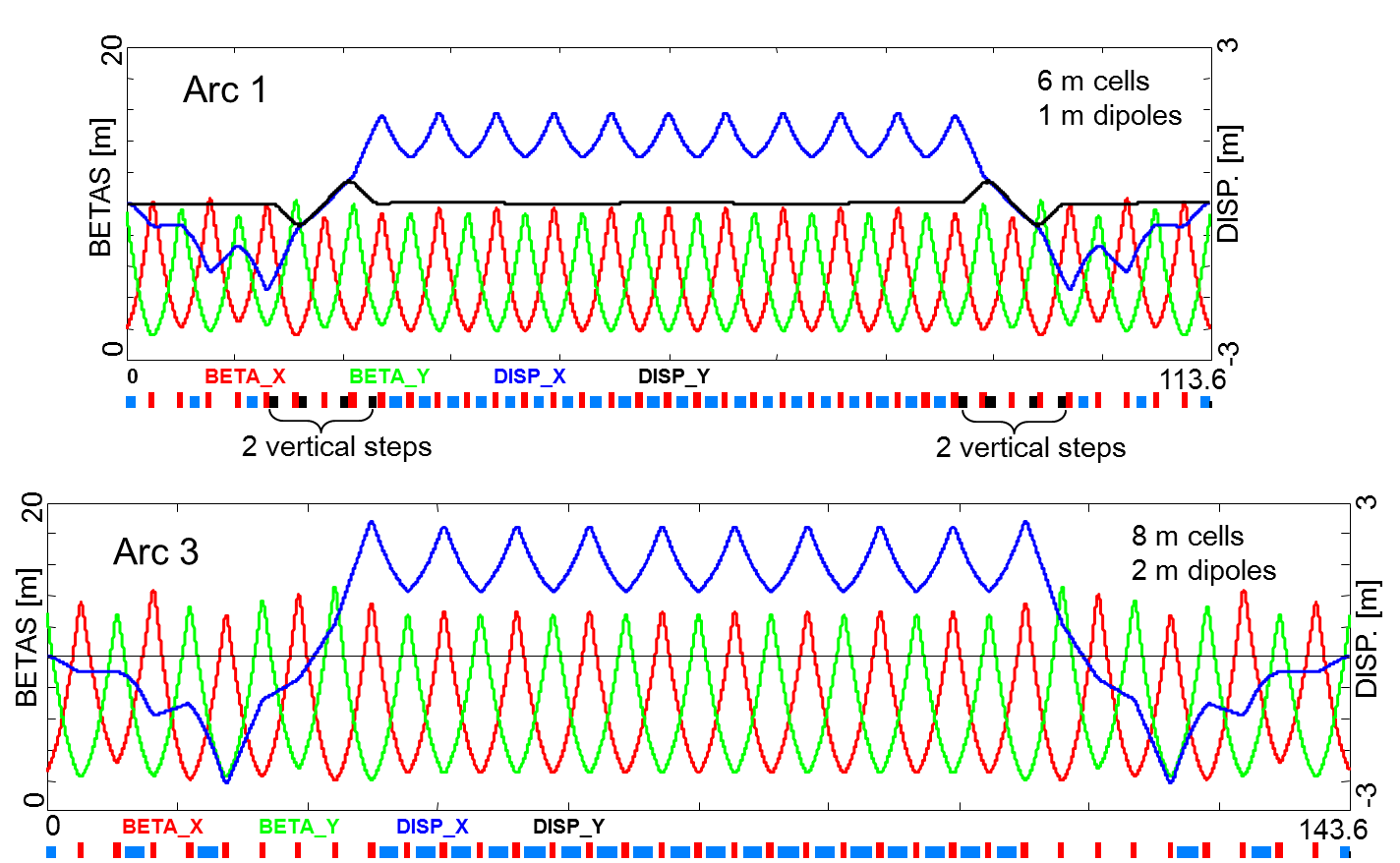}
  \caption{‘Droplet’ arc optics for a pair of arcs on one side of the ‘dogbone’; Arc 1 and Arc 3.}
  \label{fig:key_challenges:art_of_arcs:Droplet_Optics}
\end{figure}
This is partially accomplished by removing one dipole (the one furthest from the spreader) from each of the two cells following the spreader. To switch from outward to inward bending, three transition cells are used, wherein the four central dipoles are removed. The two remaining dipoles at the ends bend in the same direction as the dipoles to which they are closest. The transition region, across which the horizontal dispersion switches sign, is therefore composed of two such cells. To facilitate subsequent energy recovery following acceleration, mirror symmetry is imposed on the droplet arc optics. This puts a constraint on the exit/entrance Twiss functions for two consecutive linac passes, namely: $\beta_n^\text{out} = \beta_{n+1}^\text{in}$~and~$\alpha_n^\text{out} = -\alpha_{n+1}^\text{in}$, where $n = 0, 1, 2\ldots$ is the pass index. 
The complete droplet arc optics for the lowest-energy pair of arcs is shown in Fig.~\ref{fig:key_challenges:art_of_arcs:Droplet_Optics} 
\begin{figure}[tbh]
  \centering
  \includegraphics[width=0.7\textwidth]{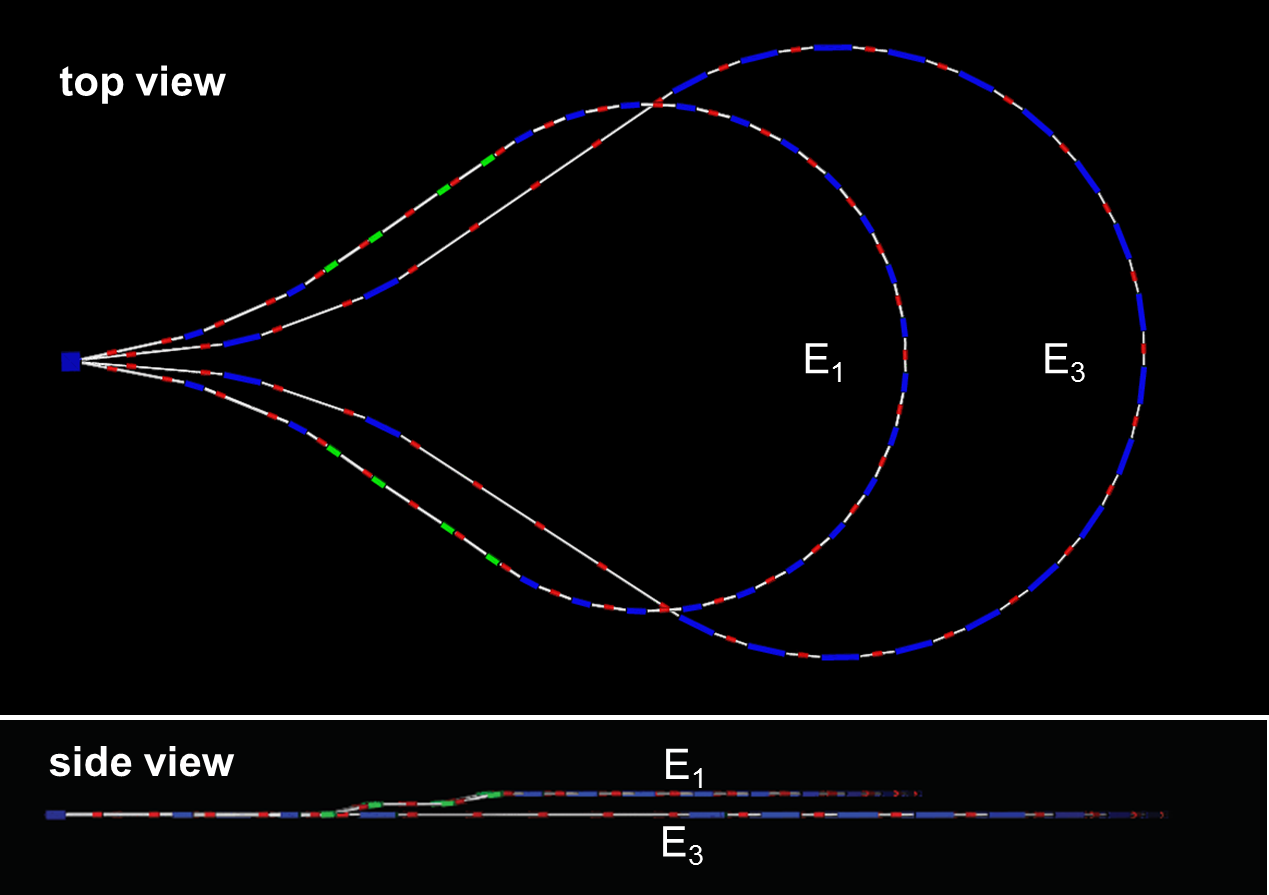}
  \caption{Layout of a pair of arcs on one side of the ‘dogbone’ RLA, Top and side views, showing the vertical two-step ‘lift’ of the middle part of the lower-energy droplet arc to avoid interference with the larger droplet (\SI{1}{\meter} vertical separation).}
  \label{fig:key_challenges:art_of_arcs:Droplet_Layout}
\end{figure}
All higher arcs are based on the same principle as Arc 1, with gradually increasing cell length (and dipole magnet length) to match naturally to the increasing beta functions dictated by the multi-pass linac. The quadrupole strengths in the higher arcs are scaled up linearly with momentum to preserve the \ang{90} FODO lattice. The physical layout of the above pair of droplet arcs is illustrated in Fig.~\ref{fig:key_challenges:art_of_arcs:Droplet_Layout}. 
One additional requirement to support energy recovery in a linac configured with elliptical twin-axis cavities is that the path-length of Arcs 1--3 has to be a multiple of the RF wavelength. Conversely, the Arc 4 path length should be a multiple plus one half of the RF wavelength to switch the beam from the ‘accelerating’ to ‘decelerating’ phase in the linac.

\subsubsection{FFA Arcs}
The ‘dogbone’ ERL can be significantly simplified by replacing a pair of single energy ‘droplet’ arcs with the proposed FFA-like arcs, which are capable of transporting different-energy beams through the same string of magnets. The multi-pass arc design has a number of advantages over separate-arc or pulsed-arc approaches. It eliminates the need for a complicated switch-yard, it reduces the total beam-line length, there is no need to accommodate multiple beam lines in the same tunnel or construct separate tunnels for individual arcs, and there is no need for vertical bypasses, which may be required for separate arcs complicating the optics. This helps to increase the number of passes through the linac, thus enhancing the top energy available with the same-size footprint.
The maximum number of passes through the RLA’s linac is often limited by design considerations for the switchyard, which first spreads the different energy passes to go into the appropriate arcs and then recombines them to align the beam with the linac axis. To reduce the complexity of the above single-energy return arcs, a recent proposal suggests a novel multi-pass arc design based on linear combined-function magnets with variable dipole and quadrupole field components, which allows two consecutive passes with very different energies (factor of two, or more) to be transported through the same string of magnets~\cite{osti_1029306}.
Such a solution combines compactness of design with all the advantages of a linear, non-scaling FFA (Fixed Field Alternating Gradient) optics, namely, large dynamic aperture and momentum acceptance essential for energy recovery, no need for complicated compensation of non-linear effects, and one can use a simpler combined-function magnet design with only dipole and quadrupole field components. The scheme utilizes only fixed magnetic fields, including those for injection and extraction.

\subsubsection{Emittance-Preserving Arc Optics}
Synchrotron radiation effects on beam dynamics, such as the transverse emittance dilution induced by quantum excitations, have a paramount impact on the collider luminosity.  
The transverse emittance dilution accrued through a given arc is proportional to the emittance dispersion function, $H$, averaged over all arc bends~\cite{Sands}: % as expressed by Eq.~\eqref{eq:Emit_dil_0}:
\begin{equation}
  \Delta \epsilon = \frac{2 \pi}{3} C_q r_0 \left<H\right> \frac{\gamma^5}{\rho^2}\,,
  \label{eq:Emit_dil_0}
\end{equation}
where
\begin{equation}
  C_q = \frac{55}{32 \sqrt{3}} \frac{\hbar}{m c} ,
  \label{eq:C_q}
\end{equation}
$r_0$ is the classical electron radius, and $\gamma$ is the Lorentz boost.
Here,
\begin{equation}
H = \frac{1+\alpha^2}{\beta} D^2 + 2 \alpha D D' + \beta D'^2 ,
\end{equation}
where $D, D'$ are the bending plane dispersion and its derivative, with $\left<\ldots\right> = \frac{1}{\pi}\int_\text{bends}\ldots~\text{d}\theta$.

Equation~\eqref{eq:Emit_dil_0} shows that emittance dilution can be mitigated through an appropriate choice of arc optics (values of $\alpha, \beta, D, D'$ at the bends). 

The optics design of each arc takes into account the impact of synchrotron radiation at different energies. At the highest energy, it is crucial to minimise the emittance dilution due to quantum excitations; therefore, one may choose the Theoretical Minimum Emittance (TME) lattice. For lower-energy arcs, the Double-Bend Achromat (DBA) may prove to be sufficient. 

\subsection{The Spreader-Arc-Recombiner as a single system}
\begin{figure}[htb]
  \centering
  \includegraphics[width=1.0\linewidth]{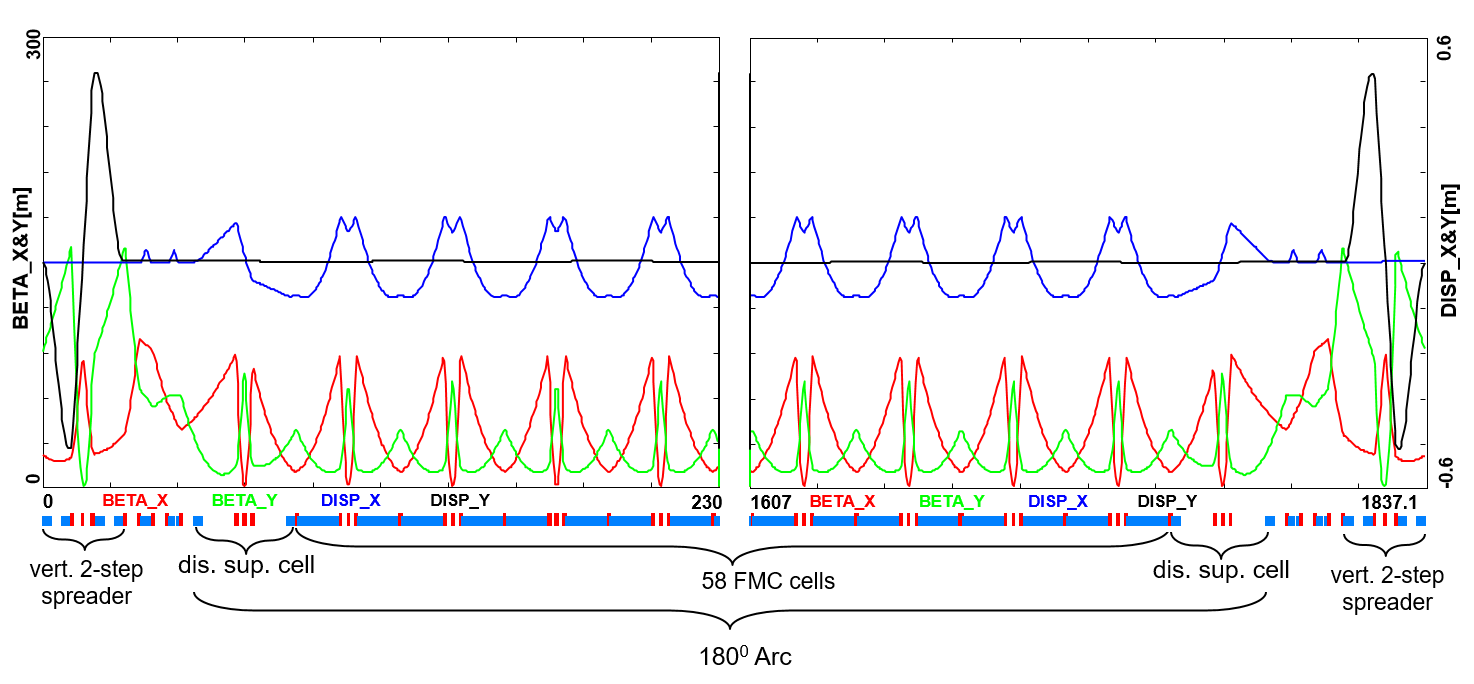}
  \caption{Complete optics for one of the LHeC ERL arcs (including switch yard); featuring: low emittance \ang{180} arc based on isochronous cells (30 cells flanked by dispersion-suppression cell with missing dipoles on each side), spreaders and recombiners with matching sections and doglegs symmetrically placed on each side of the arc proper.}
  \label{fig:key_challenges:art_of_arcs:Arc}
\end{figure}

% ---
The spreaders are placed directly after each linac to separate beams of different energies and to route them to the corresponding arcs. The recombiners facilitate just the opposite: merging the beams of different energies into the same trajectory before entering the next linac.
Each spreader starts with a vertical bending magnet, common for all three beams, that initiates the separation. The highest energy, at the bottom, is brought back to the horizontal plane with a chicane.
The lower energies are captured with a two-step vertical bending adapted from the CEBAF design.
Functional modularity of the lattice requires spreaders and recombiners to be achromats (both in the horizontal and vertical plane). To facilitate that, the vertical dispersion is suppressed by a pair of quadrupoles located between vertical steps; they naturally introduce strong vertical focusing, which needs to be compensated by the middle horizontally focusing quad. Following the spreader, there are four matching quads to \emph{bridge} the Twiss function between the spreader and the following \ang{180} arc (two betas and two alphas). 
Combined spreader-arc-recombiner optics feature a high degree of modular functionality to facilitate momentum compaction management, as well as orthogonal tunability for both the beta functions and dispersion, as illustrated in Fig.~\ref{fig:key_challenges:art_of_arcs:Arc}. 

%\subsection{Multi-turn Operational Experience}
%\subsubsection{JLab FEL Two-Turn}
%\begin{figure}[tbh]
% \centering
%  \includegraphics[width=0.7\textwidth]{figures/JLab3pass.PNG}
%    \caption{Three beams in the JLab IR-DEMO JLAB-TN-01-043.}
%     \label{fig:JLab3pass}
%
%\end{figure}
%
%\subsubsection{S-DALINAC}
%\subsubsection{CBETA}
%\section{Multi-turn Operation and the Art of Arcs}
% CEBAF, cBETA
%Alex Bogacz, Peter Williams

%
\section{ERL Operation Challenges}\label{sec:key_challenges:operation_challenges}
%brightness, halo losses, micro bunch instability, bbu, 
%
%Chris Tennant

%\subsection{Introduction}
In instances where high beam power is required, the concept of energy recovery presents an attractive solution.
Energy-recovery linacs (ERLs) are a class of novel accelerators which are uniquely qualified to meet the demands for a wide variety of applications by borrowing features from traditional architectures to generate linac-quality beams with near storage-ring efficiency~\cite{tennant_erl_chapter}.
Historically, nearly all ERLs built and operated were used to drive a free-electron laser (FEL).
The requirement for high peak current bunches necessitated bunch compression and handling the attendant beam-dynamical challenges.
In recent years, ERLs have turned from being drivers of light sources toward applications for nuclear physics experiments, Compton backscattering sources, and strong electron cooling.
Unlike an FEL, these latter use cases require long, high-charge bunches with small energy spread.
Where once a short bunch length was the key performance metric, now there is a premium on maintaining a small correlated energy spread (with a commensurately long bunch).

\subsection{Challenges}
Energy-recovery linacs are not without their own set of challenges.
In the following sections, a brief survey of some of the most relevant ones is given.
These include collective effects such as space charge, the multipass beam breakup (BBU) instability, coherent synchrotron radiation (CSR), and the microbunching instability (MBI); beam-dynamic issues such as halo and the interaction of the beam with the RF system and other environmental impedances; as well as issues related to common transport lines. 

\subsection{Space Charge}

The role of space-charge forces (both transverse and longitudinal) often dictates many operational aspects of the machine.
Maintaining beam brightness during the low-energy injection stage is vitally important.
In addition to the low energy, ERL injectors must also preserve beam quality through the merger system that directs the beam to the linac axis.
Once injected into the linac, the beam energy at the front end is often still low enough that space-charge forces cannot be neglected.
Just as important is the longitudinal space-charge (LSC) force, which manifests itself by an energy spread asymmetry about the linac on-crest phase~\cite{tennant_pac09}.
The LSC wake acts to accelerate the head of the bunch while decelerating the tail.
Operating on the rising part of the waveform leads to a decrease in the correlated energy spread, while accelerating on the falling side leads to an increase.
These observations inform where acceleration, and how the longitudinal match, is performed.

\subsection{Beam Breakup Instability}

The beam breakup instability is initiated when a beam bunch passes through an RF cavity off-axis, thereby exciting dipole higher-order modes (HOMs).
The magnetic field of an excited mode deflects following bunches traveling through the cavity.
Depending on the details of the machine optics, the deflection produced by the mode can translate into a transverse displacement at the cavity after recirculation.
The recirculated beam induces, in turn, an HOM voltage which depends on the magnitude and direction of the beam displacement.
Thus, the recirculated beam completes a feedback loop which can become unstable if the average beam current exceeds the threshold for stability~\cite{PhysRevSTAB.7.054401}.
Beam breakup is of particular concern in the design of high-average-current ERLs utilizing superconducting RF (SRF) technology.
If not sufficiently damped by the HOM couplers, dipole modes with quality factors several orders of magnitude higher than in normal-conducting cavities can exist, providing a threat for BBU to develop.
For single-pass ERLs, beam-optical suppression techniques---namely, interchanging the horizontal and vertical phase spaces to break the feedback loop between the beam and the offending HOM---are effective at mitigating BBU~\cite{PhysRevSTAB.9.064403}.\par
%% Below inserted by Peter Williams on 24th June 2021
Recently it has been realized that for a multi-pass ERL a judicious choice of filling pattern, the order in which recirculating bunches are arranged with respect to each other, can positively impact the BBU threshold. In \cite{PhysRevAccelBeams.24.061003} an example is given where a factor of 5 increase can be achieved by this method. The next generation of ERL facilities should include such studies in their experimental programs due to the potentially large benefit.

\subsection{Coherent Synchrotron Radiation}

Coherent synchrotron radiation poses a significant challenge for accelerators utilizing high-brightness beams.
When a bunch travels along a curved orbit, fields radiated from the tail of the bunch can overtake and interact with the head.
Rather than the more conventional class of head-tail instabilities where the tail is affected by the actions of the head, CSR is a tail-head instability.
The net result is that the tail loses energy while the head gains energy, leading to an undesirable redistribution of particles in the bunch.
Because the interaction takes place in a region of dispersion, the energy redistribution is correlated with the transverse positions in the bend plane and can lead to projected emittance growth.
While there has been much progress in recent years to undo the effects of CSR in the bend plane with an appropriate choice of beam optics~\cite{PhysRevLett.110.014801}, it is more difficult to undo the gross longitudinal distortion caused by the CSR wake.
This is particularly true in applications where the intrinsic energy spread is small and/or where the effect can accumulate over multiple recirculations.
One possible mitigation is shielding the CSR wake using an appropriately sized beam pipe~\cite{yakimenko_napac11}. 

\subsection{Microbunching Instability}

Microbunching develops when an initial density modulation, either from shot noise or from the drive laser, is converted to energy modulations through short-range wakefields such as space charge and CSR.
The energy modulations are then transformed back to density modulations through the momentum compaction of the lattice.
Danger arises when positive feedback is formed and the initial modulations are enhanced.
This phenomenon has been studied extensively, both theoretically and experimentally, in bunch compressor chicanes~\cite{PhysRevSTAB.5.064401, PhysRevSTAB.5.074401}.
Only recently has there been a concerted effort to study the microbunching instability in recirculating arcs~\cite{Di_Mitri_2015, PhysRevAccelBeams.19.114401, PhysRevAccelBeams.20.024401}.
Because the beam is subject to space charge and/or CSR throughout an ERL, density modulations can be converted to energy modulations.
And because of the native momentum compaction of the lattice (in arcs, spreaders/recombiners, chicanes, etc.) those energy modulations may be converted back to density modulations.
Therefore, ERLs offer potentially favorable conditions for seeding the microbunching instability, which requires careful attention in the early design stages.

\subsection{Halo}

Halo is defined as the relatively diffuse and potentially irregularly distributed components of beam phase space that can reach large amplitudes.
Numerous sources contribute to the halo.
Operational experience at various laboratories suggest that the biggest culprits are: stray light striking the photocathode, photocathode emission effects, field emission/dark current from the gun, beam dynamics during beam formation and evolution, and field emission/dark current in SRF cavities. It is of concern because ERL beams are manifestly non-Gaussian and can have beam components of significant intensity beyond the beam core~\cite{douglas_tn_2012}.
As a consequence, even a straightforward measurement of the beam size requires high-dynamic-range imaging techniques to see the core as well as the diffuse, large amplitude components of the distribution. 

Though sampling large amplitudes, halo responds to the external focusing of the accelerator transport system in a predictable manner.
It is therefore not always at large spatial amplitude, but it will at some locations instead be small in size yet strongly divergent.
Halo can therefore present itself as \enquote{hot spots} in a beam distribution and thus may be thought of as a lower-intensity, co-propagating beam that is mismatched to the core beam focusing, timing, and energy.
Beam loss due to halo scraping is perhaps the major operational challenge for higher-power ERLs.
Megawatt-class systems must control losses at unshielded locations to better than 100 parts per million to stay within facility radiation envelopes.
Scaling to \SI{100}{\mega\watt} suggests that control must be at the part-per-million level.
This has been demonstrated---but only at specific locations within an ERL~\cite{PhysRevLett.111.164801}.

\subsection{RF Transients}

Dynamic loading due to incomplete energy recovery is an issue for all ERLs~\cite{powers_erl2007}.
In some machines it is due to unintentional errors imposed on the energy-recovered beam; for instance, path-length errors in large-scale systems.
In other machines, such as high-power ERL-based FEL drivers, it is done intentionally.
In cases where there is the potential for rapid changes in the relative phase of the energy-recovered beam, dynamic loading would be difficult to completely control using fast tuners.
In such cases adequate headroom in the RF power will have to be designed into the system.
These transient beam-loading phenomena are widely unrecognized and/or neglected.
RF drive requirements for an ERL are often viewed as \enquote{minimal} because in steady-state operation the recovered beam notionally provides RF power for acceleration.
It has however been operationally established that RF drive requirements for ERLs are defined not by the steady-state but rather by beam transients, including the filling pattern~\cite{PhysRevAccelBeams.23.072002}, and environmental/design factors such as microphonics~\cite{Powers:SRF2017-FRXBA04}.
As a result, the RF power required for stable ERL operation can differ dramatically from naïve expectations.

\subsection{Wakefields and Interaction of Beam with Environment}

As with other system architectures intended to handle high-brightness beams, ERLs can be performance-limited by wakefield effects.
Not only can beam quality be compromised by interaction of the beam with environmental impedances, there is also significant potential for localized power deposition in beamline components.
Resistive-wall and RF heating have proven problematic during ERL operation in the past~\cite{4440128}.
Extrapolation of this experience to higher bunch charges and beam powers leads to serious concern regarding heating effects.
Careful analysis and management of system component impedances is required. 

\subsection{Magnet Field Quality}

Inasmuch as they rely on the generation of specific phase-energy correlations in order to bunch and/or energy-recover the beam, ERL transport systems are essentially time-of-flight spectrometers.
As a consequence, they generally require magnets with spectrometer-grade field quality to avoid performance limitations during energy recovery.
An often overlooked aspect of ERL design---and one with significant implications for system performance---is magnetic field quality. The necessary transverse-longitudinal coupling required for energy compression in high-power FEL drivers also creates the means by which magnetic field errors can generate energy errors.
Poor field quality leads to transverse steering errors, which, due to the non-zero $M_{52}$ of the recirculator, leads to path length errors (or equivalently, phase shifts).
Such phase shifts, in turn, increase the energy spread of the bunch and can lead to an unmanageably large energy spread at the dump~\cite{douglas_biw2010}.

\subsection{Multi-turn, Common Transport}

Future systems must evolve to utilize multiple turns; it is a natural cost optimization method~\cite{powers_srf2013}, and multi-turn systems can in principle provide performance equal to that of 1-pass up/down ERLs at significantly lower cost.
In addition to the use of multiple turns, cost control motivates the use of extended lengths of common transport, in which both accelerated and recovered passes are handled simultaneously using the same beam lines.
This presents unique challenges for high-energy ERLs, like LHeC in particular, where energy loss due to synchrotron radiation cannot be ignored and causes an energy mismatch for common transport lines.
A lower-energy fixed-field alternating-gradient (FFAG) optics solution for 4-pass up/down operation was recently demonstrated at the Cornell-BNL ERL Test Accelerator (CBETA)~\cite{PhysRevLett.125.044803}, see Section~\ref{sec:current_facilities:ongoing:cbeta}.
Continuing to address these challenges will open up exciting new opportunities for ERLs.
In addition to CBETA, LHeC, and PERLE, a multi-turn ERL design from Daresbury illustrates the manner in which the cost/complexity optimum lies toward shorter linacs, more turns, and multiple beams in fewer beam lines~\cite{Williams:FLS2018-THA1WA04}.
This also drives the use of multiple turns in stacking rings for hadron cooling~\cite{Benson:IPAC2018-MOPMK015}. 

Recently, a comprehensive study of the possible longitudinal matches for both collider and FEL applications utilising both common and separate transport has been published~\cite{PhysRevAccelBeams.25.021003}, highlighting the additional restrictions imposed to achieve self-consistency, particularly where SR losses are significant.

\subsection{A summary of critical open issues}
%%
%% text of Pavel from rodamap
%%
%%\subsubsection*{Critical open issues}
Operational experience of the ERLs which ran with the highest average beam power so far (1.2\,MW at IR-Upgrade, at Jefferson Lab), showed that beam halo, which is a fraction of the phase-space distribution with large amplitude and small intensity, was one of the critical operational issues.
The dynamic range of transverse beam profile measurements that are standard now and were used at IR-Upgrade is about \num{e3}.
Such measurements are made in tune-up beam mode with very low average beam current, causing some small-intensity, large-amplitude parts of the beam to be essentially invisible for the measurements, whereas in high-current mode, they would contribute to the limits of the facility in terms of practically possible average current.
If this remains unchanged, the average beam current and power at the next generation of ERLs, which are envisioned to operate with $\sim\SI{10}{\mega\watt}$ of beam power, might become limited to well below the design value.
Moreover, the transverse beam optics setup at PERLE with six arcs, five of which will need to operate with two beams simultaneously, and with 3-pass acceleration and 3-pass deceleration could prove to be far more complex than that of the IR Upgrade. Thus, transverse beam profile measurements with a dynamic range far beyond \num{e3} appear to be mandatory for the next generation of ERLs.

Having two beams in one beam line not only complicates the beam optics setup but also presents the beam diagnostics with an additional problem: When standard, invasive (intercepting) beam profile measurements are used, the beam is intercepted on the first pass. A beam viewer as thin as practically possible does not stop the beam but does cause enough multiple Coulomb scattering, such that when the beam arrives at the same location on the second, decelerating pass, its phase space and, correspondingly, the transverse profile are enlarged far beyond their original sizes. Thus, although beam viewers offer convenient 2D beam profile measurements, they present a very serious problem for the second-pass beam profile measurements, so either completely non-invasive or much less invasive methods must be considered.

To preserve the small transverse emittance, precise measurements and setup of the lattice functions will be needed. The well-established approach to such measurements is the differential orbit measurement made with the help of a beam position monitor (BPM) system. Here, having multiple beams in one beam line presents another challenge: implementing a BPM system which can measure the positions of multiple beams independently. Moreover, the BPM system will need to be used both in tune-up mode and in high-current mode with CW beam so that the stability of the beam orbit and the lattice functions can be ensured.
The following strategy for beam diagnostics at ERLs has proven to be necessary and very productive. First, there needs to be a set of beam diagnostics which allows detailed measurements of all relevant beam parameters when operating in tune-up mode. Then, there also must be a set of non-invasive measurement systems to monitor for changes of beam parameters beyond the tune-up mode, i.e., during the current ramp-up and high-power operation. Unlike the first set of diagnostics used in tune-up mode, this second set of monitors is not required to provide absolute measurements but is required to work regardless of beam mode, thus making a bridge between the tune-up and high-current modes. From this strategic perspective, some of the necessary transverse beam profile monitors are missing. These are relevant to the transverse match and emittance preservation. Practically, only synchrotron radiation (SR) monitors are available for transverse beam size monitoring with high beam currents. Monitoring the beam size in the ERL injector, where the initial beam quality (emittance) is defined, is not possible.

For efficient energy recovery and stable maintenance of beam energy and energy spread, the arrival of the beam at the linac cavities must be set up and monitored precisely. At PERLE, with six beams simultaneously in each linac at a very high repetition rate, this presents another challenge. The beam arrival measurement system would need to work in all beam modes and have picosecond or better resolution. Beam arrival measurements with such characteristics have not been demonstrated so far and need to be developed and implemented.
%\section{ERL Operation Challenges}
%brightness, halo losses, micro bunch instability, bbu, 
%Chris Tennant
%
\section{Interaction Region}\label{sec:key_challenges:interaction_region}
%Kurt Aulenbacher, Steve Benson

% MB 11/30/21: Rough edit wrt language & typesetting up to line 43.
% MB 03/06/22: Removed section "Colliders" per Andrew's email:
% "There has never been an energy recovered collider, so there is nothing to say other than it will be difficult."

%In an ideal case, the luminosity in an experiment is limited by the properties of the detection system and not by  properties  of the accelerator. In this section the situation is analyzed for ERL-based experiments. 

The interaction of an ERL beam in the region of the experiment can be categorized in three groups, those with an electromagnetic field, fixed targets and collider mode. Each of them has its specific aspects which depend on the design of the individual experiment. 
For the first two, a dilution of the beam emittance  may lead to partial  beam loss, which automatically creates radiation protection issues.
We discuss an example below. 

For colliders,  beam-beam interaction can trigger collective  instabilities in addition, which set a hard limit for the luminosity if they occur.
These implications have to be analyzed for each specific case.

\subsection{Interaction with electromagnetic fields} 
This case is relevant for applied research facilities such as FELs or undulator-based synchrotron radiation sources, but it also has to be taken into account for laser/ERL interaction in Compton-based gamma sources with high flux.
In such cases,  considerable energy spread of the electron beam may occur.
It must either be absorbed with suitable collimators, or the  recovery line needs  sufficient  acceptance as demonstrated at the JLab FEL \cite{PhysRevLett.84.662}. 

\subsection{Fixed targets}
In fixed-target experiments, Coulomb scattering plays an important role by creating an unavoidable source of beam loss.
A small beta function at the target position is advantageous, but this can conflict with the requirements of the experiment. 

Particle loss creates   beam loading. However, it turns out that radiation protection rather than installed RF power is the crucial aspect here. 

To give a concrete example, we refer to investigations done for the MAGIX experiment at MESA  \cite{Ledroit2021}. MAGIX uses a windowless gas jet target  \cite{Schlimme2021}.
GEANT-4 simulations were performed to calculate the outgoing particle distribution from different targets.
This input was used to calculate the power losses in the decelerating beamline using the tracking software BDSIM \cite{Nevay2020}.
In order to minimize  the losses in the decelerating beamline and, in particular, the cryomodules,  a collimator system between target and decelerator is helpful.
The losses were calculated for  the geometry of the  MAGIX spectrometer setup and for a   hydrogen target areal density of \SI{1 E 19}{\per\square\centi\meter} at  a beam energy of $E_0=\SI{105}{\mega\electronvolt}$.
The losses predicted at the collimator and along the beamline are \num{e-4} and \num{2e-7}, respectively.
For the given target density and nuclear charge, multiple scattering is negligible. 
FLUKA simulations of the persistent radioactivity around the MAGIX experiment indicate that a total power loss of \SI{100}{\watt} is acceptable since it happens almost completely  in the collimator, which can be locally shielded and does not need frequent interventions.
From this assumptions, we can infer a  practical luminosity limit for hydrogen of  $L_0 \approx \SI{7 E 35 }{\per\square\centi\meter\per\second}$.

For fixed  parameters,  the  beam power lost  will be proportional to the product of cross section and incoming beam power.
For Coulomb scattering, the cross section is roughly $\propto Z^2\cdot E^{-2}$ with $Z$ being the nuclear charge of the target and $E$ the beam energy, and the beam power at a given beam current  is  proportional to $E$.
For a fixed loss budget, this will lead to a luminosity limit that is  $L_\text{max}=L_0 Z^{-2} E/E_0$. 

However, in practice, one will have to take into account additional processes such as neutron production.
For beam energies in the GeV region, high-energy neutron production from $\Delta$-resonance excitation---which also takes place in  a pure hydrogen target---becomes  the most penetrating radiation component  and therefore frequently determines size and cost of shielding against direct radiation \cite{Fasso1991}.
Therefore, the $\propto E$ scaling has to be taken with care.

\subsection{Colliders}
% By Andrew 3/7/22

There have not been any ERL colliders built to date, but there have been several design exercises carried out.
Regrettably, the details of the interaction region have not been addressed in detail in any of these studies.
The interaction region of any collider is the complex result of many compromises, both theoretical and practical.
The starting point of an ERL collider interaction region would be the interaction region of a circular \positron\electron{} collider, whose designs ensure that the leptons can be accepted back into the collider ring after collision; in other words, they are minimally disrupted.
This is in contrast to a linear collider where the disruption coefficient is high, providing the maximum luminosity at the expense of a massively disturbed bunch, which must  then be immediately dumped.

In a circular collider, the bunches are brought to a tight focus at the interaction point by low-beta quadrupoles.
The chromatic effects of this strong focusing requires careful correction in dedicated regions near the interaction region and often distributed though the arcs as well.
In this respect, an ERL collider will be similar. The transverse emittances of the bunches in an ERL will be considerably smaller than in a collider ring, which enhances the beam-beam effect while producing more luminosity.
For a circular collider, the vertical emittance is due to coupling and vertical dispersion and is minimized by careful tuning; so, the beams are always extremely flat, with a horizontal-to-vertical emittance ratio of up to 1000.

In an ERL collider, the emittance ratio is either defined by the gun (nominally, equal transverse emittances) or by a damping ring, for which the emittance ratio can be tuned between 1 and $\sim 1000$.
The focusing of the low-beta quadrupoles favors unequal transverse beta values at the collision point.
Given these constraints, it is unclear whether equal or unequal emittances provide the highest luminosity for the least disruption of the beams.
This will require simulation of different options to determine the best.

The longitudinal bunch length in an ERL collider will be shorter than in a circular collider.
This may well obviate crab crossing or any of the other techniques required in a circular collider when the bunches do not collide head-on.
Again, detailed simulation will be required to confirm this hypothesis. 

Clearly, the detailed layout of a prototypical ERL interaction region is a high priority in the next few years.
This will require a small group of accelerator and detector experts working in close coordination to demonstrate what performance can be expected.
%\section{Interaction Region}
%Kurt Aulenbacher, Steve Benson
%
% Power to ERLs removed, covered by Sustainability (per Andrew)
%\input{s47.tex}
%\section{Power to ERLs}
% description of available systems and forthcoming innovation (FRT eg) 
%Erk Jensen, Nick Shipman

% Cryogenics removed, covered by Sustainability (per Andrew)
%\input{s48.tex}
%\section{Cryogenics}
%Patxi Duthil, 

%%
\chapter{Energy and Intensity Frontier Physics}\label{sec:frontier}

\section{High-Energy Colliders}\label{sec:frontier:high_energy_colliders}
%
% "Cobbled together" by Andrew from the Short Panel Report; email 3/2/2022.
%
ERLs are an extremely efficient technique for accelerating high-average-current electron beams.
In high-energy physics, the interest is in an intense, low-emittance \electron{} beam for colliding against hadrons (eh), positrons (\positron\electron) or photons ($\mathrm{e}\photon$).
Experiments rely on the provision of high electron currents (of up to \SI{100}{\milli\ampere}).
It is remarkable that following the LHeC design from 2012~\cite{AbelleiraFernandez:2012cc} (updated in 2020~\cite{LHeC:2020jxs}),
all the possibilities have been pursued: for \photon\photon{} collisions~\cite{bogacz2012sapphire} using the LHeC racetrack,
further for eh with the FCC-eh in 2018~\cite{FCC:2018bfs},
for \positron\electron{} in 2019 with an ERL concept for FCC-ee, termed CERC~\cite{Litvinenko:2019txu},
in 2021 with an ERL version of the ILC, termed ERLC~\cite{telnov2021highluminosity_v3}, and---very recently---also with a concept for the generation of picometer-emittance muon pairs through high-energy, high-current $\mathrm{e}\photon$ collisions~\cite{EXMParxiv}.

A common task for these colliders is precision SM Higgs boson measurements dealing with a small cross section (of \SI{0.2}{\pico\barn} / \SI{1}{\pico\barn} in charged-current ep interactions at LHeC/FCC-eh and similarly
of \SI{0.3}{\pico\barn} in Z-Higgsstrahlung at \positron\electron{}).
This makes maximising the luminosity a necessity to profit
from the clean experimental conditions and to access rare decay channels while limiting power.
High luminosity and energy are expected to lead beyond the SM and are essential for precision measurements
at the corners of phase space.

\subsection{LHeC and FCC-eh}
%Max Klein, Alex Bogacz, Bernhard Holzer Yannis Papaphilippou

The Large Hadron electron Collider (LHeC)
concept had been reviewed in considerable detail prior to the publication of the first design report in 2012~\cite{AbelleiraFernandez:2012cc}.
The physics results of almost a decade of LHC operation, technology developments and a more ambitious luminosity goal (related to the Higgs discovery, the brilliant LHC performance, and the capability of ERLs) have led to the recent publication of another very detailed report, written again by representatives of about $150$ institutions~\cite{agostini2020lhec}. It thus is extremely well documented. The combination of a high-energy hadron beam with a largely independent ERL configuration has also been applied to the FCC-eh. This concept has been documented in the recently published FCC design reports, Vol 1~\cite{FCC:2018xdg}
and 3~\cite{FCC:2018bfs}, and covered to considerable extent also in~\cite{agostini2020lhec}. Due consideration, both for 
LHeC and FCC-eh, was given to the electron-ion physics and
machine aspects. Since HERA missed an eA phase (collisions of electrons and hadrons other than protons), the
extension in range is phenomenal, and a complete change of
our understanding of nuclear structure and parton dynamics in nuclei is in reach with these energy-frontier EICs.

Here, a brief introduction to these future energy-frontier ep and eA colliders is given with emphasis on the choice of parameters, especially the electron beam energy and luminosity, from which basic demands on the ERL development arise.
 Parameters and components are summarised, and updated considerations are
sketched on the interaction region and on a synergetic use of the LHeC racetrack as an injector to the FCC-ee. This part concludes with the recommendations given by an 
International Advisory Committee as to the focus of work for the coming years. It is no surprise that ERL developments are of key importance.

\subsubsection{Deep Inelastic Scattering}
Deep Inelastic Scattering (DIS) was established in 1968 with the discovery of a partonic substructure of the proton in the famous SLAC-MIT experiment at the 2-mile linac at Stanford. An electron-proton scattering experiment was at the foundation of the Standard Model. DIS is 
a particularly clean process, theoretically through the operator expansion of structure functions, asymptotic freedom, and the 
evolution of parton density functions (PDFs) with the resolving
strength determined by the negative square four-momentum  transfer, $Q^2$, between the incoming lepton and the interacting parton.
Experimentally, DIS is clean because of the combination of an electromagnetic and weak probe with a strongly interacting target
such that the final state is clearly defined, free of colour reconnections, the neutral or charged
current type of interaction prescribed by the leptonic vertex, the
kinematics determined redundantly from the hadronic and the 
leptonic final state and pile-up, even at the FCC-eh, non-existent.

With high luminosity and energy, the next DIS collider has a massive
physics programme, to resolve parton interaction dynamics, to
develop Quantum Chromodynamics, for precision measurements of the Higgs boson characteristics, for finding new physics such as in the massive neutrino sector, for finding new dynamics at small Bjorken $x$, for the understanding of the Quark-Gluon Plasma phenomenon 
in heavy and possibly light-ion interactions, top and electroweak physics, all or some of which most likely leading to surprise observations without which particle physics will hardly proceed beyond where it now stands. High-energy DIS colliders are the cleanest microscopes for
resolving the substructure of matter, and it is inconceivable 
that particle physics advances without realising the next lepton-proton collider, following HERA, for which the LHeC is the singular opportunity given the success and long lifetime of the LHC. 

\subsubsection{High Energy}
The energy frontier in DIS has been held by the first ep collider 
HERA~\cite{Wiik:1985sb}, which collided protons of $E_\text{p}=\SI{920}{\giga\electronvolt}$ energy
and electrons of $E_\text{e}=\SI{27.6}{\giga\electronvolt}$, corresponding to a
centre-of-mass energy of $\sqrt{s}=2 \sqrt{E_\text{e} E_\text{p}} \simeq \SI{0.3}{\tera\electronvolt}$.
These values could be exceeded with a new LHC-based collider coupled with
an ERL of about twice the HERA electron beam energy.  
There are several fundamental reasons to illustrate why one would want to exceed the HERA parameters:
\begin{itemize}
\item HERA established the rise of the  quark and gluon densities towards
small Bjorken $x= Q^2/sy$, where $y \leq 1$ is the inelasticity related to the 
energy transfer in DIS. There are expectations, back to the seminal work
of Lev Lipatov and colleagues, that at high densities there occur non-linear 
parton-parton interactions which damp this rise and replace the known
linear DGLAP $Q^2$ evolution equations by another evolution law. Such
a discovery of new dynamics  at small $x$ would not only be a major breakthrough in QCD but
practically alter all predictions for pp collisions at future hadron colliders. 
HERA's kinematic reach was too small to resolve that 
question~\footnote{
Establishing new parton behaviour requires to measure in the genuinely deep inelastic region, $Q^2$ larger than the proton mass squared, and to cover
a minimum range to about \SI{10}{\square\giga\electronvolt}. Measuring in a region of large
$y \simeq 1 - E'/E_\text{e}$ is very difficult and requires $E_\text{e}$ to be large, with \SI{50}{\giga\electronvolt} being a suitable value, for the scattered electron energy $E'$ not to need to be too small, as is required to stay away from large hadronic backgrounds.
Thus, typically, the minimum $x$ value one can hope to cover in a precision
determination of especially the dominating gluon density is $x_\text{min} \simeq 10/{s 0.5}$ with $s$ in GeV$^2$. This value is equal to \num{1e-3} at the EIC, \num{2e-4} at HERA, \num{1.4e-5} at the LHeC and \num{1.6e-6} at the FCC-eh.
One indeed recognises in the modern PDF determinations
that the uncertainty for $xg$ substantially enlarges below $x\simeq 10^{-4}$.
With HERA not having observed any indication for a departure from DGLAP
evolution at $x \sim 10^{-3}$, its best region of coverage, it is obvious that
the EIC  energy, chosen for spin physics and adapted to the existence of RHIC, is far too low to establish a different evolution law in ep scattering.}.
\item The primary role of DIS is to determine the parton density structure of hadrons.
The ep inclusive cross section is sensitive to the sums of up and down quark and
anti-quark distributions, i.e., the four combinations $U$, $D$, $\bar{U}$ and $\bar{D}$.
These can only be disentangled with neutral (NC) and and charged (CC) current
cross section measurements.  For trigger and identification reasons, a CC measurement in ep, against the overwhelming low $Q^2$ backgrounds, is limited to the large-$Q^2$ region, 
$Q^{2}_\text{min} \simeq \SI{100}{\square\giga\electronvolt}$ at HERA. The $x$ range for CC measurement is
therefore limited to $x \geq Q^2_\text{min} / s$. It needs the LHeC to indeed utilise
the CC for the disentanglement of the quark sea. Furthermore, with
regard to predictions for the LHC or FCC, it is of importance to measure the PDFs in and near the kinematic range
where they are used in order not to depend on evolution over orders of magnitude in $Q^2$.
\item 
%New Particle Production (Top, Higgs and BSM)
%
For ep colliders to be of interest beyond QCD, the energy should be high since
cross sections of heavy particles---such as the top quark, Higgs boson, and BSM particles---become 
sizeable only when the energy is large. This is illustrated for heavy quarks 
in~\cite{AbelleiraFernandez:2012cc}. 
For the Higgs boson, this is illustrated in Fig.\,\ref{fig:Hsec}.
At the LHeC, an ab$^{-1}$ of integrated luminosity provides a sample of $\mathcal{O}(10^5)$
Higgs bosons, a base for precision Higgs physics comparable to that of the 
ILC, where the cross section of Z Higgs-Strahlung and the luminosity expectations are
similar.  Combining ep with pp, all SM Higgs decay channels can be reconstructed 
with a sum measured to within 1\,\% accuracy. The complementary measurements of
the Higgs boson characteristics in future pp, ep and \positron{}\electron{} scattering
have the potential to verify the SM Higgs mechanism to the necessary extent and
to explore whether it indeed is a window to BSM physics.  
\begin{figure}[th]
%\centerline{\includegraphics[width=100mm]{figlhec}}
\centering
\includegraphics[width=0.99\textwidth]{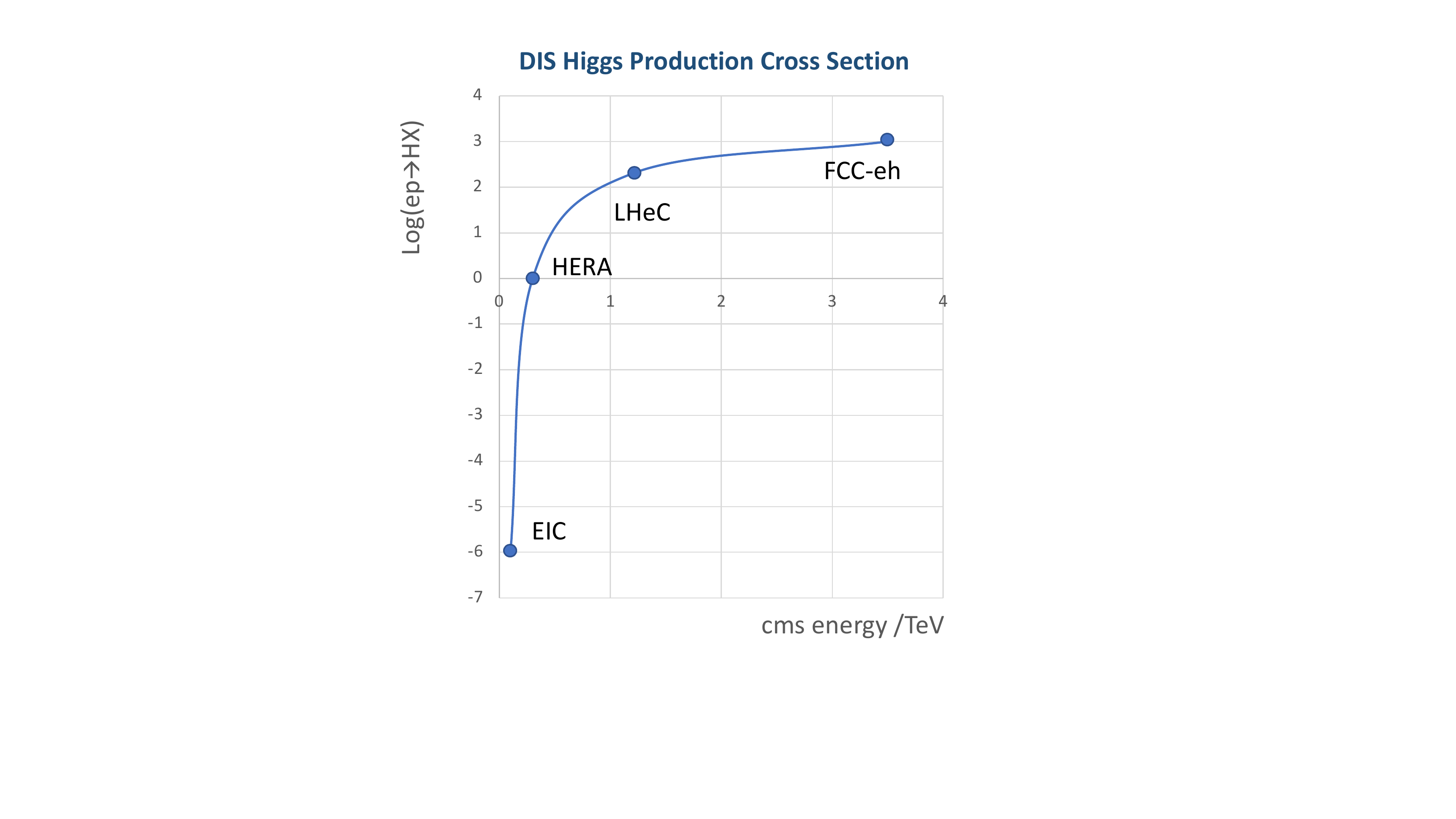}
\vspace{-2cm}
\caption{Calculation of the inclusive Higgs cross section $\sigma$, plotted
as $\log{(\sigma /\si{\femto\barn})}$, as a function of the cms $\text{e}\text{p}$ scattering
energy, $\sqrt{s} = 2 \sqrt{E_\text{e}E_\text{p}}$ in TeV. One sees that a TeV-energy DIS collider reaches the \SI{100}{\femto\barn} cross-section range, which is comparable to the $\text{e}^+\text{e}^- \to \text{H} \text{Z}$ cross section in 
electron-positron scattering.
With a thousand times its luminosity,
HERA would have had a chance to discover the Higgs boson, while
that is beyond the reach of lower-energy $\text{e}\text{p}$ colliders such as the EIC.   
}
\label{fig:Hsec}
\end{figure}
The LHeC and FCC-eh have a striking BSM discovery potential, such as right-handed
neutrinos, triplet fermions, lepto-quarks and other so far exotic particles as
has been discussed in~\cite{LHeC:2020oyt}. The discovery potential is determined
by the energy reach.
%\item BSM
%
\end{itemize}
For both the classic DIS programme, for competitive Higgs and top physics, and
for the discovery of new physics in QCD and the electroweak sector, a cms energy
beyond a TeV is crucial. With the LHC at $E_\text{p}=\SI{7}{\tera\electronvolt}$, one can reach $\sqrt{s} \geq \SI{1}{\tera\electronvolt}$ for $E_\text{e} \geq \SI{35}{\giga\electronvolt}$, while the FCC is leading much further.

\subsubsection{High Luminosity}
\begin{figure}[th]
%\centerline{\includegraphics[width=100mm]{figlhec}}
\centering
\includegraphics[width=0.7\textwidth]{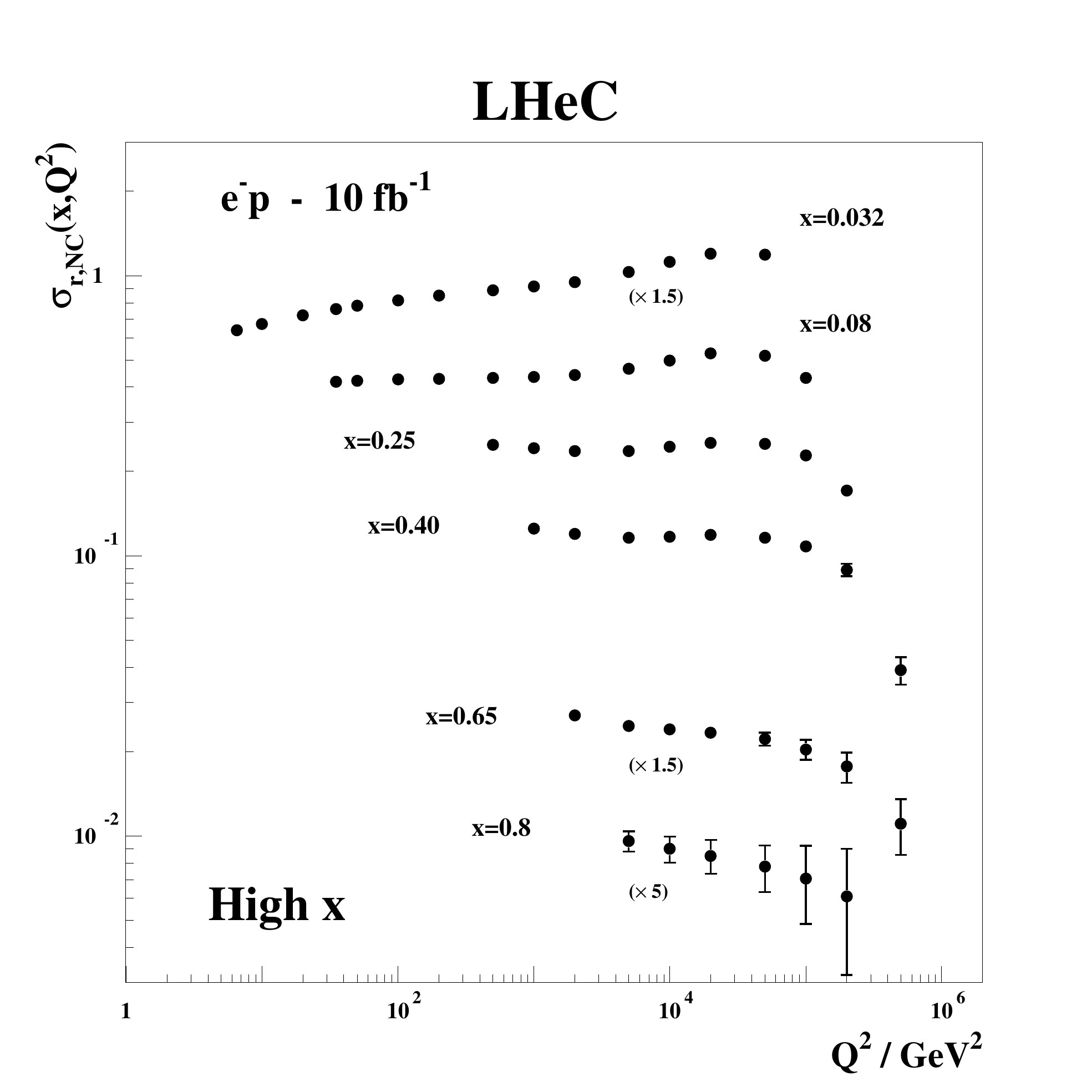}
\vspace{-0.3cm}
\caption{Simulated NC DIS cross section measurement at the LHeC for $\text{e}^-\text{p}$ scattering
as a function of $Q^2$ for different intervals of large $x$. The error bars include the 
systematic uncertainty but are dominated by statistics where visible.
Note the scaling of the cross section values at high $x$. The statistics assumed here
is ten times that of HERA  but falls visibly short of covering the `near 1 region' of Bjorken $x$
nor the 'near $s$ region' of $Q^2$.
}
\label{fig:shighx}
\end{figure}

Besides the energy, the second crucial parameter which determines the maximum
value of $Q^2 \leq s$ is naturally the luminosity. The DIS NC and CC cross sections
decrease $\propto (1-x)^3$ as $x$ approaches 1, i.e., the probability for one parton to
carry all of the available momentum is very small. The region $x$ near $1$, however, is of 
great importance for it determines the predictions for high-mass Drell-Yan scattering at the LHC.
So far, it has been explored unsatisfactorily for statistics,
 nuclear, and higher-twist-correction reasons, which will change with the simultaneous measurement of NC and CC cross sections  at high $x$ and $Q^2$ at the LHeC and FCC-eh. Much of new physics
 is expected to reside at very high $Q^2 = s x y$, which can only be accessed at high $x$ (and $y$).
This situation is illustrated in Fig.\,\ref{fig:shighx}, which shows the simulated NC DIS cross-section measurement, illustrating
the demand to have significantly smaller uncertainties. Inspection of 
top and BSM production, and especially the Higgs $\text{e}^-\text{p}$ CC production cross section
have set a goal of $\mathcal{O}(1)$\,ab$^{-1}$ of integrated luminosity, which---in reasonable operation 
times of about a decade---can only be achieved with a luminosity in the order of \SI{e34}{\per\square\centi\meter\per\second}.
This is a very high and demanding goal for $\text{e}\text{p}$ collisions, exceeding 
the HERA I value about 1000-fold.

After the Higgs discovery and having observed the LHC to operate better than 
had \enquote{ultimately} been expected, inspection of the main parameters showed
that a value near the desired goal was in reach\,\cite{Zimmermann:2013aga}.
The LHeC luminosity $\mathcal{L}$ is roughly determined as
\begin{equation}
\label{LLR}
%L = \frac{N_e N_p n_p f_{rev} \gamma_p}{4 \pi \epsilon_p \beta^*} \cdot \prod_{i=1}^3{H_i},
\mathcal{L} = \frac{N_\text{e} N_\text{p} n_\text{p} f_\text{rev} \gamma_\text{p}}{4 \pi \epsilon_\text{p} \beta^*},
\end{equation}
where $N_{\text{e}\;(\text{p})}$ is the number of electrons (protons) per bunch, $n_\text{p}$ the number of proton bunches in the LHC, 
$f_\text{rev}$ the revolution frequency in the LHC,
%[the bunch spacing in a batch 
%is given by $\Delta$, equal to $25$\,ns for protons in the LHC] a
and $\gamma_\text{p}$ the relativistic factor of the proton beam.
Further,  $\epsilon_\text{p}$ denotes the normalised proton transverse beam emittance 
and $\beta^*$  the proton beta function at the IP, assumed to be equal in 
$x$ and $y$. With the parameters listed below and the product
of hourglass, pinch, and filling factors about 1, one finds that 
a value close to \SI{e34}{\per\square\centi\meter\per\second} is a possible target. 
Since the electron current is determined as the product of $e N_\text{e} f$, $f$ being the repetition frequency of the electron linac, there
follows the LHeC design goal of a $20$\,mA current provided by
a $500$\,pC source with the $40$\,MHz frequency.
A 3-turn ERL configuration, as is the LHeC default, requires 
high-quality cavities ($Q_0 \geq \num{2e10}$)
to be able to deal with 120\,mA current altogether. These so-called default design parameters
for the  LHeC, and similarly for FCC-eh (possibly with a higher current at a later time),
have been adopted as characteristics for PERLE (see Section~\ref{sec:new_facilities:perle}),
the ERL development facility at IJCLab Orsay, for which a large International
Collaboration has recently been formed.

\subsubsection{The ERL Configuration of LHeC and FCC-eh}
The original design concept of the LHeC considered both a ring-ring and a linac-ring 
configuration~\cite{AbelleiraFernandez:2012cc} as an alternative.
The former had the problem of interfering at various crossing 
and the interaction points with existing LHC hardware installations. The latter, an ERL, has since
been adopted as the default. It consists of a racetrack configuration tangential to
the LHC proton ring with collisions foreseen at IP2 as the heavy-ion programme
had been officially declared nominally to end with LS4 while the HL-LHC is 
dedicated to maximise the luminosity for ATLAS, CMS and nowadays also LHCb.
Such an arrangement has the peculiarity of transforming the LHC to a 3-beam facility
with simultaneous pp and ep operation at different IPs, owing to the
feeble effect the electron beam has on the proton beams. This means that conceptually,
the electron-hadron operation at the LHeC and, similarly, FCC-eh is not costing
integrated luminosity to the main hh programme.

Following a careful cost evaluation as presented in~\cite{LHeC:2020oyt} it was decided to
reduce the electron beam energy from $60$ to $50$\,GeV. This economised 
almost 400 million CHF and brought the total LHeC cost, without detector, back to 
the target value of about 1 billion CHF.
It also reduced the effort as the time for building
the LHeC will become an important factor for its possible endorsement.
A possible transition from the $60$\,GeV to the $50$\,GeV configuration
 of the LHeC was already envisaged  in 2018, and presented in the  
paper submitted to the European Strategy~\cite{Bruning:2019scy}.

\begin{figure}[th]
%\centerline{\includegraphics[width=100mm]{figlhec}}
\centering
\includegraphics[width=0.7\textwidth]{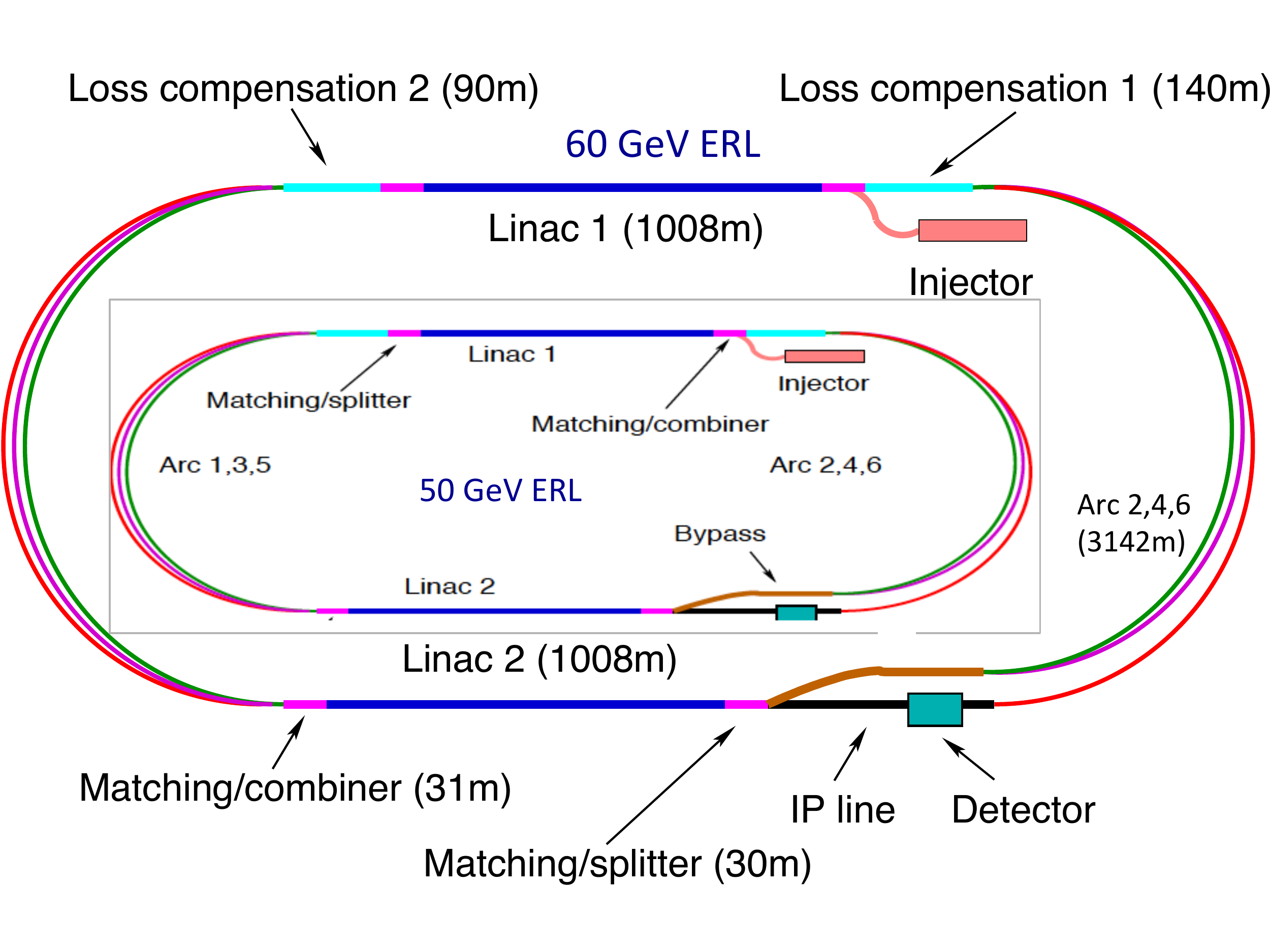}
\caption{Schematic view of the three-turn LHeC configuration with two oppositely positioned 
electron linacs and three arcs housed in the same tunnel. Two configurations are 
shown: Outer: $E_\text{e} = 60$\,GeV, as for FCC-eh, 
with linacs of about $1$\,km length and $1$\,km arc radius
leading to an ERL circumference of about $9$\,km, or $1/11$ of the FCC length.
Inner: $E_\text{e} = 50$\,GeV, as for LHeC, with linacs of about $0.8$\,km length and $0.55$\,km 
arc radius leading to an ERL circumference of $5.4$\,km, 
or $1/5$ of the LHC length, which is smaller than the size of the SPS.
The $1/5$ circumference configuration is flexible: it 
entails the possibility to stage the project as funds or physics dictate by using 
only partially equipped linacs, and it also permits upgrading to somewhat higher 
energies if one admits increased synchrotron power losses and operates at higher gradients.}
\label{fig:lhecconf}
\end{figure}

Fig.\,\ref{fig:lhecconf} illustrates the ERL configurations for the LHeC, inner
racetrack, and the FCC-eh, outer one, which has been kept at the nominal $60$\,GeV.
Obviously, such choices are for design and feasibility studies, while physics, technology,
funds, and time
will eventually lead to alterations from the parameters chosen so far, which are summarised
in Table\,\ref{tab:lumip} adapted from the recent LHeC CDR update\,\cite{LHeC:2020oyt}.
\begin{table}[ht]
  \centering
  \footnotesize
  \begin{tabular}{lcccccccc}
    %\hline
    %item & & LHeC & & FCC-eh & \\
%\hline
    \toprule
    Parameter & Unit & \multicolumn{4}{c}{LHeC} & & \multicolumn{2}{c}{FCC-eh}  \\
    \cmidrule{3-6} \cmidrule{8-9}
     & & CDR  & Run 5 & Run 6 & Dedic. & & $E_\text{p}=\SI{20}{\tera\electronvolt}$ & $E_\text{p}=\SI{50}{\tera\electronvolt}$ \\
    \midrule
%    $E_\text{p}$ & TeV & 7 & 7 & 7 & 7 & & 20 & 50 \\
    $E_\text{e}$ & GeV & 60 & 30  & 50  &  50 & & 60  &  60 \\
    $N_\text{p}$ & $10^{11}$ & 1.7 & 2.2  & 2.2 & 2.2  && 1 & 1 \\
    $\epsilon_\text{p}$ & \si{\micro\meter} & 3.7 & 2.5 & 2.5 & 2.5  && 2.2 & 2.2 \\
    $I_\text{e}$ & mA & 6.4 & 15  & 20 & 50  && 30 & 30 \\
    $N_\text{e}$ & $10^9$ & 1 & 2.3  & 3.1 & 7.8  && 3.1 & 3.1 \\
    $\beta^*$ & cm & 10 & 10 & 7 & 7  && 12 & 15 \\
%    \addlinespace
    $\mathcal{L}$ & \SI{e33}{\per\square\centi\meter\per\second} & 1 & 5 & 9 & 23& & 18 & 22 \\ 
    \bottomrule
  \end{tabular}
  \caption{Summary of luminosity parameter values for the LHeC and FCC-eh. Left: CDR 
    from 2012; Middle: LHeC in three stages, an initial run, possibly during Run\,5 of the LHC,
 the $50$\,GeV operation during Run\,6, both concurrently with the LHC, and a final,
 dedicated, stand-alone ep phase; Right: FCC-eh with a 20
 and a 50\,TeV proton beam, in synchronous operation.}
  \label{tab:lumip}
\end{table}

\subsubsection{Parameters and Components}
The main racetrack, the linac, arcs, spreaders and combiners as well as the
optics, synchrotron radiation, beam-beam interactions etc, simulations and hardware descriptions, are
all provided in~\cite{agostini2020lhec} and
partially go back to~\cite{AbelleiraFernandez:2012cc}.
For completeness, the main parameters of the LHeC ERL are listed in Table~\ref{tab:ERLparameters}.
\begin{table}[!ht]
  \centering
  \small
  \begin{tabular}{lcc} 
    \toprule
    Parameter & Unit & Value  \\
    \midrule
    Injector energy & \si{GeV} & 0.5 \\
    Total number of linacs & & 2 \\
    Number of acceleration passes & & 3 \\
    Maximum electron energy & \si{GeV} & 49.19 \\
    Bunch charge & \si{pC} & $499$ \\
    Bunch spacing & \si{ns}	& 24.95 \\
    Electron current & \si{mA}	& 20 \\
    Transverse normalized emittance & \si{\micro\meter} & 30 \\
    Total energy gain per linac & \si{GeV} & 8.114\\
    Frequency & \si{MHz} & 801.58  \\
    Acceleration gradient & \si{MV/m} & 19.73 \\
    Cavity iris diameter & \si{\milli\meter} & 130 \\
    Number of cells per cavity & & 5 \\
    Cavity length (active/real estate) & \si{\meter} & 0.918 / 1.5 \\
    Cavities per cryomodule &  & 4  \\
    Cryomodule length & \si{m} & 7 \\
    Length of 4-CM unit (group of 4 cryomodules) & \si{m} & 29.6 \\
    Acceleration per 4-CM unit & \si{MeV} & 289.8 \\
    Total number of 4-CM units per linac & & 28 \\
    Total linac length (with spr/rec matching) & \si{m} & 828.8 (980.8) \\
    Return arc radius (length) & \si{m} & 536.4 (1685.1) \\
    Total ERL length & \si{km} & 5.332 \\
    \bottomrule
  \end{tabular}
  \caption{Parameters of the LHeC Energy-Recovery Linac.}
  \label{tab:ERLparameters}
\end{table}
The default 50\,GeV LHeC main loop uses 
dipole magnets as listed in 
Table\,\ref{tab:DipolesComponents}. The field values are between $0.1$ and $0.5$\,T for
the arcs, while the spreader and combiner
need a somewhat larger field strength, up to $1.6$\,T.
The quadrupole and cavity characteristics are summarised
in Table\,\ref{tab:QuadRFComponents}, also taken from~\cite{agostini2020lhec}. The total number of cryomodules
per linac is 112 (grouped into units of 4, referred to as 4-CM units), which corresponds to 448 five-cell 802\,MHz cavities and a total of
896 for two linacs. These are the cost drivers of the LHeC as has been provisionally estimated in~\cite{agostini2020lhec}. Their number is an order of magnitude below that of the default ILC design. A total of 72 cavities at twice the base frequency, i.e., 1604\,MHz, is part of the configuration for compensation of synchrotron radiation losses.

\begin{table}[!ht]
  \centering
  \small
  \begin{tabular}{l@{\hspace{2em}}c@{\hspace{0.8em}}c@{\hspace{0.8em}}c@{\hspace{0.8em}}c@{\hspace{0.8em}}c@{\hspace{0.8em}}c@{\hspace{0.8em}}c@{\hspace{0.8em}}c@{\hspace{0.8em}}c@{\hspace{0.8em}}c@{\hspace{0.8em}}c@{\hspace{0.8em}}c@{\hspace{0.8em}}c@{\hspace{0.8em}}c} 
    \toprule
    & \multicolumn{4}{c}{Arc dipoles (horiz.)} & & \multicolumn{4}{c}{Spr/Rec dipoles (vert.)} & & \multicolumn{4}{c}{\emph{Dogleg} dipoles (horiz.)} \\
    \cmidrule{2-5}  \cmidrule{7-10}  \cmidrule{12-15} 
    Section & $N$ & $B$ & $g/2$ & $l$ &
    & $N$ & $B$ & $g/2$ & $l$ &
    & $N$ & $B$ & $g/2$ & $l$ \\
    \midrule
    Arc 1 & 352 & 0.087 & 1.5 & 3 & & 8 & 0.678 & 2 & 3 & & 16 & 1  & 1.5 & 1\\
    Arc 2 & 352 & 0.174 & 1.5 & 3 & & 8 & 0.989 & 2 & 3 & & 16 & 1 & 1.5 & 1\\
    Arc 3 & 352 & 0.261 & 1.5 & 3 & & 6 & 1.222 & 2 & 3 & & 16 & 1  & 1.5 & 1\\
    Arc 4 & 352 & 0.348 & 1.5 & 3 & & 6 & 1.633 & 2 & 3 & & 16 & 1 & 1.5 & 1\\
    Arc 5 & 352 & 0.435 & 1.5 & 3 & & 4 & 1.022 & 2 & 3 & &  &  &  & \\
    Arc 6 & 352 & 0.522 & 1.5 & 3 & & 4 & 1.389 & 2 & 3 & &  &  &  & \\
    \midrule
    Total & 2112 &  &  &  &  & 36 &  &  &  &  & 64 &  &  &\\
    \bottomrule
  \end{tabular}
  \caption{\SI{50}{GeV} ERL -- Dipole magnet count along with basic magnet parameters: Magnetic field $(B)$ [T], Half-Gap $(g/2)$ [cm], and Magnetic length $(l)$ [m].}
  \label{tab:DipolesComponents}
\end{table}

\begin{table}[!ht]
\centering
  \small
  \begin{tabular}{l@{\hspace{2.5em}}c@{\hspace{0.8em}}c@{\hspace{0.8em}}c@{\hspace{0.8em}}c@{\hspace{0.8em}}c@{\hspace{0.8em}}c@{\hspace{0.8em}}c@{\hspace{0.8em}}c@{\hspace{0.8em}}c} 
    \toprule
    & \multicolumn{4}{c}{Quadrupoles} &  & \multicolumn{4}{c}{RF cavities} \\
    \cmidrule{2-5}  \cmidrule{7-10} 
    Section & $N$ & $G~[\text{T/m}]$ & $a~[\text{cm}]$ & $l~[\text{m}]$ & &  $N$ & $f~[\text{MHz}]$ & cell & $G_\text{RF}~[\text{T/m}]$ \\
    \midrule
    Linac 1 &  29 &  1.93 & 3   & 1 & & 448 & 802 & 5 & 20 \\
    Linac 2 &  29 &  1.93 & 3   & 1 & & 448 & 802 & 5 & 20 \\
    Arc 1   & 255 &  9.25 & 2.5 & 1 & &   &   &   &   \\
    Arc 2   & 255 & 17.67 & 2.5 & 1 & &  &  &  &  \\
    Arc 3   & 255 & 24.25 & 2.5 & 1 & & 6 & 1604 & 9 & 30 \\
    Arc 4   & 255 & 27.17 & 2.5 & 1 & & 12 & 1604 & 9 & 30 \\
    Arc 5   & 249 & 33.92 & 2.5 & 1 & & 18 & 1604 & 9 & 30 \\
    Arc 6   & 249 & 40.75 & 2.5 & 1 & & 36 & 1604 & 9 & 30 \\
        \midrule
    Total & 1576 &  &  &  & & 968 &  &  & \\
    \bottomrule
  \end{tabular}
  \caption{\SI{50}{GeV} ERL---Quadrupole magnet and RF cavities count along with basic magnet/RF parameters: Magnetic field gradient $G$, Aperture radius $a$, Magnetic length $l$, Frequency $f$, Number of cells in RF cavity (cell), and RF gradient $G_\text{RF}$.}
  \label{tab:QuadRFComponents}
\end{table}

\subsubsection{Interaction Region of LHeC and FCC-eh}
%Max Klein, Alex Bogacz, Bernhard Holzer

\begin{figure}[!htb]
   \centering
   \includegraphics[width=.9\columnwidth]{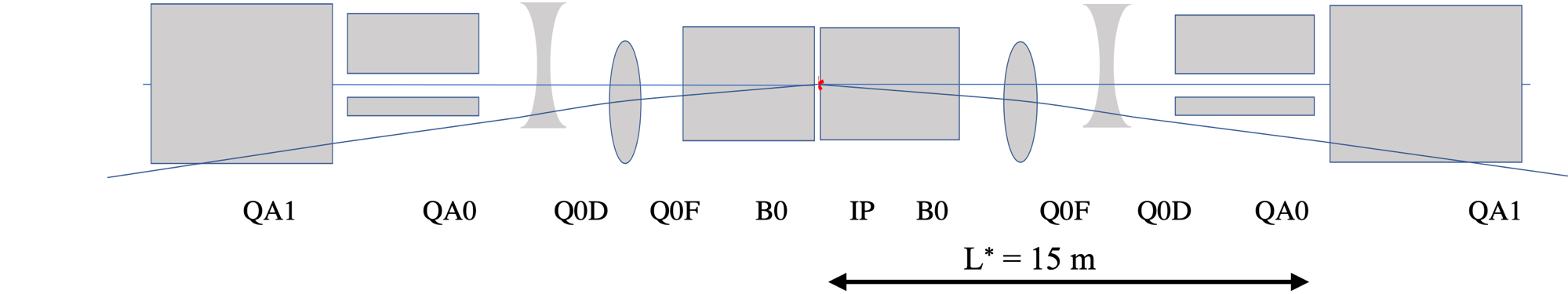}
   \caption{Schematic layout of the IR region with the electron mini-beta quadrupoles acting as combined-function magnets to separate the beams.}
   \label{fig:IR_schema}
\end{figure}

The interaction region between the electron beam of the ERL and the proton beam of the LHC or FCC is one of the most challenging parts of the design, as several aspects have to be considered at the same time. 
The required luminosity of the LHeC requests  beta functions in the order of \SI{10}{cm} at the Interaction point with matched beam sizes of electrons and protons at the IP in both planes: 
%\begin{equation}
$\sigma_{xe} =\sigma_{xp}$, $
\sigma_{ye} =\sigma_{yp}$.
%\end{equation}
Given the  considerable difference in beam energies, the electrons and protons have to be focused independently; therefore, after the IP, an efficient beam separation scheme has to separate the electron beam from the proton design orbit. Finally, the synchrotron radiation emitted during the beam separation in the vicinity of the particle detector has to be limited as much as possible.
Figure~\ref{fig:IR_schema} shows an optimized principle layout of the IR.
The requirements of small beam size, head-on collisions and efficient separation are fulfilled by combining a weak dipole
inserted in 
the  spectrometer dipole and electron mini-beta quadrupoles (off-centered with respect to the electron beam) to create a quasi-constant separation field from the IP up to the location of the first superconducting proton quadrupole QA1 at a position of $L^* = \SI{15}{m}$. At the same time, the electron quadrupoles provide an early focus to limit the electron beam size, and accordingly the separation needs. The first proton magnet, QA0, designed as a half-quadrupole,  further reduces the required horizontal distance between the two design orbits.  
%Fig. \ref{fig:IR_sy-li} summarizes the optimization process that compares the emitted light for a simple dipole based separation scheme with the different optimization steps. 
% For  more details see \cite{DIS_Proc}.

\begin{figure}[htb]
   \centering
   \includegraphics[width=.6\columnwidth]{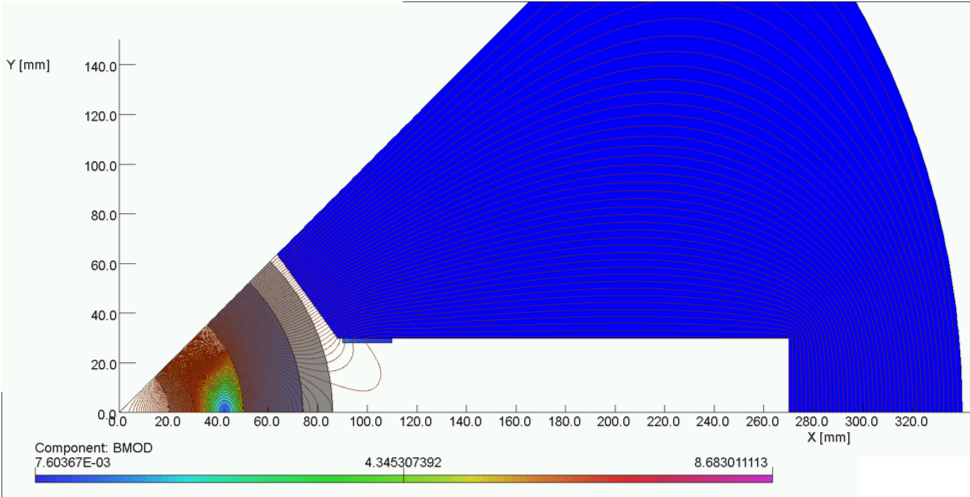}
   \caption{Cutplane view of a simulated field map of the proton mini-beta quadrupole (first design concept). The field in the area occupied by the electron beam is three orders of magnitude below the maximum value (blue area).}
   \label{fig:QA1.png}
\end{figure}

%\begin{figure}[!htb]
%   \centering
%   \includegraphics[width=.6\columnwidth]{figures/Sy_li_opti_1.pdf}
%   \caption{Optimization of the synchrotron light, emitted in different scenarios of the beam separation scheme}
%   \label{fig:IR_sy-li}
%\end{figure}
Special effort is needed in the design of the superconducting quadrupole QA1: Positioned right after the electron mini-beta quadrupoles, it has to provide sufficient aperture and gradient to re-match the proton optics towards the arc structure. Moreover, a field-free region inside the cryostat is needed  for the outgoing electron beam. Figure  \ref{fig:QA1.png} shows a first layout of the magnet. The field calculations for both apertures are determined using the magnet design code ROXIE \cite{Roxie} with special emphasis on minimizing the remaining quadrupole field in the electron aperture: located at a distance of \SI{106}{mm} from the proton design orbit, it has to be low enough not to distort the electron optics. Following the first layout and field calculations as described, some  R\&D will be needed towards 
%leading to a serious design and construction of 
a prototype magnet to further study the feasibility of the technical concept.

\subsubsection{The LHeC Racetrack as an Injector to FCC-ee}
\begin{figure}[htb]
   \centering
   \includegraphics[width=.9\columnwidth]{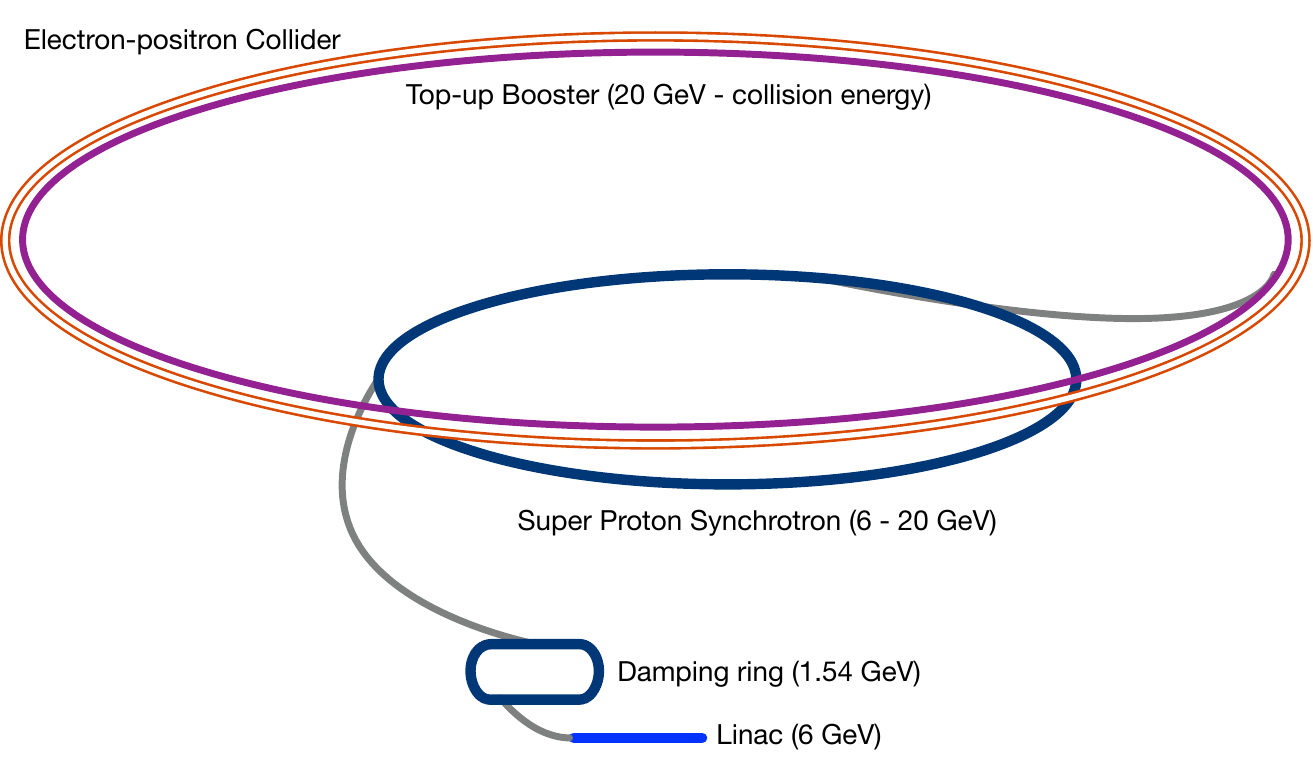}
   \caption{Schematic layout of the FCC-ee injector complex, with the SPS serving as PBR~\cite{Abada2019}.}
   \label{fig:injector_layout}
\end{figure}

The injector complex of the FCC-ee  
comprises an e\textsuperscript{+}e\textsuperscript{--} linac (for energies up to \SI{6}{GeV}),  a pre-booster synchrotron ring (PBR), accelerating from 6 to \SI{20}{GeV}, and a full-energy booster synchrotron ring (BR), integrated in the collider tunnel. A  schematic layout of the injector complex can be seen in Fig.~\ref{fig:injector_layout}.

Table~\ref{tab:FCC_inj_table1} contains a list of parameters for the injection schemes for the different collider energies and filling modes (top-up or initial filling). The baseline parameters are established assuming an SLC/SuperKEKB-like linac~\cite{154130,Miura:IPAC2014-MOPRO001} (C-band \SI{5.7}{GHz} RF system)
with 1 or 2 bunches per pulse and a repetition rate of 100 or \SI{200}{Hz}. The full filling for Z running is the most demanding with respect to the number of bunches, bunch intensity and therefore injector flux. It requires a linac bunch intensity of \num{2.13e10} particles for both species. The electron linac used for positron production should provide around a factor of two higher bunch charge, i.e., \num{4.2e10} electrons, allowing for a \SI{50}{\percent} conversion efficiency. The bunch intensity requirements include a comfortable \SI{80}{\percent} transfer efficiency throughout the injection complex (from the source to the collider).

In the current baseline, the SPS is considered as the PBR, using a scheme similar to the one used for injection into LEP~\cite{Abada2019}.
The PBR cycle length is dominated by the injection plateau and includes a fast ramp of \SI{0.2}{\second} up to \SI{20}{GeV} and a minimum fast extraction flat top of \SI{0.1}{s}.
The total number of bunches required (48 to 16640 bunches) is transferred to the main booster in at most 10 PBR cycles. 
If the SPS serves as PBR, 
the fraction of overall machine time that needs to be dedicated to filling the booster (the duty factor quoted in Table~\ref{tab:FCC_inj_table1}) varies between \SI{8}{\percent} for the $\text{t}\overline{\text{t}}$ mode and \SI{84}{\percent} on the Z pole. 
Accelerating a larger number of bunches per linac pulse or injecting more bunches per PBR cycle would provide additional time for other parallel SPS beam users.
Alternative injector options studied, which would not impact SPS fixed-target operation, include a more compact \enquote{green-field}  PBR or an extension of the linac to reach an energy of \SI{20}{\giga\electronvolt} for direct injection into the main booster. 

The bunch trains from the PBR can be directly injected into the bunch structure required by the collider, within the \SI{400}{MHz} RF. The bunches are then accelerated with a maximum ramp time of \SI{2}{\second}, and a maximum total cycle length of up to \SI{51.7}{\second}, dominated again by the long injection flat bottom, corresponding to the Z running.
Due to the short beam lifetimes of 40 to \SI{70}{\minute}, which depend on the parameter sets and running energies, continuous top-up injection from the BR is required. For the initial filling, the bunches are accumulated in the collider in less than \SI{20}{\minute}. At other times, the beam is used to top up the current, to maintain the collider beam lifetime limits within the $\pm \SI{3}{\percent}$ current drop ($\pm \SI{5}{\percent}$ for the Z). The filling of the two particle species in the machine is interleaved and is able to accommodate the current bootstrapping~\cite{Ogur:2713264, Ogur:IPAC2018-MOPMF001}.

The overall flux requirement for the FCCee at the most demanding Z pole is around 50 times smaller then the one provided by the LHeC ERL. In fact, an ERL at \SI{20}{GeV} can be a very efficient first-stage injector to FCCee. At the same time, it can minimize the injection time and be quite versatile to the requirements with respect to bunch structure for the collider.

\begin{table}\centering\scriptsize
\caption{FCC-ee injector parameters.}%
\label{tab:FCC_inj_table1}%
\begin{tabular}{lccccccccc}
\toprule
Parameter & Unit & \multicolumn{2}{c}{Z} & \multicolumn{2}{c}{W} & \multicolumn{2}{c}{H} & \multicolumn{2}{c}{$\text{t}\overline{\text{t}}$}\\
\midrule
Beam energy & GeV  & \multicolumn{2}{c}{45.6} & \multicolumn{2}{c}{80} & \multicolumn{2}{c}{120} & \multicolumn{2}{c}{182.5}  \\
Type of filling & & Initial & Top-up & Initial & Top-up & Initial & Top-up & Initial & Top-up \\
Linac bunches/pulse & & \multicolumn{2}{c}{2} & \multicolumn{2}{c}{2} & \multicolumn{2}{c}{1} & \multicolumn{2}{c}{1} \\
Linac repetition rate & Hz & \multicolumn{2}{c}{200} & \multicolumn{2}{c}{100} & \multicolumn{2}{c}{100} & \multicolumn{2}{c}{100} \\
Linac RF frequency & GHz & \multicolumn{2}{c}{2.8} & \multicolumn{2}{c}{2.8} & \multicolumn{2}{c}{2.8} & \multicolumn{2}{c}{2.8} \\
Bunch population & \num{e10} & 2.13 & 1.06 & 1.88 & 0.56 & 1.88 & 0.56 & 1.38 & 0.83 \\
No.~of linac injections & & \multicolumn{2}{c}{1040} & \multicolumn{2}{c}{1000} & \multicolumn{2}{c}{328} & \multicolumn{2}{c}{48} \\
%SPS/BR bunch spacing [MHz] & \multicolumn{8}{c}{400} \\\
PBR min.~bunch spacing & ns & \multicolumn{2}{c}{10} &	\multicolumn{2}{c}{10} & \multicolumn{2}{c}{70} & \multicolumn{2}{c}{477.5}	\\
No.~of PBR cycles & & \multicolumn{2}{c}{8} & \multicolumn{2}{c}{1} & \multicolumn{2}{c}{1} & \multicolumn{2}{c}{1} \\
No.~of PBR bunches & & \multicolumn{2}{c}{2080} &	\multicolumn{2}{c}{2000} & \multicolumn{2}{c}{328} & \multicolumn{2}{c}{48}	\\
PBR cycle time & s & \multicolumn{2}{c}{6.3}  & \multicolumn{2}{c}{11.1}  & \multicolumn{2}{c}{4.4}  & \multicolumn{2}{c}{1.6}\\
%PBR bunch population [$10^{10}$] & 0.83 & 0.03 & 0.75 & 0.15 & 0.77 & 0.20 & 0.81 & 0.44 \\
PBR duty factor & & \multicolumn{2}{c}{0.84} & \multicolumn{2}{c}{0.56}  & \multicolumn{2}{c}{0.36} & \multicolumn{2}{c}{0.14} \\
No.~of BR/collider bunches & & \multicolumn{2}{c}{16640} & \multicolumn{2}{c}{2000} & \multicolumn{2}{c}{328}  & \multicolumn{2}{c}{48} \\
No.~of BR cycles & & 10 & 1 & 10 & 1& 10 & 1 & 20 & 1 \\ 
%Transfer efficiency & \multicolumn{8}{c}{0.8} \\
Filling time (both species) & s & 1034.8 & 103.5 & 266 & 26.6 & 151.6 & 15.2 & 251.2 & 12.6\\
%Injected bunch population [$10^{10}$] & 3.3 & 0.07 & 6.0 & 0.12 & 8.0 & 0.16 & 17.4 & 0.35 \\ 
\bottomrule
\end{tabular}
\end{table}

\subsubsection{Preparations}
A decision on the LHeC has not been made yet. It exists as a possible future collider option
based on the LHC. The mid-term future of the LHC depends on the success of the coming
run, which is to begin in 2022. The long-term future of the LHC is related to the strategic, global
developments of high-energy physics and to the plans and support of CERN in particular. New physics
discoveries at the LHC or elsewhere may  alter the direction of particle physics. 
The investments for post-LHC \positron{}\electron{}
and hh colliders are of the order of ten billion CHF while the LHeC cost is an order of magnitude lower. In this context, for the next years, it has been suggested
by the International Advisory Committee of the LHeC, chaired by  
Herwig Schopper, to
``i)  further develop the ERL based ep/A scattering plans, both at LHC
and FCC, as attractive options for the mid and long term programme of CERN, resp. Before
a decision on such a project can be taken, further development work is necessary, and should
be supported, possibly within existing CERN frameworks (e.g., development of SC cavities and
high-field IR magnets).
ii) to intensify the development of the promising high-power beam-recovery technology ERL in Europe. This could be done mainly in national laboratories, in particular with the
PERLE project at Orsay. To facilitate such a collaboration, CERN should express its interest
and continue to take part.
iii) to keep the LHeC option open until further decisions have been taken.
An investigation should be started on the compatibility between the LHeC and a new heavy ion
experiment in Interaction Point 2, which is currently under discussion.''\,\cite{LHeC:2020oyt}.

The present paper is on the development of energy-recovery linacs, for which the LHeC (and FCC-eh)
are prime, high-power applications.

\subsection{CERC: FCC-ee as an ERL}\label{sec:frontier:fcc-ee}
% Kayran, Litvinenko

The Circular Energy Recovery Collider (CERC) is an alternative approach to current designs for high-energy electron-positron colliders either based on two storage rings with \SI{100}{\kilo\meter} circumference or two large linear accelerators.
Using energy-recovery linacs located in the same-size \SI{100}{\kilo\meter} tunnel would allow a large reduction of the beam energy losses, and therefore a reduction of the power consumption, while providing higher luminosity.
It also opens a path for extending the center-of-mass (CM) energy to \SI{500}{\giga\electronvolt}, which would enable double-Higgs production, and even to \SI{600}{\giga\electronvolt} for $\text{t}\overline{\text{t}}\text{H}$ production and measurements of the top Yukawa coupling.
Furthermore, this approach would allow one to recycle not only the energy but also the particles. This feature opens the possibility for colliding fully polarized electron and positron beams. 

\subsubsection{The CERC concept}

\begin{figure}[htb]\centering
\includegraphics[clip, trim=0cm 3cm 1cm 3cm, width=.8\linewidth]{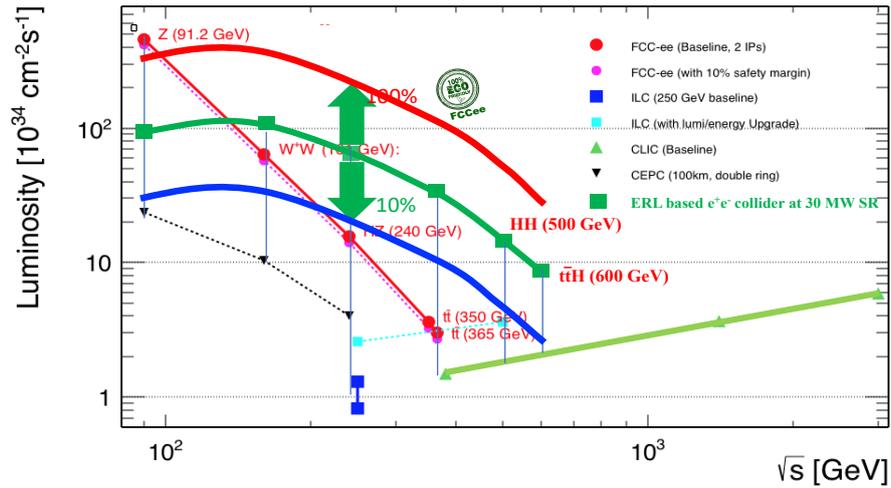}
\caption{Luminosities for various options for a high-energy $\text{e}^{+}\text{e}^{-}$ collider. The plot was taken from \cite{fccee-design-study} and had three CERC luminosity curves in blue, green, and red added to it for levels of synchrotron radiation power of \SI{10}{\mega\watt}, \SI{30}{\mega\watt}, and \SI{100}{\mega\watt}, respectively.}%
\label{fig:frontier:fcc-ee:luminosity}
\end{figure}

This section describes an update to the original proposal \cite{LITVINENKO2020135394, Litvinenko:2019txu} to recycle both energy and particles in a future polarized electron-position collider in order to expand the CM energy reach up to \SI{600}{\giga\electronvolt} while increasing the attainable luminosity; it concentrates on developments since the last publications.
The performance of the CERC is shown in Fig.~\ref{fig:frontier:fcc-ee:luminosity} for different values of RF power consumption.
The luminosity of the CERC scales proportionally with the consumed RF power.

The CERC design is based on ERLs and two damping rings that are also used for particle recycling.
It would consume about one third of the power while providing significantly higher luminosity when compared to the SR $\positron\electron$ collider, with the only exception at the Z-pole. It will also extend the CM energy reach to \SI{600}{\giga\electronvolt}, required for double-Higgs production in the ZHH channel as well as $\text{t}\overline{\text{t}}\text{H}$ production. Even with the energy consumption reduced to \SI{30}{\percent}, the integrated luminosity per year would be about \SI{1.5}{\per\atto\barn} at a center-of-mass energy of \SI{500}{\giga\electronvolt}.

In the CERC design, the electron and positron beams are accelerated to the collision energy in a 4-path ERL.
Most of the energy of the used beams is recovered by delaying them by half of an RF oscillation period and decelerating them.
The electron and positron beams are then reinjected into a damping ring, where they are cooled to low emittance prior to repeating the trip.
The small amount of beam lost during the process, e.g., due to scattering from residual gas or burn-off in the collisions, can easily be replaced by adding particles from a linear injector into the electron and positron damping rings.
Table~\ref{tab:frontier:fcc-ee:parameters} lists the main parameters for the CERC operating at various beam energies. 

\begin{table}[htb]\centering\footnotesize
\caption{Main parameters of an ERL-based $\positron\electron$ collider with a total synchrotron radiation power of \SI{30}{\mega\watt}.}%
\label{tab:frontier:fcc-ee:parameters}
\begin{tabular}{lcccccc}
\toprule
Mode of operation & Z & W & H (HZ) & $\text{t}\overline{\text{t}}$ & HH & $\text{H}\text{t}\overline{\text{t}}$ \\
\midrule
Circumference (km) & 100 & 100 & 100 & 100 & 100 & 100 \\
Beam energy (GeV) & 45.6 & 80.0 & 120.0 & 182.5 & 250.0 & 300 \\
Norm.~emittance $\epsilon_x$ (\si{\micro\meter\radian}) & 3.9 & 3.9 & 6.0 & 7.8 & 7.8 & 7.8 \\
Norm.~emittance $\epsilon_y$ (\si{\nano\meter\radian}) & 7.8 & 7.8 & 7.8 & 7.8 & 7.8 & 7.8 \\
Bend magnet filling factor & 0.9 & 0.9 & 0.9 & 0.9 & 0.9 & 0.9 \\
IP beta function $\beta_x$ (m) & 0.5 & 0.6 & 1.75 & 2 & 2.5 & 3 \\
IP beta function $\beta_y$ (mm), matched & 0.2 & 0.3 & 0.3 & 0.5 & 0.75 & 1 \\
RMS bunch length (mm) & 2 & 3 & 3 & 5 & 7.5 & 10 \\
Bunch charge (nC) & 13 & 13 & 25 & 23 & 19 & 19 \\
Electrons per bunch ($10^{11}$) & 0.78 & 0.78 & 1.6 & 1.4 & 1.2 & 1.2 \\
Bunch frequency (kHz) & 297 & 270 & 99 & 40 & 16 & 9 \\
Beam current (mA) & 3.71 & 3.37 & 2.47 & 0.90 & 0.31 & 0.16 \\
Luminosity (\SI{e35}{\per\square\centi\meter\per\second}) & 6.7 & 8.7 & 7.8 & 2.8 & 1.3 & 0.9 \\
Particle energy loss (GeV) & 4.0 & 4.4 & 6 & 17 & 48 & 109 \\
Radiated power (MW), per beam & 15.0 & 14.9 & 14.9 & 15.0 & 16.8 & 16.9 \\
Total ERL linac voltage (GV) & 10.9 & 19.6 & 29.8 & 46.5 & 67.4 & 89 \\
Disruption parameter $D_x$ & 2.2 & 1.9 & 0.8 & 0.5 & 0.3 & 0.3 \\
Disruption parameter $D_y$ & 503 & 584 & 544 & 505 & 459 & 492 \\
\bottomrule
\end{tabular}
\end{table}

The two main effects that affect the luminosity---or quality of collisions---in linear colliders, beamstrahlung and beam disruption, have been studied for the CERC concept.
It is expected that large disruption parameters would result both in pinching of the beam sizes as well as in transverse emittance growth.
We conducted preliminary studies of these effects in strong-strong collision simulations and showed that the growth of the vertical emittance is limited to about 4-fold for selected disruption parameters.

The main challenge of extending the CERC operation to energies above \SI{182.5}{\giga\electronvolt} is the low-energy tail in the recirculating electron beam generated by beamstrahlung, i.e., high-energy photons generated during collisions.
The critical energy of the beamstrahlung photons can reach \SI{1}{\giga\electronvolt} for CERC operation with \SI{300}{\giga\electronvolt} beams.
The energy recovery and damping ring systems were selected to have an energy acceptance that exceeds the energy of the beamstrahlung photons by a factor of 10.
Such a choice guarantees that beam losses will be less than 1\,ppm.
The electron and positron bunches are decompressed---up to 15-fold for \SI{300}{\giga\electronvolt} CERC operaions---in the low-energy pass of the ERL prior to injection into the damping rings.
The operating energy of the damping rings depends on the CERC top beam energy: it is \SI{2}{\giga\electronvolt} for CERC beam energies up to \SI{120}{\giga\electronvolt}, \SI{3}{\giga\electronvolt} for \SI{182.5}{\giga\electronvolt}, \SI{4.5}{\giga\electronvolt} for \SI{250}{\giga\electronvolt}, and \SI{8}{\giga\electronvolt} for \SI{300}{\giga\electronvolt} beams.
The combination of bunch decompression with $\pm\SI{5}{\percent}$ energy acceptance of the damping rings would ensure nearly perfect recovery of the circulating particles.
We plan to do Monte-Carlo simulations to identify the exact number of particles that could be lost.

\subsubsection{Studies of the collider lattice}

With diffusion caused by quantum fluctuations of the synchrotron radiation scaling as the seventh power of the beam energy, preservation of the transverse emittance in the accelerating beams is most challenging for the highest proposed energy of operation of \SIrange{250}{300}{\giga\electronvolt}.
We found that using a FODO lattice with a \SI{16}{\meter} period (e.g., two \SI{8}{\meter} combined-function magnets) and a phase advance of \ang{90} can satisfy the requirements specified in Table~\ref{tab:frontier:fcc-ee:parameters}~\cite{Litvinenko:2019txu}.
The conditions for the lower collision energies can also be satisfied.  

\begin{figure}[htb]\centering
\tikzsetnextfilename{s512_optics}
\begin{tikzpicture}
\draw (0, 0) node (topleft) {\includegraphics[height=5cm]{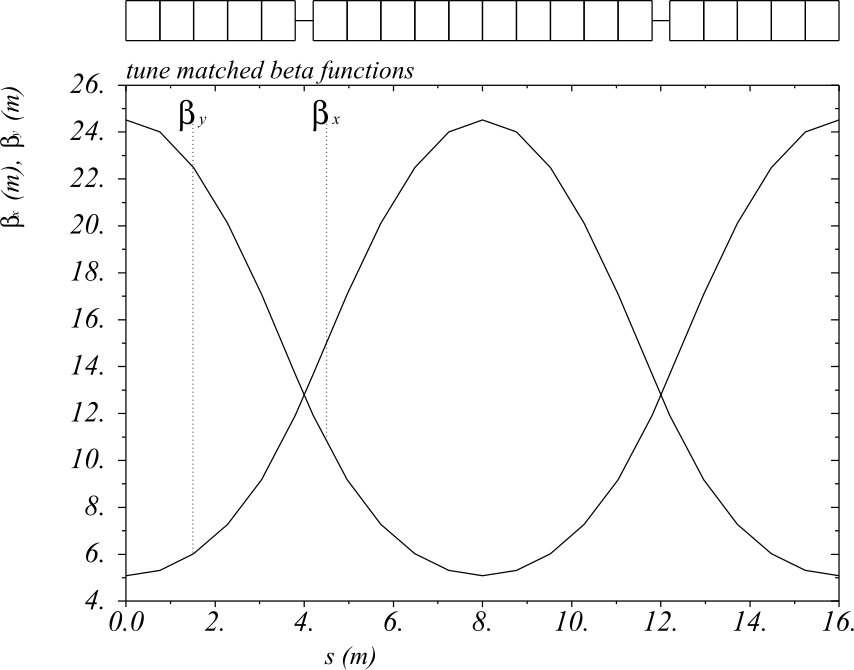}};
\draw (topleft.east) ++(1, 0) node[anchor=west] (topright) {\includegraphics[height=5cm]{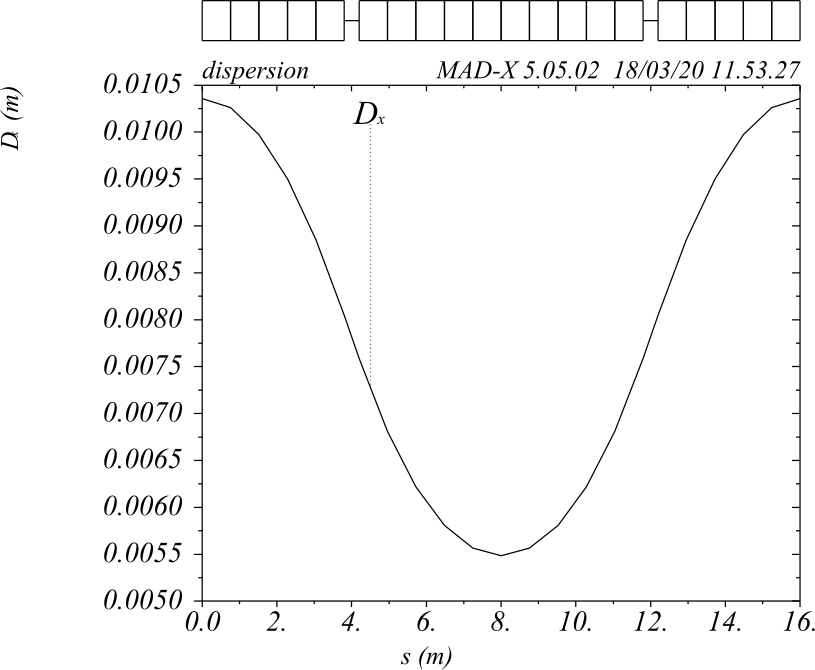}};
\draw (topleft.south) ++(0, -1) node[anchor=north] (bottomleft) {\includegraphics[height=5cm]{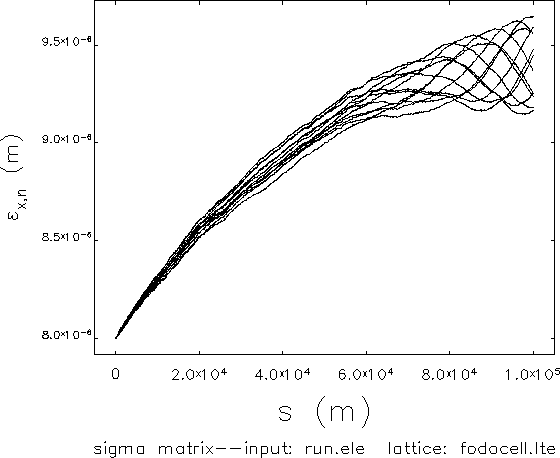}};
\draw (topright.south) ++(0, -1) node[anchor=north] (bottomright) {\includegraphics[height=5cm]{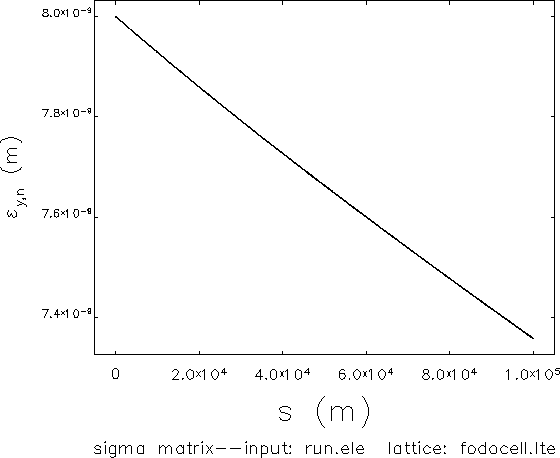}};
\end{tikzpicture}
\caption{Beta (top left) and dispersion (top right) functions of the regular ERL lattice; Evolution of normalized horizontal (bottom left) and vertical (bottom right) emittances in the \SI{100}{\kilo\meter} pass at the top energy.}%
\label{fig:frontier:fcc-ee:optics}
\end{figure}

The lattice of each path around the \SI{100}{\kilo\meter} circumference is comprised of 6250 FODO cells with combined-function (dipole, quadrupole, and sextupole) magnets and zero chromaticity.
The cell is comprised of two 7.6-meter-long magnets separated by \SI{0.4}{\milli\meter} drifts.
At the top energy, the combined magnets have a magnetic field of \SI{0.0551}{\tesla}, field gradients of $\pm\SI{32.24}{\tesla\per\meter}$, and sextupole components $S_F=\SI{267}{\tesla\per\square\meter}$ and $S_D=\SI{-418}{\tesla\per\square\meter}$.
The magnets have an aperture of $\pm\SI{0.75}{\centi\meter}$ and pole-tip fields of about 1\,kG, which is convenient for magnetic steel.
The optical functions and emittance evolution of this lattice are shown in Fig.~\ref{fig:frontier:fcc-ee:optics}.

The electron and positron beams need to undergo compression during the first pass around the tunnel as well as decompression during the last pass prior to reinjection into the damping rings.
Long bunches, which require subsequent compression, have a relatively low peak current in the damping rings, which minimizes Intra-Beam Scattering (IBS).
Similarly, the decompression will reduce the energy spread accumulated by the bunches during the acceleration/collision/deceleration cycle, allowing them to fit into the energy acceptance of the damping rings.
Using the low-energy passes of the ERL for the compression and decompression will provide for a large value of the longitudinal dispersion $R_{56}$ while maintaining low emittance growth.
This process will require additional RF gymnastics, such as chirping beam energy and compensating the energy chirp after the bunch compression/decompression.
We are pursuing a detailed design of the entire accelerator, including compressing and decompressing arcs and the SRF linac with splitters and combiners.
Details of this studies will be published elsewhere.

One of the advantages of the CERC is its capability of colliding polarized beams.
Preliminary studies using the ZGOUBI code confirmed that the proposed lattice can preserve the polarization.
The simulations include effects of synchrotron radiation, beam emittances, and orbit misalignments \cite{erl-erl-fcc-ee-polarization-transport}.
The conclusions of these studies is that beam depolarization does not exceed \SI{0.1}{\percent} per path and that collisions of highly polarized beams in such a collider might be feasible.

The impact of different polarization values for electrons and positrons on the production cross section for ZH, ZHH and $\text{t}\overline{\text{t}}\text{H}$ has been estimated with Madgraph \cite{madgraph5} and is shown in Table~\ref{tab:frontier:fcc-ee:polarization}.
The proper combination of polarization for electrons and positrons will significantly enhance the production cross section or suppress it.

\begin{table}[htb]\centering
\caption{Impact of polarization on the ZH, ZHH, and $\text{t}\overline{\text{t}}\text{H}$ production cross sections.}%
\label{tab:frontier:fcc-ee:polarization}
\begin{tabular}{ccccc}
\toprule
\multicolumn{2}{c}{Polarization} & \multicolumn{3}{c}{Scaling factor} \\
\midrule
\electron & \positron & ZH (\SI{240}{\giga\electronvolt}) & ZHH (\SI{500}{\giga\electronvolt}) & $\text{t}\overline{\text{t}}\text{H}$ (\SI{600}{\giga\electronvolt}) \\
\midrule
\multicolumn{2}{c}{Unpolarized} & 1 & 1 & 1 \\
$-70$ & 0 & 1.15 & 1.15 & 1.23 \\
$-70$ & $+50$ & 1.61 & 1.61 & 1.87 \\
$-70$ & $-50$ & 0.69 & 0.69 & 0.73 \\
$-70$ & $+70$ & 1.78 & 1.79 & 2.07 \\
$-70$ & $-70$ & 0.51 & 0.51 & 0.51 \\
$-50$ & $+50$ & 1.47 & 1.47 & 1.69 \\
$+50$ & $-50$ & 1.03 & 1.03 & 0.82 \\
$+70$ & $0$   & 0.85 & 0.85 & 0.69 \\
$+70$ & $+50$ & 0.60 & 0.60 & 0.56 \\
$+70$ & $-50$ & 1.09 & 1.09 & 0.83 \\
$+70$ & $+70$ & 0.51 & 0.51 & 0.51 \\
\bottomrule
\end{tabular}
\end{table}

\subsubsection{Key technical details and assumptions of the concept}

This section provides key technical details to illustrate the feasibility of such a collider.
First, the assumptions for the CERC linac are based on the shunt impedance of the operational \SI{703}{\mega\hertz} 5-cell cavity (so called BNL-3 design) and progress at FNAL in reaching quality factors of $Q_0=\num{e11}$ using novel doping techniques and precise demagnetization of cavity prior to cool-down \cite{posen_srf, Grassellino_2013, Trenikhina:SRF2015-MOPB055}.

\begin{figure}[htb]\centering
\includegraphics[height=5cm]{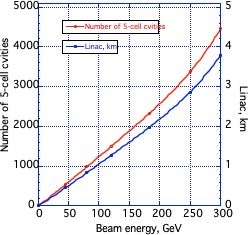}
\caption{Number of required 5-cell cavities and length of each SRF linac as a function of the energy of the colliding beams.}%
\label{fig:frontier:fcc-ee:linacs}
\end{figure}

The BNL-3 5-cell SRF cavity unit has a length of \SI{1.58}{\meter} with about \SI{1}{\meter} of accelerating structure.
We propose to use 16-meter-long cryostats housing 10 five-cell cavities.
We assume that cryomodules will be separated by \SI{1}{\meter}, which corresponds to a \SI{58.8}{\percent} filling factor for the accelerating field.
Room-temperature HOM couplers installed at the ends of each of the cavities will dissipate the majority of the HOM power.
The very-high-frequency components of the HOM power will propagate through the large apertures of the 5-cell linacs and will be absorbed by ferrite-type room-temperature HOM absorbers installed between cryomodules.
Figure~\ref{fig:frontier:fcc-ee:linacs} shows the number of 5-cell cavities as function of CERC energy.

The BNL-3 cavity has a HOM loss factor of \SI{0.16}{\volt\per\pico\coulomb} and \SI{0.12}{\volt\per\pico\coulomb} for electron bunches with an RMS bunch length of \SI{30}{\milli\meter} and \SI{50}{\milli\meter}, respectively.
It means that for a maximum charge per bunch of \SI{25}{\nano\coulomb} and \SI{30}{\milli\meter} RMS bunch length, particles will lose \SI{4}{\kilo\electronvolt} per cavity.
Each particle passes through each cavity 8 times: 4 times on the way up and 4 times on the way down in energy.
The total loss of the particle’s energy into HOM modes ranges from \SI{1.1}{\mega\electronvolt} for the lowest beam energy to \SI{10.6}{\mega\electronvolt} for the top beam energy of the CERC.
The total HOM power losses in the absorbers from the 8 passes by both electron and positions beams do not exceed \SI{250}{\kilo\watt}, and the maximum power per HOM absorber does not exceed \SI{160}{\watt}.

At the top energy for FCC-ee of \SI{182.5}{\giga\electronvolt}, the CERC will require two \SI{2}{\kilo\meter}-long SRF linacs and a \SI{20}{\mega\watt} cryo plant. The HOM losses will be at the \SI{100}{\kilo\watt} level.
Naturally, going above \SI{182.5}{\giga\electronvolt} will require longer linacs and a more powerful cryo plant.

For briefness, we do not describe second-harmonic cavities, which can compensate for synchrotron radiation losses, and other harmonic cavities.
It is estimated that these cavities will contribute \SIrange{10}{20}{\percent} to the cryo-plant power and the HOM losses. 
\SI{30}{\mega\watt} of RF power will be needed for the second-harmonic and damping-ring RF systems to compensate for the energy lost to synchrotron radiation.

Microphonics represent an additional challenge for modern SRF ERLs by requiring additional RF power to keep cavities in compliance.
Typically, a \SI{20}{\mega\volt} 5-cell cavity may require a transmitter with a power of \SI{20}{\kilo\watt} for this purpose.
For the CERC operating with \SI{182.5}{\giga\electronvolt} beams, it would require an additional \SI{46}{\mega\watt} of RF power, which would greatly reduce the attractiveness of this concept.
Fortunately, CERN successfully tested ferro-electric tuners for SRF cavities, which reduce the power requirements to about \SI{200}{\watt} \cite{shipman_frt}.
This technology is assumed to be available for the CERC; thus, \SI{3}{\kilo\watt} of RF power are allocated to compensate for microphonics.

Because the vertical beam size in the CERC is measured in \si{\micro\meter}, we propose to use magnets with a small gap of about \SI{15}{\milli\meter} \cite{Litvinenko:2019txu}.
The power consumption of the magnet is proportional to its gap and magnetic field squared, which means that 16 beam lines required for the CERC electron and position beams will consume the same amount of power as a single storage ring with a typical gap of about \SI{5}{\centi\meter}.
This is why we assumed that the CERC magnetic system will consume about \SI{50}{\percent} of the power required for the collider storage ring magnets in the FCC-ee.

\subsubsection{Summary of CERC power consumption}

Finally, Table~\ref{tab:frontier:fcc-ee:power_consumption} summarizes an estimate of the CERC power consumption.
We are assuming \SI{1.25}{\kilo\watt} of cryo-plant power per \SI{1}{\watt} loss at \SI{1.8}{\kelvin} in the SRF linac.
This includes a \SI{25}{\percent} overhead related to the cryogenic facility and liquid He transport system.
We are also using a ratio of AC power to RF power for the RF amplifiers of 1.66.
For the damping rings, we would use permanent magnets as is being done now for light sources. The same would be done for the transfer lines to and from the damping rings.

Note that the electric power consumption of the CERC is lower than that of the FCC-ee by about \SI{100}{\mega\watt} over its energy range with much higher luminosities at the higher energies.
There is also a possibility of extending the center-of-mass energy up to \SI{600}{\giga\electronvolt} without an excessive rise in power consumption.
The electric power consumption could be further reduced with focused R\&D.

\begin{table}[htb]\centering
\caption{Estimation of the AC power consumption of the CERC.}%
\label{tab:frontier:fcc-ee:power_consumption}
\begin{tabular}{ccccccc}
\toprule
Mode & Z & W & HZ & $\text{t}\overline{\text{t}}$ & HHZ & $\text{t}\overline{\text{t}}\text{H}$ \\
Beam energy (GeV) & 45.6 & 80 & 120 & 182.5 & 250 & 300 \\
\midrule
Synchrotron radiation (MW) & 30 & 30 & 30 & 30 & 30 & 30 \\
Microphonics (MW) & 1.6 & 2.9 & 4.5 & 7.0 & 10.1 & 13.4 \\
Higher-order modes (MW) & 0.1 & 0.2 & 0.3 & 0.2 & 0.1 & 0.0 \\
Total RF power (MW) & 31.7 & 33.1 & 34.8 & 37.2 & 40.2 & 43.4 \\
\midrule
Magnets (MW) & 2.0 & 6.2 & 13.9 & 32.0 & 60.1 & 86.6 \\
\SI{1.8}{\kelvin} cryo load (kW) & 5 & 10 & 15 & 23 & 34 & 45 \\
Cryo plant AC power (MW) & 6.25 & 12.5 & 18.75 & 28.75 & 42.5 & 56.25 \\
\midrule
Total AC power (MW) & 61 & 74 & 90 & 123 & 169 & 215 \\
\bottomrule
\end{tabular}
\end{table}

\subsubsection{Conclusion}

A novel concept has been proposed for a
%We did not find any showstoppers preventing our design from working as a
next-generation, high-energy, polarized $\positron\electron$ collider
which deserves 
%We continue the
detailed in-depth study to fully validate this ERL-based concept.

%Authors would like to thank Dr. Frank Zimmerman and FCC-design team at CERN for opportunity to present and discuss this option at their working meeting. Vladimir Litvinenko would like to acknowledge support by NSF grant PHY-1415252 “Center for Science and Education at Stony Brook University”.
%\subsection{FCC-ee as an ERL}
%Dimitry Kayran, Vladimir Litvinenko, 

\subsection{ERLC: ILC as an ERL}%
\label{sec:frontier:erlc}

% MB 06/03/2022: The plots are included as JPG files. We have them as PDF, but the author won't let us use them because of an alleged display issue. He also opposes redrawing them in TikZ even though I would have done all the work. OK... They aren't the worst 

Linear \positron\electron{} colliders (LC) have  been actively developed since the 1970s as a way to reach higher energies.
Their main advantage over storage rings is the absence of synchrotron radiation during acceleration, making it possible to achieve much  higher energies.
Their main weak point is the one-pass use of beams.
In storage rings, the same beams are used many millions of times, whereas in LCs they are sent  to beam dumps after a single collision.
This inefficient use of electricity results in a low collision rate and therefore low luminosity.

While there were many LC projects in the 1990s, since 2004 only two remain: ILC~\cite{ILC} and CLIC~\cite{Aicheler:1500095}.
The ILC is based on superconducting (SC) Nb technology (in the footsteps of the TESLA), while the CLIC  uses Cu cavities and operates at room temperature.
Both colliders work in one-pass mode; the difference is only  in the length of the bunch trains; the luminosities and wall plug powers are very similar.
In fact, the use of superconducting technology does not provide significant luminosity benefits to the ILC.

The main potential advantage of superconducting technology is the possibility of energy recovery, where the beam, after passing the interaction point (IP), is decelerated in the opposing linac and thus returns its energy to the accelerator.
This opportunity was noticed originally and discussed in the very first publications on linear colliders by M.~Tigner~\cite{Tigner:1965wf}, A.~Skrinsky~\cite{amaldi}, and U.~Amaldi~\cite{AMALDI1976313}.
The scheme of the LC  suggested by H.~Gerke and K.~Steffen in 1979~\cite{Gerke} assumed not only energy recovery but also multiple use of the electron and positron beams.
However, their  scheme gave a luminosity even lower than in one-pass schemes. This happened for two main reasons:
\begin{itemize}
\item The quality factor of SC cavities at that time was $Q_0 \sim 2\times 10^9$, which was not enough for a continuous mode of operation. Removal of the heat from cryogenic structures requires a lot of energy; therefore, a duty cycle of $\tfrac{1}{30}$ was adopted.
\item In order to exclude parasitic collisions inside the linac, which are a cause of beam instability, only one bunch is present at any one moment in each  half linac, which limits the collision rate to $f=\SI{30}{\kilo\hertz}$. With a duty cycle of $\tfrac{1}{30}$, the average rate would be a mere \SI{1}{\kilo\hertz}.
\end{itemize}
As a result, the estimated luminosity was $\mathcal{L}=\SI{3.6e31}{\per\square\centi\meter\per\second}$, which is too low to be of interest.
Since the 1980s, LC energy-recovery schemes have no longer been considered.
This is because the collision rate at a single-pass LC is similar to that at an ERL collider (as discussed above), and the luminosity per collision can be much higher at a single-pass LC due to the larger permissible disruption of the beams.

\subsubsection{Superconducting twin linear collider with energy recovery (ERLC)}

To solve the problem of parasitic collisions, Valery Telnov recently  proposed the concept of a \emph{twin} linear collider~\cite{Telnov_2021} with energy recovery and multiple use of beams, which can increase the luminosity by four orders of magnitude. What follows is a short version of this article.

\begin{figure}[htb]\centering
\tikzsetnextfilename{telnov_hutton_schematic}
\begin{tikzpicture}
[
	electron/.style={thick, blue},
	positron/.style={thick, green!50!black},
	annotation/.style={black, font=\footnotesize, align=center},
	colored annotation/.style={font=\footnotesize, align=center},
	linac/.style={violet, fill=violet!10!white},
	decompressor/.style={red, fill=red!10!white},
	compressor/.style={brown, fill=brown!10!white},
	wiggler/.style={red, thick, decorate, decoration={snake}},
	cavity/.style={},
]
\pgfmathsetmacro{\CavLen}{2.2}
\pgfmathsetmacro{\CavWidth}{0.2}
\pgfmathsetmacro{\CavDist}{0.5}
\pgfmathsetmacro{\CavCellRadius}{0.1}
\pgfmathsetmacro{\CavityPipeLength}{{0.5 * (\CavLen - 9 * 2 * \CavCellRadius)}}

% These are for the small elements
\pgfmathsetmacro{\BoxLen}{1}
\pgfmathsetmacro{\BoxWidth}{0.25}
% Wigglers
\pgfmathsetmacro{\WigglerLen}{1.5}
% Drift length between elements
\pgfmathsetmacro{\SegmentLen}{0.25}

\pgfmathsetmacro{\SmallBendingRadius}{0.75}
\pgfmathsetmacro{\LargeBendingRadius}{\SmallBendingRadius + \CavDist/2}
\pgfmathsetmacro{\SmallBendingRadiusP}{1.0}
\pgfmathsetmacro{\LargeBendingRadiusP}{\SmallBendingRadiusP + \CavDist/2}

% Left cavity pair
\begin{scope}[shift={(-2, 0)}]
\draw [cavity]
    (\CavLen/2, \CavWidth/2+\CavDist/2)
    -- ++(-\CavityPipeLength, 0)
    \foreach \i in {1,...,9}{
        arc (0:180:\CavCellRadius)
    }
    -- ++(-\CavityPipeLength, 0)
    -- ++(0, -\CavWidth)
    -- ++(\CavityPipeLength, 0)
    \foreach \i in {1,...,9}{
        arc (-180:0:\CavCellRadius)
    }
    -- ++(\CavityPipeLength, 0)
    -- cycle;
\draw [cavity]
    (\CavLen/2, \CavWidth/2-\CavDist/2)
    -- ++(-\CavityPipeLength, 0)
    \foreach \i in {1,...,9}{
        arc (0:180:\CavCellRadius)
    }
    -- ++(-\CavityPipeLength, 0)
    -- ++(0, -\CavWidth)
    -- ++(\CavityPipeLength, 0)
    \foreach \i in {1,...,9}{
        arc (-180:0:\CavCellRadius)
    }
    -- ++(\CavityPipeLength, 0)
    -- cycle;
\foreach \i in {-4,-3,...,4}{
	\pgfmathsetmacro{\x}{2*\CavCellRadius*\i}
	\draw[thick] (\x, -\CavDist/2+\CavWidth/2+\CavCellRadius) -- (\x, \CavDist/2-\CavWidth/2-\CavCellRadius);
}
\path (-\CavLen/2, -\CavDist/2) coordinate (bot left cav outer);
\path (-\CavLen/2, \CavDist/2) coordinate (top left cav outer);
\path (\CavLen/2, -\CavDist/2) coordinate (bot left cav inner);
\path (\CavLen/2, \CavDist/2) coordinate (top left cav inner);
\end{scope}

% Right cavity pair
\begin{scope}[shift={(2, 0)}]
\draw[cavity]
    (\CavLen/2, \CavWidth/2+\CavDist/2)
    -- ++(-\CavityPipeLength, 0)
    \foreach \i in {1,...,9}{
        arc (0:180:\CavCellRadius)
    }
    -- ++(-\CavityPipeLength, 0)
    -- ++(0, -\CavWidth)
    -- ++(\CavityPipeLength, 0)
    \foreach \i in {1,...,9}{
        arc (-180:0:\CavCellRadius)
    }
    -- ++(\CavityPipeLength, 0)
    -- cycle;
\draw [annotation, above] (0, \CavWidth/2+\CavDist/2+\CavCellRadius)
    node {acceleration};
\draw [cavity]
    (\CavLen/2, \CavWidth/2-\CavDist/2)
    -- ++(-\CavityPipeLength, 0)
    \foreach \i in {1,...,9}{
        arc (0:180:\CavCellRadius)
    }
    -- ++(-\CavityPipeLength, 0)
    -- ++(0, -\CavWidth)
    -- ++(\CavityPipeLength, 0)
    \foreach \i in {1,...,9}{
        arc (-180:0:\CavCellRadius)
    }
    -- ++(\CavityPipeLength, 0)
    -- cycle;
\draw [annotation, below] (0, -\CavWidth/2-\CavDist/2-\CavCellRadius)
    node {deceleration};
\foreach \i in {-4,-3,...,4}{
	\pgfmathsetmacro{\x}{2*\CavCellRadius*\i}
	\draw[thick] (\x, -\CavDist/2+\CavWidth/2+\CavCellRadius) -- (\x, \CavDist/2-\CavWidth/2-\CavCellRadius);
}
\path (-\CavLen/2, -\CavDist/2) coordinate (bot right cav inner);
\path (-\CavLen/2, \CavDist/2) coordinate (top right cav inner);
\path (\CavLen/2, -\CavDist/2) coordinate (bot right cav outer);
\path (\CavLen/2, \CavDist/2) coordinate (top right cav outer);
\end{scope}

% Linacs
\begin{scope}[shift={($(top right cav outer)+(\BoxLen/2, 0)+(\SegmentLen, 0)$)}]
\draw[linac] (-\BoxLen/2, -\BoxWidth/2) rectangle +(\BoxLen, \BoxWidth);
\draw (0, \BoxWidth/2) node[anchor=south, annotation] {linac\\($\Delta E$)};
\path (-\BoxLen/2, 0) coordinate (right linac inner);
\path (\BoxLen/2, 0) coordinate (right linac outer);
\end{scope}
\begin{scope}[shift={($(top left cav outer)+(-\BoxLen/2, 0)+(-\SegmentLen, 0)$)}]
\draw[linac] (-\BoxLen/2, -\BoxWidth/2) rectangle +(\BoxLen, \BoxWidth);
\path (-\BoxLen/2, 0) coordinate (left linac outer);
\path (\BoxLen/2, 0) coordinate (left linac inner);
\end{scope}

% decompressors
\begin{scope}[shift={($(bot right cav outer)+(\BoxLen/2, 0)+(\SegmentLen, 0)$)}]
\draw[decompressor] (-\BoxLen/2, -\BoxWidth/2) rectangle +(\BoxLen, \BoxWidth);
\draw (0, -\BoxWidth/2) node[anchor=north, annotation] {decomp-\\ressor};
\path (-\BoxLen/2, 0) coordinate (right decompressor inner);
\path (\BoxLen/2, 0) coordinate (right decompressor outer);
\end{scope}
\begin{scope}[shift={($(bot left cav outer)+(-\BoxLen/2, 0)+(-\SegmentLen, 0)$)}]
\draw[decompressor] (-\BoxLen/2, -\BoxWidth/2) rectangle +(\BoxLen, \BoxWidth);
\path (-\BoxLen/2, 0) coordinate (left decompressor outer);
\path (\BoxLen/2, 0) coordinate (left decompressor inner);
\end{scope}

% compressors
\begin{scope}[shift={($(right linac outer)+(\BoxLen/2, 0)+(\SegmentLen, 0)$)}]
\draw[compressor] (-\BoxLen/2, -\BoxWidth/2) rectangle +(\BoxLen, \BoxWidth);
\draw (0, \BoxWidth/2) node[anchor=south, annotation] {comp-\\ressor};
\path (-\BoxLen/2, 0) coordinate (right compressor inner);
\path (\BoxLen/2, 0) coordinate (right compressor outer);
\end{scope}
\begin{scope}[shift={($(left linac outer)+(-\BoxLen/2, 0)+(-\SegmentLen, 0)$)}]
\draw[compressor] (-\BoxLen/2, -\BoxWidth/2) rectangle +(\BoxLen, \BoxWidth);
\path (-\BoxLen/2, 0) coordinate (left compressor outer);
\path (\BoxLen/2, 0) coordinate (left compressor inner);
\end{scope}

% electron beam
\draw[electron]
	(right compressor inner)
	-- (right linac outer)
	(right linac inner)
	-- (top right cav outer)
	-- (top right cav inner);
\draw[electron, postaction={decorate, decoration={markings, mark=at position .3 with {\arrow{latex}}}}]
	(top right cav inner)
	-- ++(-0.1, 0)
	node[anchor=south east, colored annotation] {$\text{e}^{-}$}
	-- ($(bot left cav inner) + (0.1, 0)$)
	-- ++(-0.1, 0);
\draw[electron, -latex]
	(bot left cav inner)
	-- (bot left cav outer)
	-- (left decompressor inner);
\draw[electron, postaction={decorate, decoration={markings, mark=at position .5 with {\arrow{latex}}}}]
	(left decompressor outer)
	arc (90:270:\SmallBendingRadius)
	coordinate (electron wiggler outer);
\draw[wiggler]
	(electron wiggler outer)
	-- ++(\WigglerLen, 0)
	coordinate (electron wiggler inner);
\draw[electron, postaction={decorate, decoration={markings, mark=at position .4 with {\arrow{latex}}}}]
	(electron wiggler inner)
	-- ($(electron wiggler inner -| right compressor outer)$)
	coordinate (right bend)
	node[above, pos=0.4, colored annotation] {$\text{e}^{-}$}
	node[anchor=south east, pos=0.35, annotation] {$E\approx\SI{5}{\giga\electronvolt}$};
\draw[electron, postaction={decorate, decoration={markings, mark=at position .5 with {\arrow{latex}}}}]
	(right bend)
	arc(270:450:\LargeBendingRadius);
\draw[electron, dashed, -latex] (right bend) -- ++(1, 0)
	node[annotation, below] {beam dump};

% positron beam
\draw[positron]
	(left compressor inner)
	-- (left linac outer)
	(left linac inner)
	-- (top left cav outer)
	-- (top left cav inner);
\draw[positron, postaction={decorate, decoration={markings, mark=at position .3 with {\arrow{latex}}}}]
	(top left cav inner)
	-- ++(0.1, 0)
	node[anchor=south west, colored annotation] {$\text{e}^{+}$}
	-- ($(bot right cav inner) + (-0.1, 0)$)
	-- ++(0.1, 0);
\draw[positron, -latex]
	(bot right cav inner)
	-- (bot right cav outer)
	-- (right decompressor inner);
\draw[positron, postaction={decorate, decoration={markings, mark=at position .5 with {\arrow{latex}}}}]
	(right decompressor outer)
	arc (90:-90:\SmallBendingRadiusP)
	coordinate (positron wiggler outer);
\draw[wiggler]
	(positron wiggler outer)
	-- ++(-\WigglerLen, 0)
	coordinate (positron wiggler inner)
	node[below, pos=0.5, annotation] {wiggler\\$\Delta E\sim\SI{0.025}{\giga\electronvolt}$};
\draw[positron, postaction={decorate, decoration={markings, mark=at position .4 with {\arrow{latex}}}}]
	(positron wiggler inner)
	-- ($(positron wiggler outer -| left compressor outer)$)
	coordinate (left bend)
	node[below, pos=0.4, colored annotation] {$\text{e}^{+}$};
\draw[positron, postaction={decorate, decoration={markings, mark=at position .5 with {\arrow{latex}}}}]
	(left bend)
	arc(-90:-270:\LargeBendingRadiusP);
\draw[positron, dashed, -latex] (left bend) -- ++(-1, 0);

\path (top left cav inner) -- (top right cav inner) coordinate[pos=0.5] (above ip);
\draw (above ip) ++(0, 0.5)
	node[anchor=south, annotation] {$\sim$ head-on\\collision};
\path (above ip) -- ($(above ip |- positron wiggler inner)$) coordinate (below ip on positron orbit);

% Injection from DR
\draw[positron, postaction={decorate, decoration={markings, mark=at position .2 with {\arrow{latex}}}}]
    (below ip on positron orbit) ++(-1, -1)
    .. controls +(-1.5, 1) .. ++(-3, 1);
\draw[electron, postaction={decorate, decoration={markings, mark=at position .2 with {\arrow{latex}}}}]
    (below ip on positron orbit) ++(1, -1)
    .. controls +(1.5, 1.5) .. ++(3, 1.5);
\draw (below ip on positron orbit) ++(0, -1)
    node[annotation, anchor=north] {from DRs};
    
\end{tikzpicture}
\caption{The layout of the SC twin linear collider.}
\label{fig:frontier:erlc:twin-scheme}
\end{figure}
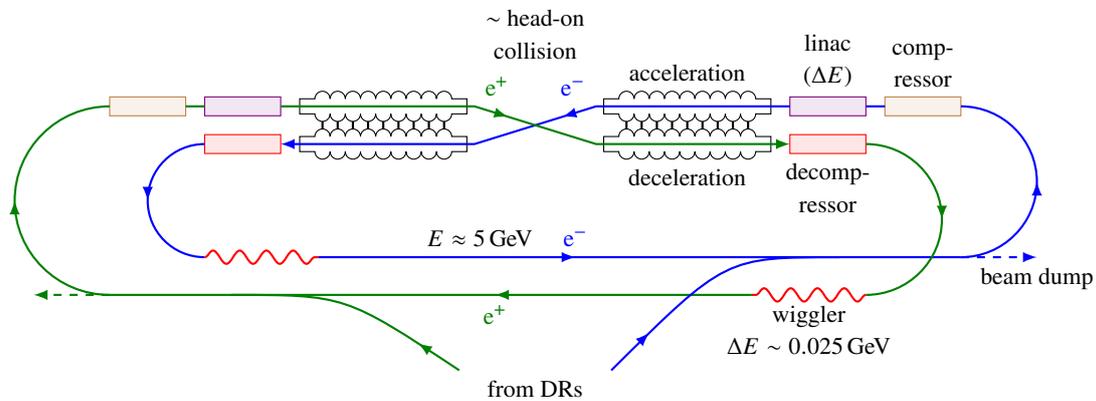

In the twin linear collider, the beams are accelerated and then decelerated down to $E\approx \SI{5}{\giga\electronvolt}$  in separate parallel linacs with coupled RF systems, see Fig.~\ref{fig:frontier:erlc:twin-scheme}.
The RF power is always divided equally among the linacs.
RF energy is transferred to the beams  both from an external RF source and from the beam being decelerated.
The linacs can be either two separate SC linacs connected by RF couplers at the ends of multi-cell cavities (9-cell TESLA cavity) or one linac consisting of twin (dual) cavities with  axes for two beams.
Such cavities have been designed and tested for XFELs~\cite{noguchi,wang2007,Park:LINAC2016-THPLR037,PhysRevAccelBeams.20.103501}.

The collider is assumed to operate at an energy of $2E_0 \approx \SI{250}{\giga\electronvolt}$ (with a possible increase up to \SI{500}{\giga\electronvolt})  in semi-continuous mode with a duty cycle: collisions for about 10 seconds (or more, see below), then a break to cool the cavities.
In one cycle, the beams make about \num{50000} revolutions.
Continuous (CW) operation is more attractive and may be an option after additional R\&D (discussed below).

The beams are prepared in damping rings, as usual.
In continuous mode, only the lost particles need to be replaced; in duty-cycle mode, the bunches are prepared anew each cycle.
The number of bunches in the ERLC is large, but there is enough time (injection time up to \SIrange{1}{2}{\second}).
The required average production rate is an order of magnitude lower than at the ILC.

During collisions, beams get an additional energy spread that is damped by wigglers installed in the  return pass at an energy of $E \approx \SI{5}{\giga\electronvolt}$.
The relative energy loss in wigglers is about $\delta E/E \sim 1/200$.
We require that the steady-state  equilibrium energy spread at the IP due to beamstrahlung is better than $\sigma_E/E_0 \sim 0.2\,\%$, the same as at the ILC and CLIC prior to collision. Such a spread would be sufficient for beam focusing.
When the beam is decelerated down to \SI{5}{\giga\electronvolt}, its relative energy spread increases by a factor of $E_0/E\sim 25$ to $\sigma_E/E\sim 5\,\%$.
To avoid losses in the subsequent arcs, the energy spread is reduced by 10--15 times with the help of the bunch (de)compressor; then, the relative energy spread in the arcs is less than 0.5\,\%.
The beam lifetime will be determined by the tails in beamstrahlung radiation.
This loss should not exceed about 1\,\% after \num{10000} revolutions. The IP energy spread, the beam instability and beam particle losses (due to single beamstrahlung) determine the IP beam parameters, and hence the luminosity.

An important question is the injection and extraction of the beams.
During the injection/extraction of the beams, normal energy exchange does not occur until the bunches fill the entire orbit, so the external RF system must work at full power.
To reduce the required RF system power, one must first inject bunches (or, better, short trains) with a large interval and then (at subsequent revolutions) add new trains between the trains that are already circulating.
The optimum distance between the bunches is $d=\lambda_\text{RF} \approx \SI{23}{\centi\meter}$ (for $f_\text{RF} = \SI{1.4}{\giga\hertz}$), which is too small for working with individual bunches, while the use of trains with gaps of \SIrange{1}{2}{\meter} makes it possible to manipulate trains using impulse deflectors.
The removal of the beams is done in reverse order.
\subsubsection{Luminosity restrictions due to collision effects}
\begin{figure}[htb]\centering
\includegraphics[width=.8\linewidth]{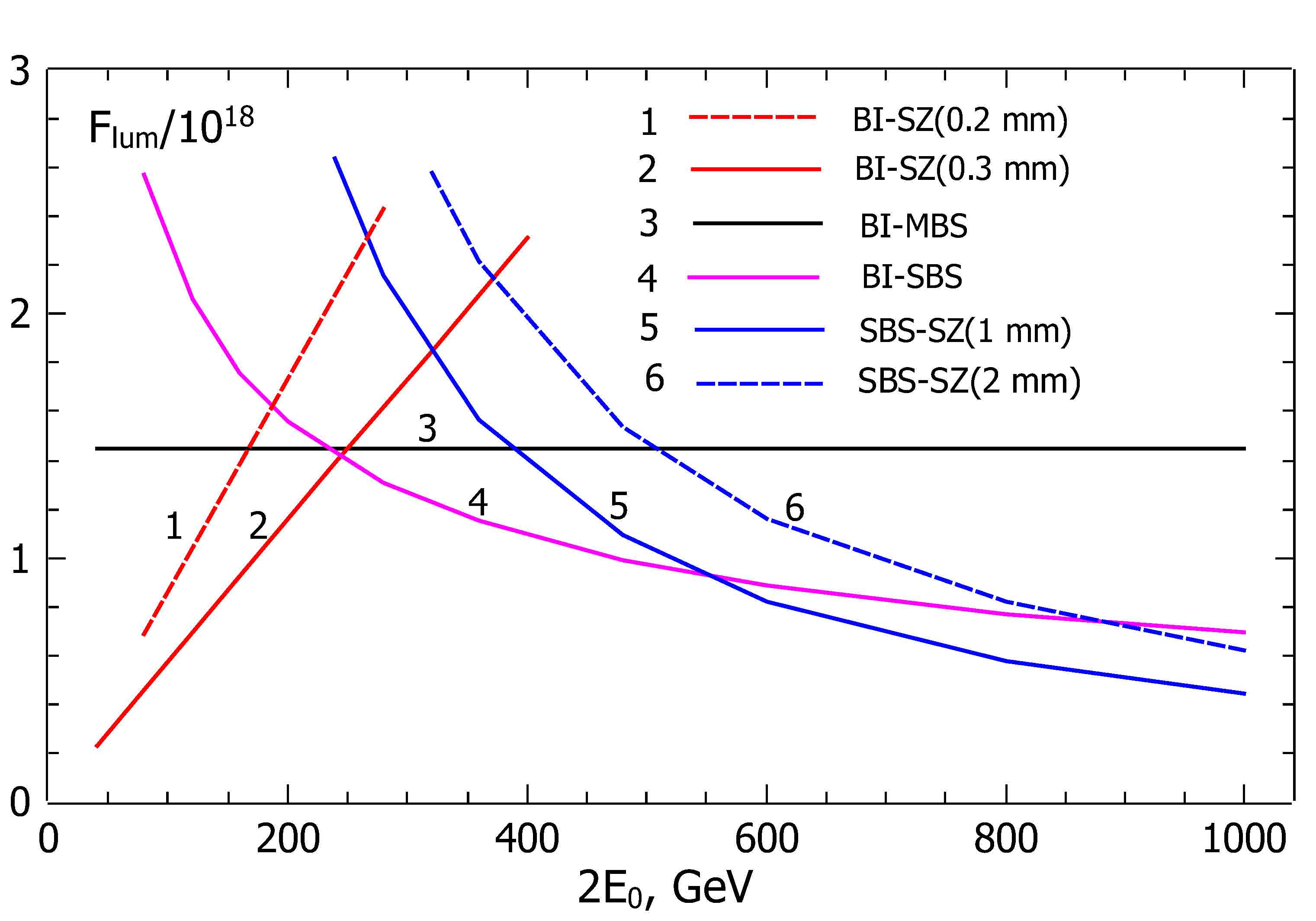}
\caption{Luminosity restrictions due to various collision effects (see text). The maximum luminosity is $\mathcal{L}=Nf\times F_\text{lum}$, where the limit is given by the lowest $F_\text{lum}$ at each energy.}
\label{fig:frontier:erlc:lum1234}
\end{figure}

Collision effects were considered in detail in \cite{Telnov_2021}; what follows is a short summary.
The beam parameters are determined by the following effects: (BI) Beam Instability (beam-beam parameter $\xi_y$); (MBS) Multiple Beamstrahlung ($\sigma_E$ at the IP); (SBS) Single Beamstrahlung (beam lifetime); (SZ) Bunch length, a) $\sigma_z> \sigma_{z, \text{min}}$, b) $\sigma_z < \sigma_{z, \text{max}}$.
With increasing energy, the luminosity is limited by the following combinations of constraints: BI-SZ(a) $\Rightarrow$ BI-MBS $\Rightarrow$ BI-SBS $\Rightarrow$ SBS-SZ(b).  The maximum luminosity due to these effects can be written as
\[ \mathcal{L}=(Nf)\,  F(E) \times D, \quad f=c/d, \quad F(E)= \min{F_i}(E), \]
where $D$ represents the duty cycle, $N$ the number of particles per bunch, and $c$ the speed of light.
The individual effects $F_i(E)$ are shown in Fig.~\ref{fig:frontier:erlc:lum1234}; which one of them is the most important depends on the energy.
For $d=\lambda_\text{RF}= \SI{23}{\centi\meter}$, $\mathcal{L}=1.3\times 10^{18} (N/10^9) F(E) \times D$.

For $2E_0 = \SIrange{250}{500}{\giga\electronvolt}$, the luminosity is limited by BI-SBS (beam instability and single beamstrahlung).
For $D=1$,  it is equal to~\cite{Telnov_2021}
\[ \mathcal{L} \approx \frac{0.58 Nf \xi^{1/3}}{\epsilon_{ny}^{1/3}r_e^{5/3}\gamma^{2/3}\Lambda^{2/3}}, \quad \Lambda= \ln{\frac{120}{(E_0/\si{\giga\electronvolt})/125}}\,. \]
For $\xi=0.1$ and $\epsilon_{ny}=\SI{3e-8}{\meter}$:
\[ \mathcal{L}\approx \num{4.15e35} \frac{(N/10^{9})}{d/\si{\meter}} \left(\frac{\ln{120}}{\ln{(120(125/(E_0/\si{\giga\electronvolt})))}}\right)^{2/3}\left(\frac{125}{E_0/\si{\giga\electronvolt}}\right)^{2/3}. \]

For $d=\SI{23}{\centi\meter}$ in CW mode, $\mathcal{L} = 1.81\times 10^{36}(N/10^9)\,\si{\per\square\centi\meter\per\second}$ at $2E_0 = \SI{250}{\giga\electronvolt}$ and $\mathcal{L} = 1.27\times 10^{36}(N/10^9)\,\si{\per\square\centi\meter\per\second}$ at $2E_0 = \SI{500}{\giga\electronvolt}$.
The average current is $I=0.21(N/10^9)\,\si{\ampere}$.
The optimum values of $N$ and $D$ are discussed below;  $N/10^9\sim\numrange{0.5}{2}$ for various types of SC cavities.
\subsubsection{RF losses in cavities}\label{sec:frontier:erlc:rf}

The main problem of SC linear accelerators operating in continuous mode, such as the one currently being developed  for the XFEL LCLS-II  at SLAC~\cite{LCLSII-Final,Raubenheimer:FLS2018-MOP1WA02}, is heat removal from the low-temperature SC cavities.
The energy dissipation in one (multi-cell) cavity is $P_\text{RF,dis}= V_\text{acc}^2/((R/Q) Q_0)$,
where $V_\text{acc}$ is the operating voltage, $R/Q$ is the fundamental-mode shunt impedance,  and $Q_0$ is the cavity quality factor.
The \SI{1.3}{\giga\hertz} TESLA--ILC  cavity has $R/Q=\SI{1036}{\ohm}$ and a length of \SI{1.04}{\meter}.
For an accelerating gradient $G=\SI{20}{\mega\electronvolt\per\meter}$ and $Q_0 =3\times 10^{10}$, the thermal power is $P_\text{dis}=\SI{13.5}{\watt\per\meter}$, or \SI{680}{\watt\per\giga\electronvolt}.
For the LCLS-II linac, $P_\text{RF,dis} \sim \SI{1}{\kilo\watt\per\giga\electronvolt}$ is expected.
Higher-order-mode (HOM) losses will further contribute to the dissipation of the cavities (see section~\ref{sec:frontier:erlc:hom}).

The overall heat transfer efficiency from temperature $T_2 \approx \SI{1.8}{\kelvin}$ to room temperature $T_1\sim \SI{300}{\kelvin}$ is $\eta=\varepsilon T_2/(T_1-T_2) \approx 0.18\times 1.8/300 = 1/900$~\cite{Posen17}.
The required refrigeration power for the twin \SI{250}{\giga\electronvolt} collider is
\[ P_\text{RF-refr, 2\,K}\approx 2 \times \SI{250}{\giga\electronvolt} \times 900 \times \SI{0.68}{\kilo\watt\per\giga\electronvolt} = \SI{306}{\mega\watt}, \]
accounting only for RF losses; under optimal conditions (see below), a similar amount of power is required for removal of HOM losses.

In recent years, great progress has been made both in increasing the maximum accelerating voltage and in increasing the quality factor $Q_0$. In the ILC project, it is assumed that $Q_0=10^{10}$ and $G=\SI{31.5}{\mega\electronvolt\per\meter}$.
For continuous operation, it is advantageous to work at $G \approx \SI{20}{\mega\electronvolt\per\meter}$, where $Q_0\sim 3\times 10^{10}$ is within reach now; one can hope for a reliable value of $Q_0=8\times 10^{10}$ at $T=\SI{1.8}{\kelvin}$~\cite{Padamsee_2017}.
An even more promising way to reduce the cooling power is to use superconductors with a higher operating temperature, such as \ce{Nb3Sn}~\cite{Posen17}.
Then, at $T_2 = \SI{4.5}{\kelvin}$, the technical efficiency of heat removal is about \SI{30}{\percent}, and the total efficiency (with Carnot) about $\tfrac{1}{220}$, i.e., about 4 times higher than that at \SI{1.8}{\kelvin}.

\subsubsection{Higher-order-mode losses (HOM)} \label{sec:frontier:erlc:hom}

Bunches traveling in a linac lose energy to higher-order modes (HOM).
The energy lost by one electron per unit length is almost independent of the distance between diaphragms and bunch length and is given by the simple formula
\begin{equation}
\frac{\mathrm{d}E}{\mathrm{d}z} \approx -2e^2N/r_a^2,
\end{equation}
where $r_a$ is the inner radius of the diaphragms~\cite{Palmer-Wakefield}.
This equation is supported by detailed numerical calculations~\cite{Novokhatsky}.
For TESLA--ILC accelerating structures ($r_a=\SI{3.5}{\centi\meter}$), $-\mathrm{d}E/\mathrm{d}z\approx 2.2(N/10^{9})\,\si{\kilo\electronvolt\per\meter}$; for $N=10^{9}$, these HOM losses are  $\sim \SI{0.01}{\percent}$ of the accelerating gradient $G\sim \SIrange{20}{30}{\mega\electronvolt\per\meter}$.
For $2E=\SI{250}{\giga\electronvolt}$ and $G=\SI{20}{\mega\electronvolt\per\meter}$, the total power of HOM energy losses (twin collider, both beams) is
\[ P_\text{HOM}=2.65 (N/10^{9})^2/{(d/\si{\meter})}\,\si{\mega\watt}. \]
For $N=10^{9}$ and $d=\SI{0.23}{\meter}$, $P_\text{HOM}=\SI{11.5}{\mega\watt}$, which is close to the synchrotron radiation power in the damping wigglers, equal to \SI{10.4}{\mega\watt}.
These numbers are for continuous operation.
For $N=10^9$, this power is about 35 times greater than the RF power dissipation in the cavities.
Fortunately, most of this energy can be extracted from the SC cavities in  two ways:
\begin{itemize}
\item using HOM couplers which dissipate the energy at room temperature;
\item with the help of special HOM absorbers located between the cavities. The latter are maintained at an intermediate temperature around \SI{80}{\kelvin} where refrigeration systems operate at much higher efficiencies.
\end{itemize}
However, some small part (\SI{1.5}{\percent} assumed in \cite{Telnov_2021}) of the HOM energy is dissipated in the walls of the SC cavities.

With this assumption (see details in \cite{Telnov_2021}), the refrigeration power needed for removal of HOM losses dissipated in HOM absorbers at a liquid nitrogen temperature of \SI{77}{\kelvin} is $P_\text{HOM,refr,77\,K} \approx 10P_\text{HOM}$.
The refrigeration power for removal of HOM heat at \SI{2}{\kelvin} is $P_\text{HOM,refr,2\,K} \approx 9 P_\text{HOM}$.
The electric power needed for compensation of beam energy losses (assuming \SI{50}{\percent} efficiency) is about $2P_\text{HOM}$.
The total electric power consumption from the wall plug (w.p.)~due to HOM losses for a collider with $2E_0=\SI{250}{\giga\electronvolt}$ in continuous mode of operation is $P_\text{HOM, w.p.} \approx 21 P_\text{HOM} \approx 240 (N/10^9)^2\,\si{\mega\watt}$.

\subsubsection{Pulse duration in duty-cycle mode}
The duration of continuous operation is determined by the heat capacity $c_p$ of the liquid He that surrounds the cavity and can be estimated as
$\Delta t= c_p m \Delta T/P_\text{dis} \sim \SI{20}{\second}$,
where $c_p(\ce{He})= \SI{2.8}{\joule\per\gram\per\kelvin}$ at $T=\SI{1.8}{\kelvin}$, $m$ is the mass of liquid He per one TESLA cavity (we take \SI{0.02}{\cubic\meter} or \SI{2.9}{\kilo\gram}), $P_\text{dis}\sim \SI{20}{\watt}$, $\Delta T \sim \SI{0.05}{\kelvin}$.
$\Delta t = \SI{10}{\second}$ is therefore a safe choice.

More promising for the ERLC are superconductors working at $T \approx \SI{4.5}{\kelvin}$ (temperature of boiling He).
Given that the heat capacity per unit volume, $\rho c_p$ ($\rho$ denoting mass density), for this temperature is 1.6 times larger and $\Delta T$ can be larger by a factor of 3, $\Delta t = \SI{20}{\second} \times 1.6 \times 3 = \SI{96}{\second}$.
In this case, the active cycle time  can last about \SI{1}{\minute}.

The duration of the break is described by the duty cycle $D$, which depends on the available power; the optimization is given below.

\subsubsection{Optimum values of $N$,  $d$,  and  $D$}

There are three main energy consumers (numbers correspond to $2E_0=\SI{250}{\giga\electronvolt}$, Nb cavities, \SI{1.3}{\giga\hertz}, $T = \SI{1.8}{\kelvin}$, $d=\SI{23}{\centi\meter}$):
\begin{enumerate}
\item radiation in the damping wigglers: $P_\text{SR}/\varepsilon=D \times 20.8(N/10^{9})$\,MW at $\varepsilon=\SI{50}{\percent}$;
\item power for cooling of the RF losses in cavities (at \SI{2}{\kelvin}): $P_\text{RF-refr, 2\,K} = D \times \SI{305}{\mega\watt}$;
\item power due to HOM losses $P_\text{HOM, w.p.} \approx D \times 240(N/10^9)^2$\,MW.
\end{enumerate}
The total power (only main contributions) amounts to
\[ P_\text{tot}= D \left(20.8(N/10^{9})+ 305+240(N/10^{9})^2 \right) \,\si{\mega\watt}. \]
Most of this power is spent on removal of RF and HOM losses.

The dependence of the luminosity on the bunch distance $d$ allows one to find the optimum values of $N$ and $D$.
Neglecting power losses in wigglers, the power in operation with a duty cycle $D$ is
\[ P= \left(a + b N^2/d\right)  D, \]
with $N$ in units of $10^9$; the coefficients  $a$ and $b$ describe RF and HOM losses, respectively, and are both proportional to the collider length (or $E_0$).
The luminosity is
\[ \mathcal{L} \propto (N/d) D = P(N/d)/(a+ bN^2/d). \]
The maximum luminosity $\mathcal{L}\propto P/\sqrt{abd}$  is reached at $N=\sqrt{ad/b}, \; D=P/2a$.
The luminosity reaches its maximum when the amounts of energy spent for removal of RF and HOM losses are equal (only valid for $D<1$).
We see that $\mathcal{L}\propto 1/\sqrt{d}$, so the distance between bunches $d$ should be as small as possible.
The coefficient $a \propto 1/(\epsilon Q_0 T)$, where $T$ is the temperature of the SC cavities, $\epsilon$ is the technical efficiency (0.18 at $T=\SI{1.8}{\kelvin}$, 0.3 at $T > \SI{4}{\kelvin}$); therefore, $\mathcal{L} \propto \sqrt{\epsilon Q_0 T}$.

The optimum number of particles in the bunch does not depend on $P$ or beam energy (because both $a$ and $b$ are proportional to $E$).
According to the above power estimates for $2E_0=\SI{250}{\giga\electronvolt}$,  $a=\SI{305}{\mega\watt}$ and $b/d=240/(10^9)^2\,\si{\mega\watt}$; therefore, the optimum number of particles per bunch is $N/10^9 \approx \sqrt{305/240}=1.13$.

For CW operation, the consideration is similar:
Here, $P= (a + bN^2/d), \; \mathcal{L} \propto N/d$, which gives
\[ N=\sqrt{(P-a)d/b}, \quad \mathcal{L}\propto \sqrt{(P-a)/bd}. \]
Again, $\mathcal{L}\propto 1/\sqrt{d}$.
The minimum power for CW operation is $P=a \propto 1/(\epsilon Q_0 T)$.
The number of particles per bunch in CW mode depends on the available power: $N/10^9 \approx \sqrt{(P-a)d/b} \sim \sqrt{(P(125/E_0) -305)/240}$.
The luminosity in CW mode is proportional to $\sqrt{P-a}$; at $P=2a$, it becomes equal to the maximum luminosity with a duty cycle.
It makes sense to work at $P$ exceeding the threshold power only by about \SI{35}{\percent} when $\mathcal{L}_\text{CW}/\mathcal{L}_\text{DC, max}=0.85$.
In this case,  $N/10^9 \sim 0.67$, and the power required for the Higgs factory in CW mode is $P=\SI{410}{\mega\watt}$, which is too much.

Luminosities calculated numerically (accounting for the neglected SR term) are shown in Fig.~\ref{fig:frontier:erlc:luminosity} and Table~\ref{tab:frontier:erlc:Table1}. For the Higgs factory with $2E=\SI{250}{\giga\electronvolt}$, there are, for example, two options:
\begin{itemize}
\item duty-cycle mode, $P=\SI{120}{\mega\watt}$, $\mathcal{L}=\SI{0.39e36}{\per\square\centi\meter\per\second}$, $D \approx 0.19$;
\item continuous mode, $P=\SI{410}{\mega\watt}$, $\mathcal{L}=\SI{1.13e36}{\per\square\centi\meter\per\second}$.
\end{itemize}
The continuous mode is attractive, but the required power is too high.
The above numbers are for $Q_0=3\times 10^{10}$ and $T=\SI{1.8}{\kelvin}$.
If $T=\SI{4.5}{\kelvin}$ (\ce{Nb3Sn} or other) and $Q_0$ is the same, then $\epsilon T$ is 4 times larger, making $\mathcal{L}=\SI{0.6e36}{\per\square\centi\meter\per\second}$ available in CW mode at $P=\SI{100}{\mega\watt}$ instead of \SI{410}{\mega\watt}.

\subsubsection{Ways to reduce power consumption}

The largest part of the power is spent on heat removal from RF and HOM losses; it can be reduced in two ways:
\begin{itemize}
\item by using a superconductor with a higher operating temperature;
\item by decreasing the RF frequency $f \equiv f_\text{RF}$ (in this section).
\end{itemize}
A promising material is \ce{Nb3Sn}, which operates at a temperature of \SI{4.5}{\kelvin}, where the efficiency of heat removal is about 4 times higher than for Nb at $T = \SI{1.8}{\kelvin}$~\cite{Posen17}.
Its thermal conductivity is about 1000 times lower than that of niobium, so it is used in the form of a thin film on a material with high thermal conductivity, such as niobium or copper.
Cavities with \ce{Nb3Sn} reach the same high $Q_0$ values as those made of pure Nb, although the technology is not reliable enough yet.
As for Nb, the value of $Q_0 \propto 1/R_\text{s}$ is limited by the BCS surface conductivity $R_\text{BCS}\propto f^2$; therefore, it is advantageous to lower the RF frequency.

The transition from $T = \SI{1.8}{\kelvin}$ to $T = \SI{4.5}{\kelvin}$ increases the efficiency of heat removal by a factor of $\epsilon T = 4$.
The RF power loss in cavities per unit length scales as $P_\text{RF} \propto R_\text{s} f^{-1} \propto f$ (if $R_\text{s}=R_\text{BCS}$).
In addition, a decrease of $f$ leads to a decrease in HOM losses (per unit length): $P_\text{HOM} \propto 1/r_a^2 \propto f^2$.
The minimum distance between bunches is $d \propto 1/f$.
As a result, $a \propto f/(\epsilon T),\, b\propto f^2,\, d \propto 1/f$.
The luminosity for duty-cycle operation is
\[ \mathcal{L} \propto P/\sqrt{abd} \propto P\sqrt{\epsilon T}/f, \quad \mathcal{L}/P\propto \sqrt{\epsilon T}/f. \]
With $\epsilon T = 4$ and a 2-fold decrease of $f$, we obtain a 4-fold increase of $\mathcal{L}$ at the same power.

For continuous operation, $\mathcal{L}\propto \sqrt{(P-a)/(bd)}$; the threshold power is $P\propto a \propto f/(\epsilon T)$.
With $\epsilon T=4$ and a 2-fold decrease of $f$, this power decreases by a factor of 8.
The required power for CW operation at \SI{250}{\giga\electronvolt}, which was earlier stated as \SI{410}{\mega\watt}, will therefore decrease to \SI{50}{\mega\watt}, which is already acceptable.
In CW mode, $\mathcal{L} \propto \sqrt{a/bd} \propto 1/\sqrt{\epsilon T}$ at $P \sim a \propto f/(\epsilon T)$, which gives $\mathcal{L}/P= \sqrt{\epsilon T}/f$,  the same gain as in duty-cycle mode.

Thus, a transition from $T=\SI{1.8}{\kelvin}$ to $T = \SI{4.5}{\kelvin}$ and halving the RF frequency increases the luminosity by a factor of 4 and decreases the threshold power for continuous operation by a factor of 8.
This is in the ideal case when the surface conductivity  $\sigma_\text{s} \approx \sigma_\text{BCS}$.
Such collider variants are also presented in the summary tables and figures.

\subsubsection{From 250 to 500 GeV}

Above, the main focus was on the high-luminosity \SI{250}{\giga\electronvolt} Higgs factory.
However, there is also great interest in higher energies: $2E_0=\SI{360}{\giga\electronvolt}$ (the top-quark threshold), or $2E_0=\SI{500}{\giga\electronvolt}$ (the Higgs self-coupling).

Beside collision effects, there is also the problem of emittance dilution in the beam delivery system where horizontal dispersion is required for chromatic correction at the final focus.
In the ERLC, the beams pass this region about 400 times during the damping time.
Some increase in length will be needed to solve this problem.

\begin{figure}[htb]\centering
\includegraphics[width=.6\linewidth]{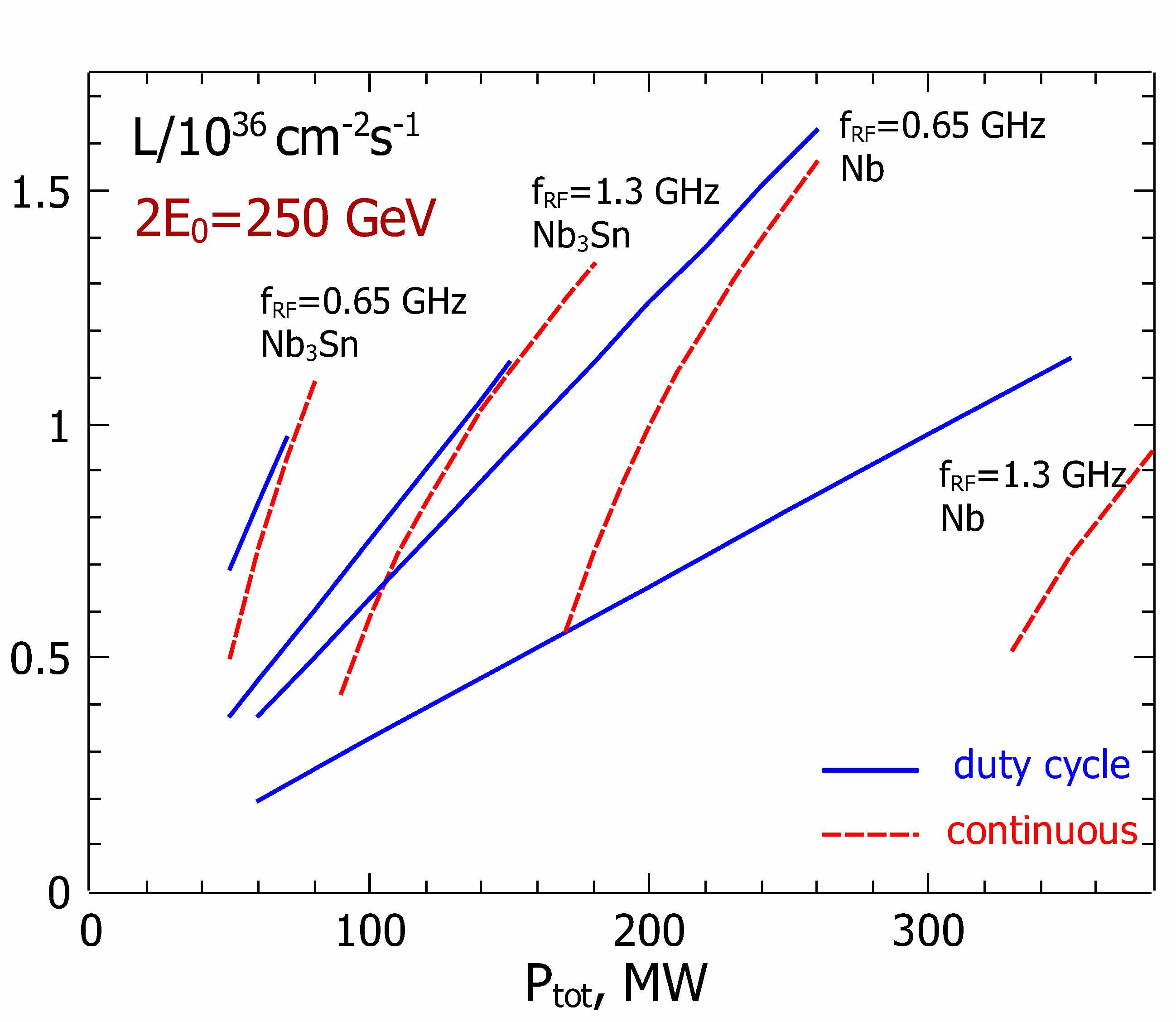}\\
\includegraphics[width=.6\linewidth]{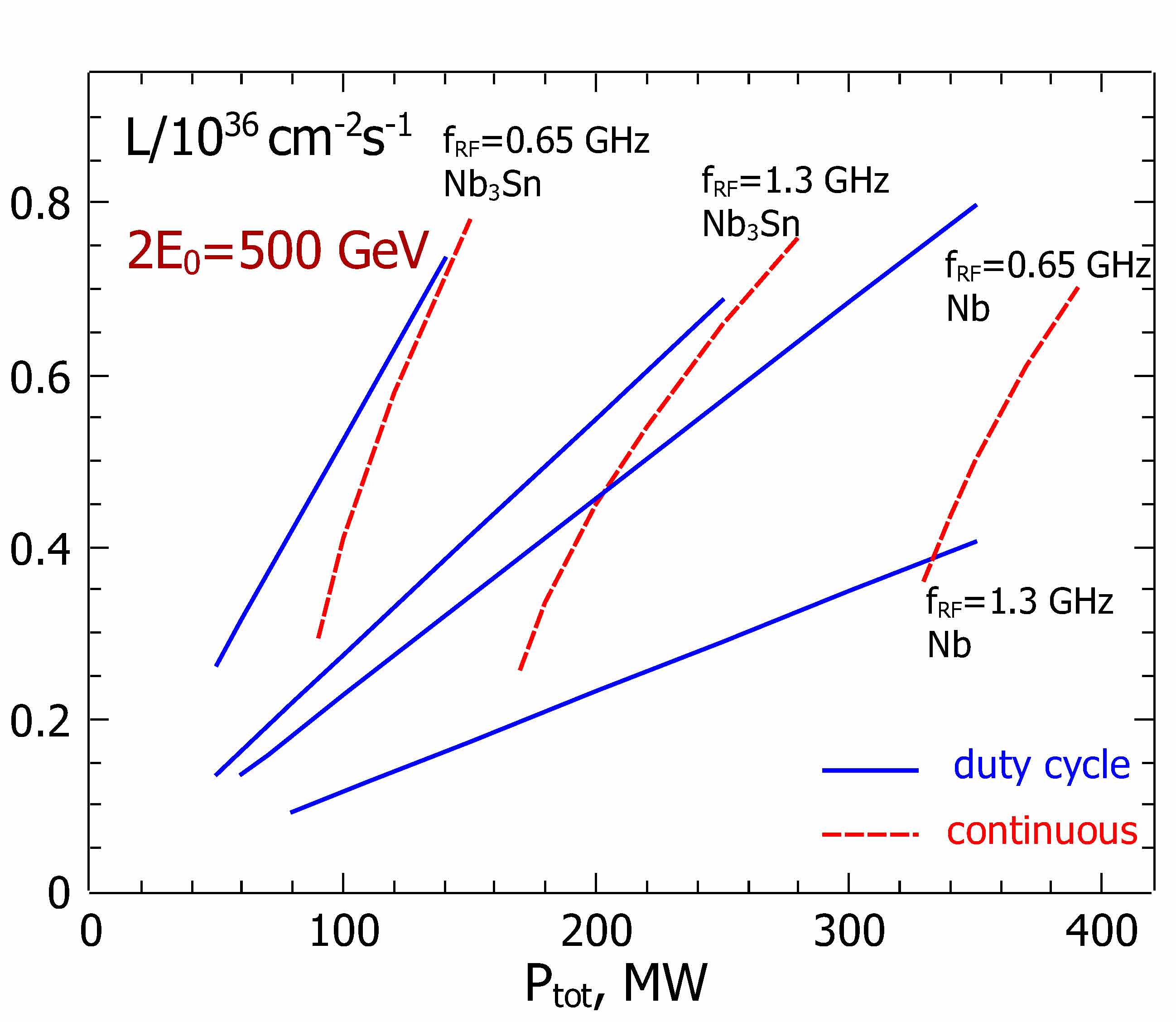}
\caption{Dependence of the luminosity on the total power for $2E_0=\SI{250}{\giga\electronvolt}$ (top) and $2E_0=\SI{500}{\giga\electronvolt}$ (bottom); blue (solid) line for optimum duty cycle operation, red (dashed) curves for continuous (CW) operation.}
\label{fig:frontier:erlc:luminosity}
\end{figure}

Continuous operation requires twice as much threshold power as at $2E_0=\SI{250}{\giga\electronvolt}$; this mode will be realistic in case of success with \ce{Nb3Sn} cavities.
The expected luminosities at $2E_0=\SI{500}{\giga\electronvolt}$ are given in Table~\ref{tab:frontier:erlc:Table2} and Fig.~\ref{fig:frontier:erlc:luminosity}; they are about 3 times lower than at $2E_0=\SI{250}{\giga\electronvolt}$ for the same total powers.

\clearpage

\subsubsection{Summary tables}
The article on the novel linear collider scheme~\cite{Telnov_2021} considered only the most important problems affecting luminosity;
there are many other issues that require careful consideration by accelerator experts.
The preliminary parameters of the collider with an energy of $2E_0 = \SI{250}{\giga\electronvolt}$ and \SI{500}{\giga\electronvolt}, respectively,  are presented in Tables~\ref{tab:frontier:erlc:Table1} and \ref{tab:frontier:erlc:Table2}.
Each table contains four ERLC options and the ILC\footnote{Note that for the ILC, the total power is given in the tables. The linac itself consumes about $\tfrac{1}{3}$ at $2E_0=\SI{250}{\giga\electronvolt}$ (sources and damping rings contribute the largest part) and $\tfrac{2}{3}$ at $2E_0=\SI{500}{\giga\electronvolt}$.}.
The dependence of the luminosities on the total power for various options is shown in Fig.~\ref{fig:frontier:erlc:luminosity}.
Table~\ref{tab:frontier:erlc:Table3} lists the contributions of the main energy consumers for two ERLC options at $2E_0=\SI{250}{\giga\electronvolt}$.

\begin{table}[htb]\centering\scriptsize
\caption{Parameters of ERLC and ILC, $2E_0=\SI{250}{\giga\electronvolt}$.}
\begin{tabular}{ l  l  c  c  c  c  c }
\toprule
                        & unit           &                      ERLC  & ERLC     &ERLC     & ERLC    &ILC \\
Beam mode               &                  &                   pulsed & pulsed   & contin. & contin.  &     \\
Cavity material         &                  &                   Nb     & Nb      & \ce{Nb3Sn} & \ce{Nb3Sn} & Nb   \\
Cavity temperature      & K                 &                  1.8  & 1.8   & 4.5    &  4.5  & 1.8    \\
RF frequency            & GHz                 &                   1.3 & 0.65 & 1.3   &  0.65 & 1.3  \\
\midrule
Energy $2E_0$           &GeV             &                      250   &  250     & 250     &250      &250      \\
Luminosity $\mathcal{L}_\text{tot}$ &\SI{e36}{\per\square\centi\meter\per\second} &         0.39   &  0.75    & 0.83    & 1.6     & 0.0135\\
$P$ (wall) (collider)   &MW              &                    120     &120       & 120     & 120     & 129 (tot.)\\
Duty cycle, $D$        &                &                     0.19  &  0.37    &   1     &  1      & n/a\\
Accel.~gradient, $G$    &MV/m            &          20                & 20       &20       &20       & 31.5 \\
Cavity quality, $Q$     & $ 10^{10}$     &                    3       &  12      & 3       & 12      & 1     \\
Length $L_\text{act}/L_\text{tot}$ &km     &                     12.5/30&12.5/30   & 12.5/30 & 12.5/30 & 8/20 \\
$N$ per bunch           &$10^{9}$       &                     1.13    & 2.26     & 0.46    & 1.77    & 20\\
Bunch distance          & m              &                   0.23     &0.46      & 0.23    & 0.46   & 166 \\
Rep.~rate, $f$          & Hz             &          \num{2.47e8} & \num{2.37e8} & \num{1.3e9} & \num{6.5e8} & 6560 \\
$\epsilon_{x,\,n}$/$\epsilon_{y,\,n}$ & \si{\micro\meter} &        10/0.035 &10/0.035   & 10/0.035 &10/0.035  & 5/0.035 \\
$\beta^*_x$/$\beta_y$ at IP           & cm    &        2.7/0.031     & 10.8/0.031 &0.46/0.031&6.8/0.031 & 1.3/0.04 \\
$\sigma_x$ at IP       &  \si{\micro\meter}        &                     1.05    &2.1      &0.43     &1.66     & 0.52\\
$\sigma_y$ at IP       & nm             &                   6.2       & 6.2      &6.2      &6.2      & 7.7\\
$\sigma_z$ at IP       & cm             &                   0.03      &0.03      &0.03     &0.03     & 0.03\\
$ (\sigma_E/E_0)_\text{BS}$ at IP  & $\%$        &             0.2      &0.2       &0.2      &0.2      & $\sim 1$ \\
\bottomrule
%\hline
\end{tabular}
\label{tab:frontier:erlc:Table1}
\end{table}

\begin{table}[htb]\centering\scriptsize
\caption{Parameters of ERLC and ILC, $2E_0=\SI{500}{\giga\electronvolt}$.}
\begin{tabular}{ l  l  c  c  c  c  c }
\toprule
                        & unit           &                      ERLC  & ERLC     &ERLC     & ERLC    &ILC \\
Beam mode               &                  &                   pulsed & pulsed   & pulsed  & contin.  &     \\
Cavity material         &                  &                   Nb     & Nb      & \ce{Nb3Sn} & \ce{Nb3Sn} & Nb   \\
Cavity temperature      & K                 &                   1.8  & 1.8   & 4.5     &  4.5  & 1.8     \\
RF frequency            & GHz                 &                   1.3 & 0.65 & 1.3   &  0.65 & 1.3  \\
\midrule
Energy $2E_0$           &GeV             &                      500   &  500     & 500     & 500      &500      \\
Luminosity $\mathcal{L}_\text{tot}$ & \SI{e36}{\per\square\centi\meter\per\second} &         0.174   &  0.342    & 0.412    & 0.78    & 0.018\\
$P$ (wall) (collider)   &MW              &                    150     &150       & 150     & 150     & 163 (tot.) \\
Duty cycle, $D$        &                &                     0.121  &  0.237   &  0.47   &  1      & n/a\\
Accel. gradient, $G$    &MV/m            &          20                & 20       &20       &20       & 31.5 \\
Cavity quality, $Q$     & $ 10^{10}$     &                    3       &  12      & 3       & 12      & 1     \\
Length $L_\text{act}/L_\text{tot}$ &km     &                     25/50  & 25/50    & 25/50   & 25/50   & 16/31 \\
$N$ per bunch           &$10^{9}$       &                     1.13    & 2.26     & 0.685    & 1.23    & 20\\
Bunch distance          & m              &                   0.23     &0.46      & 0.23     &  0.46   & 166 \\
Rep. rate, $f$          & Hz             &           \num{1.57e8} & \num{1.54e8} & \num{6.1e8} & \num{6.5e8} & 6560 \\
$\epsilon_{x,\,n}$/$\epsilon_{y,\,n}$    &   \si{\micro\meter} &  $10/0.035$  &$10/0.035$  & $10/0.035$ & $10/0.035$  & $10/0.035$ \\
$\beta^*_x$/$\beta_y$  at IP      & cm               &     $7.7/0.089$ & $31/0.089$ &$2.85/0.089$ &$9.4/0.089$ & $1.1/0.04$ \\
$\sigma_x$ at IP       & \si{\micro\meter}        &                     1.26    &2.5      &0.76     &1.38     & 0.47\\
$\sigma_y$ at IP       & nm             &                   7.4       & 7.4      &7.4      &7.4      & 5.9\\
$\sigma_z$ at IP       & cm             &                   0.089      &0.089    &0.089    &0.089    & 0.03\\
$ (\sigma_E/E_0)_\text{BS}$ at IP  & $\%$        &             0.1      &0.1       &0.1      &0.1      & $\sim 1$ \\
\bottomrule
\end{tabular}
\label{tab:frontier:erlc:Table2}
\end{table}

\begin{table}[htb]\centering\scriptsize
\caption{Power consumption of two ERLC options at $2E_0=\SI{250}{\giga\electronvolt}$, as presented in the first and fourth columns of Table~\ref{tab:frontier:erlc:Table2}. All numbers in MW.}
\begin{tabular}{ l   c  c }
\toprule
                        & Nb, \SI{1.3}{\giga\hertz}, $T=\SI{1.8}{\kelvin}$                   & \ce{Nb3Sn}, \SI{0.65}{\giga\hertz}, $T=\SI{4.5}{\kelvin}$         \\
                        & $\mathcal{L}=\SI{0.39e36}{\per\square\centi\meter\per\second}$     & $\mathcal{L}=\SI{1.6e36}{\per\square\centi\meter\per\second}$      \\
                        & $N=\num{1.13e9}$, $D=0.19$           & $N=\num{1.77e9}$, $D=1$   \\
\midrule
Beam generation         & small  & small  \\
Radiation in wigglers   & 4.45                                      & 18.4  \\
HOMs, beam energy       & 5.5                                     & 9  \\
HOM cooling at \SI{1.8}{\kelvin} / \SI{4.5}{\kelvin}     & 24.8                                    & 10  \\
HOM cooling at \SI{77}{\kelvin}      & 27.6                                    & 44.7   \\
RF diss.~cooling at \SI{1.8}{\kelvin} / \SI{4.5}{\kelvin}   & 57.6                                    & 38  \\ \hline
$P_\text{tot}$           & 120                                     & 120  \\
\bottomrule
\end{tabular}
\label{tab:frontier:erlc:Table3}
\end{table}

\subsubsection{Conclusion}
 
Superconducting technology makes it possible to build a high-energy linear \positron\electron{} collider (LC) with energy recovery and reusable beams.
The problem of parasitic collisions inside the linacs can be solved using a twin (dual) LC.
The achievable luminosity is limited by collision effects and available power.
Such a collider can operate in duty-cycle or continuous modes depending on available power.
With current SC Nb technology ($T=\SI{1.8}{\kelvin}$, $f_\text{RF}=\SI{1.3}{\giga\hertz}$, used for ILC) and with a power of $P= \SI{100}{\mega\watt}$, a luminosity $\mathcal{L} \sim \SI{0.33e36}{\per\square\centi\meter\per\second}$ is possible at the Higgs factory with $2E_0=\SI{250}{\giga\electronvolt}$.
Using superconductors operating at \SI{4.5}{\kelvin} with high $Q_0$ values, such as \ce{Nb3Sn}, and $f_\text{RF}=\SI{0.65}{\giga\hertz}$, the luminosity can reach  $\mathcal{L} \sim \SI{1.4e36}{\per\square\centi\meter\per\second}$ at $2E_0=\SI{250}{\giga\electronvolt}$ (with $P=\SI{100}{\mega\watt}$) and  $\mathcal{L} \sim \SI{0.8e36}{\per\square\centi\meter\per\second}$ at $2E_0=\SI{500}{\giga\electronvolt}$ (with $P=\SI{150}{\mega\watt}$), which is almost two orders of magnitude greater than at the ILC, where the beams are used only once.
This technology requires additional efforts to obtain the required parameters and reliable operation.
Such a collider would be the best machine for precision Higgs studies, including the measurement of Higgs self-coupling.

%\subsection{ILC as an ERL}
%Valery Telnov, Andrew Hutton

\subsection{Photon-Photon Collider}\label{sec:frontier:high_energy_colliders:photon_photon}
%Frank Zimmermann, Atoosa Meseck 

\begin{figure}[htbp]\centering
\includegraphics[width=0.75\textwidth]{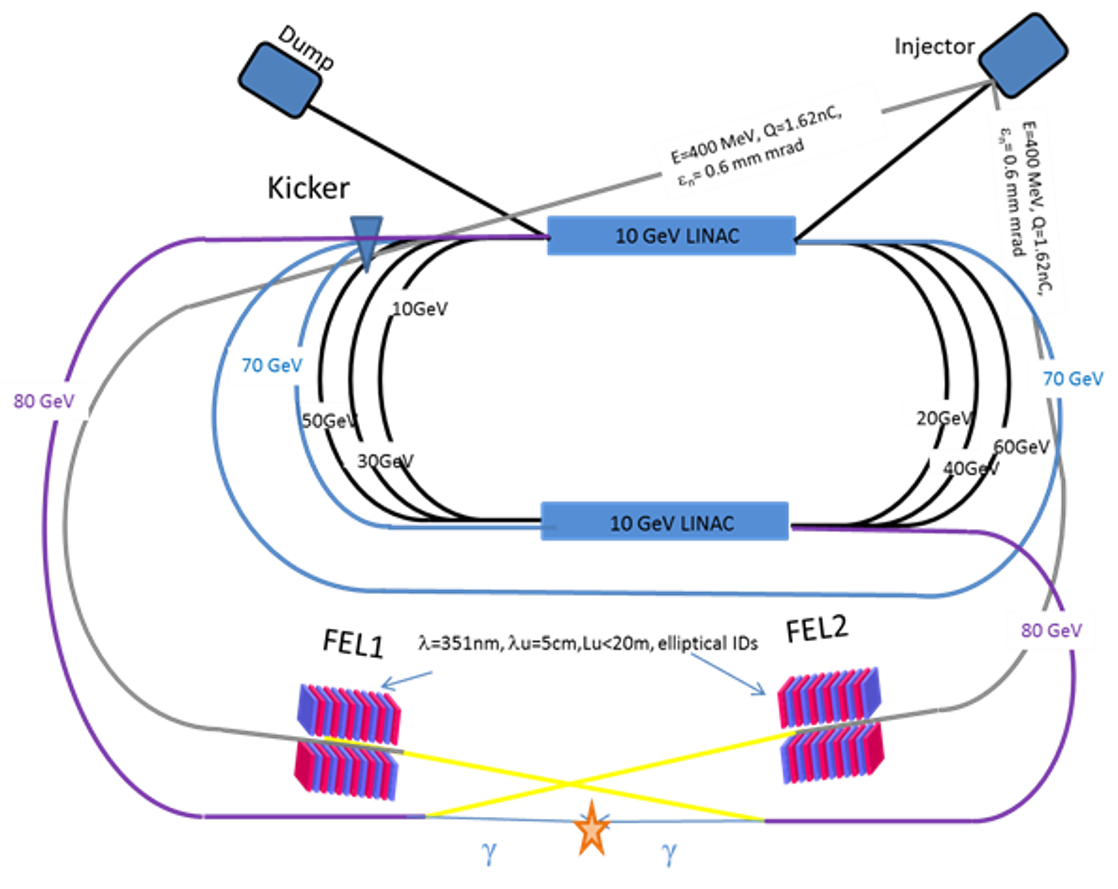}
\caption{Sketch of a layout for a \photon\photon~collider Higgs factory, 
\enquote{SAPPHiRE,} based on the LHeC recirculating SC linacs with Compton scattering of \SI{80}{\giga\electronvolt} electron bunches off \SI{3.5}{\electronvolt} photon beams; 
FEL-based configuration with fast kicker 
proposed by A.~Meseck~\cite{atoosa}.}
\label{fig:frontier:high_energy_colliders:photon_photon:lhec}
\end{figure}

A dedicated \photon\photon~Higgs factory, called \enquote{SAPPHiRE}~\cite{bogacz2012sapphire},  
could be realized by slightly reconfiguring the LHeC recirculating linacs, 
which would, in this case, be operated without energy recovery as the electrons
are consumed by Compton scattering off either a high-power laser or an FEL photon beam.   
The standard LHeC employs a pair of recirculating linacs 
capable of increasing the \electron~energy by $\sim\SI{10}{\giga\electronvolt}$ in each pass. 
The \photon\photon~Higgs factory would require an electron beam energy  
of $\sim\SI{125}{\giga\electronvolt}/0.8/2 \sim \SI{80}{\giga\electronvolt}$, where the
factor 2 arises from the centre-of-mass energy for two colliding beams, 
and the factor 0.8 approximates the 
peak of the \photon\photon~luminosity energy spectrum 
as fraction of the \electron\electron~energy, considering typical
Compton backscattering parameters.
In SAPPHiRE, the required electron energy could be achieved 
via four passes through two superconducting   
recirculating linacs, as is illustrated in Fig.~\ref{fig:frontier:high_energy_colliders:photon_photon:lhec}. 
For the configuration presented, compared to the LHeC,
two additional arcs are required on either side, 
corresponding to beam energies of 70 and \SI{80}{\giga\electronvolt}, 
respectively.  A fast kicker is used at \SI{70}{\giga\electronvolt}
to by-pass one of the linacs for  half of the bunches~\cite{atoosa}, 
which avoids the need to circulate bunches in both directions, 
as had been foreseen in the original SAPPHiRE proposal.  
The \photon\photon~(\electron\electron)
collision point is located between the two \SI{80}{\giga\electronvolt} arcs.   
The FEL \SI{3.5}{\electronvolt} photon source can be  
driven by separate low-energy beams, provided by 
the same injector.
Each FEL line delivers more than \num{e16} photons 
per pulse. Strongly focused beams are mandatory 
for inverse Compton scattering, with rms beam sizes 
in the order of \SI{300}{\nano\meter}.
Seeding and frequency up-conversion 
would be possible FEL operation modes. 

The additional high-energy 
arcs, for beam energies of 70 and \SI{80}{\giga\electronvolt}, respectively,  
can be placed inside the \enquote{existing} LHeC ERL arc tunnel,
resulting in a total energy loss from synchrotron radiation over all 8 arcs 
of \SI{3.9}{\giga\electronvolt} (about \SI{5}{\percent} of the final beam energy), which is considered acceptable.  
Alternatively, for SAPPHiRE, the LHeC linacs could be 
operated in pulsed mode at a 
\SI{33}{\percent} higher cavity gradient of \SI{26.7}{\mega\volt\per\meter}
to reach an electron energy of \SI{80}{\giga\electronvolt} in 3 passes without the need for 
additional arcs.   
Table~\ref{tab:frontier:high_energy_colliders:photon_photon:gamma} compiles a list of example parameters, 
which would meet the SAPPHiRE luminosity target of 
${\cal L}_{\photon\photon} \sim \SI{6e32}{\per\square\centi\meter\per\second}$ above \SI{125}{\giga\electronvolt} 
(or, equivalently, ${\cal L}_{\electron\electron} \sim \SI{2e34}{\per\square\centi\meter\per\second}$).
The Compton IR with an integrated optical cavity and the production of
the required photon beam using a laser or FEL still require R\&D effort.
Also, the fast kicker system needs detailed 
studies, as does the removal of the spent electron 
beams.

\begin{table}[htbp]\centering
\caption{Example parameters for a \photon\photon~collider Higgs factory, \enquote{SAPPHiRE,} based on the LHeC~\cite{Zimmermann:2013aga}.} 
\label{tab:frontier:high_energy_colliders:photon_photon:gamma}
\begin{tabular}{lcc}
\toprule
parameter & symbol & value \\
\midrule
total electric power &
$P$ &  \SI{200}{\mega\watt} \\
beam energy & $E$ &  \SI{80}{\giga\electronvolt} \\
beam polarization & $P_\text{e}$ & 0.80 \\
bunch population & $N$ &  $10^{10}$ \\
% Number of bunches per train & $n_{b}$ & (cw) \\
% Number of trains per rf pulse & $n_{t}$ & $2-11$ &  --- \\
repetition rate & $f_\text{rep}$ &  cw \\
bunch frequency & $f_\text{bunch}$ & \SI{200}{\kilo\hertz} \\ 
average beam current & $I_\text{beam}$ & \SI{0.32}{\milli\ampere} \\
rms bunch length & $\sigma_{z}$ & \SI{30}{\micro\meter} \\
crossing angle & $\theta_\text{c}$ &  $\ge \SI{20}{\milli\radian}$ \\
horizontal emittance & 
$\gamma \epsilon_{x}$  & \SI{5}{\micro\meter} \\
vertical emittance & 
$\gamma \epsilon_{y}$ & \SI{0.5}{\micro\meter} \\
horizontal interaction-point beta function &
$\beta_{x}^{\ast}$ & 5\,mm \\
vertical interaction-point beta function  &
$\beta_{y}^{\ast}$ & 0.1\,mm \\
rms horizontal~interaction-point spot size &
$\sigma_{x}^{\ast}$ &  400\,nm \\
rms vertical interaction-point spot size &
$\sigma_{y}^{\ast}$ &  18\,nm \\
rms horizontal~conversion-point spot size &
$\sigma_{x}^{C, \ast}$ & 400\,nm \\
rms vertical conversion-point spot size &
$\sigma_{y}^{C, \ast}$  & 180\,nm \\
\electron\electron~geometric luminosity & ${\cal L}$ & \SI{2e34}{\per\square\centi\meter\per\second} \\
\bottomrule
\end{tabular}
\end{table}

%\subsection{Photon-Photon Collider}
%Frank Zimmermann

\subsection{Electrons and X-rays to Muon Pairs (EXMP)}\label{sec:frontier:high_energy_colliders:exmp}

%Camilla Curatolo, Luca Serafini

% MB 07/20/2021
% Rough edit:
% Some minor issues with grammar and word choices
% Consistent use of siunitx
% Relabeled figures, tables, and equations
% Correct (upright) typesetting of particles

% The figures are not publication quality and should be redone using TikZ.
% One even has a spelling error in it (ondulator).

EXMP is a muon source based on electron-photon collisions at ultra-high luminosity, capable of reaching muon fluxes up to a few \num{e11} muon pairs per second at an outstanding normalized transverse emittance of a few \si{\nano\meter\radian}, with muon beam energies peaked at \SI{50}{\giga\electronvolt}.
In order to sustain such a large muon flux despite the very small cross section of the muon photoproduction on electrons, it is envisaged to employ two primary colliding beams with a large average current (\SI{200}{\milli\ampere}) and Free-Electron Lasers (FEL) with very large photon fluxes (the maximum achievable with \SI{200}{\milli\ampere}, associated to the beam peak brightness requested by an X-ray FEL).
The use of FELs as the source of the primary photon beam in the X-ray photon energy range is mandatory for two reasons: the maximum efficiency in transforming electron kinetic energy into photons (each electron driving the FEL can emit thousands of photons) and the coherence of the radiation, which allows maximum focusability.
Since the electron-photon collisions transfer only a very small amount of power from the primary beams into the secondary beams, efficient energy recovery must be implemented in the scenario to reduce the amount of beam power loss from \SI{100}{\giga\watt} of beam power stored in primary beams at collision down to the level of hundreds of MW.
This is the main challenge of such a muon source, together with challenging beam collision spot sizes (in the range of tens of nanometers) and handling the extremely large FEL photon beam power.
The scheme is based on a twin array of linacs arranged face-to-face, providing  both the primary electron beam and the FEL driving beam. 
Further studies on the feasibility of this scenario are necessary to assess the achievable luminosity of a muon collider based on this muon source, depending on the kind of accumulator scheme to combine with such a muon source scenario.
The promise is to achieve the requested luminosity using much lower muon beam currents with respect to other schemes, which would significantly alleviate the issue of muon-beam-induced background and neutrino radiation: the very low emittance can be achieved thanks to the absence of a target (no target handling issues, no cooling needed) and the very tight focus allowed by a linear machine not subject to beamstrahlung (no significant beam-beam interaction).

\subsubsection{Collider scheme}
\begin{figure}[ht]\centering
	\includegraphics [scale=.5] {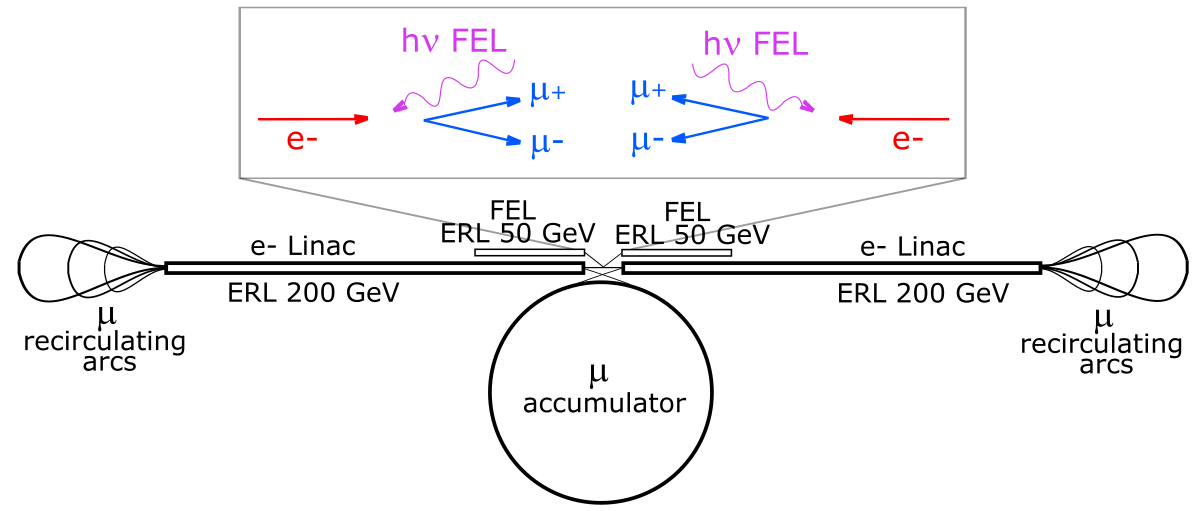}
	\caption{Twin Linac scheme. Primary \electron{} accelerated up to \SI{200}{\giga\electronvolt} collide with the counterpropagating FEL producing $\text{\textmu}^\pm$. Both \electron{} (primary and FEL) are decelerated in the opposite linacs and the energy recovered. A selected fraction of $\text{\textmu}^\pm$ are injected in the opposite linac and accelerated before storage and collision in a ring.}\label{fig:frontier:high_energy_colliders:exmp:twin}
\end{figure}

The final goal of generating a suitable beam of muons to be used in a TeV-scale muon collider is accomplished by the twin-linac scenario based on PERLE-like ERLs~\cite{Angal_Kalinin_2018} and the International Linear Collider~\cite{ILC07} schematics as sketched in Fig.~\ref{fig:frontier:high_energy_colliders:exmp:twin}.
The scheme is composed of a twin \SI{200}{\giga\electronvolt} ERL system coupled to a twin \SI{50}{\giga\electronvolt} FEL ERL system; the properties of the primary beams are listed in Table \ref{tab:frontier:high_energy_colliders:exmp:twin:parameters}.

\begin{table}[ht]	\centering
	\caption{Parameters of the primary beams.}\label{tab:frontier:high_energy_colliders:exmp:twin:parameters}
	\begin{tabular}{lcr}
		\toprule
		Electron beam energy & GeV & 200    \\
		Bunch charge & pC & 250    \\
		Electrons per bunch &   & \num{1.6e9} \\
		Repetition rate & MHz & 800 \\
		Average Current & mA & $2 \times 200$  \\
		Nominal beam power & GW & $2 \times 40$  \\
		Beam power recovery & \% &   99.9   \\
		Beam power loss & MW  & $2 \times 40$   \\
		Bunch length & ps  & 0.3    \\
		$\epsilon^\text{n}_{x,y}$ & \si{\micro\meter\radian} & 0.4\\
		$\beta_{x,y}$ & mm    & 0.2    \\
		$\sigma_{x,y}$ & nm   &  14     \\
		\midrule
		FEL photon energy & keV &    150    \\
		Photons per pulse & & \num{5e12} \\
		Repetition rate & MHz & 800 \\
		$\epsilon_{x,y}$ & \si{\pico\meter\radian} & 0.6\\
		Focal spot size & nm  & 14  \\
		FEL beam power & MW     & $2\times 100$  \\
		FEL efficiency (tapering)  &  &   \SI{1}{\percent} \\
		FEL \electron{} beam av.~curr. & mA &  200    \\
		FEL \electron{} beam energy & GeV   & 50     \\
		FEL \electron{} beam power & GW  & $2 \times 10$  \\
		Beam power recovery & \% & 99.9  \\
		Beam power loss & MW  & $2 \times 10$   \\
		\bottomrule
	\end{tabular}
\end{table}
The primary electron beam parameters listed in the table are quite consistent with the present state of the art for electron beams, with a rms normalised transverse emittance (\SI{0.4}{\milli\meter\milli\radian}, round beam) and an accelerated bunch charge of \SI{250}{\pico\coulomb}.
The value chosen for the beta function at the collision point (\SI{0.2}{\milli\meter}) is also very close to state-of-the-art performance.
That allows to match the spot size of the FEL photon beam down to \SI{14}{\nano\meter}.
No significant synchrotron radiation is expected; therefore, the primary beam power loss is set by the ERL efficiency and expected to be at most $2 \times 40 = \SI{80}{\mega\watt}$.

The photon beam needed to achieve an ultra-high luminosity in the collider is unique: it must carry an outstanding number of photons per pulse at the same repetition rate as the primary electron beam.
It must also match the ultra-tight focus spot size at collision of the primary electron beam, set by its very small beta function at the focus (hundreds of \si{\micro\meter}), in the range of a few tens of \si{\nano\meter}.
The only radiation source able to meet these demanding requirements is an FEL driven by a dedicated ERL and operated in SASE mode with tapering, as illustrated in~\cite{emma}.
An efficiency in the range of a few percent is achievable, yielding a number of photons per pulse as listed in Table~\ref{tab:frontier:high_energy_colliders:exmp:twin:parameters}, according to photon energy.
The partial coherence of the amplified FEL radiation also makes it possible to focus its photon beam down to nanometer spot sizes, as discussed in~\cite{Matsuyama2018}.
With this second crucial property of FELs, a luminosity of up to $2 \times \SI{2.5e41}{\per\square\centi\meter\per\second}$ can be envisioned.
The FEL beams considered in this study carry an impressive amount of photon beam power: running at \SI{800}{\mega\hertz} in CW mode, the number of photons per second exceeds \num{e21}.
With photon energies in the range of tens to hundreds of keV, that means up to \SI{100}{\mega\watt} of radiation beam power.
Since the FEL efficiency is about \SI{1}{\percent} considering the special mode of FEL operation, the power carried by the electron beam driving the FEL must be of the order of $\SI{100}{\mega\watt} / 0.01 = \SI{10}{\giga\watt}$, as listed in Table~\ref{tab:frontier:high_energy_colliders:exmp:twin:parameters}.
Therefore, the power budget for the linac driving the FEL is made up of twice the recovery beam power loss of \SI{10}{\mega\watt} and twice \SI{100}{\mega\watt} of radiated power, in total \SI{220}{\mega\watt}.

Summing up the power budget of the primary electron beam with that of the FEL-driving electron beam, the total beam power loss is \SI{300}{\mega\watt}.
An expected beam-to-plug efficiency not smaller than \SI{20}{\percent}, actually in the range \SIrange{20}{40}{\percent}, would set the AC power bill in the range of \SI{350}{\mega\watt} to \SI{1}{\giga\watt}.

\subsubsection{Simulation results for muon beams}

Assuming a head-on collision of the $E_\text{e}=\SI{200}{\giga\electronvolt}$ electron and the $h\nu=\SI{150}{\kilo\electronvolt}$ incident photon, the center-of-mass energy is $E_\text{CM}\simeq\sqrt{4E_\text{e}h\nu+M_\text{e}^{2}}= \SI{346}{\mega\electronvolt}$ ($M_\text{e}=\SI{0.511}{\mega\electronvolt}/c^2$ is the electron mass and natural units are used, i.e., $c=1$). Besides Muon Pair Production (MPP: $\electron+\photon \rightarrow \electron + \text{\textmu}^+/\text{\textmu}^-$), the other predominant reactions at the mentioned CM energy are Triplet Pair Production (TPP: $\electron+\photon \rightarrow \electron + \positron/\electron$) and Inverse Compton Scattering (ICS: $\electron+\photon \rightarrow \electron+\photon$).
The number of emitted muon pairs is given by
\begin{equation}
\mathcal{N}_{\text{\textmu}^{\pm}}= 2 \times\mathcal{L}\cdot \sigma^\text{MPP}_\text{tot}(E_\text{CM}) =2 \times\frac{N_\text{e}\,N_\text{ph}\,r}{4\pi \, {\sigma_{x}^2}}\cdot \SI{216}{nb} = 2 \times \SI{5.4e10}{\per\second}
\end{equation}
with $N_\text{e}, N_\text{ph}$ the number of electron and photons per bunch, $r$ the repetition rate of the collisions, and $\sigma_{x}$ the transverse dimensions of the electron and the photon beams.
Similarly, using the total cross sections as in \cite{motz} and \cite{ics}, it can be computed that $\mathcal{N}_{\text{e}^{\pm}}=2 \times \SI{4.8e15}{\per\second}$ and $\mathcal{N}_\text{ICS}=2 \times \SI{3.7e12}{\per\second}$.
The transverse normalized emittance of the produced muons is determined by the intrinsic thermal contribution of the reaction and by the features of the incoming electron beam. It can be described by
\begin{equation}
\epsilon_{\text{\textmu}}^\text{n}\simeq\frac{2}{3}\sigma_{x}\frac{\sqrt{E_\text{e} \, h\nu}-M_{\text{\textmu}}}{M_{\text{\textmu}}}+\frac{\langle\gamma_{\text{\textmu}}\rangle \epsilon_\text{e}^\text{n}}{\gamma_\text{e}}
\label{eq:frontier:high_energy_colliders:exmp:em}
\end{equation}
where $\epsilon_\text{e}^\text{n}$ is the transverse normalized emittance of the incoming electron beam, $\langle\gamma_{\text{\textmu}}\rangle$ the mean energy of the muon beam, and the first addendum represents the normalized thermal emittance of the muon beam.
In the case of this study, the emittance of the incoming electron beam has basically no impact on that of the muon beam: the second part of Eq.~\ref{eq:frontier:high_energy_colliders:exmp:em} is negligible because $\langle\gamma_{\text{\textmu}}\rangle \ll \gamma_\text{e}$. 

\begin{figure}[ht]\centering
\tikzsetnextfilename{mpp_hist_e}
\begin{tikzpicture}
\begin{groupplot}
[
    scale only axis,
    group style={
        group size=2 by 1,
        horizontal sep=2cm,
    },
    width=.35\linewidth,
    height=.25\linewidth,
]
\nextgroupplot[
    xlabel={$E$ (\si{\giga\electronvolt})},
    ylabel={\# events/bin ($\times \num{1000}$)},
    ymin=0,
    xmin=-2,
    xmax=202,
]
\addplot+[const plot mark left, mark=none, thick] table[x index=0, y expr={\thisrowno{1}/1e3}] {figures/exmp/mpp_hist_e.dat};
\draw (rel axis cs:1, 1) node[yshift=-1ex, xshift=-1ex, anchor=north east, draw, fill=white] {(a)};
\nextgroupplot[
    xlabel={$\theta$ (\si{\milli\radian})},
    ylabel={\# events/bin ($\times \num{1000}$)},
    ymin=0,
    xmin=-0.05,
    xmax=2.5,
]
\addplot+[const plot mark left, mark=none, thick] table[x expr={\thisrowno{0}*1e3}, y expr={\thisrowno{1}/1e3}] {figures/exmp/mpp_hist_theta.dat};
\draw (rel axis cs:1, 1) node[yshift=-1ex, xshift=-1ex, anchor=north east, draw, fill=white] {(b)};
\end{groupplot}
\end{tikzpicture}
\\[2ex]
\tikzsetnextfilename{mpp_hist2d}
\begin{tikzpicture}
\begin{axis}
[
    scale only axis,
    width=.35\linewidth,
    height=.25\linewidth,
    xlabel={$x$ (\si{\nano\meter})},
    ylabel={$p_x$ ($\si{\giga\electronvolt}/c$)},
    xmin=-40,
    xmax=40,
    ymin=-0.15,
    ymax=0.15,
    point meta min=0,
    point meta max=6.15,
    axis on top,
    colorbar,
    colormap/YlOrBr,
    colorbar style={ylabel={\# events/bin ($\times 100$)}},
]
%\addplot[
%    matrix plot*,
%    mesh/cols=50,
%    point meta=explicit,
%] table[x index=0, y index=1, meta index=2] {mpp_hist2d_emit.dat};
\addplot graphics[xmin=-40, xmax=40, ymin=-0.15, ymax=0.15] {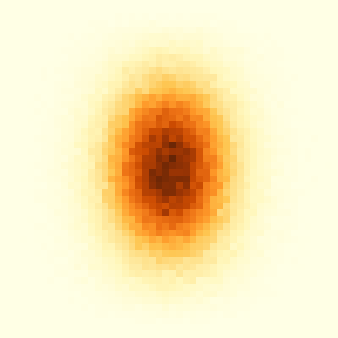};
\draw (rel axis cs:1, 1) node[yshift=-1ex, xshift=-1ex, anchor=north east, draw, fill=white] {(c)};
\path (rel axis cs:0.5, 1) coordinate (labelpos);
\end{axis}
\draw (labelpos) node[anchor=south] {$\epsilon_{x_\text{\textmu}}^{\text{n}} = \SI{0.0046}{\milli\meter\milli\radian}$};
\end{tikzpicture}
\caption{Histograms of muon beam parameters. (a) Energy distribution. (b) Angle distribution. (c) Transverse emittance. All histograms were taken from a sample of \num{2e5} particles; (a) and (b) with 100 bins, (c) with 50 bins per dimension.}
	\label{fig:frontier:high_energy_colliders:exmp:tl1}
\end{figure}
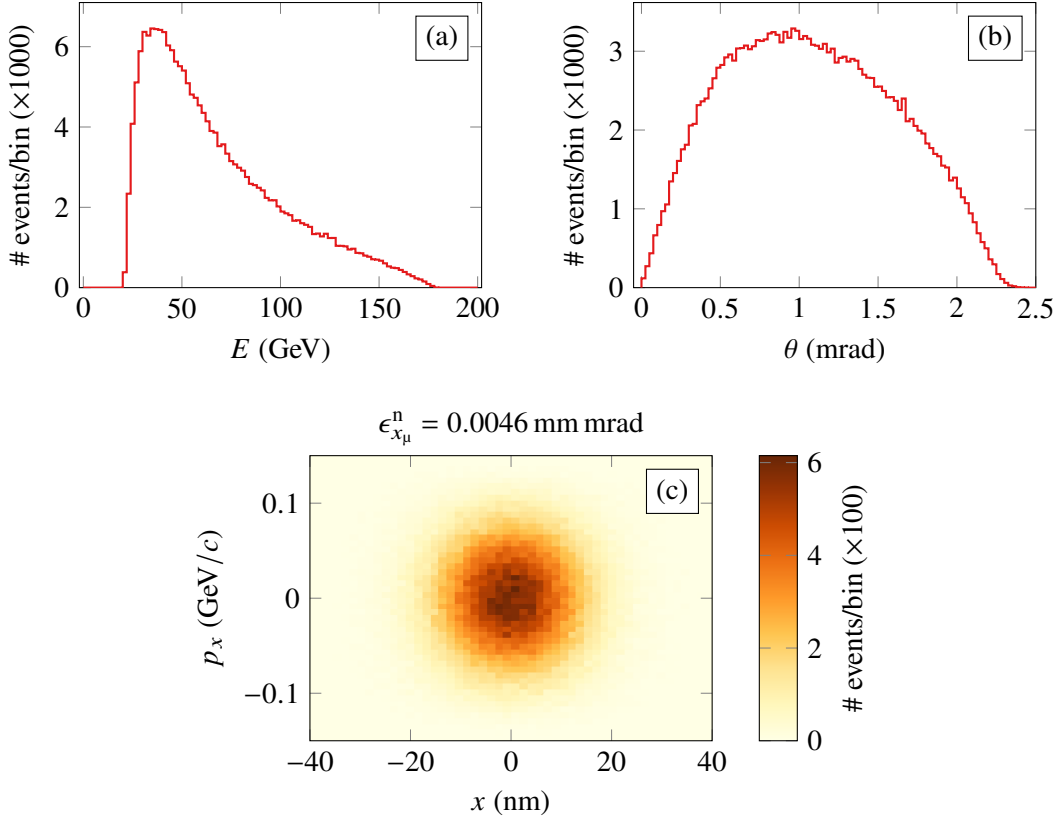

MPP has been simulated by means of the Whizard event generator~\cite{whizard}, taking into account the properties of the incoming beams.
The features of the emitted muon beam are displayed in Fig.~\ref{fig:frontier:high_energy_colliders:exmp:tl1}. The normalized transverse emittance of the muon beam is \SI{4.6}{\nano\meter\radian}: this value compares to the analytical prediction of Eq.~\ref{eq:frontier:high_energy_colliders:exmp:em} giving \SI{5.9}{\nano\meter\radian}.
The outstanding value of the normalized transverse emittance combined with the number of muon pairs per second (important figure of merit analysed in Ref.~\cite{Zimmermann:IPAC2018-MOPMF065}) gives a value of $\mathcal{N}_{\text{\textmu}^{\pm}}/\epsilon^\text{n}_{x_{\text{\textmu}}}=2 \times \SI{1.17e19}{\per\meter\per\second}$.
Considering only the muons around the energy distribution peak of \SI{50}{\giga\electronvolt} corresponding to a \SI{10}{\percent} rms relative energy spread, \SI{20}{\percent} of the produced muons are selected (with a longitudinal emittance value of $\sim\SI{4.5}{\milli\meter}$).
The aforementioned coefficient corresponding to this selection is $\mathcal{N}_{\text{\textmu}^{\pm}}/\epsilon^\text{n}_{x_{\text{\textmu}}}=2 \times \SI{2.34e18}{\per\meter\per\second}$.
TPP, which is the most probable collateral reaction, would involve  $2 \times \num{4.8e15}$ primary electrons, still a small fraction of the total (\SI{2e18}{\per\second}).

\subsubsection{Further considerations}
One should note that the power transferred from the primary colliding electron/photon beams into the secondary beams of muon pairs, photons, and electron-positron pairs is quite negligible compared to the power stored in the colliding beams at the collision point.
As a matter of fact, the muon-pair beams (\SI{e11}{\per\second} at an average energy of \SI{50}{\giga\electronvolt}) are taking out only \SI{1}{\kilo\watt} of beam power.
The power taken by the back-scattered Compton gammas (\SI{8e12}{\per\second} at an average energy of \SI{200}{\giga\electronvolt}) is \SI{250}{\kilo\watt}, and the power taken by the electron-positron pairs produced (\SI{e16}{\per\second} at an average energy of \SI{1}{\giga\electronvolt}) is about \SI{1.6}{\mega\watt}.
All these numbers are quite negligible with respect to the total power loss quoted in Table~\ref{tab:frontier:high_energy_colliders:exmp:twin:parameters} (\SI{300}{\mega\watt}), which is dominated by ERL efficiency and FEL photons.

A useful characteristic of the twin-linac layout shown in Fig.~\ref{fig:frontier:high_energy_colliders:exmp:twin} is the possibility to accelerate muons in the same linacs used to accelerate the primary electron beams.
This is possible thanks to the very small emittance of the muons and the lack of interaction between muon beams and electron beams.
Beam optics could rely on RF focusing effective both on muons and electrons, as further analysed in a future work.
Linac acceleration of muons would allow to bring them up rapidly to the TeV kinetic energy range requested by muon collider physics, just in a few passes (each twin-linac pass is \SI{400}{\giga\electronvolt} energy gain) through the twin-linac system, using proper muon recirculation arcs.

Concerning the possibility to generate polarised muon beams, it is noted that that the FEL photon beam is linearly polarised by nature (the FEL
radiation being transversely coherent and basically single-mode TEM-00), so by using a polarised primary electron beam, the muon-pair beam will
be clearly polarised to a very large extent.
How to preserve and maintain such a polarisation during acceleration and accumulation of the muons
is a subject of future studies.

A \SI{500}{\giga\electronvolt} working point for the twin-linac scheme and a proof-of-principle experiment are described and discussed in \cite{app12063149}.
%An alternative scenario described in~\cite{EXMParxiv} and sketched in Fig.~\ref{fig:frontier:high_energy_colliders:exmp:fcctwin} has been considered: adopting the future FCC tunnel, its planned radius of curvature being \SI{15}{\kilo\meter}, so as to host the two Linacs generating the primary electron beam and the FEL driving electron beam in a slightly modified race-track geometry.
%This scenario is limited in performance by the huge amount of power lost due to synchrotron radiation emission in the arcs.
%
%\begin{figure}[ht]\centering
	%\includegraphics [scale=.5] {figures/EXMPFCCTwin.png}
	%\caption{FCC-twin Linac scheme, FCC ring with ERL insertions. Primary \electron{} accelerated up to \SI{100}{\giga\electronvolt} collide with the counterpropagating FEL, producing $\text{\textmu}^\pm$. The energy of both \electron{} beams (primary and FEL) is recovered. $\text{\textmu}^\pm$ are accumulated and accelerated in a separate ring.}\label{fig:frontier:high_energy_colliders:exmp:fcctwin}
%\end{figure}

%\subsection{Electrons and X-rays to Muon Pairs (EXMP)}
%Camilla Curatolo, Luca Serafini

\section{Low-Energy Particle Physics}\label{sec:frontier:low_energy_particle}

The unique beam properties of ERLs---high intensity and small emittance---enable substantial experimental
advances for a variety of physics at lower energies.

Form factors of nucleons and nuclei are traditionally accessed via elastic electron scattering. Recently,
the low-$Q^2$ form factor of the proton was the focus of increased scrutiny because of the proton
charge radius puzzle~\cite{RevModPhys.94.015002}, a more than $5\sigma$ difference in the charge radius
extracted from muonic spectroscopy and all other determination methods.
The determination of the proton form factors is limited by experimental systematics stemming from target-related background.
The high beam current available at ERLs allows us to employ comparatively thin targets, for example cluster jets~\cite{Schlimme:2021gjx}, which minimise this
background, paving the way for a new generation of experiments.
In a similar vein, the relatively high
luminosity and typically small energies at facilities like MESA allow us to measure the magnetic form
factor, only accessible at backward angles at low $Q^2$, with substantially increased precision in a $Q^2$ range
highly relevant for the magnetic and Zeeman radii and where the current data situation is especially dire.
Further electron scattering experiments include dark sector searches like DarkLight@ARIEL, aiming at
masses of a couple of (tens of) MeV.

In photon backscattering, the luminosity available exceeds that of the Extreme Light Infrastructure project (ELI)~\cite{book} by a few orders
of magnitude, paving the way to nuclear photonics, a development area possibly comparable with the
appearance of lasers in the sixties.
For example, the intensities achievable at an ERL allow nuclear parity
mixing to be accessed.
Photonuclear reactions test the theory for nuclear matrix elements relevant for
the neutrino mass determination from neutrinoless double beta decay.
They can be used to study key reactions for stellar evolution.
Ab-initio calculations of light nuclei (e.g.,~\cite{Friman21}) are advanced and need to
be tested with precision measurements.

A further fundamental interest regards the exploration of unstable nuclear matter with intense
electron beams of about \SI{500}{\mega\electronvolt} energy, as is planned for PERLE and envisaged for GANIL in
France.
This follows the recognition of the field by NuPECC in their strategic plan in 2017: ``Ion-electron
colliders represent a crucial innovative perspective in nuclear physics to be pushed forward in
the coming decade. They would require the development of intense electron machines to be installed at
facilities where a large variety of radioactive ions can be produced''.

\subsection{Elastic Electron-Hadron Scattering}
% Jan Bernauer
 
In our current understanding, Quantum Chromodynamics (QCD) in the non-per\-tur\-ba\-tive regime describes the physics inside nucleons and nuclei. Elastic electron-hadron scattering allows us to determine fundamental properties of QCD systems, validating our current theoretical understanding and providing crucial input for other fields, from astrophysics to atomic physics. Precision measurements are benchmarks for lattice calculations and might open a portal to the dark sector.

\subsubsection{Proton targets}
Using a proton as a target, the elastic form factors $G_E$ and $G_M$ are accessible via elastic scattering, either using the Rosenbluth separation technique to extract the from factors from cross section measurements, or using polarization degrees of freedom to determine the form factor ratio. 

The form factors are connected to the distribution of charge and currents inside the proton, and might show emergent behaviors like the existence of a meson cloud around a bare nucleon. In first order, the elastic electron-proton scattering cross section can be rewritten as  
\begin{equation}
    \frac{\text{d}\sigma}{\text{d}\Omega}=\frac{1}{\epsilon(1+\tau)}\left[\epsilon G_E^2+\tau G_M^2\right]\left(\frac{\text{d}\sigma}{\text{d}\Omega}\right)_\textrm{Mott}\label{eq:redcross},
\end{equation}
where the photon polarization $\epsilon$ varies between 0 and 1 for backward to forward scattering, and $\tau=Q^2/(4m_\textrm{proton}^2)$ is proportional to the exchanged four-momentum squared.  From the extracted form factors, one can then extract critical parameters, e.g., the proton charge and magnetic radius, given by
\begin{equation}
    \left<r_{E/M}\right>^2=-\frac{6\hbar^2}{G_{E/M}(0)}\left.\frac{\text{d}G_{E/M}}{\text{d}Q^2}\right|_{Q^2=0}.
\end{equation}

The low-$Q^2$ regime has attracted intensive interest in the last decade through the so-called proton radius puzzle, originally established in 2010 by a \SI{4}{\percent} difference between extractions using muonic spectroscopy \cite{pohlSizeProton2010} ($r_\text{p}=\SI{0.84184(67)}{\femto\meter}$) and both the results of the Mainz high-precision form factor experiment \cite{bernauerHighprecisionDeterminationElectric2010} ($r_\text{p}=0.879(5)_\text{stat}(6)_\text{syst}\,\si{\femto\meter}$)  and the CODATA value \cite{mohrCODATARecommendedValues2008} ($r_\text{p}=\SI{0.8768(69)}{\femto\meter}$), based on a series of normal hydrogen spectroscopy measurements and radius extractions from earlier scattering data. 

\begin{figure}[htb]
    \centering
    \includegraphics[width=\textwidth]{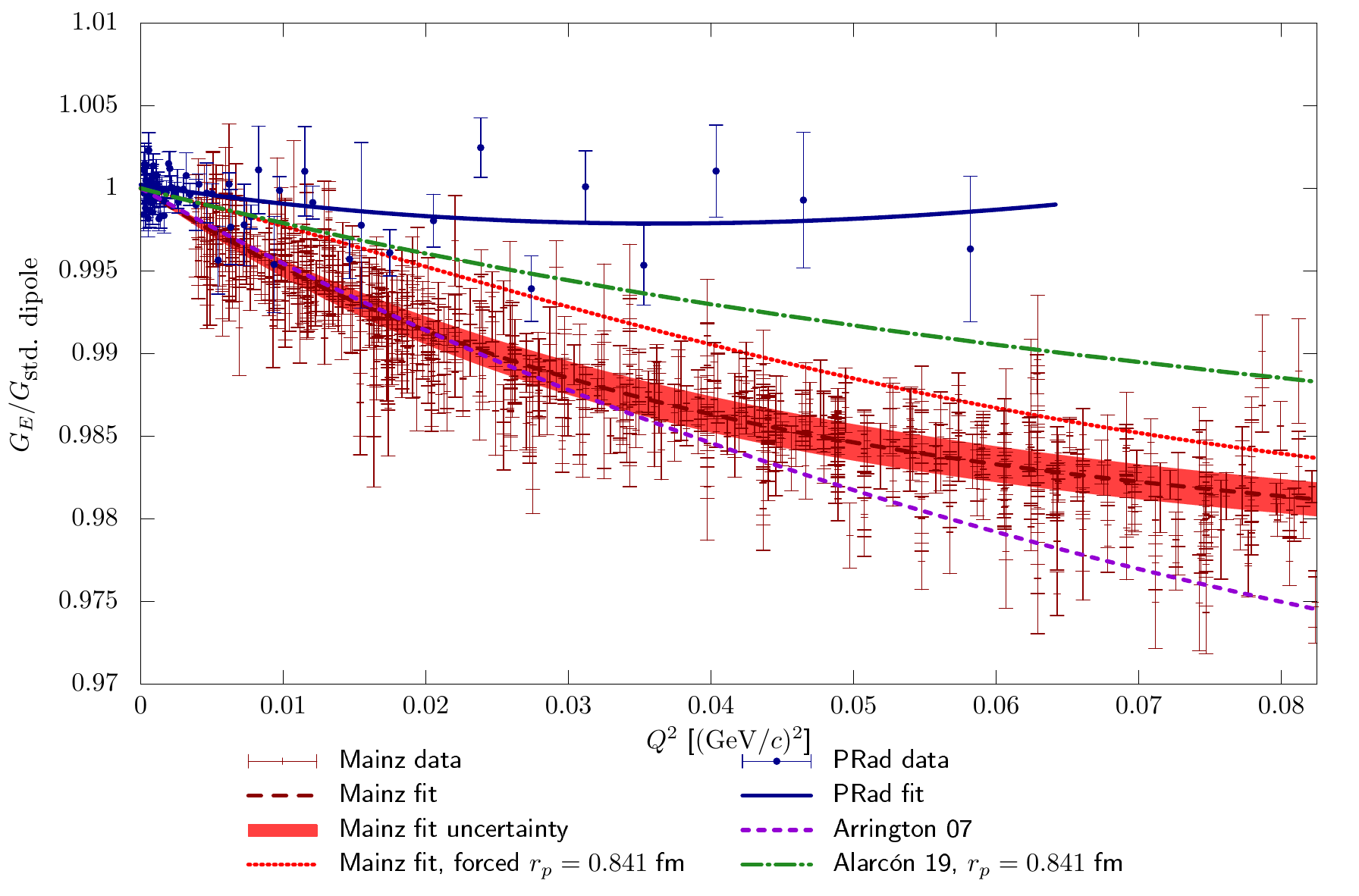}
    \caption{Recent results on the proton electric form factor $G_E$, normalized to the standard dipole, for the overlap region between the PRad and the Mainz data. Red data points are the Mainz data \cite{bernauerHighprecisionDeterminationElectric2010} with fit and error band. Additionally, a fit to the Mainz data with a radius forced to the small value from muonic spectroscopy is shown (red, dotted). The PRad data and fit \cite{Xiong:2019umf} are shown in blue. The lilac dashed line labeled \enquote{Arrington 07} is a fit  to pre-Mainz data \cite{arrington2007}. The green dot-dashed line is from a dispersively improved
    effective-field-theory calculation with only the radius as a parameter, chosen here to be the small radius from muonic spectroscopy. The general agreement between the Mainz and earlier data, the PRad data, and this calculation is poor, raising questions about the reliability of all existing form factor measurements.}
    \label{fig:gewhatweknow}
\end{figure}

While newer measurements, especially in spectroscopy, show a trend to the smaller radii, some new measurements prefer the bigger radius. The situation is especially interesting on the scattering side, where the recent PRad \cite{Xiong:2019umf} result gives the smaller result but shows a clearly different result from all earlier measurements, putting our knowledge about the proton form factors not only at small $Q^2$ into question. The current situation is shown in Fig.~\ref{fig:gewhatweknow}.
A series of experiments are planning to illuminate different aspects of this puzzle, and ERL-based experiments will play a key role.

\begin{figure}[htb]
    \centering
    \includegraphics[width=\textwidth]{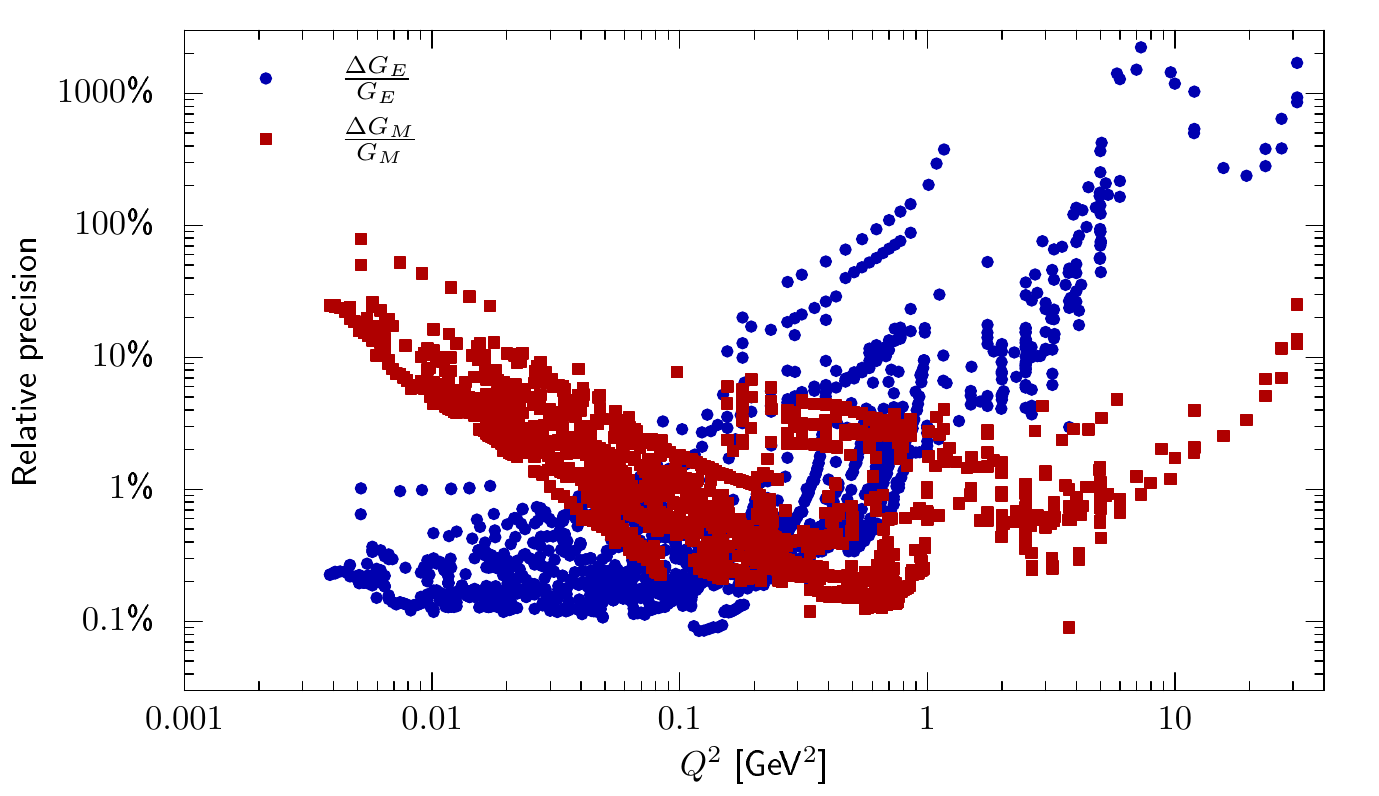}
    \caption{The impact of available cross section data on the determination of $G_E$ and $G_M$, symbolized by translating the statistical error of the data on the cross section to effective errors of the form factors (assumed the other form factor is known). Data set is the same as in \cite{PhysRevC.90.015206}; the newer PRad data is not sensitive to $G_M$.}
    \label{fig:wwkgm}
\end{figure}
The proton radius puzzle steered the focus onto the electric form factor; however, the situation for the magnetic form factor at small $Q^2$ is worse.
As can be seen from Eq.~\ref{eq:redcross}, the magnetic contribution is strongly suppressed for small $Q^2$.

In Fig.~\ref{fig:wwkgm}, the effective uncertainty of $G_E$ and $G_M$, calculated from the uncertainty of $d\sigma$ for the existing data sets, is shown.
As can be seen, for smaller $Q^2$, the uncertainty of $G_M$ is many orders worse than that of $G_E$.
In fact, most magnetic radius extractions use rather stiff fits to extrapolate from the data at larger $Q^2$. Indeed, the Mainz fit \cite{bernauerHighprecisionDeterminationElectric2010}, which is more flexible, finds some structure at $Q^2=0.03\,(\mathrm{GeV}/c)^2$, while other fits with less flexibility do not resolve this structure. This leads to large variations in the extracted radii.

ERLs allow us to improve on current experiments in several ways.
The possible large currents at excellent beam qualities make it possible to run with thin targets with sufficient luminosity, even at backward angles, to improve the sensitivity to $G_M$.
The possibility of a thin target has additional benefits; as part of the MAGIX program, the A1 and MAGIX collaborations are developing a cluster-jet target \cite{Schlimme:2021gjx} which would realize such a thin target.
Here, a jet of frozen hydrogen clusters forms when hydrogen is forced into vacuum with high pressure through a thin, cryogenically cooled laval nozzle.
The clusters are then collected in a catcher.
This realizes a pure hydrogen target without any cell walls which could interact with the electron beam.
The jet is very narrow, simplifying the reconstruction of the scattered particle trajectory and of the detector acceptance.
Further, the thin target minimizes external radiative effects and makes it possible to detect the recoiling proton.

\subsubsection{Light nuclei}
Similar to the experimental program for a proton target, ERLs enable experiments to measure the equivalent properties for other nuclei. For gaseous targets, a similar jet target can be used.
If the waste gas of the required pumping is collected, cleaned, and compressed, it is possible to build a closed loop with minimal losses so that even expensive gas targets like $^3$He are feasible. 

New, precise measurements of the deuterium, $^3$He and $^4$He form factors, and with that, of their radii, would allow additional comparisons with muonic and electronic spectroscopy measurements. 

%\subsubsection{Pion electroproduction}
%Leaving pure v

%Using virtual photon tagging,  it is possible to study confinement-scale QCD. In photo- production, 
%the photon tagger sets the rate limit and only a small fraction of the tagged photons interact with 
%the target, leading to low data-taking efficiency.  At forward angles, the virtual photons are 
%almost real, so that a forward scattering electron tagger can be used as a highly efficient 
%substitute.  Because of the high efficiency and high beam currents, it is possible to use pure, 
%thin targets and detect low energy recoil particles which would not
%escape traditional, thick targets. It is thus possible to measure the reactions γ p → π0 p, π+n,
%γn → π0n, π−p and γD → π0D.  Coherent π0  production in D and ³He measure relative signs of the γ p 
%→ π0 p,γn → π0n amplitudes.

%\subsection{Elastic Electron-Hadron Scattering}
% Jan Bernauer

\subsection{Weak Interaction at Low Energy}
% Hubert Spiesberger, Kurt Aulenbacher, Maarten Boonekamp
%\\ 

At high energies, testing the Standard Model (SM) of particle 
physics is generally linked to the search for new particles. 
In contrast, at low energies, one hopes to explore the 
frontier to new physics beyond the SM by searching for 
deviations from SM predictions, performing measurements 
at the highest possible precision. Any deviation from SM 
predictions could be pointing towards an incomplete 
understanding of the SM and a sign of new physics. 
Many different measurements will have to be combined 
in order to infer from indirect signals the type 
of new particles or new interactions. 

High precision in scattering experiments requires high 
rates and can, therefore, be obtained usually only with charged 
particles. The background from the well-understood electromagnetic 
interaction described by Quantum Electrodynamics is often 
overwhelming. In order to isolate effects from the weak 
interaction and, possibly, new physics, one has to focus 
on rare processes, or to filter out properties which are 
specific to the weak interactions, like parity-violation 
or effects violating the charge symmetry of QED.

\subsubsection{Effective low-energy Lagrangean}

The interaction of electrons with matter, i.e., with electrons 
or with quarks inside hadrons, can be described by an 
effective Lagrangean for the neutral-current interaction 
at low energies 
\begin{equation}
L_{NC}^{ef} 
= 
\frac{G_F}{\sqrt{2}} \sum_f 
\left[ 
g_{VV}^{ef} \bar e\gamma^\mu e \bar f\gamma_\mu f 
+ 
g_{AV}^{ef} \bar e\gamma^\mu\gamma_5 e \bar f\gamma_\mu f 
+  
g_{VA}^{ef} \bar e\gamma^\mu e \bar f\gamma_\mu \gamma_5 f 
+ 
g_{AA}^{ef} \bar e\gamma^\mu \gamma_5 e \bar f\gamma_\mu \gamma_5 f \right], 
\label{eqs522:L}
\end{equation}
where $G_F$ is the Fermi constant. At lowest order (tree level) 
in the SM, the four-fermion couplings are products of the fermion 
couplings $g_{V,A}^f$ to the $Z_0$ boson,
\begin{equation}
g_{AV}^{ef} = 2 g_A^e g_V^f \ , 
\quad 
g_{VA}^{ef} = 2 g_V^e g_A^f \ , 
\quad
g_{AA}^{ef} = - 2 g_A^e g_A^f \ , 
\label{eqs522:gav} 
\end{equation}
with
\begin{equation}
g_A^f = T_3^f \ , 
\quad 
g_V^f = T_3^f - 2 Q^f \sin^2 \theta_W 
\end{equation}
where $\theta_W$ is the weak mixing angle, $T_3^f$ the third 
component of the weak isospin and $Q^f$ the charge of the 
fermion. 
The vector times vector term proportional to $g_{VV}^{ef}$ 
is important for phenomenology only at high energies since 
at low energies it is overwhelmed by QED effects. 
The $g_{AV}^{ef}$ and $g_{VA}^{ef}$ terms in Eq.~\ref{eqs522:L} 
induce parity violation. They can be identified by measuring 
the cross-section asymmetry between left- and right-handed 
electrons scattering off unpolarized targets. The coupling 
$g_{AV}^{eq}$ was also determined in atomic parity violation 
experiments. The terms involving $g_{AA}^{ef}$ do not 
violate parity, but they can be accessed by comparing cross sections 
of electron to positron scattering \cite{Zheng:2021hcf}. 
Electron couplings, i.e., the special case with $f=e$, are 
accessible in Møller scattering. For the $g_{AV}^{ef}$ 
couplings, because of the conservation of the vector current, 
one can define in the limit of vanishing momentum transfer 
the weak charge of a nucleus, $Q_W^{Z,N}$, composed of $Z$ 
protons and $N$ neutrons: 
\begin{equation}
Q_W^{Z,N} = 
-2 \left( Z g_{AV}^{ep} + N g_{AV}^{en} \right) 
\label{eqs522:qw-lo}
\end{equation}
with 
\begin{equation}
g_{AV}^{ep} = 2 g_{AV}^{eu} + g_{AV}^{ed} \,  
\quad \text{and} \quad 
g_{AV}^{en} = g_{AV}^{eu} + 2 g_{AV}^{ed} \, . 
\end{equation} 
Coincidentally, the weak charge of the proton is small in the SM,
\begin{equation}
Q_\text{W}(\text{p}) 
= 
Q_W^{1,0} 
= 
1 - 4 \sin^2 \theta_\text{W} 
\label{eqs522:QWp_tree_level}
\end{equation}
since $\sin^2 \theta_W$ is close to $1/4$. Its measurement is 
therefore expected to be particularly sensitive to new physics. 

New physics beyond the SM can be described by additional terms 
of perturbations of the SM Lagrangian (Eq.~\ref{eqs522:L}), that 
is by replacements of the form,
\begin{equation}
\frac{G_F}{\sqrt{2}} g_{ij}^{ef} 
\rightarrow 
\frac{G_F}{\sqrt{2}} g_{ij}^{ef} 
+ \eta_{ij}^{ef}\frac{4\pi}{\left(\Lambda_{ij}^{ef}\right)^2}\ ,
\label{eqs522:ciqmodified}
\end{equation}
where $ij = AV,\, VA,\, AA$ and $\eta_{ij}^{ef} = \pm 1$. 
By convention, one usually assumes that new physics is 
strongly coupled with a coupling $g^2 = 4\pi$. Then, new 
physics is described by a mass scale $\Lambda$, up to 
which the SM is valid and beyond which new particles 
could exist. Different targets and different observables 
probe different combinations of $\Lambda_{ij}^{ef}$.

\subsubsection{P2 at MESA}

\begin{figure}[htb]
\centering
\includegraphics[width=0.7\textwidth]{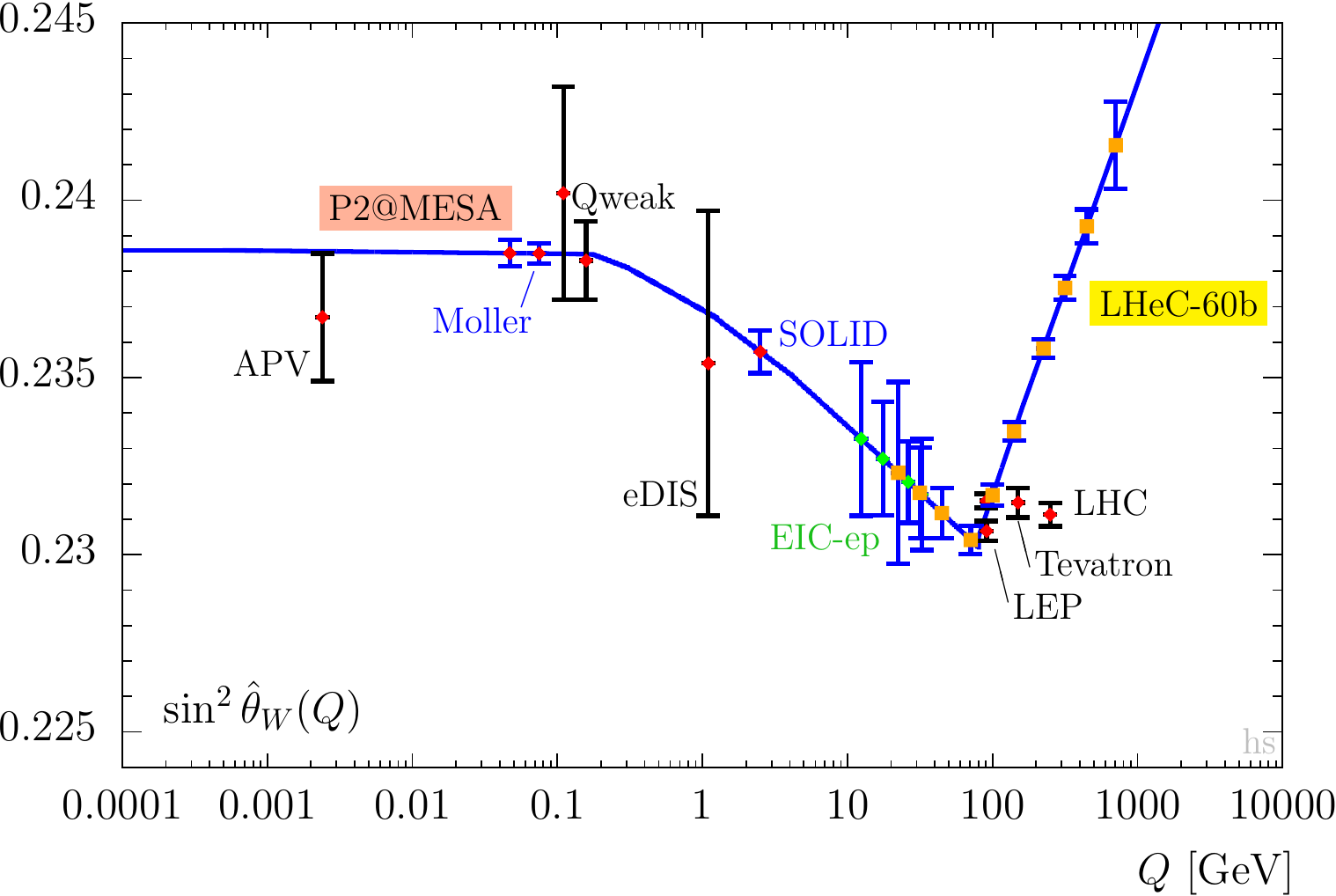}
\caption{%
   The running weak mixing angle compared with existing (red 
   points with black error bars) and planned measurements (blue 
   error bars).%
}
\label{fig522:sw2running}
\end{figure}

The P2 experiment \cite{Becker:2018ggl} at MESA plans to 
determine the weak charge of the proton by measuring the 
polarization asymmetry in elastic electron-proton scattering, 
\begin{equation}
A^\text{PV} = 
\frac{\mathrm{d}\sigma_\text{ep}^+ -
      \mathrm{d}\sigma_\text{ep}^-}%
      {\mathrm{d}\sigma_\text{ep}^+ +
      \mathrm{d}\sigma_\text{ep}^-} \, .
\label{eqs522:Apv_differential_form}
\end{equation}
In this equation, $\mathrm{d}\sigma_\text{ep}^\pm$ is the 
differential cross section for the elastic scattering of 
electrons with helicity $\pm 1/2$ off unpolarized protons.
The parity-violating helicity asymmetry is predicted to be 
\begin{equation}
A^\text{PV} 
= 
\frac{-G_\text{F} Q^2}{4\pi\alpha_\text{em}\sqrt{2}} 
\left[Q_\text{W}(\text{p}) - F(E, Q^2) \right] .
\label{eqs522:Apv}
\end{equation}
Here, $\alpha_\text{em}$ is the electromagnetic fine structure 
constant, $E$ the beam energy and $Q^2$ the invariant square 
of the 4-momentum transfer. At present, the setup at P2 assumes 
a beam energy of $E=155$\,MeV and scattering angles in the range 
of $\ang{25} < \theta < \ang{45}$, corresponding to $Q^2 = \SI{0.0045}{\square\giga\electronvolt}$.
This beam energy is the maximum that can be reached 
at MESA with three passages through the accelerating cryomodule. 
A low beam energy is advantageous since corrections due to 
$\text{\textgamma}\mathrm{Z}$-box graphs are suppressed proportionally to $E$. The 
scattering angles cover a range where the combination of 
statistical and systematic uncertainties is minimized 
\cite{Becker:2019gvs}. Such a measurement requires the highest 
possible luminosity, i.e., also a target with high density. 
At P2, therefore, the measurement with the required high statistics 
will be performed in the EB-mode 
(see Section~\ref{sec:new_facilities:mesa}). 

The form factor contribution $F(E, Q^2)$ can be decomposed 
into well-known electromagnetic proton and neutron form factors, 
but contains also less well-known strangeness and axial-vector 
parts. The strangeness form factor has been determined recently 
from lattice QCD with good precision \cite{Djukanovic:2019jtp} 
and further improved results are expected. Remaining uncertainties 
due to the axial form factor can be reduced by an auxiliary 
measurement at backward scattering angles. 

Provided that $F(E, Q^2)$ is known with sufficiently high 
precision, one can determine the weak mixing angle at very 
high precision: a relative uncertainty of $0.15\,\%$ for 
$\sin^2 \theta_\text{W}$ can be expected for a $1.4\,\%$ 
measurement of the asymmetry. The expected data point is 
shown in Fig.~\ref{fig522:sw2running}, compared with the SM 
prediction \cite{Erler:2017knj} as well as with existing 
measurements (taken from the PDG \cite{Zyla:2020zbs}) and 
possible measurements at the future experiments MOLLER  
\cite{Benesch:2014bas} and SoLID \cite{Souder:2012zz}, the EIC \cite{Boughezal:2022pmb}, or 
at the LHeC \cite{Britzger:2020kgg}. Assuming, in turn, 
that the value of the weak mixing angle is known with high 
precision, one can determine exclusion limits for new physics. 
Mass scales up to 50\,TeV are possible, competitive with and 
complementary to ongoing measurements at the high-energy 
frontier, i.e., at the LHC experiments. 

%\subsection{Weak Interaction at Low Energy}
% Hubert Spiessberger, Kurt Aulenbacher, Maarten Boonekamp

% Dark Photons deleted per Andrew
%\input{s523.tex}
%\subsection{Dark Photons}
% Oliver Fischer, Steve Benson, Jan Bernauer

\section{Low-Energy Electron-Ion Scattering}\label{sec:frontier:low_energy_nuclear}
% David Verney
\subsection{Introduction, physical and historical contexts}
Nuclear physics of the 21st century faces two challenges: It must try to go beyond the discovery frontier on the one hand (the challenge being the synthesis of new and more exotic nuclei) and the precision frontier on the other hand (the challenge being the cross-accumulation of ever richer and more precise observables to understand the nuclear structure).
The stakes attached to these two ambitions are often divergent, requiring the use of varied and complementary techniques, instruments, and facilities (a specificity of research in nuclear structure).
On the road to the precision frontier, nuclear physicists have an ally of choice: the electromagnetic probe.
Because the electromagnetic interaction is perfectly known, this probe allows to obtain information about the nucleus independently of nuclear models and to get rid of the assumptions and approximations inherent to the incredible complexity of the description of the nuclear interaction involved in the use of a hadronic probe.

The range of use of electromagnetic interaction to probe the nucleus extends, at very low energy, from the radioactive ion manipulation with electromagnetic fields in traps (e.g., high-precision mass measurements), through the interaction of the nucleus with the hyperfine field (laser spectroscopy, nuclear orientation), gamma-spectroscopy techniques (e.g., using advanced electronically segmented HPGe crystal arrays for gamma tracking) to electron scattering. In the first three cases, many examples of very efficient devices for the study of nuclei far from stability exist in the world, but in the last case, electron scattering off exotic nuclei, no example can be cited (except a pioneering, demonstration SCRIT device in RIKEN, Japan \cite{SCRIT17}). Contrary to any of the other techniques, which can only give access to integrated quantities (mean square radii, electromagnetic transition probabilities), the scattering of electrons of several MeV energy offers a unique access to spatially-dependent distributions (radial charge density, charge transition density, magnetic current distributions), i.e., an access to the interior of the nucleus. From the point of view of theory, detailed densities are much more demanding than integrated quantities and encapsulate different correlation effects. As such, they offer an unprecedented test bench for state-of-the-art nuclear structure models. Their availability over a wide range of unstable isotopes would thus systematically provide model-independent constraints very complementary to information from other probes like proton scattering.

The European Nuclear Physics community’s interest for electron beams of \SIrange{400}{800}{MeV} providing the ideal spatial resolution scale of about 0.5\,fm to study charge distributions of unstable nuclei has already been put forward in the framework of the NuPECC long-range plan perspective in 2016--2017 \cite{LRP}. Conclusions and recommendation of the community were written as follows: ``Ion-electron colliders represent a crucial innovative perspective in nuclear physics to be pushed forward in the coming decade. They would require the development of intense electron machines to be installed at facilities where a large variety of radioactive ions can be produced''.

More recently, during its national foresight exercise, the French low-energy Nuclear Physics community has set among its priority objectives for the future of its national facility, GANIL (Caen, France), the launch of an ambitious program to measure electron scattering off radioactive ions. The scenario retained for these experiments would be that of a fixed target consisting of a cloud of trapped ions interacting with an electron beam of energy of the order of \SI{500}{\mega\electronvolt}. This idea has been endorsed by a committee of international experts led by M.~Spiro and mandated by the GANIL funding agencies. This committee recommends the construction of an electron radioactive ion (e-RI) collider for possible operation in the 2040s. It is clear that such an ambitious undertaking requires a preliminary extensive research and development phase. The committee further underlines: \enquote{feasibility studies and prototyping may start as soon as possible elsewhere than at GANIL as part of international/EU initiative in Accelerator R\&D}. The realization of an ERL like PERLE in Orsay (see Section \ref{sec:new_facilities:perle}) would represent a unique opportunity in this sense. The objective of DESTIN \textit{[DEep STructure Investigation of (exotic) Nuclei]} that the IJCLab Nuclear Physics community is pushing forward is to seize this opportunity.

\subsection{The Luminosity challenge}
The insight one can get into density distributions depends on the accuracy of the measured form factor and the range of momentum transfer covered. This translates into luminosity constraints to access different structure observables.
In order to fix the ideas on the orders of magnitude of luminosities necessary to carry out this type of experiments one can base oneself on the many works which have been carried out on stable targets from the 50s until the end of the 90s.
\begin{table}[htb]
\centering
\begin{tabular}{ccccc}
\hline\hline
 & $E_\text{e}$ & $I_\text{e}$ & target thickness & Luminosity \\
\hline
Hofstadter's era & 150\,MeV & $\sim \SI{1}{\nano\ampere} $ & $\sim \SI{e19}{\per\square\centi\meter}$ & $\sim \SI{e28}{\per\square\centi\meter\per\second}$ \\
(1950's) & & ($\sim \SI{e9}{\per\second}$) & & \\
\hline
JLAB & 6\,GeV & $\sim \SI{100}{\micro\ampere}$ &  $\sim \SI{e22}{\per\square\centi\meter}$ & $\sim \SI{e36}{\per\square\centi\meter\per\second}$ \\
&             &   ($\sim \SI{e14}{\per\second}$)    && \\
\hline\hline
\end{tabular}
\caption{}
\label{tab:frontier:low_energy_nuclear:1}
\end{table}
 A luminosity of \SI{e36}{\per\square\centi\meter\per\second}, corresponding to the order of magnitude needed for the most complete studies at high momentum transfer (3--4\,fm$^{-1}$) required for the charge density extraction, would be obtained with an electron beam of 1\,mm$^2$ size, $I_\text{e}=\SI{20}{\micro\ampere}$ impinging on a \SI{100}{\milli\gram\per\square\centi\meter} target of high-Z elements ($\sim 3 \times 10^{20}$ atoms/cm$^2$). We can immediately realize the distance that separates us from this type of luminosity by considering Table~\ref{tab:frontier:low_energy_nuclear:1} and the fact that a target formed by the accumulation of radioactive ions within a trap cannot exceed ion populations of $\approx 10^{8}$, in order of magnitude, in the interaction region (according to our current know-how \cite{GANIL2}). While an ERL machine like PERLE would deliver beam intensities larger than the presently available ones by two orders of magnitude, gains of one or two additional orders of magnitude will still be required on the ion-capture efficiency side to reach ${\cal L}\approx \SI{e29}{\per\square\centi\meter\per\second}$.
 This also poses challenges in the production of very short-lived radioactive ions ($\approx100$\,ms and lower), and only facilities capable of providing such radioactive beams with intensities greater than $10^7$\,pps will be able to accommodate such a measurement program.
 
 \subsubsection{Nuclear structure observables and required luminosities}
 Orders of magnitudes of required luminosities for different structure observables are given in Table~\ref{tab:frontier:low_energy_nuclear:2} (adapted from \cite{LRP, GANIL2}), with the following ranges of nuclear mass: light, $Z^2 \leq 100$; medium $100 < Z^2 \leq 1024$ ($=32^2$); heavy $Z^2 >1024$.
\begin{table}[htb]
\centering
\begin{tabular}{cccc}
\hline\hline
 \textbf{Observables} & \textbf{Reactions} & \textbf{target mass} & \textbf{Required}  \\
 \textbf{deduced quantities} &  & & \textbf{luminosity}\\
 &&& ${\cal L}$ (\si{\per\square\centi\meter\per\second}) \\
\hline
r.m.s.~charge radii & (e,e) elastic & Light & $\sim 10^{24}$ \\
 & at small q  && \\
\hline
Charge density distribution & (e,e) $1^\text{st}$ min. in & Light & $\sim10^{28}$ \\
$\rho_c$ with 2 parameters & elastic form factor & Medium  & $\sim10^{26}$ \\
&& Heavy & $\sim10^{24}$ \\
\hline
Charge density distribution & (e,e) $2^\text{nd}$ min. in & Medium & $\sim10^{29}$ \\
$\rho_c$ with 3 parameters & elastic form factor & Heavy  & $\sim10^{26}$ \\
\hline
$F_L$, $F_T$ magnetic form factors & (e,e) $2^\text{nd}$ min. in  & Medium & $\sim10^{30}$ \\
p and n transition densities & elastic form factor & Heavy & $\sim10^{29}$ \\
(direct access to neutron skin) &(odd nuclei)&  &  \\ 
\hline
Energy spectra, width, strength, & (e,e') & Medium & $\sim10^{28\text{--}29}$\\
decays, collective excitations && to Heavy & \\
\hline
Extraction of $\rho_c$ using functionals & (e,e) and (e,e') & Light & $\sim10^{30\text{--}31}$ \\
(series of Fourier-Bessel functions...) && Medium & $\sim10^{29\text{--}30}$ \\
 && to Heavy & \\ 
\hline
spectral function, & (e,e'p) && $\sim10^{30\text{--}31}$ \\
correlations  &&& \\
\hline\hline
\end{tabular}
\caption{Required luminosities for different structure observables and target mass regions: light, $Z^2 \leq 100$; medium $100 < Z^2 \leq 1024$ ($=32^2$); heavy $Z^2 >1024$.}
\label{tab:frontier:low_energy_nuclear:2}
\end{table}
A preliminary analysis carried out within the framework of the \enquote{Spiro mission}~\cite{GANIL2} has allowed to dimension the essential constraints and to highlight the main technological challenges. This study clearly shows that whatever the target/ultimate/ideal (to reach ${\cal L}> \SI{e29}{\per\square\centi\meter\per\second}$) electron machine design would be, a key point is the ion capture efficiency. The more efficient the capture is, the less electron intensity is needed. An intermediary step is crucial to study and understand all processes involved, and develop and optimize an original ion trapping system that needs to be tested on a high-performance electron machine to fully explore the ion efficiency by varying some key parameters like the electron beam size. More precisely, if one is to demonstrate the ion capture efficiency, one would need benchmark tests, done at an electron machine which can deliver a beam size smaller than \SI{0.1}{\milli\meter} (or similar to the target one), a sufficiently high average current (to achieve the saturation in the ion trap) and sufficiently high energy. One also needs to have enough place to host the trap plus a detector.\par
If these conditions are met, and they could well be at PERLE, the installation of a nuclear physics device next to this machine, including an intense radioactive ion source (produced by the ISOL method, Isotope Separation On Line) would allow, in a first step, and based only on trapping technologies well established to date, to open a program of e-RI elastic scattering (e,e) measurements with short lived targets in a luminosity regime from $10^{26}$ to \SI{e28}{\per\square\centi\meter\per\second}. It is clear that this would be a resounding world and historical first. This is the goal of the DESTIN@PERLE initiative.
\subsubsection{Selected physics case examples for an e-RI elastic (e,e) scattering program}
\paragraph{Light systems} Few-body nuclear systems (lighter than carbon) are very appealing ones for studying exotic forms of nuclear correlations (clustering, halo phenomena, etc.) and new emergent phenomena related to their behavior as open quantum systems because their description is now achievable via fully microscopic, first-principle-based, ab-initio theoretical approaches. Charge radii of halo nuclei $^6$He and $^8$He have been measured more than a decade ago with laser spectroscopy techniques \cite{Wang04,Mue07}, unveiling nontrivial effects of neutron-proton correlations. Information on the charge density distribution in these systems would provide additional insight and shed light on the properties of alpha clustering. The picture on the neutron-rich side could be complemented by charge radii measurements of certain systems that are up to now not accessible with laser spectroscopy, like $^{12}$Be or $^{17}$C. On the neutron-deficient side, attention has been recently attracted on systems believed to exhibit an extended proton distribution such as $^{8}$B, $^{14}$O or $^{17,18}$Ne. At present, only the charge radius of the latter has been measured.

\paragraph{Bubble nuclei} From empirical considerations and as supported by state-of-the-art theories \cite{Grasso09,Duguet17} $^{34}$Si (sometimes coined ``bubble nucleus'')   is believed to exhibit a pronounced central depletion, but this conjecture can only be experimentally verified by unambiguous measurement of its proton density distribution. An elastic scattering measurement (eventually with other nuclei in the vicinity) would indicate which theoretical assumptions are consistent or not with the measured charge densities. Similarly, charge densities for Sn and Xe isotopes could be obtained from (e,e) scattering measurements and compared to ab-initio calculations.

\paragraph{Symmetry energy of the nuclear equation of state} The study of neutron skins in neutron-rich isotopes would greatly benefit from the combination of electron and proton scattering data on unstable nuclei having different neutron-proton ratios. As exemplified for $^{208}$Pb \cite{Roca11}, this observable helps to shed light on the density dependence of the nuclear symmetry energy of the nuclear equation of state (EoS). Differences in charge radii and densities of proton-rich mirror nuclei will also be directly measurable and helpful to characterize the isospin dependence of the EoS \cite{Brown17,Reinhard16}. These new constraints would thus contribute to improve our understanding of nuclear matter, a necessary step to model different neutron star systems \cite{Baus19}, such as mergers recently identified by the detection of gravitational waves \cite{GW17} and sites of r-process nucleosynthesis \cite{Goriely_2011}.

%\section{Low Energy Nuclear Physics}
% David Verney
%  e- isotope scattering, ..

\section{Photonuclear Physics}\label{sec:frontier:photonuclear}
%Norbert Pietralla, Geoff Krafft
 
 Photonuclear reactions are a prolific tool for investigating 
 complex atomic nuclei in great detail. 
 Since the electromagnetic interaction is fully understood and can 
 be treated perturbatively to any desired precision, photonuclear reactions 
 allow for a separation of the details of the reaction 
 mechanism from the nuclear response under investigation. 
 They, hence, offer insight into the properties and 
 the dynamics of the complex nuclear system with high precision. 
 This makes photonuclear reactions very valuable ingredients for 
 \begin{itemize} 
 \item supporting the usage of complex atomic nuclei for investigating 
 fundamental symmetries of nature, 
 \item testing and further developing  state-of-the-art nuclear modeling 
 in terms of chiral effective-field theories for the nuclear forces, 
 \item uncovering new phenomena of nuclear motion and dynamics, 
 \item providing fundamental data for the modeling of stellar evolution 
 and cosmic events, 
 \item understanding the origin of chemical elements in the universe, or 
 \item medical diagnosis and treatment, commercial applications, and 
 nuclear-waste management. 
 \end{itemize} 
 
\begin{figure}[htb]
\centering
\includegraphics[scale=0.45]{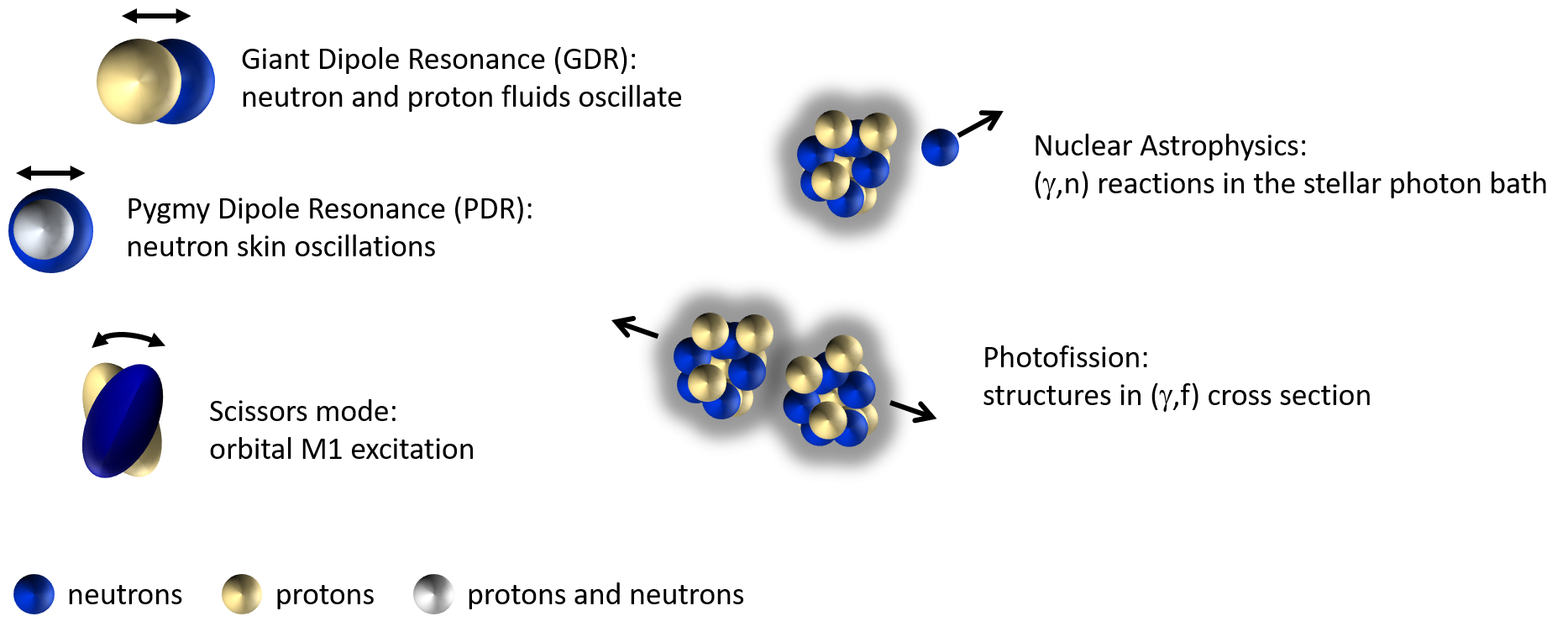}
\caption{Schematic view of some phenomena in nuclei induced by photons. 
From Ref.~\cite{Zilg_PPNP2022}.}
\label{fig:frontier:photonuclear:phenomena}
\end{figure}

 The discovery potential of photonuclear reactions has steadily grown 
 alongside the development of ever more brilliant sources of MeV-ranged 
 photon beams. 
 Comprehensive review articles \cite{knei96, Zilg_PPNP2022} 
 document the vitality of the field of nuclear 
 research and applications using photonuclear reactions. 
 Figure \ref{fig:frontier:photonuclear:phenomena} indicates a few nuclear modes that are 
 predominantly accessible by photonuclear reactions. 

ERL technology can provide cost-efficient, quasi-monoenergetic 
\textgamma-ray beams of unprecedented brilliance from the inverse 
Compton-scattering process of laser beams on the ERL electron beam. 
The technology of SRF-ERLs combines the advantages of highly repetitive  electron beams with large currents with optimum beam emittance 
from a linac and is described in more detail in section \ref{sec:applications:ics}. 
It clearly offers a leap beyond the capabilities of leading present-day 
infrastructure, such as the High-Intensity \textgamma-ray Source 
(HI\textgamma{}S) at Duke Univ., or facilities under construction, 
such as the VEGA system at the European Extreme Light 
Infrastructure---Nuclear Physics (ELI-NP). 
 
The discovery potential of a 4th-generation \textgamma-ray source, 
i.e., an ICS source driven by a high-current ERL, has been reviewed 
recently \cite{Zilg_PPNP2022, Howell_2021}. 
Here, we therefore emphasize only a few complementary research 
opportunities by using shortened adaptations of prior work.
For more details, the readers are referred to the original literature. 
   
\subsection{Testing Fundamental Symmetries}%
\label{sec:frontier:photonuclear:funsym}
 
While the strong force conserves parity, the effective nuclear forces 
violate parity due to contributions of the weak interaction to the
effective nucleon–nucleon interaction. 
At the current stage, the weak meson–nucleon coupling constants deduced 
from various experiments are not consistent.  
Various theoretical and experimental approaches have been employed to investigate parity violation in nuclei. 
In particular, studies of parity doublets $J^\pm$ are well suited to 
the observation of parity violation in nuclei. 

Beller \textit{et al.} \cite{Beller2014} have employed photonuclear reactions 
using linearly and circularly polarized MeV-ranged photon beams from the 
Laser-Compton backscattering (LCB) mechanism at the HI\textgamma{}S facility to characterize the $1^\pm$ 
parity doublet of $^{20}$Ne at \SI{11.26}{\mega\electronvolt} excitation energy. 
They found a small energy separation of \SI{3}{\kilo\electronvolt} and a nuclear enhancement 
factor of \SI{1.4}{\per\kilo\electronvolt} for the study of parity mixing in this doublet. 
This is the largest nuclear enhancement factor known today. 
Their analysis shows that a measurement of the nuclear parity mixing will 
be achievable, provided that brilliant MeV-ranged photon beams at 
intensities two orders of magnitude higher than today will be available. 
An LCB source at a high-current ERL as described here will provide 
the needed intensities. 
 
In addition, considerable international effort is being put into experiments 
to detect neutrino-less double-beta (0\textnu\textbeta\textbeta) decays. 
For a final determination of the neutrino mass, the nuclear matrix 
element (NME) ${\cal M}^{(0\text{\textnu})}$ needs to be calculated sufficiently 
precisely from nuclear structure theory. 
While photonuclear reactions cannot provide information on 
(0\textnu\textbeta\textbeta) decay reactions themselves, they can be used 
instead to assess the 
required nuclear modeling to the desired accuracy. 
Photonuclear reactions on the nucleus $^{154}$Gd, which is 
the double-beta decay product of $^{154}$Sm and the $N=90$ isotone 
of the double-beta emitter $^{150}$Nd, have revealed \cite{Beller2013} 
that the proper modeling of the state-dependent nuclear deformation 
in the initial and final states of 0\textnu\textbeta\textbeta\ decay reactions 
is mandatory for a reliable calculation of the required NMEs. 
In particular, $M1$ decays of the $J^\pi = 1^+$ nuclear scissors mode 
to lower-lying $J^\pi = 0^+$ states with different amounts of quadrupole 
deformation have been measured in photonuclear reactions \cite{Beller2013}. 
The data have led to a significant improvement of our understanding 
for the detailed modeling of 0\textnu\textbeta\textbeta\ NMEs.

\subsection{Constraining Nuclear Models}%
\begin{figure}[htb]
\centering
\includegraphics[scale=0.66]{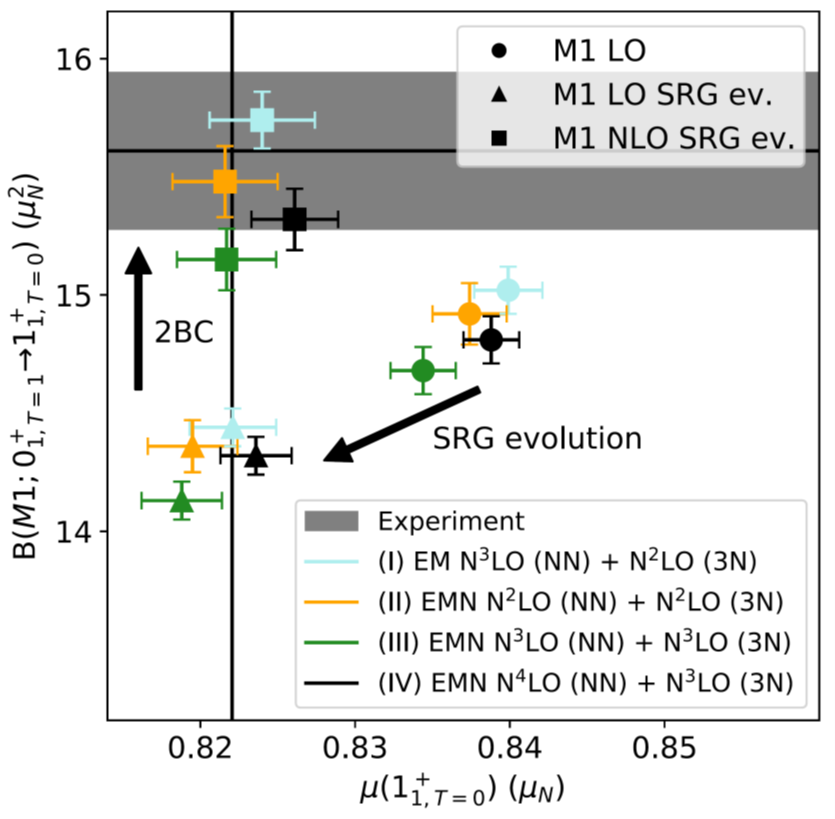}
\caption{Experimental $B(M1)$ value and magnetic moment of $^6$Li in 
comparison to importance-truncated no-core shell model results based 
on chiral effective-field theory interactions at various orders.  
A satisfactory description of the data at the level of precision achieved 
experimentally by photonuclear reactions is only obtained by including 
Similarity Renormalization Group (SRG) evolution in the modeling of the 
nuclear wave functions and 2BC in the $M1$ operator. 
From Ref.~\cite{Friman21}.}
\label{fig:frontier:photonuclear:test2BC}
\end{figure}

\label{sec:frontier:photonuclear:nuclmodels}
Nuclear structure physics has entered an era of precision studies, both in experiment and theory. 
For light nuclei, \textit{ab initio} theory based on interactions from chiral effective-field theory is reaching an accuracy at which corrections to electromagnetic (EM) operators that emerge naturally in the chiral 
expansion become relevant. 
Friman-Gayer \textit{et al.} \cite{Friman21} have recently pioneered the 
method of photonuclear Relative Self-Absorption (RSA) to model-independently 
measure the isovector $M1$ excitation strength of $^6$Li to a precision 
of \SI{2}{\percent}, thereby quantifying the contributions of two-body currents (2BC) 
to the formulation of the $M1$ transition operator at the given resolution 
scale of the model. 
Figure \ref{fig:frontier:photonuclear:test2BC} provides the details of the nuclear model analysis. 
Simultaneous description of the ground state's magnetic moment and the 
$M1$ excitation strength requires the inclusion of 2BC in the $M1$ operator. 
 
The demonstrated experimental constraint to nuclear modeling was made 
possible by the RSA measurement of the half-life of the first excited 
$J^\pi = 0^+$ of $^6$Li to an accuracy at the attosecond scale. 
The powerful RSA method, however, requires a very high luminosity of 
about $10^4$ $\text{\textgamma} / (\si{\second} \Gamma_0$), which is at the limit of 
present-day technology and applicable today only to very peculiar 
cases with extraordinarily large excitation widths $\Gamma_0$ such 
as $^6$Li.
Fourth-generation gamma-ray sources, such as LCB beams from high-current 
ERLs, may provide brilliant MeV-ranged photon beams two orders of magnitude 
beyond what is available today. 
They will make similarly precise tests of nuclear theory possible 
at a much larger variety of nuclear key properties and will thereby 
support the corresponding development of nuclear theory in the future. 
    
\subsection{New Phenomena of Nuclear Collective Modes}
\label{sec:frontier:photonuclear:collmodes}
 Collective nuclear dipole and quadrupole excitations, such as the 
 isovector Giant Dipole Resonance (GDR), the Scissors Mode (ScM), 
 and the proton-neutron symmetric and mixed-symmetry quadrupole-phonon 
 excitations, belong to the fundamental building blocks of nuclear 
 structure. 
 Photonuclear reactions are particularly well suited for precision 
 studies of these modes \cite{Zilg_PPNP2022}. 
 
 While the GDR has been known for almost a century, very little 
 information is available on  
 its \textgamma-decay to the nuclear ground state and or even to excited 
 states. 
 For a few nuclei the GDR has been observed to decay by \textgamma-ray 
 emission on the order of about \SI{1}{\percent} 
 relative to the emission of neutrons. 
 Photonuclear reactions with quasi-monochromatic beams in the energy 
 range between about 10 and \SI{25}{\mega\electronvolt} will enable nuclear physicists to 
 measure the re-emission of \textgamma-rays from the GDR in an energy-resolved 
 manner. 
 Pioneering experiments by Kleemann \textit{et al.} have recently been 
 performed at the HI\textgamma{}S facility on the \textgamma-decay of the 
 GDR of the spherical nucleus $^{140}$Ce and the deformed nucleus 
 $^{154}$Sm. 
 Decay transitions to the first rotational state of the deformed 
 ground state of the latter have been observed for the first time 
 and its \textgamma-decay branching ratio measured. 
 The data challenge the assignment of pure $K$ quantum numbers to the 
 two humps of the GDR in axially quadrupole deformed nuclei and 
 demonstrate the discovery potential of this new approach.

 While the nuclear GDR corresponds to translational out-of-phase 
 oscillations of the proton and neutron fluids, orbital 
 out-of-phase oscillations of a coupled two-component 
 many-body quantum system are generally called scissors modes (ScM). 
 It has initially been discovered in the deformed nuclide $^{156}$Gd. 
 While the nuclear ScM occurs due to the quadrupole deformation of 
 the proton and neutron subsystems, its signature is the electromagnetic 
 coupling to the ground-state band via strong magnetic dipole ($M1$) 
 transitions caused by the predominant isovector character of its 
 decay transitions to low-energy nuclear states with proton-neutron 
 symmetry. 
 Despite its quadrupole-collective origin, the electric quadrupole-decay 
 ($E2$) properties of the ScM were unknown until recently. 
 Utilizing intense quasi-monochromatic MeV-ranged photon beams from the 
 HI\textgamma{}S facility enabled Beck \textit{et al.} \cite{beck17} to not only 
 measure the $E2/M1$ multipole mixing ratio, and hence the $E2$ strength, 
 of the $1^+_\text{sc} \to 2^+_1$ transition unambiguously for the first 
 time, but also to identify the $2^+_\text{sc}$ first rotational excitation 
 of the ScM rotational band. 
 The data agree with early predictions of very small isovector $E2$ 
 transition rates from the ScM but indicate also its surprisingly large 
 rotational moment of inertia. 
 More precision studies of $M1$ excitations of deformed nuclei, including 
 those with $\Delta K = 0$ \cite{beck20}, will be possible with 
 brilliant MeV-ranged photon beams from the next generation of LCB sources.

\subsection{Key Reactions for Stellar Evolution and Cosmic Nucleosynthesis}
\label{sec:frontier:photonuclear:stellarevol}
 
The potential of photonuclear reactions for supporting various 
aspects of contemporary nuclear astrophysics has recently been reviewed 
in a white paper \cite{Howell_2021}: 
Nuclear reactions generate the energy in stars and are responsible for 
the synthesis of the elements. 
When stars eject part of their matter through various means, 
they enrich the interstellar medium with their nuclear ashes and 
thereby provide the building blocks for the birth of new stars, 
of planets, and of life itself. 
Element synthesis and nuclear energy generation in stars are the 
two primary research topics in nuclear astrophysics. 
 
 The beams available at the next-generation LCB sources will enable 
 measurements that contribute to a better understanding of 
 stellar evolution, of the extreme matter in neutron stars, and 
 on the cosmic nucleosynthesis \cite{Howell_2021}. 
 The main opportunities are for cross-section measurements of nuclear 
 resonance florescence (NRF) processes and (\textgamma, particle) reactions. 
 The NRF measurements provide important information for determining photon strength functions (PSFs), electromagnetic transition probabilities, 
 and nuclear structure spectroscopic information, all of which are inputs 
 to nuclear astro-physics reaction-network calculations. 
 The (\textgamma, particle) reaction measurements provide data that are 
 important input for \textgamma-ray-induced reactions on stable nuclei 
 in stars and also for the time reverse of particle capture on 
 unstable nuclei. 
 Of utmost importance will be energy-resolved studies of photofission 
 reactions in the mass range $A \approx 250$ where the $r$-process 
 of nucleosynthesis in binary neutron star mergers is conjectured 
 to terminate and to initiate the $r$-process fission cycle which is 
 crucial for the robustness of the $r$-process elemental abundance pattern.

\subsection{Technological and Commercial Applications}%
\label{sec:frontier:photonuclear:applications}
% by Geoff Krafft 
 Narrowband sources of MeV-scale photons have been proposed in several 
 applications, utilizing their photonuclear reactions on specific 
 target nuclei. 
 Wavelength-tunable sources, for example via inverse Compton scattering, 
 have been proposed as a means of nondestructive radionuclide assay  
 \cite{Haj2008,HAYAKAWA2010695,Hajima_2014}
 and detecting hidden nuclear material  \cite{BERTOZZI2005820,Kikuzawa_2009,Hayakawa_2009}.
 
\begin{figure}[htb]\centering
\includegraphics[width=8cm]{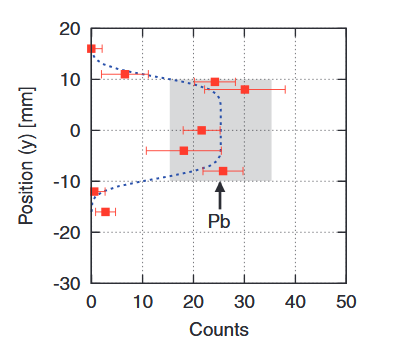}
\caption{Counts detected as a function of vertical gamma beam position during a scan of a small lead block. From~\cite{Kikuzawa_2009}.}
\label{fig:frontier:photonuclear:Pb208nonde}
\end{figure}
 
 An early example of a successful experiment is shown in Figure \ref{fig:frontier:photonuclear:Pb208nonde}. 
 A small lead block, \SI{52}{\percent} $^{208}$Pb, was hidden inside an iron box. 
 A laser-Compton-generated gamma beam was vertically scanned through the box. 
 Scattered photons were detected at ninety degrees to the incident beam 
 direction, resulting in a one-dimensional image of the enclosed lead. 
 The presence and distribution of the lead isotope in the container 
 was clearly detected. 
 As emphasized in the reference, having sufficient photonuclear cross 
 section in the target nucleus is important to allow such remote detections 
 to occur.
%\section{Photo-Nuclear Physics}
% Norbert Pietralla, Geoff Krafft
%

\chapter{Applications}\label{sec:applications}
The first CW ERLs were at Jefferson Lab, funded by the Office of Naval Research (ONR) to develop a high-power infrared (IR) laser, intended to protect ships against cruise missiles.
The FEL was also intended to be for industrial applications and had an active Laser Processing Consortium.
In the area of polymer surface processing, amorphization to enhance adhesion, fabric surface texturing, enhanced food packaging, and induced surface conductivity were being evaluated.
In micromachining, applications were ultrahigh density CD-ROM technology, surface texturing; micro-optical components, Micro-Electrical Mechanical Systems (MEMS), and six-axis micromachining was to be used for producing micro-satellites.
In metal surface processing, proposed applications were laser glazing for corrosion resistance and adhesion pre-treatments.
In electronic materials processing, large-area processing (flat-panel displays) and a laser-based \enquote{cluster tool} for combined deposition, etching, and in-situ diagnostics were being developed.

This highly interdisciplinary range of applications led to several significant successes: the high-power IR light was shown to preferentially heat fat in skin-and-fat samples of pig tissue, leaving the other portions relatively unaffected.
The result could one day be a safe, effective treatment of acne, cellulite, and even heart disease.
The work on boron nitride nanotubes led to a spin-off company, which is now a thriving company with a full order book, producing the purest boron nitride nanotubes in the world.

While it is regrettable that much of the application development specific to the Jefferson Lab FEL was cut short when the program lost ONR funding, this did not lead to the extinction of ERL-driven FELs: The Recuperator in Novosibirsk, which was the first multi-pass CW ERL, is now being used to drive three separate FELs at different energies, providing a wide range of wavelengths.
Future ERL-driven FELs will be discussed in detail in Section \ref{sec:applications:highpower_fel}.
Additionally, a joint effort between Jefferson Lab and ASML was initiated to develop ERL-driven FELs for semiconductor lithography.
This early work was followed up at other laboratories, notably Daresbury and KEK.
This will be discussed in detail in Section \ref{sec:applications:euv-fel}.

Cornell proposed using an ERL to produce high-brightness synchrotron radiation in CESR, with a single turn of the electrons in the storage ring, maintaining the small emittance of the photo-injector.
However, the advent of multi-bend achromats, pioneered at MAX in Lund, enabled storage rings to compete with the brightness of an ERL-driven ring, and the project was not funded, nor has any other laboratory followed this direction. 

A different source of photons is Compton back-scattering from an ERL.
In this case, the use of an ERL provides high brightness because of the high currents that are possible.
This was the initial motivation for the cERL facility at KEK, Japan.
The recent BRIX proposal in Milan is an Inverse Compton Source (ICS) using a modified ERL layout to achieve high flux and brightness.
The applications intended for BRIX are medically oriented research/investigations, mainly in the radiodiagnostics and radiotherapy fields, exploiting the unique features of monochromatic X-rays, as well as in microbiological studies, and, within this mainstream, material studies, crystallography and museology for cultural heritage investigations.
ICS gamma sources will be discussed in detail in Section \ref{sec:applications:ics}.

\section{ERL-Driven High-Power FEL}\label{sec:applications:highpower_fel}
% Frank Zimmermann, Zafer Nergiz, Avni Aksoy, Najmeh Mirian, Demin Zhou 

The high-current ERL of the LHeC would provide the opportunity for driving a Free Electron Laser (FEL) \cite{schopper, Zimmermann:IPAC2019-TUPRB076}. 
Though the LHeC is designed for energy frontier electron-hadron scattering experiments at the LHC, it is conceivable that the ERL program can be temporarily redefined, independently of electron-hadron operation, as, for example, during the decade in which the LHC may possibly be reconfigured to double its hadron beam energy within the High Energy LHC (HE-LHC) proposal~\cite{HELHC}, 
and during which no lepton-hadron collisions would take place.
In light of the performance expected from the LHeC-FEL,
the construction of a dedicated ERL-based X-ray FEL user facility could---
and, perhaps, should---be considered as well.

For the LHeC proper, the electron-beam emittance is not critical, since the proton-beam 
emittance is quite large.
Incoherent synchrotron radiation significantly increases 
the normalized RMS 
emittance during the arc passages at 40 and 50 GeV beam energy,
by about \SI{7}{\micro\meter} \cite[Table 7.14]{LHeCdesign}.
 However, in order to obtain coherent 
 X-rays at low wavelengths in FEL operation, the beam emittance must be sufficiently small. 
Partly because of this emittance requirement, for the FEL operation,  
the electron beam energy is chosen as \SI{40}{GeV}  or lower \cite{Zimmermann:IPAC2019-TUPRB076}, 
depending on the X-ray wavelength desired, rather than the nominal LHeC beam energy of \SI{60}{GeV}.
% Accordingly the beam is accelerated (or decelerated) 
% on either one or two turns, instead of three.

The beam energy of \SI{40}{GeV} can be attained after two passes through 
the two \SI{10}{GeV} linacs, instead of the three passes of the standard LHeC operation. 
The subsequent deceleration would also happen 
during two additional passes. An energy of \SI{20}{GeV}
would already be achieved after a single pass through the two linacs, 
again followed by another pass of deceleration.
Beam energies of 10 and \SI{30}{GeV} are also readily obtained after one or two turns,
with appropriate linac voltages and phasing. 

The possible performance was simulated \cite{nergiz2020}, 
taking into account
 linac wake fields, incoherent synchrotron radiation, and 
coherent synchrotron radiation (CSR) including shielding, 
using the codes CSRZ \cite{zhoujjap,zhouipac12}, 
ELEGANT \cite{elegant}, and GENESIS \cite{genesis}.
Resistive-wall wake fields were not included in the simulations
but only estimated analytically.

\begin{figure}[htb]\centering
\tikzsetnextfilename{s61_fel_power_g}
\begin{tikzpicture}
\begin{axis}
[
    scale only axis,
    width=.8\linewidth,
    height=.5\linewidth,
    xlabel={$z$ (m)},
    ylabel={$P$ (W)},
    ymode=log,
    ymin=1e4,
    ymax=1e12,
    xmin=-5,
    xmax=150,
    legend pos=south east,
    cycle multiindex* list={
        mark list\nextlist
        Dark2-4\nextlist
        mylinestyles\nextlist
    },
]
\addplot+[mark=none, thick] table[x index=0, y index=1] {figures/s61/s61_fel_power_g_pmax_z10.dat};
\addlegendentry{\SI{765}{\pico\meter}}
\addplot+[mark=none, thick] table[x index=0, y index=1] {figures/s61/s61_fel_power_g_pmax_z20.dat};
\addlegendentry{\SI{203.3}{\pico\meter}}
\addplot+[mark=none, thick] table[x index=0, y index=1] {figures/s61/s61_fel_power_g_pmax_z30.dat};
\addlegendentry{\SI{98.25}{\pico\meter}}
\addplot+[mark=none, thick] table[x index=0, y index=1] {figures/s61/s61_fel_power_g_pmax_z40.dat};
\addlegendentry{\SI{50.45}{\pico\meter}}
\end{axis}
\end{tikzpicture}
\caption{Growth of photon pulse power  at 
\SI{7.6}{\angstrom} (solid green), \SI{2}{\angstrom} (dashed orange), \SI{1}{\angstrom} (dotted blue) 
and \SI{0.54}{\angstrom} (dot-dashed magenta) for an LHeC electron beam of energy
10, 20, 30 and 40\,GeV, respectively,  
passing through the undulator FEL line with period 
$\lambda_u=\SI{39}{\milli\meter}$, as simulated with the code GENESIS \protect\cite{nergiz2020}.%
}%
\label{fig:FEL_powerG}
\end{figure}
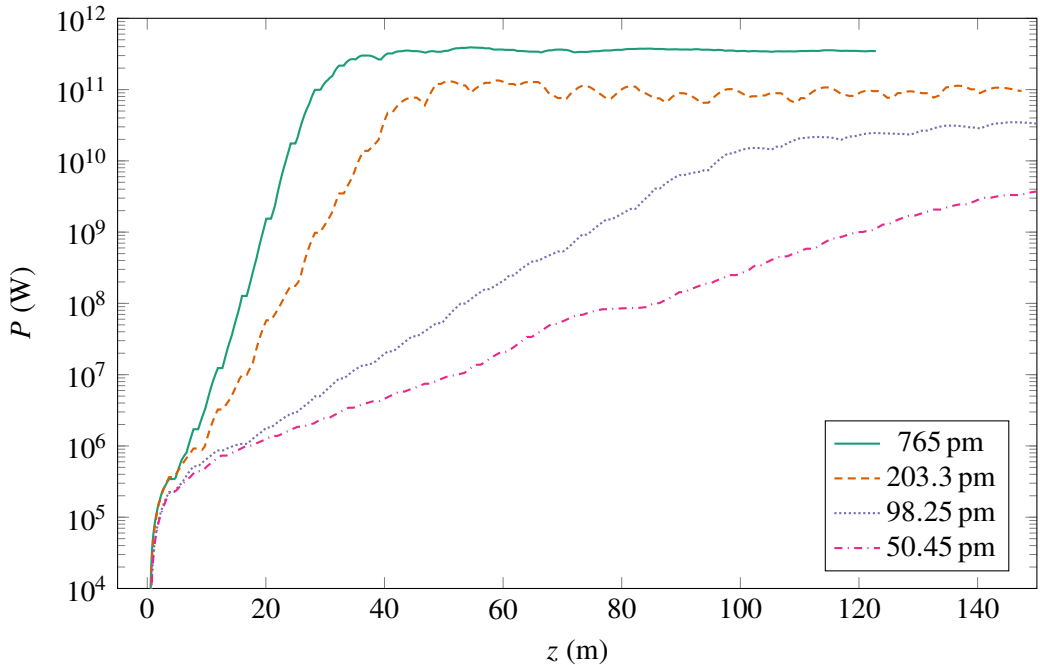

Figure~\ref{fig:FEL_powerG}
shows the simulated power growth at different FEL wavelengths 
generated by electron beams of the corresponding  energies. 
Depending on the wavelength, saturation occurs after a 
distance varying between \SI{30}{\meter} and about \SI{120}{\meter}.

\begin{figure}[htb]\centering
\tikzsetnextfilename{s61_fel_bril}
\begin{tikzpicture}
\begin{groupplot}
[
    group style={
        group size=1 by 2,
        vertical sep=2cm,
    },
    scale only axis,
    width=.8\linewidth,
    height=.5\linewidth,
    xlabel={Energy (keV)},
    ymode=log,
    legend style={font=\footnotesize},
    ylabel style={align=center},
    legend cell align={left},
]
\nextgroupplot[
    cycle multiindex* list={
        mark list\nextlist
        Dark2-8\nextlist
        mylinestyles\nextlist
    },
    xmin=0.05,
    xmax=1000,
    xmode=log,
    ymin=1e18,
    ymax=1e36,
    ylabel={Peak brilliance, \SI{0.1}{\percent} BW\\(\si{\per\second\per\square\milli\meter\per\square\milli\radian})},
]
\addplot+[mark=none, thick] table[x expr={\thisrowno{0}*1e-3}, y index=1] {figures/s61/brilliance_peak/lhec_blist.txt};
\addlegendentry{LHeC ERL}
\addplot+[mark=none, thick] table[x expr={\thisrowno{0}*1e-3}, y index=1] {figures/s61/brilliance_peak/xfel2.txt};
\addlegendentry{XFEL}
\addplot+[mark=none, thick] table[x expr={\thisrowno{0}*1e-3}, y index=1] {figures/s61/brilliance_peak/lcls.txt};
\addlegendentry{LCLS}
\addplot+[mark=none, thick] table[x expr={\thisrowno{0}*1e-3}, y index=1] {figures/s61/brilliance_peak/swissFel.txt};
\addlegendentry{SwissFEL}
\addplot+[mark=none, thick] table[x expr={\thisrowno{0}*1e-3}, y index=1] {figures/s61/brilliance_peak/petra3.txt};
\addlegendentry{Petra III}
\addplot+[mark=none, thick] table[x expr={\thisrowno{0}*1e-3}, y index=1] {figures/s61/brilliance_peak/esrf.txt};
\addlegendentry{ESRF}
\addplot+[mark=none, thick] table[x expr={\thisrowno{0}*1e-3}, y index=1] {figures/s61/brilliance_peak/bessy.txt};
\addlegendentry{BESSY}
\addplot+[mark=none, thick] table[x expr={\thisrowno{0}*1e-3}, y index=1] {figures/s61/brilliance_peak/als.txt};
\addlegendentry{ALS}
\nextgroupplot[
     cycle multiindex* list={
         mark list\nextlist
         Dark2-4\nextlist
         mylinestyles\nextlist
     },
    xmin=-1,
    xmax=25,
    ymin=1e18,
    ymax=1e30,
    ylabel={Average brilliance, \SI{0.1}{\percent} BW\\(\si{\per\second\per\square\milli\meter\per\square\milli\radian})},
    legend pos=south east,
]
\addplot+[mark=none, thick] table[x expr={\thisrowno{0}*1e-3}, y index=1] {figures/s61/brilliance_avg/lhec_blist.txt};
\addlegendentry{LHeC ERL}
\addplot+[mark=none, thick, forget plot] table[x expr={\thisrowno{0}*1e-3}, y index=1] {figures/s61/brilliance_avg/lcls2.txt};
\addplot+[mark=none, thick] table[x expr={\thisrowno{0}*1e-3}, y index=1] {figures/s61/brilliance_avg/lcls2_2.txt};
\addlegendentry{LCLS-II HE}
\addplot+[mark=none, thick] table[x expr={\thisrowno{0}*1e-3}, y index=1] {figures/s61/brilliance_avg/xfel2.txt};
\addlegendentry{XFEL}
\addplot+[mark=none, thick] table[x expr={\thisrowno{0}*1e-3}, y index=1] {figures/s61/brilliance_avg/lcls.txt};
\addlegendentry{LCLS (\SI{120}{\hertz})}
\end{groupplot}
\end{tikzpicture}
\caption{Comparison of FEL peak (top)
and average brilliance (bottom) for the LHeC-FEL
with several existing or planned hard X-ray FEL and SR sources \protect\cite{nergiz2020}.}
\label{fig:applications:highpower_fel:bril}
\end{figure}
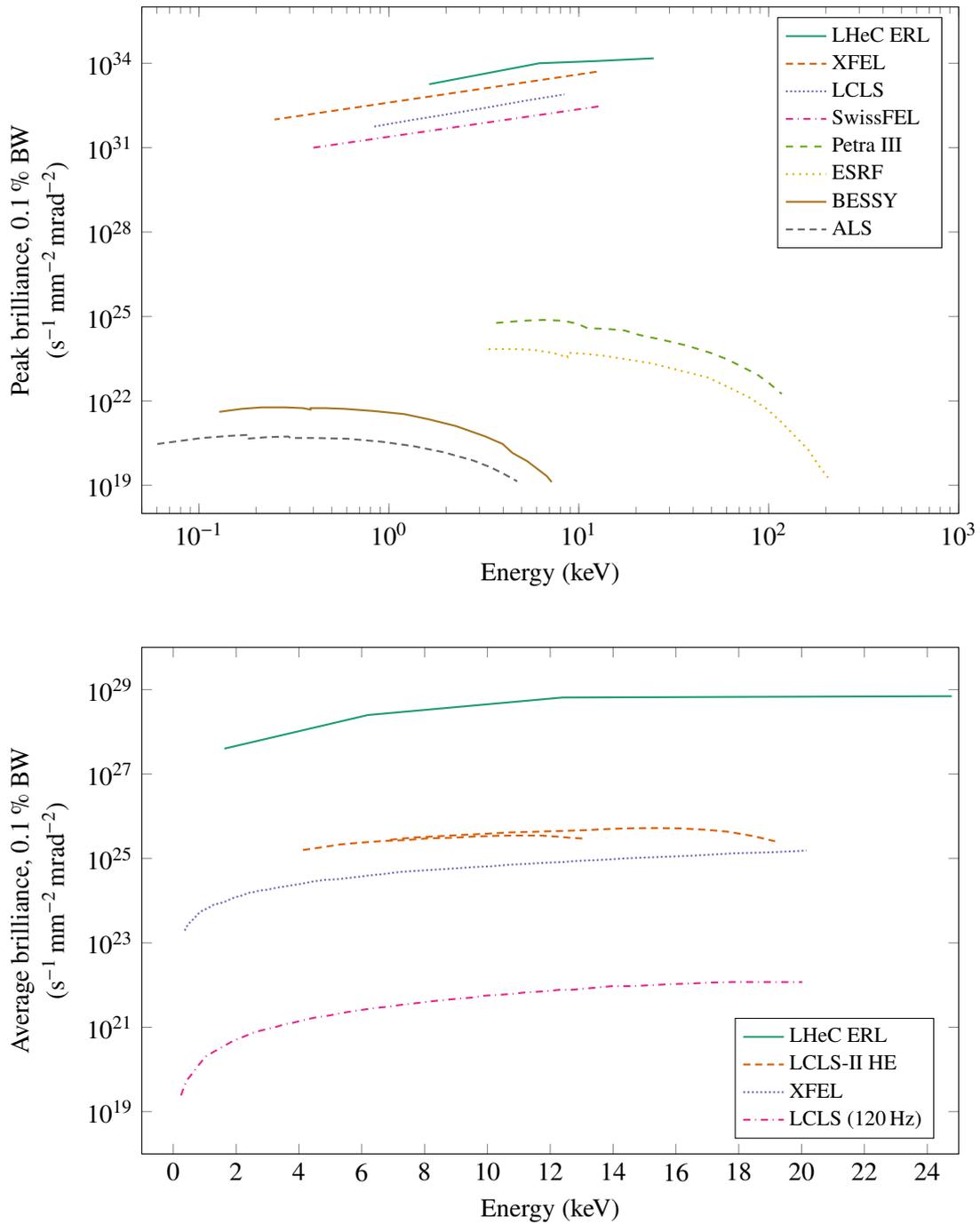

A comparison of the LHeC ERL-FEL with
a few existing and planned hard X-ray sources 
\cite{LCLS2,brachmann,SLSdesign,XFEL,LCLS2_2}  
is presented in Fig.~\ref{fig:applications:highpower_fel:bril}.
These figures demonstrate 
that the peak brilliance of the LHeC ERL-FEL is as high as 
that of the European XFEL, while the average 
brilliance is orders of magnitude higher, thanks to the high
average beam current, enabled by energy recovery.

\begin{figure}[htb]\centering
\includegraphics[width=0.9\textwidth]{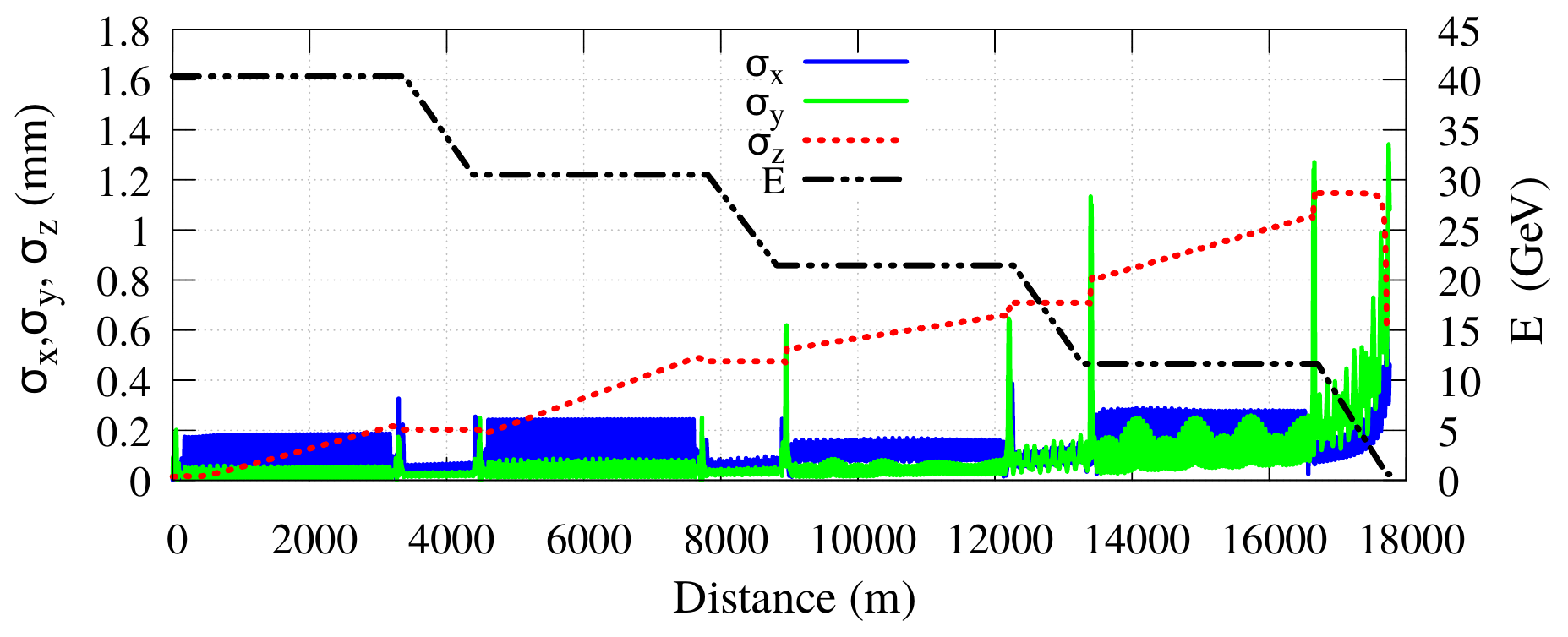} 
\caption{Beam energy and beta functions for the deceleration of the spent beam, after lasing at \SI{0.5}{\angstrom},  
over two complete LHeC turns starting from \SI{40}{GeV} \protect\cite{nergiz2020}.}
\label{fig:applications:highpower_fel:twiss}
\end{figure}

To demonstrate the feasibility of
energy recovery during FEL operation, not only the acceleration process, but also the deceleration
process was simulated from the maximum beam energy about \SI{40}{GeV}
down to about \SI{0.5}{GeV},
starting with the beam distribution exiting the undulator, 
after lasing at a wavelength of \SI{0.5}{\angstrom}.
This distribution, modelled by \num{8e5} macroparticles representing a single bunch, was obtained from
the GENESIS FEL simulation \cite{nergiz2020}. 
The simulation code ELEGANT 
%  PLACET \cite{placet2,placet}
was used to track the \num{3e5}
macroparticles through the exact optics \cite{LHeCdesign,bogaczarc}
for the last two decelerating turns (four arcs and four linac passages)
of the LHeC, composed of 16,000 beam-line elements. 
As also for the acceleration,  both 
the linac wake fields 
and the shielded CSR in the arcs were 
taken into account. The energy
spread and bunch length during deceleration 
were controlled by adjusting 
the bunch arrival phase
in the linacs. Figure~\ref{fig:applications:highpower_fel:twiss}
shows the simulated beam size, bunch length and beam energy during the 
deceleration process. In the simulation,  
not a single macroparticle is lost. 
The final RMS beam size, which is of the order of \SI{1}{\milli\meter},
is much smaller than   
the linac RF cavity iris radius of \SI{7}{\centi\meter} \cite{Calaga:2020926}.  
Deceleration is 
also possible, and even 
easier, for the \SI{20}{GeV} 
single-turn ERL operation.

%\section{ERL Driven High Power FEL}
% Frank Zimmermann
 
\section{EUV-FEL Semiconductor Lithography}
\label{sec:applications:euv-fel}
%Peter Williams, George Neil, Hiroshi Kawata

The industrial process of producing semiconductor chips comprises the placing of electronic components of nanometre scale onto a substrate or wafer via photolithography. Light transfers a pattern from a photomask to a light-sensitive photoresist deposited on the substrate. Chemical treatments then etch this pattern of exposure producing the integrated circuits. The wavelength of the light used for this exposure must be close to the required component size in the final circuit. The power of the light source determines the rate at which wafers can be scanned, which in turn determines the cost of production. Standard lithography of semiconductor chips currently utilises argon-fluoride laser-produced \SI{193}{\nano\meter} radiation. However, since 2016, this is gradually being augmented/replaced by the finer patterning enabled by shorter-wavelength illumination at \SI{13.5}{\nano\meter}, referred to in the industry as EUV. This is the main contemporary technique allowing the industry to keep up with Moore's Law, namely the societal demand for computational processing power to double roughly every $18$ months. Within commercially available photolithography scanners EUV is produced through the excitation of a tin or xenon plasma with a carbon-dioxide laser. Commercial scanners can now produce EUV with a few \SI{100}{\watt} average power. An alternative method would be the deployment at chip fabrication plants of an EUV Free-Electron Laser (FEL) capable of operating at multi-kW average power. An EUV-FEL would have the advantage of scale-ability to both higher power and shorter wavelength than laser-plasma methods. Such a possibility has been investigated in detail by a leading company in lithography apparatus, but as yet the low level of technical maturity of high-average-power FELs has prohibited commercial commitment.\par
In order to produce the necessary EUV power to make deployment of FELs feasible for industrial photolithography with acceptable operating costs, the FEL must be driven by a superconducting ERL. An ERL with electron beam energy of $\sim \SI{1}{\giga\electronvolt}$ would enable multi-kW production of EUV. This would benefit the global semiconductor industry by allowing study of FEL capabilities at an industrial output level, and developing and proving kW-capable EUV optical elements/beamlines for photon transport to chip scanners. Such a flexible, industrial-research-led tool would also facilitate investigations beyond the current industry state-of-the-art. The superior flexibility and controllability of FEL photons as compared to ``brute-force'' laser-plasma produced photons would have far-reaching implications for many diverse aspects of lithography. Examples of this could include shorter wavelengths, or indeed variation of wavelength (two-colour exposures), variation of polarisation or transverse coherence in differentiating patterning layers, or generation of orbital angular momentum photons to manipulate helicity within a wafer. There is clear potential for novel semiconductor devices to be developed utilising such source capability.\par
An EUV lithography ERL-FEL would be the first deployment of a large scale particle accelerator in an industrial, rather than research, setting. The most stringent challenge in making this translation will be the operational reliability. This because one machine is likely to provide light to an entire chip fabrication factory comprising many individual scanners. As such the cost of source downtime is likely to run into millions of Euros per hour. This entry barrier points to the initial possibility of a hybrid research / industrial development platform that would enable the application of systems engineering techniques to make incremental reliability improvements to all components. This has been the successful model employed within the semiconductor industry over past decades.\par

%\textbf{PW to approach ASML again to seek permission to disclose a few more details?}

\subsection{Example ERL-FEL high-power EUV light source for lithography}
An ERL-FEL based EUV light source has been designed by KEK using available technologies to assess the feasibility of generation of EUV with average power more than \SI{10}{\kilo\watt}. For industrialization, high availability is essential as well as high power, and reduction of the light source size is preferred. In the following, a brief outline of the designed ERL-FEL based EUV light source for semiconductor lithography as well as some considerations and developments for obtaining high availability and size reduction of the machine are given~\cite{Nakamura2018ERLFELBH}.
\begin{figure}[htb]
\centering
\includegraphics[scale=0.33]{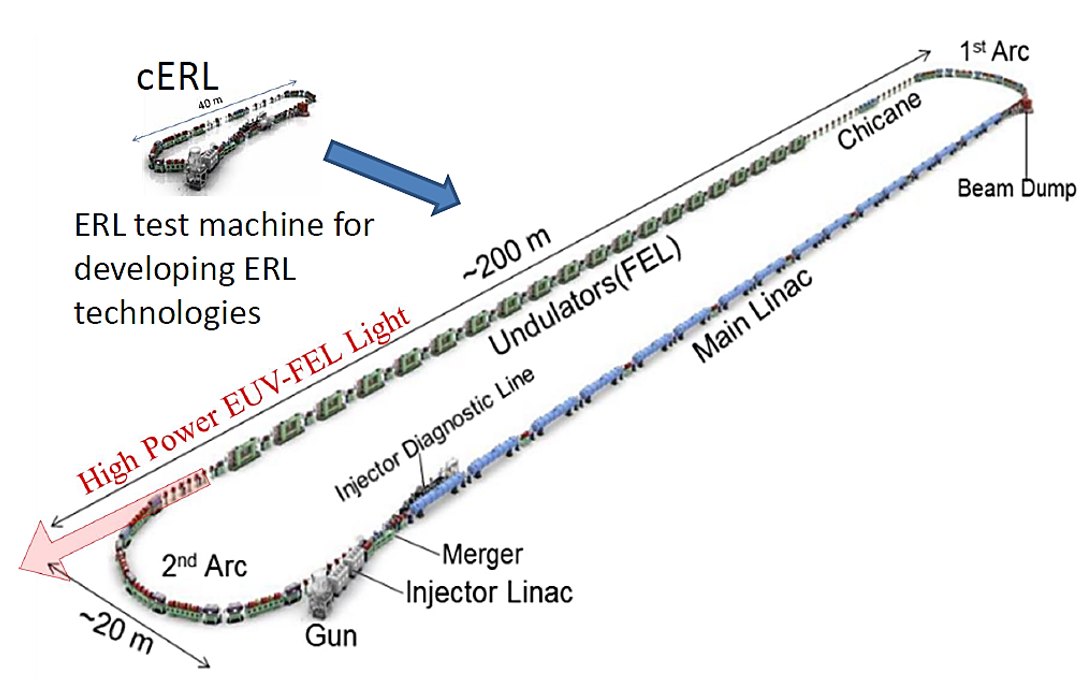}
\caption{Prototype design of the EUV-FEL.}
\label{fig:EUV}
\end{figure}

Figure~\ref{fig:EUV} shows a schematic EUV light source~\cite{Kawata:18}, and Table~\ref{tab:applications:euv-fel:cerltab} shows the design parameters. The electron beam is injected at \SI{10.5}{MeV} and accelerated to \SI{800}{MeV} by the main linac.
It is then magnetically compressed within the first arc to obtain the high peak current required for SASE-FEL lasing. Then it is passed through the undulator, producing \SI{13.5}{\nano\meter} radiation with an output of \SI{10}{\kilo\watt} or higher average power.
Following lasing, it is decompressed within the second arc, then decelerated in the main linac for energy recovery, and discarded in the beam dump. With a bunch charge of \SI{60}{\pico\coulomb} and a bunch repetition frequency of \SI{162.5}{MHz}, the average current is $\sim \SI{10}{\milli\ampere}$. The discarded beam power is reduced from \SI{8}{\mega\watt} to \SI{100}{\kilo\watt} by the energy recovery process.

%%%%%%%%%%%%%%%%%%%% TABLE starts here %%%%%%%%%%%%%%%%%%%%

\begin{table}[htb]\centering
\footnotesize
\caption{Specification of the EUV-FEL and proposed demonstration steps using the KEK cERL (see Section~\ref{sec:cERL}).}
\begin{tabular}{lccc}
\toprule
   & EUV-FEL & cERL-MIR-FEL & cERL-MIR-FEL \\
   & & (Target) & (Preliminary) \\
 \midrule
Beam energy	& \SI{800}{MeV} &	\SIrange{17.5}{20.0}{MeV} &	\SIrange{17.5}{20.0}{MeV} \\
Beam current  (ave.) &	\SI{10}{mA} &	\SI{5}{mA} & Burst mode \\
Bunch charge &	\SI{60}{pC} &	\SI{60}{pC} &	\SI{60}{pC} \\
Bunch length (FWHM) & \SI{0.1}{ps} & \SIrange{0.5}{2}{ps} & \SIrange{3}{5}{ps} \\
Normalized emittances &	$~\SI{0.7}{\milli\meter\milli\radian}$ &	$~\SI{3}{\milli\meter\milli\radian}$	& $\SIrange{3}{10}{\milli\meter\milli\radian}$ \\
Energy spread &	\SI{0.03}{\percent} & \SI{0.1}{\percent} & \SI{0.3}{\percent} \\
Repetition rate & 162.5 MHz & 81.25 MHz & 81.25 MHz \\
Undulator type & APPLE II & Planar & Planar \\
Length ($\text{period}\times\text{number}$) & \SI{5}{\meter} ($\SI{28}{\milli\meter} \times 175$) & \SI{3}{\meter} ($\SI{24}{\milli\meter} \times 124$) & \SI{3}{\meter} ($\SI{24}{\milli\meter} \times 124$) \\
Number of units & 17 & 2 & 2 \\
FEL wavelength & \SI{13.5}{\nano\meter} & \SIrange{15}{20}{\micro\meter} & \SIrange{11}{20}{\micro\meter} \\
Output power (ave.) & $>\SI{10}{\kilo\watt}$ & \SI{1}{\watt} & several--tens mW \\
\bottomrule
\end{tabular}
\label{tab:applications:euv-fel:cerltab}
\end{table}

%%%%%%%%%%%%%%%%%%%%%%%%%%%%%%%%%%%%%

A stepwise development has been proposed utilising upgrades to the cERL to realize an EUV-FEL light source~\cite{Kato:20}. This also builds upon scanner R\&D work undertaken on dedicated EUV beamlines at the NewSUBARU storage ring adjacent to SPring-8:
\begin{enumerate}
  \item Development of the feasible technologies;
  \item Establishment of the EUV-FEL Lithography system;
  \item International Development Center on the processing of EUV-FEL lithography.
\end{enumerate}
The first step would include a Proof of Concept (PoC) machine for ERL-based SASE-FEL. Thus, ERL-based SASE-FEL light production guaranties the high-power requirement, even though the FEL wavelength is not EUV wavelength. Even at the mid-infrared (MIR) light sources, there is to date no such high-repetition-rate and high-power light source without a mirror system in which the wave length is tune-able. The JLab FELs were oscillators rather than SASE. In 2019, KEK started to contribute another project to develop high average power mid-infrared FEL with an \SI{81.25}{MHz} repetition rate. The details of the project have been presented at several conferences \cite{Nakamura:2021} and has led to publications in collaboration with industry investigating known areas in which R\&D will be required~\cite{Brynes2018BeyondRadiation,Brynes_2021,Brynes2020MicrobunchingPulses,Di_Mitri_2020,Brynes:2020kkl}. This FEL could also develop into a proof-of-concept machine for an EUV-SASE-FEL. 

\subsection{Future prospects}
A high-power EUV light source using an ERL-FEL could be an epoch-making light source that supplies a large number of semiconductor lithography scanners with $\sim$~kW class EUV. Developments in the reliability of accelerator components are required for such large scale industrial deployment, and necessary developments are under investigation. For example, in order to improve uptime, it is important to remotely control the preparation and replacement work of the electron gun cathode~\cite{PhysRevAccelBeams.22.053402}, improve the trip rate of the superconducting cavity~\cite{umemori}, develop an in-situ recovery method for field emission~\cite{PhysRevAccelBeams.22.022002}, and design a redundant system.
Remote control of cathode preparation / replacement is not a major technical problem, and the trip rate of the superconducting cavity is not serious from the operation of cERL. Pulse processing is used to suppress field emission, but other methods should be also considered. In order to reduce the size of the light source, high acceleration gradient of the acceleration cavity and multi-pass recirculation are being considered. In the former, the development of clean assembly technology and nitrogen doping technology for cavities and cryomodules are underway. For the latter, the challenge is to advance the design research of the double-loop structure. Reducing beam energy is not effective in reducing size, as it results in a significant reduction in EUV output.
%\section{EUV-FEL Semiconductor Lithography}
%Peter Williams, George Neil

\section{ICS Gamma Source}
\label{sec:applications:ics}
%Peter Williams

\begin{figure}[htb]
    \centering
    \includegraphics[width=0.47\textwidth]{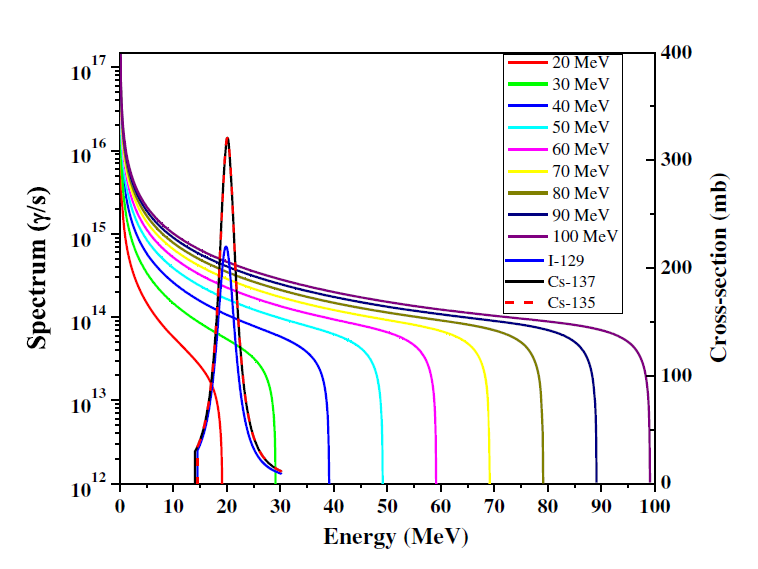}
    \includegraphics[width=0.5\textwidth]{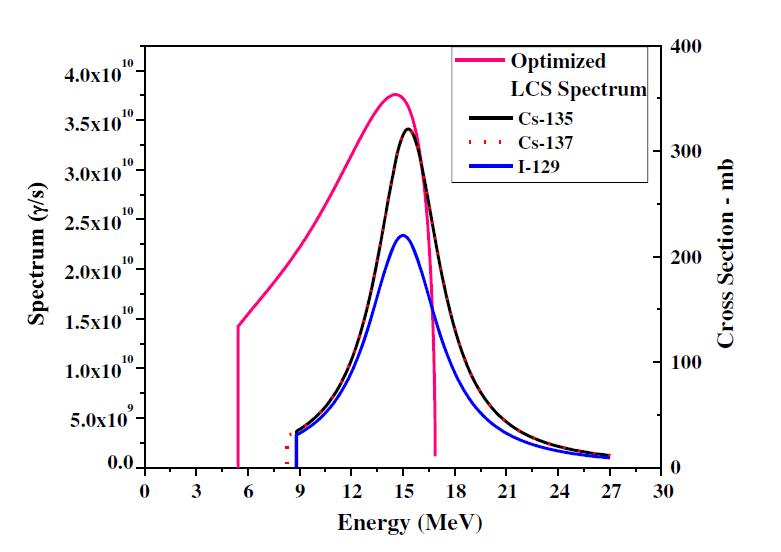}
    \caption{Bremsstrahlung (left) and collimated ICS (right) gamma spectra compared to photonuclear dipole resonances of $^{135}\text{Cs}$ and $^{129}\text{I}$ (right only)~\cite{https://doi.org/10.1002/er.3904}}
    \label{fig:applications:ics:brem}
\end{figure}

A \SIrange{1}{2}{\giga\electronvolt} superconducting ERL producing high average electron current in the \SIrange{10}{100}{\milli\ampere} range would enable a high-flux, narrowband gamma source based on inverse Compton scattering (ICS) of the electron beam with an external laser within a high-finesse recirculating laser cavity. 
The production of \SIrange{1}{100}{\mega\electronvolt} gammas via ICS results in properties of the gamma beam  fundamentally improved with respect to standard bremsstrahlung generation. Bremsstrahlung is broadband emission peaked at low energy with a cut-off at the electron energy.
ICS has a correlation between the angle of emission and energy, therefore in combination with angular collimation, narrowband (or \enquote{monoenergetic}) gamma beams can be produced.
A comparison is shown in Figure~\ref{fig:applications:ics:brem}.
In addition, ICS preserves the polarisation of the incident laser. Presently the worlds brightest narrowband gamma source is the High Intensity Gamma Source (HI\textgamma{}S) at Duke University~\cite{WELLER2009257}. In the construction phase is the EU funded Extreme Light Infrastructure – Nuclear Physics (ELI-NP) Variable Energy Gamma System (VEGA) in Magurele, Romania~\cite{book}.\par
To quantify the potential of a GeV-scale ICS source driven by an SC-ERL, calculations of the source properties have been performed based on the methods used for a similar proposal on CBETA~\cite{PhysRevAccelBeams.24.050701} with input electron beam (Table~\ref{TAB:DIANA_electron_bunch}) and laser beam (Table~\ref{TAB:DIANA_laser_pulse}) are shown in Table~\ref{tab:DIANA_spectral_output}.
For the interaction point we assume a Fabry-Perot bow-tie like resonator cavity as demonstrated at KEK~\cite{Akagi:2017blh}.
Shown are parameters achievable in each of three recirculation turns with the top \textgamma~energy of \SI{20}{\mega\electronvolt}.

\begin{table}[htbp]
\centering\footnotesize
\begin{threeparttable}
\begin{tabular}{lccccc}
\toprule
Parameter & \multicolumn{3}{c}{Quantity} & Unit \\
\midrule
Turn number & 1 & 2 & 3  \\
Injection Energy, $E_{\mathrm{inj}}$ & \multicolumn{3}{c}{7} & MeV\\
\tnote{$\dagger$}~Electron kinetic energy, $E_\text{e}$ & 362 & 717 & 1072 & MeV\\
Bunch repetition rate, $f_\text{b}$ & \multicolumn{3}{c}{125} & MHz\\
Bunch charge, $e N_\text{e}$ & \multicolumn{3}{c}{100} & pC \\
Beam current, $I$ & \multicolumn{3}{c}{12.5} & mA \\
Transverse norm.~\textit{rms} emittance, $\epsilon_\text{N}$ & \multicolumn{3}{c}{0.5} & \si{\milli\meter\milli\radian}\\
\textit{rms} bunch length, $\Delta \tau$ & \multicolumn{3}{c}{0.9 (3)} & mm (ps)\\
Bunch spacing, $t_\text{b}$ & \multicolumn{3}{c}{8} & ns \\
RF frequency, $f_\text{RF}$ & \multicolumn{3}{c}{750} & MHz \\
\tnote{*}~Absolute energy spread, $\Delta E_\text{e}$ & \multicolumn{3}{c}{$\sim\num{10}$} & keV \\ 
\tnote{*}~Relative energy spread, $\left(\Delta E_\text{e}/E_\text{e}\right)$ & \multicolumn{3}{c}{$\sim\num{e-5}$} & \\
\midrule
\multicolumn{5}{c}{Baseline Parameters} \\
\midrule
$\beta$-functions at the IP, $\beta_{x}^{*}$/$\beta_{y}^{*}$ & 0.2/0.2 & 0.2/0.2 & 0.2/0.2 & m \\
Electron bunch spot size, $\sigma_{\text{e},x}$/$\sigma_{\text{e},y}$ & 11.87/11.87 & 8.44/8.44 & 6.90/6.90 & \si{\micro\meter}\\
\midrule
\multicolumn{5}{c}{Optimised \SI{0.5}{\percent} \textit{rms} Bandwidth} \\
\midrule
$\beta$-functions at the IP $\beta_{x}^{*}$/$\beta_{y}^{*}$ & 1.33/0.298 & 2.62/0.587 & 3.90/0.874 & m \\
Electron bunch spot size, $\sigma_{\text{e},x}$/$\sigma_{\text{e},y}$ & 30.62/14.49 & 30.54/14.46 & 30.48/14.43 & \si{\micro\meter}\\
Collimation Angle, $\theta_{\mathrm{col}}$ & 0.180 & 0.091 & 0.061 & mrad \\ 
\bottomrule
\end{tabular}
\begin{tablenotes}
\item[*]{Estimated values.}
\item[$\dagger$]{Electron beam energies to accomplish $E_{\text{\textgamma}}^{\mathrm{max}}$ = \SI{20}{\mega\electronvolt} \textgamma-rays. $\Delta E_{\mathrm{turn}} = \SI{355}{\mega\electronvolt}$.}
\end{tablenotes}
\end{threeparttable}
\caption{Electron bunch properties used for calculation of output shown in Table~\ref{tab:DIANA_spectral_output}}
\label{TAB:DIANA_electron_bunch}
\end{table}

\begin{table}[htbp]
\centering
\begin{tabular}{lcc}
\toprule
Parameter & Quantity & Unit \\
\midrule
Wavelength, $\lambda_\textrm{laser}$ & 1064 & nm\\
Photon energy, $E_\textrm{laser}$ & 1.17 & eV\\
Pulse energy, $E_\text{pulse}$  & 100 & \si{\micro\joule}\\
Number of photons, $N_{\textrm{laser}}$ & \num{5.34e14}\\ 
Repetition rate, $f$ & 125 & MHz\\
Spot size at the IP, $\sigma_\textrm{laser}$ & 25 & \si{\micro\meter}\\
Crossing angle, $\phi$ & 5 & deg \\
Pulse length, $\tau_{\mathrm{laser}}$  & 10 & ps\\
Relative energy spread, $\Delta E_\textrm{laser}/E_\textrm{laser}$ & \num{6.57e-4} &   \\
 % 0.7nm error on 1064nm
\bottomrule
\end{tabular}
\caption{External incident laser pulse properties used for calculation of Table~\ref{tab:DIANA_spectral_output}.}
\label{TAB:DIANA_laser_pulse}
\end{table}

\begin{table}[htbp]
\centering\footnotesize
\begin{tabular}{lcccc}
\toprule
 & \multicolumn{3}{c}{Electron Kinetic Energy (MeV)} & \\
 \cline{2-4}
 & 362 & 717 & 1072 & \\
\midrule
\textgamma-ray peak energy  & 2.33 & 9.06 & 20.11 & MeV\\
Source Size ($x$/$y$)  & 10.72/10.72 & 8.00/8.00 & 6.65/6.65 & \si{\micro\meter} \\
Uncollimated flux  & 5.77$\times 10^{10}$ & 6.02$\times 10^{10}$ & 6.08$\times 10^{10}$ & ph/s\\
Spectral density  & 2.48$\times 10^{5}$ & 6.65$\times 10^{4}$ & 3.03$\times 10^{4}$ & ph/s eV\\
Average brilliance  & 5.64$\times 10^{12}$ & 2.05$\times 10^{13}$ & 4.45$\times 10^{13}$ & ph/s mm$^{2}$ mrad$^{2}$ \SI{0.1}{\percent} bw\\
Peak brilliance  & 5.60$\times 10^{17}$ & 2.22$\times 10^{18}$ & 4.99$\times 10^{18}$ & ph/s mm$^{2}$ mrad$^{2}$ \SI{0.1}{\percent} bw\\
\midrule
 & \multicolumn{3}{c}{\SI{0.5}{\percent} \textit{rms} bandwidth} & \\
\midrule
Source Size ($x$/$y$) & 19.36/12.54 & 19.35/12.52 & 19.33/12.50 & \si{\micro\meter} \\ 
Collimated flux  & 1.30$\times 10^{9}$ & 1.29$\times 10^{9}$ & 1.29$\times 10^{9}$ & ph/s \SI{0.5}{\percent} bw \\
\bottomrule
\end{tabular}
\caption{Calculated spectral properties of output \textgamma~radiation from an example 3-pass recirculating \SI{1}{\giga\electronvolt}-scale ERL-driven inverse Compton scattering source. Based on methods from \cite{PhysRevAccelBeams.24.050701}.}
\label{tab:DIANA_spectral_output}
\end{table}

\begin{figure}[h]
    \centering
    \includegraphics[width=0.6\textwidth]{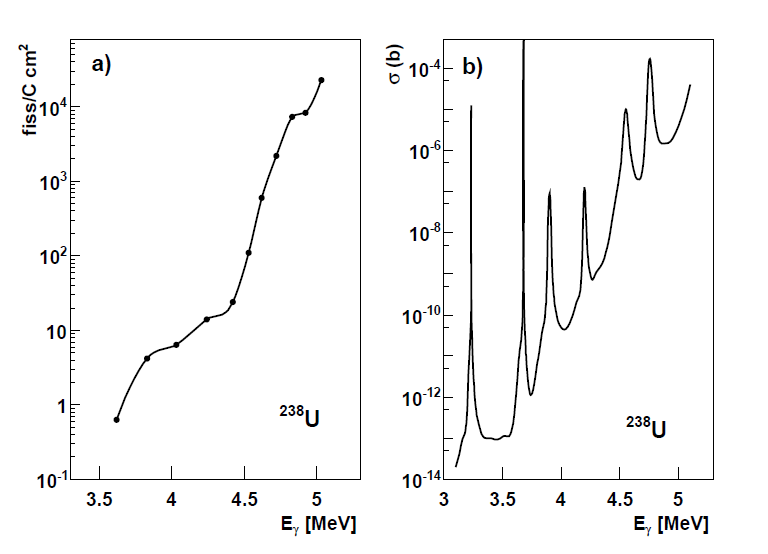}
    \caption{Left: observed, bremsstrahlung-induced photo-fission Giant Dipole Resonance (GDR) of $^{238}\text{U}$. Right: predicted \enquote{hidden} resonances within GDR revealed by an ICS narrowband gamma source~\cite{refId0}.}
    \label{fig:applications:ics:resonances}
\end{figure}

This corresponds to at least two orders of magnitude greater spectral energy density than HI\textgamma{}S, and likely similarly exceed the performance of ELI-NP-GBS. It should be noted that this calculation uses conservative, previously demonstrated parameters as input, and therefore should be viewed as a lower bound. Narrowing the energy bandwidth below \SI{0.5}{\percent} whilst retaining significant flux will enable the resolution of individual nuclear excitations according to the nuclear shell model.
This is because the source bandwidth would be less than predicted typical spacings between adjacent energy levels.
The simulated improvement in the knowledge of an example photonuclear cross section is shown in Fig.~\ref{fig:applications:ics:resonances}.
This would lead to the establishment / consolidation of a new field of science, Nuclear Photonics, named by analogy to the field of atomic photonics opened up by lasers from the 1960s onwards.
This is because the ICS source would be a step change in high-flux, tune-able, narrowband gamma production.
\subsection{Non-destructive assay via radiography, nuclear resonance fluorescence and photofission}

Radiographic imaging ($<\SI{10}{\mega\electronvolt}$) of assemblies (e.g.~shipping container contents) can be conducted with broadband (bremsstrahlung) systems, but a near-monoenergetic ICS source would significantly simplify the deconvolution of detector and filter responses during the post-processing of radiographic data whilst reducing dose to the object.
Tune-ability of the source (including below \SI{1}{\mega\electronvolt}) may permit some degree of discrimination between materials. \SIrange{1}{5}{\mega\electronvolt} photons with around few percent bandwidth are expected to enable development of verification techniques for non-proliferation security through nuclear resonance fluorescence (NRF) computed tomography.
This would be a source of high-quality underpinning data for future, smaller systems deployable in the field (e.g.~at ports of entry, or at other locations where there is a requirement to rapidly detect and diagnose materials of concern, including possible contraband).
Specifically, nuclear cross-sections and the resonance characteristics of materials could be acquired across the photon energy range of operational interest.
The narrow bandwidth and tune-ability of the source are also key, in that specific resonances can be targeted with the aim of demonstrating materials identification from the resulting fluorescence signature.
Furthermore, the narrowband nature of the ICS source ensures that the input dose to the target can be significantly reduced compared to existing broadband sources.
This fact is also important in the context of the non-destructive assay of components within the nuclear industry, for example spent fuels and unknown legacy wastes.
The \textgamma-rays from the ICS system would allow fast, high-resolution, isotopically sensitive probing of waste packages, allowing nuclear assay even through shielded flask walls~\cite{jaea_report,ANGELL201511}.
From a security perspective, the ICS source could be used to demonstrate the feasibility of using induced photofission signatures ($\SI{5}{\mega\electronvolt} < E < \SI{10}{\mega\electronvolt}$) for low-dose identification of specific isotopes~\cite{osti_1376659}.
As for NRF, underpinning materials data (in this case, nuclear photofission cross-sections) acquired using the ICS source would be central for developing this new detection methodology.

\subsection{Nuclear waste management via photonuclear transmutation}
Combining NRF with irradiation at higher photon energy (\SIrange{5}{40}{\mega\electronvolt}) and narrower bandwidth (\SIrange{0.1}{0.01}{\percent}) is expected to enable development of techniques to possibly selectively transmute specific isotopes~\cite{ZHU2016109,doi:10.1080/18811248.2007.9711592}.
This would allow investigation of improved nuclear waste management techniques via induced photofission of actinides and long-lived fission products.
Crucially, a high value source or product oxide could be purified without the need for wet chemical partitioning, thus allowing purification of a mixture of isotopes or, alternatively, selective destruction of a contaminant that has grown into a material.
For example, ingrowth of $^{241}\text{Am}$ in a can of PuO\textsubscript{2} greatly increases operator dose when this is recycled as MOX fuel.
This could be mitigated if the $^{241}\text{Am}$ could be eliminated in situ.
For future reprocessing plants, new reprocessing strategies are being investigated to segregate actinides in differing combinations, to improve long term waste handling options.
It may be practical to selectively destroy a specific high-hazard actinide or alternatively a fission product waste such as $^{129}\text{I}$ or $^{99}\text{Tc}$.
To be economically viable, the waste volume would need to be small and specific long-term waste storage and waste stabilisation costs be high.
Major economic benefit may therefore arise in, for example, reconfiguring the long-term management of current and future stockpiles of problematic legacy wastes.

\subsection{Medical radionuclide production}
The high spectral flux available may make the production of novel medical radioisotopes economically viable, including by harvesting them from legacy material currently considered as waste. The pencil nature of the gamma beam implies potential for very high specific activity in the material produced. Applications would include new medical diagnostic techniques, such as gamma-PET (photonuclear generation of $^{44}\text{Ti}$, $^{195\text{m}}\text{Pt}$, $^{117\text{m}}\text{Sn}$, $^{44}\text{Sc}$)~\cite{Habs:2010nn}.

\subsection{Detector calibration}
For example, calibrated gas Cherenkov detectors (GCD) are routinely used to monitor burn characteristics in Inertial Confinement Fusion (ICF) schemes~\cite{doi:10.13182/FST15-173}.
The monochromatic but tune-able characteristics of the ICS source are ideal for determining detector response as a function of gamma photon energy~\cite{doi:10.1063/1.4812572}.

%\section{ICS Gamma Source}
%Peter Williams

\chapter{ERLs and Sustainability}\label{sec:sustainability}
%Andrew Hutton, Erk Jensen, Olga Tanaka, Nick Shipman

% MB 01/19/2022: Merged edits from Andrew per today's email.

%\section{Introduction}%
%\label{sec:sustainability:intro}

In any new accelerator proposal, sustainability issues will be heavily scrutinized, be that in electricity and water use, the overall efficiency of the facility, including reusing the heat for other purposes (space heating, biogas production, etc.) or energy recycling.

The European Spallation Source (ESS) is exemplary in this regard.
The facility includes a wind farm in the Baltic Sea, on-site solar systems, and a biomass conversion facility that also produces fertilizer.
The power consumption of the facility (270~GWh/year) is about half of the original estimates for the energy required for the facility.
The reduction came from reusing the heat, including heat storage, sending hot water to the Lund distribution system for heating houses and industrial buildings, and improved insulation for the buildings.

ESS had an advantage as it was built on a green-field site (literally), so some of the solutions adopted would be harder to implement in an urban environment such as CERN.
Nevertheless, minimizing, if not reducing to zero, the carbon footprint of a new facility will be a requirement.
This has been recognized by the ICFA Panel on Sustainability, which was established to address all of the issues required to meet the new sustainability expectations of society.

These aspects are important for all new facilities, but ERLs bring a new dimension.
Directly returning the energy of an unused beam into RF that can be used for acceleration with practically no losses is a unique feature of ERLs.
While not all of the energy can be recovered, the overall efficiency of the process is extremely high.
This advantage starts with a reduction in the RF power needed for acceleration, which translates into smaller RF sources and their associated power transformers (reducing the resources needed for their production), and less electric power and water cooling required (reduced operating costs as well as a reduced carbon footprint). 
ERLs have a special place in future colliders, given the importance of sustainability and the reduction of electrical power and water consumption.
While the technology improvements in component efficiency can improve all future colliders, only ERLs also recover energy from the beam in the form of RF power.
This results in a smaller investment in the power required from the RF sources.
For klystrons, the electrical power requirements are only determined by the maximum power, and so the smaller the power source, the less energy is required.  
The R\&D that will be needed in the coming years has to focus on every aspect of the collider, including operation of superconducting cavities at a higher temperature (\SI{4.5}{K}, where the efficiency is three times better than at \SI{2}{K}).

Given the inherent advantages of ERLs, it is to be expected that their sustainability profile will eclipse other colliders with similar physics potential. 

\section{Power consumption}%
\label{sec:sustainability:intro:power_consumption}

Energy efficiency and sustainability have received a lot of attention over recent years, and society’s concern about climate change and global warming must also be taken seriously by the accelerator community.
Improving the energy efficiency has to be an integral part of any strategy for future accelerator facilities, which must be included in the more usual cost optimization strategy.
For new facilities, improving the energy efficiency can really make a difference in the overall project cost, as well as in the operating costs, and will have a large impact on their carbon footprint.
These aspects will be more heavily scrutinized when seeking approval for future accelerators, and convincing solutions to minimize the environmental impact will be a prerequisite to such approval. 

Even more, the accelerator community drives research and development at the cutting edge for a wider range of applications than just making the next accelerator better.
Society expects a return from the investment in this research, which includes other applications of accelerators, for example for medicine, but also other spin-offs (like the touchscreen technology or the World-Wide Web in the past).
The same is true for the development of concepts to optimize energy efficiency.
As described in \cite {JThomas2020GreenAccelerator}, work on energy-efficient accelerators has started and is making good progress. 

How can the energy efficiency of an accelerator be optimized?
Using the example of an accelerator for high-energy physics, the beam energy and the luminosity are parameters required by physics, so the question is: can we build an accelerator which can reach the same physics parameters but consumes less primary energy? 

The first element of the answer is to make the components of the accelerator better in terms of their individual efficiencies, such as power converters, magnets, or RF amplifiers.
The next step will be the conversion efficiency of RF power to beam power.
For cryogenic systems, there is also the requirement to minimize cryogenic losses (better thermal insulation) and to look for the optimum operating temperature.
The final element of the answer is the recovery of otherwise \enquote{lost} energy.
Here, \enquote{recovery} means both conversion to a higher-quality form of energy than heat, as well as the recovery of the energy in the low-grade heat. 

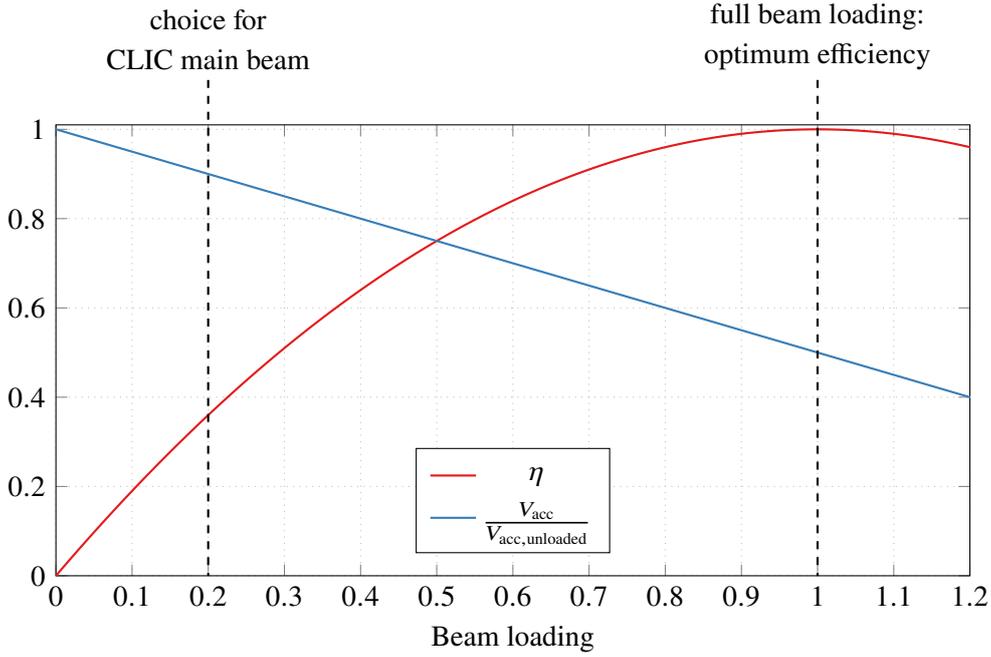
\begin{figure}[htb]\centering
\tikzsetnextfilename{sustainability_fullbl}
\begin{tikzpicture}
\begin{axis}
[
	width=.9\linewidth,
	height=.5\linewidth,
	xlabel={Beam loading},
	legend style={at={(rel axis cs: 0.5, 0.05)}, anchor=south, row sep=0.5ex, inner sep=1ex},
	xmin=0,
	xmax=1.2,
	ymin=0,
	ymax=1.01,
	xmajorgrids=true,
	ymajorgrids=true,
	major grid style={dotted},
]
\addplot+[thick, mark=none, samples=1000] {1 - (x-1)^2};
\addlegendentry{$\eta$}
\addplot+[thick, mark=none, samples=2] {1 - x/2};
\addlegendentry{$\frac{\mathstrut{}V_\mathrm{acc}}{\mathstrut{}V_\mathrm{acc, unloaded}}$}
\path (axis cs:0.2, 0.5) coordinate (clic);
\path (axis cs:1.0, 0.5) coordinate (full);
\path (rel axis cs:0, 0) coordinate (bottom);
\path (rel axis cs:0, 1.1) coordinate (top);
\end{axis}
\draw[thick, dashed] ($(top -| clic)$) -- ($(bottom -| clic)$)
    node[pos=0, anchor=south, align=center] {choice for\\CLIC main beam};
\draw[thick, dashed] ($(top -| full)$) -- ($(bottom -| full)$)
    node[pos=0, anchor=south, align=center] {full beam loading:\\optimum efficiency};
\end{tikzpicture}
\caption{RF-to-beam power conversion efficiency ($\eta$) and accelerating voltage ($V_\text{acc}$) as a function of beam loading}
\label{fig:sustainability:intro:power_consumption:full_bl}
\end{figure}

Concerning the RF-to-beam efficiency, it was shown with the CLIC Test Facility 3 \cite{Corsini04}, which uses a traveling-wave accelerating structure, that the RF-to-beam efficiency can be increased to close to \SI{100}{\percent} (``full beam loading'') at the expense of reducing the accelerating voltage by half compared to the unloaded case.
Figure~\ref{fig:sustainability:intro:power_consumption:full_bl} demonstrates this trade-off: When the beam current is increased to a value $V_\text{acc}/R_\text{shunt}$, where $V_\text{acc}$ is the unloaded accelerating voltage and $R_\text{shunt}$ is the shunt impedance, the beam-induced voltage completely compensates the unloaded voltage. The total accelerating voltage would be zero; in Fig.~\ref{fig:sustainability:intro:power_consumption:full_bl} this corresponds to a beam loading of 2.
At half this current, the RF-to-beam efficiency is maximum.
In the normal-conducting drive beam accelerator of CTF3, the RF-to-beam efficiency was demonstrated experimentally to be \SI{90}{\percent} initially; later, even \SI{96}{\percent} was achieved.

For completeness, the chosen operating point for the CLIC main accelerator is marked in Fig.~\ref{fig:sustainability:intro:power_consumption:full_bl}. A beam loading of 0.2 leads to a reduction of the unloaded accelerating gradient by \SI{10}{\percent}, but at the expense of an RF-to-beam efficiency of only \SI{36}{\percent}.
Accepting an accelerating gradient \SI{20}{\percent} below the unloaded case (beam loading 0.4) increased this efficiency to \SI{84}{\percent} \cite{Corsini09}.

\par
In order to reach high luminosity at high energy in physics colliders, the power in the particle beams is necessarily very high, so the question arises: can one recover the energy from the particle beam? For illustration, consider a linear collider consisting of two counter-running linacs. One way to write the luminosity of such a collider \cite{DELAHAYE1999369} is 
\begin{equation*}
    \mathcal{L} \propto \frac{N}{\sigma_x \sigma_y E_\text{b}} P_\text{b}
\end{equation*}
where  $N$ is the number of particles per bunch, $\sigma_x$ and $\sigma_y$ are the transverse beam sizes at collision, $E_\text{b}$ is the beam energy and $P_\text{b}$ is the beam power. With $E_\text{b}$ given and $\sigma_x$, $\sigma_y$ and $N$ limited by technology, it is clear that $P_\text{b}$ should be maximized to maximize the luminosity $\mathcal{L}$.
The typical numbers presented in \cite{DELAHAYE1999369} for the linear collider projects in 1999 are average beam powers ranging from \SIrange{4.8}{16.4}{\mega\watt}, resulting in AC powers for RF generation ranging from \SIrange{57}{153}{MW}.
The values assumed in 1999 for the conversion of AC power to RF power were ranging from \SI{30}{\percent} to \SI{40}{\percent}.
At first glance, it seems that recovery of beam power will save only \SI{10}{\percent} at best, even if one can recover all beam power.
If, however, the AC-to-RF conversion efficiency and the RF-to-beam efficiency can be increased, the beam power should also be recovered, making a real ``green'' accelerator.
% MB 3/7/22: Reference to chapter removed; the chapter is gone.
%We describe the power needs for ERLs in chapter \ref{sec:key_challenges:power}.

Looking again at the above equation, it is important to question to what level $\sigma_x$ and $\sigma_y$ can be further reduced---the typical argument here uses the tolerable beam-beam tune shifts $\xi_{x,y}$. Of course, squeezing the beams more in the final focus will lead to a larger luminosity (assuming good alignment and stabilization); however, it will also lead to a stronger pinch effect caused by the space charge of the opposite bunches, which in turn will lead to more curved particle trajectories and consequently synchrotron radiation and energy spread. 

For the sake of illustration, assume for a moment that \SI{100}{\percent} beam energy recovery can be reached. In this case, one could allow an increase of $P_\text{b}$ and at the same time relax on $\sigma_x$ and $\sigma_y$ in such a way that pinch, beam-beam tune spread, and energy spread would be reduced.
In other words, the assumptions generally accepted for colliders without beam energy recovery change completely if the beam energy can be recovered. 

\section{Beam Energy Recovery}

Even if the overall energy conversion efficiency from AC power to beam power may still be limited by technical concerns today, let us separate these limitations from the actual recovery of the beam energy.
Here, the CLIC study provides quantitative results.
The concept of CLIC is a two-beam scheme, where the energy of the drive beam is completely recovered to serve as a power source for the main beam.
With the nominal parameters of CLIC, the drive beam is decelerated from \SI{2.38}{\giga\electronvolt} to about \SI{238}{\mega\electronvolt}, i.e., \SI{90}{\percent} of the drive beam energy is recovered.

An ERL works by placing the accelerated beam on the crest of the RF and the decelerating beam exactly \ang{180} out of phase.
Under these conditions, the energy of the decelerating beam is fully extracted. There are two corrections to this ideal scenario.

Both the accelerating and decelerating bunches create higher-order modes (HOMs) in the superconducting RF cavities, which add rather than cancel.
The primary design criterion for the cavity is to minimize the surface losses for a given gradient in order to reduce the cryogenic load.
With this constraint, it is difficult to simultaneously minimize the energy lost to HOMs.
However, considerable effort is made to avoid resonant excitation of the HOMs which can cause beam-break-up instabilities.

The HOMs also cause heating of the cavity surface, so it is important to extract the HOM energy to higher temperature than the \SI{2}{K} used for the cavity---ideally, this would be at room temperature, but it also creates a heat leak into the cryomodule, so the resulting temperature is a balance of these two effects.
Bringing the HOM energy out to the shield temperature of $\sim\SI{50}{K}$ is the usual configuration that results from this compromise.
It is worth noting that even with careful optimization, \SI{7}{\percent} of the HOM heat is deposited at \SI{2}{K} in the ILC, and it may be hard to do better.

In low-energy ERLs, the decelerated beam is run off-crest in order to minimize the energy spread in the last cavities.
This leads to partial energy recovery equal to $1-\cos\theta$, where $\theta$ is the phase angle.
In high-energy colliders, this effect is likely to be small and limited to a small section of the booster cavities.

\section{Technology and Infrastructure}

Many of the technologies required to reduce the energy consumption are common to all modern accelerators.
However, the two main areas that are different in ERLs are: the RF power sources and delivery, as well as cavity heat deposition and the cryogenics.

High-efficiency klystrons are being developed, while the use of injection-locked magnetrons is considered as well.
Klystrons draw the same electric power regardless of whether they are producing RF power or not.
If the klystron is not producing RF power, the drive beam ends up in the collector where the energy is converted to heat.
While this may be acceptable for CW machines, it makes klystrons a bad choice for pulsed operation.
Prototype solid-state amplifiers (SSAs) with efficiency over \SI{80}{\percent} and good turn-down efficiency have been developed, but these are not in the market place yet.
SSAs appear to be comparable with CW klystrons in cost per watt, efficiency, and footprint, but while klystrons are getting more expensive, the SSAs are falling in price, so they are clearly the way of the future.
Magnetrons are an order of magnitude cheaper and have an efficiency above \SI{80}{\percent} but have yet to be adopted for SRF applications, although they are widely used for copper linacs.
As they are oscillators and naturally somewhat noisy, special measures are necessary to control them for SRF applications.
Injection phase locking and adaptive power supply control are needed to achieve this, which have been demonstrated on the bench but not yet in full tests with beam.
SSAs are a safe option for modest power, but if cost is the main concern, magnetrons offer the best potential cost-benefit ratio.
One work-around to reduce start-up power in ERLs is to ramp up the average current.
It is better to ramp up the bunch rate rather than charge per bunch to maintain the injector chain in a constant state, and if the ramp time is a bit longer than the recirculation time, it will smooth out the start-up transient. 

For ERLs, where the energy absorbed by the beams is small (principally replacing HOM losses), the most important factor is properly matching the power source to the cavity under the different operating conditions.
For fundamental power couplers, the match is either set during cryomodule fabrication for fixed couplers or variable couplers are used.
For ERLs, variable couplers would be needed, as power must be delivered to the cavity during start-up, which requires a large coupling coefficient. During steady-state operation, the power required by the cavity drops to a small value, so a small coupling constant is preferred.
Basically, variable couplers are used to avoid large reflected (and, therefore, wasted) power.

Traditionally, tuners that squeeze the cavity are used to match the frequency of superconducting cavities to the power source.
These are relatively slow (seconds), but piezo-electric tuners have been used to make faster adjustments (up to $\sim\SI{3}{\kilo\hertz}$).
The recent development of Fast Reactive Tuners (FRTs) is a big step forward for ERLs.
The change in tuning is achieved using an external magnetic field to change the permittivity of a special ceramic.
There are therefore no moving parts, and the FRT can respond to fast transients (up to \SI{10}{\mega\hertz}).
For ERLs, this capability is a game-changer as simulations show that the reflected RF power can almost be reduced to zero.
This will be tested in a cryomodule at bERLinPro.

If the facility is operating with pulsed RF systems, the Lorentz forces change the resonant frequency of the superconducting cavity, which then vibrates (microphonics).
The use of piezoelectric tuners has been pioneered at DESY for the XFEL; this method reduces the impact of the Lorentz forces using a feed-forward adaptive algorithm.
It is to be expected that the use of the faster FRTs will improve the precision of the response.

The RF surface heating of superconducting cavities depends on several factors. First, the fundamental RF frequency; the lower the frequency, the thicker the skin depth and the lower the losses.
However, the lower the frequency, the larger the surface area of the cavity.
The optimum RF frequency is in the region of \SIrange{600}{800}{\mega\hertz} for the high currents required for ERLs~\cite{Marhauser:2014fya}.

Next is the cavity shape; if the irises are small, that decreases the fundamental RF losses but increases the production of HOMs.
In addition, small irises tend to trap the HOMs so that they cannot be coupled out of the cavity.
This leads to HOM power being deposited in the cavity, which causes beam-break-up instabilities and increased surface heating.
For ERLs, a rather large iris diameter is chosen to avoid instabilities, but even so, only \SI{93}{\percent} of the HOM power in the ILC is extracted to be dissipated at $\sim\SI{50}{\kelvin}$, while \SI{7}{\percent} of the HOM power is deposited in the cavity and dissipated at \SI{2}{\kelvin}.
Clearly, R\&D to improve the HOM couplers is still needed.

The surface treatment of the superconducting cavities has been the subject of intense research over the last decade, resulting in a breakthrough with nitrogen doping of the niobium.
In particular, the specification for the LCLS-II cavities at \SI{1300}{\mega\hertz} is a $Q_0$ of \num{3e10}, an improvement of at least a factor of 4 over previous cavities.
At this time, the improvement in \SIrange{600}{800}{\mega\hertz} cavities has not been so dramatic, so this is another area where R\&D should be spent for ERLs.

\begin{figure}[htb]\centering
\tikzsetnextfilename{sustainability_cop}
\begin{tikzpicture}
[
    annotation/.style={font=\footnotesize, fill=white, inner sep=1pt},
    datapoint/.style={red, fill, circle, inner sep=0, minimum size=5pt},
]
\begin{axis}
[
	width=.9\linewidth,
	height=.5\linewidth,
	xlabel={$T_\text{refr}$ (\si{\kelvin})},
	ylabel={COP (W/W)},
    legend style={row sep=0.5ex},
    legend cell align={left},
	xmin=1.4,
	xmax=5,
	ymin=0,
	ymax=1400,
	xmajorgrids=true,
	ymajorgrids=true,
	major grid style={dotted},
	/pgf/number format/1000 sep={},
]
\addplot+[thick, mark=none, samples=100, domain=1.4:5] {293.0/x - 1};
\addlegendentry{Carnot: $\frac{\mathstrut{}T_\mathrm{ambient}}{\mathstrut{}T_\mathrm{refr}} - 1$}
\addplot+[thick, mark=none, samples=100, domain=1.4:5] {(293.0/x - 1) / (0.278744 - 0.038811*(4.5-x))};
\addlegendentry{Real (installed at LHC)}
\draw (axis cs:1.8, 930) node[datapoint] (p1) {};
\draw (axis cs:2.0, 801) node[datapoint] (p2) {};
\draw (axis cs:4.2, 260) node[datapoint] (p3) {};
\draw (axis cs:4.5, 230) node[datapoint] (p4) {};
\path (rel axis cs:0, 0) coordinate (bottom);
\path (rel axis cs:0, 1) coordinate (top);
\end{axis}
%\draw[dashed] (p1) -- ($(bottom -| p1)$)
\draw (p1) node[anchor=west, rotate=90, xshift=5pt, annotation] {\SI{1.8}{\kelvin}: 930};
% \draw[dashed] (p1) -- ($(bottom |- p1)$);
% \draw[dashed] (p2) -- ($(bottom -| p2)$)
\draw (p2) node[anchor=west, rotate=90, xshift=5pt, annotation] {\SI{2.0}{\kelvin}: 801};
% \draw[dashed] (p2) -- ($(bottom |- p2)$);
% \draw[dashed] (p3) -- ($(bottom -| p3)$)
\draw (p3) node[anchor=west, rotate=90, xshift=5pt, annotation] {\SI{4.2}{\kelvin}: 260};
% \draw[dashed] (p3) -- ($(bottom |- p3)$);
% \draw[dashed] (p4) -- ($(bottom -| p4)$)
\draw (p4) node[anchor=west, rotate=90, xshift=5pt, annotation] {\SI{4.5}{\kelvin}: 230};
% \draw[dashed] (p4) -- ($(bottom |- p4)$);
\end{tikzpicture}%
\caption{Coefficient of performance (COP) as a function of temperature of a cryogenic system, here using the example of the LHC [thanks to P.~Lebrun]. To extract $P$ at $T_\text{refr}$, one needs $\mathrm{COP} \cdot P$ at $T_\text{ambient}$.}
\label{fig:sustainability:technology:cop}
\end{figure}
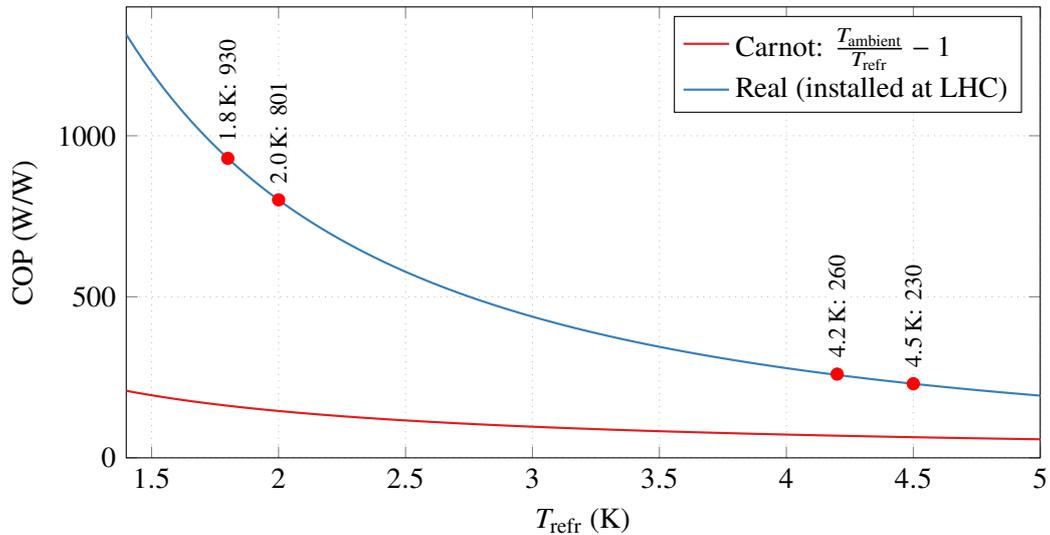

Finally, the cryogenic efficiency has been the subject of enormous improvements over the last few years, but it is unlikely that there will be any more breakthroughs.
The only parameter left to improve the energy consumption is the operating temperature.
Operating the cavities at temperatures above \SI{2}{\kelvin}, nominally \SI{4.5}{\kelvin}, improves the energy consumption of the cryoplant considerably.
Niobium is no longer the best superconductor at these higher temperatures, so a coating of a superconducting material with a higher transition temperature is preferred.
The coating that has been the subject of the most R\&D has been $\text{Nb}_3\text{Sn}$, and this seems likely the first to be incorporated in a cavity.
The improvement to be expected in the cryogenic efficiency is shown in Fig.~\ref{fig:sustainability:technology:cop}.
For the installed LHC plants, the cryogenic efficiency is \SI{801}{\watt} at room temperature per watt at \SI{2}{\kelvin}, and \SI{230}{\watt} at room temperature per watt at \SI{4.5}{\kelvin}, an improvement of a factor 3.5.
Note that these numbers are for the cryoplant itself; in general, it is necessary to add about \SI{15}{\percent} to cover the shields, distribution, and electrical efficiency.

The \enquote{Ganni Cycle} is a recent improvement in the operation of cryogenic plants~\cite{doi:10.1063/1.3422267}.
A cryogenic plant operates using many stages of: compression of the gas; removal of the heat; and decompression of the gas, which lowers the temperature.
Each of these intermediate stages has an input pressure and an output pressure.
The conventional wisdom was that these input and output pressures should be fixed and the compressors/decompressors should be optimized for these fixed pressures.
In the Ganni Cycle, the pressures at the interfaces between the different stages of the cool-down are allowed to float, which increases the plant efficiency because optimizing each stage is less efficient than optimizing the overall system.
What was initially surprising was that the intermediate pressures would naturally stabilize at the optimum values; this is the so-called \enquote{floating pressure} principle.
The full Ganni Cycle incorporates other efficiency improvements, which together allow the efficiency of the plant to remain at the same high level from \SI{100}{\percent} down to $\sim\SI{30}{\percent}$.
Since a large cryoplant must be sized for the maximum load plus a safety margin, most existing plants are operated with reduced efficiency; this is not the case for the Ganni Cycle.
At this time, the Ganni cycle is patented~\cite{patent_7409834, patent_7278280} and licensed to Linde, one of the two European Cryoplant constructors; the other, Air Liquide, declined.

Theoretically, the Ganni Cycle would make it possible to operate the cryogenic plant in a \enquote{gated} mode (e.g., two seconds on, four seconds off), but this has never been demonstrated.
This is another area for R\&D because if the same number of interactions can be squeezed into one third of the operating cycle, the power required by the cryogenic plant would be drastically reduced.
The dynamic load would go down to one third, but the static load (which is usually smaller than the dynamic load) would remain unchanged.

\chapter{Conclusions}\label{sec:conclusions}
%Max Klein, Andrew Hutton

\begin{figure}[htb]\centering
\tikzsetnextfilename{hype_cycle}
\begin{tikzpicture}
[
    annotation/.style={font=\footnotesize},
]
\begin{axis}
[
    width=.95\linewidth,
    height=.45\linewidth,
    xlabel={Time},
    ylabel={Expectations},
    xmin=0,
    xmax=9,
    ymin=-0.1,
    ymax=1.3,
    ytick=\empty,
    xtick=\empty,
]
\addplot+[thick, mark=none] table[x index=0, y index=1] {figures/hype_cycle.dat};
\draw (axis cs:0.45, 0.02) node[annotation, anchor=west, align=left] {Innovation\\trigger};
\draw (axis cs:2, 1) node[annotation, anchor=south, align=center] {Peak of Inflated\\Expectations};
\draw (axis cs:4.3, 0.18) node[annotation, anchor=north, align=center] {Trough of\\Disillusionment};
\draw (axis cs:6, 0.38) node[annotation, anchor=north, align=center] {Slope of\\Enlightenment};
\draw (axis cs:8, 0.53) node[annotation, anchor=north, align=center] {Plateau of\\Productivity};
\draw (axis cs:5, 0.33) node[red, anchor=south, rotate=20, align=center] (erl) {ERL};
\end{axis}
\draw[red, thick, -latex] (erl) -- ++(20:1);
\end{tikzpicture}
\caption{The Hype Cycle used by the American research advisory and information technology firm Gartner Inc.~to represent the maturity, adoption and social application of specific technologies \cite{gartner_hypecycle}.}
\label{fig:hype_cycle}
\end{figure}
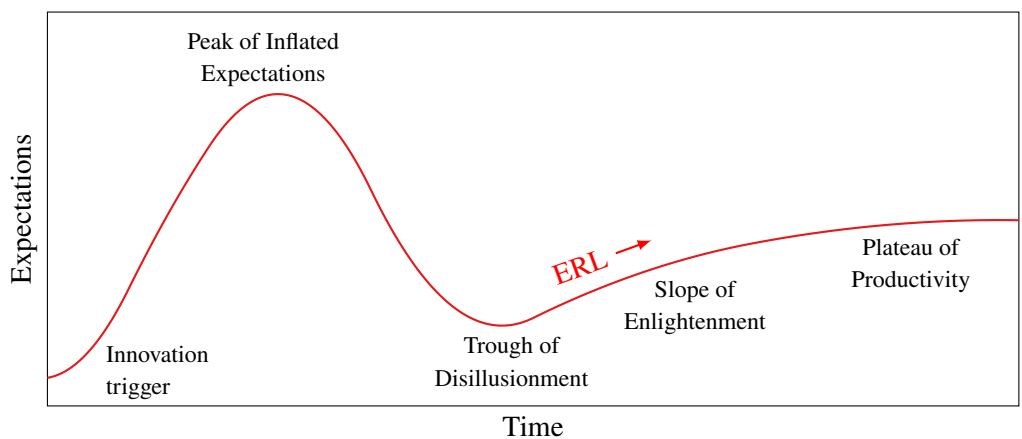

ERLs have followed the typical curve for the development of technologies (Fig.~\ref{fig:hype_cycle}), with the difference that the initial adoption was slow because the SRF technology required was not yet ready.
ERLs really took off at Jefferson Lab in the USA in the 1990s with several upgrades leading to the world-leading results shown in Fig.~\ref{tab:appendix:facilities:completed} in Appendix~\ref{sec:appendix:facilities}.
This success triggered the construction of ALICE at STFC in England and cERL at KEK in Japan.
However, with the termination of the Jefferson Lab FEL, the peak of ERL enthusiasm was passed. The successful construction of CBETA at Cornell in the USA took place during the phase of decreasing expectations, which resulted in the lack of funding to fully complete the commissioning.
ALICE ceased operation, and the cERL was also paused in Japan.

ERLs are currently having a resurgence around the world as enlightenment sets in.
In Japan, the cERL is being repurposed for industrial use, an ERL has been adopted for electron cooling in the EIC Project at Brookhaven in the USA, MESA in Mainz (Germany) has been funded and is under construction, while the S-DALINAC in Darmstadt (Germany) has continued operation without interruption and recently achieved two-pass operation as an ERL. 

It is in this context that this report has been written, as ERLs are transitioning to the ``Plateau of Productivity.'' The proposed completion of bERLinPro in Germany, and the construction of PERLE at Orsay in France are evidence of this new era in Europe.
The ERL roadmap for this decade that enables this bright future comprises three interlinked elements:
\begin{enumerate}
\item The continuation and development of the various facility programs, for which no funds are needed from the particle physics field. For Europe, these are S-DALINAC in Darmstadt and MESA in Mainz (both in Germany). For Japan, this is the continuation of the cERL. For the US, this is the 5-pass CEBAF experiment and the EIC electron cooler.
\item A number of key technologies to be developed:
\begin{enumerate}
\item High-current electron sources reaching the \SI{100}{\milli\ampere} electron current regime; 
\item High-current superconducting RF cryomodules;
\item Development of superconducting cavities operating at \SI{4.4}{\kelvin}; 
\item Fast reactive tuners (FRTs);
\item Monitoring and beam instrumentation;
\item Simulation and training in ERLs.
\end{enumerate}
Some of these, such as electron sources of high brightness, FRTs, and---for longer term---the development of an \SI{802}{\mega\hertz}, \SI{4.4}{\kelvin} cavity-cryomodule have been explicitly integrated into the plans for bERLinPro and PERLE.
There are two other, aspirational items of strategic importance which deserve separate support: HOM damping at high temperature; and the development of twin cavities.
\item The timely upgrade of bERLinPro and construction of PERLE at Orsay as necessary steps to move ERLs forward to their introduction to collider developments, possibly mid-term and long-term.
\end{enumerate}
The present status of all of the recent ERLs as well as those being built, planned, or envisaged was presented in this report.
It should be clear that the field of ERL accelerators is booming and that only a rather small investment would enable the few remaining technology challenges to be addressed so that the large-scale ERL-based electron-hadron machines can be confidently proposed.
The goal of being able to build a high-energy electron-ion collider at CERN using an ERL (the LHeC or FCC-eh) with all of the technologies demonstrated now seems within reach.
Similarly, very interesting prospects have been described and evaluated for future electron-positron colliders, with the vision, as presented in the European roadmap, of a \SI{500}{\giga\electronvolt} machine of extremely high luminosity to precisely map the vacuum structure of the Higgs field.
A new idea has also been presented to produce high-intensity muon beams with an ERL arrangement at the initial step of a future possible muon collider.
The prospects for energy- and intensity-frontier colliders for particle physics and the needs to reduce the energy consumption of such machines have revived the field of energy-recovery linacs and technology, also leading to possibly revolutionary industrial applications and, through the \SI{4}{\kelvin} technology, to unprecedented innovations of the field of superconducting RF.

\appendix
\chapter{Overview on ERL Facilities}\label{sec:appendix:facilities}
%
%Andrew
%
% These tables were converted from Andrew's Excel spreadsheets. They still need some work.

\begin{table}[ht]\centering\scriptsize
\caption{Completed facilities.}%
\label{tab:appendix:facilities:completed}
\begin{tabular}{lcccc}
\toprule
& & ALICE & JLab FEL & CEBAF 1-pass \\
\midrule
Top energy	& MeV	& 27.5		& 165		& 1045		\\
Beam power	& kW		& 0.18		& 1300		& 104.5	 \\
\midrule
Injector / Dump \\
\midrule
Gun energy		& keV	& 350		& 100	& 100	 \\
Bunch charge		& pC	& 60		& 270		& 0.06	 \\
Current			& mA	& 5		& 8.5		& 0.1	 \\
Polarization		&	& No		& No	& No \\
Beam energy		& MeV	& 6.5		& 9		& 25 / 45	\\
Emittance (norm.)	& \si{\micro\meter}	& 2.5	& 8		& 0.05	\\
Dump energy     & MeV   & $< 10$ & 11 & 25 / 45 \\
Dump power		& kW		& 30		& 100	& 4.5	\\
\midrule
Acceleration \\
\midrule
Energy gain per linac	& MeV	& 20	& 156	& $2\times 500$	 \\
RF frequency		& MHz	& 1300		& 1497		& 1497 \\
Bunch repetition rate 	&	MHz &	81.25	&		&		\\
Total linac current	& mA	& 10 (peak)	& 17		& 0.2	\\
Macropulse length	& 	& \SI{100}{\micro\second}	& CW		& CW	\\
Emittance (norm.)	& \si{\micro\meter}	& 3		& 10		& 0.05	\\
Average gradient	& MV/m	& 11		& 12		& 12 \\
Quality factor		& $\times \num{e10}$ & 0.5	& 1	& 1	\\
RF controls		& 	& analog / DLLRF & analog & analog \\
Beam loss & 	& not measured	&	\SI{100}{\nano\ampere}	&	\\
\midrule
Arcs \\
\midrule
Passes		&	& 1	& 1		& 1 \\
Optics design		&	& 	& Bates bends		& achromatic, isochronous	\\
Beam loss & & \SI{2}{\percent} & $< \num{1e-4}$ & \\
\midrule
Interaction region \\
\midrule
$\beta_x$ / $\beta_y$	& cm	& 20	& $\approx 6$	& n/a \\
Beam size		& \si{\micro\meter}	& 50		&	50 & n/a \\
\bottomrule
\end{tabular}
\end{table}

\begin{table}[ht]\centering\scriptsize
\caption{Ongoing facilities (parameters achieved).}%
\label{tab:appendix:facilities:ongoing}
\begin{tabular}{lcccccc}
\toprule
& & S-DALINAC & bERLinPro & cERL & Recuperator & CBETA \\
& & 1-pass / 2-pass & & & & 1-pass / 4-pass \\
\midrule
Top energy	& MeV	& 22.5 / 41	& 25		& 17.6 & 40 & 42 / 150		\\
Beam power	& kW		& 0.027 / 0.3		& 150		& 20 & 200 & 2.9 / 0.3	 \\
\midrule
Injector / Dump \\
\midrule
Gun energy		& keV	& 125 / 250		& $< 3500$	& 500 & 300 & 350	 \\
Bunch charge		& pC	& & 77		& 0.77		& 1500 & 5	 \\
Current			& mA	& 0.0012 / 0.007		& 100		& 0.9 & 10 & 0.07 / \num{2e-6}	 \\
Polarization		&	& Yes / No		& No	& No & No & No \\
Beam energy		& MeV	& 2.5 / 4.5		& 6		& 3 & 1.5 & 6	\\
Emittance (norm.)	& \si{\micro\meter}	& & $\leq 1$ & 0.29 / 0.26		& 20 & 0.3	\\
Dump energy     & MeV   & 2.5 / 4.5 & 6 & 3 & 1.5 &  \\
Dump power		& kW    & 0.003 / 0.032 & $< 60$ & 2.7    & 15	& 0.42 / 0.012	\\
\midrule
Acceleration \\
\midrule
Energy gain per linac	& MeV	& 20 / 18 & 43 & 14.6	& 10	& 6	 \\
RF frequency		& MHz	& 2997		& 1300		& 1300 & 180 & 1300 \\
Bunch repetition rate 	&	MHz & 2997 &	&	& 5.6 / 7.5 / 3.8 & 	\\
Total linac current	& mA	& 0.0024 / 0.028	& 200 & 1.8 & 10 / 30 / 70	& 0.28 / \num{8e-6}	\\
%Harmonic frequency	& MHz	& n/a		& n/a		& n/a & n/a & n/a	\\
Macropulse length	& 	& CW & CW	& CW		& CW, copper	& CW \\
Emittance (norm.)	& \si{\micro\meter}	& & $\leq 1$		& 0.42 / 0.26 & 20		& 0.3	\\
Average gradient	& MV/m	& $< 5$ & 18		& \numrange{5.8}{8.3}		& 0.4 & 16 \\
Quality factor		& $\times \num{e10}$ & 0.3 & 1 & \numrange{0.25}{0.6} 	& \num{4e-6}	& \\
%RF controls		& 	& analog / DLLRF & analog & analog \\
Beam loss & 	& &	$<\num{1e-5}$	&	$<\num{1e-4}$ &	$<\num{1e-2}$ & \\
\midrule
Arcs \\
\midrule
Passes		&	& 1 / 2	& 1		& 1 & 1 / 2 / 4 & 1 / 4\\
Optics design		&	& MBA	& DBA	& \ang{180} achromatic 	& \ang{180} achromatic & FFAG	\\
Beam loss & & & & & $< \num{1e-4}$ & $< \SI{1}{\percent}$ \\
\bottomrule
\end{tabular}
\end{table}

\begin{table}[ht]\centering\scriptsize
\caption{Facilities in progress (target values).}%
\label{tab:appendix:facilities:future}
\begin{tabular}{lccccc}
\toprule
& & MESA & PERLE & CEBAF 5-pass & EIC CeC \\
\midrule
Top energy	& MeV	& 105		& 500		& 7584		& 22.3 / 54.1 / 150 \\
Beam power	& MW		& 0.21		& 10		& 0.758		& 2.2 / 5.3 / 14.7 \\
\midrule
Injector / Dump \\
\midrule
Gun energy		& keV	& 100		& 350 / 200	& 100		& 400 \\
Bunch charge		& pC	& 1		& 500		& 0.06		& 1000 \\
Current			& mA	& 2		& 20		& 0.1		& 98.5 \\
Polarization		&	& Yes		& Yes / No	& Yes		& No \\
Beam energy		& MeV	& 5		& 7		& 84		& 5.6 \\
Emittance (norm.)	& \si{\micro\meter}	& $< 1$		& 6		& 0.05		& $< 3$ \\
Dump power		& kW		& 5		& 140	& 8.4		& 551.6	\\
\midrule
Acceleration \\
\midrule
Energy gain per linac	& MeV	& $2 \times 25$	& $2\times 82$	& $2\times 750$	& 17.3 / 49.1 / 145 \\
RF frequency		& MHz	& 1300		& 801.58	& 1497		& 591 \\
Bunch repetition rate 	&	MHz &		&		&		& 98.5 \\
Total linac current	& mA	& 8		& 120		& 1		& 197 \\
Harmonic frequency	& MHz	& n/a		& n/a		& n/a		& 1773 \\
Macropulse length	& 	& CW		& CW		& CW		& CW \\
Emittance (norm.)	& \si{\micro\meter}	& $< 1$		& 6		& 0.05		& $< 3$ \\
Average gradient	& MV/m	& 12.5		& 20		& \numrange{12}{17.5} & 20 \\
Quality factor		& $\times \num{e10}$ & $> 1.25$	& $> 1$	& 1	& \\
RF controls		& 	& MTCA (digital) & & analog/digital & TBD \\
Beam loss & 	& \num{e-5} &		&		& TBD \\
\midrule
Arcs \\
\midrule
Passes		&	& 2	& 3		& 5	& 3 \\
Optics design		&	& MBA		& flexible		& achromatic,		& $R_{56}$-canceling \\
                    &   &           & momentum      & isochronous       & bending, Bates \\
                    &   &           & compaction    &                   & \\
Beam loss & 	& \num{e-3} 	& 		&		& TBD \\
\midrule
Interaction region \\
\midrule
$\beta_x$ / $\beta_y$	& cm	& $\approx 100$	& 	& n/a		& 40 / 40 \\
Beam size		& \si{\micro\meter}	& 100		&	& n/a		& 1330 / 550 / 200 \\
Beam divergence		& \si{\micro\radian}	& 100		&	& n/a		& 4 \\
\bottomrule
\end{tabular}
\end{table}

\begin{table}[ht]\centering\scriptsize
\caption{ERL-based energy-frontier collider projects.}%
\label{tab:appendix:facilities:hep}
\begin{tabular}{lccccccc}
\toprule
& & LHeC & FCC-eh & CERC\textsubscript{Z} & CERC\textsubscript{H} & ERLC & EXMP \\
\midrule
Top energy	& GeV	& 50		& 60		& 45.6		& 120 & 250 & 500 \\
Beam power	& MW		& 1000		& 1200		& 48	& ?	& \num{40000} & \num{100000} \\
Luminosity estimate & \si{\per\square\centi\meter\per\second} & \num{e34} & \num{e34} & \num{7e35} & \num{8e35} & \num{5e35} & \num{1.25e42} \\
\midrule
Injector / Dump \\
\midrule
Gun energy		& keV	& \numrange{220}{350}		& \numrange{220}{350}	& n/a		& n/a & ? & n/a \\
Bunch charge		& pC	& 500		& 750		& \num{12500}	& \num{25000}	& 800 & 250 \\
Current			& mA	& 20		& 30		& 3.7	& 2.5	& 98.5 & 200 \\
Beam energy		& MeV	& 500		& 500	& 2000	& 2000	& 5000 & 5 \\
Dump power		& kW		& 100		& 150	& n/a	& n/a	& $\approx 100$	& 1000 \\
\midrule
Acceleration \\
\midrule
RF frequency		& MHz	& 801.58		& 801.58	& 750	& 750	& 1300 & 801.58 \\
Total linac current	& mA	& 120		& 180		& 29.6	& 19.8	& 320 & 400 \\
Emittance (norm.)	& \si{\micro\meter}	& 30		& 30		& 4 / 0.008	& 6 / 0.008	& 20 / 0.035 & 0.4\\
Average gradient	& MV/m	& 20		& 20		& 20 & 20 & 20 & 20 \\
Quality factor		& $\times \num{e10}$ & $> 1$	& $> 1$	& $\approx 1$ & $\approx 1$	& 3 & 1 \\
\midrule
Arcs \\
\midrule
Passes		&	& 3	& 3		& 4	& 4 & 1 & 1 \\
\midrule
Interaction region \\
\midrule
$\beta_x$ / $\beta_y$	& cm	& \numrange{7}{10}	& \numrange{7}{10}	& 15 / 0.08	 & ?	& 25 / 0.03 & 0.023\\
Beam size		& \si{\micro\meter}	& 6		&	6 & 	&	& 4.5 / 0.006 & 0.01 \\
\bottomrule
\end{tabular}
\end{table}

\chapter{ERL High-Energy \positron\electron{} sub-Panel Report}
\label{sec:appendix:electronpositron}

\section{Introduction}

While the Panel was collecting information, an ERL concept was put forward to possibly build a superconducting linear collider as an energy-recovery twin collider, termed ERLC \cite{telnov2021highluminosity_v1},
with the prospect of a large increase of the instantaneous $\positron\electron$ luminosity as compared to the ILC.
This caused the formation, in agreement with the LDG, of a sub-Panel to evaluate the prospects (primarily luminosity), involved R\&D, and the schedule and cost consequences for the ERLC.
The sub-Panel was also asked to evaluate a concept, developed recently, for a high-luminosity circular energy-recovery collider, CERC \cite{LITVINENKO2020135394}, with the same criteria.

A sub-Panel with wide experience in accelerator design, construction, and operation was formed with the following members:
\begin{center}
\begin{tabular}{cc}
Chris Adolphsen (SLAC)	& Reinhard Brinkmann (DESY) \\
Oliver Brüning (CERN) & Andrew Hutton (Jefferson Lab)---Chair \\
Sergei Nagaitsev (Fermilab)	& Max Klein (Liverpool) \\
Peter Williams (STFC, Daresbury) & Akira Yamamoto (KEK) \\
Kaoru Yokoya (KEK) & Frank Zimmermann (CERN) \\
\end{tabular}
\end{center}
For each concept, the sub-Panel looked at the published information and provided questions to the authors.
The authors were invited to give a one-hour presentation to the sub-Panel, followed by thirty minutes for questions.
This meeting was followed up by another thirty-minute question and answer session after the sub-Panel had had time to go through the material provided.
The draft report was then sent to the authors of the two concepts for comment.
Both sets of project parameters were modified in response to the sub-Panel’s comments, but we did not have time to fully evaluate the new parameter sets, limiting ourselves to indicating the effect the changes would have on our conclusions.

This report is divided into two main sections; one on the CERC, the other on the ERLC, with an identical format for the two concepts.  

\section{Executive Summary}

A group of experts (the ERL $\positron\electron$ sub-Panel) was invited to study two concepts to re-imagine the FCC-ee and the ILC as ERLs, called CERC and ERLC, respectively.
The sub-Panel spent several months evaluating these concepts, an indication of the importance that the sub-Panel members felt they deserved, not least for the excitement that the proposals had generated.
The overall, unanimous conclusion of the sub-Panel members was that neither concept is currently at a stage where it could be developed into a project in the near future.
As might have been expected for concepts developed by really small teams (the CERC had three authors and the ERLC only one), both of the proposals were incomplete, and many details remained to be worked out. 

However, with these ideas, the prospects for a new era of using ERLs for high-energy colliders have now been looked at in a new light, and we expect that other ideas will soon be forthcoming.
The sub-Panel therefore focused on identifying the new hardware and beam physics topics that were identified during the evaluation, rather than focusing on missing details of the concepts.
In this context, the sub-Panel was able to identify several areas for R\&D that would positively impact not only ERLs for high-energy $\positron\electron$ colliders but the whole field of ERLs.
Most notably, the potential is striking for future projects that would be unlocked by operation at \SI{4.5}{\kelvin} with a high $Q_0$; for example, a next-generation ERL-based $\positron\electron$ linear collider to study the Higgs vacuum potential.  

\section{CERC---Vladimir Litvinenko, Thomas Roser, Maria Chamizo-Llatas}

\subsection{General}

The concept was presented in \cite{LITVINENKO2020135394} and was modified in a presentation by V.~Litvinenko to the sub-Panel [unpublished] to respond to a criticism from V.~Telnov regarding the bunch length \cite{telnov_talk_lcws21}.
The damping ring energy was also changed to \SI{8}{\giga\electronvolt} in the presentation.
Since neither parameter set was fully self-consistent, we concentrated on the presentation made to the sub-Panel but also looked at changes in the published version (notably the effect of different bunch lengths).

\subsection{Findings}

The concept is still in the early stages and is therefore difficult to compare with the FCC-ee design, which is the result of many years’ work by a large group.
This became evident in the discussions because the concept was being modified during the evaluation to respond to problems that had been identified.
The sub-Panel decided to focus on a review of the published concept but collected the improvements proposed by the authors and the sub-Panel members in Section \ref{sec:appendix:electronpositron:cerc:performance}.

The drivers for the proposal were to make the collider more ``sustainable,'' and to increase the luminosity, particularly at higher energies.
These areas were therefore a particular focus of the sub-Panel.
However, the cost is always an important factor in choosing a design philosophy, so this was also evaluated.  

\subsection{Performance}
\label{sec:appendix:electronpositron:cerc:performance}

This section describes the luminosity issues and the emittance problems throughout the facility.  

\subsubsection{Luminosity}

There were a few design-parameter sets of CERC presented in the sub-Panel meeting.
Here, we mainly discuss the ``updated parameter'' set in \cite{troser166} with the long bunch choice.

The luminosity is defined somewhat differently for CERC and FCC-ee.
Collisions in the CERC occur in only one of the up to three interaction regions at a time, so the luminosity is the total facility luminosity for the three interaction regions.
In the FCC-ee, the luminosity is per interaction region, so to obtain the total facility luminosity, it should be multiplied by the number of installed detectors, two or four, with a slight dependence on the number of detectors.
This is based on the operating experience at LEP, which had four detectors, each with a beam-beam tune shift of $\sim 0.1$.

\subsubsection{Beam-Beam Interaction}

One of the major issues that drive the whole parameter set is the beam-beam interaction.
Let us take the collision parameter set for $\text{t}\overline{\text{t}}$:

\[
\begin{array}{llr}
N =\num{1.4e11} & & \text{(number of particles per bunch)} \\
\sigma_z = \SI{50}{\milli\meter} & & \text{(RMS bunch length)} \\
\beta_x^* = \SI{1}{\meter} & \beta_y^* = \SI{2}{\milli\meter} & \text{(IP beta function)} \\
\sigma_x^* = \SI{4.7}{\milli\meter} & \sigma_y^* = \SI{6.6}{\nano\meter} & \text{(IP beam size)} \\
D_x = 5.0 & D_y = 3500 & \text{(disruption parameter)}
\end{array}
\]

It is assumed the beams are kept focused during the collision by the focusing force of the opposite bunch owing to the matching of the $\beta_y^*$ and the space-charge beta function $\beta_\text{SC}$ in spite of the long bunch $\sigma_z = 25 \beta_y^*$.
This extremely long bunch has been chosen so that the low-energy tail due to beamstrahlung can be accepted by the deceleration beamline and the damping ring.

Whether or not the extreme choice of the collision parameters ($D_x$, $D_y$, $\sigma_z /\beta_y^*$) is realistic drives the entire scenario.
In addition, compared with the original parameter set \cite{LITVINENKO2020135394}, the horizontal disruption parameter $D_x$ in the updated parameter set is significantly larger than 1 (from $D_x = 22$ for Z-pole to 4.4 for \SI{300}{\giga\electronvolt}).
Hence, the horizontal beam size will also change during the collision. Accurate simulations are required, and, obviously, the simulations must take into account the horizontal force to the same level of precision as the vertical.

A shorter bunch ($\sim \SI{10}{\milli\meter}$) is suggested in Section \ref{sec:appendix:electronpositron:cerc:linac_design} from the point of view of RF acceleration.
In that case, the critical energy of the beamstrahlung would increase by a factor of 5, which makes the design of the decompressor and the damping ring more demanding.

\subsubsection{Final focus system}

The emittance increase due to the beam-beam interaction is large.
The authors expect a factor $\sim 5$ increase for $D_y \sim 100$.
The increase must be re-evaluated for the larger $D_x$ and $D_y$.
The quality of the beams before and after the collision is significantly different.
The present design adopts a head-on collision and uses the same optics for the beams after collision.
We suggest adopting a finite crossing angle, crab crossing, and different optics for accepting and focusing the disrupted beams after collision.
Taking into account the energy spread induced by beamstrahlung, the transverse emittance obtained after passing the disrupted beam through the outgoing final focus should be confirmed by simulations.

\subsubsection{Damping rings}

The vertical/horizontal emittance ratio $\sim 1000$ is large compared with existing linear collider parameters ($\sim 100$ for CLIC, $\sim 200$ for ILC) but not compared with modern light sources.
This seems to be feasible.
On the other hand, the normalized vertical emittance of \SI{8}{\nano\meter} is much smaller than those of ILC (\SI{35}{\nano\meter}) and FCC-ee (\SI{1000}{\nano\meter} at \SI{182.5}{\giga\electronvolt}).
The tolerances should be carefully examined.
To keep this value up to the collision energy can be a very difficult problem, which will be discussed later. 

Because the vertical emittance \SI{8}{\nano\meter} is small and the bunch charge is very high (\SIrange{13}{25}{\nano\coulomb}), the effects of intrabeam scattering should be evaluated (the effect is already visible at ILC with \SI{20}{\nano\meter} and \SI{3.2}{\nano\coulomb}, though at a lower energy of \SI{5}{\giga\electronvolt}).
The electron cloud instability in the positron damping ring must also be examined since the beam current is rather high (up to $\sim\SI{5}{\ampere}$), and the bunch spacing is much shorter than in the ILC.

We would need more concrete design parameters of the damping rings (beam energy, circumference, damping time, beam dwell time, etc.), together with those of the decompressor, for a more detailed assessment.

The CLIC design considers a factor 3 vertical emittance increase between damping ring and collision point (\SI{20}{\nano\meter} normalized), while for CERC the normalized vertical damping ring emittance of \SI{8}{\nano\meter} is assumed to be preserved on the way to the IP, which seems optimistic.

\subsubsection{Arcs}

Most problems related to the arcs come from the orbit length: $\sim\SI{400}{\kilo\meter}$ each for acceleration and deceleration. 

There is a proposed focusing system for the arcs (combined function, with dipole, quadrupole, and sextupole components, \SI{16}{\meter}-period FODO, and a phase advance per cell of \ang{90}). Presumably, weaker focusing (lower phase advance) would be better for the arcs with lower-energy beams.

The increase of the horizontal emittance due to synchrotron radiation has been estimated and found to be acceptable.
The most important issue is the preservation of the small vertical emittance of \SI{8}{\nano\meter} over the \SI{400}{\kilo\meter}-long orbit in the presence of strong focusing magnets.
Emittance growth comes both from the misalignment of the combined-function magnets and the ground motion, and tolerances are normally tighter for stronger focusing.
It should be easy to estimate the tolerance of alignment jitter (though this can presumably be cured by the feedback system).
The next step would be to estimate the vertical emittance growth with both misalignment and ground motion.
The orbit correction algorithm must be studied (the dispersion-free method, in which the beam energy is changed, cannot be used).

For these purposes, the CLIC studies will be very helpful (ILC is somewhat different because of the large aperture and weak focusing system). In the case of CLIC, a strong focusing system is required due to the strong wake field of the high-frequency cavities.
The CERC arcs do not have cavities, but a strong focusing system is needed for the synchrotron radiation.
The combined-function magnets, containing dipole, quadrupole, and sextupole components, are likely to complicate beam-based alignment. 

The effects of the wake fields should be studied, in particular because of the small beam pipe (\SI{15}{\milli\meter} radius) and the high bunch charge.
The long bunch may also be sensitive to transverse wake fields, even though there are no RF cavities.
The resistive-wall wake field should also be studied because of the long orbit.

Another issue may be beam-gas scattering by the residual gas because pumping will be difficult in the small-bore magnets. The fast ion instability from residual gas ionization will probably not be an issue because of the long inter-bunch distance. 

We do not yet know the impact on the vertical emittance preservation due to the vertical orbit displacements before and after the RF, which are needed for sharing the linacs by the different-energy beams.

\subsubsection{Linac beam dynamics}

Cumulative beam breakup due to the deflecting HOMs should be studied in the linacs because the beam current is high (sixteen times greater than in the arcs), but presumably it is acceptable.
The effects of the short-range wake will define the alignment tolerance of the linac components (cryomodules and quadrupole magnets).
Definite quantitative conclusions cannot be made since the important parameters---the RF frequency, the bunch length, and the focusing system---are unknown.
However, even with favorable choices for these parameters, the tolerance would be tighter than for the ILC ($\sim\SI{200}{\micro\meter}$), even if the target vertical emittance is the same as in the ILC.

\subsection{Linacs}

For the linac design and operating parameters, we consider here just the $\text{t}\overline{\text{t}}$ physics case where the energy of each beam at the IP is \SI{182.5}{\giga\electronvolt}, the bunch charge is \SI{22.5}{\nano\coulomb}, and the collision rate is \SI{45}{\kilo\hertz}.
Each bunch makes eight passes through two linacs: four to reach the IP energy, and four to de-accelerate to the \SI{2}{\giga\electronvolt} injection energy.
The beam is segmented into trains of eight bunches that have a spacing equal to the RF period of the main ERL cavities.
The trains are separated by \SI{22.2}{\micro\second}, and with the two counter-rotating beams, the average linac current is \SI{16.2}{\milli\ampere}.

\subsubsection{Linac design}
\label{sec:appendix:electronpositron:cerc:linac_design}

The linac design was not specified in the proposal, so we had to make assumptions about the CW SRF cavities that would be used.
The BNL ERL 5-Cell, \SI{704}{\mega\hertz} design was chosen as the main (energy-recovery) cavity assuming that short bunches (\SI{2}{\milli\meter} RMS) would be accelerated as originally proposed (the feasibility of longer bunches is discussed in subsequent sections).
Each cavity is housed in a single cryomodule with an in-line, room-temperature, ferrite HOM absorber attached at each end: the flange-to-flange length of the cryomodule with absorbers is \SI{5.0}{\meter}.
The loss factor for \SI{2}{\milli\meter} RMS bunches is \SI{3.2}{V/pC} excluding the fundamental mode, which requires the two HOM loads to absorb \SI{1.2}{\kilo\watt} of broad-band HOM power---this is comparable to what the 2006 BNL ERL design assumed.  

The cavity has an active length of \SI{1.52}{\meter} and an $R/Q$ of \SI{404}{\ohm}.
The gradient is assumed to be \SI{16}{\mega\volt\per\meter} with a $Q_0$ of \num{2e10} at \SI{2}{\kelvin}, which was achieved in prototype cavity measurements.
At this gradient and $Q_0$, the \SI{2}{\kelvin} niobium heat load is \SI{73.2}{\watt}.
The cavity $Q_\text{ext}$ was chosen to be \num{3.5e7} (half BW = \SI{10}{\hertz}) so the RF power is minimized with \SI{10}{\hertz} detuning, which is a rough guess of the maximum value based on the CW performance of other SRF cavities.
With \SI{2}{\hertz} average detuning from microphonics, the RF power required per cavity with no beam loading is \SI{10.8}{\kilo\watt}. 

The linacs also have to restore the beam energy loss from synchrotron radiation in the arcs---in this case, \SI{10.8}{\giga\electronvolt} per beam---which produces about \SI{20}{\mega\watt} of radiated power.
To do this, the bunches could be run off-crest by \ang{23} and the cavities detuned by \SI{18.4}{\hertz} to cancel the out-of-phase beam loading.
A more straight-forward approach, which is assumed here, is to include second-harmonic (\SI{1408}{\mega\hertz}) cavities that would accelerate the beams during both energy ramp-up and ramp-down.
The disadvantages of using such cavities are the higher wake fields and larger RF-induced bunch energy spread.
This approach is also about \SI{6}{\percent} less efficient in terms of the total RF power required. 

Assuming there would be two second-harmonic cavities per cryomodule, the length of the cryomodule plus absorber is \SI{5.2}{\meter}.
Based on a scaled cavity design, the cavity parameters are: loss factor without fundamental mode $= \SI{4.2}{V/pC}$, $R/Q = \SI{340}{\ohm}$, and active length $= \SI{0.76}{\meter}$.
The \SI{2}{\kelvin} niobium heat load is \SI{21.7}{\watt} per cryomodule for a $Q_0$ of \num{2e10} and a gradient of \SI{16}{\mega\volt\per\meter}, the HOM power is \SI{3.1}{\kilo\watt} per cryomodule, and the required RF power is \SI{393}{\kilo\watt} per cryomodule, which mostly ($\sim\SI{90}{\percent}$) goes to accelerating the beams.
To achieve the required \SI{46.5}{\giga\electronvolt} energy gain per pass, 1856 cryomodules with \SI{704}{\mega\hertz} cavities and 56 cryomodules with \SI{1408}{\mega\hertz} cavities are required in the two linacs, which would span \SI{9.6}{\kilo\meter} in total without interleaved magnets, etc.

\subsection{Bunch length}

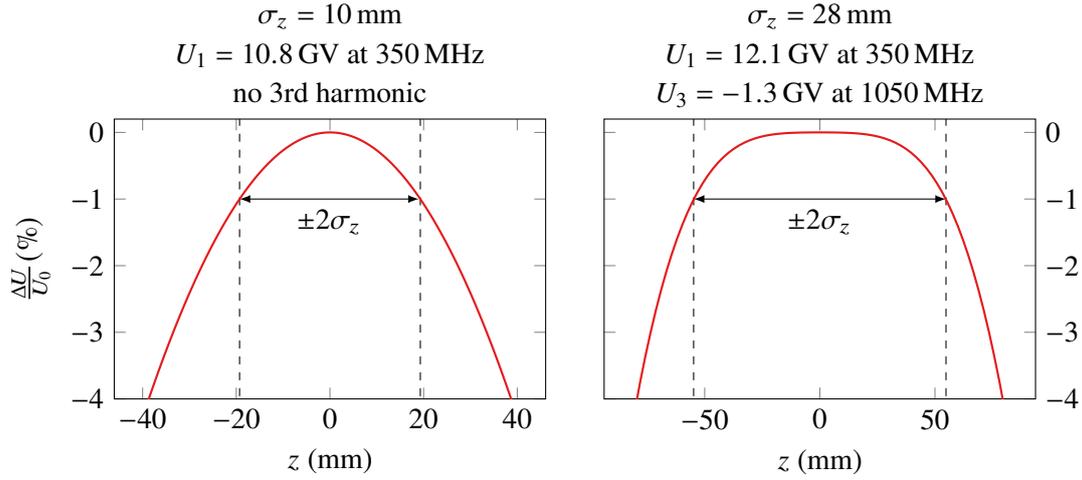
\begin{figure}[htb]\centering
\tikzsetnextfilename{appendix_electronpositron_cerc_bunchlength1}
\begin{tikzpicture}
\begin{groupplot}
[
    group style={
        group size=2 by 1,
        horizontal sep=2em,
        %vertical sep=1.5cm,
    },
    width=.48\linewidth,
    height=.35\linewidth,
    legend style={
        cells={align={left}},
        anchor=south,
    },
    %ymajorgrids=true,
    %xmajorgrids=true,
    %major grid style={dotted},
    ymin=-4,
    ymax=0.2,
]
\nextgroupplot[
    xlabel={$z$ (\si{\milli\meter})},
    ylabel={$\frac{\Delta U}{U_0} (\si{\percent})$},
]
%\draw[draw=none, fill=yellow, fill opacity=0.2] (axis cs:-20, 100) rectangle (axis cs:20, -100);
\draw[dashed] (axis cs:-19.3, 100) -- (axis cs:-19.3, -100);
\draw[dashed] (axis cs:19.3, 100) -- (axis cs:19.3, -100);
\path (axis cs:-19.3, 0) coordinate (left1);
\path (axis cs:19,3, 0) coordinate (right1);
%\path (rel axis cs:0.5, 0.4) coordinate (y1);
\path (axis cs:0, -1) coordinate (y1);
\draw[latex-latex] ($(left1 |- y1)$) -- ($(right1 |- y1)$)  node[anchor=north, pos=.5] {$\pm 2\sigma_z$};
\addplot+[thick, mark=none] table[x expr={\thisrowno{0}*1e3}, y index=1] {figures/cerc_bunchlength_no3rd.txt};
\path (rel axis cs:0.5, 1) coordinate (title1);
\nextgroupplot[
    xlabel={$z$ (\si{\milli\meter})},
    ylabel near ticks, yticklabel pos=right,
]
\draw[dashed] (axis cs:-54.8, 100) -- (axis cs:-54.8, -100);
\draw[dashed] (axis cs:54.8, 100) -- (axis cs:54.8, -100);
\path (axis cs:-54.8, 0) coordinate (left2);
\path (axis cs:54.8, 0) coordinate (right2);
\path (axis cs:0, -1) coordinate (y2);
%\path (rel axis cs:0.5, 0.4) coordinate (y2);
\draw[latex-latex] ($(left2 |- y2)$) -- ($(right2 |- y2)$)  node[anchor=north, pos=.5] {$\pm 2\sigma_z$};
\addplot+[thick, mark=none] table[x expr={\thisrowno{0}*1e3}, y index=1] {figures/cerc_bunchlength_with3rd.txt};
\path (rel axis cs:0.5, 1) coordinate (title2);
\end{groupplot}
\draw (title1)  node[anchor=south, align=center] {$\sigma_z = \SI{10}{\milli\meter}$\\$U_1 = \SI{10.8}{\giga\volt}$ at \SI{350}{\mega\hertz}\\no 3rd harmonic\vphantom{fg}};
\draw (title2)  node[anchor=south, align=center] {$\sigma_z = \SI{28}{\milli\meter}$\\$U_1 = \SI{12.1}{\giga\volt}$ at \SI{350}{\mega\hertz}\\$U_3 = \SI{-1.3}{\giga\volt}$ at \SI{1050}{\mega\hertz}\vphantom{fg}};
\end{tikzpicture}
\caption{Relative bunch energy profile caused by RF curvature. Left: \SI{350}{\mega\hertz} system. Right: Added third-harmonic RF. In both cases, the allowable bunch length $\sigma_z$ corresponds to a maximum relative RF voltage deviation of \SI{1}{\percent}.}
\label{fig:appendix:electronpositron:cerc:bunchlength1}
\end{figure}

Lengthening of the bunch in order to reduce beamstrahlung becomes a potential problem for the RF system due to the non-linearity of the RF potential.
A low-frequency system is advantageous in this context, and an improvement of the RF field flatness can be achieved by adding a higher-harmonic system with a voltage opposite to the main ERL RF.
For the example of a \SI{350}{\mega\hertz} 1st-harmonic system, the situation is depicted in Fig.~\ref{fig:appendix:electronpositron:cerc:bunchlength1} (left).
Assuming that the FFS requires an energy deviation from the on-crest particle of not more than \SI{1}{\percent} for the core of the beam ($\pm 2\sigma_z$) in order to avoid an increase of the beam size at the IP (this could be more relaxed with a wide-band FFS), the maximum tolerable bunch length is \SI{10}{\milli\meter}.
When adding a 3rd-harmonic system with \SI{11}{\percent} of the voltage of the 1st-harmonic system, a bunch length up to \SI{28}{\milli\meter} would be tolerable, as shown in Fig.~\ref{fig:appendix:electronpositron:cerc:bunchlength1} (right).
Note that in that case, the 1st-harmonic voltage also has to be increased by \SI{11}{\percent} to maintain the full beam energy.

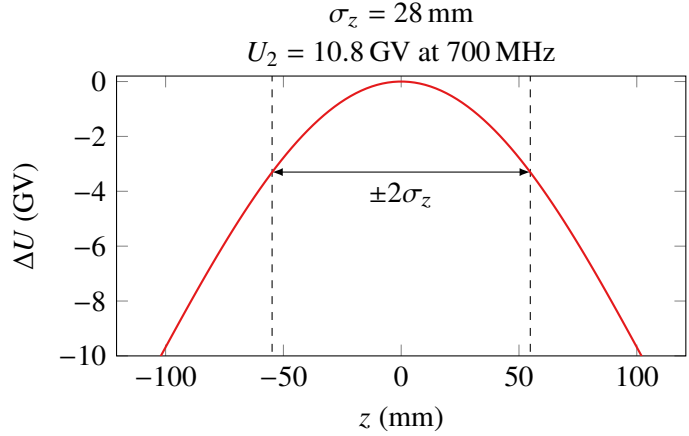
\begin{figure}[htb]\centering
\tikzsetnextfilename{appendix_electronpositron_cerc_bunchlength2}
\begin{tikzpicture}
\begin{axis}
[
    width=.6\linewidth,
    height=.35\linewidth,
    ymin=-10,
    ymax=0.2,
    xlabel={$z$ (\si{\milli\meter})},
    ylabel={$\Delta U$ (\si{\giga\volt})},
]
\draw[dashed] (axis cs:-54.8, 100) -- (axis cs:-54.8, -100);
\draw[dashed] (axis cs:54.8, 100) -- (axis cs:54.8, -100);
\path (axis cs:-54.8, 0) coordinate (left2);
\path (axis cs:54.8, 0) coordinate (right2);
\path (axis cs:0, -3.3) coordinate (y2);
%\path (rel axis cs:0.5, 0.4) coordinate (y2);
\draw[latex-latex] ($(left2 |- y2)$) -- ($(right2 |- y2)$)  node[anchor=north, pos=.5] {$\pm 2\sigma_z$};
\addplot+[thick, mark=none] table[x expr={\thisrowno{0}*1e3}, y index=1] {figures/cerc_bunchlength_700mhz.txt};
\path (rel axis cs:0.5, 1) coordinate (title2);
\end{axis}
\draw (title2)  node[anchor=south, align=center] {$\sigma_z = \SI{28}{\milli\meter}$\\$U_2 = \SI{10.8}{\giga\volt}$ at \SI{700}{\mega\hertz}};
\end{tikzpicture}
\caption{Bunch energy profile caused by RF curvature in the 2nd-harmonic system.}
\label{fig:appendix:electronpositron:cerc:bunchlength2}
\end{figure}

The energy spread in the bunch caused by the RF curvature on the accelerating branch of the ERL is completely removed on the decelerating branch so that it does not contribute to the issue of large energy spread of the beam at the damping ring energy.
However, the 2nd-harmonic system, which is designed to compensate for the radiative energy loss, creates a non-linear energy profile which is not compensated.
The bunch energy profile for a \SI{700}{\mega\hertz} system with a voltage of \SI{10.8}{\giga\volt} is shown in Fig.~\ref{fig:appendix:electronpositron:cerc:bunchlength2} for a bunch length of \SI{28}{\milli\meter}.
The energy deviation with respect to the on-crest particle amounts to about \SI{-3.5}{\giga\electronvolt}
at $\pm 2\sigma_z$, which is obviously a serious problem for the damping ring energy and energy acceptance.
One could also consider adding a 3rd-harmonic system here, but at \SI{2.1}{\giga\hertz} it will be very difficult to handle HOM problems; furthermore, the energy variation in the tails of the longitudinal bunch distribution becomes too large.
Therefore, a bunch length of much more than about \SI{10}{\milli\meter} does not seem to be realistically feasible even with such a rather low-frequency system.

\subsection{Polarisation}

Two cases must be considered, namely making use of the Sokolov-Ternov radiative polarization in the damping rings, and the injection and acceleration of pre-polarized beams. 

The polarization time of the Sokolov-Ternov effect at \SI{8}{\giga\electronvolt} is of the order of an hour, very roughly speaking.
The cycle time of DR–Arc–IP–Arc–DR is of the order of \SI{10}{\milli\second}.
Therefore, a particle makes $\sim\num{3e5}$ cycles during a polarization time.
So, Sokolov-Ternov polarization does not work if the depolarization in one cycle exceeds $\sim 1/\num{3e5} \approx \SI{3}{ppm}$.
Estimating the depolarization to this level is not easy, but presumably it is not fatal (beam-beam depolarization must be carefully simulated including the downstream FFS).
Also note that the probability of spin-flip transition by beamstrahlung by one crossing is of order \num{e-4} in the case $\sigma_z=\SI{2}{\milli\meter}$.
The most difficult problem is the beam loss due to beamstrahlung, since obviously the polarization lifetime cannot exceed the beam lifetime.
Suppressing the beam loss to the level of a few parts per million seems to be very difficult.
For a \SI{2}{\giga\electronvolt} damping ring, the polarization times become too long for this mechanism.

The depolarization in one cycle does not impose a tight condition for the case of a pre-polarized beam.
The number of particles per second at the IP ranges from \num{2e16} for Z to \num{1e15} for \SI{300}{\giga\electronvolt} (with \num{6e15} for $\text{t}\overline{\text{t}}$).
The polarized electron source for ILC is designed to produce \SI{1.3e14}{\per\second}.
Hence, if the beam loss per cycle is $\ll\SI{1}{\percent}$, the loss can be replenished by a polarized source (top-up injection in DR).
Of course, producing polarized positrons is hard, but, in principle, the ILC baseline scheme (an undulator with $>\SI{100}{\giga\electronvolt}$ electrons) can be employed for CERC (but would require more investment). 

In the FCC-ee, an accurate energy measurement can be obtained by using polarization: the method uses resonance with the spin tune and the frequency of the depolarizer.
However, for CERC, the spin tune (number of precessions in one cycle IP-Arc-DR-Arc-DR) is not well-defined due to the long stay in the damping ring.
Moreover, even if the spin tune is defined, its relationship with the beam energy at the IP is undefined.
Hence, the beam energy cannot be measured by using polarization.

Note that, if the main contribution to the radiation damping in the damping rings comes from the (normal, non-asymmetric) wigglers rather than the arc magnets, a high degree of polarization cannot be expected by the Sokolov-Ternov effect.

\subsection{Cost estimate}

The cost of the CERC is compared to the cost of the accelerator design presented in the FCC-ee CDR, which foresees \SI{47}{\percent} of the total cost for civil engineering, \SI{17}{\percent} for technical infrastructure, \SI{27}{\percent} for the injector complex and collider for operation at beam energies between 45.6 and \SI{120}{\giga\electronvolt}, and about \SI{9}{\percent} for the upgrade to \SI{182.5}{\giga\electronvolt} (essentially an \SI{800}{\mega\hertz} SRF system).

Since details of the CERC design are not known or not yet decided (arc corrector magnets, arc vacuum system (coating?), the presence or lack of magnet movers, energy and size of damping rings\ldots) and/or likely to evolve, some of the CERC cost estimates have large uncertainties.  

\subsubsection{Civil engineering and technical infrastructure}

It is assumed that the costs of the main tunnels, access shafts, surface sites, etc.~and of the technical infrastructure are the same as for FCC-ee.  

\subsubsection{Damping rings}

A rough cost estimate for CERC is based on \SI{1}{\kilo\meter}-circumference rings operating at \SI{2}{\giga\electronvolt} beam energy and with \SI{1.3}{\ampere} beam current (\SI{365}{\giga\electronvolt} collider case).
With a damping time of \SI{2}{\milli\second} (requiring $\sim\SI{150}{\meter}$ of \SI{2}{\tesla} damping wigglers), the energy loss per turn amounts to about \SI{3}{\mega\electronvolt}; the RF power that must be supplied to the beam is about \SI{4}{\mega\watt}---even higher in proportion to the beam current for the lower center-of-mass (cryomodule) energy options.
The cost for two rings, including the \SI{1}{\kilo\meter} tunnel (for both rings) and technical infrastructure, beam sources, pre-accelerators, transfer lines, and injection/extraction systems is estimated at \SI{3}{\percent} of the total FCC-ee cost.
Should it be necessary to increase the damping ring energy (e.g., to \SI{8}{\giga\electronvolt} as suggested by the authors in response to questions by the sub-Panel) to be able to accommodate the large longitudinal emittance of the decelerated beam from the ERL, then costs for the magnets and RF will be higher, but fewer damping wigglers would be required.
Very roughly, the cost would double to about \SI{6}{\percent} of the total FCC-ee cost.

These costs will increase significantly if additional tunnels, possibly with lengths of a few km or even tens of km, are required to connect damping rings and collider tunnels.

\subsubsection{Collider arc magnet \& vacuum systems}

Starting from a cost estimate for eRHIC magnets and taking into account a possible cost reduction due to mass production of \SI{30}{\percent}, the total arc magnet cost for 16 beam lines has been estimated at about \SI{13}{\percent} of the total FCC-ee cost.
Taking into account additional costs for vacuum systems, beam diagnostics, precision alignment, and corrector elements, we are multiplying this by a factor of 3, which still appears optimistic.
We then arrive at \SI{39}{\percent} of the total FCC-ee cost. 

\subsubsection{Collider SRF system and cryogenics}

The fundamental RF system consists of two \SI{23.3}{\giga\volt} linacs.
Scaling from the LHeC SRF and cryogenic cost estimate for a \SI{20}{\giga\volt} linac, we obtain \SI{24}{\percent} of the total FCC-ee cost for these two linacs.
Alternatively, we can also estimate this cost by extrapolation from the FCC-ee upgrade from 240 to \SI{365}{\giga\electronvolt}, which requires \SI{15}{\giga\volt} \SI{800}{\mega\hertz} SRF and associated cryogenics.
Scaling from \SI{16}{\giga\volt} to \SI{46.5}{\giga\volt} and slightly reducing the resulting price (as less RF power will be required) provides a value that is consistent with the \SI{24}{\percent} estimated from LHeC.

The energy loss from synchrotron radiation (about \SI{15}{\giga\electronvolt} per beam at \SI{365}{\giga\electronvolt} center-of-mass energy) can be compensated by a higher-harmonic RF system.
Considering a \SI{1.3}{\giga\volt} RF linac at higher frequency (8 passes through this linac make up for \SI{10.8}{\giga\electronvolt} energy loss in the arcs) and including the high RF power required, we assign \SI{1}{\percent} of the FCC-ee total cost.

In this cost estimate, a possible higher-harmonic RF system for flattening the RF potential (in case of long bunches) is not included; it would cause an additional cost contribution.

\subsubsection{Other elements}

Several further elements will contribute to the total cost, such as the straight-section and final-focus magnets, the interaction region, transfer \& bypass lines, survey and alignment systems for the final focus, corrector magnet systems, interaction-region beam diagnostics, accelerator control systems, etc.
We assign \SI{4}{\percent} of the total FCC-ee cost to these remaining items. 

\subsubsection{Total cost estimate}

The relative costs are summarized in Table~\ref{tab:appendix:electronpositron:cerc:relative_costs}.

\begin{table}[htb]\centering
\caption{Cost estimate for the CERC accelerator relative to FCC-ee for the case of \SI{365}{\giga\electronvolt} ($\text{t}\overline{\text{t}}$ machine).}
\label{tab:appendix:electronpositron:cerc:relative_costs}
\begin{tabular}{p{.7\linewidth}r}
\toprule
Item & Estimated fraction of\\
& total FCC-ee cost \\
\midrule
Civil engineering (if no new straight tunnels are required to accommodate the longer RF sections) & \SI{47}{\percent}\\
Technical infrastructure & \SI{17}{\percent}\\
Damping rings (\SI{2}{\giga\electronvolt} / \SI{8}{\giga\electronvolt}) and injector & \SI{3}{\percent} / \SI{6}{\percent}\\
Collider arcs (16 beam lines) & \SI{39}{\percent} *\\
Main RF system \& cryogenics & \SI{24}{\percent}\\
Harmonic RF system \& cryogenics & \SI{1}{\percent}\\
Other (straight sections, final focus, IR, survey \& alignment, IR beam diagnostics, controls \ldots & \SI{4}{\percent}\\
\midrule
Total & \SI{138}{\percent} \\
\bottomrule
\end{tabular}
* possibly still optimistic
\end{table}

\subsection{Staging and Upgradability}
\label{sec:appendix:electronpositron:cerc:staging}

The CERC is projected to have a higher performance (luminosity for a given power footprint) than the FCC-ee for center-of-mass energies higher than about \SI{150}{\giga\electronvolt}.
The relative performance advantage becomes more pronounced the higher the center-of-mass collision energy; for center-of-mass collision energies below \SI{150}{\giga\electronvolt}, the FCC-ee appears to offer better performance reach than the CERC.

An appealing operation scenario would therefore be to start the physics program for the Z and WW physics programs with the FCC-ee Synchrotron configuration and to move to the CERC ERL configuration once the required center-of-mass collision energies surpass \SI{150}{\giga\electronvolt}.
We refer to this scenario as Staging or Upgradability of the FCC-ee configuration. 

The FCC-ee Design Report describes four operation phases for the machine with an increasing number of RF cryomodules required at each phase.
The first two phases are dedicated to the Z and W physics program.
The Z physics program requires 26 cryomodules in the main ring and up to 13 cryomodules in the booster.
The W physics program phase requires 52 cryomodules in the main ring and up to 34 cryomodules in the booster.
For higher collision energies, the number of required SRF cryomodules more than doubles again, and it might be appealing to explore the high-energy regime in the FCC in an ERL configuration where the SRF cryomodules are passed more than once. 

The CERC features three interaction points placed in between two SRF linacs with a length of approximately \SI{4.8}{\kilo\meter} each, placed symmetrically on either side of the central interaction zone.
Note that the sub-Panel estimates the length of the linacs to be twice this (Section~\ref{sec:appendix:electronpositron:cerc:linac_design}).
The CERC layout is required to minimize the synchrotron radiation losses in the arcs.
The FCC-ee layout, on the other hand, envisions two to four interaction points and features several \SIrange{2.1}{2.8}{\kilo\meter}-long SRF sections distributed around the ring.
Implementing the CERC configuration inside the FCC-ee tunnel would require a redesign of the FCC tunnel layout with sufficient space for the CERC linacs next to the central interaction point.
However, the required caverns for the detector placement are not compatible with the experimental caverns envisioned in the FCC-ee layout.
The extent to which such a design iteration affects the FCC-ee performance reach and cost would need to be assessed.

Furthermore, the FCC-ee configuration requires the installation of a booster ring inside the FCC tunnel (unless one plans on using an LHeC-type ERL in Recirculating Linac mode as an injector, which would be possible for Z-pole operation, and possibly still at the WW threshold), while the CERC requires two damping rings for the electron and positron beams in order to reduce the beam emittances of the spent beams before reuse for a new collision cycle.
Staging the FCC-ee and CERC configurations therefore requires both the booster ring (or LHeC-type Recirculating Linac injector) and the two damping rings.
The exact additional cost of this infrastructure still needs to be assessed. 

Upgradability of the FCC-ee configuration to the CERC configurations therefore appears not to be straightforward, and the exact cost and performance benefits can only be assessed once the implications of the above design iterations (long straight section in the FCC tunnel for the SRF, experimental caverns and added cost for Booster and damping rings) have been fully assessed.

\subsection{Earliest possible time for implementation}

It is appropriate to compare the possible time line for a project like the CERC to the envisaged timescale of FCC-ee.
As discussed in Section \ref{sec:appendix:electronpositron:cerc:staging}, there are different scenarios to consider, namely a stand-alone construction of CERC or construction of FCC-ee followed by an upgrade to the CERC.
There are two separate considerations; the time needed for R\&D that would put the CERC at the same risk level as the FCC-ee, and the actual construction time.  

For the R\&D, a precursor machine seems almost obligatory.
This could be a variable-energy CBETA-type machine with 4 passes with small-aperture magnets to show that the magnets can be installed and aligned cheaply.
We think this will require a multiple-magnet innovation as radical as was developed for Max IV or the permanent magnets developed for CBETA.
It would need at least one cryomodule ($\sim\SI{100}{\mega\electronvolt}$) so four passes would give \SI{400}{\mega\electronvolt}.
The same beam lines can be used for both accelerating and decelerating beams if the bending radius is sufficiently large so that there is minimal synchrotron radiation loss.
A five-year development would seem reasonable, but this should be done prior to project approval. 

Other topics to be studied:
\begin{enumerate}
\item SRF development in the areas of HOM damping and extraction, and towards higher quality factor. A ten-year development would seem reasonable, which could run in parallel to the test facility design, construction and testing.
\item Spreader and recombiner design without the enormous complexity of CBETA.
\item Using lower-energy beams over a shorter distance as a stand-in for high-energy beams over a long distance. Simulation should be able to compare them, but this does not sound unreasonable.
\item The interactions between the ERL system and the damping rings.
\end{enumerate}

For both the FCC-ee and CERC, the critical path will be tunnel construction, which is estimated as eight years.
For the CERC, a significant upfront design effort will be required to fit the envisaged sixteen beam lines into the FCC-ee tunnel cross-section.
Without an engineering design, it is difficult to estimate the increase in installation time required; nevertheless, an additional year seems reasonable.
For the FCC-ee, a two-year period would be required after tunnel completion to complete installation, so we assume three years for the CERC.
The preparatory work should not increase from the FCC-ee estimate of seven years.
This makes a total of eighteen years for the CERC.

\subsection{Power consumption}

For a power consumption evaluation, we consider the $\text{t}\overline{\text{t}}$ physics case, for which the linac design was discussed previously.
Using the LCLS-II design assumptions as guidance, the cryoplant efficiencies are assumed to be \SI{920}{\watt} per watt of heat removed at \SI{2}{\kelvin}, and that an additional \SI{20}{\percent} of power is required for the cryomodule thermal shield and power coupler cooling at \SI{5}{\kelvin} and \SI{40}{\kelvin}.
For simplicity, we assume that the \SI{2}{\kelvin} load is just from the RF heating of the cavity niobium (\SI{138}{\mega\watt} total) and ignore the static and dynamic heat losses at low temperature related to the HOM absorber connections on each end of the cryomodules, which are not straightforward to estimate but could be significant.
In this case, the AC power required for the cryoplants is \SI{153}{\mega\watt}. 

For the RF source AC power, a \SI{50}{\percent} AC-to-RF efficiency is assumed.
With the \SI{8.1}{\mega\watt} RF power required for the DRs and \SI{41.9}{\mega\watt} for the linac SRF cavities, the AC power consumption is \SI{100}{\mega\watt} (the RF power required to make up for HOM and beam-beam losses are relatively small and ignored for this analysis).
Note that a higher damping ring energy would increase the power consumption. 

Assuming all of the AC power ends up being transferred to air in the facility cooling towers, and that it takes \SI{1}{\watt} of AC power per \SI{4}{\watt} of heat dissipated, the utility power for the cryoplants and RF sources is \SI{63}{\mega\watt}.
Summing the AC power for the cryoplants, RF sources, and the associated heat removal gives a total of \SI{316}{\mega\watt}.
The actual AC power would be larger when including that required for the magnets, HVAC, controls, etc.

\subsection{Comments and suggestions for improvements}

In principle, the ERL is an excellent concept to overcome beam-beam limitations of a conventional circular collider by introducing single-pass collisions and the power consumption limitations of a conventional linear collider layout by recovering the beam energy of the spent beam and re-cycling it to accelerate the subsequent beam in the same accelerator system.
It therefore aims at combining the best parts of both collider concepts.
However, it is very important to minimize additional power/energy consumption to keep the advantages of this feature and to avoid canceling out the energy saving in the total wall plug-power balance including RF, cryogenics, magnets, and general services.
It is suggested to confirm the energy balance to be emphasized in addition to the reduction in synchrotron radiation.
For example, the wall-plug power of the required injector and damping ring complex and the RF and cryogenics for the SRF-ERL should be clearly discussed, as additional balance.

\subsection{R\&D required}

The CERC concept relies on several technologies which should be studied before a detailed proposal can be prepared.
Obviously, a detailed proposal would require a lot of simulations, and some of the required codes exist already and do not need to be developed.
The R\&D topics (many of which were pointed out by the authors) are as follows:  
\begin{enumerate}
\item High-order mode (HOM) damping of the SRF cavities to avoid the transverse beam-breakup instability.  
\item Absolute beam energy measuring systems with accuracy $\sim\num{e-5}$ at the Interaction Regions.
\item High-repetition-rate ejection and injection kickers for the \SI{2}{\giga\electronvolt} damping rings.
\item Compression and de-compression of the electron and positron bunches to match the energy acceptance of the damping rings.
\item Development of small-gap magnets with integrated pumping and beam position monitors, which can be easily aligned (in particular, reaching a tight roll tolerance).
\item Development of SRF technology towards higher quality factor and/or for \SI{4}{\kelvin} operating temperature to reduce power consumption from cryogenic load.
\item Demonstration of high HOM power absorption avoiding any cryogenic load into \SI{2}{\kelvin} level.
\item A beam-beam simulation code much more accurate than existing ones must be developed to treat the disruption parameter $\sim 3000$.
\end{enumerate}

\subsection{Updated parameters}

As this report was being finalized, the authors proposed an updated set of operating parameters and gave specific choices for the linac cavity design, voltage gain, and quality factor, which had not been provided in the initial proposal.
These new parameters are reported in Section \ref{sec:frontier:fcc-ee}.
We had assumed a $Q_0$ of \num{3e10}, the present state of the art.
The authors assumed that the $Q_0$ would be \num{e11} as a result of future R\&D.
The gradient was also reduced by a factor of 2. Taken together, these values would significantly lower the machine electrical power requirements from our assessment in the $\text{t}\overline{\text{t}}$ case but would roughly double the number of linac cavities.
Our simple cost model is not adequate to accurately assess these changes although an overall decrease in the cost is likely.
However, the new parameters reduce the luminosity by a factor of three and do not change the large beamstrahlung-induced bunch energy spread that brings into question the viability of this approach.
With the new parameters, the CERC would still be significantly more expensive than the FCC-ee.

\subsection{Recommendations}

The sub-Panel supports the idea of designing a collider based on an ERL to reduce the energy footprint of the facility, and the CERC is an excellent first attempt.
While the present proposal has several flaws due to the limited effort that the authors were able to devote to the design, the sub-Panel chose to look for ways that the design could be improved rather than focus on the problem areas.
\begin{enumerate}
\item We strongly recommend the development of a self-consistent set of parameters with associated preliminary simulations to fully demonstrate that the idea is viable.
\item The bunch length is a critical parameter: too short and the beamstrahlung becomes excessive; too long and the energy spread from the RF curvature becomes excessive. It will be necessary to carefully optimize the choice.
\item The energy requirements of the damping rings must be integrated in the design.  
\item We recommend R\&D on high-$Q_0$ cavities operating at \SI{4.5}{\kelvin}, which would reduce both the cost and the power consumption.
\end{enumerate}

\section{ERLC---Valery Telnov}

\subsection{General}

The basic concept is described in Chapter~\ref{sec:frontier:erlc} and has since been published in \cite{Telnov_2021}\footnote{It should be noted that the sub-Panel report is based on assumptions and design parameters from an earlier version of the ERLC publication~\cite{telnov2021highluminosity_v1}. While the general concept remains the same, some inconsistencies between Chapter~\ref{sec:frontier:erlc} and the statements made here are to be expected.}.
The concept promises extremely high luminosity, which has excited the future user community.
The sub-Panel therefore focused on this aspect as well as cost and sustainability.

\subsection{Performance}

\begin{table}[htb]\centering
\caption{Beam parameters of ERLC, based on \cite{telnov2021highluminosity_v1}. The parameters marked with $\dagger$ are derived.}
\label{tab:appendix:electronpositron:erlc:parameters}
\begin{tabular}{lr}
\toprule
Ring circumference & $\sim\SI{60}{\kilo\meter}$ \\
Electrons per bunch & \num{0.5e10} \\
Bunch distance & \SI{1.5}{\meter} \\
Number of stored positrons & $\sim\num{2e14}$ \\
Beam current (peak in pulse) & \SI{160}{\milli\ampere} \\
Horizontal normalized emittance at IP & \SI{20}{\micro\meter} \\
Vertical normalized emittance at IP & \SI{35}{\nano\meter} \\
Damping ring beam energy & \SI{5}{\giga\electronvolt} \\
Energy loss in DR & \SI{25}{\mega\electronvolt} \\
$\sigma_z$ at IP and in the linacs & \SI{0.3}{\milli\meter} \\
$\sigma_z$ in the \SI{5}{\giga\electronvolt} section & \SI{3}{\milli\meter} \\
$\sigma_x^*$ & \SI{4.5}{\milli\meter} \\
$\sigma_y^*$ & \SI{6.5}{\nano\meter} \\
Beam-beam tune $\xi_x$~$^\dagger$ & 0.112 \\
Beam-beam tune $\xi_y$~$^\dagger$ & 0.093 \\
Disruption parameter $D_x$~$^\dagger$ & 0.002 \\
Disruption parameter $D_y$~$^\dagger$ & 1.17 \\
Luminosity & \SI{4.8e35}{\per\square\centi\meter\per\second} \\
\bottomrule
\end{tabular}
\end{table}

The parameters used here are listed in Table~\ref{tab:appendix:electronpositron:erlc:parameters}.
The ring circumference is an estimate:
\begin{equation}
    2 \times \left[ \frac{\SI{250}{\giga\volt}}{\SI{20}{\mega\volt\per\meter}} \times \frac{1.1}{0.7} + \SI{4}{\kilo\meter} + \SI{1}{\kilo\meter}\right] + 2\times \SI{3}{\kilo\meter} \approx \SI{60}{\kilo\meter} ,
\end{equation}
where 0.7 is the cavity packing factor, 1.1 accounts for the space for a HOM absorber at every cavity, \SI{4}{\kilo\meter} = two FFS, \SI{1}{\kilo\meter} is the space needed for the bunch compressor/decompressor, \SI{3}{\kilo\meter} = arc at the end, and everything is doubled because of the return line.
If the cryomodule assumed in Section \ref{sec:appendix:electronpositron:erlc:power} is used, the circumference will be $\sim\SI{10}{\kilo\meter}$ longer.

\subsubsection{Longitudinal beam dynamics}

The entire machine is a storage (damping) ring with an unusual insertion from the bunch compressor to the decompressor consisting of the acceleration linac, Final Focus system (FFS), Interaction point (IP), FFS, and the deceleration linac.
The longitudinal dynamics can be somewhat different from a normal storage ring due to this long insertion.
If the bunch compressor and the decompressor are both of the simplest type, i.e., combination of an RF linac and a chicane, the total transformation in the longitudinal phase space from the compressor to the decompressor would be $z \rightarrow -z$ and $\epsilon \rightarrow -\epsilon$ (with $\epsilon = \frac{\Delta E}{E}$).
This will make the synchrotron tune close to a half integer, which means the head and the tail of a bunch are swapped every turn. The longitudinal beam dynamics might be quite unusual (the transverse may also be affected).
This situation can be avoided by a proper choice of the compressor/decompressor such that the total effect is an identity transformation.
This new configuration needs careful study as it is likely to be a configuration used in other, future ERL concepts.  

The energy loss (HOM loss) due to the longitudinal wake in the acceleration and deceleration linacs with its compensation scheme (normally off-crest compensation) is a large perturbation of the longitudinal dynamics.
This effect must be carefully studied.

The energy spread due to beamstrahlung is one of the key issues of the collider and is treated properly in the proposal.  

\subsubsection{Transverse emittance}

The vertical emittance in the parameter table is the same as in ILC.
However, since the proposed transverse damping time corresponds to $\sim400$ turns, some types of the emittance increase are accumulated for $\sim400$ turns to reach equilibrium, in contrast to the case of single-pass colliders such as the ILC.
Various stochastic effects belong in this category.

\subsubsection{Synchrotron radiation}
The effects of synchrotron radiation can be easily estimated, but if a future energy upgrade is envisaged, the required space must be carefully considered in the first stage, in particular the high-energy part (note that the increase of the normalized emittance by synchrotron radiation for a given orbital geometry is proportional to $E^6$).

For example, the FFS contains a long bending section to produce the dispersion for chromaticity correction.
The ILC dogleg can accept \SI{500}{\giga\electronvolt} beam (with some magnets omitted in the first stage).
However, the reserved space is far from enough if the emittance increase is multiplied by 400.
The FCC-ee will have such a bending section, but the beam energy is limited to $\sim\SI{180}{\giga\electronvolt}$, and the damping will be faster in the high energy region (40 turns), hence less accumulation.

The vertical layout of the tunnel might be similar to the ILC case, following the earth's curvature in the main linac and laser-straight in the FFS.
This means there is effective vertical bending in the main linac, and a vertical kick at the joint (\SI{0.3}{\milli\radian} in ILC).
The radiation effects must be taken into account for a future upgrade.
There will be no problem at \SI{250}{\giga\electronvolt} (center-of-mass), but this must be checked for higher energies.

\subsubsection{Linac misalignment and wake}
Much more complex is the emittance increase in the main linac (and FFS) due to the misalignment and the wake field.
ILC expects $\sim\SI{10}{\nano\meter}$ increase of the vertical normalized emittance in a single pass.
Obviously, the major components of the emittance increase are coherent turn by turn, but they do not accumulate over multiple turns.

However, some of them may be cumulative. Even \SI{0.1}{\nano\meter} out of \SI{10}{\nano\meter} single-pass effect can exceed the design emittance if multiplied by 400.
The increase in the deceleration linac must also be added. The possible source of the cumulative components may be a combination of the above effects (misalignment and wake field) with the chromaticity, which cannot be compensated in the linacs, unlike the case of ring colliders.

Another possible source of cumulative components may be the components of the ground motion faster than $\sim\SI{5}{\kilo\hertz}$ (the revolution time is $\sim\SI{0.2}{\milli\second}$).

\subsubsection{Beam-beam interaction}
A similar increase can come from the beam-beam interaction. How large is the cumulative component? The vertical disruption parameter is of order one as shown in the parameter table above.
The combination of the beamstrahlung and the disruption should also be studied.

\subsubsection{Transverse higher-order modes}
One of the problems in the transverse dynamics is the deflecting HOMs in the linac.
The beam current in the pulse is \SI{160}{\milli\ampere} compared with \SI{9}{\milli\ampere} in ILC (luminosity upgrade case).
Though there is still a safe margin on this issue in ILC, the effect in ERLC must be checked.

\subsubsection{Other issues related to the luminosity}
The collimators upstream of the interaction point suppress possible background coming from the beam halos, but they also create secondary particles, in particular muons, which become another background source.
The average beam current in ERLC is $\sim 2500$ times higher than in ILC, whereas the luminosity is ``only'' 35 times higher.
This means the number of secondary muons normalized by the luminosity is $\sim 70$ times higher than in ILC. Hence, a much more efficient system of collimator and muon shielding must be developed for the ERLC.

\subsubsection{RF}
\label{sec:appendix:electronpositron:erlc:rf}
For the linac design and operating parameters, we considered the HZ physics case, where the energy of each beam at the IP is \SI{125}{\giga\electronvolt} and the bunch charge is \SI{0.8}{\nano\coulomb}.
To achieve the ERLC design duty factor of one-third, the proposal suggests alternating \SI{2}{\second} of beam operation with \SI{4}{\second} periods with the RF off. However, this seems impractical as \SI{2}{\kelvin} cryoplants cannot handle large ($> \SI{10}{\percent}$) swings in the helium gas return flow rate on this time scale, and the ramp-up to high luminosity with a narrow-band energy-recovery linac would likely take more than one second.
For example, a \SI{300}{\milli\second} RF ramp to the nominal gradient, frequency, and phase, followed by a \SI{400}{\milli\second} beam current ramp at \SI{100}{\micro\ampere} per turn (to limit the required RF power) and then at least \SI{300}{\milli\second} for the feedback systems to tune up the luminosity.
A more realistic (but more expensive) approach is to operate CW with three times the bunch spacing, which maintains the desired luminosity.
In this case, the bunch spacing is \SI{15}{\nano\second}, the bunch rate is \SI{66.7}{\mega\hertz}, and the beam current is \SI{53.3}{\milli\ampere}.
This CW scenario is assumed in the discussion below.

The linac design was not specified in the proposal, so assumptions were made about the CW SRF cavities that would be used.
A CBETA-like cryomodule concept was chosen, but with dual cavities, that is, side-by-side \SI{1.3}{\giga\hertz} cavities with niobium cross connections so power can flow from one cavity to its neighbor as required for energy recovery.
The cryomodule would be \SI{9.8}{\meter} long, contain 6 dual cavities and have silicon carbide HOM absorbers between the cavities and at the ends of the two cavity strings.
For these cavities, the loss factor for the proposed \SI{0.3}{\milli\meter}-long bunches (RMS) would be \SI{10.1}{\volt\per\pico\coulomb} excluding the fundamental mode contribution.
This would require that each HOM load absorb \SI{430}{\watt} of broad-band HOM power, which is close to the maximum assumed in the CBETA design.

Each cavity would have an active length of \SI{0.81}{\meter}, and the pair would have an $R/Q$ of \SI{385}{\ohm}.
Assuming a gradient of \SI{20}{\mega\volt\per\meter} and a $Q_0$ of \num{3e10} at \SI{2}{\kelvin} (similar to the LCLS-II cavities), the \SI{1.8}{\kelvin} niobium heat load would be \SI{22.7}{\watt} per cavity pair.
The cavity pair coupler $Q_\text{ext}$ was chosen to be \num{6.5e7} (half BW $=\SI{10}{\hertz}$) so the RF power would be minimized with \SI{10}{\hertz} detuning (like for LCLS-II).
With \SI{2}{\hertz} average detuning from microphonics, the RF power required per cavity pair with no beam loading would be \SI{2.7}{\kilo\watt}.

The linacs would also have to restore the beam energy loss (\SI{0.1}{\percent}) from HOM generation, which would produce a total of \SI{12.8}{\mega\watt} of radiated power in the two linacs.
To do this, the decelerated beams could be run off-crest by \ang{2.6} and the cavities detuned by \SI{37}{\hertz} to cancel the out-of-phase beam loading.
The required RF power per cavity pair in this case would increase to \SI{3.2}{\kilo\watt}.
These values are beam-current-dependent, so they would need to be varied as the beam current was ramped up.

The huge steady-state loading (\SI{1.6}{\giga\volt\per\meter}) from each \SI{53}{\milli\ampere} beam makes the cavity fields very sensitive to imperfect loading cancelation (i.e., partial energy recovery). In particular, the relative timing of the \electron{} and \positron{} bunches at the cavities may vary due to slow tunnel temperature changes that move the cryomodules longitudinally.
For example, a \SI{1}{\milli\meter} displacement would produce a \ang{3} relative bunch phase difference and require a \SI{25}{\percent} increase in RF power and \SI{43}{\hertz} cavity detuning to compensate.
Instead, the cavities and/or cryomodules, which effectively act as RF interferometers, could be mounted on movers and their position adjusted to keep the average RF power constant.
Larger, long-range relative motion could perhaps be compensated by changing the reference RF frequency to keep the machine a fixed number of RF wavelengths long. Relative beam current changes would also be an issue.

To achieve a \SI{250}{\giga\electronvolt} center-of-mass energy, about 2500 cryomodules would be required, and the length of each linac would be about \SI{12}{\kilo\meter} (the CBETA cryomodules include space for quadrupole magnets, etc.).
The total required RF power would be \SI{47}{\mega\watt}, and with a \SI{40}{\percent} AC-to-RF efficiency, the AC power consumption would be \SI{117}{\mega\watt}.
The total \SI{2}{\kelvin} heat load from niobium RF heating is \SI{340}{\kilo\watt}, which would require \SI{281}{\mega\watt} of AC power to cool, assuming the expected LCLS-II cryoplant efficiency at \SI{2}{\kelvin}.
Based on a heat load analysis that was done for the Cornell ERL, \SI{180}{\mega\watt} of AC power would be required to cool the HOM absorbers at \SI{100}{\kelvin}, and another \SI{93}{\mega\watt} to remove the static heat load.
The static and dynamic heat loads between the cavities and HOM absorbers require simulations to estimate, but they are likely significant.
Without these losses, the total AC power related to cooling and RF generation is more than \SI{700}{\mega\watt} with \SI{100}{\percent}-duty-factor operation.

\subsubsection{Pulsed operation}
Pulsed operation with a duty factor of 1/3 is foreseen in order to avoid an unreasonably excessive cryogenic load, unless \SI{4.5}{\kelvin} operation becomes available.
In principle, CW operation would be very beneficial for stable operation of the ERLC.
Careful investigation is needed to determine whether cryogenic stability can be achieved in pulsed mode, how the stability of the RF field can be guaranteed during ramp-up, etc.
Furthermore, the beams have to be generated from scratch with every pulse, which causes a significant challenge for the positron source.
The production of \num{2e14} positrons must be done in a time much shorter than \SI{2}{\second}, say $\sim\SI{100}{\milli\second}$ (or \SI{400}{\milli\second} as in Section \ref{sec:appendix:electronpositron:erlc:rf}).
The ILC positron source can produce \num{2e14} positrons in \SI{1.5}{\second}.
On the positive side, the pulsed mode could be scaled to a lower duty factor, reducing the capital investment in cryoplants and power consumption, while still maintaining a very high luminosity proportional to the duty factor.

\subsection{Cost and schedule}
\label{sec:appendix:electronpositron:erlc:cost}

A comparison of the ILC-250 and ERLC cost estimates is summarized in Table~\ref{tab:appendix:electronpositron:erlc:cost_estimates}.
The cost estimate comparisons are normalized with the ILC total cost to be \SI{100}{\percent}.
The breakdown foresees \SI{67}{\percent} of the total cost for accelerator system (with \SI{44}{\percent} for Main Linac (ML) \& SRF and \SI{23}{\percent} for other sub-systems) and \SI{33}{\percent} for the sum of civil engineering (site, tunnel, and building) and technical infrastructure (electrical, cooling, ventilation, etc.).
Since details of the ERLC design and cost estimates are not known, we are simply adding cost estimates for unknown sub-systems (such as sources and low-energy beamlines), referring to the corresponding ILC cost estimates.
It is still difficult to accurately evaluate the large cryogenic capacity and the cost related to significant HOM losses, which are inevitable in CW-mode operation, with a one-third duty factor, and which must be efficiently extracted to higher temperature ($>\SI{100}{\kelvin}$).

% Massive fudge, may come up with a better solution later
%\newcommand{\subsubsubsection}[1]{\vskip1.5ex\noindent\textbf{#1}\\}
\subsubsection{Main linac and SRF cost}

We assume the ERLC energy recovery linac will use a twin-SRF-cavity string with a gradient of \SI{20}{\mega\volt\per\meter}, about 2/3 of the \SI{31.5}{\mega\volt\per\meter} gradient for the ILC.
It requires a sophisticated twin-aperture cavity configuration requiring a larger cross section of the cryomodule (with more cost for the larger cross section), and \SI{57}{\percent} more cryomodules, resulting in a more than \SI{57}{\percent} longer linac.
In addition, HOM absorbers need to be installed at the end of each cavity to extract the significant HOM loss directly to a much higher temperature ($\geq\SI{100}{\kelvin}$). This requires an increase in cryomodule length, and therefore linac length, by at least \SI{10}{\percent}.
The HOM loss in CW operation, even in the case of a duty factor of 1/3, would have a big impact on the cryogenic capacity, and therefore the construction cost. From past experience, we assume that the cryogenic cost would follow a power scaling law of $(C1/C2)^{0.65}$.
%The relative cost comparison is summarized in Table~\ref{tab:appendix:electronpositron:erlc:cost_estimates}.
In summary, we estimate the ERLC relative cost for the main linac and SRF to be more than twice as high as that of the ILC (the details are shown for Case A in Table~\ref{tab:appendix:electronpositron:erlc:cost_estimates}).

\subsubsection{Other accelerator sub-systems}
The electron and positron source would cost less for the ERLC because of the partial CW operation with a duty factor of 1/3.
The ERLC low-energy beam transport includes beam injection into the main linac for acceleration and recirculation after deceleration, combined with beam compression and decompression, and wigglers for beam damping.
We consider that the low-energy transport cost would be similar to the ILC low-energy beam lines consisting of the Damping Ring and the RTML beam line.
The ERLC beam delivery system (BDS) needs two beam lines in each side, which would cost about \SI{50}{\percent} more.
The cost of the interaction region (IR) should be similar in both cases.
The main beam dump would cost much less in the ERLC because of the lower beam energy and the operational duty factor.
In summary, the overall cost for the other accelerator systems would be similar between the ILC and the ERLC, except the BDS, which would be more expensive.

\subsubsection{Civil engineering and technical infrastructure}
We assume that the cost of the ERLC tunnel would be roughly proportional to the main linac cavity and cryomodule cost, except for the cross-section effect, which would be slightly lower. This causes some over-estimate because no damping ring tunnel is required for the ERLC, but this would be a small correction. 

\begin{table}[htb]\centering
\caption{ILC-250 and ERLC (Case A) cost estimate comparisons, assuming the HOMs directly extracted to $\geq\SI{100}{\kelvin}$, at each ILC-type 9-cell, but twin-cavity end, with 9 cavities/cryomodule, duty factor = 1/3.}
\label{tab:appendix:electronpositron:erlc:cost_estimates}
\begin{tabular}{p{.35\linewidth}rrr}
\toprule
Item & Relative ILC cost & Factor & relative ERLC cost \\
& & & (normalized to ILC) \\
\midrule
Civil engineering & 19\,\% & $\times 2$ & 38\,\% \\
Technical infrastructure & 14\,\% & $\times 2$ & 28\,\% \\
\midrule
Accelerator ML \& SRF: & & & \\
~~~~Cavity \& cryomodule & 27\,\% & $\times 2.6$ & 70\,\% \\
~~~~RF & 7\,\% & $\times 1$ & 7\,\% \\
~~~~Cryogenics & 8\,\% & $\times 6$ & 54\,\% \\
Other accelerator sub-systems: & & & \\
~~~~Sources, DR, and LE beams & 21\,\% & $\times 1$ & 21\,\% \\
~~~~BDS & 4\,\% & $\times 1.5$ & 6\,\% \\
\midrule
Sum & 100\,\% & & 224\,\% \\
\bottomrule
\end{tabular}
\\[2ex]
Notes for relative cost factor breakdown:
\begin{enumerate}
    \item CE \& CF: $2 = 1.2 \times (1.57 \times 1.1)$ for `cavity \& cryomodule cross section' $\times$ `Length (gradient and frequent HOM absorbers)'.
    \item ML \& SRF:
    \begin{itemize}
        \item Cavity \& cryomodule: $2.6 = 1.5 \times (1.57 \times 1.1)$ for `cavity \& cryomodule cross section' $\times$ `Length (gradient and frequent HOM absorbers)'. 
        \item RF: similar
        \item Cryogenics: $6 = (273/15.7)^{0.65}$, the scaling law \cite{cryogenic_scaling_law} referring to the ratio of AC power between ERLC and ILC 
    \end{itemize}
    \item Other accelerator systems:
    \begin{itemize}
        \item Sources, DR, and LE beams: $\times 1$, similar overall, although BDS more expensive and \electron/\positron{} sources less expensive in ERLC
        \item BDS: $\times 1.5$, twice the beam lines for ERLC
    \end{itemize}
\end{enumerate}
\end{table}

\subsection{Staging and Upgradability}
The ERLC concept has the potential to exceed the performance projections of the ILC by over an order of magnitude, but still requires fundamental R\&D efforts for the design of fully coupled SRF systems.
One appealing scenario could therefore be to start the physics program with the baseline ILC configuration and to look at the ERLC as a future upgrade option of the collider.
Noting from the above cost estimate table that the Main Linac and SRF system amount to approximately \SI{45}{\percent} of the total ILC budget, one can conclude further that such an upgrade of the ILC implies an additional investment of at least half of the total ILC budget.
While this clearly represents a significant cost item, it might still be an interesting option for the long-term exploitation of the ILC if one considers the potential increase of the collider performance by over one order of magnitude and the extension of the ILC exploitation period by perhaps another decade.
This approach assumes that the ERLC cryostats are compatible with the main tunnel dimensions and that the Interaction Region design of the ERLC is designed to be compatible with the ILC Interaction Region.

\subsection{Earliest possible time for implementation}

The results of the cost estimate (Section~\ref{sec:appendix:electronpositron:erlc:cost}) and power consumption estimate (Section~\ref{sec:appendix:electronpositron:erlc:power}) performed by the sub-Panel imply that, to realise the benefits of ERLC in comparison to ILC, the superconducting cavities should differ from those of the ILC in three regards:

Firstly, a twin-aperture configuration within a common cryomodule is preferable.
Twin aperture would allow counter-propagating beams without the serious disruption that would be caused by parasitic crossovers within the main linac.
A common cryomodule will maximise the ER efficiency by enabling RF power transfer between the apertures with minimal loss.

Secondly, the geometry, number and location of HOM absorbers should be optimised, and the HOMs should be extracted to as high a temperature as practicable.
This will be a major determining factor in the overall power consumption of the facility.

Thirdly, the fundamental frequency of the cavities should be lower than \SI{1300}{\mega\hertz} to mitigate HOM excitation.

Therefore, the cavities and cryomodules should be very different to those developed by the Tesla Technology Collaboration (TTC) and those derived from them.
This will form the critical path in any schedule to realise an ERLC.
The TTC recently celebrated its 30-year anniversary, which gives an indication of the time to perfect an SRF implementation.
Of course, the experience gained over that timeframe will help future development of a newly optimised ERLC cavity / cryomodule / cryosystem.
On balance, the panel estimates 15 years to develop from pre-concept to a fully industrialised production of proven designs.

In summary, our expectation is that the earliest time for implementation of an ERLC is 2037.

\subsection{Power consumption}
\label{sec:appendix:electronpositron:erlc:power}

The ILC and ERLC estimated AC (wall-plug) power is compared in Table~\ref{tab:appendix:electronpositron:erlc:power}.
The calculations assume the same basic cavity and cryomodule design as the ILC; nine \SI{1.3}{\giga\hertz}, 9-cell elliptical cavities in each cryomodule, but with a HOM absorber at the end of each cavity. Also, there are two parallel strings of such cavities with niobium cross-connects to allow energy recovery from the decelerated beam.

The AC power estimate for the ERLC requires significant AC (wall-plug) power consumption, mainly for the cryogenics load due to the cavity RF dynamic loss and the HOM load caused by the beam current.
The AC power for HOM heating would be extremely high if an ILC-like cryomodule were used with HOM absorbers only between cryomodules. To significantly reduce the AC power, the ERLC cryomodule needs to be designed with:
\begin{itemize}
    \item HOM power directly extracted to a higher temperature (\SI{100}{\kelvin}),
    \item HOM absorber/damper at each cavity end (instead of each cryomodule end).
\end{itemize}
For the Cornell ERL Main Linac design, as discussed earlier, the HOM load was optimized to extract the HOMs to \SI{100}{\kelvin} at each cavity end to minimize the cryogenics load at \SI{1.8}{\kelvin} and the overall AC wall plug power.
This requires increasing the length of the cryomodules and the linacs, leading to additional construction costs.

\begin{table}[htb]\centering\footnotesize
\caption{Wall-plug power comparisons of ILC with ERLC, including HOM load extraction at \SI{100}{\kelvin}, and DF 1/3 (Case A) and DF 1 (Case B), the latter two having been estimated by the sub-Panel.}
\label{tab:appendix:electronpositron:erlc:power}
\begin{tabular}{p{.35\linewidth}crrrr}
\toprule
 & Unit & ILC & \multicolumn{3}{c}{ERLC} \\
\midrule
& & ILC-250 & Proposal & Case A & Case B \\
& & estimate & estimates & & \\
\midrule
Operating parameters: \\
Electric field gradient & \si{\mega\volt\per\meter} & 31.5 & 20 & 20 & 20 \\
$Q_0$ at operating temperature & \num{e10} & 1 & 3 & 3 & 3 \\
Beam current & \si{\milli\ampere} & $\left<0.021\right>$ & $\left<53.3\right>$ & $\left<53.3\right>$ & 53.3 \\
Duty factor (for beam) & & 0.0037 & $1/3$ & $1/3$ & 1 \\
Number $N$ of $\text{e}^+$ / $\text{e}^-$ per bunch & \num{e9} & 20 & 5 & 5 & 5 \\
Distance between bunches & \si{\meter} & 166 & 1.5 & 1.5 & 4.5 \\
HOM absorber temperature & \si{\kelvin} & 2, 5, 60 & (300) & 100 & 100 \\
\midrule
Linac AC power: $^{(1)}$ \\
RF systems $^{(2)}$:\\
~~~~RF to keep cavity gradient & \si{\mega\watt} & 24 & --- & 26 & 79 \\
~~~~HOM energy-loss compensation & \si{\mega\watt} & --- & 30 & 23 & 23 \\
Cryogenic loads: & \si{\mega\watt} & & 92 & & \\
~~~~Cavity dynamic $^{(3)}$ & \si{\mega\watt} & 5.1 & --- & 96 & 289 \\
~~~~HOM dynamic $^{(4)}$ & \si{\mega\watt} & 0.7 & --- & 134 & 134 \\
~~~~Power coupler dynamic \& static & \si{\mega\watt} & 1.6 & --- & $\sim 7$ & $\sim 7$ \\
~~~~HOM static & \si{\mega\watt} & (small) & --- & 33 & 33 \\
~~~~Other static loads $^{(5)}$ & \si{\mega\watt} & 8.3 & --- & 30 & 30 \\
%\midrule
Utilities $^{(6)}$ & \si{\mega\watt} & 10.5 & n/a & 86 & 147 \\
\midrule
Linac AC power totals & \si{\mega\watt} & 50 & 122 & 428 & 734 \\
(Total collider AC power) $^{(7)}$ & & (111) & ($130 + \text{n/a}$) & & \\
\midrule
Luminosity & $10^{34}\,\si{\per\square\centi\meter\per\second}$ & 1.35 & 48 & 48 & 48 \\
\bottomrule
\end{tabular}
\\[2ex]
Notes:
\begin{enumerate}
\item The ILC SRF design (9-cell cavity and 9-cavities/cryomodule) is commonly applied for ILC and ERLC in this comparison. The results are consistent with the evaluation using the Cornell-ERL-MLC design as discussed in the RF section.
\item Power required for beam acceleration (ILC) or compensation of energy loss during circulation (ERLC).
\item loaded by RF loss: $P_\text{RF-loss}= \frac{V^2}{(R/Q) Q_0}$,
\item The HOM power lost by the beam is $P_\text{HOM} = k_\text{L} e N L I_\text{av}$, where $eN$ is the bunch charge, $L$ the active length of the linacs, and $I_\text{av}$ the average current. The loss parameter $k_\text{L}$ is about \SI{11}{\volt\per\pico\coulomb\per\meter} for TESLA cavities assuming a bunch length of \SI{0.3}{\milli\meter} (fundamental mode excluded).
\item assuming HOM static = $1/4$ of HOM dynamic, and others static loads are constant.
\item assuming $1/4$ of AC power is used for cooling towers, etc.~to remove the heat to air.
\item including: (ILC) sources, DR, RTML, BDS, IR, and campus, or (ERLC) LE beamline systems.
\item COP: 900 at \SI{1.8}{K}, 790 at \SI{2}{K}, 208 at \SI{4.5}{K}, 21 at \SI{80}{K}, 11.9 at \SI{100}{K}.
\end{enumerate}
\end{table}

In Table~\ref{tab:appendix:electronpositron:erlc:power}, the AC power estimates for the ILC-250 design and the original ERLC proposal are compared with two estimates: Case A (DF 1/3 and $d=\SI{1.5}{\meter}$) and Case B (DF 1 and $d=\SI{4.5}{\meter}$). Case B corresponds to the more realizable \SI{100}{\percent}-duty-factor approach discussed in Section \ref{sec:appendix:electronpositron:erlc:rf}.

It should be noted that the basic cavity and cryomodule design is kept with the ILC-style \SI{1.3}{\giga\hertz}, 9-cell cavity, and 9 cavities per cryomodule, for both ILC and ERLC in these power estimates.
The results are consistent within \SI{10}{\percent} to the values discussed for the Cornell ERL-MLC-based design in Section~\ref{sec:appendix:electronpositron:erlc:rf}.

The thermal load due to RF losses in the cavities plus the beam-generated HOM power is significantly larger than that in the ILC and the original ERLC proposal. However, the luminosity normalized to AC power would be much higher in the ERLC than in the ILC. 

In either case, a long-term R\&D program will be required to establish the technology and to develop a project with practical AC plug power consumption, twin-aperture cavities with much higher $Q_0$ and higher-temperature operation, more efficient HOM load extraction, thermal and cryogenic design optimization, etc.

\subsection{Updated parameters}
The author developed an update to the published parameters with a reduced distance between bunches (\SI{23}{\centi\meter} instead of \SI{1.5}{\meter}) with an equivalent reduction in the number of particles per bunch (\cite{telnov2021highluminosity_v3}), which reduces the HOMs by the same factor.
The luminosity is kept the same by adopting a smaller horizontal beam size at the IP (keeping the same vertical beam-beam tune shift).
The new parameter set considers full CW operation, and the author estimates that the electrical power for the beams is \SI{250}{\mega\watt}.
This assumes that the cryogenic efficiency is equal to 0.3 times the ideal Carnot efficiency ($1/550$).
We estimate this efficiency to be $1/900$ (the value obtained at LCLS-II), to which \SI{25}{\percent} should be added for shield cooling and to dissipate the cryoplant heat loads in cooling towers.
Adding the site power requirements gives a total of over \SI{600}{\mega\watt}, which the sub-Panel considers unacceptable.
We also believe that the closer bunch spacing in the ERLC would require a crossing angle at the interaction region, with the added complexity of including a bend that returns the bunches to the decelerating linac after collision.

\subsection{Comments and suggestions for improvements}
Without a significant reduction in cryogenic power, this proposal is interesting but not compelling. If \SI{4.5}{\kelvin} operation with high $Q_0$ can be developed, this would be an extremely interesting proposal, with high luminosity being the primary driver rather than high energy.
This could be a stand-alone machine (preferable) or a future upgrade to the ILC.

\subsection{R\&D Required}
This is the first proposal that integrates a linac and a damping ring, which brings a new set of theoretical problems.
We recommend that this be studied in detail as this may well not be the only project that will want to use this layout. Obviously, the development of \SI{4.5}{\kelvin} operation with a high $Q_0$ would make this a much more attractive project.

\subsection{Recommendations}
The sub-Panel approves the idea of designing a linear collider based on an ERL to reduce the energy footprint of the facility, and the ERLC is an excellent first attempt. The present proposal was developed by a single author and is therefore incomplete in many details.
Therefore, the sub-Panel chose to look for ways that the design could be improved as part of a more detailed study. 
\begin{enumerate}
    \item We recommend a study of the new beam-dynamics problems inherent in the integration of a linac and a damping ring. 
    \item We recommend R\&D on high-$Q_0$ cavities operating at \SI{4.5}{\kelvin}, which would reduce both the cost and the power consumption. 
    \item If the ERLC is envisioned as an ILC upgrade, then the ERLC cryostats should be compatible with the main tunnel dimensions, and the Interaction Region design must be compatible with the ILC geometry.
\end{enumerate}

\section{Overall Conclusions}
The sub-Panel was presented with two extremely interesting ideas to evaluate.
While neither is ready to be adopted now, they point to the future in different ways. The CERC aims for multiple passes in a tunnel with an extremely large bending radius to minimize the synchrotron radiation loss.
The ERLC proposes a single acceleration and deceleration, separating the two beams by using twin-axis cavities.
Both of these ideas provide an indication of the variety of different ERL layouts that might be developed in the future.

A particularly interesting prospect is to design an energy-efficient, ultra-high-luminosity ERL-based electron-positron collider at \SI{500}{\giga\electronvolt}, which would enable the exploration of the Higgs vacuum potential with a measurement of the tri-linear Higgs coupling in \positron\electron.

The most important R\&D activity that would make this kind of development viable is to be able to operate at \SI{4.5}{\kelvin} with high $Q_0$.
We strongly recommend R\&D on this topic as it would also allow universities to adopt small superconducting accelerators for inverse Compton back-scattering, FELs, isotope production, etc.
Apart from the societal aspect, this would provide a steady product line for SRF cavity and cryomodule production by industry, which would in turn benefit future HEP colliders.

\acknowledgments

This paper provides a status report on ERLs---past, present, and future---around the world and could not have been written without the enthusiastic involvement of the entire ERL community.
In addition,
the authors most gratefully acknowledge information, insight, and guidance they received in discussions
with Roy Aleksan, Jean-Luc Biarotte, Phil Burrows, Dimitri Delikaris, Grigory Emereev, Eric Fauve,
Rao Ganni, Frank Gerigk, Karl Jakobs, Jan Lüning, Eugenio
Nappi, Sam Posen, Guillaume Rosaz, Hiroshi R Sakai, Herwig Schopper, Mike
Seidel, Alexander Starostenko, and many other colleagues.

Work partially supported by the U.S.~Department of Energy, Office of Science, Office of Nuclear Physics under Contract No.~DE-AC05-06OR23177.

%For Jan Bernauer:
Work partially supported by US National Science Foundation, grant No.\ 2012114.

CBETA construction was supported by NSF award DMR-0807731, DOE grant DE-SC0012704, and NYSERDA agreement number 102192.

Work of V.~I.~Telnov was supported by the Russian Foundation for Basic Research grant RFBR 20-52-12056.

Work partially supported by the German Research Council (DFG) within
the research training group, GRK 2128 \enquote{AccelencE} (Project ID 264883531),
by the German Ministry for Education and Research (BMBF) under grant
No.~05H21RDRB1, and by the State of Hesse under the grant \enquote{Nuclear Photonics}
within the LOEWE program and within the Hessian Research Cluster ELEMENTS
(Project ID 500/10.006).

%\printbibliography
\bibliographystyle{JHEP}
\bibliography{biberl}

%%%%%%%%%%%%%%%%%%%%%%%%%%%%%%%%%%%%%%%%%
\end{document}